\documentclass[a4paper,oneside,12pt]{book}

\newcommand{\thesistitle}{Scattering Amplitudes in the Yang-Mills sector of Quantum Chromodynamics} 
\newcommand{\degree}{Doctor of Philosophy} 
\newcommand{\authorname}{Author: Hiren Kakkad} 
\newcommand{\supervisor}{dr hab. Piotr Kotko} 
\newcommand{\keywords}{this, that, more} 
\newcommand{\school}{\href{https://www.agh.edu.pl/en/}{AGH University of Kraków,}} 

\newcommand{\department}{\href{http://www.ftj.agh.edu.pl/}{Faculty of Physics and Applied Computer Science}} 

\usepackage[T1]{fontenc} 
\usepackage[utf8]{inputenc}
\usepackage[polish,english]{babel}
\usepackage{bigints}
\usepackage{esint}
\usepackage{enumitem}
\usepackage{amssymb}
\usepackage{enumitem}
\usepackage{multirow}
\usepackage[]{cite}
\usepackage[a4paper,top=2.54cm,bottom=2.54cm,left=2.54cm,right=2.54cm,headheight=16pt]{geometry}
\usepackage{amsmath}
\usepackage{hhline}
\usepackage{bbold}
\usepackage[autostyle=true]{csquotes} 
\usepackage{graphicx}
\usepackage[colorinlistoftodos]{todonotes}
\usepackage[colorlinks=true, allcolors=black]{hyperref}
\usepackage{hyperxmp}
\usepackage{caption} 
\usepackage{sfmath} 
\usepackage[parfill]{parskip}    
\usepackage{setspace} 
\usepackage{enumerate} 
\usepackage{booktabs} 
\usepackage{fancyhdr}
\usepackage{xcolor} 
\numberwithin{equation}{chapter} 
\definecolor{tcd_blue}{RGB}{5, 105, 185}

\usepackage{titlesec}
\titleformat{\chapter}[hang]{\normalfont\huge\bfseries\color{tcd_blue}}{\thechapter}{1cm}{}{}

\title{\thesistitle}
\author{\authorname}

\hypersetup{
   pdftitle=\thesistitle, 
   pdfauthor=\authorname, 
   pdfkeywords=\keywords, 
   pdfsubject=\degree, 
   pdfinfo={
     pdfsupervisor=\supervisor, 
   }
}
\newcommand\Tr{\mathrm{Tr}}

\usepackage{ragged2e}

\frontmatter
\begin{document}
\begin{titlepage}

\center 



\includegraphics[width=8cm]{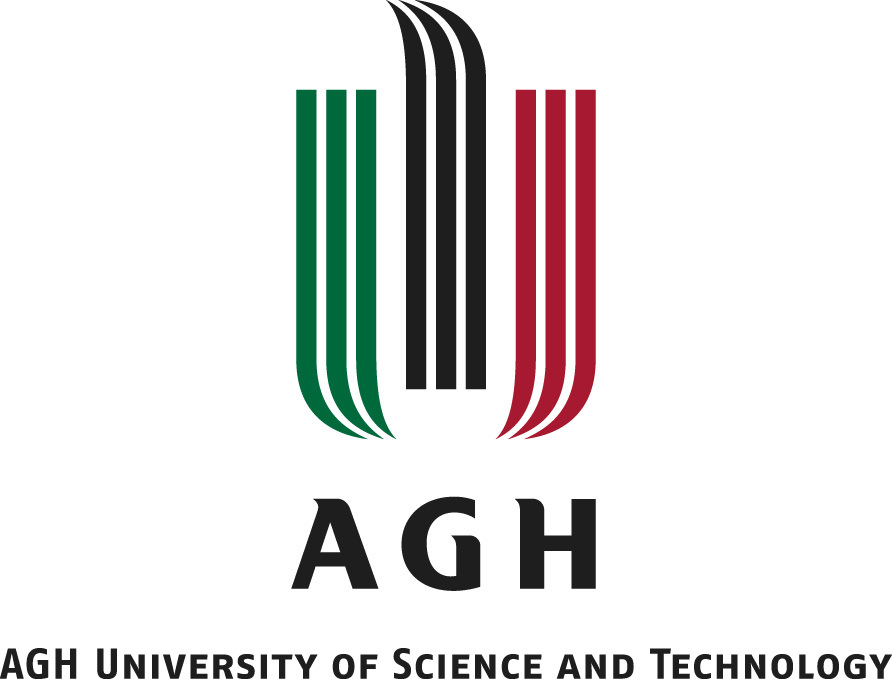}\\
\vspace{1.5cm}
\Large\textbf{FIELD OF SCIENCE: NATURAL SCIENCES}\\
\large{SCIENTIFIC DISCIPLINE: PHYSICAL SCIENCES}\\
\vspace{1.5cm}
\Huge\textbf{DOCTORAL DISSERTATION}\\
\vspace{1.5cm}

\Large\textbf{\textit{Scattering Amplitudes
in the Yang-Mills sector of
Quantum Chromodynamics}}
\vspace{1cm}


\ifdefined\authorid
\authorname\\ 
\authorid\\ 
\else
\authorname\\ 
\fi


Supervisor: \supervisor\\[2cm] 
\ifdefined\cosupervisor
Cosupervisor: \cosupervisor\\[2cm] 
\fi


\Large \school\\ 
\ifdefined\department
\Large \department\\[1.5cm] 
\Large{Kraków, 2023} 


\end{titlepage}

\pagenumbering{roman}
\justifying








\chapter*{Abstract}

 Scattering amplitudes are one of the most crucial objects in Quantum Field Theories because they are the building blocks for computing cross sections measured at the particle colliders. Traditionally, amplitudes were computed using the Feynman diagram technique. However, in field theories that involve self-interactions, like the Yang-Mills sector of Quantum Chromodynamics (QCD) which describes pure gluonic interactions, this technique becomes impractical because the number of diagrams grows factorially with the number of external gluons. 

We derive a new classical action for the pure gluonic sector of QCD that implements new interaction vertices (local in the light-cone time) with at least four legs and fixed helicities, which makes it efficient in calculating tree-level pure gluonic scattering amplitudes. We demonstrated this by computing several tree-level amplitudes up to 8-point and the maximum number of planar diagrams we got was 13 (in the Feynman diagram technique, an 8-point tree amplitude requires tens of thousands of diagrams). The new action was obtained by performing a canonical field transformation on the light-cone Yang-Mills action such that it eliminates both the triple point interaction vertices from the latter. The transformation replaces the fundamental gluon fields of the Yang-Mills theory with the Wilson line degrees of freedom -- geometric objects facilitating parallel transport of vectors in curved spaces. The new action extends the so-called MHV action, i.e. the action implementing the Cachazo-Svrcek-Witten method that utilizes the Maximally Helicity Violating (MHV) scattering amplitudes as interaction vertices.

At the loop level, both the MHV action and the new classical action turn out to be incomplete because the eliminated triple gluon vertices with helicity $(+ + -)$ and $(+ - -)$ contribute to loops. To systematically develop loop corrections to the MHV action first, we used the one-loop effective action approach where we start with constructing it for the Yang-Mills action and then perform the field transformation to obtain the classical MHV action plus loop contributions. We verified that there are no missing loop contributions by computing  four-point one-loop amplitudes where all the gluons have plus helicity $(+ + + +)$ and where one of them has a minus helicity $(+ + + -)$. These could not be computed in the MHV theory.  A major advantage of this approach is that the transformation accounts for all the tree level connections one could make by connecting the $(+ + -)$ triple gluon vertex with the external legs of the loop contributions. As a result, the number of diagrams required to compute one-loop amplitudes is way less when compared with the one-loop effective Yang-Mills action.

Next, we extend this approach to develop loop corrections for our new Wilson line-based action. We start with the one-loop effective Yang-Mills action and then perform the transformation to obtain the new classical action plus loop corrections. In this case, the number of diagrams required to compute a higher multiplicity one-loop amplitude is lesser than in the one-loop effective MHV action. This is because the transformation accounts for all the tree level connections one could make by connecting both the $(+ + -)$ and $(+ - -)$ triple gluon vertices with the external legs of the loop contributions. To validate the one-loop action we computed one-loop amplitudes with helicities $(+ + + +)$, $(+ + + -)$, $(+ + - -)$, $(- - - -)$ and $(- - - +)$.

Although one-loop complete (no missing loop contribution), a major drawback of this approach is that it uses Yang-Mills vertices inside the loop and the new efficient vertices of our action outside the loops only. 
We, therefore, re-derive the one-loop actions via a different approach, where we first perform the canonical transformation to the Yang-Mills action transforming also the current dependent terms, and then integrate the field fluctuations to derive the one-loop effective actions. This way the new interaction vertices of our action are explicit in the loop. We test this approach first for the MHV action and demonstrate that the one-loop effective MHV action, derived this way, is both one-loop complete and has MHV vertices explicit in the loop. Since there are "bigger" interaction vertices when compared with the Yang-Mills vertices in the loop, the efficiency of computing higher multiplicity one-loop amplitude further increases in this approach. We finally extend the new approach to our action and derive the one-loop action such that the interaction vertices of our action are manifest in the loop. Since our interaction vertices are even bigger when compared to the MHV vertices, computing one-loop amplitude requires even fewer diagrams.

 The research done in this thesis provides a new field-theory action-based method to efficiently calculate pure gluonic scattering amplitudes up to one loop.

\selectlanguage{polish} 
\newpage
\chapter*{Abstrakt}


Amplitudy rozpraszania są jednymi z najważniejszych obiektów w kwantowej teorii pola, ponieważ stanowią podstawę obliczeń przekrojów czynnych mierzonych w akceleratorach cząstek. Tradycyjnie, amplitudy rozpraszania były obliczane za pomocą diagramów Feynmana. Jednakże, w teoriach pola zawierających samooddziaływania, takich jak teoria Yanga-Millsa, będąca częścią Chromodynamiki Kwantowej (QCD), opisująca oddziaływania gluonów, technika ta staje się niepraktyczna, ponieważ liczba diagramów rośnie jak silnia liczby zewnętrznych gluonów.

W niniejszej pracy, wyprowadzamy nowe klasyczne działanie dla czysto gluonowego sektora QCD, które wprowadza nowe wierzchołki oddziaływań (lokalne w czasie na stożku świetlnym), posiadające co najmniej cztery nogi i ustalone skrętności. Dzięki temu metoda ta umożliwia efektywne obliczanie gluonowych amplitud rozpraszania na poziomie drzewiastym. Demonstrujemy to obliczając kilka drzewiastych amplitud, aż do 8 zewnętrznych nóg, gdzie maksymalna liczba diagramów planarnych wyniosła 13 (w technice diagramów Feynmana, drzewiasta amplituda 8-punktowa wymagałaby dziesiątek tysięcy diagramów). Nowe działanie uzyskano wykonując kanoniczną transformację pól działania Yanga-Millsa na stożku świetlnym, eliminując trójpunktowe wierzchołki oddziaływania. Transformacja ta zastępuje podstawowe pola gluonowe teorii Yang-Millsa liniami Wilsona - geometrycznymi obiektami umożliwiającymi transport równoległy wektorów w zakrzywionych przestrzeniach. Nowe działanie rozszerza tzw. działanie MHV, czyli działanie realizujące metodę Cachazo-Svrcek-Witten, która wykorzystuje amplitudy rozpraszania Maximally Helicity Violating (MHV) jako wierzchołki oddziaływań.

Na poziomie pętlowym, zarówno działanie MHV, jak i nowa działanie okazują się niekompletne, ponieważ wyeliminowane potrójne wierzchołki gluonowe o skrętnościach $(+ + -)$ i $(+ - -)$ dają przyczynki do diagramów pętlowych. Aby systematycznie wprowadzić poprawki pętlowe w pierwszej kolejności do działania MHV, użyliśmy jednopętlowego działania efektywnego, w którym rozpoczynamy od jego konstrukcji dla teorii Yang-Millsa, a następnie wykonujemy transformację pól, aby uzyskać działanie MHV wraz z wkładami pętlowymi. Zweryfikowaliśmy, że nie ma brakujących przyczynków pętlowych, poprzez obliczenie 4-punktowych amplitud na poziomie jednej pętli dla gluonów o dodatniej skrętności $(+ + + +)$ oraz gdzie jeden z nich ma skrętność ujemną $(+ + + -)$. Takie obliczenia nie były możliwe w teorii MHV. Główną zaletą tego podejścia jest to, że transformacja uwzględnia wszystkie drzewiaste połączenia, otrzymane łącząc wierzchołek trój-gluonowy $(+ + -)$ z zewnętrznymi nogami wkładów pętlowych. W rezultacie, liczba diagramów wymaganych do obliczenia amplitud jednopętlowych jest znacznie mniejsza w porównaniu działaniem Yanga-Millsa.

Następnie rozszerzamy to podejście, w celu opracowania poprawek pętlowych w nowym działaniu opartym na liniach Wilsona. Rozpoczynamy od jednopętlowego działania efektywnego Yanga-Millsa, a następnie wykonujemy transformację pól, aby uzyskać nowe działanie wraz z poprawkami pętlowymi. W tym przypadku, liczba diagramów wymaganych do obliczenia jednopętlowej amplitudy o wielu zewnętrznych nogach okazuje się mniejsza niż w jednopętlowym działaniu efektywnym MHV. Wynika to z faktu, że transformacja uwzględnia wszystkie drzewiaste połączenia, tworzone łącząc zarówno trójgluonowy wierzchołek $(+ + -)$ jak i $(+ - -)$, z zewnętrznymi nogami wkładów pętlowych. Aby zweryfikować nowe jednopętlowe działanie efektywne, obliczyliśmy amplitudy jednopętlowe dla skrętności gluonów $(+ + + +)$, $(+ + + -)$, $(+ + - -)$, $(- - - -)$ i $(- - - +)$.

Mimo że nowe działanie jest kompletne na poziomie pętli, główną jego wadą jest to, że wciąż używa wierzchołków Yang-Millsa wewnątrz pętli, zaś nowe efektywne wierzchołki są obecne jedynie na zewnątrz pętli. W związku z tym, wyprowadzamy jednopętlowe działanie efektywne za pomocą innego podejścia, w którym najpierw wykonujemy kanoniczną transformację pól działania Yang-Millsa, przekształcając również człony zależne od zewnętrznych prądów, a następnie całkujemy fluktuacje kwantowe aby uzyskać działanie efektywne. W ten sposób nowe wierzchołki oddziaływań naszej akcji są jawnie obecne w pętli. Testujemy to nowe podejście najpierw dla działania MHV i pokazujemy, że otrzymane w ten sposób jednopętlowe działanie efektywne MHV,  jest zarówno jednopętlowo-kompletne, jak i ma wierzchołki MHV w pętli. Ponieważ w pętli występują "większe" wierzchołki oddziaływań w porównaniu z wierzchołkami Yanga-Millsa, wzrasta efektywność obliczania amplitud jednopętlowych o wyższej krotności. Wreszcie,  rozszerzamy to podejście na nasze nowe działanie oparte na liniach Wilsona i wyprowadzamy jednopętlowe działanie efektywne, tak, aby wierzchołki oddziaływań nowego działania były jawne w pętli. Ponieważ nowe wierzchołki oddziaływań są „jeszcze większe” w porównaniu do wierzchołków MHV, obliczanie jednopętlowych amplitud wymaga jeszcze mniej diagramów.

Badania będące przedmiotem tej rozprawy dostarczają nowej, opartej na teorio-polowym działaniu, metody umożliwiającej efektywne obliczanie czysto gluonowych amplitud rozpraszania na poziomie jednej pętli.

\selectlanguage{english}
\chapter*{Publications}

The research work presented in this thesis resulted in the publications of the following  3 research articles
\begin{itemize}
\item H. Kakkad, P. Kotko and A. Stasto, \textit{One-Loop effective action approach to quantum MHV theory,
JHEP} 11 (2022) 132.
\item H. Kakkad, P. Kotko and A. Stasto, \textit{A new Wilson line-based action for gluodynamics,
JHEP} 07 (2021) 187. 
\item H. Kakkad, P. Kotko and A. Stasto, \textit{Exploring straight infinite Wilson lines in the
self-dual and the MHV Lagrangians, Phys. Rev. D} 102 (nov, 2020) 094026.
\end{itemize}
All the results presented in Chapter 5 and some results in Chapter 3 are completely new and have not been published yet. The manuscript for those results is under preparation.

In addition, I also contributed to the following two research projects exploring phenomenological aspects of scattering amplitudes in Quantum Chromodynamics researched in this project. These results are not included in this thesis. 

\begin{itemize}
\item H. Kakkad, A. K. Kohara and P. Kotko, \textit{Evolution equation for elastic scattering of
hadrons, The European Physical Journal C} 82 (sep, 2022).
\item M. A. Al-Mashad, A. van Hameren, H. Kakkad, P. Kotko, K. Kutak, P. van Mechelen
et al., \textit{Dijet azimuthal correlations in p-p and p-Pb collisions at forward LHC
calorimeters, JHEP} 12 (2022) 131.
\end{itemize}

\section*{Conference Proceedings}

\begin{itemize}
\item  H. Kakkad, A. K. Kohara and P. Kotko, \textit{Energy Evolution Equation for Elastic
Scattering Amplitudes of Hadrons in b-space, Acta Phys. Polon. Supp.} 16 (2023) 3.
    \item H. Kakkad, P. Kotko and A. Stasto, \textit{A new Wilson line-based classical action for gluodynamics, SciPost Phys. Proc.,} page 011, 2022.
\item H. Kakkad, P. Kotko and A. Stasto, \textit{Exploring straight infinite Wilson lines towards formulating a new classical theory for gluodynamics, PoS, EPS-HEP} 2021:720, 2022.
\end{itemize}


\newpage
\onehalfspacing\raggedright 

\section*{\Huge\textcolor{tcd_blue}{Acknowledgements}}
\justifying

I would like to express my sincere gratitude to all those who have contributed to the completion of my PhD thesis. Their support, guidance, and encouragement have been invaluable throughout this journey.

First and foremost, I am deeply grateful to my supervisor, dr hab. Piotr Kotko, for his unwavering guidance, expertise, and patience. His insightful advice, constant support, and constructive feedback have been instrumental in shaping the direction and quality of my research. I am truly fortunate to have had the opportunity to work under his supervision. I would like to extend my appreciation to Professor Anna Stasto. I am truly grateful for the countless hours of discussions, brainstorming sessions, and fruitful collaborations we have shared. 

I am indebted to Lance Dixon, Jacob Boujaily, and  Radu Roiban who generously gave their time for interesting discussions. I am also thankful to the staff and administrators at SLAC, Stanford University, and Pennsylvania State University for their assistance with administrative matters, access to resources, and logistical support during my research visit.

I would like to extend my appreciation to the members of my thesis committee, for their valuable insights, critical evaluation, and valuable suggestions. Their expertise and feedback have greatly enriched the quality of my research and thesis.

I am grateful to the faculty members and researchers at the Faculty of Physics and Applied Computer Science, AGH UST, for creating a stimulating academic environment. A special thanks to Dr. Anderson Kendi Kohara for the knowledge-sharing, intellectual discussions, and collaboration opportunities that have played a significant role in shaping my research skills and expanding my horizons. I would like to thank my colleagues and fellow graduate students for their camaraderie, support, and valuable discussions. Their friendship and shared experiences have made this journey memorable and enjoyable. 

Last but not least, I want to express my heartfelt appreciation to my family, friends, and especially my sister Dr Kruti Magdani, for their unwavering love, understanding, and encouragement. Their continuous support and belief in my abilities have been my driving force during challenging times.

My research is supported by the National Science Centre, Poland grant no. 2021/41/N/ ST2/ 02956.

To everyone mentioned above and to those who have supported me in ways beyond words, I extend my deepest gratitude. This accomplishment would not have been possible without your contributions. Thank you all for being a part of my academic and personal growth.

\tableofcontents






\newpage
\section*{\Huge\textcolor{tcd_blue}{Notation}}

Below we summarize the notations that we will be using in the text. There are some notations that we may adapt temporarily in the text. We define them in the text itself.

Let us begin with the contravariant four-vector $x^\mu$
\begin{equation}
    x^\mu = (x^0, x^1, x^2, x^3 )\,,
\end{equation}
in the four dimensional space-time with the metric: $g^{00}=1$, $g^{ii}=-1$ with $i = 1,2,3$ and $g^{ij}=0, \forall i\neq j$. 

In the majority of the text, we will use the so-called "double null" coordinates defined as
\begin{gather}
v^{+}=v\cdot\eta\,,\,\,\,\, v^{-}=v\cdot\widetilde{\eta}\,,\label{eq:plusmindef}\\
v^{\bullet}=v\cdot\varepsilon_{\bot}^{+}\,,\,\,\,\, v^{\star}=v\cdot\varepsilon_{\bot}^{-}\,,\label{eq:zzbardef}
\end{gather}
where ${\eta}$ and $\widetilde{\eta}$ are the two light-like  basis four-vectors defined as follows
\begin{gather}
\eta^\mu=\frac{1}{\sqrt{2}}\left(1,0,0,-1\right)\,,\,\,\,\,\widetilde{\eta}^\mu=\frac{1}{\sqrt{2}}\left(1,0,0,1\right)\, ,\label{eq:etavec}
\end{gather}
and $\varepsilon_{\bot}^{\pm}$ are the two space-like complex four-vectors spanning the transverse plane. These read
\begin{equation}
\varepsilon_{\perp}^{\pm}=\frac{1}{\sqrt{2}}\left(0,1,\pm i,0\right)\,.\label{eq:epsPlMin}
\end{equation}
Thus, the four-vector has the following explicit form
\begin{equation}
v=v^{+}\widetilde{\eta}+v^{-}\eta-v^{\star}\varepsilon_{\perp}^{+}-v^{\bullet}\varepsilon_{\perp}^{-}\,.
\end{equation}
Using the above definitions, one can lower the indices to obtain the covariant components. These read
\begin{gather}
v_{+}=v^{-},\,\,\, v_{-}=v^{+}\,,\\
v_{\bullet}=-v^{\star},\,\,\, v_{\star}=-v^{\bullet}\,.
\end{gather}
Consider two four-vectors $u$, $w$. Their dot product reads
\begin{equation}
u\cdot w=u^{+}w^{-}+u^{-}w^{+}-u^{\bullet}w^{\star}-u^{\star}w^{\bullet}\,.\label{eq:scalarprod}
\end{equation}

Consider a surface where $x^+$ is held fixed. On this surface, the position of a point is defined using the following three vector

\begin{equation}
\mathbf{x}\equiv\left(x^{-},x^{\bullet},x^{\star}\right)\, ,
\end{equation}
and the measure reads
\begin{equation}
d^{3}\mathbf{x}=dx^{-}dx^{\bullet}dx^{\star}\,.
\end{equation}

We introduce similar notations for the three vectors in the momentum space. These read
\begin{equation}
\mathbf{p}\equiv\left(p^{+},p^{\bullet},p^{\star}\right)\, ,
\end{equation}
and similarly, the measure reads
\begin{equation}
d^{3}\mathbf{p}=\frac{dp^{+}dp^{\bullet}dp^{\star}}{\left(2\pi\right)^{3}}\,.
\end{equation}
Throughout the text, we reserve the bold-faced alphabet for the position and momentum three vectors.  

In our notation, the Fourier transformation (on the surface of constant $x^+$) reads
\begin{equation}
A\left(x^{+},\mathbf{x}\right)=\int d^{3}\mathbf{p}\, e^{-i\mathbf{x}\cdot\mathbf{p}}\widetilde{A}\left(x^{+},\mathbf{p}\right)\,.\label{eq:FT_def}
\end{equation}
We use the tilde to represent the momentum space version for a given quantity
\begin{equation}
   {A}(x^+;\mathbf{x})\,, \quad\,\widetilde{A}(x^+;\mathbf{p})\,.
\end{equation}

Consider now a pair of four-momenta $p$ and $q$. For these we define
\begin{equation}
v_{qp}=\frac{q^{\star}}{q^{+}}-\frac{p^{\star}}{p^{+}},\;\;\;\; v_{qp}^{\star}=\frac{q^{\bullet}}{q^{+}}-\frac{p^{\bullet}}{p^{+}}\,.\label{eq:v_def}
\end{equation}
The above definition is anti-symmetric. We additionally define a non-antisymmetric version as follows
\begin{equation}
\widetilde{v}_{qp}=q^{+}v_{pq}=-q^{\star}+q^{+}\frac{p^{\star}}{p^{+}} \, ,\,\label{eq:vtild_def}
\end{equation}
and similarly $\widetilde{v}^{\star}_{qp}$ as well. Interestingly, these variables can be expressed in terms of the polarization vector $\varepsilon_{p}^{\pm}$ as follows
\begin{equation}
\widetilde{v}^{\star}_{qp}=q\cdot\varepsilon_{p}^{+},\,\,\,\,\widetilde{v}_{qp}=q\cdot\varepsilon_{p}^{-},\,
\end{equation}
where the polarization vector $\varepsilon_{p}^{\pm}$ itself reads
\begin{equation}
\varepsilon_{p}^{\pm}=\varepsilon_{\perp}^{\pm}-\frac{p\cdot\varepsilon_{\perp}^{\pm}}{p^{+}}\eta\,.\label{eq:PolarizationVect}
\end{equation}

For momentum sum, we use
\begin{equation}
p_{1\dots i}=\sum_{k=1}^{i}p_{k}\,.\label{eq:momentum_sum_def}
\end{equation}
Thus, in general, we will use
\begin{equation}
\widetilde{v}_{p_{i}p_{j}}\equiv\widetilde{v}_{ij},\,\,\,\quad\widetilde{v}_{\left(p_{1\dots i}\right)\left(p_{1\dots j}\right)}\equiv\widetilde{v}_{\left(1\dots i\right)\left(1\dots j\right)}\,.
\end{equation}
These symbols are related to the conventionally defined spinor products as follows
\begin{equation}
\left\langle ij\right\rangle =-\sqrt{\frac{2p_{j}^{+}}{p_{i}^{+}}}\,\widetilde{v}_{ij},\;\;\;\left[ij\right]=-\sqrt{\frac{2p_{j}^{+}}{p_{i}^{+}}}\,\widetilde{v}_{ij}^{*}\,,\label{eq:spinor_prod}
\end{equation}
where
\begin{equation}
 \langle \lambda_i \lambda_j \rangle  = 
\epsilon_{\alpha\beta}\, (\lambda_i)^\alpha \,(\lambda_j)^\beta \equiv \langle{i}{j}\rangle \, ,\quad \quad
 [\widetilde{\lambda}_i \widetilde{\lambda}_j] =-
\epsilon_{\dot\alpha\dot\beta}\, (\widetilde\lambda_i)^{\dot\alpha} \,(\widetilde\lambda_j)^{\dot\beta} \equiv \left[i j \right] \,
\, .
\end{equation}

Finally, for the Lie algebra valued quantities we use
\begin{equation}
\hat{A}^{\mu}=A_{a}^{\mu}t^{a}\,,
\end{equation}
 where $t^{a}$ are the generators in the fundamental representation satisfying
\begin{equation}
\left[t^{a},t^{b}\right]=i\sqrt{2}f^{abc}t^{c}\,,
\end{equation}
and
\begin{equation}
\mathrm{Tr}\left(t^{a}t^{b}\right)=\delta^{ab}\,.
\end{equation}
In the above normalization, we re-scale the coupling constant 
\begin{equation}
g \longrightarrow \frac{g}{\sqrt{2}}\,,
\end{equation}
to account for the additional factors of $\sqrt{2}$.


\mainmatter
\chapter{Introduction}
\label{Chap1}
\justifying

We live in the era of particle colliders like the Large Hadron Collider (LHC) and the planned Electron-Ion Collider (EIC) where particles are collided at ultra-relativistic speeds with a primary aim to probe the sub-atomic realm so as to determine the fundamental constituents of nature and the laws that govern them. The signatures of these fundamentals are encoded in the physical quantities measured at the detectors after the collisions. One such physical quantity, which is of prime importance, is the cross section. It determines the probability of a process to occur. At present, the theoretical predictions for the cross sections are made using  Quantum Field Theories (QFTs), like the Standard Model of particle physics, which describes all the elementary particle and their interactions. However, it is well known that the Standard Model is incomplete. As a result, it is crucial that the theoretical predictions made for things we already know are precise so that the difference in the predictions and measurements could point towards potential new physics.

Strong interactions dominate inelastic hadronic collisions at the LHC. Quantum Chromodynamics (QCD) provides the necessary framework for the theoretical description of strong interactions between quarks (matter particles) mediated by the gluons (gauge bosons). The latter originates from the principle of local gauge invariance of strong interactions under the $\mathrm{SU(3)}$ gauge group \cite{YangPGI}. QCD is thus a non-abelian gauge theory. Since gluons, just like quarks, have a color charge, they can undergo self-interactions. Due to this, QCD has a sub sector -- the Yang-Mills sector -- which describes only pure gluonic interactions.

A fundamental property of QCD is color confinement which, given the conditions at the particle colliders, prohibits color charges to exist freely \footnote{Under certain conditions, like those in the Quark–gluon plasma (QGP), these color charges deconfine and can therefore exist freely.}. Due to this, the end products of the collisions are again hadrons. Therefore, in order to compute cross sections in QCD, a bridge is required between the fundamental constituents of the hadrons --  partons (quarks and gluons) -- which undergo collision and the final state hadrons. This bridge and therefore the computation of the cross sections is  facilitated by the factorization theorems which expresses the cross section as a convolution of parton distribution function (PDFs) and the partonic cross section (see \cite{Collins:2011zzd} for a review on factorization theorems). The PDFs describe the composition of the hadrons in terms of the partons. It is a non-perturbative object and is determined either experimentally or from Lattice QCD or effective models of QCD (see reviews \cite{pPDF1,pPDF2}, and the references therein). The partonic cross section, on the other hand, is a perturbative object and is proportional to the square of scattering amplitudes of the partonic interactions.  The latter can be computed in QCD. This brings scattering amplitudes at the intersection between theory and experiment. In this thesis, we approach this conjunction along the theoretical front.

The partonic interactions stated above include interactions involving both the quarks and gluons. Of all these interactions, the pure gluonic ones are the most difficult to compute in perturbative QCD. In this thesis, we, therefore, focus only on the pure gluonic interactions described using the Yang-Mills theory with the primary aim of developing an efficient approach to compute scattering amplitudes.  The traditional approach to computing the pure gluonic amplitudes is using the Feynman diagram technique where we start with the gauge-fixed Yang-Mills action (or Lagrangian), develop the Feynman rules for the propagator and the interaction vertices, draw all possible Feynman diagrams that can contribute to a given scattering process and then sum them over. Finally, if the diagrams involve loops, we need to integrate over the loop momenta. In four dimensions, this integration is, however, plagued with divergences: infrared at low energies and ultraviolet at high energies. Taming these divergences requires regularization and further lead to the renormalization of the theory (see standard textbooks \cite{Peskin:1995ev, Srednicki:2007qs}). 

From the above discussion, the problem of computing scattering amplitudes seems solved. The reality is, however, far from this. The problem is that the number of diagrams quickly gets out of hand as the number of external gluons increases: 2485 for tree-level and 227,585 for the one-loop seven-point amplitude, for instance.  The final result, on the other hand, often turns out to be compact analytic expressions, in fact, a single term for the case of maximally helicity violating (MHV) tree amplitudes \cite{Parke:1986gb}, when expressed in terms of appropriate variables (helicity spinors). The simplicity of these results questions the complexity of the procedure based on Feynman diagrams with fundamental QCD vertices. Over the past couple of decades, this complexity has been analyzed in different ways which in turn have led to a number of different approaches to bypassing it (see reviews \cite{SC1,SC2,SC3}). 

One reason for this complexity is gauge redundancy. When computing an on-shell (on-mass-shell) amplitude, each Feynman diagram is not gauge invariant and depends on off-shell degrees of freedom. The final amplitude, on the other hand, is gauge invariant and relies only on the on-shell information, implying that the complexity stems from the baggage of un-physical information that the procedure for computing amplitudes depends on. As a result, we get complicated analytic expressions for each diagram that undergoes remarkable cancellation resulting in simple analytic amplitude. This realization has led to the development  of on-shell approaches like the Britto-Cachazo-Feng-Witten (BCFW) on-shell recursion relations \cite{Britto:2004ap,Britto:2005fq}. In this approach, one computes a higher point on-shell tree-level amplitude using lower point on-shell amplitudes where the momentum of some of the external legs of each of the lower point amplitude is shifted to the complex domain. Since the building blocks are on-shell amplitudes, they are gauge invariant. Following the same spirit at the loop level, unitarity-based techniques \cite{Bern1994,Bern2007,Brandhuber2008,Perkins2009,Bern2011, DU1, DU2, DU3, DU4, DU5, DU6} were developed which utilize tree amplitudes to compute the loop amplitudes. This approach provided a remarkable simplicity in computing one-loop amplitudes. Following the Passarino-Veltman reduction \cite{Passarino}, it is well known that the one-loop integral can be reduced in terms of box, triangle, and bubble diagrams with rational coefficients - the integral basis \cite{LR1,LR2,LR3}. Comparing the integral obtained from the unitarity cuts (putting the cut loop momenta on-shell) of the one-loop amplitude with those from the integral basis allows one to systematically determine the coefficients.

The on-shell approaches have also been explored geometrically following the realization that the entire kinematic data of an amplitude consisting of the momentum and the helicities of all the external gluons (specified in terms of 2-component spinors), constitutes positive Grassmannian \cite{Arkani-Hamed_book_2016}. This, later, led to the discovery of Amplituhedron \cite{Arkani-Hamed2014} whose volume determines the scattering amplitudes in $\mathcal{N}=4$ super Yang-Mills (SYM). The latter has its own rich history in the world of pure gluonic scattering amplitudes (see reviews \cite{SYM1, SYM2, SYM3}). In fact, most of the abstract novel reformulations of scattering amplitudes, ranging from amplitudes in twistor space \cite{Witten2004} to Amplituhedron \cite{Arkani-Hamed2014}, that have unraveled hidden structures and symmetries underlying particle interactions originated from investigations in $\mathcal{N} = 4$
SYM. This is because owing to the huge amount of symmetries, computing amplitudes in $\mathcal{N}=4$ SYM is way simpler than any other theory. For instance, the analytic expressions for tree amplitudes, used for numerical predictions of cross sections in QCD, were first obtained in SYM using the Yangian symmetry. At loop level, although the amplitudes in SYM and QCD are not equal, the former provides a simple arena to test new ideas and techniques which could later be used to perform similar calculations in QCD. 

Most of the above approaches focus on computing on-shell amplitudes evading the need for the Yang-Mills action (or Lagrangian) completely. In this thesis, however, our focus is on developing a new action, equivalent to the Yang-Mills action, that allows for efficient computation of pure gluonic amplitudes. The starting point is to realize that one of the major reasons for the complexity of the Feynman diagram technique, discussed above, is that it relies on "very small" building blocks: three and four-point interaction vertices. As a result, the number of diagrams required to compute amplitudes grows rapidly. One can, on the other hand, use a different action (or Lagrangian), equivalent to the Yang-Mills action, consisting of bigger building blocks thereby allowing for efficient computation of the amplitudes. 

The "MHV action", for instance, is an explicit realization of this idea \cite{Mansfield2006,Ettle2006b, Gorsky_2006}\footnote{MHV action is just an example. In this thesis, we will focus on yet another action that we developed.}. The Feynman rules for computing amplitudes in the MHV action correspond to the Cachazo, Svrcek, and Witten (CSW) rules or the MHV rules \cite{Cachazo2004}, which state that the MHV amplitudes continued off-shell can be used as vertices. Gluing them together using a scalar propagator, one can compute any tree-level pure gluonic amplitudes with considerably fewer diagrams. 

Although the MHV rules originated geometrically from the twistor space representation of amplitudes \cite{Witten2004} (and later shown to be a special case of BCFW recursion \cite{Risager2005}), Paul Mansfield showed that the MHV action, on the other hand, could be derived from the light-cone Yang-Mills action via a canonical field redefinition \cite{Mansfield2006}. This sparked a new line of research aimed at re-writing the Yang-Mills action in a form much more suitable for computing amplitudes. In fact, similar actions have also been developed for supersymmetric gauge theories in \cite{Morris_2008, Feng2009, Fu_2010}.  Furthermore, later in \cite{Kotko2017, Kakkad2020}, it was shown that the new fields in the MHV action are certain types of Wilson lines. This led to the realization that although the gluons are considered fundamental, they are probably not the most appropriate degrees of freedom when computing amplitudes. Wilson lines, on the other hand, might be. Similar realizations of using such collective degrees of freedom, instead of gauge fields, have been made in the past, for instance, in high energy QCD \cite{Lipatov:1995pn,Gelis2010} and lattice QCD \cite{Polyakov1980}, to name some. Motivated by these ideas, in \cite{Kakkad:2021uhv}, we derived a new Wilson line-based action which allows for even more efficient computation of amplitudes as compared to the MHV action. 

Despite the success of such actions in computing tree-level amplitudes, there are issues when computing loop amplitudes. Consider first the MHV action. It is well known that using just the MHV vertices one cannot compute amplitudes where either all gluons have plus helicity, the so-called all-plus $(+ + \dots +)$ amplitudes, or one of the gluons has a minus helicity, the so-called single minus $(- + \dots +)$ amplitudes \cite{Brandhuber2007a,Fu_2009, Boels_2008, Ettle2007,Brandhuber2007,Elvang_2012}. This implies that they are zero in the MHV action approach. In QCD, these amplitudes are zero at the tree level but non-zero in general; at one loop they are given by a finite, rational function of the spinor products \cite{APOL1, APOL2,APOL3,Bern:1991aq,Kunszt_1994}. The origin of these issues is intertwined with the canonical transformation that derives the MHV action. Since our new action is derived in a similar fashion, it has missing loop contributions as well. In fact, our action has more missing loop contributions as compared to the MHV action. Therefore, in order to develop quantum corrections systematically such that there are no missing loop contributions, recently in \cite{Kakkad_2022} we used the One-loop Effective Action approach to develop quantum corrections to the MHV action. The main motivation for using the one-loop effective action approach was that it separates the classical action from one-loop contributions. Thus the effect of the field transformation can be explored separately on each set of contributions: the tree and the loops. We demonstrated that when developing quantum corrections to the MHV action this way, there are no missing loop contributions. In fact, to validate this claim, we computed $(+ + + +)$  and $(- + + +)$ one-loop amplitudes and found agreement with the known results \cite{Bern:1991aq,Kunszt_1994}. 

The above mentioned approach is fairly generic and can be used to systematically develop quantum corrections to any action related to the Yang-Mills action via canonical field transformation. Therefore, we extend it to derive the loop contributions in our new action. There is, however, one drawback to this idea. The interaction vertices participating in the loop formation are still the Yang-Mills vertices, and not the MHV or the interaction vertices of our new action. We remedy this by deriving the one-loop effective action in a different way such that the new interaction vertices are explicit in the loop. 

\section{Outline}
\label{sec:outline}

This thesis is a culmination of the work done in \cite{Kakkad2020, Kakkad:2021uhv, Kakkad_2022}. The results presented in Chapter \ref{QZth-chapter} are new and have not been published yet. The manuscript is under preparation. 

Below we present the outline of the entire thesis.

In Chapter \ref{Chap1}, the present chapter, we will review the basics which in turn will lay the foundation for the discussions in the following chapters. In Section \ref{sec:YM_Theory}, we review the basics of Quantum Chromodynamics (QCD). This can be found in standard textbooks on quantum field theories as well. After that, since the focus of this thesis is on pure gluonic scattering amplitudes, in Section \ref{sec:SAmp_YM} we review the textbook approach of computing scattering amplitudes from the Yang-Mills partition function following the Lehman-Symanzik-Zimmermann(LSZ) reduction \cite{LSR_red}. We then discuss the traditional Feynman diagram approach of computing scattering amplitudes. Here we also intend to motivate the reader about why the interaction vertices of the Yang-Mills are not suitable building blocks for computing higher multiplicity amplitudes. In the remaining of Section \ref{sec:SAmp_YM} we review the two important tools: color decomposition which allows one to re-express the full-color amplitudes in terms of color-ordered amplitudes; and Helicity Spinors which provides the appropriate set of variables to develop compact expressions for the color-ordered amplitudes and then highlight the well known results for some helicity tree amplitudes. Finally, Section \ref{sec:MHV_rules} focuses on the Cachazo, Svrcek, and Witten (CSW) rules or the MHV rules for computing tree-level pure gluonic amplitudes with an intention to highlight that the complexity in the Feynman diagram approach of computing scattering amplitudes can be greatly reduced choosing a different set of building blocks. 

In Chapter \ref{Chapt2}, we begin with a review of the derivation of the MHV action in Section \ref{sec:Re-MHV} following the work of Paul Mansfield \cite{Mansfield2006}. The central idea is that the MHV action can be derived from the Yang-Mills action on the light cone via a canonical field redefinition. This action provides a field theory realization of the MHV rules and thus lays the foundation of our work. After that in Section \ref{sec:EX_SOL_MT} we explore the physical interpretation of the solutions of the transformation. In Subsection \ref{subsec:SDYM_EOM} we show that the solution for the plus helicity fields in the transformation is intimately related to the Self-Dual sector of the Yang-Mills theory. Later in Subsections \ref{subsec:Bbul[A]_WL}-\ref{subsec:Bstar[A]_WL} we demonstrate that the fields in the MHV action are in fact straight infinite Wilson lines in position space. This generates the motivation towards the idea that Wilson lines might be a better degree of freedom instead of gauge fields in the context of computing scattering amplitudes. Finally, in Section \ref{sec:MHV_vert_TS}, we explore the MHV vertices geometrically using the fact that the fields in the vertex are straight infinite Wilson lines in the complexified Minkowski space. The majority of the calculations or derivations related to this chapter have been put in Appendices \ref{sec:app_A1}-\ref{sec:app_A4}.

In Chapter \ref{WLAc-chapter}, we go beyond the MHV action to derive a new Wilson line-based action that allows for even more efficient computation of the tree level pure gluonic amplitudes. We begin by explaining the main motivation, in Section \ref{sec:Zac_mot}, that led us to develop this new action. Following this we explicitly derive the new action in Section \ref{sec:Z_ACTION}. This section is further divided into three parts. In Subsection \ref{subsec:gen_id}, we discuss the outline of the entire derivation, skipping the details,  starting first with the structure of the transformation that derives the new action from the Yang-Mills action. Then we consider the structure of the solutions of this transformation using which we finally outline the structure of the new action and elaborate on its properties. Then, in Subsection \ref{subsec:Zac_der}, we work out all the details associated with the derivation. We begin by demonstrating that the transformation eliminates both the triple gluon vertices in the Yang-Mills action but there are two ways of executing this. Either of the two ways can be used to derive the new action. Then we move on to developing the solutions of the transformation and its inverse both in the position and momentum space. In position space, the solution turns out to be "Wilson lines of Wilson lines". Finally, using the inverse of these solutions we derive the explicit content of the vertices in the new action. After that, in Subsection \ref{subsec:Zac_FR} we discuss the Feynman rules for computing amplitudes in the new action. These are then used in Section \ref{sec:Zac_TR_AMP} to compute several split-helicity tree level amplitudes up to 8 points. Interestingly, the number of diagrams we get follows a well known number series called the Delannoy number. In Subsection \ref{subsec:del_num}, we explore the connection of these numbers with the computation of amplitudes in our action. After that, in Section \ref{sec:Zac_TS}, we engage in an intuitive discussion about the vertices of our action in twistor space. In Appendix \ref{sec:app_A5}, we present the details of the elimination of both the triple gluon vertices from the Yang-Mills action using our transformation.

In Chapter \ref{QMHV-chapter}, we develop quantum corrections to the MHV action. We begin by highlighting the problem of missing loop contributions in the MHV action which makes computing loop amplitudes of certain types like the all-plus and single-minus impossible. The source of these missing loop contributions is the Self-Dual vertex which gets eliminated via Mansfield's transformation. Thus, we use the on-loop effective action approach. In Section \ref{sec:SDYM}, we review the one-loop effective action approach using the simple example of the Self-Dual Yang-Mills theory. After that, in Section \ref{sec:Yang-Mills_OLEA}, we derive the one-loop effective action for the Yang-Mills theory. This is the starting point of our systematic approach to developing quantum corrections to the MHV action. In Section \ref{sec:MHV_OLEA}, we apply Mansfield's transformation to the Yang-Mills one-loop effective action. We then demonstrate, using the example of the 2-point $(+ -)$ amputated connected Green's function, that the new action we obtain consists of the classical MHV action plus all the one-loop corrections (MHV as well as non-MHV). Finally, to verify that there are indeed no missing one-loop contributions we compute one-loop amplitudes in Section \ref{sec:One-loop-OLEAMHV}. We first begin by reviewing the 4D world-sheet regularization scheme that we employ during the computation in Subsection \ref{subsec:CQTreview}. Then in Subsection~\ref{sub:4plus}, we compute the leading trace 4-point $(+ + + +)$ one-loop amplitude using the one-loop effective MHV action. After that, in Subsection \ref{sub:3plus-minus} we compute the leading trace 4-point $(+ + + -)$ one-loop amplitude. Both the results agree with the known results confirming that there are indeed no missing one-loop contributions this way. In Appendices \ref{sec:app_A6}-\ref{sec:app_A8}, we present the technical details of some of the selected calculations/derivations related to the present chapter.

In Chapter \ref{QZth-chapter}, we begin with a straightforward extension of the one-loop effective action approach developed in Chapter \ref{QMHV-chapter} to systematically develop quantum correction to the new Wilson line-based action (also called the Z-field action) in Section \ref{sec:Zth_OLEA}. After that, in Section \ref{sec:Issue_MHV_Zac}, we highlight a major issue associated with the one-loop effective MHV as well as the Z-field action derived via the approach developed in Chapter \ref{QMHV-chapter}. The issue is that the vertices entering the loop formation in these one-loop effective actions are the Yang-Mills vertices and not the bigger interaction vertices of the MHV action or the Z-field action. We remedy this issue in Section \ref{sec:OLEA_rev} where we use a new way of deriving the two one-loop effective actions. In this approach, we first perform the canonical transformations that derive the MHV and Z-field action from the Yang-Mills action and then proceed to obtain the corresponding one-loop effective actions. We employ this in Subsection \ref{subsec:OLEA_MHVrev} to re-derive the one-loop effective MHV action and then in Subsection \ref{subsec:OLEA_ZACrev} to re-derive the one-loop effective Z-field action. In both cases, we demonstrate that the actions are one-loop complete and are identical (up to a factor redundant for amplitude computation) to the one-loop effective action derived via the previous approach developed in Chapter \ref{QMHV-chapter}. Finally, to validate the one-loop effective Z-field action, in Section \ref{sec:loopamp_ZAC} we compute  the leading trace 4-point $(+ + + +)$, $(+ + + -)$, $(- - - -)$, $(- - - +)$ and $(+ + - -)$ one-loop amplitudes. The results agree with the known results confirming that there are indeed no missing one-loop contributions. In Appendices \ref{sec:app_A9}-\ref{sec:app_A10}, we present the technical details of some of the selected calculations/derivations related to the present chapter.
\section{Quantum Chromodynamics (QCD)}
\label{sec:YM_Theory}

Straightforwardly put, Quantum Chromodynamics (QCD) is a quantum field theory describing strong interactions between quarks and gluons in the Standard Model of particle physics. It is a non-abelian gauge theory, meaning that this theory is invariant under space-time dependent (local) special unitary transformations \cite{YangPGI}. 

Let us elaborate on the above statement starting first with the general philosophy. Theoretically, physical systems are represented by a scalar object called the action functional (or simply \textit{action}); denoted by S. For gauge theories, this action is invariant under local transformations $\mathrm{U(x)}$, the so-called  gauge transformation \cite{YangPGI}. The set of all these transformations forms a group which is commonly referred to as the gauge group. For non-abelian gauge theories, this group is $\mathrm{SU(N)}$ which, in fundamental representation, consists of a set of $\mathrm{N}\times \mathrm{N}$ unitary matrices ($\mathrm{U(x)}\mathrm{U(x)}^{\dagger} = \mathrm{U(x)}^{\dagger}\mathrm{U(x)} =\mathbb{1}$)\footnote{The transpose conjugate $\mathrm{U(x)}^{\dagger}$ of the matrix $\mathrm{U(x)}$ is it's inverse.} satisfying: $\det(\mathrm{U(x))} = 1$. The group operation being matrix multiplication. Finally, this group has an associated Lie algebra $\mathfrak{su}(\mathrm{N})$ which is a vector space of infinitesimal transformations equipped with a lie bracket $[\cdot,\cdot]:\mathfrak{g}\times \mathfrak{g} \longrightarrow \mathfrak{g}$ which measures the non-commutativity of two transformations. Owing to the non-commutative nature of the transformations, the gauge group is termed as a non-abelian group, and the corresponding gauge theory as a non-abelian gauge theory. 

For QCD, the gauge group is $\mathrm{SU(3)}$, which in fundamental representation consists of $3\times 3$ unitary matrices $\mathrm{U(x)}$ satisfying: $\det(\mathrm{U(x))} = 1$. Owing to the latter, there are only 8 linearly independent matrices in this matrix space. These correspond to the 8 generators $t^a$, where $a$ is the so-called adjoint index and runs from 1 to 8, defining the linearly independent directions of the $\mathfrak{su}(\mathrm{3})$ Lie algebra vector space. The rows and columns of these generator matrices can further be enumerated using the so-called fundamental indices $j,k\dots$ which run from 1 to 3. In terms of $t^a$, any generic element of the group can be expressed as
\begin{equation}
    \mathrm{U(x)} = \exp\left[i\, \sum_{a=1}^{8} \phi_a (x) t^a \right]\,.
    \label{eq:gen_GT}
\end{equation}
In what follows, we shall use the notation $\hat{\phi}(x
) = \phi_a (x) t^a $ for the Lie algebra valued quantities. Finally, the generators $t^a$ are Hermitian traceless matrices and we use the following normalizations for these
\begin{equation}
    \left[t^{a},t^{b}\right]=i\sqrt{2}f^{abc}t^{c}\,, \quad \quad \mathrm{Tr}(t^{a}t^{b}) = \delta^{ab}\,.
    \label{eq:su3_sc_id}
\end{equation}
Above, $f^{abc}$ are the totally anti-symmetric structure constants. $t^{a}$ are in fundamental representation. 

Now, we have the basic ingredients to elaborate more on QCD as a non-abelian gauge theory. Traditionally, the starting point of this discussion is the free (no interactions) Dirac action
\begin{equation}
    S_{\mathrm{Dirac}} = \int d^{4}x\, \Big\{\overline{\psi}_{f,j} \left[i\, \gamma^{\mu} (\partial_{\mu})_{jk} - m_f \delta_{jk} \right] \psi_{f,k}\Big\}\,,
    \label{eq:Dirac_free}
\end{equation}
where the integral is over the 4D space-time. Note, however, that we suppressed the space-time coordinates in the integrand for simplicity. Above, all the repeated indices are summed over. The fermionic matter fields -- $\psi_{f,k}$ -- are complex valued. In QCD, these represent the quarks of flavor $f$ and mass $ m_f$, and $k$ is the fundamental color index associated with the quark. Since $k$ runs from 1 to 3, there are three quarks.  $\gamma^{\mu}$ are the Dirac $\gamma$-matrices which in 4D satisfies the Clifford algebra
\begin{equation}
    \left\{\gamma^{\mu}, \gamma^{\nu} \right\} = 2 g^{\mu \nu}\,.
\end{equation}
Finally, $\partial_{\mu}= \partial/\partial x^{\mu}$ is the space-time derivative with $(\partial_{\mu})_{jk} = \partial_{\mu} \delta_{jk}$ and $\overline{\psi}_{f,j} = {\psi}_{f,j}^{\dagger} \gamma^{0}$.

Under "global" (space time independent, $\phi_a (x) \equiv \phi_a$ in Eq.~\eqref{eq:gen_GT}) gauge transformation, the quark fields transform as 
\begin{equation}
    \psi_{f,k}(x) \longrightarrow \Big(\exp[i {\phi}_a t^a]\Big)_{kl}\psi_{f,l}(x) \,, \quad \quad \psi_{f,k}^{\dagger} (x) \longrightarrow \psi_{f,l}^{\dagger}(x)\Big(\exp[i {\phi}_a t^a]\Big)^{\dagger}_{lk} \,.
    \label{eq:psi_GGT}
\end{equation}
Owing to the above relations we see that the free Dirac action Eq.~\eqref{eq:Dirac_free}, describing free quarks, is invariant under global gauge transformations. However, if we now impose the transformations to be local, that is
\begin{align}
    \psi_{f,k}(x) \longrightarrow \Big(\exp[i {\phi}_a(x) t^a]\Big)_{kl}\psi_{f,l}(x) \,, &\quad \quad \psi(x)\longrightarrow   \mathrm{U(x)} \psi(x)\,, \nonumber
    \\ \psi_{f,k}^{\dagger} (x)\longrightarrow \psi_{f,l}^{\dagger}(x)\Big(\exp[i {\phi}_a (x) t^a]\Big)^{\dagger}_{lk} \,, &\quad \quad \psi^{\dagger}(x)\longrightarrow  \psi^{\dagger}(x) \mathrm{U(x)}^{\dagger}\,,
    \label{eq:psi_LGT}
\end{align}
where on R.H.S. we introduced compact notation by suppressing the indices for the sake of simplicity, we see that the mass term in the free Dirac action Eq.~\eqref{eq:Dirac_free} is still invariant. But the space time derivative term is not. In fact, since the derivative subtracts fields defined at two different space time points, there is no appropriate transformation rule for $\partial_{\mu}\psi$. This is overcome by introducing the so-called \textit{comparator}: $\mathcal{W} (z, y)$\footnote{$\mathcal{W} (z, y)$ satisfies: $\mathcal{W} (y, y) = 1$ and $\mathcal{W} (z, y)\mathcal{W} (y, x) = \mathcal{W} (z, x)$.} -- also called a Wilson line
\begin{equation}
    \mathcal{W} (z, y) = \mathbb{P}\exp\left[ig\int_{0}^{1}\! ds\, \frac{dx^{\mu}}{ds} \hat{A}_{\mu}\left(x(s)\right)\right]\, ,
    \label{eq:WL_compa}
\end{equation}
where $\mathbb{P}$ represents path-ordering with respect to the increase of the parameter s and $x(s)$ represents the contour with $x(0)=y$ and $x(1)=z$. $g$ is the coupling constant, and owing to normalization Eq.~\eqref{eq:su3_sc_id}, we re-scale the coupling constant as $g\rightarrow g/\sqrt{2}$ to accommodate for the additional factors of $\sqrt{2}$. Finally, $\hat{A}^{\mu}$ is the spin-1 gauge field, also known as the \textit{connection}. It is Lie algebra-valued and can be expanded in terms of the generators as follows
\begin{equation}
    \hat{A}^{\mu}=A_a^{\mu}t^a \quad \Leftrightarrow \quad \mathrm{Tr}(t^{a}\hat{A}^{\mu}) = A_a^{\mu} \,.
\end{equation}

Using the comparator $\mathcal{W} (z, y)$, one replaces the space time derivative $\partial_\mu$ with the so-called covariant derivative $(D_{\mu})_{jk}$ defined as follows
\begin{equation}
    (D_{\mu})_{jk} = \partial_{\mu} \delta_{jk} - i g{A}_{\mu a}(t^a)_{jk} \,.
\end{equation}

Under the local gauge transformation Eq.~\eqref{eq:psi_LGT}, $\mathcal{W} (z, y)$ transforms as 
\begin{equation}
    \mathcal{W} (z, y) \longrightarrow \mathrm{U(z)}\mathcal{W} (z, y) \mathrm{U(y)}^{\dagger}\,.
\end{equation}
Following this, one can show that
\begin{equation}
    \hat{A}^{\mu} \longrightarrow \mathrm{U(x)}\left[\hat{A}^{\mu} + \frac{i}{g} \partial^{\mu} \right]\mathrm{U(x)}^{\dagger}\,, \quad \quad D_{\mu}\longrightarrow \mathrm{U(x)}\,D_{\mu}\,\mathrm{U(x)}^{\dagger}\,.
    \label{eq:gauge_trans1}
\end{equation}
Therefore we see that a new action defined as
\begin{equation}
    S = \int d^{4}x\, \Big\{\overline{\psi}_{f,j} \left[i\, \gamma^{\mu} (D_{\mu})_{jk} - m_f \delta_{jk} \right] \psi_{f,k}\Big\}\,,
    \label{eq:Dirac_int}
\end{equation}
is invariant under the local gauge transformations. 

Note, demanding the local gauge invariance transformed the Dirac action from Eq.~\eqref{eq:Dirac_free} to Eq.~\eqref{eq:Dirac_int}. The new action has two important differences as compared to the old one. First, it contains 8 new spin-1 gauge fields ${A}^a_{\mu}$. In QCD these are the so-called gluons. Secondly, the action is no longer free. It has a triple-point interaction vertex consisting of two quarks and one gluon. Thus, in some sense, the "existence" of the gauge fields (gluons) as well as interactions are deeply intertwined with the principle of local gauge invariance.

What is missing in Eq.~\eqref{eq:Dirac_int}, so as to obtain the full QCD action, is the description of gluons. This requires the Yang-Mills action, which reads
\begin{equation}
S_{\mathrm{YM}}=\int d^{4}x\,\mathrm{Tr}\left\{ -\frac{1}{4}\hat{F}_{\mu\nu}\hat{F}^{\mu\nu}\right\} \,.
\label{eq:YM_cov}
\end{equation}
Above, $\hat{F}^{\mu\nu}$ is the Lie algebra-valued field strength tensor
\begin{equation}
\hat{F}^{\mu\nu}=\partial^{\mu}\hat{A}^{\nu}-\partial^{\nu}\hat{A}^{\mu}-ig\left[\hat{A}^{\mu},\hat{A}^{\nu}\right]\,,
\label{eq:FST_c}
\end{equation}
where $\hat{F}^{\mu\nu}= {F}_a^{\mu\nu} t^a$. Under local gauge transformation, using Eq.~\eqref{eq:gauge_trans1} one can see that $\hat{F}^{\mu\nu}$ transforms as follows
\begin{equation}
    \hat{F}^{\mu\nu}\longrightarrow \mathrm{U(x)}\,\hat{F}^{\mu\nu}\,\mathrm{U(x)}^{\dagger}\,.
    \label{eq:gauge_trans}
\end{equation}
Although, $\hat{F}^{\mu\nu}$ transforms covariantly under the local gauge transformations, the Yang-Mills action Eq.~\eqref{eq:YM_cov} is invariant owing to the trace. 

Substituting Eq.~\eqref{eq:FST_c} to Eq.~\eqref{eq:YM_cov} we can express the Yang-Mills action explicitly in terms of the gauge field $\hat{A}^{\mu}$ as shown below
\begin{multline}
S_{\mathrm{YM}}=\int d^{4}x\,\mathrm{Tr}\Bigg\{-\frac{1}{2}\left(\partial^{\mu}\hat{A}_{\nu}\right)^{2}+\frac{1}{2}\partial^{\mu}\hat{A}^{\nu}\partial_{\nu}\hat{A}_{\mu}+ig\partial^{\mu}\hat{A}^{\nu}\left[\hat{A}_{\mu},\hat{A}_{\nu}\right]\\
+\frac{1}{4}g^{2}\left[\hat{A}^{\mu},\hat{A}^{\nu}\right]\left[\hat{A}_{\mu},\hat{A}_{\nu}\right]\Bigg\}\,.\label{eq:YM_action_cov}
\end{multline}
Above, there are three types of terms: quadratic, cubic, and quartic in the gauge fields. The quadratic term represents the kinetic term whereas the cubic and quartic terms represent self-interactions. Thus, even without coupling any matter Eq.~\eqref{eq:Dirac_int}, the Yang-Mills theory represents an interaction theory of spin-1 fields. At this point, if we change the gauge group from $\mathrm{SU(3)}$ to $\mathrm{U(1)}$, the commutator in Eq.~\eqref{eq:FST_c} and Eq.~\eqref{eq:YM_action_cov} will vanish and we will get an abelian gauge theory  action with just the kinetic term and no interactions. This action would describe a photon propagating in space-time. In order to have interactions, we need to include matter fields with electric charge and this would give rise to the theory of Quantum Electrodynamics (QED).

Note, however, the Yang-Mills action Eq.~\eqref{eq:YM_action_cov} is incomplete. There are two terms missing. These are the gauge-fixing and the ghost terms. We discuss their origin as well as their systematic inclusion to the Yang-Mills action via the Fadeev-Popov approach \cite{Faddeev:1967fc} in the following section (in the same section we also explain the name \textit{ghosts}). Here we simply quote the commonly used terms. For gauge fixing, we have the $\xi$ dependent term as shown below
\begin{equation}
    S_{\mathrm{GF}} = -\int d^{4}x\,\frac{1}{2\xi} \left(\partial^{\mu} A_{\mu}^a \right)^2\,,
    \label{eq:gauge_fix}
\end{equation}
where "GF" stands for gauge fixing and $\xi$ is the gauge parameter. Setting $\xi = 1$, for instance, we get the Feynman-'t Hooft gauge. Finally, the ghost term reads
\begin{equation}
    S_{\mathrm{Ghosts}} = -\int d^{4}x\, \overline{c}^a\partial^\mu D^{ab}_\mu c^b \,,
    \label{eq:FPgho}
\end{equation}
where $c$ and $\overline{c}$ are the Lie algebra valued anticommuting fields.

Putting everything together we get the full QCD action
\begin{equation}
    S_{\mathrm{QCD}} = \int d^{4}x\, \Big\{\overline{\psi}_{f,j} \left[i\, \gamma^{\mu} (D_{\mu})_{jk} - m_f \delta_{jk} \right] \psi_{f,k}\Big\} - \mathrm{Tr}\left\{\frac{1}{4}\hat{F}_{\mu\nu}\hat{F}^{\mu\nu}\right\} + S_{\mathrm{GF}}+ S_{\mathrm{Ghosts}}\,.
    \label{eq:S_QCD}
\end{equation}
The discussion so far was focused on the full QCD action. Hereafter, we shall focus only on the Yang-Mills action Eq.~\eqref{eq:YM_action_cov} describing pure-gluonic interactions (also dubbed as \textit{Gluodynamics}).

\section{Scattering amplitudes in Yang-Mills theory}
\label{sec:SAmp_YM}

In particle physics, as stated previously, interactions are studied experimentally through particle collisions at the colliders. In such collisions, cross sections of various scattering processes are one of the most crucial measured physical quantities. Theoretically, these are calculated via scattering amplitudes which are probability amplitudes for a given scattering process. The cross section is proportional to the square of scattering amplitudes. These are, therefore, at the intersection between theoretical predictions and experimental measurements. Below, we review the general procedure for computing these objects for pure gluonic interactions (i.e., for the Yang-Mills theory Eq.~\eqref{eq:YM_action_cov}) which shall be the focus of the rest of the text.

Consider an $n$-point pure gluonic scattering process with $n=n_i +n_f$, where $n_i$ ($n_f$) represents the number of gluons in the initial (final) state $|I\rangle$ ($|F\rangle$) at $t=-\infty$ ($t=+\infty$). The scattering amplitude ${A}_n $ for this process is given by the S-matrix element $\langle F| S |I\rangle$ (here S must not be confused with the action, it is in fact the time-ordered exponential of the interaction part of the action in the interaction picture description). Assuming the asymptotic initial and final states to be non-interacting, we can express these as the direct product of the single-particle states $| \left\{ p_i, h_i, a_i \right\}\rangle$, which for on-shell gluons depend on the on-shell momenta $p_i^2 = 0$, polarization $\varepsilon_{p_i}^{\pm}$ represented by the helicities $h_i = \pm 1$ and the color charge $a_i$. Using these we can write,
\begin{equation}
    {A}_n \left( \left\{ p_i, h_i, a_i\right\}\right) = \langle \left\{ p_{n}, h_{n}, a_{n} \right\}, \dots, \left\{ p_{n_i + 1}, h_{n_i + 1}, a_{n_i + 1} \right\}| S |\left\{ p_{n_i}, h_{n_i}, a_{n_i} \right\}\dots \left\{ p_1, h_1, a_1 \right\}\rangle \,.
\end{equation}
Above, R.H.S can be computed using the Lehman-Symanzik-Zimmermann(LSZ) reduction \cite{LSR_red} which basically states that the amplitudes can be obtained by amputating the momentum space connected Greens function $G^{\mu_1 \dots \mu_n}(p_1, \dots, p_n)$ with external states on-shell $p_i^2 = 0$. Let us see how it's done. The Greens function, in position space, is computed using the generating functional for the Greens function or the partition function $Z[J]$ which reads
\begin{equation}
    Z[J]= \mathcal{N}\int[dA]\, e^{i\left(S_{\mathrm{YM}}[A] + \int\!d^4x\, \Tr \hat{J}_{\mu}(x) \hat{A}^{\mu}(x)\right) } \,,
    \label{eq:gen_YM_cov}
\end{equation}
where 
\begin{equation}
    [dA] = \prod_x \prod_{a, \mu}d A_a^{\mu}(x) \,,
    \label{eq:A_mes}
\end{equation}
and $S_{\mathrm{YM}}[A]$ is the Yang-Mills action Eq.~\eqref{eq:YM_action_cov} coupled with an external source $\hat{J}_{\mu} (x)$. The source is an arbitrary function. $\mathcal{N}$ is the overall normalization. The integral is over all the configurations of the gauge field, denoted compactly as $[dA]$. The Greens function are obtained as follows
\begin{align}
    G^{\mu_1 \dots \mu_n}_{a_1, \dots, a_n}(x_1, \dots, x_n) =& \frac{\mathcal{N}\int[dA]\,{A}^{\mu_1}_{a_1}(x_1)\, \dots {A}^{\mu_n}_{a_n}(x_n)\, e^{i\left(S_{\mathrm{YM}}[A] \right) }}{\mathcal{N}\int[dA]\, e^{i\left(S_{\mathrm{YM}}[A]\right) }}\,, \label{eq:GF_1}\\
    =& (-i)^n \frac{1}{Z[J]} \left. \frac{\delta^n Z[J]}{\delta J_{\mu_1 a_1}(x_1) \dots \delta J_{\mu_n  a_n}(x_n)} \right|_{J=0}\,. \label{eq:GF_2}
\end{align}

The R.H.S of Eq.~\eqref{eq:GF_1} involves integrals (both in the numerator and denominator) that cannot be computed exactly. However, it can be computed order by order perturbatively. To see this, consider the integral in the numerator. The action is first split into a free part and an interaction part as $S_{\mathrm{YM}}[A] = S_{\mathrm{free}}[A] + S_{\mathrm{int}}[A]$, where the former contains the kinetic term and the latter contains the cubic and quartic interaction vertices which are proportional to the coupling constant $g$. Assuming $g$ to be small, one expands the exponential of $S_{\mathrm{int}}[A]$ into a series
\begin{equation}
 \sim \mathcal{N}\int[dA]\,{A}^{\mu_1}_{a_1}(x_1)\, \dots {A}^{\mu_n}_{a_n}(x_n)\, e^{i\,S_{\mathrm{free}}[A] } \left( 1 + \sum_{n=1} \frac{\left(i\,S_{\mathrm{int}}[A]\right)^n}{n!}\right)\,,
 \label{eq:pert_exp}
\end{equation}
where $S_{\mathrm{int}}[A] \sim \int\!d^4x \, [\mathcal{V}_3 (x) + \mathcal{V}_4 (x)]$ (the cubic and the quartic interaction vertices in Eq.~\eqref{eq:YM_action_cov}). The above trick reduces the problem into a series of integrals. Since the kinetic term (or equivalently the $S_{\mathrm{free}}[A]$) is quadratic in the gauge fields, each term in the series is a Gaussian-type integral which can be evaluated exactly. The integration for each term gives rise to a set of diagrams each of which represents a certain way of connecting all the fields, external as well as those originating from the cubic and quartic interaction vertices, via the \textit{propagator} (inverse of the kinetic term, also referred to as the two-point Greens function of the free theory). At this point, one is only left with the integration over the vertex positions. Fourier transforming everything, one obtains the Greens function $G^{\mu_1 \dots \mu_n}_{a_1,\dots, a_n}(p_1, \dots, p_n)$ in momentum space where the integral over vertex positions gives rise to momentum conserving delta at each vertex plus integrals over momentum of each line. Following this procedure for the integral in the denominator of Eq.~\eqref{eq:GF_1} gives rise to "vacuum bubbles" (diagrams with no external legs) which can be factored out from the numerator as well and thus they get canceled explicitly. As a result, the R.H.S of Eq.~\eqref{eq:GF_1} consists of only two types of terms: connected where each external point is connected to the others via a set of propagators and interaction vertices, and disconnected. Only the former contributes to the computation of the scattering amplitudes therefore we define a subset $\{G_{a_1,\dots, a_n}^{\mu_1 \dots \mu_n}(p_1, \dots, p_n)_{\mathrm{con}}\}$ consisting of only the connected contributions. In order to obtain amplitudes, one needs to contract $\{G_{a_1,\dots, a_n}^{\mu_1 \dots \mu_n}(p_1, \dots, p_n)_{\mathrm{con}}\}$ with appropriate polarization $\varepsilon_{p_i}^{\pm}$ of all the external gluons, 
 amputate the propagators on the external legs and finally take the on-shell limit for all the external gluons. That is
 \begin{equation}
    {A}_n \left( \left\{ p_i, h_i, a_i\right\}\right) = \sum_{\mathrm{all}} \mathrm{Amputated}\, \left\{  \varepsilon_{p_1}^{\pm} \dots \varepsilon_{p_n}^{\pm}\left. \{G_{a_1,\dots, a_n}^{\mu_1 \dots \mu_n}(p_1, \dots, p_n)_{\mathrm{con}}\}\right|_{p_i^2 = 0\, \forall\, i\, \in\, \left\{1, \dots, n\right\} }\, \right\} \,.
    \label{eq:amp_pert}
\end{equation}
There is, however, a caveat in this entire discussion. The partition function Eq.~\eqref{eq:gen_YM_cov} in its current form is an ill-defined quantity and therefore requires a fixing. Recall, the partition function is integral over all the configurations of the gauge field Eq.~\eqref{eq:A_mes}. There are infinitely many configurations that are related to each other via the gauge transformation Eq.~\eqref{eq:gauge_trans}. This introduces an additional divergence to the integral. In order to properly define the partition function, one must distribute the field configurations into classes that are not related via gauge transformation and then consider only one configuration from each class in the integral. This is systematically done via the Fadeev-Popov approach \cite{Faddeev:1967fc} where one introduces
\begin{equation}
    1 = \int \,[d\phi]\,\, \delta(G[A^{\phi}])\, \det \left( \frac{\delta G[A^{\phi}]}{\delta \phi}\right)\,, \quad \mathrm{where} \quad [d\phi] = \prod_{x,a}d\phi^a(x) \,,
    \label{eq:fad_pop}
\end{equation}
into the definition of the partition function. Above $\hat{A}^{\phi}$ is the gauge transformed field 
\begin{equation}
    (\hat{A}^{\mu })^{\phi} = e^{i \hat{\phi}(x)}\left[\hat{A}^{\mu} + \frac{i}{g} \partial^{\mu} \right]e^{-i \hat{\phi}(x)}\,,
    \label{eq:gaf_gt}
\end{equation}
and $G[A]$ is the so-called gauge fixing functional. For any choice of gauge, one obtains two terms. First, the gauge fixing term originates from $\delta(G[A^{\phi}])$. Second, the determinant can be rewritten as an integral over un-physical fields - \textit{ghosts}. These two terms introduce corrections to the action $S_{\mathrm{YM}}[A]$ in the partition function. To see this, let us begin by including Eq.~\eqref{eq:fad_pop} to Eq.~\eqref{eq:gen_YM_covFP} as shown below
\begin{equation}
    Z[J]= \mathcal{N}\int[dA]\,\int \,[d\phi]\,\, \delta(G[A^{\phi}])\, \det \left( \frac{\delta G[A^{\phi}]}{\delta \phi}\right)  e^{i S_{\mathrm{YM}}[A]  } \,.
    \label{eq:gen_YM_covFP}
\end{equation}
For the determinant term above, the infinitesimal version for the transformation Eq.~\eqref{eq:gaf_gt} is more suitable. It reads
\begin{equation}
    ({A}_a^{\mu })^{\phi} = \left[{A}_a^{\mu} + \frac{1}{g} D^{\mu} {\phi}_a\right]\,.
    \label{eq:gaf_gt1}
\end{equation}
Now, consider the following gauge condition
\begin{equation}
    G[A] = \partial_{\mu}{A}_a^{\mu} - \Lambda_a(x)\,,
    \label{eq:gauge_cho}
\end{equation}
where $\Lambda_a(x)$ is arbitrary. This breaks the Lorentz invariance but it can be restored by integrating it using a weight function. The usual choice is the Gaussian weight. With this we have
\begin{equation}
    Z[J]= \mathcal{N}\int[d\Lambda]\,\int [dA]\,\int \,[d\phi]\,\, \delta(G[A^{\phi}])\, \det \left( \frac{1}{g}\partial^{\mu}D_{\mu}\right)  e^{-\frac{i}{2\xi}\int d^{4}x\,\Lambda_a(x)\Lambda^a(x) + i S_{\mathrm{YM}}[A]  } \,.
    \label{eq:gen_YM_covFP1}
\end{equation}
Recall, the action $S_{\mathrm{YM}}[A]$ is invariant under the transformation Eq.~\eqref{eq:gaf_gt}. Furthermore, since this transformation is a combination of a linear shift followed by unitary rotation, the measure is also invariant under this transformation. As a result, we can replace $A\longrightarrow A^{\phi}$ everywhere. This makes it a dummy variable and we can rewrite the entire expression Eq.~\eqref{eq:gen_YM_covFP1} as follows
\begin{equation}
    Z[J]= \mathcal{N}\left(\int \,[d\phi]\,\right) \int[d\Lambda]\,\int [dA]\,\, \delta(G[A])\, \det \left( \frac{1}{g}\partial^{\mu}D_{\mu}\right)  e^{-\frac{i}{2\xi}\int d^{4}x\,\Lambda_a(x)\Lambda^a(x) + i S_{\mathrm{YM}}[A]  } \,,
    \label{eq:gen_YM_covFP2}
\end{equation}
where the integral over $\phi$ factors out and it can be absorbed into the normalization. Substituting the gauge choice Eq.~\eqref{eq:gauge_cho} in the delta, we can perform the integral with respect to~$\Lambda$. By doing this we get
\begin{equation}
    Z[J]= \mathcal{N} \int [dA]\,\, \det \left( \frac{1}{g}\partial^{\mu}D_{\mu}\right)  e^{-\frac{i}{2\xi}\int d^{4}x\,\left( \partial_{\mu}{A}_a^{\mu}\right)^2 + i S_{\mathrm{YM}}[A]  } \,.
    \label{eq:gen_YM_covFP3}
\end{equation}
In the Fadeev-Popov procedure, the determinant is rewritten as follows
\begin{equation}
    \det \left( \frac{1}{g}\partial^{\mu}D_{\mu}\right) = \int [dc]\,[ d \overline{c}]e^{-i\int d^{4}x\, \overline{c}^a\partial^\mu D^{ab}_\mu c^b } \,,
\end{equation}
where the factor of $1/g$ has been absorbed in the normalization of the  fields $c$ and $\overline{c}$. These are Lie algebra-valued anti-commuting scalar fields and they satisfy the Fermi statistics. These are, therefore, fictitious fields with no physical meaning. Hence, they are termed as the \textit{ghosts}. Putting all of this together we get
\begin{equation}
    Z[J]= \mathcal{N} \int [dA]\,[dc]\,[ d \overline{c}]\, e^{-\frac{i}{2\xi}\int d^{4}x\,\left( \partial_{\mu}{A}_a^{\mu}\right)^2 -i\int d^{4}x\, \overline{c}^a\partial^\mu D^{ab}_\mu c^b  + i S_{\mathrm{YM}}[A]  } \,.
    \label{eq:gen_YM_covFP4}
\end{equation}
Notice, the first two terms in the exponent are exactly the gauge fixing and the ghost terms we discussed in Eq.~\eqref{eq:gauge_fix} and Eq.~\eqref{eq:FPgho}, respectively.

In the rest of the text, we will work in the light-cone gauge $G[A] = {\hat A}^{\mu}\cdot\eta^{\mu} = 0$ where $\eta^{\mu} = (1,0,0,-1)/\sqrt{2}$ is a light-like four-vector. In this gauge, $G[A^{\phi}]$ and therefore both the above terms originating from it become independent of the gauge field. As a result, the integral Eq.~\eqref{eq:fad_pop} can be factored out of the partition function and can be absorbed in the overall normalization $\mathcal{N}$. Hence, we do not get any extra terms.

The above procedure, summarized in Eq.~\eqref{eq:amp_pert}, gives rise to the perturbative approach for computing scattering amplitudes ${A}_n \left( \left\{ p_i, h_i, a_i\right\}\right)$, squaring which one obtains the calculated cross section for a process which could be compared against the measured values in the experiments. Note, however, that the accuracy of the computed cross section depends on how many terms one computes in the series Eq.~\eqref{eq:pert_exp}. The higher the number of terms, the higher the accuracy.

\subsection{Feynman diagrams}
\label{subsec:Fey_dia}

\begin{figure}
    \flushleft
    \includegraphics[width=15.9cm]{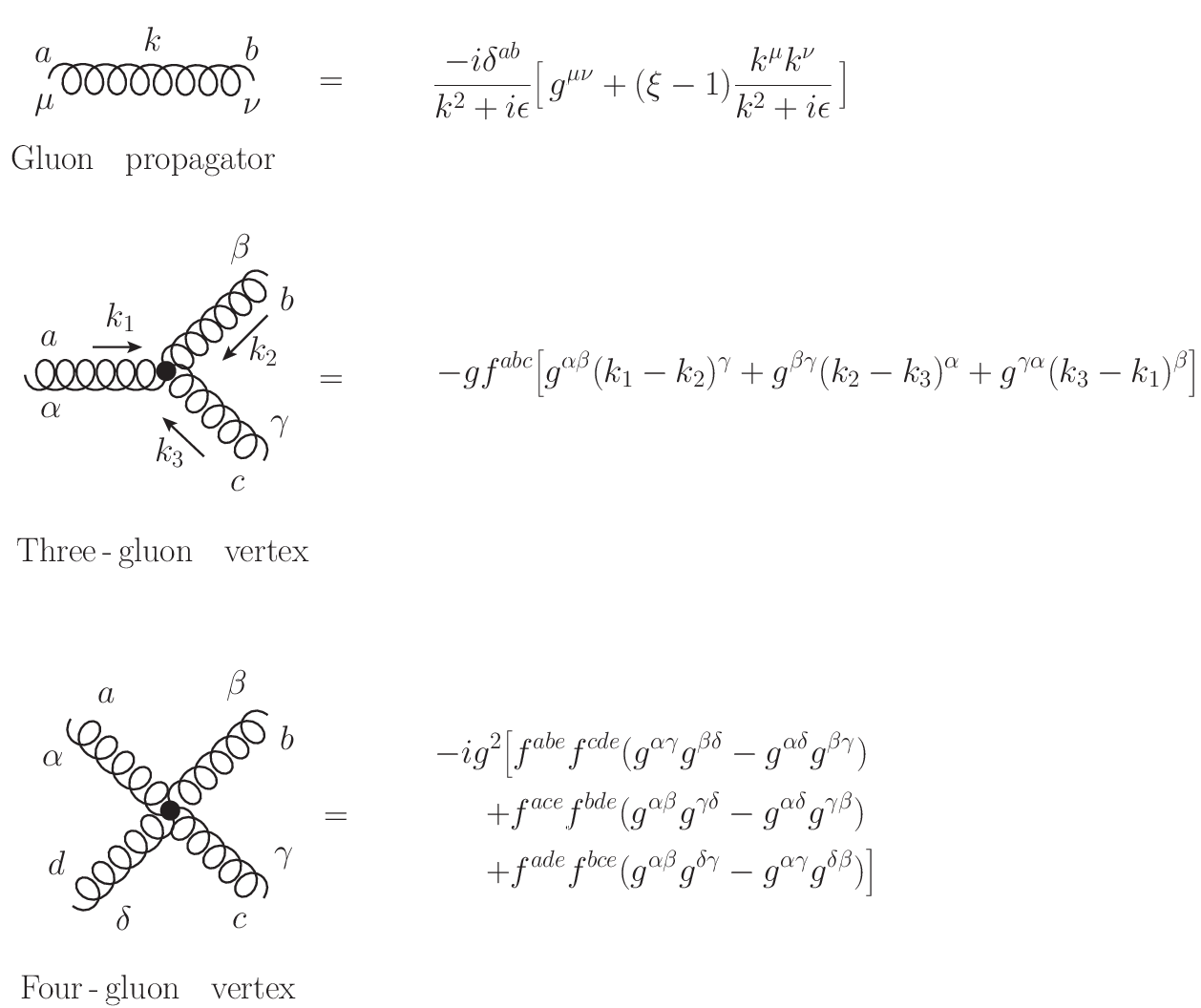}

    \vspace{1cm}
    
   \hspace*{0.2cm} \includegraphics[width=7cm]{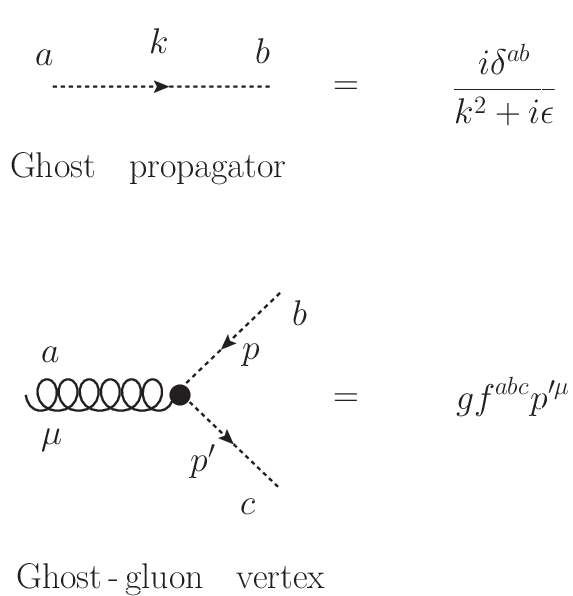}
    \caption{\small The momentum space Feynman rules derived using Eq.~\eqref{eq:gen_YM_covFP4}.
}
    \label{fig:FR_AG}
\end{figure}

The perturbative approach for computing scattering amplitudes ${A}_n \left( \left\{ p_i, h_i, a_i\right\}\right)$ analytically, discussed in the previous section, can be systematically reproduced using the well known \textit{Feyman diagram} technique. To do this, one first starts by identifying the building blocks for developing amplitudes -- the so-called Feynman rules, see Figure \ref{fig:FR_AG}. Although we work with the light-cone gauge throughout the rest of the thesis in which we can ignore the ghosts, just for the sake of completeness we show the Feynman rules for the full Yang-Mills Eq.~\eqref{eq:gen_YM_covFP4}, including the gauge fixing term (which contributes to the gluon propagator) and the ghosts, in Figure \ref{fig:FR_AG}. These include 

\begin{itemize}
    \item  \textit{Propagators}: Inverse of the kinetic term in the action. There are two propagators in Eq.~\eqref{eq:gen_YM_covFP4}. First is the gluon propagator. It is the inverse of the kinetic term originating both from the $S_{\mathrm{YM}}[A]$ Eq.~\eqref{eq:YM_action_cov} plus the gauge fixing term $S_{\mathrm{GF}}[A]$ Eq.~\eqref{eq:gauge_fix}. It represents a gluon propagating in space-time without interacting and can be used to connect gauge fields defined at two different space-time points.  The second is the Ghost propagator. It is the inverse of the kinetic term in the ghost action $S_{\mathrm{Ghosts}}[A]$ Eq.~\eqref{eq:FPgho}. The momentum space expression for both these propagators is shown in Figure \ref{fig:FR_AG}.
    \item \textit{Interaction vertices}: The cubic and the quartic pure gluonic interaction vertices in the action $S_{\mathrm{YM}}[A]$ Eq.~\eqref{eq:YM_action_cov} plus the gluon-ghost interaction vertex in  $S_{\mathrm{Ghosts}}[A]$ Eq.~\eqref{eq:FPgho}. These represent local (at a given space-time point) interactions and the fields in these vertices can be connected with other fields either from the same vertex or other vertices via the propagators. The momentum space expressions for the these vertices are shown in Figure \ref{fig:FR_AG}.
\end{itemize}

Consider an $n$-gluon scattering process where each of these external gluons is characterized by three indices $\left\{ p_i, h_i, a_i \right\}$ which are held fixed during the computation of the amplitude ${A}_n \left( \left\{ p_i, h_i, a_i\right\}\right)$. Draw all possible connected Feynman diagrams using the building blocks mentioned above such that each of the external legs belongs to an interaction vertex. This requirement follows from the LSZ reduction where one needs to amputate the external propagators. There can be infinitely many Feynman diagrams contributing to a given process. All of these diagrams can be segregated into two classes: \textit{tree} and \textit{loop} diagrams. In the former, each interaction vertex is connected to any other vertex via a single propagator representing an internal gluon that is not on-shell. On the other hand, for loop diagrams, the vertices may be connected through more than one propagator. This forms the loop. Such diagrams can be further divided based on the number of loops. By doing this we get
\begin{equation}
    {A}_n \left( \left\{ p_i, h_i, a_i\right\}\right) =\sum_{L=0} {A}_n^L \left( \left\{ p_i, h_i, a_i\right\}\right)\,,
    \label{eq:pert_amp_loop}
\end{equation}
where ${A}_n^L \left( \left\{ p_i, h_i, a_i\right\}\right)$ represents the sum of all the Feynman diagrams with $L$ loops and therefore  corresponds to $L$-loop $n$-gluon amplitude ($L=0 \longrightarrow \mathrm{tree}\,\mathrm{amplitude}$). The expression Eq.~\eqref{eq:pert_amp_loop} represents a perturbative expansion of an $n$-gluon amplitude in terms of the number of loops (and also in terms of the power of coupling constant $g$: ${A}_n^{L =0} \left( \left\{ p_i, h_i, a_i\right\}\right) \sim g^{n-2}$, ${A}_n^{L =1} \left( \left\{ p_i, h_i, a_i\right\}\right) \sim g^{n}$ and so on) where the accuracy of the computed amplitude depends on the number of the terms ${A}_n^L \left( \left\{ p_i, h_i, a_i\right\}\right)$ computed. Each term on the R.H.S. of the series Eq.~\eqref{eq:pert_amp_loop} can, in principle (hopefully), be computed explicitly by converting each of the diagrams contributing to it into an analytic expression and then summing over all the contributions. This is achieved by replacing the propagators and interaction vertices in the diagram with corresponding analytic expressions derived from Eq.~\eqref{eq:gen_YM_covFP4}, shown in Figure \ref{fig:FR_AG}. Furthermore, owing to the integration over the vertex positions in each diagram, in momentum space we get momentum conserving delta and integration over all of the loop momenta (if there are loops in the diagram). 

From the above discussion, it appears that the Feynman diagram technique for computing amplitudes is a systematic approach and all one needs to do is to compute all the diagrams for a given loop-order to obtain the corresponding amplitude ${A}_n^L \left( \left\{ p_i, h_i, a_i\right\}\right)$. Although true, the caveat is that the number of diagrams explodes already at the tree level as the number of external gluons (multiplicity) increases. Consider $ g + g \longrightarrow n\,g$ scattering process. The number of tree diagrams for this process are given below in Table \ref{Tab:diagmult}\cite{Mangano}.
 \begin{table}[tbh]
 \centering
    \begin{tabular}{|c||r|r|r|r|r|r|r|} \hline
       $ n $  &     2   &  3  &  4  &  5  &    6   &  7     &    8  
\\  \hline                                              
$ \# $ of diagrams    
	&   4	& 25 &  220 & 2485 & 34300 & 559405 & 10525900 
\\ \hline                       
	\end{tabular}
 \caption{\small Number of tree-level Feynman diagrams for a pure gluonic $ g + g \longrightarrow n\,g$ scattering process.}
 \label{Tab:diagmult}
\end{table}  

The final result ${A}_n^{L=0} \left( \left\{ p_i, h_i, a_i\right\}\right)$, however, for quite a few cases turns out to be very simple, in fact, a single rational term as we will see below. This simplicity does not seem to be manifest in the above approach and has led to the development of a lot of techniques for computing pure-gluonic amplitudes, most of which are on-shell techniques evading the need for the Yang-Mills action Eq.~\eqref{eq:YM_action_cov} completely. Our focus, on the other hand, will be on a field theory approach where we derive an equivalent action that allows for efficient computation of these amplitudes. The motivation stems from the realization that the number of diagrams shown in Table \ref{Tab:diagmult} are large also due to the fact that the building blocks used for obtaining the Feynman diagrams are very "small". One could instead use a different action consisting of a new set of "bigger" building blocks. The Cachazo-Svrcek-Witten (CSW) method \cite{Cachazo2004} is an explicit example of this. Since this method is also the starting point of this thesis, it is worth recalling the details but before we do that let us introduce the last set of two tools (\textit{Color Decomposition} and \textit{Helicity Spinors}) that allow for further simplification in the computation of amplitudes and also provide appropriate variables to express them compactly.
\subsection{Color Decomposition}
\label{subsec:Col_dec}

In Color Decomposition, the central idea \cite{CO1,CO2} is that the entire color dependence of a pure gluonic  amplitude can be disentangled and factored out. To see this, for simplicity, consider  $n$-point tree-level amplitude ${A}_n^{L=0} \left( \left\{ p_i, h_i, a_i\right\}\right)$. From the Feynman diagram technique, we know that this amplitude consists of a set of tree diagrams each of which is made up of propagators and interaction vertices from the Yang-Mills action Eq.~\eqref{eq:YM_action_cov}, shown in Figure \ref{fig:FR_AG}. The entire color dependence of the cubic and the quartic interaction vertices is encapsulated in a single structure constant $f^{b_1 b_2b_3}$ and a contracted product of structure constants $f^{b_1 b_2 a} f^{a b_3b_4}$ respectively (\emph{cf.} Figure \ref{fig:FR_AG}). As a result, the color dependence of a given Feynman diagram is just a contracted product of structure constants
\begin{equation}
    \sim  \dots f^{e\, b_{i+1}\, a} \, f^{a\, b_{i+2} \,c}\, f^{c\, b_{i+3} \, d}\dots
\end{equation}
Using the identity $i\sqrt{2}\,f^{b_1 b_2b_3}=\mathrm{Tr}( t^{b_1} t^{b_2} t^{b_3})-\mathrm{Tr}(t^{b_1} t^{b_3} t^{b_2})$ (it follows from the two expressions in Eq.~\eqref{eq:su3_sc_id}), one can trade each of the structure constants in terms of the trace of color generators. By doing this for all the structure constants, one obtains a sum of terms where each term consists of the product of traces of the type
\begin{equation}
    \sim  \dots  \mathrm{Tr}( t^{a} t^{b_i} \dots t^{b_j} t^{c})\mathrm{Tr}(t^{c} t^{b_k} \dots t^{b_l}t^{d})\dots
\end{equation}
which can be combined into a single trace $\sim  \dots  \mathrm{Tr}( t^{a} t^{b_i} \dots t^{b_j} t^{b_k} \dots t^{b_l}t^{d}) \dots$ plus sub-leading contributions using the $\mathrm{SU(N)}$ \textit{Fierz} identity
\begin{equation}
    \left(t^{a}\right)^{j_1}_{i_1}\left(t^{a}\right)^{j_2}_{i_2} = \delta^{j_2}_{i_1}\delta^{j_1}_{i_2} -\frac{1}{\mathrm{N}}\delta^{j_1}_{i_1}\delta^{j_2}_{i_2}\,.
\end{equation}
The sub-leading contributions originating from the $1/\mathrm{N}$ term vanishes owing to the photon decoupling theorem. This term is required only when quarks are present either externally or in the loop.  Due to this, for the current discussion, the product of traces reduces to a single trace.

All the trace products can be, similarly, combined into a single trace. Repeating this procedure for all the diagrams, the tree-level amplitude ${A}_n^{L=0} \left( \left\{ p_i, h_i, a_i\right\}\right)$ can be expressed as
\begin{equation}
    {A}_n^{L=0} \left( \left\{ p_i, h_i, a_i\right\}\right) =
    \!\!\sum_{\sigma \in S_n/Z_n}
 \mathrm{Tr}\left(t^{b_{\sigma_1}}\dots t^{b_{\sigma_n}}\right)
 \mathcal{A}_n^{L=0} \left( \left\{ p_{\sigma_1}, h_{\sigma_1}\right\},\dots ,\left\{ p_{\sigma_n}, h_{\sigma_n}\right\}\right)\,,
\end{equation}
where $S_n$ represents the set of all permutations of $n$ objects and $Z_n$ represents the subset of all the cyclic permutations. The latter preserves the trace. Thus $S_n/Z_n$ represents the set of all non-cyclic permutations of $n$ objects. Therefore, the sum is over distinct (cyclically inequivalent) orderings of the $n$-gluons. In later chapters, instead of the above, we shall be using the following notation
\begin{equation}
    {A}_n^{L=0} \left( \left\{ p_i, h_i, a_i\right\}\right) =
    \!\!\sum_{\underset{\text{\scriptsize permutations}}{\text{noncyclic}}}
 \mathrm{Tr}\left(t^{b_1}\dots t^{b_n}\right)
 \mathcal{A}_n^{L=0} \left( \left\{ p_1, h_1\right\},\dots ,\left\{ p_n, h_n\right\}\right)\,,
\end{equation}
to convey the same idea. The above expression represents the color-decomposed form for the amplitude ${A}_n^{L=0} \left( \left\{ p_i, h_i, a_i\right\}\right)$ where, on the R.H.S., the entire color dependence of the amplitude factors out as a single trace of color generators. The object $\mathcal{A}_n^{L=0} \left( \left\{ p_i, h_i\right\}\right)$ is commonly referred to as the \textit{color-ordered} amplitude. It contains the kinematic information of the amplitude and is gauge invariant. Furthermore, it follows the same symmetries as the trace of the generators associated with it. Given that the trace is invariant under cyclic permutations of the generators, only the diagrams that are cyclically related to each other for a fixed ordering contribute to a given color-ordered amplitude.

The above procedure can also be extended to the loop amplitudes ${A}_n^{L} \left( \left\{ p_i, h_i,a_i\right\}\right)$. However, in that case, one also obtains multi-trace terms. For instance, the color decomposition of the one-loop amplitude ${A}_n^{L=1} \left( \left\{ p_i, h_i,a_i\right\}\right)$ reads
\begin{multline}
    {A}_n^{L=1} \left( \left\{ p_i, h_i, a_i\right\}\right) =
    \!\!\sum_{\underset{\text{\scriptsize permutations}}{\text{noncyclic}}} \Big[ \mathrm{N}\,
 \mathrm{Tr}\left(t^{b_1}\dots t^{b_n}\right)
 \mathcal{A}_{n; 1}^{L=1} \left( \left\{ p_1, h_1\right\},\dots ,\left\{ p_n, h_n\right\}\right) \\
 + \sum_{j=2}^{[n/2]+1}  \mathrm{Tr}\left(t^{b_1}\dots t^{b_j-1}\right) \mathrm{Tr}\left(t^{b_j}\dots t^{b_n}\right)\mathcal{A}_{n; j}^{L=1} \left( \left\{ p_1, h_1\right\},\dots ,\left\{ p_n, h_n\right\}\right)\Big]\,,
\end{multline}
where $[j]=$ greatest integer of $j$. Above, $\mathcal{A}_{n; 1}^{L=1} \left( \left\{ p_1, h_1\right\},\dots ,\left\{ p_n, h_n\right\}\right)$ is the single-trace color-ordered one-loop amplitude and $\mathcal{A}_{n; j>1}^{L=1} \left( \left\{ p_1, h_1\right\},\dots ,\left\{ p_n, h_n\right\}\right)$ is the double-trace contribution. The latter can also be constructed via permutations of the former $\mathcal{A}_{n; 1}^{L=1}$  \cite{CO3,CO4} and therefore these are not independent. Consequently, one can simply focus on deriving the single-trace amplitudes. This is what we will do by imposing the 't Hooft large $\mathrm{N}$ limit ($\mathrm{N}\longrightarrow \infty$) \cite{tHooft:1973alw}. In this limit the single-trace amplitudes dominate. 

One of the major advantages of using color decomposition is that when computing a color-ordered amplitude, the only contributions originate from planar Feynman diagrams (no non-planar) in which the external gluons follow the chosen color ordering. One can also derive the so-called color-ordered Feynman rules which could be used to compute the color-ordered amplitudes. We will, however, avoid doing it here.

\subsection{Spinor Helicity formalism}
\label{subsec:spi_hel}

Color decomposition reduces the problem of computing a full-color amplitude to a color-ordered amplitude, where the latter is a function of the two on-shell degrees of freedom: momentum, and polarization, of the external massless gluons. The Spinor Helicity formalism provides a systematic way of uniformly expressing these two quantities, for each gluon, in terms of two-component objects known as the Helicity Spinors. 

Consider first the four momentum $p_i$ of the $i^{th}$ external gluon. This four-vector can be mapped to a $2\times2$ matrix using the Pauli matrices as shown below
 \begin{equation}
   p_i^\mu \longrightarrow p_i^\mu (\sigma_\mu)^{\dot\alpha \alpha} = p_i^{\dot\alpha \alpha}= 
\left(\begin{matrix} p_i^0 + p_i^3 & p_i^1 - i p_i^2 \\
               p_i^1 + i p_i^2 & p_i^0 - p_i^3 \\
\end{matrix}\right)\, ,
\label{eq:bi-spin_mom}
\end{equation}
where $(\sigma_\mu)^{\dot\alpha \alpha} = (\mathbb{1}, \sigma)$. Since the external gluons are on-shell $p_i^\mu p_{i \mu} =0$, the determinant of the $2\times2$ matrix on the R.H.S. of the expression above is zero; $\det (p_i^{\dot\alpha \alpha}) = 0$. This implies that the rank of this matrix is one and it can, therefore, be factorized into a product of 2 two-component objects
\begin{equation}
 p_i^{\dot\alpha \alpha}  =   (\widetilde\lambda_i)^{\dot\alpha}(\lambda_i)^\alpha\, \quad \quad \mathrm{where} \,\,\, \dot\alpha\,, \alpha = 1,2\,.
\end{equation}
These objects $\left\{(\lambda_i)^\alpha,  (\widetilde\lambda_i)^{\dot\alpha} \right\}$, given the explicit form of the matrix in Eq.~\eqref{eq:bi-spin_mom}, are the solutions of the left-handed and the right-handed Weyl equations. Therefore, they have left-handed and right-handed helicity and are thus commonly referred to as helicity spinors. Furthermore, they are related as $[(\lambda_i)^\alpha]^* = \pm(\widetilde\lambda_i)^{\dot\alpha}$, via the complex conjugation $*$, if the components of the four momentum $p_i^\mu$ are real. However, if the components are complex, which is what we will have, they are independent. 

In order for the helicity spinors to be useful, it should be possible to translate not just the four-vector $p_i^\mu$ but also the other useful operations associated with them like the lowering and raising of indices and defining the dot product. The spinor indices $\left\{\alpha, \dot\alpha\right\}$ can be lowered using the $\mathrm{SL}(2,\mathbb{C})$ invariant Levi Civita tensors $\left\{ \epsilon_{\alpha \beta}, \epsilon_{\dot\alpha\dot\beta}\right\}$ respectively, as shown below
\begin{equation}
   \lambda_{i \alpha} =  \epsilon_{\alpha \beta} \lambda_i^\beta\,,\quad \quad  \widetilde\lambda_{i \dot\alpha} = \epsilon_{\dot\alpha\dot\beta} \widetilde\lambda_i^{\dot\beta}\,,\quad \mathrm{where} \quad \epsilon_{\alpha \beta} = \epsilon_{\dot\alpha\dot\beta} = \left(\begin{matrix}
       0 & -1 \\
               1 & 0 \\
   \end{matrix}\right)\,,
\end{equation}
and these could be raised using the inverse of $\left\{ \epsilon_{\alpha \beta}, \epsilon_{\dot\alpha\dot\beta}\right\}$ as shown below
\begin{equation}
   \lambda_{i}^{\alpha} =  \epsilon^{\alpha \beta} \lambda_{i \beta}\,,\quad \quad  \widetilde\lambda_{i}^{\dot\alpha} = \epsilon^{\dot\alpha\dot\beta} \widetilde\lambda_{i \dot\beta}\,,\quad  \mathrm{where} \quad \epsilon^{\alpha \beta} = \epsilon^{\dot\alpha\dot\beta} = \left(\begin{matrix}
       0 & 1 \\
               -1 & 0 \\
   \end{matrix}\right)\,.
\end{equation}
Using the Levi-Civita tensors, one can define the Lorentz invariant spinor products as follows
\begin{equation}
  \langle \lambda_i \lambda_j \rangle  = 
\epsilon_{\alpha\beta}\, (\lambda_i)^\alpha \,(\lambda_j)^\beta \equiv \langle{i}{j}\rangle \, ,\quad \quad
 [\widetilde{\lambda}_i \widetilde{\lambda}_j] =-
\epsilon_{\dot\alpha\dot\beta}\, (\widetilde\lambda_i)^{\dot\alpha} \,(\widetilde\lambda_j)^{\dot\beta} \equiv \left[i j \right] \,
\, .
\label{eq:spi_prod_def}
\end{equation}
Above, $\langle{i}{j}\rangle$ and $\left[i j \right]$ represent the commonly used compact notation for these products. From above, it follows that $\langle{i}{j}\rangle = - \langle{j}{i}\rangle$ and $\left[i j \right] = - \left[j i \right]$ implying that $\langle{i}{i}\rangle = \left[i i \right] = 0$.

Just like the four momentum $p_i^\mu$, the polarization vector $\varepsilon_{p_i}^{\pm}$ of the $i^{th}$ external gluon can also be expressed in terms of these helicity spinors. In fact, the origin of the spinor helicity techniques is associated with the work \cite{SPH1, SPH2, SPH3, SPH4, SPH5} where the authors showed that the polarization of a massless vector particle can be expressed using spinors as follows
\begin{equation}
 \varepsilon_{+ i}^{\alpha\dot\alpha}  = -\sqrt{2}\, \frac{(\widetilde\lambda_i)^{\dot\alpha}(\chi_i)^\alpha }{\langle \lambda_i \chi_i  \rangle } \, ,\quad \quad \varepsilon_{- i}^{\alpha\dot\alpha}  =\sqrt{2} \,\frac{(\lambda_i)^\alpha (\widetilde\chi_i)^{\dot\alpha}}{[\widetilde{\lambda}_i \widetilde{\chi}_i ]}\,,
 \label{eq:pol_spinor}
\end{equation}
where $\varepsilon_{+ i}^{\alpha\dot\alpha}$ and $\varepsilon_{- i}^{\alpha\dot\alpha}$ are the $2\times 2$ matrix representation (in terms of the spinor indices) of the polarization of the $i^{th}$ gluon of helicity $+1$ and $-1$ respectively. The quantities $\langle \cdot \rangle$ and $[\cdot]$ in the denominator are the spinor products Eq.~\eqref{eq:spi_prod_def}. The spinors $(\chi_i)^\alpha$ and $(\widetilde\chi_i)^{\dot\alpha}$ are reference spinors and are associated with a reference momentum $q_i^\mu \longrightarrow q_i^{\alpha\dot\alpha}  =  (\chi_i)^\alpha (\widetilde\chi_i)^{\dot\alpha}$, which could be chosen independently for each external gluon. This freedom is an outcome of the local gauge invariance and the resulting Ward identity. To see this, consider the effect on $\varepsilon_{+}^{\alpha \dot{\alpha}}$ of an infinitesimal change in the reference spinors $\chi^\alpha \longrightarrow \chi^\alpha + \delta \chi^\alpha$
\begin{align}
\delta \varepsilon_{+}^{\alpha \dot{\alpha}} & =-\sqrt{2}\left(\frac{\widetilde{\lambda}^{\dot{\alpha}} \delta \chi^\alpha}{\langle\lambda \chi\rangle}-\widetilde{\lambda}^{\dot{\alpha}} \chi^\alpha \frac{\langle\lambda \delta \chi\rangle}{\langle\lambda \chi\rangle^2}\right)\, \nonumber\\
& =-\sqrt{2} \frac{\widetilde{\lambda}^{\dot{\alpha}}}{\langle\lambda \chi\rangle^2} \left(\delta \chi^\alpha\langle\lambda \chi\rangle-\chi^{\dot{\alpha}}\langle\lambda \delta \chi\rangle\right)\, \nonumber\\
& =p^{\alpha \dot{\alpha}}\left(\sqrt{2} \frac{\langle\delta \chi \chi\rangle}{\langle\lambda \chi\rangle^2}\right)\,.
\label{eq:var_POLA}
\end{align}
Above, in going from the second line to the third, we used the Schouten identity, see below Eq.~\eqref{eq:Sch_id}. Recall, in order to compute amplitude, we contract the amputated connected Greens function with the polarizations of the external gluons. Due to the result Eq.~\eqref{eq:var_POLA}, the variation in the amplitude due to the change of reference spinor in the polarization is proportional to the momentum of the given gluon contracted with the amputated connected Green's function which owing to the Ward identity is zero in the on-shell limit. 

Finally, from Eq.~\eqref{eq:pol_spinor} we see that the polarization is transverse both to the external momenta $(p_i \cdot \varepsilon_{p_i}^{\pm} \sim \langle{\lambda} \lambda \rangle\, \mathrm{or}\, \left[{\lambda} \lambda \right] =0)$ and the reference momentum $(q_i \cdot \varepsilon_{p_i}^{\pm} \sim \langle{\chi} \chi \rangle\, \mathrm{or}\, \left[{\chi} \chi \right] =0)$.

After switching over to the spinor variables, the color-ordered amplitudes are now functions of these, and the helicities associated with each external gluon $\mathcal{A}_n^{L} \left( \left\{ (\lambda_i)^\alpha,  (\widetilde\lambda_i)^{\dot\alpha}, h_i\right\}\right)$. In general, an amplitude consists of some incoming and outgoing states. Since helicity is defined via the projection of spin on the 3-momentum of the particle, exchanging the incoming states with the outgoing or vice versa may result in a switch of helicity. Therefore, in order to have a uniform description of the amplitudes with respect to the crossing symmetry, it is preferable that  all the particles are considered either incoming or outgoing. From now on we will stick to the latter and associate helicities accordingly. Finally, using momentum conservation in this convention, one obtains the following identity
\begin{equation}
    \sum_{i=1}^n p_i^{\mu} = \sum_{i=1}^n (\widetilde\lambda_i)^{\dot\alpha}(\lambda_i)^\alpha\, =0\,, \quad \implies \quad \sum_{i=1}^n \langle{k}{i}\rangle \left[i l \right] = 0\,.
    \label{eq:spi_MC}
\end{equation}
Identities like the above are very useful when computing amplitudes explicitly. Below we list two more identities for the spinor products
\begin{itemize}
    \item Mandelstam invariants $s_{ij}$
    \begin{equation}
        s_{ij} = (p_i+p_j)^2 = 2 p_i \cdot p_j = \langle{i}{j}\rangle \left[j i \right]\,.
    \end{equation}
    \item Schouten identity 
    \begin{equation}
        \langle{i}{j}\rangle \langle{k}{l}\rangle - \langle{i}{k}\rangle \langle{j}{l}\rangle = \langle{i}{l}\rangle \langle{k}{j}\rangle \,.
        \label{eq:Sch_id}
    \end{equation}
    A similar identity exists for the square brackets as well.
\end{itemize}
\subsection{Helicity tree amplitudes}
\label{sub:Hel_TA}

Now, we are ready to discuss the well known results for some amplitudes. Recall from Table~\ref{Tab:diagmult} that the number of Feynman diagrams required for computing tree-level amplitudes was growing factorially with the number of external legs. Using the above tools, Tomasz Taylor and Stephen Parke in \cite{Parke:1986gb} reported that the tree-level amplitudes where either all the external gluons have the same helicity or if one of them has a different helicity vanish. That is
\begin{equation}
    \mathcal{A}_n^{L=0} \left( 1^+,2^+,\dots,n^+\right) = 0 \,,\quad \mathcal{A}_n^{L=0} \left( 1^+,\dots,{(j-1)}^+,j^-,{(j+1)}^+,\dots,n^+\right) = 0\,,
    \label{eq:plus_tree_amp}
\end{equation}
\begin{equation}
    \mathcal{A}_n^{L=0} \left( 1^-,2^-,\dots,n^-\right) = 0 \,,\quad \mathcal{A}_n^{L=0} \left( 1^-,\dots,{(j-1)}^-,j^+,{(j+1)}^-,\dots,n^-\right) = 0\,,
    \label{eq:minus_tree_amp}
\end{equation}
where we use $i^{\pm}= \left\{ (\lambda_i)^\alpha,  (\widetilde\lambda_i)^{\dot\alpha}, h_i=\pm\right\}$. The first non-zero result is when two gluons have different helicity as compared to the rest. Furthermore, these have a single-term expression, in terms of the spinor products, irrespective of the number of external gluons as shown below
\begin{equation}
    \mathcal{A}_n^{L=0} \left(\dots,j^-,\dots,l^-,\dots \right) = i (-g)^{n-2}\delta^4(p_1+\dots+p_n)\frac{ {\langle jl\rangle}^4 }{ \langle 12\rangle \langle 23\rangle \cdots \langle n1\rangle }\,.
    \label{eq:MHV_ampl}
\end{equation}
Above the dots represent the plus helicity legs. These are known as the \textit{maximally helicity violating} (MHV) amplitudes \footnote{Conservation of helicity states that the sum of helicities of all the external legs should be zero. Thus this conservation is maximally violated when all the gluons have plus helicity. But these amplitudes are zero at the tree level Eq.~\eqref{eq:plus_tree_amp}-\eqref{eq:minus_tree_amp}. So the next case where it is maximally violated is when one of the gluons has minus helicity. These too vanish Eq.~\eqref{eq:plus_tree_amp}-\eqref{eq:minus_tree_amp}. Thus the next candidate is where two gluons have negative helicities as compared to the rest. This amplitude is non-zero Eq.~\eqref{eq:MHV_ampl} hence the name "maximally helicity violating".}. Notice, it depends only on one type of spinor product $\langle \cdot \rangle$ and is therefore "holomorphic". The parity conjugate of this amplitude, obtained by flipping the helicities of each leg, reads 
\begin{equation}
    \mathcal{A}_n^{L=0} \left(\dots,j^+,\dots,l^+,\dots \right) = i (g)^{n-2} \delta^4(p_1+\dots+p_n)\frac{\left[j l \right]^4}{\left[12 \right] \left[23 \right]\cdots \left[n 1 \right]}\,.
    \label{eq:MHVbar_ampl}
\end{equation}
These are the so-called $\overline{\mathrm{MHV}}$ amplitudes. The other tree-level amplitudes, in which the number of negative helicity legs $n_{-}$ varies from $3\leq n_{-} \leq n-3$, do not have a single term expression. In fact, the $(- - - + \dots +)$ tree-level amplitudes, also known as the Next-to-MHV (NMHV), are the next simplest, and then the Next-to-Next-to-MHV (NNMHV) $(- - - - + \dots +)$ and so on. 

The simplicity of the final results Eq.~\eqref{eq:MHV_ampl} is oblivious in the perturbative approach, discussed earlier, involving the cubic and the quartic interaction vertices in the Yang-Mills action Eq.~\eqref{eq:YM_action_cov}. Furthermore, since the number of Feynman diagrams grows factorially, computing these via brute force is also out of question. In the next section, we will discuss a new perturbative approach that uses the MHV amplitudes Eq.~\eqref{eq:MHV_ampl} as interaction vertices to efficiently compute any tree-level pure gluonic amplitudes.

\section{MHV rules} 
\label{sec:MHV_rules}

In \cite{Cachazo2004}, Cachazo, Svrcek, and Witten (CSW) developed a new perturbative approach for computing tree-level pure gluonic amplitudes. The central idea is to use the MHV amplitudes $({A}_n^{\mathrm{MHV}}; n\geq 3)$ Eq.~\eqref{eq:MHV_ampl} as interaction vertices instead of the cubic and the quartic vertices of the Yang-Mills action Eq.~\eqref{eq:YM_action_cov}. In fact, these are the only interaction vertices in this approach. No other amplitudes, not even the $\overline{\mathrm{MHV}}$ amplitudes Eq.~\eqref{eq:MHVbar_ampl} are required. They are rather computed using the MHV vertices. We elaborate on the motivation for this new approach a bit later towards the end of this section.

Recall, all the external legs in the MHV amplitudes Eq.~\eqref{eq:MHV_ampl} are on-shell $p_i^2=0$. In order to use them as interaction vertices, one requires a prescription to perform an off-shell continuation of these because the internal gluons that connect these vertices are off-shell. In \cite{Cachazo2004}, CSW offered the following prescription. Consider a leg in the  MHV amplitude Eq.~\eqref{eq:MHV_ampl} with momentum $P$. The spinor $(\lambda_P)^\alpha$ associated with this leg is re-defined as
\begin{equation}
    (\lambda_P)^\alpha = \eta_{\dot\alpha} P^{\dot\alpha \alpha}\,,
    \label{eq:CSW_offshell}
\end{equation}
where $\eta_{\dot\alpha}$ is a reference spinor and can be chosen arbitrarily. However, once chosen, it should be the same for the off-shell continuation of all the legs in all the Feynman diagrams. Substituting Eq.~\eqref{eq:CSW_offshell} back to the MHV amplitude Eq.~\eqref{eq:MHV_ampl} results in the MHV vertex. These vertices can then be connected via a scalar propagator $\sim 1/P^2$. The propagator connects legs of opposite helicities $(+ -)$. Although the analytical expression for each diagram depends on the reference spinor $\eta_{\dot\alpha}$, the final results are independent of it (we show this in the example considered below). This was demonstrated explicitly through amplitude computation in \cite{Cachazo2004}.

\begin{figure}
    \centering
    \includegraphics[width=15.9cm]{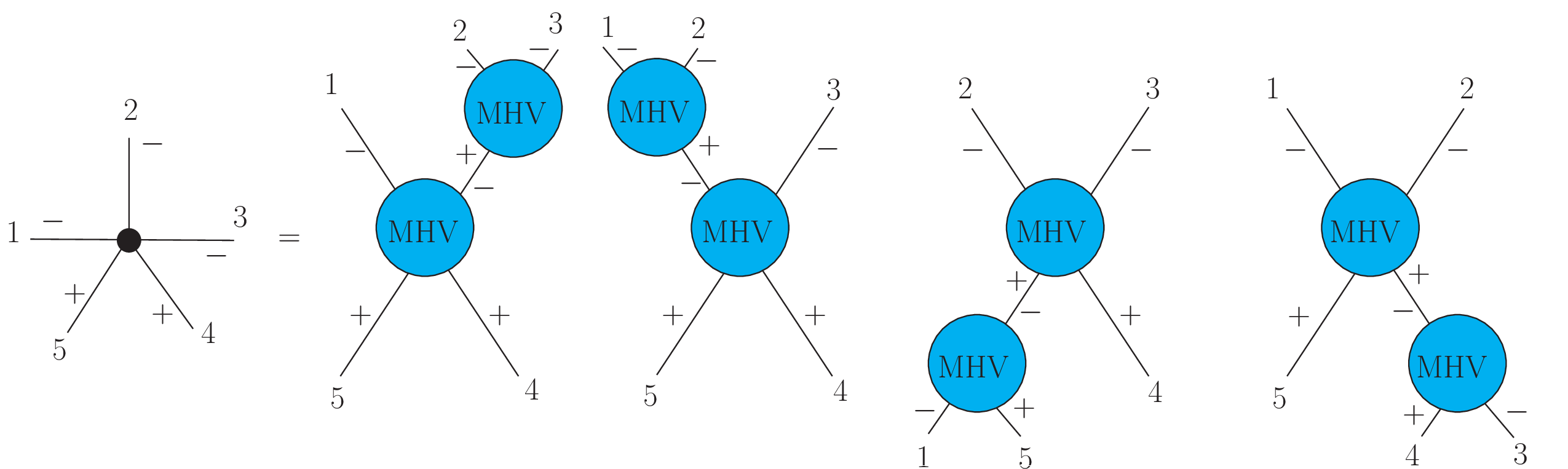}
    \caption{\small 
    LHS: 5-point $\overline{\text{MHV}}$ $(---++)$ amplitude. RHS: The contributions to the color-ordered amplitude in the MHV rules.
    }
    \label{fig:5mhvbar}
\end{figure}

Using these new building blocks, one can compute any tree-level pure gluonic amplitude. All one has to do is draw all possible Feynman diagrams contributing to a given tree-level color-ordered amplitude, where each diagram consists of the MHV vertices, defined above, connected via a scalar propagator. Consider, for instance, 5-point $\overline{\mathrm{MHV}}$ $(- - - + +)$ amplitude. The contributing diagrams are shown in Figure \ref{fig:5mhvbar}.
Note, there are just 4 contributions. We assign $\{b_i, p_i \}$ as the color and the momentum, respectively, to the external legs $(1^-,2^-,3^-,4^+,5^+)$ of each diagram in a clockwise fashion. For the sake of simplicity, let us focus on the color-ordered amplitude $\mathcal{A}(1^-,2^-,3^-,4^+,5^+)$ with $\Tr(t^{b_1} t^{b_2} t^{b_3} t^{b_4} t^{b_5})$ color ordering. The analytic expression for the first contribution on the R.H.S. reads
\begin{align}
 \mathcal{D}_1(1^-,2^-,3^-,4^+,5^+)= &g^3\frac{ {\langle P 1\rangle}^4 }{ \langle 15\rangle \langle 54\rangle \langle 4 P\rangle \langle P1\rangle } \frac{1}{P^2}\frac{ {\langle 32\rangle}^4 }{ \langle 2P\rangle \langle P3\rangle \langle 32\rangle }\,, \nonumber\\
    =&g^3\frac{ {(\phi_2\langle 2 1\rangle + \phi_3\langle 3 1\rangle)^3} }{ \langle 15\rangle \langle 54\rangle (\phi_2\langle 4 2\rangle+ \phi_3\langle 4 3\rangle ) }\frac{ -1 }{[32]\phi_3 \phi_2  }\,,
    \label{eq:u1_MHV}
\end{align}
where for each of the two MHV vertices in the diagram we used Eq.~\eqref{eq:MHV_ampl}. In the first line above, $P$ represents the momentum of the off-shell leg joining the two MHV vertices. For this leg, we use the CSW off-shell prescription Eq.~\eqref{eq:CSW_offshell}
\begin{equation}
    (\lambda_P)^\alpha = \eta_{\dot\alpha} P^{\dot\alpha \alpha}\,,
    \label{eq:csw_cal}
\end{equation}
where $\eta_{\dot\alpha}$ is the reference spinor. Owing to momentum conservation for each vertex, we have $P=p_2+p_3$. Substituting this to Eq.~\eqref{eq:csw_cal} (and using $ p_i^{\dot\alpha \alpha}  =   (\widetilde\lambda_i)^{\dot\alpha}(\lambda_i)^\alpha$) we get
\begin{equation}
    (\lambda_P)^\alpha = \phi_2 (\lambda_2)^\alpha + \phi_3 (\lambda_3)^\alpha\, \quad\quad \mathrm{where} \quad \phi_i=\eta_{\dot\alpha}(\widetilde\lambda_i)^{\dot\alpha}\,.
    \label{eq:phi_def}
\end{equation}
We used the above and the identity $P^2=(p_2+p_3)^2 = \langle 23\rangle[32]$ to go from the expression in the first line of Eq.~\eqref{eq:u1_MHV} to the second line. Notice, the power of $\phi$ in the numerator is exactly the same as the denominator. Both are $\sim \phi^3$. From Eq.~\eqref{eq:phi_def}, we see that $\phi \propto \eta$ (suppressing the indices). Thus the $\eta$ dependence cancels out of this contribution.

In a similar fashion, the color-ordered expression for the remaining three contributions can be computed. These read
\begin{equation}
    \mathcal{D}_2(1^-,2^-,3^-,4^+,5^+)= g^3 \frac{-1}{\phi_1 \phi_2 [21]}\frac{(\phi_2\langle 3 2\rangle + \phi_1\langle 3 1\rangle)^3}{(\phi_1\langle 1 5\rangle + \phi_2\langle 2 5\rangle) \langle 5 4\rangle \langle 4 3\rangle}\,, 
\end{equation}
\begin{equation}
    \mathcal{D}_3(1^-,2^-,3^-,4^+,5^+)= g^3 \frac{\phi_5^3}{\phi_1 [51]}\frac{\langle 3 2 \rangle^3}{(\phi_1\langle 2 1 \rangle + \phi_5\langle 2 5\rangle)(\phi_1\langle 1 4 \rangle + \phi_5\langle 5 4 \rangle)\langle 4 3\rangle}\,, 
\end{equation}    
\begin{equation}
    \mathcal{D}_4(1^-,2^-,3^-,4^+,5^+)= g^3 \frac{\phi_4^3}{\phi_3 [34]}\frac{\langle  21 \rangle^3}{(\phi_4\langle 54 \rangle + \phi_3\langle 53 \rangle)(\phi_4\langle  4 2\rangle + \phi_3\langle 32 \rangle)\langle 15\rangle}\,.
\end{equation}
Again notice, the $\eta$ dependence cancels out of each of the above contributions. After a bit of tedious algebra, the sum of these reproduces the result Eq.~\eqref{eq:MHVbar_ampl} \cite{Cachazo2004}. This example clearly establishes the fact that the complexity of the Feynman diagram technique for computing amplitudes perturbatively greatly depends on the building blocks being used for computing them. Making a wise choice can provide remarkable simplicity. In the original work \cite{Cachazo2004}, the authors also computed six and seven-point amplitudes. In each case, they just got a handful of diagrams, and the computed amplitudes agreed with the known results.  

Notice, finally, that each diagram contributing to $(- - - + +)$ amplitude in Figure \ref{fig:5mhvbar} consists of precisely two MHV vertices. This observation generalizes to higher-point amplitudes as follows. Consider a diagram consisting of $k$ MHV vertices connected to form a tree diagram via $k-1$ propagators $(+ -)$. The total negative helicity legs of such a diagram are $2k-(k-1)= k+1$. Therefore, an $n$-point amplitude with $n_{-}$ negative helicity legs, in this approach, can be computed from diagrams made up of  exactly $(n_{-} - 1)$ MHV vertices. This implies that amplitudes of the type $(+ + \dots +)$, $(- + \dots +)$ with $n_{-} = 0, 1$, respectively, cannot be computed using MHV vertices and are thus zero in this approach which is consistent with the results Eq.~\eqref{eq:plus_tree_amp}. 

\begin{figure}
    \centering
    \includegraphics[width=11cm]{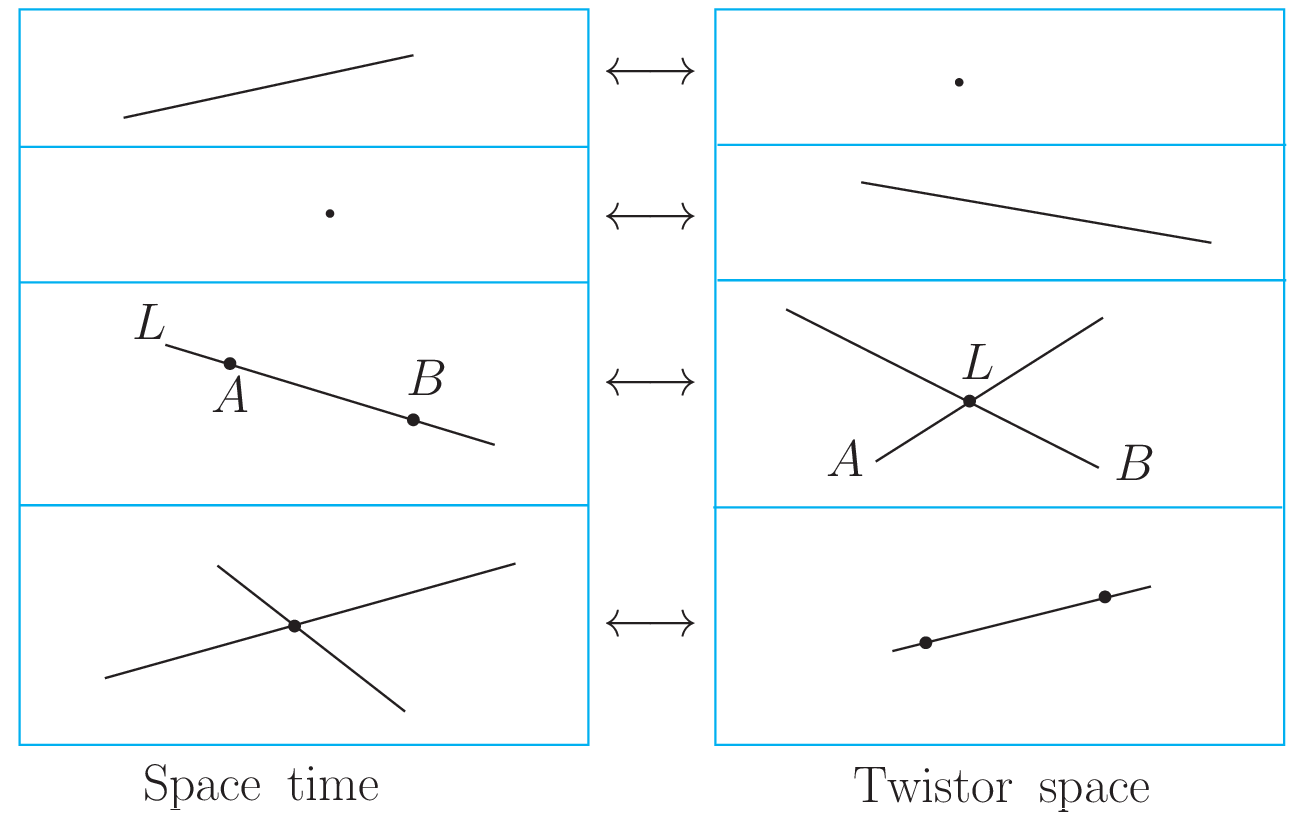}
    \caption{\small
    LHS: The real Minkowski space $\mathbb{M}$. RHS: The twistor space  $\mathbb{PN}$. $\longleftrightarrow$ represents the correspondence between the geometric objects in one space to those in the other.
Explicitly, the first correspondence states that a line in the real Minkowski space $\mathbb{M}$ corresponds to a point in the twistor space  $\mathbb{PN}$ and vice versa. Second states a point $\mathbb{M}$ corresponds to a line in $\mathbb{PN}$ and vice versa. Third states that a pair of co-linear points in $\mathbb{M}$ corresponds to a pair of intersecting lines in $\mathbb{PN}$ and vice versa. Finally, the fourth states that a pair of intersecting lines in $\mathbb{M}$ corresponds to a pair of co-linear points in $\mathbb{PN}$ and vice versa. }
    \label{fig:inc_rel}
\end{figure}

All of the above discussion simply reiterates how remarkable the CSW approach for computing tree-level pure gluonic scattering amplitudes is. But none of it points towards the origin of the MHV rules and to the fact of how amplitudes, a seemingly non-local object, can be used as a local interaction vertex. The origin of the MHV rules stems from the work of Witten \cite{Witten2004} where he demonstrated that the simplicity of the gauge theory amplitudes can be understood from the  geometric structures underlying these in the \textit{twistor space}.  In a nutshell, twistor space $\mathbb{PT}$ is a 3-dimensional complex projective space described using homogenous coordinates $Z^A$. These coordinates can be expressed in terms of a pair of left and right-handed Weyl spinors as $Z^A = (  \mu^{\dot\alpha}, \lambda_\alpha )$ and are identified up to a scaling $(  \mu^{\dot\alpha}, \lambda_\alpha ) \sim ( c \mu^{\dot\alpha},  c\lambda_\alpha )$ where $c \in \mathbb{C}^* = \mathbb{C}/\{0\}$ is a non-zero complex number.  The central role in all of these is played by the \textit{incidence relations} which define a mapping between the points $Z^A$ in twistor space $\mathbb{PT}$ to those in complexified Minkowski space $\mathbb{M}_\mathbb{C}$ \footnote{The complexification of the Minkowski space implies allowing for the co-ordinates $x^\mu$ to take complex values while keeping the metric $g^{\mu \nu}$ holomorphic. The latter simply means that there is no dependence of the metric on the complex conjugate ${\overline x}^\mu$. Owing to this complexification, there is no meaning of the signature of $\mathbb{M}_\mathbb{C}$. However, taking different real slices we retrieve back the real Minkowski space-time $\mathbb{M}$ $(+ - - -)$, the $\mathbb{R}^4$ with Euclidean signature $(+ + + +)$, and also the $\mathbb{R}^{2,2}$ with split signature $(+ + - - )$.}. They read
\begin{equation}
    \mu^{\dot\alpha}= x^{\dot\alpha \alpha} \lambda_\alpha\,,
    \label{eq:inc_rel}
\end{equation}
where $x^{\dot\alpha \alpha}$ is the representation ($2\times 2$ matrix representation in terms of spinor indices, to be precise) of a point in $\mathbb{M}_\mathbb{C}$. Using the above relation for a given point $x^{\dot\alpha \alpha}$ in $\mathbb{M}_\mathbb{C}$ we get a set of two equations that define a complex holomorphic (no dependence on the complex conjugate) line $X \cong \mathbb{CP}^1$ in the twistor space $\mathbb{PT}$. This implies that the mapping Eq.~\eqref{eq:inc_rel} is non-local since a point in complexified Minkowski space $\mathbb{M}_\mathbb{C}$ maps to an extended object (a line) in twistor space $\mathbb{PT}$. In a similar fashion, using the incidence relations, one can discover what geometric objects in one space map to in the other. In Figure \ref{fig:inc_rel}, we show this geometric correspondence between the real Minkowski space $\mathbb{M}$ and its corresponding twistor sub-space  $\mathbb{PN} \subset \mathbb{PT}$. The twistor space  $\mathbb{PN}$ is defined as
\begin{equation}
    \mathbb{PN}= \left\{ Z^A \in \mathbb{PT}| Z^A\cdot \overline{Z}^A = 0 \right\} \,,
    \label{eq:def_PN}
\end{equation}
where $\overline{Z}^A$ represents the twistor dual to ${Z}^A$ such that the complex conjugation of points in ${Z}^A = (\mu^{\dot\alpha}, \lambda_\alpha )$ correspond to the point $({\overline \lambda}_{\dot\alpha}, {\overline \mu}\,^{\alpha} ) \in \overline{Z}^A $. And, the inner product is defined as
\begin{equation}
    Z^A\cdot \overline{Z}^A = [\mu \overline{\lambda}] + \langle \overline{\mu} \lambda \rangle\,.
\end{equation}
The condition in Eq.~\eqref{eq:def_PN} defines a subset $\mathbb{PN} \subset \mathbb{PT}$ such that the points $x^{\dot\alpha \alpha}$ obtained via the incidence relations belong to the real Minkowski space $\mathbb{M}$ and vice versa.

Coming back to the amplitudes, Witten in \cite{Witten2004} discovered that the tree-level pure gluonic amplitudes localize on algebraic curves in twistor space. The degree $d$ of these curves is related to the number of negative helicity legs in the amplitude via $d = n_{-} - 1$. This implies that the MHV amplitudes localize on degree-one curves in twistor space which is a line. And since the latter corresponds to a point in Minkowski space, although seemingly non-local, the MHV amplitudes would localize on a point in Minkowski space and could therefore be treated as a vertex. Furthermore, he showed that a generic amplitude living on a $d = n_{-} - 1$ curve in twistor space could be computed using a set of $n_{-} - 1$ degree-one curves connected via propagator. In the language of amplitudes, this means a generic amplitude with $ n_{-}$ negative helicity legs could be computed using $n_{-} - 1$ MHV amplitudes as vertices connected together via a suitable propagator (which they showed was $\sim 1/P^2$) thus giving rise to the MHV rules (also known as the CSW prescription).


\chapter{Wilson lines in the MHV action}
\label{Chapt2}

In this chapter, our focus will be on two important things. First, reviewing the derivation of the "MHV action" following the work of Paul Mansfield \cite{Mansfield2006} where he demonstrated that the Yang-Mills action on the light-cone can be canonically transformed into a new action (the MHV action) consisting of MHV vertices and a scalar propagator connecting opposite helicity fields. The Feynman rules for computing tree-level scattering amplitudes in the MHV action correspond to the MHV (or the CSW) rules, discussed in Section \ref{sec:MHV_rules}, thus providing a field theory realization of an approach motivated geometrically via the twistor space representation of gauge theory amplitudes.

Second, exploring the physical interpretation of the solutions of the field transformation (Mansfield's transformation) that derives the MHV action. This line of research was initiated in \cite{Kotko2017}, where the authors showed that the plus helicity fields in the MHV action correspond to a certain type of straight infinite Wilson line. Later in \cite{Kakkad2020}, we found that the minus helicity fields are also given by a similar Wilson line indicating that Wilson lines are probably better degrees of freedom as compared to the gauge fields in the context of computing scattering amplitudes. These Wilson lines are intimately connected to the Self-Dual sector of the Yang-Mills theory. In fact, the kernels of the inverse of the Wilson line (this will be defined later in the chapter) for the plus helicity fields satisfy the light-front versions of the Berends-Giele currents. 

\section{Recap of the MHV action}
\label{sec:Re-MHV}

The MHV action is an action whose Feynman rules for computing tree-level scattering amplitudes correspond to the MHV rules of Section \ref{sec:MHV_rules}. Recall, the latter utilizes the MHV amplitudes Eq.~\eqref{eq:MHV_ampl} with two minus and the remaining plus helicity gluons $(- - + \dots +)$ continued off-shell as vertices which in turn are glued together using a scalar propagator connecting the opposite helicity fields. Therefore, in order to derive the MHV action from the Yang-Mills action, the fully covariant form is not a suitable starting point. This is because, firstly, the Yang-Mills propagator is not scalar (\emph{cf.} Figure \ref{fig:FR_AG}). Secondly, the gluon helicities are not explicit. In \cite{Mansfield2006}, Mansfield started with the Yang-Mills action on the light cone \cite{Scherk1975} which reads 
\begin{multline}
S_{\mathrm{YM}}\left[A^{\bullet},A^{\star}\right]=\int dx^{+}\int d^{3}\mathbf{x}\,\,\Bigg\{ 
-\mathrm{Tr}\,\hat{A}^{\bullet}\square\hat{A}^{\star}
-2ig\,\mathrm{Tr}\,\partial_{-}^{-1}\partial_{\bullet} \hat{A}^{\bullet}\left[\partial_{-}\hat{A}^{\star},\hat{A}^{\bullet}\right] \\
-2ig\,\mathrm{Tr}\,\partial_{-}^{-1}\partial_{\star}\hat{A}^{\star}\left[\partial_{-}\hat{A}^{\bullet},\hat{A}^{\star}\right]
-2g^{2}\,\mathrm{Tr}\,\left[\partial_{-}\hat{A}^{\bullet},\hat{A}^{\star}\right]\partial_{-}^{-2}\left[\partial_{-}\hat{A}^{\star},\hat{A}^{\bullet}\right]
\Bigg\}
\,,\label{eq:YM_LC_action}
\end{multline}
where $x^{+}=x\cdot\eta$ is the light-cone time Eq.~\eqref{eq:zzbardef},
and
\begin{equation}
\mathbf{x}\equiv\left(x^{-},x^{\bullet},x^{\star}\right)\,,
\end{equation}
represents the position on the constant light-cone time surface. We work with the so-called "double-null" coordinates
\begin{gather}
v^{+}=v\cdot\eta\,,\,\,\,\, v^{-}=v\cdot\widetilde{\eta}\,,\\
v^{\bullet}=v\cdot\varepsilon_{\bot}^{+}\,,\,\,\,\, v^{\star}=v\cdot\varepsilon_{\bot}^{-}\,,
\end{gather}
where ${\eta}$ and $\widetilde{\eta}$ are the two light-like  basis four-vectors defined as follows
\begin{gather}
\eta^\mu=\frac{1}{\sqrt{2}}\left(1,0,0,-1\right)\,,\,\,\,\,\widetilde{\eta}^\mu=\frac{1}{\sqrt{2}}\left(1,0,0,1\right)\, ,
\end{gather}
and $\varepsilon_{\bot}^{\pm}$ are the two space-like complex four-vectors spanning the transverse plane. These read
\begin{equation}
\varepsilon_{\perp}^{\pm}=\frac{1}{\sqrt{2}}\left(0,1,\pm i,0\right)\,.
\end{equation}
In these coordinates, the dot product of two vectors $v$ and $w$ reads
\begin{equation}
u\cdot w=u^{+}w^{-}+u^{-}w^{+}-u^{\bullet}w^{\star}-u^{\star}w^{\bullet}\,.
\end{equation}
Thus, we have
\begin{equation}
    \square=\partial^\mu \partial_\mu = 2(\partial_+\partial_- - \partial_{\bullet}\partial_{\star})\,,
\end{equation}
and $\partial_-^{-1}$ in Eq.~\eqref{eq:YM_LC_action} is realised as an antiderivative \cite{Scherk1975}. Notice, there are only two field components in the action Eq.~\eqref{eq:YM_LC_action}. These read
\begin{gather} 
 {\hat A}^{\bullet}={\hat A}\cdot\varepsilon_{\bot}^{+}=-\frac{1}{\sqrt{2}} \left({\hat A}^1+i{\hat A}^2\right) \,, \\
 {\hat A}^{\star}={\hat A}\cdot\varepsilon_{\bot}^{-}=-\frac{1}{\sqrt{2}} \left({\hat A}^1-i{\hat A}^2\right) \,.
\end{gather}

In Appendix \ref{sec:app_A1}, we re-derive the action Eq.~\eqref{eq:YM_LC_action} starting with the fully covariant form of the Yang-Mills action. Stating briefly, we begin by re-writing the latter using the double-null coordinates, in which the field components are 
\begin{equation}
    {\hat A}^{+}={\hat A}\cdot\eta\,,\quad\, {\hat A}^{-}={\hat A}\cdot\widetilde{\eta}\,, \quad\, {\hat A}^{\bullet}={\hat A}\cdot\varepsilon_{\bot}^{+}\,,\quad\, {\hat A}^{\star}={\hat A}\cdot\varepsilon_{\bot}^{-}\,.
    \label{eq:Afi_com}
\end{equation}
We, then, impose the light-cone gauge
\begin{equation}
    {\hat A}^+ = 0\,.
\end{equation}
This reduces the unphysical degrees of freedom by one. Furthermore, in this gauge, the Fadeev-Popov determinant becomes field independent and can therefore be absorbed into the normalization of the path integral. The ${\hat A}^{-}$ field appears quadratically and is integrated out \cite{Scherk1975}, as a result, the action is expressed in terms of just the two physical degrees of freedom corresponding to the two transverse components $\left( {\hat A}^{\bullet}, {\hat A}^{\star} \right)$ as shown in Eq.~\eqref{eq:YM_LC_action}. In the on-shell limit, these components $\left( {\hat A}^{\bullet}, {\hat A}^{\star} \right)$ correspond to the two physical transverse polarization states of gluons. To see this, recall from Eq.~\eqref{eq:PolarizationVect} that the polarization vector $\varepsilon_{p}^{\pm}$ has the following explicit form
\begin{equation}
\varepsilon_{p}^{\pm}=\varepsilon_{\perp}^{\pm}-\frac{p\cdot\varepsilon_{\perp}^{\pm}}{p^{+}}\eta\,.\label{eq:PolarizationVectre}
\end{equation}
Thus, we see that
\begin{align}
{\hat A}\cdot\varepsilon_{p}^{\pm} = &{\hat A}\cdot\varepsilon_{\perp}^{\pm}-\frac{p\cdot\varepsilon_{\perp}^{\pm}}{p^{+}}{\hat A}\cdot\eta\,, \\
 = &{\hat A}\cdot\varepsilon_{\perp}^{\pm} \,.
\end{align}
In going from the first expression to the second, we used the light-cone gauge condition ${\hat A}\cdot\eta ={\hat A}^+ = 0$. Finally, from Eq.~\eqref{eq:Afi_com} we have ${\hat A}^{\bullet}={\hat A}\cdot\varepsilon_{\bot}^{+} $ and $ {\hat A}^{\star}={\hat A}\cdot\varepsilon_{\bot}^{-}$. Therefore, ${\hat A}\cdot\varepsilon_{p}^{+} = {\hat A}^{\bullet}$ and ${\hat A}\cdot\varepsilon_{p}^{-} = {\hat A}^{\star}$. This leads to a natural identification of these components $\left( {\hat A}^{\bullet}, {\hat A}^{\star} \right)$ with the "plus" and "minus" helicity gluons. The action, Eq.~\eqref{eq:YM_LC_action}, can therefore be compactly re-expressed as
\begin{equation}
S_{\mathrm{YM}}\left[A^{\bullet},A^{\star}\right]=\int dx^{+}\left(\mathcal{L}_{+-}+\mathcal{L}_{++-}+\mathcal{L}_{+--}+\mathcal{L}_{++--}\right)\,,\label{eq:actionLC}
\end{equation}
where
\begin{equation}
\mathcal{L}_{+-}\left[A^{\bullet},A^{\star}\right]=-\int d^{3}\mathbf{x}\,\mathrm{Tr}\,\hat{A}^{\bullet}\square\hat{A}^{\star}\,,\label{eq:YM_kinetic_A}
\end{equation}
\begin{equation}
\mathcal{L}_{++-}\left[A^{\bullet},A^{\star}\right]=-2ig\,\int d^{3}\mathbf{x}\,\mathrm{Tr}\,\partial_{-}^{-1}\partial_{\bullet} \hat{A}^{\bullet}\left[\partial_{-}\hat{A}^{\star},\hat{A}^{\bullet}\right]\,,
\end{equation}
\begin{equation}
\mathcal{L}_{+--}\left[A^{\bullet},A^{\star}\right]=-2ig\,\int d^{3}\mathbf{x}\,\mathrm{Tr}\,\partial_{-}^{-1}\partial_{\star}\hat{A}^{\star}\left[\partial_{-}\hat{A}^{\bullet},\hat{A}^{\star}\right]\,,
\end{equation}
\begin{equation}
\mathcal{L}_{++--}\left[A^{\bullet},A^{\star}\right]=-2g^{2}\int d^{3}\mathbf{x}\,\mathrm{Tr}\,\left[\partial_{-}\hat{A}^{\bullet},\hat{A}^{\star}\right]\partial_{-}^{-2}\left[\partial_{-}\hat{A}^{\star},\hat{A}^{\bullet}\right]\, .
\end{equation}
In our convention, $\bullet$ corresponds to plus and $\star$ to minus helicity. Notice, firstly, the kinetic term $\mathcal{L}_{+-}$ gives rise to a scalar propagator connecting opposite helicity fields, and secondly, the helicities in the Yang-Mills action are explicit as evident from Eq.~\eqref{eq:actionLC}. Thus, the Yang-Mills action on the light cone is most suitable to derive the MHV action.

The major differences in the Feynman rules for the Yang-Mills action Eq.~\eqref{eq:actionLC} and the MHV rules include, firstly, the $\mathcal{L}_{++-}$ triple gluon vertex. This vertex doesn't have the MHV-type helicity $(- - + \dots +)$ configuration, as opposed to the remaining two vertices $\mathcal{L}_{+--}$ and $\mathcal{L}_{++--}$, which do obey the trend. Here we are, obviously, overlooking the kinematic content of these vertices for the moment. Secondly, higher point interaction vertices (beyond four-point) of the type $\mathcal{L}_{+\dots +--}$ are missing. In \cite{Mansfield2006}, Mansfield proposed the following transformation
\begin{equation}
    \mathcal{L}_{+-}\left[A^{\bullet},A^{\star}\right] +\mathcal{L}_{++-}\left[A^{\bullet},A^{\star}\right] \rightarrow \mathcal{L}_{+-}\left[B^{\bullet},B^{\star}\right] \,,
    \label{eq:Man_Transf1}
\end{equation}
that is
\begin{equation}
\mathrm{Tr}\,\hat{A}^{\bullet}\square\hat{A}^{\star}
+2ig\,\mathrm{Tr}\,\partial_{-}^{-1}\partial_{\bullet} \hat{A}^{\bullet}\left[\partial_{-}\hat{A}^{\star},\hat{A}^{\bullet}\right]
\,\, \longrightarrow \,\,
\mathrm{Tr}\,\hat{B}^{\bullet}\square\hat{B}^{\star}
\,,
\label{eq:Man_Transf2}
\end{equation}
where $\hat{B}^{\bullet} $ and $ \hat{B}^{\star}$ are the new fields. It maps the kinetic term and the "unwanted" triple gluon vertex $(- + +)$ of the Yang-Mills action to a new kinetic term. The idea was that this transformation would not only get rid of the $(- + +)$ vertex but also generate the missing higher point interaction terms via the remaining two vertices in the Yang-Mills action. Note, the transformation Eq.~\eqref{eq:Man_Transf1} is over the constant light-cone time $x^+$, implying that the new fields ${\hat B}^{\bullet}(x^+;\mathbf{y}), {\hat B}^{\star}(x^+;\mathbf{y})$ as well as the old ones ${\hat A}^{\bullet}(x^+;\mathbf{x}), {\hat A}^{\star}(x^+;\mathbf{x})$ have the same $x^+$.

The transformation was further required to be canonical satisfying
\begin{equation}
B^{\bullet}_a(x^+;\mathbf{x})=B^{\bullet}_a\left[A^{\bullet}\right](x^+;\mathbf{x})\,, \quad \quad
\partial_{-}A_{a}^{\star}(x^+;\mathbf{x})=\int d^{3}\mathbf{y}\,\frac{\delta B_{c}^{\bullet}(x^+;\mathbf{y})}{\delta A_{a}^{\bullet}(x^+;\mathbf{x})}\partial_{-}B_{c}^{\star}(x^+;\mathbf{y})\,,
\label{eq:AtoB_CT_def}
\end{equation}
where the first expression on the left of Eq.~\eqref{eq:AtoB_CT_def} implies that the ${\hat B}^{\bullet}$ field is a functional of only ${\hat A}^{\bullet}$ field and vice versa. The expression on the right of Eq.~\eqref{eq:AtoB_CT_def} relates the canonically conjugate momenta. From Eq.~\eqref{eq:YM_kinetic_A}, it is straightforward to see that $\partial_{-}{\hat A}^{\star}$ is the momentum canonically conjugate to ${\hat A}^{\bullet}$ and similarly $\partial_{-}{\hat B}^{\star}$ is conjugate to ${\hat B}^{\bullet}$. Assuming the transformation to be canonical preserves the phase space volume up to a field-independent factor. This could also be verified via the Jacobian for the transformation which reads
\begin{equation}
    \mathcal{J}_{\mathrm{MT}}  = \det \begin{vmatrix}
     \frac{\delta {\hat A}^{\bullet} (x^+;\mathbf{x})}
    {\delta {\hat B}^{\bullet} (x^+;\mathbf{y})} 
     &\mathbb{0} \\ \\
\frac{\delta {\hat A}^{\star}(x^+;\mathbf{x})}
    {\delta {\hat B}^{\bullet}(x^+;\mathbf{y})} 
     &\frac{\delta {\hat A}^{\star}(x^+;\mathbf{x})}
    {\delta {\hat B}^{\star}(x^+;\mathbf{y})} 
\end{vmatrix} \,.
\label{eq:MT_jac}
\end{equation}
Above, $\mathrm{MT}$ stands for "Mansfield's Transformation". From Eq.~\eqref{eq:AtoB_CT_def}, we see that the Jacobian is indeed field-independent. Furthermore, since the phase space measure $[d\, A^{\bullet}][d\, \partial_{-}A^{\star}]$ differs from the path integral measure $[d\, A^{\bullet}][d\, A^{\star}]$ by a field-independent factor, assuming the transformation to be canonical also preserves the functional measure in the partition function. 

Substituting Eq.~\eqref{eq:AtoB_CT_def} in Eq.~\eqref{eq:Man_Transf2} we get \cite{Mansfield2006}
\begin{multline}   \partial_{\bullet}\partial_{\star}\partial_{-}^{-1}B_a^{\bullet}(x^+;\mathbf{x}) = \int d^{3}\mathbf{y}\,\mathrm{Tr}\Big\{\left(\partial_{\bullet}\partial_{\star}\partial_{-}^{-1}\hat{A}^{\bullet}(x^+;\mathbf{y})\right.\\
  \left. +ig\left[\hat{A}^{\bullet}(x^+;\mathbf{y}),\partial_{-}^{-1}\partial_{\bullet}\hat{A}^{\bullet}(x^+;\mathbf{y})\right]\right)t^{c}\Big\}\frac{\delta B_{a}^{\bullet}(x^+;\mathbf{x})}{\delta A_{c}^{\bullet}(x^+;\mathbf{y})}
\,. \label{eq:MT_abulb_def}
\end{multline}
 We re-derived the above result in Appendix \ref{sec:CFR} just for the sake of completeness. The straightforward way to obtain the new action is to solve Eq.~\eqref{eq:MT_abulb_def} and \eqref{eq:AtoB_CT_def} for ${\hat A}^{\bullet}$ and ${\hat A}^{\star}$ fields and then substitute these to the Yang-Mills action Eq.~\eqref{eq:YM_LC_action}. In the original work \cite{Mansfield2006}, however, using the analytic properties of the transformation and the S-matrix equivalence theorem\footnote{The Equivalence Theorem states that the S-matrix, calculated via the LSZ reduction formula, is invariant under local field redefinition/reparametrization.}, the new action was demonstrated to have the following form 
 \begin{equation}
S_{\mathrm{MHV}}\left[{B}^{\bullet}, {B}^{\star}\right]=\int dx^{+}\left(
-\int d^{3}\mathbf{x}\,\mathrm{Tr}\,\hat{B}^{\bullet}\square\hat{B}^{\star} 
+\mathcal{L}_{--+}+\dots+\mathcal{L}_{--+\dots+}+\dots\right)\,,
\label{eq:MHV_action}
\end{equation}
where $\mathcal{L}_{--+\dots+}$ represents the $n$-point MHV vertex. In momentum space, it reads
\begin{multline}
\mathcal{L}_{--+\dots+}\left[B^{\bullet},B^{\star}\right]=\int d^{3}\mathbf{p}_{1}\dots d^{3}\mathbf{p}_{n}\delta^{3}\left(\mathbf{p}_{1}+\dots+\mathbf{p}_{n}\right)\,
\widetilde{\mathcal{V}}_{--+\dots+}^{b_{1}\dots b_{n}}\left(\mathbf{p}_{1},\dots,\mathbf{p}_{n}\right)
\\ \widetilde{B}_{b_{1}}^{\star}\left(x^+;\mathbf{p}_{1}\right)\widetilde{B}_{b_{2}}^{\star}\left(x^+;\mathbf{p}_{2}\right)\widetilde{B}_{b_{3}}^{\bullet}\left(x^+;\mathbf{p}_{3}\right)\dots\widetilde{B}_{b_{n}}^{\bullet}\left(x^+;\mathbf{p}_{n}\right)
\,,
\label{eq:MHV_n_point}
\end{multline}
and the vertex $\widetilde{\mathcal{V}}_{--+\dots+}^{b_{1}\dots b_{n}}\left(\mathbf{p}_{1},\dots,\mathbf{p}_{n}\right)$ , in our notation, has the following explicit form 
\begin{multline}
\widetilde{\mathcal{V}}_{--+\dots+}^{b_{1}\dots b_{n}}\left(\mathbf{p}_{1},\dots,\mathbf{p}_{n}\right)= \!\!\sum_{\underset{\text{\scriptsize permutations}}{\text{noncyclic}}}
 \mathrm{Tr}\left(t^{b_1}\dots t^{b_n}\right)\\
 \frac{(-g)^{n-2}}{(n-2)!}  \left(\frac{p_{1}^{+}}{p_{2}^{+}}\right)^{2}
\frac{\widetilde{v}_{21}^{*4}}{\widetilde{v}_{1n}^{*}\widetilde{v}_{n\left(n-1\right)}^{*}\widetilde{v}_{\left(n-1\right)\left(n-2\right)}^{*}\dots\widetilde{v}_{21}^{*}}
\,,
\label{eq:MHV_vertex}
\end{multline}
where the quantities $\widetilde{v}^{\star}_{ij}$ and similarly $\widetilde{v}_{ij}$ (introduced in \cite{Motyka2009}) are defined as follows
\begin{equation}
    \widetilde{v}_{ij}=
    p_i^+\left(\frac{p_{j}^{\star}}{p_{j}^{+}}-\frac{p_{i}^{\star}}{p_{i}^{+}}\right), \qquad 
\widetilde{v}^*_{ij}=
    p_i^+\left(\frac{p_{j}^{\bullet}}{p_{j}^{+}}-\frac{p_{i}^{\bullet}}{p_{i}^{+}}\right)\, .
\label{eq:vtilde}
\end{equation}
These are proportional to the spinor products $\left<ij\right>$, $\left[ij\right]$ as shown below
\begin{equation}
\left\langle ij\right\rangle =-\sqrt{\frac{2p_{j}^{+}}{p_{i}^{+}}}\,\widetilde{v}_{ij},\;\;\;\quad \quad \left[ij\right]=-\sqrt{\frac{2p_{j}^{+}}{p_{i}^{+}}}\,\widetilde{v}_{ij}^{*}\,.\label{eq:spinor_prod1}
\end{equation}
Note, the MHV vertex Eq.~\eqref{eq:MHV_vertex} has exactly the Parke -Taylor form Eq.~\eqref{eq:MHV_ampl}. Although in the remainder of this section, we  follow the straightforward approach (discussed above) where we first consider the explicit solutions of the transformation and then substitute them in the Yang-Mills action, let us briefly highlight the approach followed in the original work \cite{Mansfield2006}. 

Following the S-matrix equivalence theorem, the Green's functions computed using the B-fields (in the MHV action) are equivalent to those computed in terms of A-fields (in the Yang-Mills action) up to a wave-function renormalization factor. Consider computing the $n$-point MHV tree amplitude using both the actions via amputation of the connected Green's function. The result must be identical. From the latter (the Yang-Mills action), we know it should have the Parke -Taylor form Eq.\eqref{eq:MHV_ampl} in the on-shell limit. As for the former, there is just one vertex in the MHV action Eq.~\eqref{eq:MHV_action} that contributes to this amplitude. Thus in the on-shell limit, the vertex itself should give the MHV tree amplitude. But this does not constrain the form of the off-shell vertex itself in the MHV action. There could exist more generic forms for the vertex, instead of the one shown in Eq.~\eqref{eq:MHV_vertex}, such that the on-shell limit reproduces the Parke -Taylor amplitude Eq.\eqref{eq:MHV_ampl}. It turns out that the transformation Eq.~\eqref{eq:AtoB_CT_def} does not allow that. In a nutshell, given the form of the transformation, the vertices in the MHV action must not depend on $x^+$, should be holomorphic (cannot be dependent on $p^{\star}$ in momentum space), and cannot have any additional terms to those shown in Eq.~\eqref{eq:MHV_vertex}. We show this below via the straightforward approach instead.

In \cite{Ettle2006b}, the explicit solution for the ${\hat A}^{\bullet}$ and ${\hat A}^{\star}$ fields were derived in the momentum space using Eq.~\eqref{eq:MT_abulb_def} and \eqref{eq:AtoB_CT_def}. In our notation, these read
\begin{multline}
\widetilde{A}^{\bullet}_a(x^+;\mathbf{P}) = \widetilde{B}^{\bullet}_a(x^+;\mathbf{P})\\
+\sum_{n=2}^{\infty} 
    \int d^3\mathbf{p}_1\dots d^3\mathbf{p}_n \, \widetilde{\Psi}_n^{a\{b_1\dots b_n\}}(\mathbf{P};\{\mathbf{p}_1,\dots ,\mathbf{p}_n\}) \prod_{i=1}^n\widetilde{B}^{\bullet}_{b_i}(x^+;\mathbf{p}_i)\,,
    \label{eq:A_bull_solu}
\end{multline}
\begin{multline}
\widetilde{A}^{\star}_a(x^+;\mathbf{P}) = {\widetilde B}^{\star}_a(x^+;\mathbf{P})\\
+ \sum_{n=2}^{\infty} 
    \int d^3\mathbf{p}_1\dots d^3\mathbf{p}_n \, {\widetilde \Omega}_{n}^{a b_1 \left \{b_2 \cdots b_n \right \} }(\mathbf{P}; \mathbf{p}_1 ,\left \{ \mathbf{p}_2 , \dots ,\mathbf{p}_n \right \}) \widetilde{B}^{\star}_{b_1}(x^+;\mathbf{p}_1)\prod_{i=2}^n\widetilde{B}^{\bullet}_{b_i}(x^+;\mathbf{p}_i)\, ,
    \label{eq:A_star_solu}
\end{multline}
where
\begin{equation}
    {\widetilde \Psi}_{n}^{a \left \{b_1 \cdots b_n \right \} }(\mathbf{P}; \left \{\mathbf{p}_{1},  \dots ,\mathbf{p}_{n} \right \}) =- (-g)^{n-1} \,\,  
    \frac{{\widetilde v}^{\star}_{(1 \cdots n)1}}{{\widetilde v}^{\star}_{1(1 \cdots n)}} \, 
    \frac{\delta^{3} (\mathbf{p}_{1} + \cdots +\mathbf{p}_{n} - \mathbf{P})\,\,  \mathrm{Tr} (t^{a} t^{b_{1}} \cdots t^{b_{n}})}{{\widetilde v}^{\star}_{21}{\widetilde v}^{\star}_{32} \cdots {\widetilde v}^{\star}_{n(n-1)}}  
      \, ,
    \label{eq:psi_kernel}
\end{equation}
\begin{equation}
    {\widetilde \Omega}_{n}^{a b_1 \left \{b_2 \cdots b_n \right \} }(\mathbf{P}; \mathbf{p}_{1} , \left \{ \mathbf{p}_{2} , \dots ,\mathbf{p}_{n} \right \} ) = n \left(\frac{p_1^+}{p_{1\cdots n}^+}\right)^2 {\widetilde \Psi}_{n}^{a b_1 \cdots b_n }(\mathbf{P};  \mathbf{p}_{1},  \dots , \mathbf{p}_{n}) \, ,
    \label{eq:omega_kernel}
\end{equation}
with
\begin{multline}
    \widetilde{\Psi}_n^{a b_1b_2 \dots b_n}(\mathbf{P};\mathbf{p}_1,\mathbf{p}_2, \dots \mathbf{p}_n) = -\frac{(-g)^{n-1}}{n!}\delta^3(\mathbf{p}_{12 \dots n} - \mathbf{P})\\
    \!\!\sum_{\text{\scriptsize permutations}}
    \frac{{\widetilde v}^{\star}_{(1 \cdots n)1}}{{\widetilde v}^{\star}_{1(1 \cdots n)}}\frac{1}{\widetilde{v}^{\star}_{n\, n-1} \dots \widetilde{v}^{\star}_{32}\widetilde{v}^{\star}_{21}}\mathrm{Tr}(t^a t^{b_1} t^{b_2} t^{b_3})\,.
    \label{eq:psi_kernel_notation}
\end{multline}
Note the subtle difference between $\widetilde{\Psi}_n^{a\{b_1\dots b_n\}}(\mathbf{P};\{\mathbf{p}_1,\dots ,\mathbf{p}_n\})$ and $\widetilde{\Psi}_n^{a b_1b_2 \dots b_n}(\mathbf{P};\mathbf{p}_1,\mathbf{p}_2, \dots \mathbf{p}_n)$. Also, the kernels $\widetilde{\Psi}_n^{a\{b_1\dots b_n\}}(\mathbf{P};\{\mathbf{p}_1,\dots ,\mathbf{p}_n\})$ and ${\widetilde \Omega}_{n}^{a b_1 \left \{b_2 \cdots b_n \right \} }(\mathbf{P}; \mathbf{p}_{1} ,\left \{ \mathbf{p}_{2} , \dots ,\mathbf{p}_{n} \right \})$ are integrated over the momentum of the associated fields in the expansion. Therefore we can symmetrize them such that when renaming the $\widetilde{B}^{\bullet}_{b_i}(x^+;\mathbf{p}_i)$ fields, the overall expression remains unchanged. The curly braces on the kernels over the momentum and color indices of the $\widetilde{B}^{\bullet}_{b_i}(x^+;\mathbf{p}_i)$ fields represent this symmetry. In \cite{Ettle2006b}, the solutions Eq.~\eqref{eq:A_bull_solu}-\eqref{eq:A_star_solu} were derived using recursive techniques. In Appendix \ref{sec:SoFT}, we explicitly derive the first few orders of the solutions Eq.~\eqref{eq:A_bull_solu}-\eqref{eq:A_star_solu} using Eq.~\eqref{eq:MT_abulb_def} and \eqref{eq:AtoB_CT_def}, after which we generalize the results for any $n$. During the process, we also elaborate on the aforementioned symmetrization.  

\begin{figure}
    \centering
    \includegraphics[width=15.6cm]{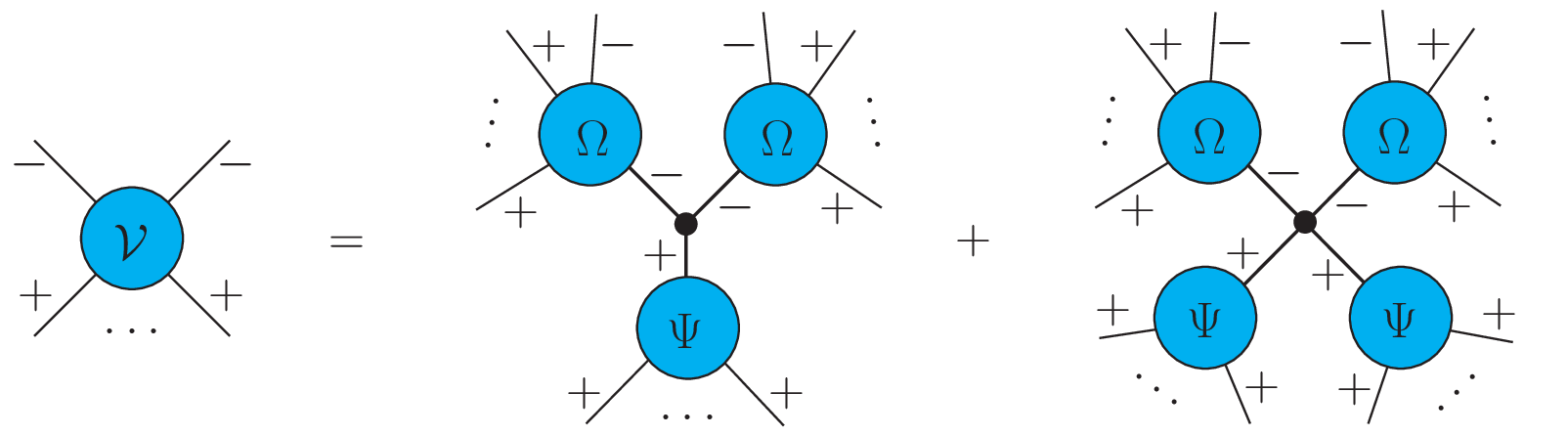}
    \caption{\small
    LHS: $n$-point color-ordered split-helicity vertex $\mathcal{V}\left(1^-,2^-,3^+,\dots,n^+\right)$ in the MHV action. RHS: The contributions originating from the substitution of the solutions for ${\hat A}^{\bullet}[B^{\bullet}]$ and ${\hat A}^{\star}[B^{\bullet}, B^{\star}]$ Eq.~\eqref{eq:A_bull_solu}-\eqref{eq:A_star_solu} to the triple gluon $(- - +)$ and the four gluon $(- - + +)$ vertices in the Yang-Mills action Eq.~\eqref{eq:YM_LC_action}.
}
    \label{fig:mhv_ver}
\end{figure}

Substituting (the position space version of) Eq.~\eqref{eq:A_bull_solu}-\eqref{eq:A_star_solu} in the Yang-Mills action Eq.~\eqref{eq:YM_LC_action} for the first two terms, following Mansfield's transformation, we have
\begin{equation}
\int d^{3}\mathbf{x}\,\mathrm{Tr}\Big\{-\hat{A}^{\bullet}\square\hat{A}^{\star}-2ig\partial_{-}^{-1}\partial_{\bullet}\hat{A}^{\bullet}\left[\partial_{-}\hat{A}^{\star},\hat{A}^{\bullet}\right]\Big\}=\int d^{3}\mathbf{x}\,\mathrm{Tr}\Big\{-\hat{B}^{\bullet}\square\hat{B}^{\star}\Big\}\,.
\end{equation}
Essentially, substituting the first order expansion of the fields (${\hat A}^{\bullet}= {\hat B}^{\bullet}$ and ${\hat A}^{\star}= {\hat B}^{\star}$) to the kinetic term on the L.H.S of the above expression, we obtain the kinetic term on the R.H.S. Whereas, the contributions originating from the substitution of the higher order expansion of ${\hat A}^{\bullet}[B^{\bullet}]$ and ${\hat A}^{\star}[B^{\bullet}, B^{\star}]$ to the kinetic term on the L.H.S. get exactly canceled (term by term) by similar contributions originating from the substitution of ${\hat A}^{\bullet}[B^{\bullet}]$ and ${\hat A}^{\star}[B^{\bullet}, B^{\star}]$ to the triple gluon vertex $(+ + -)$. As a result, we are left with just the kinetic term shown on the R.H.S above. Before we consider the substitution of ${\hat A}^{\bullet}[B^{\bullet}]$ and ${\hat A}^{\star}[B^{\bullet}, B^{\star}]$ to the remaining two vertices, i.e, $(+ - -)$ and $(+ + - -)$, of the Yang-Mills action, it is worth mentioning that as seen from Eq.~\eqref{eq:A_bull_solu}-\eqref{eq:A_star_solu}, in terms of helicity the ${\hat A}^{\star}$ is linear in ${\hat B}^{\star}$ and multiplicates the ${\hat B}^{\bullet}$ fields whereas the ${\hat A}^{\bullet}$ fields simply multiplicates the ${\hat B}^{\bullet}$ fields. Therefore the resulting action should indeed have the form shown in Eq.~\eqref{eq:MHV_action}. Let us, now, consider the $n$-point vertex $\mathcal{L}_{--+\dots+}$ in the new action. For simplicity, we color-decompose the vertex as below
\begin{equation}
\widetilde{\mathcal{V}}_{--+\dots+}^{b_{1}\dots b_{n}}\left(\mathbf{p}_{1},\dots,\mathbf{p}_{n}\right)= \!\!\sum_{\underset{\text{\scriptsize permutations}}{\text{noncyclic}}}
 \mathrm{Tr}\left(t^{b_1}\dots t^{b_n}\right)
 \mathcal{V}\left(1^-,2^-,3^+,\dots,n^+\right)
\,.
\end{equation}
In the color ordered vertex above, we use numbers to represent the momentum associated with a given leg. Furthermore, we put the helicity of that given leg explicitly on the number. The contribution to the color-ordered vertex $\mathcal{V}\left(1^-,2^-,3^+,\dots,n^+\right)$ originating from the substitution of Eq.~\eqref{eq:A_bull_solu}-\eqref{eq:A_star_solu} to $(+ - -)$ and $(+ + - -)$ vertices of the Yang-Mills action is shown in Figure~\ref{fig:mhv_ver}. 
 Using this, in Appendix \ref{sec:mhv_action} we show that  for 3-point and 4-point, the resulting vertex indeed has the MHV form Eq.\eqref{eq:MHV_vertex}. In \cite{Ettle2006b}, the authors confirmed this for the 5-point vertex as well. Thus, we see that the new action -- MHV action -- resulting from the Mansfield transformation of the Yang-Mills action has, indeed, MHV vertices and a scalar propagator. 

At this point, it is tempting to conclude that the Feynman rules for the MHV action Eq.~\eqref{eq:MHV_action} indeed correspond to the MHV rules. However, there is one last piece in this puzzle. In MHV rules, the MHV amplitudes are continued off-shell using a fixed spinor $\chi_{\dot\alpha}$ (\emph{cf.} Eq.~\eqref{eq:CSW_offshell}). Precisely, the prescription reads
\begin{equation}
    [\chi\lambda]\,\lambda_P^\alpha = \chi_{\dot\alpha} P^{\dot\alpha \alpha}\,.
    \label{eq:csw_cal1}
\end{equation}
where $P$ is the momentum of the leg which is being continued off-shell  and $\chi_{\dot\alpha}$ is the reference spinor used for the off-shell continuation (note previously we used $\eta_{\dot\alpha}$ in Eq.~\eqref{eq:CSW_offshell} for the reference spnior but now we change it to $\chi_{\dot\alpha}$ to avoid confusions because we use $\eta$ for the light-cone time direction). In the CSW prescription Eq.~\eqref{eq:CSW_offshell}, the factor of $[\chi\lambda]$ is suppressed because the tree amplitudes are invariant under rescaling of $\lambda_P^\alpha$ associated with the off-shell internal legs. For the current discussion, we keep the factor explicit. 

It is important to make sure that the off-shell continued vertices of the MHV rules are the same as the vertices in the MHV action Eq.~\eqref{eq:MHV_action}. In  \cite{Ettle2006b}, the authors showed that  the reference spinor considered in the MHV rules (to continue the amplitudes off-shell) corresponds to the direction defining the light-cone time $x^+$. To see this, recall that an on-shell momentum $q$ could always be factorized in terms of spinors as follows
\begin{equation}
 q^{\dot\alpha \alpha}  =   \widetilde{\lambda}^{\dot\alpha}\lambda^\alpha\,.
\end{equation}
This on-shell momentum $q$ can be rewritten in terms of an off-shell momentum $P$ as follows (in four vector notation)
\begin{equation}
    q^\mu = P^\mu + c \eta^\mu \,, \quad \mathrm{where}\quad  c = \frac{P^{\bullet}P^{\star}}{P^+} -P^- \, .
    \label{eq:qoffp}
\end{equation}
Above,  $\eta^\mu$ is the light-like four vector along the light-cone time ${ x}^{+}={ x}\cdot\eta$. Given the above definition, it is straightforward to see that $q^2 = 0$. Rewriting Eq.~\eqref{eq:qoffp}, using spinor indices we get
\begin{equation}
    q^{\dot\alpha \alpha} = P^{\dot\alpha \alpha} + c \eta^{\dot\alpha \alpha} = \widetilde{\lambda}^{\dot\alpha}\lambda^\alpha\,.
\end{equation}
Contracting the above with a reference spinor $\widetilde{\chi}_{\dot\alpha}$ we get
\begin{equation}
   \widetilde{\chi}_{\dot\alpha} P^{\dot\alpha \alpha} + c \widetilde{\chi}_{\dot\alpha}\eta^{\dot\alpha \alpha} = [\chi \lambda]\lambda^\alpha\,.
\end{equation}
Finally, since $\eta^{\dot\alpha \alpha}$ is light-like, let us decompose it in terms of spinors as
\begin{equation}
    \eta^{\dot\alpha \alpha} = \widetilde{\xi}^{\dot\alpha}\xi^\alpha\,.
\end{equation}
Thus we get
\begin{equation}
   \widetilde{\chi}_{\dot\alpha} P^{\dot\alpha \alpha} + c [\chi \xi] \xi^\alpha = [\chi \lambda]\lambda^\alpha\,.
\end{equation}
Now, if $\chi \propto \xi$, that is if the reference spinor is proportional to the spinor that defines the light-cone time direction, $[\chi \xi] = 0$ and we get
\begin{equation}
   \widetilde{\chi}_{\dot\alpha} P^{\dot\alpha \alpha} = [\chi \lambda]\lambda^\alpha\,.
\end{equation}
The above expression is exactly the same as the CSW off-shell prescription Eq.~\eqref{eq:csw_cal1}.
\section{Exploring solutions of Mansfield's transformation}
\label{sec:EX_SOL_MT}

In the original work \cite{Mansfield2006}, the MHV action Eq.~\eqref{eq:MHV_action}, as stated previously, was derived by exploiting the analytic properties of the transformation and the S-matrix equivalence theorem. Although later the solutions ${\widetilde A}^{\bullet}[B^{\bullet}]$ and ${\widetilde A}^{\star}[B^{\bullet}, B^{\star}]$ Eq.~\eqref{eq:A_bull_solu}-\eqref{eq:A_star_solu} to the transformation Eq.~\eqref{eq:Man_Transf2} was explicitly derived in \cite{Ettle2006b}, the physical interpretation of these  were explored in \cite{Kotko2017, Kakkad2020} where they were shown to be intimately connected with the Self-Dual sector of the Yang-Mills theory. Furthermore, in the former work \cite{Kotko2017}, the authors considered, for the first time,  the structure of ${\hat B}^{\bullet}[A^{\bullet}]$ field in the position space and found that it is given by a straight infinite Wilson line on a plane in $\mathbb{M}_\mathbb{C}$ which turns out to be "Self-Dual" in the sense explained later. Following which, in latter \cite{Kakkad2020}, we found that the ${\hat B}^{\star}[A^{\bullet}, A^{\star}]$ field is given by a similar Wilson line. In this section, we explore these Wilson line solutions to Mansfield's transformation and show how geometrically rich and intriguing these objects are.
\subsection{Self-Dual Yang-Mills and the classical EOM}
\label{subsec:SDYM_EOM}

The light-cone action for the Self-Dual sector of the Yang-Mills theory is the Chalmers-Siegel action \cite{Bardeen1996,Chalmers1996,Cangemi1997,Rosly1997,Monteiro2011}
\begin{equation}
S_{\mathrm{SD}}\left[A^{\bullet},A^{\star}\right]=\int dx^{+}\left(\mathcal{L}_{+-}+\mathcal{L}_{++-}\right)\,,
\label{eq:SDactionLC}
\end{equation}
where
\begin{gather}
\mathcal{L}_{+-}\left[A^{\bullet},A^{\star}\right]=-\int d^{3}\mathbf{x}\,\mathrm{Tr}\,\hat{A}^{\bullet}\square\hat{A}^{\star}\,,\\
\mathcal{L}_{++-}\left[A^{\bullet},A^{\star}\right]=-2ig\,\int d^{3}\mathbf{x}\,\mathrm{Tr}\,\partial_{-}^{-1}\partial_{\bullet} \hat{A}^{\bullet}\left[\partial_{-}\hat{A}^{\star},\hat{A}^{\bullet}\right]\,.
\end{gather}
The above action is essentially a truncation of the full Yang-Mills action Eq.~\eqref{eq:actionLC} including just the kinetic term and the $(+ + -)$ triple gluon vertex. Recall, these are exactly the terms participating in Mansfield's transformation Eq.~\eqref{eq:Man_Transf1}-\eqref{eq:Man_Transf2} implying that the transformation maps the Self-Dual sector of Yang-Mills to a free theory $\mathcal{L}_{+-}\left[B^{\bullet},B^{\star}\right]$. The Self-Dual Yang-Mills is known to be classically free because the tree-level amplitudes obtained using Eq.~\eqref{eq:SDactionLC} are of the type $(- + \dots +)$ which vanish in the on-shell limit (\emph{cf.} Eq.~\eqref{eq:plus_tree_amp}). However, the loop amplitudes are in general non-zero \cite{Bern_1994}. Therefore, the above transformation can result in missing loop amplitudes. We discuss this later in detail in Chapter \ref{QMHV-chapter}. Coming back to the transformation, from the above discussion, it follows that the solution of Mansfield's transformation must encode the Self-Dual sector of the Yang-Mills theory. In \cite{Kakkad2020}, we demonstrated that ${\widetilde A}^{\bullet}[B^{\bullet}]$ is related to the solution of the Self-Dual equation of motion (EOM). Below, we summarize the findings.

Using Eq.~\eqref{eq:SDactionLC}, the generating functional for the Green's function can be written as
\begin{equation}
    Z[J]=\int[dA]\, e^{i\left(S_{\mathrm{SD}}[A] + \int\!d^4x\, \Tr \hat{J}_j(x) \hat{A}^j(x)\right) } \,,
    \label{eq:Z_SDYM}
\end{equation}
where the index $j$ runs over the transverse components $\{\bullet, \star\}$ and $\hat{J}_j$ represents the auxiliary external current. From Eq.~\eqref{eq:Z_SDYM}, the classical EOM can be shown to be
\begin{equation}
   \left. \frac{\delta S_{\mathrm{SD}}[A]}{\delta {\hat A}^j(x)}\right|_{{\hat A}={\hat A}_{c}[J]}+{\hat J}_j(x)=0 \,,
    \label{eq:SDEOM0}
\end{equation}
where ${\hat A}_{c}[J]=({\hat A}_{c}^{\bullet}[J],{\hat A}_{c}^{\star}[J])$ represents the solution of the classical EOM. Out of the two equations, the one for the ${\hat A}_{c}^{\bullet}[J]$ is dynamic and is commonly referred to as the Self-Dual equation. Substituting Eq.~\eqref{eq:SDactionLC} in Eq.~\eqref{eq:SDEOM0}, we get
\begin{equation}
    \Box {\hat{A}}^{\bullet} + 2ig{\partial}_{-} \left[ ({\partial}_{-}^{-1} {\partial}_{\bullet} {\hat{A}}^{\bullet}), {\hat{A}}^{\bullet} \right] + {\hat J}^{\bullet} = 0\, .
    \label{eq:SD_EOM}
\end{equation}
The above equation, with the current ${\hat J}^{\bullet}$ put to zero, can be explicitly derived from the following Self-Duality condition after imposing the light-cone gauge ${\hat A}^+ = 0$ 
\begin{equation}
    \hat{F}^{\mu\nu} = \ast \hat{F}^{\mu\nu}\, , \quad\quad \ast \hat{F}_{\mu\nu} = -i \epsilon_{\mu\nu\alpha\beta}\hat{F}^{\alpha\beta}\,.
\end{equation}

\begin{figure}
    \centering
    \includegraphics[width=11cm]{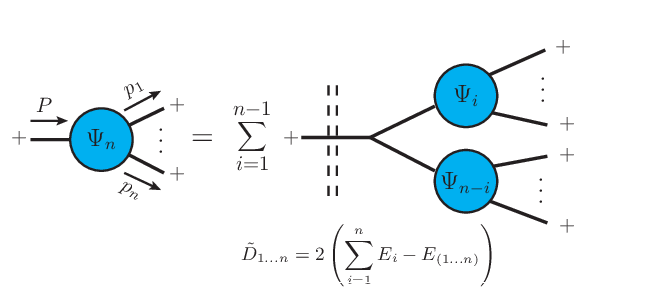}
    \caption{\small 
    The kernels $\widetilde{\Psi}_n^{a\{b_1\dots b_n\}}\left(P;\{p_1,\dots ,p_n\}\right)$ in the solution of the Self-Dual EOM satisfy a recursion relation which corresponds to the light-front version of the Berends-Giele relations. The double-dashed line represents the energy denominator $\widetilde{D}_{1\dots n}$ and $+$ represents the helicity of the associated leg.
}
    \label{fig:PSI_BG}
\end{figure}

The solution  ${\hat A}_{c}^{\bullet}[J^{\bullet}]$ to the Self-Dual equation Eq.~\eqref{eq:SD_EOM} can be obtained iteratively under the assumption that the currents have the support on the light cone. Redefining the current as 
\begin{equation}
{\hat j}^{\bullet}(x)= \square^{-1}  {\hat J}^{\bullet}(x)\,,
\end{equation}
the solution in momentum space reads \cite{Cangemi1997,Motyka2009}
\begin{equation}
    {\widetilde A}_a^{\bullet}\left[j^{\bullet}\right](P) = \sum_{n=1}^{\infty} 
    \int d^4p_1\dots d^4p_n \, {\widetilde \Psi}_n^{a\{b_1\dots b_n\}}(P;\{p_1,\dots ,p_n\})\, {\widetilde j}^{\bullet}_{b_1}(p_1)\dots {\widetilde j}^{\bullet}_{b_n}(p_n)\,,
    \label{eq:Abullet_sol1}
\end{equation}
where the kernels are
\begin{multline}
    \widetilde{\Psi}_n^{a\{b_1\dots b_n\}}\left(P;\{p_1,\dots ,p_n\}\right) =     
    - (-g)^{n-1}  
    \delta^{4} (p_{1} + \cdots + p_{n} - P) \\
    \Tr \left(t^{a} t^{b_{1}} \cdots t^{b_{n}}\right) \,  
    \frac{{\widetilde v}^{\ast}_{(1 \cdots n)1}}{{\widetilde v}^{\ast}_{1(1 \cdots n)}}
    \frac{1}{{\widetilde v}^{\ast}_{21}{\tilde v}^{\ast}_{32} \cdots {\widetilde v}^{\ast}_{n(n-1)}} \, .
    \label{eq:BG_Psi_n}
\end{multline}
The kernels $\widetilde{\Psi}_n^{a\{b_1\dots b_n\}}\left(P;\{p_1,\dots ,p_n\}\right)$ satisfy the recursion relation (shown in Figure~\ref{fig:PSI_BG}) which corresponds to the light-front version of the Berends-Giele relations \cite{Berends:1987me, Motyka2009}. Thus, physically, the kernels $\widetilde{\Psi}_n^{a\{b_1\dots b_n\}}\left(P;\{p_1,\dots ,p_n\}\right)$ represent the current of an incoming plus helicity off-shell gluon of momentum $P$ splitting into $n$ on-shell final state gluons. Amputating the incoming off-shell leg by multiplying by $P^2$ or the energy denominator 

\begin{equation}
    \widetilde{D}_{1\dots n} = 2\left(\sum_{i=1}^n E_i - E_{(1\dots n)}\right)\,, \quad\quad E_i = \frac{k_i^{\bullet}k_i^{\star}}{k_i^+} \, ,
    \label{eq:Dtilde}
\end{equation}
reduces the kernel $\widetilde{\Psi}_n^{a\{b_1\dots b_n\}}\left(P;\{p_1,\dots ,p_n\}\right)$ to the $(n+1)$-point $(- \rightarrow +\dots+)$ amplitude where the incoming negative helicity gluon is still off-shell. It is straightforward to see that in the on-shell limit, $P^2\rightarrow 0$ the amplitude $P^2\,\widetilde{\Psi}_n^{a\{b_1\dots b_n\}}\left(P;\{p_1,\dots ,p_n\}\right)$ is zero which is in agreement with the well-known result for the fully on-shell $(- \rightarrow +\dots+)$ amplitude.

The kernels $\widetilde{\Psi}_n^{a\{b_1\dots b_n\}}\left(P;\{p_1,\dots ,p_n\}\right)$ in Eq.~\eqref{eq:BG_Psi_n} appear quite similar to those in the solution ${\widetilde A}^{\bullet}[B^{\bullet}]$ of Mansfield's transformation (see Eq.~\eqref{eq:psi_kernel}). These are in fact identical except for a delta function for the minus component of the momentum as shown below
\begin{multline}
    \widetilde{\Psi}_n^{a\{b_1\dots b_n\}}\left(P;\{p_1,\dots ,p_n\}\right) = 
    \delta\left(p_1^- + \dots + p_n^- - P^-\right)
    \widetilde{\Psi}_n^{a\{b_1\dots b_n\}}\left(\mathbf{P};\{\mathbf{p}_1,\dots ,\mathbf{p}_n\}\right) \, .
    \label{eq:Psi_n_3D}
\end{multline}
The kernel in the solution of the Self-Dual equation is a function of four momenta $p_i$ whereas the kernels in ${\widetilde A}^{\bullet}[B^{\bullet}]$ is a function of three momenta $\mathbf{p}_i$ (recall Mansfield's transformation was over the constant light cone time $x^+$). However, the delta over the minus component can be Fourier transformed from Eq.~\eqref{eq:Abullet_sol1} to obtain
\begin{multline}
    \widetilde{A}_a^{\bullet}\left[j^{\bullet}\right](x^+;\mathbf{P}) = \sum_{n=1}^{\infty} 
    \int d^3\mathbf{p}_1\dots d^3\mathbf{p}_n \, \widetilde{\Psi}_n^{a\{b_1\dots b_n\}}(\mathbf{P};\{\mathbf{p}_1,\dots ,\mathbf{p}_n\}) \\ \widetilde{j}^{\bullet}_{b_1}(x^+;\mathbf{p}_1)\dots \widetilde{j}^{\bullet}_{b_n}(x^+;\mathbf{p}_n)\,.
    \label{eq:Abullet_sol2}
\end{multline}
Thus we see that the 3D solution (over constant $x^+$) of the Self-Dual equation is exactly equal to the solution ${\widetilde A}^{\bullet}[B^{\bullet}]$ of Mansfield's transformation if we reinterpret the ${\widetilde B}^{\bullet}$ field in the MHV action as the auxiliary external current ${\widetilde J}^{\bullet}$, coupled to the Self-Dual action in $Z[J]$, as follows
\begin{equation}
    B^{\bullet}_a(x) \equiv j^{\bullet}_{a}\left(x\right)\equiv \square^{-1}  {J}^{\bullet}(x)\, .
    \label{eq:B_field}
\end{equation}
An interesting question is what would the kernels of the inverse transformation $\widetilde{B}_a^{\bullet}\left[A^{\bullet}\right]$ correspond to, given that the kernels of ${\widetilde A}^{\bullet}[B^{\bullet}]$ are related to the Berends-Giele currents. This question has already been answered in \cite{Kotko2017}. In the following section, we summarize the results. 

\subsection{\texorpdfstring{${\hat B}^{\bullet}[A^{\bullet}]$}{Bbullet} as a straight infinite Wilson line}
\label{subsec:Bbul[A]_WL}

The 3D solution ${\widetilde A}^{\bullet}[B^{\bullet}]$ of the Self-Dual equation Eq.~\eqref{eq:SD_EOM} can be inverted to obtain $\widetilde{B}_a^{\bullet}\left[A^{\bullet}\right]$ and just like ${\widetilde A}^{\bullet}[B^{\bullet}]$, it is natural to expect that the latter should encode the Self-Dual sector of the Yang-Mills theory in some sense.  Unraveling this, in turn, allows us to provide a physical interpretation of the plus helicity field $B^{\bullet}$  in the MHV action. 

Let us begin by postulating the following series expansion for the inverse in momentum space
\begin{multline}
    \widetilde{B}_a^{\bullet}\left[A^{\bullet}\right](x^+;\mathbf{P}) = \sum_{n=1}^{\infty} 
    \int d^3\mathbf{p}_1\dots d^3\mathbf{p}_n \, \widetilde{\Gamma}_n^{a\{b_1\dots b_n\}}(\mathbf{P};\{\mathbf{p}_1,\dots ,\mathbf{p}_n\}) \\ \widetilde{A}^{\bullet}_{b_1}(x^+;\mathbf{p}_1)\dots \widetilde{A}^{\bullet}_{b_n}(x^+;\mathbf{p}_n)\,.
    \label{eq:Bbullet_sol1}
\end{multline}
To determine $\widetilde{\Gamma}_n^{a\{b_1\dots b_n\}}(\mathbf{P};\{\mathbf{p}_1,\dots ,\mathbf{p}_n\})$, we can substitute ${\widetilde A}^{\bullet}[B^{\bullet}]$ using Eq.~\eqref{eq:A_bull_solu} to the expression above and equate terms with same order in $B^{\bullet}$ fields. This calculation was done explicitly in \cite{Kotko2017} and therefore we do not repeat it. Here we simply recall the details necessary to assemble the final result. 

The first order is trivial $\widetilde{\Gamma}_1^{a b_1}(\mathbf{P}; \mathbf{p}_1) = 1$. For the second order, we get diagrammatically
\begin{center}
\includegraphics[width=5.5cm]{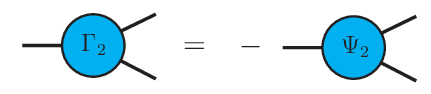}
\end{center}
\begin{align}
    \widetilde{\Gamma}_2^{a\{b_1 b_2\}}(\mathbf{P};\{\mathbf{p}_1,\mathbf{p}_2\})  =&\, -\widetilde{\Psi}_2^{a\{b_1 b_2\}}(\mathbf{P};\{\mathbf{p}_1,\mathbf{p}_2\})\,, \nonumber\\
    =&\,g\, \frac{1}{\widetilde{v}^{\star}_{21}}\frac{p_{12}^+}{p_1^+}\delta^3(\mathbf{p}_1 + \mathbf{p}_2 - \mathbf{P})\mathrm{Tr}(t^a t^{b_1} t^{b_2})\,,\nonumber\\
   =&\, -g\,\frac{1}{\widetilde{v}_{1\left(12\right)}^{\star}}\,\delta^3(\mathbf{p}_1 + \mathbf{p}_2 - \mathbf{P})\mathrm{Tr}\left(t^{a}t^{b_{1}}t^{b_{2}}\right)\,.
   \label{eq:psi2_gamma2}
\end{align}
In going from the second expression above to the third we used the following two identities
\begin{equation}
   \widetilde{v}^{\star}_{(kl)k}= \widetilde{v}^{\star}_{lk} \, \quad \mathrm{and}\quad \widetilde{v}^{\star}_{ij} = -\frac{p_i^+}{p_j^+} \widetilde{v}^{\star}_{ji}\,.
\end{equation}

For the third order, we get diagrammatically 
\begin{center}
\includegraphics[width=12.3cm]{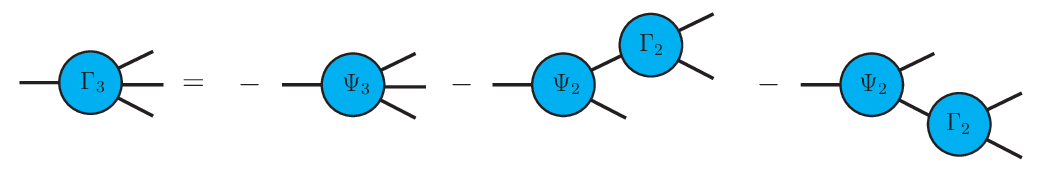}
\end{center}
To each of the terms in the diagram above (both on the L.H.S. as well as the R.H.S) we assign a momentum and color $\{\mathbf{P},a\}$, respectively, to the incoming leg and $\{\mathbf{p}_{i},b_i\}$ with $i= 1,2,3$  to the three outgoing legs in the anticlockwise fashion. Substituting for the $\widetilde{\Psi}_3$, $\widetilde{\Psi}_2$ and $\widetilde{\Gamma}_2$, after a bit of tedious algebra we get \cite{Kotko2017}
\begin{equation}
\widetilde{\Gamma}_{3}^{a\{b_{1} b_2 b_{3}\}}\left(\mathbf{P};\{\mathbf{p}_{1},\mathbf{p}_{2},\mathbf{p}_{3}\}\right)=\left(-g\right)^{2}\delta^{3}\left(\mathbf{p}_{123}-\mathbf{P}\right)\frac{1}{\widetilde{v}_{1\left(123\right)}^{\star}\widetilde{v}_{\left(12\right)\left(123\right)}^{\star}}\,\mathrm{Tr}\left(t^{a}t^{b_{1}}t^{b_{2}}t^{b_{3}}\right)\,.
\label{eq:Gamma_3}
\end{equation}
Similarly, for the fourth order, we get
\begin{center}
\includegraphics[width=16cm]{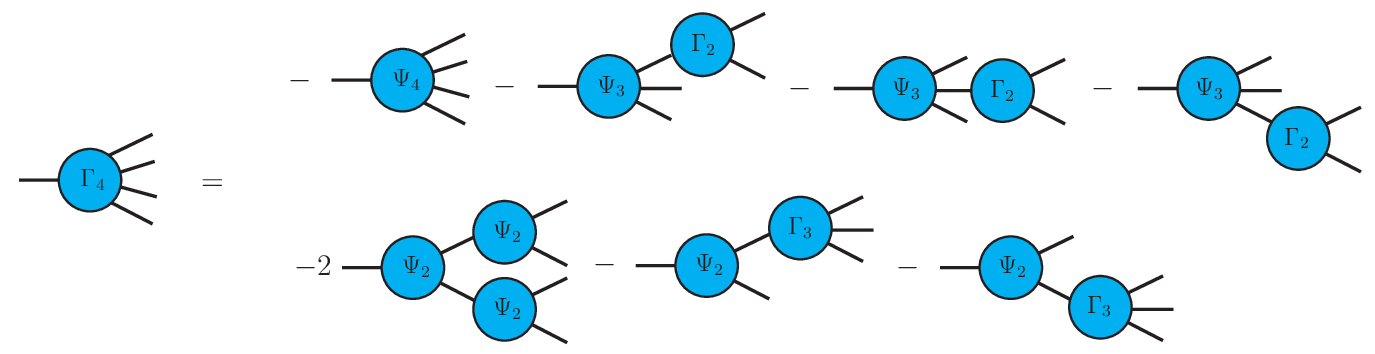}
\end{center}
\begin{multline}
\widetilde{\Gamma}_{4}^{a\{b_{1} b_2 b_{3}b_{4}\}}\left(\mathbf{P};\{\mathbf{p}_{1},\mathbf{p}_{2},\mathbf{p}_{3},\mathbf{p}_{4}\}\right)=\left(-g\right)^{3}\delta^{3}\left(\mathbf{p}_{1234}-\mathbf{P}\right)\\
\frac{1}{\widetilde{v}_{1\left(1234\right)}^{\star}\widetilde{v}_{\left(12\right)\left(1234\right)}^{\star}\widetilde{v}_{\left(123\right)\left(1234\right)}^{\star}}\,\mathrm{Tr}\left(t^{a}t^{b_{1}}t^{b_{2}}t^{b_{3}}t^{b_{4}}\right)\,.
\label{eq:Gamma_4}
\end{multline}

The results can be generalized for any $n$ as below
\begin{multline}
\widetilde{\Gamma}_{n}^{a\{b_{1}\dots b_{n}\}}\left(\mathbf{P};\{\mathbf{p}_{1},\dots,\mathbf{p}_{n}\}\right)
=\left(-g\right)^{n-1}\delta^{3}\left(\mathbf{p}_{1\dots n}-\mathbf{P}\right)\\
\frac{1}{\widetilde{v}_{1\left(1\dots n\right)}^{\star}\widetilde{v}_{\left(12\right)\left(1\dots n\right)}^{\star}\dots\widetilde{v}_{\left(1\dots n-1\right)\left(1\dots n\right)}^{\star}}\,\mathrm{Tr}\left(t^{a}t^{b_{1}}\dots t^{b_{n}}\right)\,.\label{eq:Gamma_n}
\end{multline}

At this point, one may wonder that, the substitution of the correspondence $B^{\bullet}_a(x) \equiv j^{\bullet}_{a}\left(x\right)$ to the Self-Dual equation Eq.~\eqref{eq:SD_EOM} gives
\begin{equation}
    \Box {\hat{A}}^{\bullet} + 2ig{\partial}_{-} \left[ ({\partial}_{-}^{-1} {\partial}_{\bullet} {\hat{A}}^{\bullet}), {\hat{A}}^{\bullet} \right] - \Box\, \hat{B}^{\bullet} = 0\, ,
    \label{eq:SD_EOM_j}
\end{equation}
which implies that $\widetilde{B}_a^{\bullet}\left[A^{\bullet}\right]$ should be at most quadratic. Therefore, a series solution of the type postulated above in Eq.~\eqref{eq:Bbullet_sol1} may appear contradictory because $\widetilde{\Gamma}_n^{a\{b_1\dots b_n\}}(\mathbf{P};\{\mathbf{p}_1,\dots ,\mathbf{p}_n\}) \neq 0$ for $n\geq 3$. In  \cite{Kakkad2020}, we showed that although the kernels themselves are non-vanishing $\widetilde{\Gamma}_n^{a\{b_1\dots b_n\}}(\mathbf{P};\{\mathbf{p}_1,\dots ,\mathbf{p}_n\}) \neq 0$, we have $\Box\, \hat{B}^{\bullet} = 0$  for $n\geq 3$. We repeat this in Appendix \ref{sec:app_A3}.

Notice, although the kernels $\widetilde{\Gamma}_{n}^{a\{b_{1}\dots b_{n}\}}\left(\mathbf{P};\{\mathbf{p}_{1},\dots,\mathbf{p}_{n}\}\right)$ and $\widetilde{\Psi}_{n}^{a\{b_{1}\dots b_{n}\}}\left(\mathbf{P};\{\mathbf{p}_{1},\dots,\mathbf{p}_{n}\}\right)$ differ, the above procedure indicates that the former is also made up of the Self-Dual vertex $(+ + -)$ in the Yang-Mills action. In fact, in \cite{Kotko2017}, the difference in these kernels was shown to arise exclusively from the difference in the energy denominators. Precisely, when both these kernels are expressed diagrammatically using Feynman diagrams in momentum space, each of the two kernels (for a given $n$) consists of a sum of diagrams where each diagram is made up of the Self-Dual interaction vertex $(+ + -)$ in the Yang-Mills action connected via energy denominators. In light front dynamics, we have energy denominators \cite{lff1,lff2,lff3} (associated with each intermediate state) connecting the interaction vertices instead of propagators. These denominators arise, essentially, from the integration of the minus component of the momentum $p^-$ in the propagators. The energy denominator represents the difference in the light cone energy
\begin{equation}
    E_i = \frac{k_i^{\bullet}k_i^{\star}}{k_i^+}\,,
\end{equation}
of the given intermediate state and the light cone energy of a reference state. The latter can be the initial or the final state. In the case of the $\widetilde{\Gamma}_{n}^{a\{b_{1}\dots b_{n}\}}\left(\mathbf{P};\{\mathbf{p}_{1},\dots,\mathbf{p}_{n}\}\right)$ kernels, all the intermediate state energy denominators are defined with respect to the energy of the initial state whereas for the $\widetilde{\Psi}_{n}^{a\{b_{1}\dots b_{n}\}}\left(\mathbf{P};\{\mathbf{p}_{1},\dots,\mathbf{p}_{n}\}\right)$ kernels, all the intermediate state energy denominators are defined with respect to the energy of the final state. Thus, these two could be interchanged into one another by simply changing the reference state in the energy denominators while keeping the rest of the Feynman diagrams exactly the same. The origin of that "symmetry" is still to be explored.

\begin{figure}
    \centering
    \includegraphics[width=10cm]{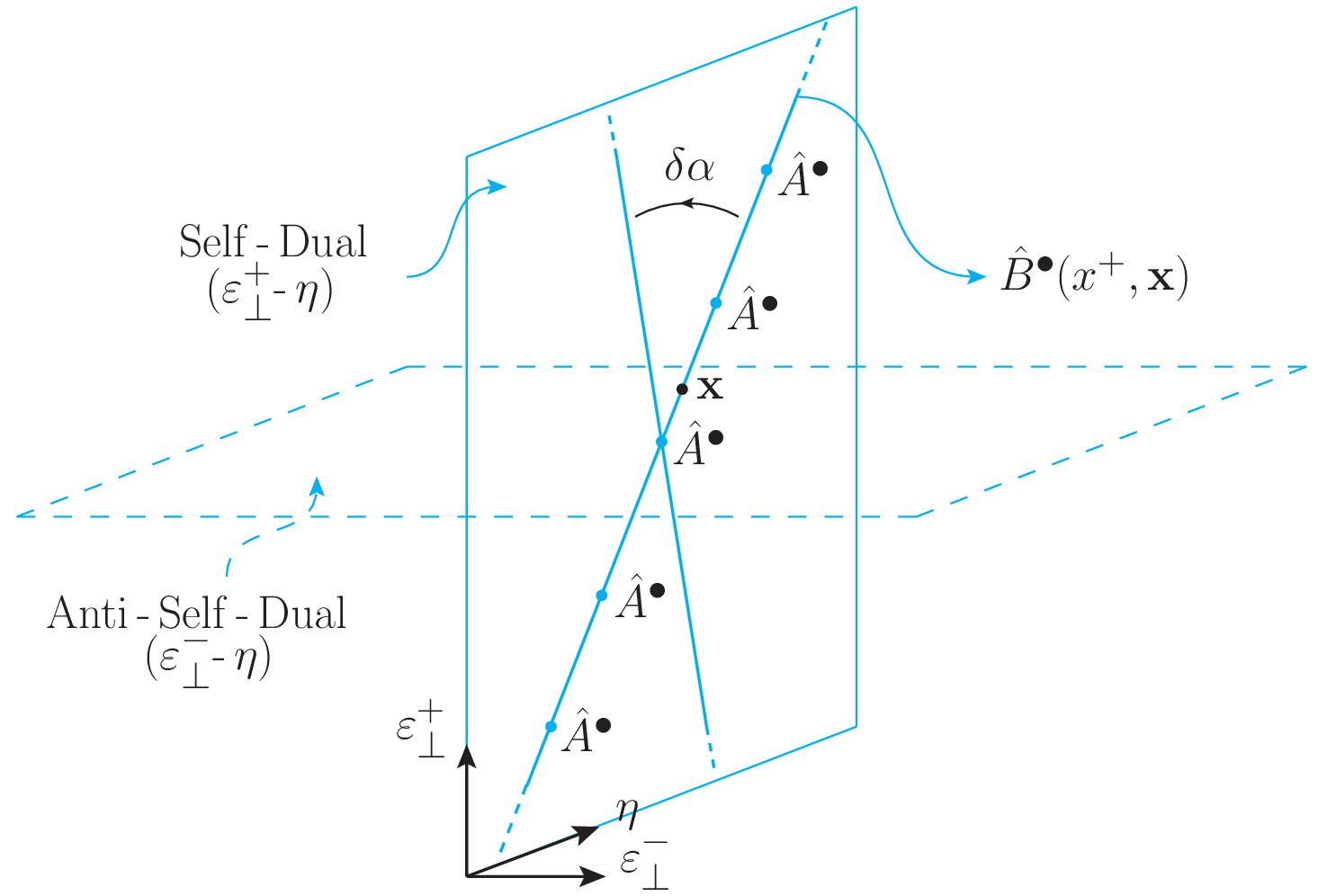}
    \caption{\small
    The plus helicity field in the MHV action $B^{\bullet}_a[A^{\bullet}](x^+,\mathbf{x})$ is a straight infinite  Wilson line on the Self-Dual plane spanned by $\varepsilon_{\perp}^{+} $, $ \eta$ along a generic vector of slope $\alpha$. $\delta \alpha$ represents the variation in the slope. It gets integrated over.
}
    \label{fig:Bbul_WL}
\end{figure}

Furthermore, in the same work \cite{Kotko2017}, the authors showed that the series solution $\widetilde{B}_a^{\bullet}\left[A^{\bullet}\right](x^+;\mathbf{P})$ given in Eq.~\eqref{eq:Bbullet_sol1} is the momentum space expression for a straight infinite Wilson line $B^{\bullet}_a[A^{\bullet}](x)=\mathcal{W}_{(+)}^a[A](x)$. We use $\mathcal{W}_{(\pm)}^a[K](x)$ to represent a generic straight infinite Wilson line of ${\hat K}$ field along the vector $\varepsilon_{\alpha}^{\pm}$ with the following explicit definition
\begin{equation}
     \mathcal{W}^{a}_{(\pm)}[K](x)=\int_{-\infty}^{\infty}d\alpha\,\mathrm{Tr}\left\{ \frac{1}{2\pi g}t^{a}\partial_{-}\, \mathbb{P}\exp\left[ig\int_{-\infty}^{\infty}\! ds\, \varepsilon_{\alpha}^{\pm}\cdot \hat{K}\left(x+s\varepsilon_{\alpha}^{\pm}\right)\right]\right\} \, .
\label{eq:WL_gen}
\end{equation}
Let us dissect the above definition. A generic Wilson line is defined as the path-ordered exponential of a gauge field Eq.~\eqref{eq:WL_compa} (for a detailed review on Wilson lines, see \cite{Cherednikov2014})
\begin{equation}
\mathbb{P}\exp\left[ig\int dz\,\cdot \hat{K}\left(z\right)\right]=\mathbb{P}\exp\left[ig\int_{-\infty}^{\infty}ds\,\varepsilon_{\alpha}^{\pm}\cdot \hat{K}\left(x+s\varepsilon_{\alpha}^{\pm}\right)\right]\,,
\end{equation}
above $dz\,\cdot \hat{K}\left(z\right) = dz_{\mu}\, \hat{K}^{\mu}\left(z\right)$ and $z^\mu$ represents the coordinates along a path. On the R.H.S we substituted the following definition for the path
\begin{equation}
z^{\mu}\left(s\right)=x^{\mu}+s\varepsilon_{\alpha}^{\pm \mu},\,\,\, s\in\left(-\infty,+\infty\right)\, ,\label{eq:path}
\end{equation}
which represents a straight line passing through some fixed point $x^{\mu}$ along $\varepsilon_{\alpha}^{\pm\, \mu}$. $s$ is a scalar parameterizing the path and $\mathbb{P}$ denotes the path ordering with respect to the increase in $s$. 
 The four vector $\varepsilon_{\alpha}^{\pm\, \mu}$ reads
\begin{equation}
    \varepsilon_{\alpha}^{\pm\, \mu} = \varepsilon_{\perp}^{\pm\, \mu }- \alpha \eta^{\mu} \, .
    \label{eq:epsilon_alpha}
\end{equation}
where $\varepsilon_{\perp}^{\pm\, \mu }$ and $\eta^{\mu}$ were defined in Eq\eqref{eq:epsPlMin} and \eqref{eq:etavec} respectively. As a result, we get
\begin{equation}
    \varepsilon_{\alpha}^{\pm}\cdot \hat{K} = \varepsilon_{\perp}^{\pm}\cdot \hat{K} - \alpha  \hat{K}^{+} = \varepsilon_{\perp}^{\pm}\cdot \hat{K}\, .
    \label{eq:eps_dot_K}
\end{equation}
Since we work in the light cone gauge, we set $\hat{K}^{+} = 0$ to obtain the final expression above. Therefore, substituting $\varepsilon_{\perp}^{+}\cdot \hat{K} = \hat{K}^{\bullet}$ Eq.~\eqref{eq:zzbardef} in Eq.~\eqref{eq:WL_gen}, we get the following explicit form for $B^{\bullet}_a[A^{\bullet}](x)$ field
\begin{equation}
     B^{\bullet}_a[A^{\bullet}](x)=\int_{-\infty}^{\infty}d\alpha\,\mathrm{Tr}\left\{ \frac{1}{2\pi g}t^{a}\partial_{-}\, \mathbb{P}\exp\left[ig\int_{-\infty}^{\infty}\! ds\, \hat{A}^{\bullet}\left(x+s\varepsilon_{\alpha}^{+}\right)\right]\right\} \, ,
\label{eq:WL_Bbul}
\end{equation}
From Eq.~\eqref{eq:epsilon_alpha}, we see, that the Wilson lines $B^{\bullet}_a[A^{\bullet}](x)$ live on a plane spanned by $\varepsilon_{\perp}^{+} $, $ \eta$ along a generic vector of slope $\alpha$ as shown in Figure \ref{fig:Bbul_WL}. The "slope" gets integrated over. 

Although intuitive, it is worth mentioning that the Wilson line $B^{\bullet}_a[A^{\bullet}](x)$ Eq.~\eqref{eq:WL_Bbul}, representing the solution of the plus helicity field in the MHV action Eq.~\eqref{eq:MHV_action}, encodes the Self-Dual sector of the Yang-Mills theory as follows. The plane spanned by the two four vectors $\varepsilon_{\perp}^{+} $ and $ \eta$ is "Self-Dual". Any plane that satisfies the following conditions is a Self-Dual plane.
\begin{itemize}
    \item Any vector $P^\mu$ tangent to the plane is light-like i.e. $P^2=0$. This implies that the plane is a null plane. Consequently, the basis vectors spanning the plane are also light-like. 
    \item Consider a pair of tangent vectors (say, $P^{\mu}, Q^{\nu}$) and define a bivector as follows
\begin{equation}
    G^{\mu\nu}=P^{\mu}Q^{\nu}-P^{\nu}Q^{\mu}\,.
\end{equation}
It should satisfy the Self-Duality condition 
\begin{equation}
    {G}^{\mu\nu} = \ast {G}^{\mu\nu}\, ,\quad \mathrm{where} \quad \ast {G}_{\mu\nu} = -i \epsilon_{\mu\nu\alpha\beta}{G}^{\alpha\beta}\,.
\end{equation}
Above, $\ast {G}_{\mu\nu}$ is the Hodge dual.
\end{itemize}
Note, if the bivector ${G}^{\mu\nu}$, considered above, satisfies the Anti-Self-Duality condition instead, then the corresponding plane is the Anti-Self-Dual plane. Thus we see that the plane spanned by $\varepsilon_{\perp}^{-} $ and $ \eta$ is Anti-Self-Dual. These planes are commonly referred to as $\alpha$, $\beta$ planes in the twistor space terminology \cite{Ward1977a}.

From above we conclude that the plane spanned by $\varepsilon_{\perp}^{+} $ and $ \eta$ is indeed Self-Dual. This, however, implies that any Wilson line defined on this plane should be a functional of a Self-Dual gauge field (or Self-Dual connection) which in the present case is $A^{\bullet}$ (recall $A^{\bullet}$ is the solution of the Self-Dual equation Eq.~\eqref{eq:SD_EOM}). This is consistent with the definition of $B^{\bullet}_a[A^{\bullet}](x)$ Eq.~\eqref{eq:WL_Bbul}. Furthermore, the structure of $B^{\bullet}_a[A^{\bullet}](x)$ Eq.~\eqref{eq:WL_Bbul} is unique in the sense that one can easily invert it to obtain the Self-Dual gauge field  $A^{\bullet}$. Since this is non-trivial, let us elaborate. A generic Wilson line $\mathcal{W}_{(+)}^a[A^{\bullet}](y, z)$ on the $\varepsilon_{\perp}^{+}-\eta$ plane will depend on the endpoints $(y, z)$ of the path thereby making the procedure of inversion non-trivial (maybe impossible). However, if we consider the path to be a straight line passing through a point $x$ with endpoints at infinity then the Wilson line $\mathcal{W}_{(+)\, \alpha}^a[A^{\bullet}](x)$ depends only on $x$ and the slope $\alpha$ of the line. Furthermore, since we integrate over the slope, the Wilson line $\mathcal{W}_{(+)}^a[A^{\bullet}](x)$ becomes independent of the path details and can now be easily inverted to derive the Self-Dual gauge field. 

Finally, one can verify that the momentum space expression for $B^{\bullet}_a[A^{\bullet}](x)$ Eq.~\eqref{eq:WL_Bbul} is indeed Eq.~\eqref{eq:Bbullet_sol1}. Since this was done explicitly in \cite{Kotko2017}, we will not repeat it. Here we highlight the main steps necessary to derive the expression. We begin by expanding the path ordered exponential as shown below
\begin{multline}
\widetilde{B}_{a}^{\bullet}(x^+;\mathbf{P})=\frac{1}{2\pi g}\int d^{3}\mathbf{x}\, e^{i\mathbf{x}\cdot\mathbf{P}}\int d\alpha\,\partial_{-} \Bigg[1+
ig\int_{-\infty}^{+ \infty}\! ds_1\,  {{A}}_a^\bullet (x^+;\mathbf{x}+s_{1}\mathbf{e}_{\alpha})\\
    +  (ig)^2\int_{-\infty}^{+ \infty} \! ds_1\,\int_{-\infty}^{s_1}\!  ds_2\, {{A}}_{b_{1}}^\bullet (x^+;\mathbf{x}+s_{1}\mathbf{e}_{\alpha}) {{A}}_{b_{2}}^\bullet(x^+;\mathbf{x}+s_{2}\mathbf{e}_{\alpha}) \mathrm{Tr}\left(t^{a}t^{b_{1}} t^{b_{2}}\right)+ \dots\\
\left(ig\right)^{n}\int_{-\infty}^{+\infty}ds_{1}\int_{-\infty}^{s_{1}}ds_{2} \dots\int_{-\infty}^{s_{n-1}}ds_{n}\,
A_{b_{1}}^{\bullet}\left(x^+;\mathbf{x}+s_{1}\mathbf{e}_{\alpha}\right)\dots A_{b_{n}}^{\bullet}\left(x^+;\mathbf{x}+s_{n}\mathbf{e}_{\alpha}\right)\,\mathrm{Tr}\left(t^{a}t^{b_{1}}\dots t^{b_{n}}\right)\\
+ \dots \Bigg]\,.
\end{multline}
Above we use 
\begin{equation}
    \mathbf{e}_{\alpha} \equiv (-\alpha, 0, -1)\,,
\end{equation}
to represent  the $(x^-, x^\bullet, x^\star)$ coordinates of  $\varepsilon_{\alpha}^{+}$. The $n$th order term in the expansion above can be rewritten as
\begin{multline}
\frac{1}{2\pi g}\left(ig\right)^{n}\int d^{3}\mathbf{x}\, e^{i\mathbf{x}\cdot\mathbf{P}}\int d^{3}\mathbf{p}_{1}\dots d^{3}\mathbf{p}_{n}\int d\alpha\,\partial_{-}\int_{-\infty}^{+\infty}ds_{1}\int_{-\infty}^{s_{1}}ds_{2}\dots\int_{-\infty}^{s_{n-1}}ds_{n}\\
e^{-i\mathbf{x}\cdot\left(\mathbf{p}_{1}+\dots+\mathbf{p}_{n}\right)}\, e^{-is_{1}\mathbf{e}_{\alpha}\cdot\mathbf{p}_{1}}\dots e^{-is_{n}\mathbf{e}_{\alpha}\cdot\mathbf{p}_{n}}\widetilde{A}_{b_{1}}^{\bullet}\left(x^+;\mathbf{p}_{1}\right)\dots \widetilde{A}_{b_{n}}^{\bullet}\left(x^+;\mathbf{p}_{n}\right)\,\mathrm{Tr}\left(t^{a}t^{b_{1}}\dots t^{b_{n}}\right)\,.
\label{eq:NthWLcoef}
\end{multline}
 For the ordered integrals we use the following identity
\begin{multline}
\int_{-\infty}^{+\infty}ds_{1}\int_{-\infty}^{s_{1}}ds_{2}\dots\int_{-\infty}^{s_{n-1}}ds_{n}\,
 e^{-is_{1}\mathbf{e}_{\alpha}\cdot\mathbf{p}_{1}}\dots e^{-is_{n}\mathbf{e}_{\alpha}\cdot\mathbf{p}_{n}} \\
=2\pi\,\delta\left(\mathbf{e}_{\alpha}\cdot\mathbf{p}_{1\dots n}\right)\frac{i^{n-1}}{\left(\mathbf{e}_{\alpha}\cdot\mathbf{p}_{2\dots n}+i\epsilon\right)\left(\mathbf{e}_{\alpha}\cdot\mathbf{p}_{3\dots n}+i\epsilon\right)\dots\left(\mathbf{e}_{\alpha}\cdot\mathbf{p}_{n}+i\epsilon\right)}\,,
\label{eq:eikonals}
\end{multline}
where $\mathbf{p}_{1\dots n} \equiv \mathbf{p}_{1} + \dots + \mathbf{p}_{n}$. Notice the delta $\delta(\mathbf{e}_{\alpha}\cdot \mathbf{p}_{1\cdots n}) \equiv  \delta(- \alpha\, {p}^+_{1\dots n} + {p}^{\bullet}_{1\dots n} )$ on the R.H.S of the expression above. Using it we can integrate out $\alpha$. Beyond the triviality of this integration, there is something interesting worth mentioning. Recall, prior to the integration over $\alpha$ the Wilson line $B^{\bullet}_a[A^{\bullet}](x)$ Eq.~\eqref{eq:WL_Bbul} is along 
\begin{equation}
    \varepsilon_{\alpha}^{+ \mu} = \varepsilon_{\perp}^{+ \mu }- \alpha \eta^{\mu} \, .
    \label{eq:varep_dis}
\end{equation}
The integration fixes the slope of the Wilson line to $\alpha\, \equiv {p}^{\bullet}_{1\dots n}/ {p}^+_{1\dots n}$. Furthermore, notice, the above four vector has the form of a generic polarization (\emph{cf.} Eq.\eqref{eq:PolarizationVect}). The integration fixes the polarization to $\varepsilon_{\alpha}^{+ \mu} \longrightarrow \varepsilon_{p_{1 \dots n}}^{+ \mu}  \equiv \varepsilon_{P}^{+ \mu}$ (we used momentum conservation $P=p_{1 \dots n}$) of the Wilson line $B^{\bullet}_a[A^{\bullet}](x)$ Eq.~\eqref{eq:WL_Bbul}. Following this, we see that each term in the denominator of Eq.~\eqref{eq:eikonals} reduces to
 \begin{equation}
    \left(p_i + \dots + p_j \right) \cdot \varepsilon^+_{p_{1 \dots n}} = -\widetilde{v}^{\star}_{(i\dots j)(1\dots n)} \, .
\end{equation}
Substituting this to Eq.~\eqref{eq:eikonals} we get 
\begin{equation}
    \left(-g\right)^{n-1}\delta^{3}\left(\mathbf{p}_{1\dots n}-\mathbf{P}\right)\\
\frac{1}{\widetilde{v}_{1\left(1\dots n\right)}^{\star}\widetilde{v}_{\left(12\right)\left(1\dots n\right)}^{\star}\dots\widetilde{v}_{\left(1\dots n-1\right)\left(1\dots n\right)}^{\star}}\,\mathrm{Tr}\left(t^{a}t^{b_{1}}\dots t^{b_{n}}\right)\,.
\end{equation}
Above we suppressed $i\epsilon$ for brevity. The above expression is exactly $\widetilde{\Gamma}_{n}^{a\{b_{1}\dots b_{n}\}}\left(\mathbf{P};\{\mathbf{p}_{1},\dots,\mathbf{p}_{n}\}\right)$ from Eq.~\eqref{eq:Gamma_n}.

\subsection{\texorpdfstring{${\hat B}^{\star}[A^{\bullet}, A^{\star}]$}{Bstar} as a straight infinite Wilson line}
\label{subsec:Bstar[A]_WL}

The Self-Dual sector is encoded in the solution ${\widetilde A}^{\star}[B^{\bullet}, B^{\star}]$ Eq.~\eqref{eq:A_star_solu} for the negative helicity field in Mansfield transformation in pretty much the same fashion as ${\widetilde A}^{\bullet}[B^{\bullet}]$. This should be already apparent from the relation between their kernels
\begin{equation}
    {\widetilde \Omega}_{n}^{a b_1 \left \{b_2 \cdots b_n \right \} }(\mathbf{P}; \mathbf{p}_{1} , \left \{ \mathbf{p}_{2} , \dots ,\mathbf{p}_{n} \right \} ) = n \left(\frac{p_1^+}{p_{1\cdots n}^+}\right)^2 {\widetilde \Psi}_{n}^{a b_1 \cdots b_n }(\mathbf{P};  \mathbf{p}_{1},  \dots , \mathbf{p}_{n}) \, .
    \label{eq:omega-Psi-pair}
\end{equation}
The two kernels differ by an overall factor of the ratio of the plus components of certain momentum and a number $n$. The striking similarities between the two sets of kernels indicate that upon inversion, the minus helicity field in the MHV action  ${\widetilde B}^{\star}[A^{\bullet}, A^{\star}]$ should have a similar structure as the plus helicity field ${\widetilde B}^{\bullet}[A^{\bullet}]$. Motivated by this, we explored ${\widetilde B}^{\star}[A^{\bullet}, A^{\star}]$ in \cite{Kakkad2020} where we found the interesting connections summarized below.

\begin{figure}
    \centering
    \includegraphics[width=10cm]{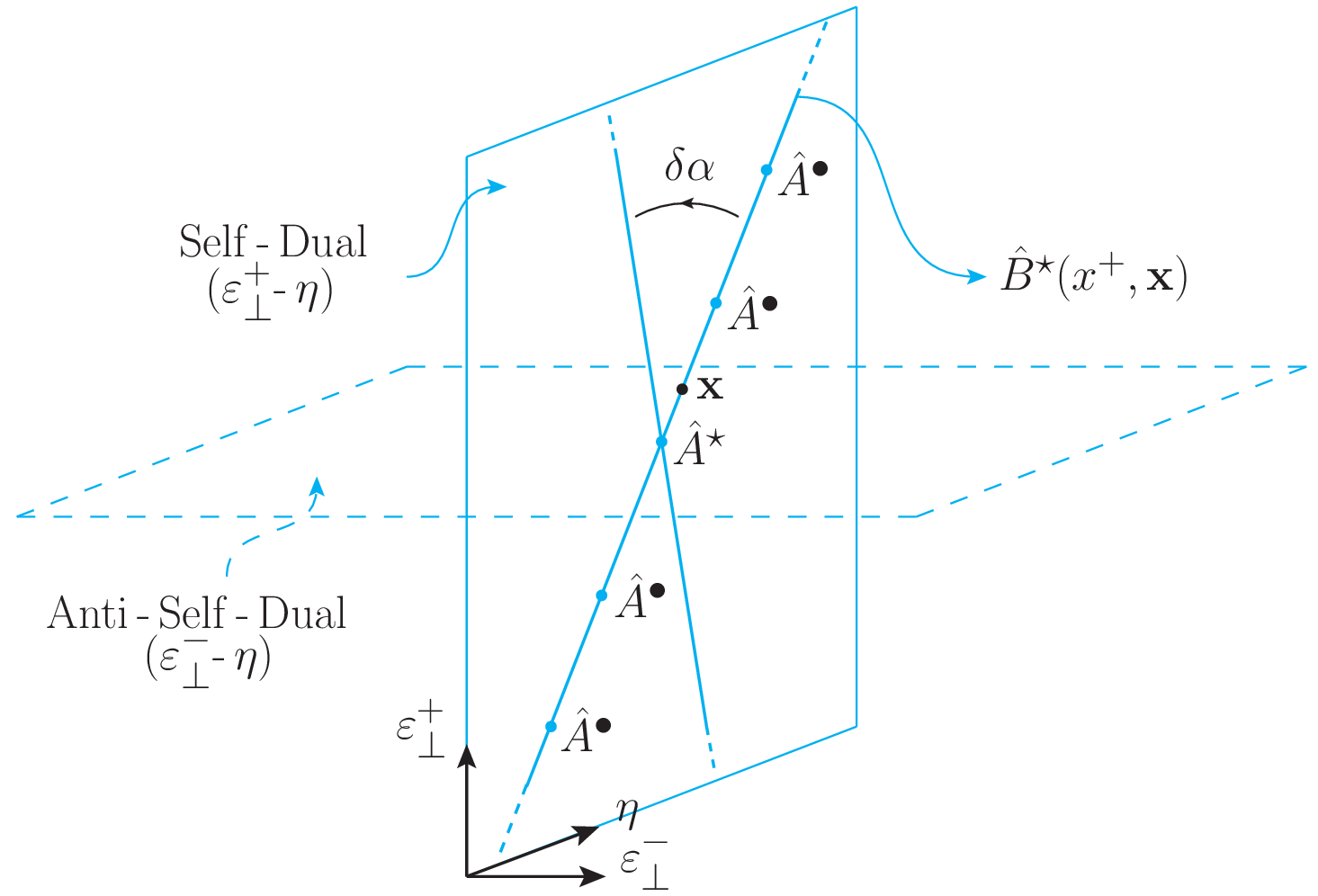}
    \caption{\small
     The minus helicity field $B^{\star}_a[A^{\bullet}, A^{\star}](x^+,\mathbf{x})$ in the MHV action is given by a similar  straight infinite Wilson line as the plus helicity field $B^{\bullet}_a[A^{\bullet}](x^+,\mathbf{x})$ (see Figure \ref{fig:Bbul_WL}) on the Self-Dual plane with an insertion of $\hat{A}^{\star}$ field somewhere on the line.
}
    \label{fig:Bstar_WL}
\end{figure}

First, we derived the momentum space expression for ${\widetilde B}^{\star}[A^{\bullet}, A^{\star}]$. The derivation can be found in Appendix \ref{subsec:app_A41}. Here we simply summarize the procedure. We start by assuming the following series expansion for  ${\widetilde B}^{\star}[A^{\bullet}, A^{\star}]$
\begin{multline}
    {\widetilde{B}}_a^\star (x^+;\mathbf{P}) = {\widetilde A}^\star_{a} (x^+;\mathbf{P}) + \sum_{n=2}^{\infty} \int\!d^{3}\mathbf{p}_{1}\cdots d^{3}\mathbf{p}_{n} {\widetilde \Upsilon}_{n}^{a b_1 \left \{b_2 \cdots b_n \right \} }(\mathbf{P}; \mathbf{p}_{1} ,\left \{ \mathbf{p}_{2} , \cdots \mathbf{p}_{n} \right \})  \\
     {\widetilde A}^\star_{b_1} (x^+;\mathbf{p}_{1}){\widetilde A}^\bullet_{b_2} (x^+;\mathbf{p}_{2}) \cdots {\widetilde A}^\bullet_{b_n} (x^+;\mathbf{p}_{n}) \, .
   \label{eq:Bstar_rep}
\end{multline}
To obtain the kernels ${\widetilde \Upsilon}_{n}^{a b_1 \left \{b_2 \cdots b_n \right \} }(\mathbf{P}; \mathbf{p}_{1} ,\left \{ \mathbf{p}_{2} , \cdots \mathbf{p}_{n} \right \})$ we substitute the expression for ${\widetilde A}^{\bullet}[B^{\bullet}]$ and ${\widetilde A}^{\star}[B^{\bullet},B^{\star}]$ using Eq.~\eqref{eq:A_bull_solu}-\eqref{eq:A_star_solu} on the R.H.S of the expression above and equate terms with equal order in fields. By doing this, we get
\begin{equation}
    {\widetilde \Upsilon}_{n}^{a b_1 \left \{b_2 \cdots b_n \right \} }(\mathbf{P}; \mathbf{p}_{1} ,\left \{ \mathbf{p}_{2} , \cdots \mathbf{p}_{n} \right \}) = n\left(\frac{p_1^+}{p_{1\cdots n}^+}\right )^2 {\widetilde \Gamma}_{n}^{a b_1 \cdots b_n }(\mathbf{P}; \mathbf{p}_{1}  \cdots \mathbf{p}_{n} ) \, ,
    \label{eq:Upsilon_n_rep}
\end{equation}
where
\begin{multline}
    \widetilde{\Gamma}_n^{a b_1b_2 \dots b_n}(\mathbf{P};\mathbf{p}_1,\mathbf{p}_2, \dots \mathbf{p}_n) = \frac{(-g)^{n-1}}{n!}\delta^3(\mathbf{p}_{12 \dots n} - \mathbf{P})\\
    \!\!\sum_{\text{\scriptsize permutations}}
    \frac{1}{\widetilde{v}_{1\left(1\dots n\right)}^{\star}\widetilde{v}_{\left(12\right)\left(1\dots n\right)}^{\star}\dots\widetilde{v}_{\left(1\dots n-1\right)\left(1\dots n\right)}^{\star}}\mathrm{Tr}(t^a t^{b_1} t^{b_2}\dots  t^{b_n})\,.
\end{multline}
Note the subtle difference between the kernels $\widetilde{\Gamma}_n^{a\{b_1\dots b_n\}}(\mathbf{P};\{\mathbf{p}_1,\dots ,\mathbf{p}_n\})$ in Eq.~\eqref{eq:Gamma_n} and $\widetilde{\Gamma}_n^{a b_1b_2 \dots b_n} (\mathbf{P};\mathbf{p}_1,\mathbf{p}_2, \dots \mathbf{p}_n)$. Furthermore, the pair of kernels $\left\{{\widetilde \Upsilon}_{n}, {\widetilde \Gamma}_{n} \right\}$ in Eq.~\eqref{eq:Upsilon_n_rep} are related in exactly the same way as the pair $\left\{{\widetilde \Omega}_{n}, {\widetilde \Psi}_{n} \right\}$ in Eq.~\eqref{eq:omega-Psi-pair}. Furthermore, given that the kernels $\widetilde{\Gamma}_n$ are momentum space coefficients of a straight infinite Wilson line, the result Eq.~\eqref{eq:Upsilon_n_rep} suggests that in position space a similar Wilson line type structure should exist for the minus helicity field ${\hat B}^{\star}[A^{\bullet}, A^{\star}]$ in the MHV action. Indeed, in \cite{Kakkad2020}, we discovered the following structure for the latter
\begin{multline}
    B_a^{\star}(x) = 
    \int_{-\infty}^{+\infty}\! d\alpha\,\, 
    \mathrm{Tr} \Big\{
    \frac{1}{2\pi g} t^a \partial_-^{-1} 
    \int\! d^4y \,
     \left[\partial_-^2 {A}_c^{\star}(y)\right]  \\
     \frac{\delta}{\delta {A}_c^{\bullet}(y)} \,
    \mathbb{P} \exp {\left[ig \int_{-\infty}^{+ \infty}\! ds \,
    \hat{A}^{\bullet}(x+s\varepsilon_{\alpha}^+)\right]  } 
    \Big\} \,.
    \label{eq:Bstar_WL}
\end{multline}
Although, the above expression seems fairly complicated, using the Wilson line expression for $B^{\bullet}_a[A^{\bullet}](x)$ Eq.~\eqref{eq:WL_Bbul}, it can be compactly re-written as
\begin{equation}
    B_a^{\star}(x) = 
    \int\! d^3\mathbf{y} \,
     \left[ \frac{\partial^2_-(y)}{\partial^2_-(x)} \,
     \frac{\delta B_a^{\bullet}(x^+;\mathbf{x})}{\delta {A}_c^{\bullet}(x^+;\mathbf{y})} \right] 
     {A}_c^{\star}(x^+;\mathbf{y})
      \, .
    \label{eq:Bstar_Bbullet}
\end{equation}
Already at this stage, one can see that the above result has the required structure to obtain the kernels in Eq.~\eqref{eq:Upsilon_n_rep}. The functional derivative facilitates the replacement of an $A^{\bullet}$ field by the $A^{\star}$ field in the definition Eq.~\eqref{eq:Bstar_Bbullet}. To see this let us rewrite the $B^{\bullet}_a[A^{\bullet}](x)$ Wilson line Eq.~\eqref{eq:WL_Bbul} as follows
\begin{equation}
     B^{\bullet}_a[A^{\bullet}](x)=\int_{-\infty}^{\infty}d\alpha\,\mathrm{Tr}\left\{ \frac{1}{2\pi g}t^{a}\partial_{-}\,\mathcal{W}(+\infty, -\infty) \right\} \, ,
\label{eq:WL_Bbul-dr}
\end{equation}
where 
\begin{equation}
    \mathcal{W}(+\infty, -\infty) = \mathbb{P}\exp\left[ig\int_{-\infty}^{\infty}\! ds\, \hat{A}^{\bullet}\left(x+s\varepsilon_{\alpha}^{+}\right)\right]\,.
\end{equation}
The arguments of $\mathcal{W}(+\infty, -\infty)$ represent the final and initial point respectively of the Wilson line. The functional derivative of a Wilson line with respect to the gauge field is given as \cite{FDWL}
\begin{equation}
    \frac{\delta \mathcal{W}(+\infty, -\infty)}{\delta {A}_c^{\bullet}(x^+;\mathbf{y})} = ig\int_{-\infty}^{\infty}\! ds\, \mathcal{W}(+\infty, s)\,  \delta^3(\mathbf{x}+s\varepsilon_{\alpha}^{+}- \mathbf{y})\, t^c\, \mathcal{W}(s, -\infty)\,.
\end{equation}
Therefore, we can re-write Eq.~\eqref{eq:Bstar_Bbullet} as
\begin{multline}
    B_a^{\star}(x) = 
    \int\! d^3\mathbf{y} \,
     \Bigg[ \frac{\partial^2_-(y)}{\partial^2_-(x)} \,
     \int_{-\infty}^{\infty}d\alpha\,\mathrm{Tr}\Bigg\{ \frac{1}{2\pi g}t^{a}\partial_{-}\, \int_{-\infty}^{\infty}\! ds\, \mathcal{W}(+\infty, s)\, \\     \delta^3(\mathbf{x}+s\varepsilon_{\alpha}^{+}- \mathbf{y})\, ig\, {A}_c^{\star}(x^+;\mathbf{y}) t^c\, \mathcal{W}(s, -\infty)\Bigg\} \Bigg] 
      \, .
    \label{eq:Bstar_Bbullet_FD}
\end{multline}
Integrating out $\mathbf{y}$ we get
\begin{multline}
    B_a^{\star}(x) =   
     \Bigg[ \frac{1}{\partial^2_-(x)} \,
     \int_{-\infty}^{\infty}d\alpha\,\mathrm{Tr}\Bigg\{ \frac{1}{2\pi g}t^{a}\partial_{-}\, \int_{-\infty}^{\infty}\! ds\, \mathcal{W}(+\infty, s)\, \\  ig\, \partial^2_-{A}_c^{\star}(x^+;\mathbf{x}+s\varepsilon_{\alpha}^{+}) t^c\, \mathcal{W}(s, -\infty)\Bigg\} \Bigg] 
      \, ,
    \label{eq:Bstar_Bbullet_FD2}
\end{multline}
Indeed, the above result shows that the functional derivative splits the Wilson line $B^{\bullet}_a[A^{\bullet}](x)$ into two semi-infinite Wilson lines, and at the point of differentiation, ${A}_c^{\star}$ field is introduced which again \textit{patches} the two semi-infinite Wilson lines into one "impure" (due to the replacement ${A}^{\bullet}\longrightarrow {A}^{\star}$) straight infinite Wilson line (shown in Figure \ref{fig:Bstar_WL}). As a result, in momentum space, we retrieve back the $\widetilde{\Gamma}_n$ kernel of the $B^{\bullet}_a[A^{\bullet}](x)$ Wilson line Eq.~\eqref{eq:WL_Bbul}. Furthermore, this replacement could be anywhere on the line. In momentum space, for $n$-th order in expansion, this introduces the factor of $n$ in $\widetilde{\Upsilon}_n$. Finally, the operator $\partial^2_-(y) / \partial^2_-(x)$ results in ratio of the plus component momenta. In Appendix \ref{subsec:app_A42}, we explicitly follow this procedure to demonstrate that the momentum space  kernels obtained from  Eq.~\eqref{eq:Bstar_Bbullet} are indeed ${\widetilde \Upsilon}_{n}^{a b_1 \left \{b_2 \cdots b_n \right \} }(\mathbf{P}; \mathbf{p}_{1} ,\left \{ \mathbf{p}_{2} , \cdots \mathbf{p}_{n} \right \})$ Eq.~\eqref{eq:Upsilon_n_rep}.

\begin{figure}
    \centering
    \includegraphics[width=10cm]{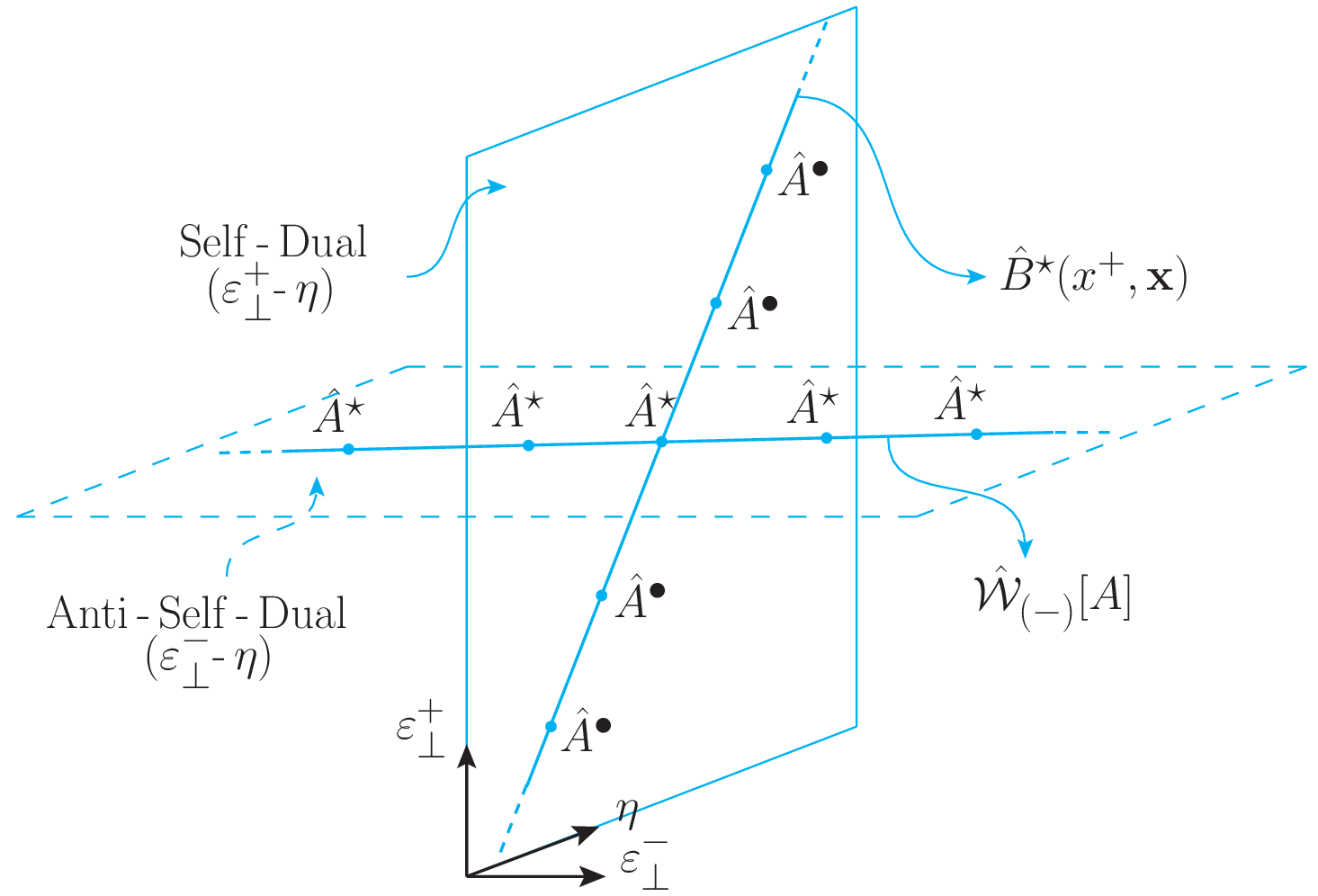}
    \caption{\small
     The $\hat{A}^{\star}$ field in the Wilson line description of $B^{\star}_a[A^{\bullet}, A^{\star}](x^+,\mathbf{x})$ can be thought as belonging to the $\mathcal{W}_{(-)}^a[A](x)$ Wilson line on the $\varepsilon_{\perp}^{-}-\eta$ plane. This leads to the realization that the $B^{\star}_a[A^{\bullet}, A^{\star}](x^+,\mathbf{x})$ field is probably a slice of a bigger geometric object spanning both the Self-Dual and the Anti-Self-Dual planes.
}
    \label{fig:Bstar_WL_BSD}
\end{figure}

Geometrically, although the minus helicity field ${\hat B}^{\star}[A^{\bullet}, A^{\star}]$ in the MHV action appears to be a similar straight infinite Wilson line as the plus helicity field ${\hat B}^{\bullet}[A^{\bullet}]$, the former is far more intriguing due to the "impurity" of an ${\hat A}^{\star}$ field. Recall, $B^{\bullet}_a[A^{\bullet}](x)=\mathcal{W}_{(+)}^a[A](x)$ Eq.~\eqref{eq:WL_Bbul}. This relation implies that one can think of the ${\hat A}^{\bullet}$ field as belonging to the Wilson line  $\mathcal{W}_{(+)}^a[A](x)$ spanning the $\varepsilon_{\perp}^{+} $, $ \eta$ plane. The question arises if the ${\hat A}^{\star}$ field could belong to the Wilson line  $\mathcal{W}_{(-)}^a[A](x)$ spanning the $\varepsilon_{\perp}^{-} $, $ \eta$ plane as follows
\begin{equation}
     \mathcal{W}^{a}_{(-)}[A](x)=\int_{-\infty}^{\infty}d\alpha\,\mathrm{Tr}\left\{ \frac{1}{2\pi g}t^{a}\partial_{-}\, \mathbb{P}\exp\left[ig\int_{-\infty}^{\infty}\! ds\, \varepsilon_{\alpha}^{-}\cdot \hat{A}\left(x+s\varepsilon_{\alpha}^{-}\right)\right]\right\} \, ,
\label{eq:WL_min}
\end{equation}
where $\varepsilon_{\alpha}^{-}\cdot \hat{A} = \hat{A}^{\star}$ (\emph{cf.} Eq.\eqref{eq:eps_dot_K}). As a result, one can imagine a geometric object consisting of a pair of Wilson lines spanning over both the $\varepsilon_{\perp}^{+}-\eta$ and $\varepsilon_{\perp}^{-}-\eta$ planes such that ${\hat B}^{\star}[A^{\bullet}, A^{\star}]$ is just a slice of this object. We demonstrate this diagrammatically in Figure \ref{fig:Bstar_WL_BSD}.
Following this geometric curiosity led us to derive a new Wilson line based action in \cite{Kakkad:2021uhv}. We discuss this in detail in Chapter \ref{WLAc-chapter}.

\section{Geometric exploration of the MHV vertices}
\label{sec:MHV_vert_TS}

In this section, we intend to summarize the very interesting geometric structure of the MHV vertices in terms of the straight infinite Wilson lines on the Self-Dual plane in the complexified Minkowski space $\mathbb{M}_\mathbb{C}$. 

In the previous Subsections \ref{subsec:Bbul[A]_WL} and \ref{subsec:Bstar[A]_WL}, following the work in \cite{Kotko2017, Kakkad2020}, we demonstrated that the plus ${\hat B}^{\bullet}[A^{\bullet}]$ and the minus ${\hat B}^{\star}[A^{\bullet}, A^{\star}]$ helicity fields  in the MHV action Eq.~\eqref{eq:MHV_action} are given by straight infinite Wilson lines Eq.~\eqref{eq:WL_Bbul}-\eqref{eq:Bstar_WL} on a the Self-Dual plane ($\varepsilon_{\perp}^{+}-\eta$) in complexified Minkowski space $\mathbb{M}_\mathbb{C}$ (see Figures \ref{fig:Bbul_WL} and \ref{fig:Bstar_WL}).

Now, consider a generic $n$-point vertex in the MHV action Eq.~\eqref{eq:MHV_action}. It has the following form in momentum space Eq.~\eqref{eq:MHV_n_point}
\begin{multline}
\mathcal{L}_{--+\dots+}\left[B^{\bullet},B^{\star}\right]=\int d^{3}\mathbf{p}_{1}\dots d^{3}\mathbf{p}_{n}\,\,\delta^{3}\left(\mathbf{p}_{1}+\dots+\mathbf{p}_{n}\right)\,
\widetilde{\mathcal{V}}_{--+\dots+}^{b_{1}\dots b_{n}}\left(\mathbf{p}_{1},\dots,\mathbf{p}_{n}\right)
\\ \widetilde{B}_{b_{1}}^{\star}\left(x^+;\mathbf{p}_{1}\right)\widetilde{B}_{b_{2}}^{\star}\left(x^+;\mathbf{p}_{2}\right)\widetilde{B}_{b_{3}}^{\bullet}\left(x^+;\mathbf{p}_{3}\right)\dots\widetilde{B}_{b_{n}}^{\bullet}\left(x^+;\mathbf{p}_{n}\right)
\,.
\label{eq:MHV_ver_mom}
\end{multline}
Recall, Mansfield's transformation Eq.~\eqref{eq:Man_Transf1} was performed over the constant light-cone time surface $x^+ = const$. As a result, all the fields in the vertex Eq.~\eqref{eq:MHV_ver_position} have the same $x^+$. However the remaining three co-ordinates in momentum space $\mathbf{p}_i\equiv\left(p^{+}_i,p^{\star}_i,p^{\bullet}_i\right)$ are in general not the same $\mathbf{p}_i \neq \mathbf{p}_j \forall i \neq j$. The interesting aspect of the vertex $\widetilde{\mathcal{V}}_{--+\dots+}^{b_{1}\dots b_{n}}\left(\mathbf{p}_{1},\dots,\mathbf{p}_{n}\right)$ is that it is \textit{holomorphic} (\emph{cf.} Eq.\eqref{eq:MHV_vertex}) i.e. it depends on only one type of spinor product. In our notation it depends only on $\widetilde{v}^{\star}_{ij}$. It is therefore a rational function of the momentum components $\left\{ p^{+}_i,p^{\bullet}_i \right\}$ and is independent of $p^{\star}_i$. 

\begin{figure}
    \centering
    \includegraphics[width=4cm]{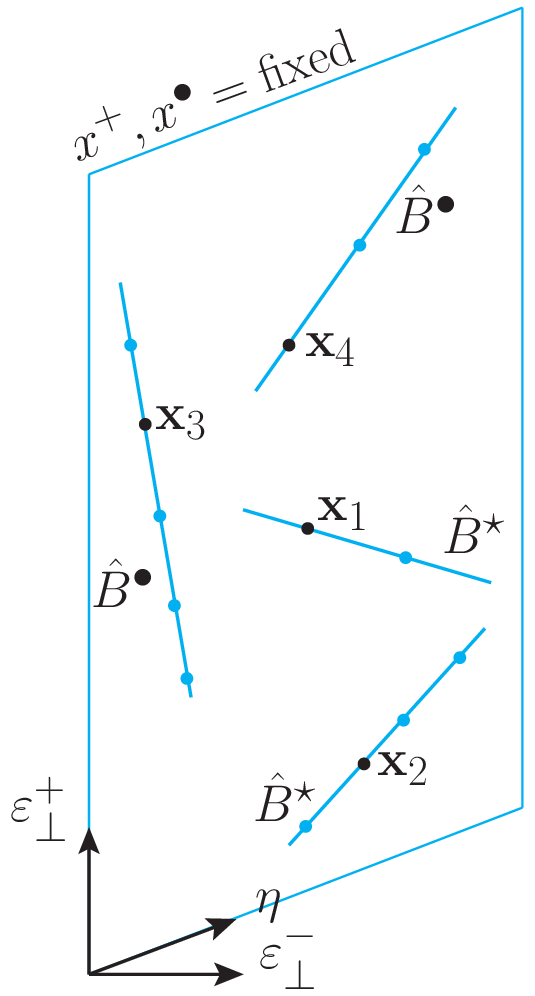}     \caption{\small
     The two plus ${B}_{b_i}^{\bullet}[A^{\bullet}](x_i)$ and the two minus helicity fields ${B}_{b_j}^{\star}[A^{\bullet}, A^{\star}](x_j)$ in the 4- point MHV vertex Eq.~\eqref{eq:MHV_ver_position} represent Wilson line on the same Self-Dual plane $\varepsilon_{\perp}^{+}-\eta$ plane with $\left\{ x^{+},x^{\bullet} \right\}$ fixed. The above representation generalizes to $n$-point MHV as well.
 }
     \label{fig:WL_origin}
 \end{figure}

Replacing the delta conserving the $p^{\star}_i$ components in Eq.~\eqref{eq:MHV_ver_mom} with the following integral ($x^{\bullet}$ is conjugate to ${p}^{\star}$)
\begin{equation}
\delta\left({p}^{\star}_{1}+\dots+{p}^{\star}_{n}\right) = \int d x^{\bullet}\,\,e^{ix^{\bullet}\,\left({p}^{\star}_{1}+\dots+{p}^{\star}_{n}\right)}\,.
\end{equation}
And then performing the Fourier transform, we get
\begin{multline}
\mathcal{L}_{--+\dots+}\left[B^{\bullet},B^{\star}\right]=\int d x^{\bullet}\int d^{2}{p}_{1}^{+, \bullet}\dots d^{2}{p}_{n}^{+, \bullet}\,\,\delta^{2}\left({p}_{1}^{+, \bullet}+\dots+{p}_{n}^{+, \bullet}\right)\,\\
\widetilde{\mathcal{V}}_{--+\dots+}^{b_{1}\dots b_{n}}\left({p}_{1}^{+, \bullet},\dots,{p}_{n}^{+, \bullet}\right)
\widetilde{B}_{b_{1}}^{\star}\left(x^+,x^{\bullet};{p}_{1}^{+, \bullet}\right)\widetilde{B}_{b_{2}}^{\star}\left(x^+,x^{\bullet};{p}_{2}^{+, \bullet}\right)\\
\widetilde{B}_{b_{3}}^{\bullet}\left(x^+,x^{\bullet};{p}_{3}^{+, \bullet}\right)\dots\widetilde{B}_{b_{n}}^{\bullet}\left(x^+,x^{\bullet};{p}_{n}^{+, \bullet}\right)
\,,
\end{multline}
 where we introduced the notation ${p}_{i}^{+, \bullet} = \left\{ p^{+}_i,p^{\bullet}_i \right\}$. 
Thus, we see that owing to the holomorphic nature of the MHV vertex,  the transverse component $x^{\bullet}$ in position space is also the same in all the fields in the MHV vertex. At this point, it is more convenient to re-write the above expression fully in position space
\begin{multline}
\mathcal{L}_{--+\dots+}\left[B^{\bullet},B^{\star}\right]=\int d^{3}\mathbf{x}_{1}\dots d^{3}\mathbf{x}_{n}\,\,
{\mathcal{V}}_{--+\dots+}^{b_{1}\dots b_{n}}\left({x}_{1}^{-, \star},\dots,{x}_{n}^{-, \star}\right)
\\ {B}_{b_{1}}^{\star}\left(x^+,x^{\bullet};{x}_{1}^{-, \star}\right){B}_{b_{2}}^{\star}\left(x^+,x^{\bullet};{x}_{2}^{-, \star}\right){B}_{b_{3}}^{\bullet}\left(x^+,x^{\bullet};{x}_{3}^{-, \star}\right)\dots{B}_{b_{n}}^{\bullet}\left(x^+,x^{\bullet};{x}_{n}^{-, \star}\right)
\,.
\label{eq:MHV_ver_position}
\end{multline}
From above we see that in position space the two components $\left\{ x^{+},x^{\bullet} \right\}$ are the same in all the fields in the MHV vertex. Only the remaining, two components $\left\{ x^{-}_i,x^{\star}_i \right\}$ (conjugate to ${p}_{i}^{+, \bullet} = \left\{ p^{+}_i,p^{\bullet}_i \right\}$ respectively) vary. This is consistent with the Wilson line description we encountered earlier. Recall, both the types of fields ${B}_{b_i}^{\bullet}[A^{\bullet}](x_i)$ and ${B}_{b_i}^{\star}[A^{\bullet}, A^{\star}](x_i)$ lie on the $\varepsilon_{\perp}^{+}-\eta$ plane along
\begin{equation}
z_i^{\mu}\left(s_i\right)=x^{\mu}_i+s_i\varepsilon_{\alpha _i}^{+ \mu},\,\,\, s_i\in\left(-\infty,+\infty\right)\, ,
\end{equation}
where $\varepsilon_{\alpha_i}^{+ \mu} = \varepsilon_{\perp}^{+ \mu }- \alpha_i \, \eta^{\mu} $.   Using $v^{-}=v\cdot\widetilde{\eta}$ and $v^{\star}=v\cdot\varepsilon_{\bot}^{-}$, we can re-write the above in the component form as
\begin{equation}
\left(z_i^ +,z^{-}_i,z^{\star}_i,z^{\bullet}_i\right)=\left(x_i^ +,x^{-}_i,x^{\star}_i,x^{\bullet}_i\right)+ \left(0,- s_i \alpha_i, -s_i,0\right)\,.
\label{eq:WL_comp}
\end{equation}
Thus we see that on the $\varepsilon_{\perp}^{+}-\eta$ plane, only the components $\left\{ x^{-}_i,x^{\star}_i \right\}$ of the Wilson lines ${B}_{b_i}^{\bullet}[A^{\bullet}](x_i)$ and ${B}_{b_i}^{\star}[A^{\bullet}, A^{\star}](x_i)$ vary whereas $\left\{ x^{+},x^{\bullet} \right\}$ remain unaltered. This indicates that all the fields in the vertex Eq.~\eqref{eq:MHV_ver_position} can be considered to localize on the same Self-Dual plane. For 4-point MHV we show this diagrammatically in Figure \ref{fig:WL_origin}

From the above discussion, we see that all the fields in a given vertex in the MHV action are geometric objects (Wilson lines to be specific) localized on a single Self-Dual plane in the complexified Minkowski space $\mathbb{M}_\mathbb{C}$. This is intertwined with the holomorphic nature of the MHV vertex. Secondly, this vertex solely results in the tree level MHV amplitude in the on-shell limit. However, the latter, owing to its holomorphic nature, localizes on a line (complex) in the twistor space as discussed in Section \ref{sec:MHV_rules}. Given these two geometric pictures, one in the complexified Minkowski space $\mathbb{M}_\mathbb{C}$ and the other in the twistor space, it is intriguing to further explore their relationship. Such an investigation goes, however, beyond the scope of the thesis and is left for future.

\chapter{A new Wilson line-based action}
\label{WLAc-chapter}

In this chapter, we summarize the work we did in \cite{Kakkad:2021uhv} where we derived a new Wilson line-based action for computing pure gluonic amplitudes. This action can be derived in two ways. First is via a direct canonical transformation of the Yang-Mills action on the light-cone. This transformation maps the kinetic term and both the triple-gluon vertices $(+ + -)$ and $(+ - -)$ of the Yang-Mills action to the kinetic term in the new action. The second way is via a pair of canonical transformations. The first transformation maps the Self-Dual part of the Yang-Mills action to a new kinetic term. This gives the MHV action (discussed in the previous chapter). The second transformation maps the  Anti-Self-Dual part of the MHV action to the kinetic term of the new action. We show that both ways give the same new action but mostly we use the latter approach to develop both the solutions of the transformation and then to derive the explicit content of the new action. 

The solutions turn out to be "Wilson lines of Wilson lines" spanning over both the Self-Dual and the Anti-Self-Dual planes. These resum both the triple-gluon vertices $(+ + -)$ and $(+ - -)$ that were eliminated by the transformation and thus account for all the tree-level contributions originating from these. Later we go on to derive the new action and show that there are no triple-point interaction vertices. The lowest interaction vertex is 4-point MHV. Higher point vertices include $\mathrm{N}^k\mathrm{MHV}$ vertices, with $1\leq k \leq n-4$ where $n$ is the total number of external legs. We validate the action by computing several tree-level amplitudes and show that the absence of triple-gluon vertices indeed results in very few diagrams when compared with the MHV action, needless to mention a lot fewer than in the Yang-Mills action.  Finally, we discuss two interesting discoveries we made recently, not published in \cite{Kakkad:2021uhv}. First is that the number of diagrams that we get when computing split-helicity tree amplitudes using our action follows a well-known number series the so-called \textit{Delannoy numbers}. This in turn not only allows us to predict the number of diagrams required to compute these amplitudes using our action but also to quantitatively compare against the number of diagrams required in other formalisms, say the MHV action. Second, the realization of the vertices in our action in twistor space. We try to argue, intuitively, that our action provides a novel prescription for computing amplitudes in terms of twistor space objects.
\section{Motivation}
\label{sec:Zac_mot}

In \cite{Kakkad:2021uhv}, we had the following two motivations to develop a new transformation that would, in turn, lead to the derivation of a new action allowing for computing pure gluonic amplitudes even more efficiently as compared to the MHV action
\begin{itemize}
    \item The CSW method established the idea that using appropriate building blocks can provide a remarkable simplicity, as compared to the Yang-Mills action, in computing on-shell tree-level pure gluonic amplitudes. The MHV action, on the other hand, demonstrated that this simplicity could also be realized for off-shell Green's function. It, however, requires a field transformation that exchanges the gauge fields in the Yang-Mills action to "inverse" Wilson lines indicating that the latter is a more appropriate degree of freedom when it comes to scattering amplitudes. For the case of MHV action, the explicit description of the plus helicity $B^{\bullet}_a[A^{\bullet}](x)$ and minus helicity $B^{\star}_a[A^{\bullet}, A^{\star}](x)$ fields in terms of $\mathcal{W}_{(+)}^a[A](x)$ Wilson line was discussed in the previous chapter. In fact, we pointed out that the $A^{\star}$ field in the latter can be thought of as belonging to $\mathcal{W}_{(-)}^a[A](x)$ Wilson line spanning the Anti-Self-Dual plane implying that $B^{\star}_a[A^{\bullet}, A^{\star}](x)$ is a cut of a bigger geometric object (intersecting $\mathcal{W}_{(\pm)}^a[A](x)$ Wilson lines) spanning both the Self-Dual and the Anti-Self-Dual planes. Recall, further that the Wilson line $B^{\bullet}_a[A^{\bullet}](x) = \mathcal{W}_{(+)}^a[A](x)$ originates from Mansfield's transformation which involves the Self-Dual vertex $(+ + -)$. Similarly, the $\mathcal{W}_{(-)}^a[A](x)$ Wilson line would originate from a transformation involving the Anti-Self-Dual vertex $(+ - -)$. This means that in order to realize the bigger structure involving both the $\mathcal{W}_{(\pm)}^a[A](x)$ Wilson lines, one needs to think of a transformation including both the triple-gluon vertices. From MHV action, we already know that introducing such degrees of freedom has been fruitful for amplitude computation.  
    \item On-shell approaches like the CSW and the BCFW recursion relations at the tree level and Unitarity-based approaches at the loop level led to the realization that the procedure for computing pure gluonic amplitudes could be greatly simplified if one uses amplitudes themselves as building blocks. For that matter, at the tree level, MHV rules are much more restrictive because they use only the MHV amplitudes. BCFW, on the other hand, uses non-zero amplitudes with all helicity configurations and therefore provides even more simplicity in computing them. This triggered us to think of developing an action whose vertices are as close to the pure gluonic scattering processes as possible (with different helicities, more in the spirit of BCFW. Note, however, BCFW does not have an "action" realization). In order to achieve this, our attention was again drawn toward the Anti-Self-Dual vertex $(+ - -)$ in the MHV action for two simple reasons. First, a triple gluon vertex, in general, is not an effective building block for computing amplitudes because it is very "small" due to which the number of Feynman diagrams grows quickly. Second, in the on-shell limit for real momenta (physical scattering), these are zero. Thus we considered eliminating it via a transformation. But the important question was, will eliminating it help us achieve the type of action we are looking for? From Mansfield's transformation, we know that the $(+ + -)$ vertex that was being eliminated in the transformation got effectively resummed in the solutions $A^{\bullet}_a[B^{\bullet}](x)$ and $A^{\star}_a[B^{\bullet}, B^{\star}](x)$. These, therefore, account for all the tree-level contributions originating from $(+ + -)$ and thus give rise to the infinite sets of MHV vertices, upon substitution to $(+ - -)$ and $(+ + - -)$ vertices in Yang-Mills action, which correspond to MHV amplitudes in the on-shell limit. Similarly, if we develop a transformation (like Mansfield's transformation) including instead both the triple gluon vertices, the solutions will account for tree-level connections originating from both the vertices and will therefore result in an action with an infinite set of interaction vertices  of different helicity configuration, not limited to a certain type.
\end{itemize}

\section{The new action}
\label{sec:Z_ACTION}

In \cite{Kakkad:2021uhv}, motivated by the above ideas, we derived a new Wilson line-based action that does not contain any triple-gluon vertices. In this section, we will discuss the step-by-step derivation of this new action from scratch starting with the general idea where we intend to provide an outline of the entire derivation and the structure of the transformation as well as the new action. Later, we will discuss the technicalities and the actual results.

\subsection{The general idea}
\label{subsec:gen_id}

The starting point is again the Yang-Mills action on the light-cone
\begin{equation}
S_{\mathrm{YM}}\left[A^{\bullet},A^{\star}\right]=\int dx^{+}\left(\mathcal{L}_{+-}+\mathcal{L}_{++-}+\mathcal{L}_{+--}+\mathcal{L}_{++--}\right)\,.\label{eq:actionLC_YM}
\end{equation}
We want a field redefinition that transforms the Yang-Mills fields into a new pair of fields
\begin{equation}
    \left\{\hat{A}^{\bullet},\hat{A}^{\star}\right\} \rightarrow \Big\{\hat{Z}^{\bullet}\big[{A}^{\bullet},{A}^{\star}\big],\hat{Z}^{\star}\big[{A}^{\bullet},{A}^{\star}\big]\Big\} \, ,
    \label{eq:general_transf}
\end{equation}
such that it maps the kinetic term $\mathcal{L}_{+-}\left[A^{\bullet},A^{\star}\right]$ as well as both the triple gluon vertices $\mathcal{L}_{++-}\left[A^{\bullet},A^{\star}\right]$ and $\mathcal{L}_{+--}\left[A^{\bullet},A^{\star}\right]$ in the Yang-Mills action to a new kinetic term $\mathcal{L}_{+-}\left[Z^{\bullet},Z^{\star}\right]$. We use $\left\{\hat{Z}^{\bullet},\hat{Z}^{\star}\right\}$ to represent the fields in the new action $S\left[Z^{\bullet},Z^{\star}\right]$. Following Mansfield's transformation \cite{Mansfield2006}, we put two more constraints. First, the transformation Eq.~\eqref{eq:general_transf} is over the constant light-cone time $x^+$. Due to this the set of new fields ${\hat Z}^{\bullet}(x^+;\mathbf{y}), {\hat Z}^{\star}(x^+;\mathbf{y})$ as well as the old ones ${\hat A}^{\bullet}(x^+;\mathbf{x}), {\hat A}^{\star}(x^+;\mathbf{x})$ have the same $x^+$. Second, the transformation is canonical. This requirement, just as in the case of Mansfield's transformation, will preserve the integral measure in the partition function up to a field-independent factor.

In order to derive the new action $S\left[Z^{\bullet},Z^{\star}\right]$ we need the explicit relations between the new $\left\{\hat{Z}^{\bullet},\hat{Z}^{\star}\right\}$ and the old $\left\{\hat{A}^{\bullet},\hat{A}^{\star}\right\}$ set of fields. To obtain these we use the generating function approach to canonical transformation. Although this exposition can be found in standard textbooks on classical mechanics, let us briefly recall the key ideas. Consider an initial set of canonical coordinates $\{q,p\}$ where $q$ is the generalized coordinate and $p$ is the momentum. Let $\mathcal{G}$ be the generating function that facilitates a canonical transformation from the initial set to a new set of canonical coordinates $\{q,p\} \longrightarrow \{Q,P\}$. The generating function $\mathcal{G}$ depends on one coordinate from the initial and one from the new set. As a result, there are four possibilities for $\mathcal{G}$ (one can also consider a mixture of these four types). Consider, for example, the generating function of the type $\mathcal{G}(q,P)$. Using this one can determine the momentum $p$ and the generalized coordinate $Q$ as follows
\begin{equation}
     p =  \frac{\partial\mathcal{G}(q,P)}{\partial q} \,, \qquad 
     Q =  \frac{\partial\mathcal{G}(q,P)}{\partial P} \,.
    \label{eq:generatingfunc_tr}
\end{equation}
Mansfield's transformation is an explicit realization of this type. To see this, recall that for the old (Yang-Mills) set of fields $\left\{\hat{A}^{\bullet},\hat{A}^{\star}\right\}$, $\partial_{-}{\hat A}^{\star}(x)$ was the momentum canonically conjugate to ${\hat A}^{\bullet}(x)$, and for the new set of fields $\left\{\hat{B}^{\bullet},\hat{B}^{\star}\right\}$,  $\partial_{-}{\hat B}^{\star}(x)$ was conjugate to ${\hat B}^{\bullet}(x)$, implying  $\{q,p\} = \{\hat{A}^{\bullet}(x), \partial_{-}\hat{A}^{\star}(x)\}$ and  $\{Q,P\} = \{\hat{B}^{\bullet}(x), \partial_{-}\hat{B}^{\star}(x)\}$. Using the generating functional  
\begin{equation}
    \mathcal{G}(q,P) \equiv  \mathcal{G}[A^\bullet, B^\star](x^+) = \int\! d^3\mathbf{x}\,\,\,\Tr\,
    \hat{\mathcal{W}}_{(+)}[A](x) \,\,
     \partial_- \hat{B}^{\star}(x) \,,
     \label{eq:Gen_func_MT}
\end{equation}
we can write the following equations (analogous to Eq.~\eqref{eq:generatingfunc_tr})
\begin{align}
    \partial_{-}A_{a}^{\star}\left(x^+;\mathbf{x}\right) = &
    \int\! d^3\mathbf{y} \,
     \left[ \,
     \frac{\delta \mathcal{W}^c_{(+)}[A](x^+;\mathbf{x})}{\delta {A}_a^{\bullet}(x^+;\mathbf{x})} \right] 
     \partial_{-}B_{c}^{\star}\left(x^+;\mathbf{y}\right)\,, \label{eq:MT_Bstar}\\
    B^{\bullet}_a[A^{\bullet}](x^+;\mathbf{x})= & \mathcal{W}_{(+)}^a[A](x^+;\mathbf{x}) \, . \label{eq:MT_Bbul}
\end{align}
The above equations are exactly Mansfield's transformation Eq.~\eqref{eq:AtoB_CT_def}. Notice, the generating function Eq.~\eqref{eq:Gen_func_MT} is defined over the constant light-cone time $x^+$ and $\hat{\mathcal{W}}_{(+)}[A](x)$ is the Wilson line defined in Eq.~\eqref{eq:WL_gen}.

For our transformation Eq.~\eqref{eq:general_transf}, on the other hand, the old and the new set of canonical coordinates read
\begin{align}
    \{q,p\} =& \{\hat{A}^{\bullet}(x), \partial_{-}\hat{A}^{\star}(x)\}\,,\\
    \{Q,P\} =& \{\hat{Z}^{\star}(x), \partial_{-}\hat{Z}^{\bullet}(x)\}\,.
\end{align}
It turns out that the generating functional for this transformation depends only on the generalized coordinates $\mathcal{G}(q,Q)$.  It reads \cite{Kakkad:2021uhv}
\begin{equation}
   \mathcal{G}(q,Q) \equiv \mathcal{G}[A^\bullet,Z^\star](x^+) =
    -\int\! d^3\mathbf{x}\,\,\,\Tr\,
     \hat{\mathcal{W}}^{\,-1}_{(-)}[Z](x)\,\,
     \partial_- \hat{\mathcal{W}}_{(+)}[A](x) \,.
    \label{eq:generatingfunc3}
\end{equation}
Above, $\mathcal{W}^{\,-1}$ is the inverse of the Wilson line $\mathcal{W}$. For a generic field $K$, it is defined as $\mathcal{W}[\mathcal{W}^{-1}[K]]=K$. The generating function is again over the constant light-cone time $x^+$. Finally, using Eq.~\eqref{eq:generatingfunc3}, we can write the following explicit relations for the conjugate momenta
\begin{align}
     p =  \frac{\partial\mathcal{G}(q,Q)}{\partial q} \,, \qquad &\implies \qquad \partial_{-}A^{\star}_a(x^+,\mathbf{y}) =  \frac{\delta \, \mathcal{G}[A^{\bullet},Z^{\star} ](x^+)}{\delta A_a^{\bullet}\left(x^+,\mathbf{y}\right)} \,, \label{eq:AtoZ_ct1}\\
     P = - \frac{\partial\mathcal{G}(q,Q)}{\partial Q} \,, \qquad &\implies \qquad \partial_{-}Z^{\bullet}_a(x^+,\mathbf{y}) = - \frac{\delta \, \mathcal{G}[A^{\bullet},Z^{\star} ](x^+)}{\delta Z_a^{\star}\left(x^+,\mathbf{y}\right)} \,. \label{eq:AtoZ_ct2}
\end{align}
Now, to derive the new action $S\left[Z^{\bullet},Z^{\star}\right]$, all that needs to be done is solve the transformations Eq.~\eqref{eq:AtoZ_ct1}-\eqref{eq:AtoZ_ct2} using the explicit form for the generating function Eq.~\eqref{eq:generatingfunc3} to obtain the solutions $\left\{\hat{A}^{\bullet}[{Z}^{\bullet},{Z}^{\star}],\hat{A}^{\star}[{Z}^{\bullet},{Z}^{\star}]\right\}$ and then substitute the latter to the Yang-Mills action Eq.~\eqref{eq:actionLC_YM}. Following Eq.~\eqref{eq:generatingfunc3}, \eqref{eq:AtoZ_ct1}-\eqref{eq:AtoZ_ct2}, the solutions $\left\{\hat{A}^{\bullet}[{Z}^{\bullet},{Z}^{\star}],\hat{A}^{\star}[{Z}^{\bullet},{Z}^{\star}]\right\}$ are assumed to have the following generic form in position space
\begin{equation}
    A_a^{\bullet}(x^+;\mathbf{x})=\sum_{n=1}^{\infty}
    \int\! d^3\mathbf{y}_1\dots d^3\mathbf{y}_n \sum_{i=1}^{n}\, \Xi_{i,n-i}^{ab_1\dots b_n}(\mathbf{x};\mathbf{y}_1,\dots,\mathbf{y}_n) \prod_{k=1}^{i}Z_{b_k}^{\bullet}(x^+;\mathbf{y}_k)
    \prod_{l=i+1}^{n}Z_{b_l}^{\star}(x^+;\mathbf{y}_l) \,,
    \label{eq:Abullet_to_Z}
\end{equation}
\begin{equation}
    A_a^{\star}(x^+;\mathbf{x})=\sum_{n=1}^{\infty}
    \int\! d^3\mathbf{y}_1\dots d^3\mathbf{y}_n \sum_{i=1}^{n}\, \Lambda_{i,n-i}^{ab_1\dots b_n}(\mathbf{x};\mathbf{y}_1,\dots,\mathbf{y}_n) \prod_{k=1}^{i}Z_{b_k}^{\star}(x^+;\mathbf{y}_k)
    \prod_{l=i+1}^{n}Z_{b_l}^{\bullet}(x^+;\mathbf{y}_l) \,,
    \label{eq:Astar_to_Z}
\end{equation}
where $\Xi_{i,n-i}^{ab_1\dots b_n}(\mathbf{x};\mathbf{y}_1,\dots,\mathbf{y}_n)$ is the kernel in the $n$-th order expansion of $A^{\bullet}$ field with $i$ representing the number of $Z^{\bullet}$ fields and $n-i$ the number of $Z^{\star}$ fields. Similarly, $\Lambda_{i,n-i}^{ab_1\dots b_n}(\mathbf{x};\mathbf{y}_1,\dots,\mathbf{y}_n)$ is the kernel in the $n$-th order expansion of $A^{\star}$ field with $i$ representing the number of $Z^{\star}$ fields and $n-i$ the number of $Z^{\bullet}$ fields. These kernels are independent of the light-cone time $x^+$. At the lowest order, the solution reads
\begin{equation}
    A_a^{\bullet}(x^+;\mathbf{x})= Z_{a}^{\bullet}(x^+;\mathbf{x})+\dots \,\,, \qquad
    A_a^{\star}(x^+;\mathbf{x})=
    Z_{a}^{\star}(x^+;\mathbf{x})+\dots \,\,.
    \label{eq:A_to_Z_zeroth}
\end{equation}
The momentum space expressions for the solutions $\left\{\widetilde{A}^{\bullet}[{Z}^{\bullet},{Z}^{\star}],\widetilde{A}^{\star}[{Z}^{\bullet},{Z}^{\star}]\right\}$ can be obtained by Fourier transforming Eq.~\eqref{eq:Abullet_to_Z}-\eqref{eq:Astar_to_Z}. 

Substituting Eq.~\eqref{eq:Abullet_to_Z}-\eqref{eq:Astar_to_Z} to the Yang-Mills action Eq.~\eqref{eq:actionLC_YM}, we get the new action $S\left[Z^{\bullet},Z^{\star}\right]$ with the following generic form
\begin{align}
S\left[Z^{\bullet},Z^{\star}\right] =\int dx^{+} \Bigg\{ & 
-\int d^{3}\mathbf{x}\,\mathrm{Tr}\,\hat{Z}^{\bullet}\square\hat{Z}^{\star} \nonumber \\
 & + \mathcal{L}_{--++}+ \mathcal{L}_{--+++}+\mathcal{L}_{--++++} + \dots \nonumber \\
& + \mathcal{L}_{---++}+ \mathcal{L}_{---+++}+\mathcal{L}_{---++++} + \dots \nonumber \\
& + \mathcal{L}_{----++}+ \mathcal{L}_{----+++}+\mathcal{L}_{----++++}+ \dots \nonumber \\
& \,\, \vdots \nonumber \\
& + \mathcal{L}_{---\dots -++}+ \mathcal{L}_{---\dots -+++}+\mathcal{L}_{---\dots -++++}+ \dots
\Bigg\}
\,.\label{eq:Z_action1}
\end{align}
Above $\mathcal{L}_{\underbrace{-\,\cdots\,-}_{m}\underbrace{+ \,\cdots\, +}_{n-m}}$  represents the $n$-point ($n\geq 4$) vertex consisting of $m$ minus helicity $Z^{\star}$ and $n-m$ plus helicity $Z^{\bullet}$ fields as shown below
\begin{equation}
    \mathcal{L}_{\underbrace{-\,\cdots\,-}_{m}\underbrace{+ \,\cdots\, +}_{n-m}}= 
   \int\!d^{3}\mathbf{y}_{1}\dots d^{3}\mathbf{y}_{n} \,\, \mathcal{U}^{b_1 \dots b_{n}}_{-\dots-+\dots+}\left(\mathbf{y}_{1},\cdots \mathbf{y}_{n}\right) 
   \prod_{i=1}^{m}Z^{\star}_{b_i} (x^+;\mathbf{y}_{i})
   \prod_{j=1}^{n-m}Z^{\bullet}_{b_j} (x^+;\mathbf{y}_{j}) \, .
   \label{eq:Z_vertex_lagr_pos}
\end{equation}

The new action $S\left[Z^{\bullet},Z^{\star}\right]$ Eq.~\eqref{eq:Z_action1}, has the following interesting features 
\begin{itemize}
    \item The smallest interaction vertex is the 4-point MHV $(- - + +)$. The action doesn't have any triple-point interaction vertices. The elimination of both the triple point interaction vertices is demonstrated in Appendix \ref{sec:app_A5}.
    \item The first row consists of the off-shell MHV interaction vertices $(- - + \dots +)$. These correspond to the MHV amplitudes in the on-shell limit. This is demonstrated in Section~\ref{sec:Zac_TR_AMP} where we compute amplitudes.
    \item The first column consists of the off-shell $\overline{\mathrm{MHV}}$ interaction vertices $(- \dots - + +)$. These correspond to the $\overline{\mathrm{MHV}}$ amplitudes in the on-shell limit. This too is demonstrated in Section \ref{sec:Zac_TR_AMP} where we compute amplitudes.
    \item There are no all-plus $(+ + \dots +)$, all-minus $(- - \dots - )$, single-minus $(- +\dots + +)$ and single-plus $(+ - \dots - -)$ type interaction vertices in the action.
    \item All of the interaction vertices in the action have a closed generic form which is easy to compute. We derive the master equation Eq.~\eqref{eq:Z_gen_ker} for any generic vertex in the action in the following Subsection.
\end{itemize}

So far, we have considered the derivation of the new action $S\left[Z^{\bullet},Z^{\star}\right]$ Eq.~\eqref{eq:Z_action1} from a generic point of view. In the following Subsection, we will discuss the details of each and every aspect of this derivation starting with the generating functional Eq.~\eqref{eq:generatingfunc3}. Precisely, we will begin by exploring the implications of this generating functional and demonstrate that the transformation Eq.~\eqref{eq:AtoZ_ct1}-\eqref{eq:AtoZ_ct2} indeed eliminates both the triple gluon vertices from the Yang-Mills action Eq.~\eqref{eq:actionLC_YM}. After that, we will consider the solutions to the transformation, both $\left\{\hat{A}^{\bullet}[{Z}^{\bullet},{Z}^{\star}],\hat{A}^{\star}[{Z}^{\bullet},{Z}^{\star}]\right\}$ as well as $\left\{\hat{Z}^{\bullet}[{A}^{\bullet},{A}^{\star}],\hat{Z}^{\star}[{A}^{\bullet},{A}^{\star}]\right\}$. For the latter we will also discuss the geometric representation in the position space. Finally, using these solutions, we will derive the generic form of any interaction vertex in the action $S\left[Z^{\bullet},Z^{\star}\right]$ Eq.~\eqref{eq:Z_action1}.
\subsection{The derivation}
\label{subsec:Zac_der}

Substituting the generating function Eq.~\eqref{eq:generatingfunc3} to Eq.~\eqref{eq:AtoZ_ct1}-\eqref{eq:AtoZ_ct2} we get the following equations
\begin{equation}
\partial_{-}A_{a}^{\star}\left(x^+;\mathbf{x}\right)=-\int d^{3}\mathbf{y}\,\, \mathcal{W}^{c\,\,-1}_{(-)}[Z](x^+;\mathbf{y})\, 
\partial_{-}\,\frac{\delta}{\delta A_{a}^{\bullet}\left(x^+;\mathbf{x}\right)} \mathcal{W}^{c}_{(+)}[A](x^+;\mathbf{y})
\,,
\label{eq:Transformation_A-_Z}
\end{equation}
\begin{equation}
\partial_{-}Z_{a}^{\bullet}\left(x^+;\mathbf{x}\right)=\int d^{3}\mathbf{y}\,\left[\frac{\delta }{\delta Z_{a}^{\star}\left(x^+;\mathbf{x}\right)}
\mathcal{W}^{c\,\,-1}_{(-)}[Z](x^+;\mathbf{y}) \right]
\,\, \partial_- \mathcal{W}^{c}_{(+)}[A](x^+;\mathbf{y}) \, .
\label{eq:Transformation_Z+}
\end{equation}
The above set of equations govern the canonical transformation Eq.~\eqref{eq:general_transf} deriving the new action $S\left[Z^{\bullet},Z^{\star}\right]$ Eq.~\eqref{eq:Z_action1} from the Yang-Mills action Eq.~\eqref{eq:actionLC_YM}. Notice, however, from Mansfield's transformation we know $ B^{\bullet}_a[A^{\bullet}](x^+;\mathbf{x})=  \mathcal{W}_{(+)}^a[A](x^+;\mathbf{x}) $ (\emph{cf.} Eq.~\eqref{eq:MT_Bbul}). Substituting this to Eq.~\eqref{eq:Transformation_A-_Z} and integrating by parts we get
\begin{equation}
    \partial_{-}A_{a}^{\star}\left(x^+;\mathbf{x}\right)=\int d^{3}\mathbf{y}\,
    \left[\partial_{-}\mathcal{W}^{c\,\,-1}_{(-)}[Z](x^+;\mathbf{y})\right]\frac{\delta B_{c}^{\bullet}\left(x^+;\mathbf{y}\right)}{\delta A_{a}^{\bullet}\left(x^+;\mathbf{x}\right)}
    \, .
    \label{eq:Z_wl_src}
\end{equation}
The above equation resembles the second equation Eq.~\eqref{eq:MT_Bstar} of Mansfield's transformation shown below
\begin{equation}
\partial_{-}A_{a}^{\star}\left(x^+;\mathbf{x}\right)=\int d^{3}\mathbf{y}\,\frac{\delta B_{c}^{\bullet}\left(x^+;\mathbf{y}\right)}{\delta A_{a}^{\bullet}\left(x^+;\mathbf{x}\right)}
\, \partial_{-}B_{c}^{\star}\left(x^+;\mathbf{y}\right) \, .
\label{eq:Transformation_A-}
\end{equation}
Equating the two we get
\begin{equation}
   B_{c}^{\star}[Z^{\star}](x) = \mathcal{W}^{c\,\,-1}_{(-)}[Z](x) \,, 
   \label{eq:B_star_Z}
\end{equation}
which could be inverted to obtain 
\begin{equation}
   Z_{c}^{\star}[B^{\star}](x) = \mathcal{W}^{c}_{(-)}[B](x) \,. 
   \label{eq:B_star_Z_1}
\end{equation}
Repeating the above procedure, using Mansfield's transformation, for Eq.~\eqref{eq:Transformation_Z+} we get
\begin{equation}
\partial_{-}Z_{a}^{\bullet}\left(x^+;\mathbf{x}\right)=\int d^{3}\mathbf{y}\,\frac{\delta B_{c}^{\star}\left(x^+;\mathbf{y}\right)}{\delta Z_{a}^{\star}\left(x^+;\mathbf{x}\right)}
\, \partial_{-}B_{c}^{\bullet}\left(x^+;\mathbf{y}\right) \, ,
\label{eq:Transformation_Z+_1}
\end{equation}
 which can also be inverted to obtain 
\begin{equation}
\partial_{-}B_{a}^{\bullet}\left(x^+;\mathbf{x}\right)=\int d^{3}\mathbf{y}\,\frac{\delta Z_{c}^{\star}\left(x^+;\mathbf{y}\right)}{\delta B_{a}^{\star}\left(x^+;\mathbf{x}\right)}
\, \partial_{-}Z_{c}^{\bullet}\left(x^+;\mathbf{y}\right) \, .
\label{eq:Transformation_Z+_2}
\end{equation}

The implication of the above exercise is that the canonical transformation generated by the function Eq.~\eqref{eq:generatingfunc3} can be broken into a two-step procedure. First is Mansfield's transformation 
\begin{equation}
B^{\bullet}_a(x^+;\mathbf{x})=\mathcal{W}_{(+)}^a[A](x^+;\mathbf{x})\,, \quad \quad
\partial_{-}A_{a}^{\star}(x^+;\mathbf{x})=\int d^{3}\mathbf{y}\,\frac{\delta B_{c}^{\bullet}(x^+;\mathbf{y})}{\delta A_{a}^{\bullet}(x^+;\mathbf{x})}\partial_{-}B_{c}^{\star}(x^+;\mathbf{y})\,,
\label{eq:AtoB_CT_defre}
\end{equation}
which converts the Yang-Mills action Eq.~\eqref{eq:actionLC_YM} into the MHV action \eqref{eq:MHV_action}. Second, a transformation generated by the following equations Eq.~\eqref{eq:B_star_Z_1}-\eqref{eq:Transformation_Z+_2}
\begin{equation}
Z_{c}^{\star}(x^+;\mathbf{x}) = \mathcal{W}^{c}_{(-)}[B](x^+;\mathbf{x}) \,, \quad \quad 
    \partial_{-}B_{a}^{\bullet}\left(x^+;\mathbf{x}\right)=\int d^{3}\mathbf{y}\,\frac{\delta Z_{c}^{\star}\left(x^+;\mathbf{y}\right)}{\delta B_{a}^{\star}\left(x^+;\mathbf{x}\right)}
\, \partial_{-}Z_{c}^{\bullet}\left(x^+;\mathbf{y}\right) \, ,
\label{eq:BtoZ_CT_def}
\end{equation}
which converts the MHV action into the new action $S\left[Z^{\bullet},Z^{\star}\right]$ Eq.~\eqref{eq:Z_action1}. 

Notice the similarities between the two set of transformations in Eq.~\eqref{eq:AtoB_CT_defre} and Eq.~\eqref{eq:BtoZ_CT_def}. In fact, the latter can be obtained from the former via two changes. First is changing the set of $B$ and $A$ fields to $Z$ and $B$ fields respectively 
\begin{equation}
    \left\{\left(\hat{B}^{\bullet},\hat{B}^{\star}\right), \left(\hat{A}^{\bullet},\hat{A}^{\star}\right)\right\} \rightarrow \left\{\left(\hat{Z}^{\bullet},\hat{Z}^{\star}\right), \left(\hat{B}^{\bullet},\hat{B}^{\star}\right)\right\}\,.
    \label{eq:field_int}
\end{equation}
The second is conjugation $\bullet\leftrightarrow\star$. This flips the helicities of the fields. It also changes the Wilson lines $\mathcal{W}_{(+)}$  to $\mathcal{W}_{(-)}$. This can also be crosschecked using  Eq.~\eqref{eq:WL_gen} from which we know that the two Wilson lines differ only in their projection along $\varepsilon_{\alpha}^+$ and $\varepsilon_{\alpha}^-$. 

Recall, however, that at the level of Lagrangian, Mansfield's transformation mapped the Self-Dual part of the Yang-Mills action to a free term
\begin{equation}
\mathcal{L}_{-+}[A^{\bullet},A^{\star}]+\mathcal{L}_{-++}[A^{\bullet},A^{\star}]
\,\, \longrightarrow \,\,
\mathcal{L}_{-+}[B^{\bullet},B^{\star}]
\,.
\label{eq:SD_MT}
\end{equation}
Applying $\left\{\left(\hat{B}^{\bullet},\hat{B}^{\star}\right), \left(\hat{A}^{\bullet},\hat{A}^{\star}\right)\right\} \rightarrow \left\{\left(\hat{Z}^{\bullet},\hat{Z}^{\star}\right), \left(\hat{B}^{\bullet},\hat{B}^{\star}\right)\right\}$ followed by conjugation $\bullet\leftrightarrow\star$ to the above mapping we get
\begin{equation}
\mathcal{L}_{-+}[B^{\bullet},B^{\star}]+\mathcal{L}_{--+}[B^{\bullet},B^{\star}]
\,\, \longrightarrow \,\,
\mathcal{L}_{-+}[Z^{\bullet},Z^{\star}]
\,,
\label{eq:BtoZtransform}
\end{equation}
which represents the Anti-Self-Dual part in the MHV action getting mapped to the kinetic term in the new action $S\left[Z^{\bullet},Z^{\star}\right]$ Eq.~\eqref{eq:Z_action1}. And, given the form of transformation in Eq.~\eqref{eq:BtoZ_CT_def}, this should indeed be the case. Furthermore, just as in the case of Mansfield's transformation, this transformation must, in fact, be a canonical transformation with a field-independent Jacobian
\begin{equation}
    \mathcal{J}  = \det \begin{vmatrix}
     \frac{\delta {\hat B}^{\star} (x^+;\mathbf{x})}
    {\delta {\hat Z}^{\star} (x^+;\mathbf{y})} 
     &\mathbb{0} \\ \\
\frac{\delta {\hat B}^{\bullet}(x^+;\mathbf{x})}
    {\delta {\hat Z}^{\star}(x^+;\mathbf{y})} 
     &\frac{\delta {\hat B}^{\bullet}(x^+;\mathbf{x})}
    {\delta {\hat Z}^{\bullet}(x^+;\mathbf{y})} 
\end{vmatrix} \,.
\label{eq:BZ_jac}
\end{equation}

\begin{figure}
    \centering
    \includegraphics[width=7cm]{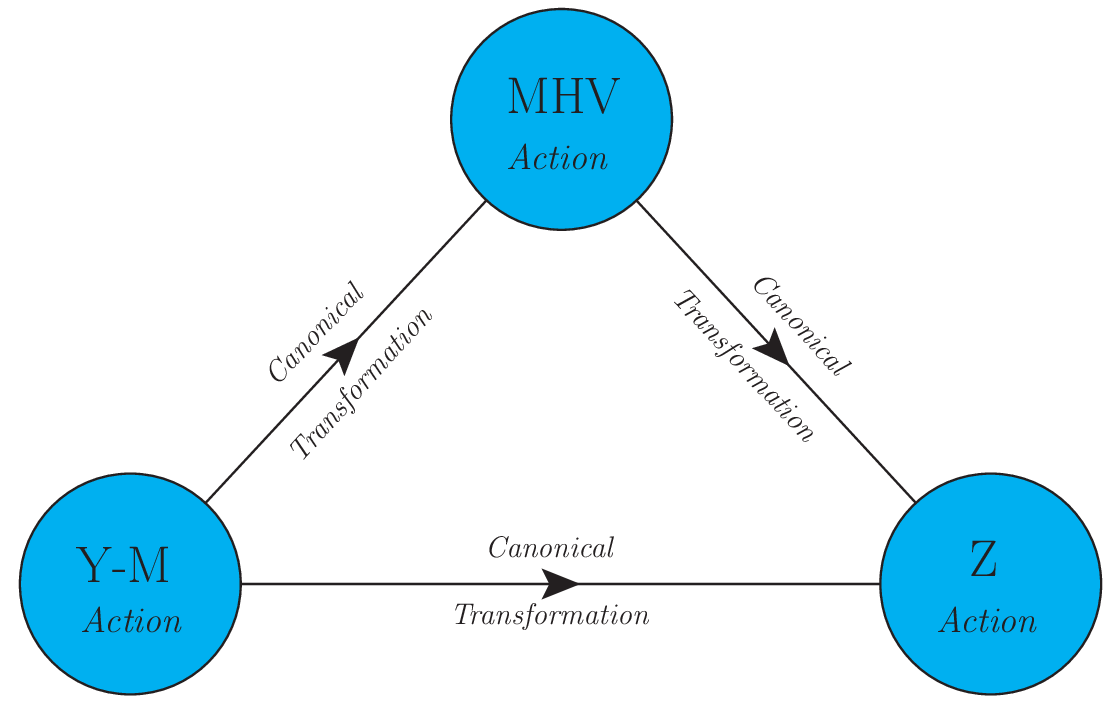}
    \caption{ 
    \small Two ways of deriving the new action $S\left[Z^{\bullet},Z^{\star}\right]$. First is the direct approach $S_{\mathrm{YM}}\left[A^{\bullet},A^{\star}\right] \longrightarrow S\left[Z^{\bullet},Z^{\star}\right]$ using the generating functional Eq.~\eqref{eq:general_transf}. The second is via two consecutive canonical transformations $S_{\mathrm{YM}}\left[A^{\bullet},A^{\star}\right] \longrightarrow S_{\mathrm{MHV}}\left[B^{\bullet},B^{\star}\right] \longrightarrow S\left[Z^{\bullet},Z^{\star}\right]$. This image was modified from our paper \cite{Kakkad:2021uhv}.
    } 
    \label{fig:CT_paths}
\end{figure}

With this, we conclude that the new action $S\left[Z^{\bullet},Z^{\star}\right]$ Eq.~\eqref{eq:Z_action1} can be derived from the Yang-Mills action Eq.~\eqref{eq:actionLC_YM} in two ways. The first is direct, via a canonical transformation using the generating functional  Eq.~\eqref{eq:general_transf}.  The second is via two consecutive canonical transformations: Yang-Mills to MHV action eliminating the  $(- + +)$ triple gluon vertex and then MHV action to the new action eliminating the $(- - +)$ triple gluon vertex. These two ways are shown in Figure \ref{fig:CT_paths}.

Owing to the above correspondence, we can obtain the solution $\left\{\hat{Z}^{\bullet}[{A}^{\bullet},{A}^{\star}],\hat{Z}^{\star}[{A}^{\bullet},{A}^{\star}]\right\}$  of our transformation from the solution $\left\{\hat{B}^{\bullet}[{A}^{\bullet}],\hat{B}^{\star}[{A}^{\bullet},{A}^{\star}]\right\}$ of the Mansfield's transformation via the two-step procedure: interchanging the fields Eq.~\eqref{eq:field_int} followed by conjugation $\bullet\leftrightarrow\star$. By doing this to the position space solutions for $\hat{B}^{\bullet}[{A}^{\bullet}]$ Eq.~\eqref{eq:WL_Bbul} and $\hat{B}^{\star}[{A}^{\bullet},{A}^{\star}]$ Eq.~\eqref{eq:Bstar_Bbullet} we get
\begin{align}
    Z_{a}^{\star}[B^{\star}](x^+;\mathbf{x}) = & \mathcal{W}^{a}_{(-)}[B](x^+;\mathbf{x}) \,, \\
    = & \int_{-\infty}^{\infty}d\alpha\,\mathrm{Tr}\left\{ \frac{1}{2\pi g}t^{a}\partial_{-}\, \mathbb{P}\exp\left[ig\int_{-\infty}^{\infty}\! ds\, \varepsilon_{\alpha}^{-}\cdot \hat{B}\left(x+s\varepsilon_{\alpha}^{-}\right)\right]\right\} \, ,
    \label{eq:Zstar_WL}
\end{align}
where $\varepsilon_{\alpha}^{-\, \mu} = \varepsilon_{\perp}^{-\, \mu }- \alpha \eta^{\mu}$ and 
\begin{equation}
    Z_a^{\bullet}[B^{\bullet},B^{\star}](x^+;\mathbf{x}) = 
    \int\! d^3\mathbf{y} \,
     \left[ \frac{\partial^2_-(y)}{\partial^2_-(x)} \,
     \frac{\delta \mathcal{W}^a_{(-)}[B](x^+;\mathbf{x})}{\delta {B}_c^{\star}(x^+;\mathbf{y})} \right] 
     {B}_c^{\bullet}(x^+;\mathbf{y})
      \, .
      \label{eq:Zbul_WL}
\end{equation}
\begin{figure}
    \centering
    \includegraphics[width=11cm]{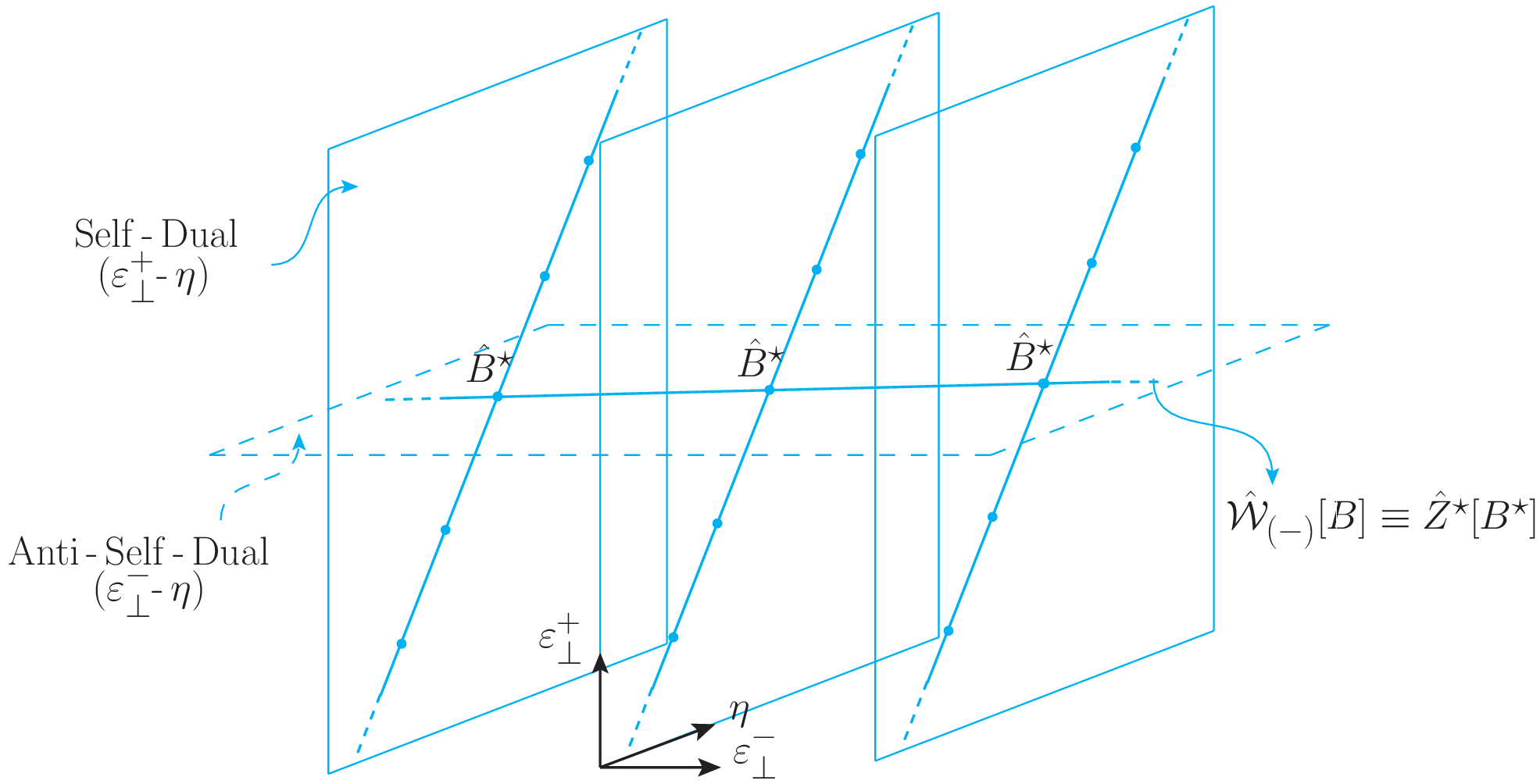}
    \caption{
    \small  ${\hat Z}^{\star}[B^{\star}]$ is a straight infinite Wilson line of $B^{\star}$ fields on the Anti-Self-Dual plane spanned by $\varepsilon_{\perp}^{-} $, $ \eta$. And, ${\hat B}^{\star}[A^{\bullet}, A^{\star}]$ itself is a straight infinite Wilson line on the Self-Dual plane spanned by $\varepsilon_{\perp}^{+} $, $ \eta$ (\emph{cf.} Figure \ref{fig:Bstar_WL})}
    \label{fig:Zstar_WL}
\end{figure}
Geometrically, $Z^{\star}[B^{\star}]$ is a straight infinite Wilson line of $B^{\star}$ fields on the Anti-Self-Dual plane spanned by $\varepsilon_{\perp}^{-} $, $ \eta$. Recall, however, that the latter itself is a Wilson line of $\{A^{\star}, A^{\bullet}\}$ fields on the Self-Dual plane spanned by $\varepsilon_{\perp}^{+} $, $ \eta$ (\emph{cf.} Figure \ref{fig:Bstar_WL}). Thus, the functional $Z^{\star}[A^{\bullet}, A^{\star}]$ is a Wilson line of Wilson lines spanning over both the Self-Dual and the Anti-Self-Dual planes (see Figure \ref{fig:Zstar_WL}). The $Z^{\bullet}[B^{\bullet}, B^{\star}]$ field, on the other hand, has the structure of an "impure" Wilson line $\mathcal{W}_{(-)}[B]$ on the Anti-Self-Dual plane where one of the $B^{\star}$ field has been replaced by a $B^{\bullet}$ field. But since both $B^{\bullet}$ and $B^{\star}$ fields are Wilson line on Self-Dual planes, overall $Z^{\bullet}[A^{\bullet}, A^{\star}]$ also has the geometric representation as a Wilson line of Wilson lines spanning over both the Self-Dual and the Anti-Self-Dual planes (see Figure \ref{fig:Zbul_WL}). With this, we see that, the new fields $\left\{\hat{Z}^{\bullet}[{A}^{\bullet},{A}^{\star}],\hat{Z}^{\star}[{A}^{\bullet},{A}^{\star}]\right\}$  are Wilson line degrees of freedom. Due to this, we call the new action $S\left[Z^{\bullet},Z^{\star}\right]$ Eq.~\eqref{eq:Z_action1} as \textit{a Wilson line-based action}. Finally, they indeed have the geometry that, as stated in Motivation (Section \ref{sec:Zac_mot}), we were intuitively looking for. 

\begin{figure}
    \centering
    \includegraphics[width=11cm]{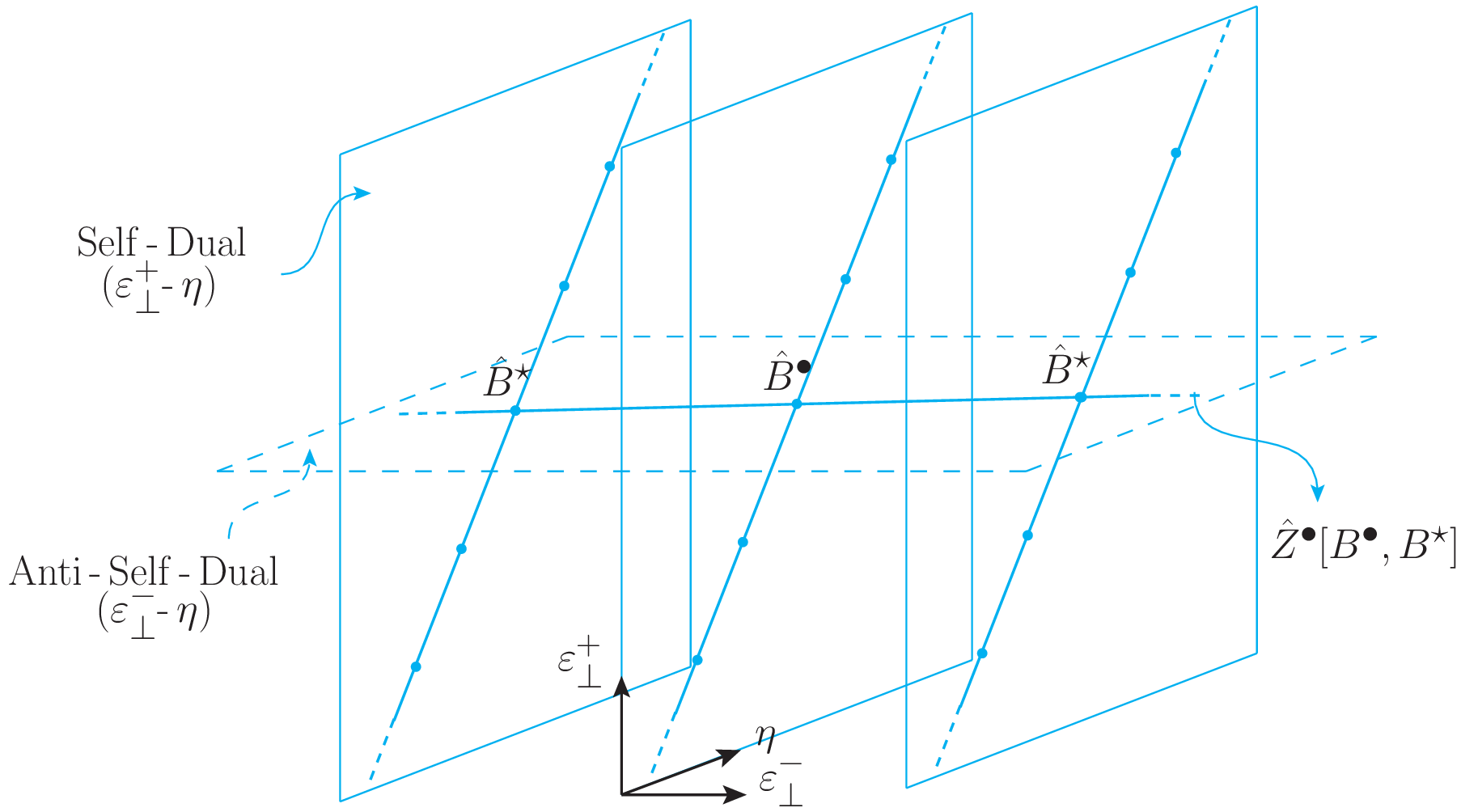}
    \caption{
    \small ${\hat Z}^{\bullet}[B^{\bullet}, B^{\star}]$ is the "impure" straight infinite Wilson line on the Anti-Self-Dual plane spanned by $\varepsilon_{\perp}^{-} $, $ \eta$. Geometrically it is similar to ${\hat Z}^{\star}[B^{\star}]$ (\emph{cf.} Figure \ref{fig:Zstar_WL}) but with one of the $\hat{B}^{\star}$ fields replaced with a $\hat{B}^{\bullet}$ field somewhere on the line. }
    \label{fig:Zbul_WL}
\end{figure}

The solution $\left\{\hat{Z}^{\bullet}[{A}^{\bullet},{A}^{\star}],\hat{Z}^{\star}[{A}^{\bullet},{A}^{\star}]\right\}$ Eq.~\eqref{eq:Zbul_WL}-\eqref{eq:Zstar_WL}, can be Fourier transformed to obtain the momentum space expressions. This calculation proceeds in the same way as for the $\left\{\hat{B}^{\bullet}[{A}^{\bullet}],\hat{B}^{\star}[{A}^{\bullet},{A}^{\star}]\right\}$ fields Eq.~\eqref{eq:WL_Bbul} -~\eqref{eq:Bstar_Bbullet} therefore we do not repeat it. We rather quote the final results
\begin{equation}
    \widetilde{Z}^{\star}_a(x^+;\mathbf{P}) = \sum_{n=1}^{\infty} 
    \int d^3\mathbf{p}_1\dots d^3\mathbf{p}_n \, \overline{\widetilde{\Gamma}}\,_n^{a\{b_1\dots b_n\}}(\mathbf{P};\{\mathbf{p}_1,\dots ,\mathbf{p}_n\}) \prod_{i=1}^n\widetilde{B}^{\star}_{b_i}(x^+;\mathbf{p}_i)\,,
    \label{eq:Z_sta_exp}
\end{equation}
\begin{equation}
    \widetilde{Z}^{\bullet}_a(x^+;\mathbf{P}) = \sum_{n=1}^{\infty} 
    \int d^3\mathbf{p}_1\dots d^3\mathbf{p}_n \, \overline{{\widetilde \Upsilon}}\,_{n}^{a b_1 \left \{b_2 \dots b_n \right \} }(\mathbf{P}; \mathbf{p_1} ,\left \{ \mathbf{p_2} , \dots ,\mathbf{p_n} \right \}) \widetilde{B}^{\bullet}_{b_1}(x^+;\mathbf{p}_1)\prod_{i=2}^n\widetilde{B}^{\star}_{b_i}(x^+;\mathbf{p}_i)\,,
    \label{eq:Z_bul_exp}
\end{equation}
where
\begin{equation}
    \overline{\widetilde{\Gamma}}\,^{a\{b_1\dots b_n\}}_n (\mathbf{P};\{\mathbf{p}_1,\dots,\mathbf{p}_n\}) = (-g)^{n-1} \frac{\delta^3\left(\mathbf{p}_1+\dots+\mathbf{p}_n-\mathbf{P}\right)\,\Tr\!\left(t^at^{b_1}\dots t^{b_n}\right)}{\widetilde{v}_{1(1\cdots n)} \widetilde{v}_{(12)(1\cdots n)} \cdots \widetilde{v}_{(1 \cdots n-1)(1\cdots n)}} \, ,
    \label{eq:overGamma_n}
\end{equation}
\begin{equation}
    \overline{{\widetilde \Upsilon}}\,_{n}^{a b_1 \left \{b_2 \cdots b_n \right \} }(\mathbf{P}; \mathbf{p_1} ,\left \{ \mathbf{p_2} , \dots ,\mathbf{p_n} \right \}) = n\left(\frac{p_1^+}{p_{1\cdots n}^+}\right )^2 \overline{{\widetilde \Gamma}}\,_{n}^{a b_1 \dots b_n }(\mathbf{P}; \mathbf{p_{1}},  \dots ,\mathbf{p_{n}} ) \, .
    \label{eq:overUpsilon_n}
\end{equation}
Above, the kernels $\left\{  \overline{\widetilde{\Gamma}}\,_n , \overline{{\widetilde \Upsilon}}\,_{n} \right\}$ (the tilde denotes Fourier transform) have exactly the same form as $\left\{  \widetilde{\Gamma}_n , {\widetilde \Upsilon}_{n} \right\}$ Eq.~\eqref{eq:Gamma_n}-\eqref{eq:Upsilon_n_rep} but with the momentum components conjugated $p^\bullet \leftrightarrow p^\star$. This exchanges the spinor products ${\widetilde v}^{\star}_{ij} \leftrightarrow {\widetilde v}_{ij}$. As a result, we conclude that the two-step procedure: interchanging the fields $\left\{\left(\hat{B}^{\bullet},\hat{B}^{\star}\right), \left(\hat{A}^{\bullet},\hat{A}^{\star}\right)\right\} \rightarrow \left\{\left(\hat{Z}^{\bullet},\hat{Z}^{\star}\right), \left(\hat{B}^{\bullet},\hat{B}^{\star}\right)\right\}$ Eq.~\eqref{eq:field_int} followed by conjugation $\bullet\leftrightarrow\star$, holds for the momentum space expressions as well. We, therefore, use this to write the momentum space expressions for the inverse 
\begin{equation}
    \widetilde{B}^{\star}_a(x^+;\mathbf{P}) = \sum_{n=1}^{\infty} 
    \int d^3\mathbf{p}_1\dots d^3\mathbf{p}_n \, \overline{\widetilde{\Psi}}\,^{a\{b_1\dots b_n\}}_n(\mathbf{P};\{\mathbf{p}_1,\dots ,\mathbf{p}_n\}) \prod_{i=1}^n\widetilde{Z}^{\star}_{b_i}(x^+;\mathbf{p}_i)\,,
    \label{eq:BstarZ_exp}
\end{equation}
with
\begin{equation}
    \overline{\widetilde \Psi}\,^{a \left \{b_1 \cdots b_n \right \}}_{n}(\mathbf{P}; \left \{\mathbf{p}_{1},  \dots ,\mathbf{p}_{n} \right \}) =- (-g)^{n-1} \,\,  
    \frac{{\widetilde v}_{(1 \cdots n)1}}{{\widetilde v}_{1(1 \cdots n)}} \, 
    \frac{\delta^{3} (\mathbf{p}_{1} + \cdots +\mathbf{p}_{n} - \mathbf{P})\,\,  \mathrm{Tr} (t^{a} t^{b_{1}} \cdots t^{b_{n}})}{{\widetilde v}_{21}{\widetilde v}_{32} \cdots {\widetilde v}_{n(n-1)}}  
      \, ,
    \label{eq:psiBar_kernel}
\end{equation}
and
\begin{equation}
    \widetilde{B}^{\bullet}_a(x^+;\mathbf{P}) = \sum_{n=1}^{\infty} 
    \int d^3\mathbf{p}_1\dots d^3\mathbf{p}_n \, \overline{\widetilde \Omega}\,^{a b_1 \left \{b_2 \cdots b_n \right \}}_{n}(\mathbf{P}; \mathbf{p_1} ,\left \{ \mathbf{p_2} , \dots ,\mathbf{p_n} \right \}) \widetilde{Z}^{\bullet}_{b_1}(x^+;\mathbf{p}_1)\prod_{i=2}^n\widetilde{Z}^{\star}_{b_i}(x^+;\mathbf{p}_i)\,,
    \label{eq:BbulletZ_exp}
\end{equation}
where
\begin{equation}
    \overline{\widetilde \Omega}\,^{a b_1 \left \{b_2 \cdots b_n \right \}}_{n}(\mathbf{P}; \mathbf{p}_{1} , \left \{ \mathbf{p}_{2} , \dots ,\mathbf{p}_{n} \right \} ) = n \left(\frac{p_1^+}{p_{1\cdots n}^+}\right)^2 \overline{\widetilde \Psi}\,^{a b_1 \cdots b_n }_{n}(\mathbf{P}; \mathbf{p}_{1},  \dots ,\mathbf{p}_{n}) \, .
    \label{eq:omegaBar_kernel}
\end{equation}
At this point, we can substitute the expressions for $\widetilde{B}^{\star}_a[{Z}^{\star}](x^+;\mathbf{P})$ and $\widetilde{B}^{\bullet}_a[{Z}^{\star}, {Z}^{\bullet}](x^+;\mathbf{P})$ Eq.~\eqref{eq:BstarZ_exp}-\eqref{eq:BbulletZ_exp} into the expressions for $\widetilde{A}^{\bullet}_a[{B}^{\bullet}](x^+;\mathbf{P})$ and $\widetilde{A}^{\star}_a[{B}^{\bullet}, {B}^{\star}](x^+;\mathbf{P})$ Eq.~\eqref{eq:A_bull_solu}-\eqref{eq:A_star_solu} to obtain the momentum space expression for $\widetilde{A}^{\bullet}_a[{Z}^{\bullet}, {Z}^{\star}](x^+;\mathbf{P})$ and $\widetilde{A}^{\star}_a[{Z}^{\bullet}, {Z}^{\star}](x^+;\mathbf{P})$. The kernels for these would correspond to the momentum space versions of the kernels $\Xi_{i,n-i}^{ab_1\dots b_n}(\mathbf{x};\mathbf{y}_1,\dots,\mathbf{y}_n)$ and $\Lambda_{i,n-i}^{ab_1\dots b_n}(\mathbf{x};\mathbf{y}_1,\dots,\mathbf{y}_n)$ considered in Eq.~\eqref{eq:Abullet_to_Z}-\eqref{eq:Astar_to_Z}. Finally, substituting $\widetilde{A}^{\bullet}_a[{Z}^{\bullet}, {Z}^{\star}](x^+;\mathbf{P})$ and $\widetilde{A}^{\star}_a[{Z}^{\bullet}, {Z}^{\star}](x^+;\mathbf{P})$ to the Yang-Mills action Eq.~\eqref{eq:actionLC_YM}, we can derive the new action $S\left[Z^{\bullet},Z^{\star}\right]$ Eq.~\eqref{eq:Z_action1}. In fact, in Appendix \ref{sec:app_A5} we follow this approach to explicitly demonstrate the cancellation of both the triple gluon vertices in the Yang-Mills action. 

However, in order to derive the new action, following Figure \ref{fig:CT_paths}, we use instead the simplest approach where we start with the MHV action and then substitute $\widetilde{B}^{\star}_a[{Z}^{\star}](x^+;\mathbf{P})$ and $\widetilde{B}^{\bullet}_a[{Z}^{\star}, {Z}^{\bullet}](x^+;\mathbf{P})$ Eq.~\eqref{eq:BstarZ_exp}-\eqref{eq:BbulletZ_exp}. Notice, the substitution of $\widetilde{B}^{\star}_a[{Z}^{\star}](x^+;\mathbf{P})$ will simply multiplicate the minus helicity fields. The $\widetilde{B}^{\bullet}_a[{Z}^{\star}, {Z}^{\bullet}](x^+;\mathbf{P})$, on the other hand, is linear in plus helicity therefore, it doesn't alter the number of plus helicity fields but it multiplicates the minus helicity fields. As a result, the substitution of $\widetilde{B}^{\star}_a[{Z}^{\star}](x^+;\mathbf{P})$ and $\widetilde{B}^{\bullet}_a[{Z}^{\star}, {Z}^{\bullet}](x^+;\mathbf{P})$ to the MHV action keeps the number of plus helicity fields unchanged whereas it multiplicates the minus helicity fields. Thus, given the cancellation of triple gluon vertices shown in Appendix \ref{sec:app_A5}, we can already verify that the structure of the new action derived this way should indeed be of the type shown in Eq.~\eqref{eq:Z_action1}.

\begin{figure}
    \centering
 \includegraphics[width=13cm]{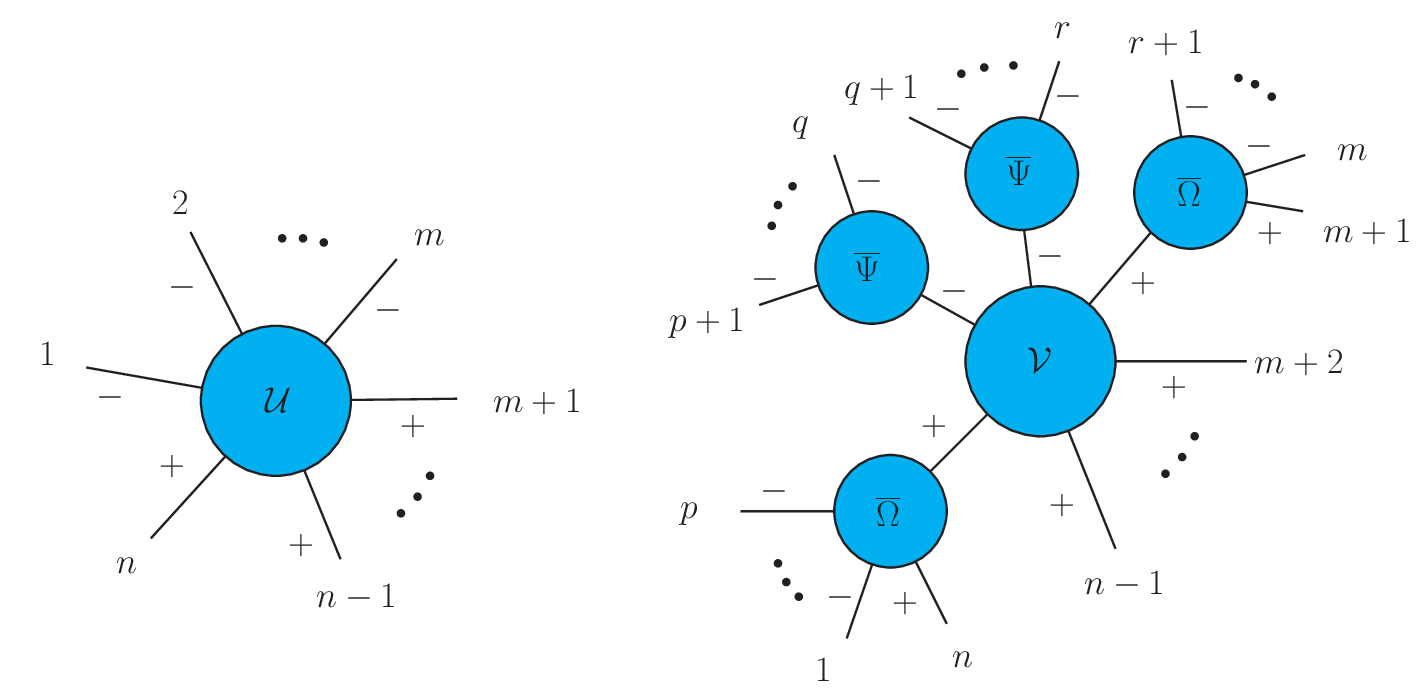}
    \caption{\small 
    Left: A generic color-ordered split-helicity vertex with $m$ minus helicity legs and $n-m$ plus helicity legs $\mathcal{U}\left(1^-,2^-,\dots,m^-,(m\!+\!1)^+,\dots,n^+\right)$ in our action. Right: A generic contribution to this vertex when deriving it via the substitution of $\widetilde{B}^{\star}_a[{Z}^{\star}](x^+;\mathbf{P})$ and $\widetilde{B}^{\bullet}_a[{Z}^{\star}, {Z}^{\bullet}](x^+;\mathbf{P})$ to the MHV action.  This image was modified from our paper \cite{Kakkad:2021uhv}.}
    \label{fig:ZTH_vertex_gen}
\end{figure}

Now, let us proceed to derive the explicit content of the new action. To this end, consider the following generic $n$ point interaction vertex in momentum space
\begin{equation}
    \mathcal{L}_{\underbrace{-\,\cdots\,-}_{m}\underbrace{+ \,\cdots\, +}_{n-m}}^{\left(\mathrm{LC}\right)}= 
   \int\!d^{3}\mathbf{p}_{1}\dots d^{3}\mathbf{p}_{n} \,\, \mathcal{U}^{b_1 \dots b_{n}}_{-\dots-+\dots+}\left(\mathbf{p}_{1},\cdots \mathbf{p}_{n}\right) 
   \prod_{i=1}^{m}Z^{\star}_{b_i} (x^+;\mathbf{p}_{i})
   \prod_{j=1}^{n-m}Z^{\bullet}_{b_j} (x^+;\mathbf{p}_{j}) \, ,
   \label{eq:Z_vertex_mom}
\end{equation}
consisting of $m$ minus helicity legs and $n-m$ plus helicity legs where $\mathcal{U}^{b_1 \dots b_{n}}_{-\dots-+\dots+}\left(\mathbf{p}_{1},\cdots \mathbf{p}_{n}\right)$ is the unknown vertex to be determined. For the sake of simplicity, we considered the so-called \textit{split-helicity} case where all the minus and all the plus helicities are together. Vertices with any other helicity configuration can be derived in exactly the same way. We will, further, restrict to the color-ordered vertex defined below
\begin{equation}
    \mathcal{U}_{-\dots-+\dots+}^{b_{1}\dots b_{n}}\left(\mathbf{p}_{1},\dots,\mathbf{p}_{n}\right)= \!\!\sum_{\underset{\text{\scriptsize permutations}}{\text{noncyclic}}}
 \mathrm{Tr}\left(t^{b_1}\dots t^{b_n}\right)
 \mathcal{U}\left(1^-,\dots,m^-,(m+1)^+,\dots,n^+\right)
\,.
\label{eq:Zvertex_color_decomp}
\end{equation}
Finally, since we derive the color-ordered vertex via the substitution of $\widetilde{B}^{\star}_a[{Z}^{\star}](x^+;\mathbf{P})$ and $\widetilde{B}^{\bullet}_a[{Z}^{\star}, {Z}^{\bullet}](x^+;\mathbf{P})$ Eq.~\eqref{eq:BstarZ_exp}-\eqref{eq:BbulletZ_exp} to the MHV action, we need color-ordered versions for the kernels as well. These read
\begin{equation}
    \overline{\widetilde{\Psi}}\,^{a\{b_{1}\dots b_{m}\}}_m\left(\mathbf{P};\{\mathbf{p}_{1},\dots,\mathbf{p}_{m}\}\right)= \!\!\sum_{\underset{\text{\scriptsize permutations}}{\text{noncyclic}}}
 \mathrm{Tr}\left(t^{b_1}\dots t^{b_m}\right)
 \overline{\Psi}\left(1^-,\dots,m^-\right)
\,,
\label{eq:PsiBar_color_decomp}
\end{equation}
and 
\begin{equation}
    \overline{\widetilde{\Omega}}\,^{ab_{1}\{b_2\dots b_{m}\}}_m\left(\mathbf{P};\mathbf{p}_{1},\{\mathbf{p}_{2},\dots,\mathbf{p}_{m}\}\right)= \!\!\sum_{\underset{\text{\scriptsize permutations}}{\text{noncyclic}}}
 \mathrm{Tr}\left(t^{b_1}\dots t^{b_m}\right)
 \overline{\Omega}\left(1^+,2^-,\dots,m^-\right)
\,.
\label{eq:OmegaBar_color_decomp}
\end{equation}
Above we use numbers in the kernel to represent the  momentum associated with the legs $i \equiv p_i$ and we assign the helicity of the leg to this number. In order to derive the color-ordered split-helicity $\mathcal{U}\left(1^-,\dots,m^-,(m+1)^+,\dots,n^+\right)$ with $n-m$ plus helicity legs, we begin with the color-ordered split-helicity MHV vertex in the B-field theory (i.e. the MHV action) with $n-m$ plus helicity legs $ \mathcal{V}\left(1^-,2^-,3^+,\dots,(n-m+2)^+\right)$ Eq.~\eqref{eq:MHV_vertex}. This is because, recall, the $\overline{\Psi}$, $\overline{\Omega}$ kernels cannot alter the number of plus helicity legs. Now, to the two negative helicity legs in the MHV action, we can substitute the $\overline{\Psi}$ kernels which, in turn, will multiplicate the number of minus helicity legs. An important point to note here is that all the momenta in the MHV vertex are outgoing whereas the momentum $\mathbf{P}$ of the minus and plus helicity leg in the kernels $\overline{\Psi}$ and $\overline{\Omega}$, respectively, is incoming. As a result upon substitution, there is no flip in the helicity. That is, the helicity flow of the entire leg in the MHV vertex undergoing substitution remains the same. Furthermore, there is no propagator connecting the MHV vertex to the kernels $\overline{\Psi}$, $\overline{\Omega}$. Finally, to the plus helicity leg of the MHV vertex, we can substitute the $\overline{\Omega}$ kernels. This, however, can be done only to the two plus helicity legs situated on either side of the two minus helicity legs of the MHV vertex. This is because substituting $\overline{\Omega}$ to any other plus helicity leg will not result in a split-helicity vertex. So, in total, there are just four consecutive legs $(+ - - +)$ to which the kernels can be substituted to get a split helicity vertex in the new action. The generic form of this is shown in Figure \ref{fig:ZTH_vertex_gen}. Note, however, there will be a sum of terms contributing to $\mathcal{U}\left(1^-,\dots,m^-,(m+1)^+,\dots,n^+\right)$, where, in each term, the kernels are expanded to different orders such that the sum of minus helicity legs is $m$. This sum reads \cite{Kakkad:2021uhv}:
\begin{multline}
    \mathcal{U}\left(1^-,2^-,\dots,m^-,(m\!+\!1)^+,\dots,n^+\right) = 
    \sum_{p=0}^{m-2}\sum_{q=p+1}^{m-1}\sum_{r=q+1}^{m}\\
    \mathcal{V}\left(\,[p\!+\!1,\dots,q]^-,[q\!+\!1,\dots,r]^-,[r\!+\!1,\dots,m\!+\!1]^+,(m\!+\!2)^+,\dots,(n\!-\!1)^+,[n,1,\dots,p]^+\right) \\
    \overline{ \Omega}\left(n^+,1^-,\dots,p^-\right) \,\,
    \overline{ \Psi}\left((p\!+\!1)^-,\dots,q^-\right) \,\, 
    \overline{ \Psi}\left((q\!+\!1)^-,\dots,r^-\right) \,\, \\
    \overline{ \Omega}\left((r\!+\!1)^-,\dots,m^-,(m\!+\!1)^+\right) \,\,.
    \label{eq:Z_gen_ker}
\end{multline}
Above we use the collective index notation $[i,i+1,\dots,j]$ to represent the momentum sum $\mathbf{p}_{i(i+1)\dots j}=\mathbf{p}_i+\mathbf{p}_{i+1}+\dots+\mathbf{p}_j$. Eq.~\eqref{eq:Z_gen_ker} represents the generic form for the color-ordered vertex $\mathcal{U}\left(1^-,2^-,\dots,m^-,(m\!+\!1)^+,\dots,n^+\right)$ in the new ation $S\left[Z^{\bullet},Z^{\star}\right]$ Eq.~\eqref{eq:Z_action1}. Although it doesn't simplify any further, in the next section we will validate the expression by using it to compute amplitudes. But before we do that, in the following Subsection, we highlight the Feynman rules for computing pure gluonic scattering amplitudes using our new action.

\subsection{The Feynman rules}
\label{subsec:Zac_FR}

In the next section, we will focus on computing color-ordered amplitudes. Therefore, below, we summarize the color-ordered Feynman rules for computing these using our new action $S\left[Z^{\bullet},Z^{\star}\right]$ Eq.~\eqref{eq:Z_action1} 
\begin{itemize}
\item \emph{Scalar propagator}: The propagator in our action is just like the one in the MHV action. That is, it is scalar and connects opposite helicity legs $(- +)$ \\
\begin{center}
\includegraphics[width=4.8cm]{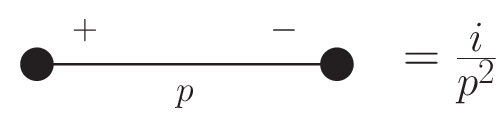}
\end{center}
\item \emph{$n$-point interaction vertex}: The interaction vertices in our action have $n\geq4$ legs. Furthermore, if $m$ is the number of minus helicity legs, then $2\leq m \leq n-2$. The general form, $\mathcal{U}\left(1^-,2^-,\dots,m^-,(m\!+\!1)^+,\dots,n^+\right)$, for this was derived in Eq.~\eqref{eq:Z_gen_ker}. \\
\begin{center}
\includegraphics[width=12cm]{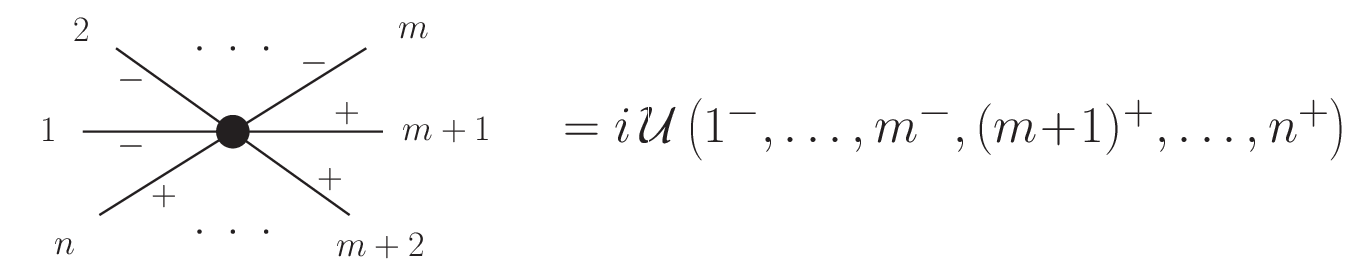}
\end{center}
MHV vertices $(- - + \dots +)$ have the least number of minus helicity legs i.e. $m=2$. After that we have Next-to-MHV (NMHV) $(- - - + \dots +)$, Next-to-Next-to-MHV (NNMHV) $(- - - - + \dots +)$ and so on. The $\overline{\mathrm{MHV}}$ vertices $(- \dots - + +)$ have the highest number of negative helicity legs i.e. $m=n-2$. 
\end{itemize}

\section{Tree amplitudes}
\label{sec:Zac_TR_AMP}

In this section, using our new action $S\left[Z^{\bullet},Z^{\star}\right]$ Eq.~\eqref{eq:Z_action1}, we will compute several color-ordered tree-level pure gluonic amplitudes with different helicity configurations ranging from MHV, $\overline{\mathrm{MHV}}$ up to NNMHV.

\subsection{4-point and 5-point amplitudes}

The possible helicity configurations for a 4-point pure gluonic amplitude are $(+ + + +)$, $(- + + +)$, $(+ - - -)$, $(- - - -)$ and $(- - + +)$. At the tree level, in the physical domain (real on-shell momenta), the first four amplitudes are zero (\emph{cf.} Eq.~\eqref{eq:plus_tree_amp}). The only non-vanishing tree level 4-point amplitude is the 4-point MHV $(- - + +)$. In our action $S\left[Z^{\bullet},Z^{\star}\right]$ Eq.~\eqref{eq:Z_action1}, no contributions can give rise to $(+ + + +)$, $(- + + +)$, $(+ - - -)$, and $(- - - -)$ tree-level amplitudes, as a result, they are naturally zero. For the 4-point MHV $(- - + +)$, there is just a single contribution  that comes from the vertex $\mathcal{L}_{--++}$. Therefore, for the color-ordered split-helicity amplitude $\mathcal{A}\left(1^-,2^-,3^+,4^+\right)$, using Eq.~\eqref{eq:Z_gen_ker}, we have
\begin{align}
    \mathcal{A}\left(1^-,2^-,3^+,4^+\right)= & \left.\mathcal{U}\left(1^-,2^-,3^+,4^+\right)\right|_{\mathrm{on-shell}} \nonumber \\
    = &   \left[\mathcal{V}\left(1^-,2^-,3^+,4^+\right)
    \overline{ \Omega}_1\,\,
    \overline{ \Psi}_1\,\, 
    \overline{ \Psi}_1\,\,
    \overline{ \Omega}_1\right]_{\mathrm{on-shell}} \,\,, \nonumber\\
    = & g^2 \left(\frac{p_{1} ^{+}}{p_{2}^{+}}\right)^{2}
\frac{\widetilde{v}_{21}^{\star 4}}{\widetilde{v}^{\star}_{14}\widetilde{v}^{\star}_{43}\widetilde{v}^{\star}_{32}\widetilde{v}^{\star}_{21} } \, ,
\label{eq:zac_MHV4}
\end{align}
where, $\overline{ \Omega}_1\,\, = \overline{ \Psi}_1\,\, \equiv 1$. The above result is in agreement with Eq.~\eqref{eq:MHV_ampl}. Finally, the conjugate of 4-point MHV $(- - + +)$ is the 4-point $\overline{\mathrm{MHV}}$ $(+ + - -)$ and it also has a single contribution from the same vertex. We checked that the above result, after a bit of algebra using the identities for  $\widetilde{v}^{\star}_{ij}$ and $\widetilde{v}_{ij}$, can be re-written as
\begin{equation}
    \mathcal{A}\left(1^-,2^-,3^+,4^+\right) = g^2 \left(\frac{p_{3} ^{+}}{p_{4}^{+}}\right)^{2}
\frac{\widetilde{v}_{43}^{4}}{\widetilde{v}_{14}\widetilde{v}_{43}\widetilde{v}_{32}\widetilde{v}_{21} } \, ,
\label{eq:zac_MHVbar4}
\end{equation}
which corresponds to the $\overline{\mathrm{MHV}}$ representation of the same amplitude Eq.~\eqref{eq:MHVbar_ampl}. Note, in both the cases Eq.~\eqref{eq:zac_MHV4}, \eqref{eq:zac_MHVbar4} we suppressed the overall momentum conserving delta.

For tree level 5-point amplitude, there are just two non-vanishing configurations: MHV $(- - + + +)$ and  $\overline{\mathrm{MHV}}$ $(+ + - - -)$. Computing the former is trivial because, just as in the case of 4-point MHV $(- - + +)$, there is only one vertex contributing to it, and this vertex is exactly the same as in the MHV action with $\overline{ \Omega}_1\,\, = \overline{ \Psi}_1\,\, \equiv 1$. Therefore, for the color-ordered amplitude in the on-shell limit, we get
\begin{equation}
    \mathcal{A}(1^-,2^-,3^+,4^+,5^+) =
 -g^3 \left(\frac{p_{1} ^{+}}{p_{2}^{+}}\right)^{2}
\frac{\widetilde{v}_{21}^{\star 4}}{\widetilde{v}^{\star}_{15}\widetilde{v}^{\star}_{54}\widetilde{v}^{\star}_{43}  \widetilde{v}^{\star}_{32}\widetilde{v}^{\star}_{21} } \, .
 \label{eq:5G_MHV_onshell}
\end{equation}

\begin{figure}
    \centering
 \includegraphics[width=16cm]{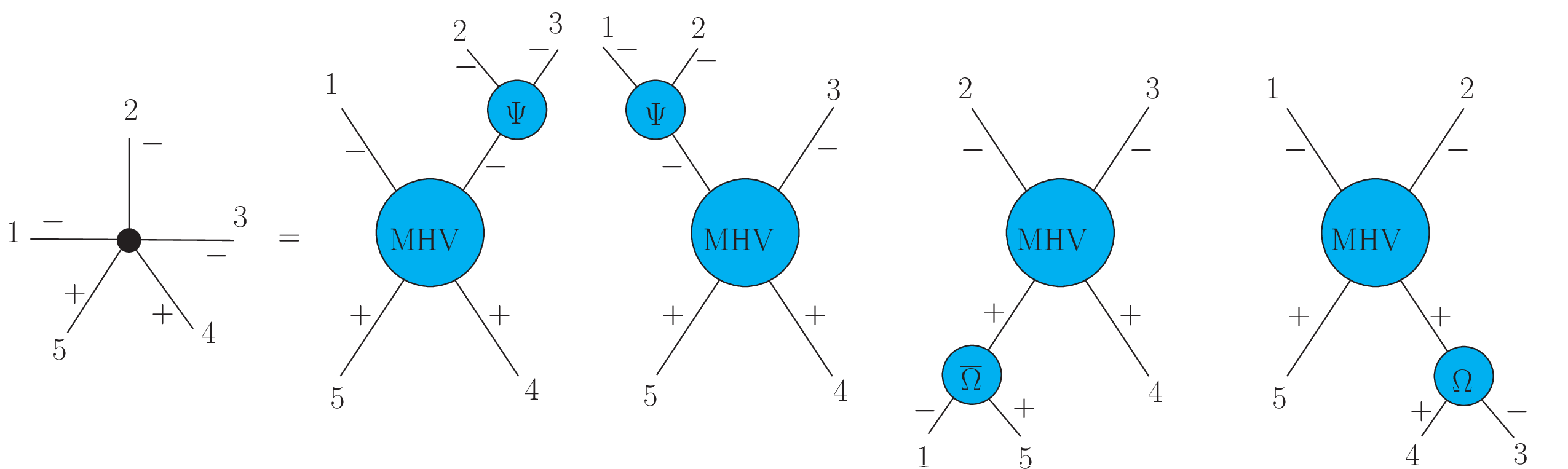}
\caption{\small 
    Left: The 5-point $\overline{\mathrm{MHV}}$ vertex $\mathcal{U}(1^-,2^-,3^-,4^+,5^+)$ in our action. Right: The contributions when deriving it via substituting $\widetilde{B}^{\star}_a[{Z}^{\star}](x^+;\mathbf{P})$ and $\widetilde{B}^{\bullet}_a[{Z}^{\star}, {Z}^{\bullet}](x^+;\mathbf{P})$ to the MHV action. This image was modified from our paper \cite{Kakkad:2021uhv}.}
    \label{fig:MHVbar5_vertex}
\end{figure}

The 5-point $\overline{\mathrm{MHV}}$ $(- - - + +)$ also receives contribution from a single vertex $\mathcal{L}_{---++}$. Notice, in comparison to the MHV action, this vertex is new in our action $S\left[Z^{\bullet},Z^{\star}\right]$ Eq.~\eqref{eq:Z_action1}. Using the general form for the $\mathcal{U}(1^-,2^-,3^-,4^+,5^+)$ vertex following Eq.~\eqref{eq:Z_gen_ker}, we can compute the color-ordered split-helicity 5-point $\overline{\mathrm{MHV}}$ $(- - - + +)$ amplitude. There are four terms in it, which we represent diagrammatically in Figure \ref{fig:MHVbar5_vertex}. These read
\begin{multline}
 \mathcal{U}(1^-,2^-,3^-,4^+,5^+) = g^3  \Bigg[  \left(\frac{p_{1} ^{+}}{p_{23}^{+}}\right)^{2}
\frac{\widetilde{v}_{(23)1}^{\star 4}}{\widetilde{v}_{15}^{\star}\widetilde{v}_{54}^{\star}\widetilde{v}_{4(23)}^{\star}  \widetilde{v}_{({23})1}^{\star} } \times  \frac{{\widetilde v}_{({23})2}}{{\widetilde v}_{32}{\widetilde v}_{2({23})}}   \\
+     \left(\frac{p_{12} ^{+}}{p_{3}^{+}}\right)^{2}
\frac{\widetilde{v}_{3({12})}^{ \star 4}}{\widetilde{v}_{({12}){5}}^{\star}\widetilde{v}_{54}^{\star}\widetilde{v}_{43}^{\star}  \widetilde{v}_{{3}({12})}^{\star} } \times  \frac{{\widetilde v}_{({12})1}}{{\widetilde v}_{21}{\widetilde v}_{1({12})}}\\
+ \left(\frac{p_{2} ^{+}}{p_{3}^{+}}\right)^{2}
\frac{\widetilde{v}_{32}^{\star 4}}{\widetilde{v}_{2({15})}^{\star}\widetilde{v}_{({15})4}^{\star}\widetilde{v}_{43}^{\star}  \widetilde{v}_{32}^{\star} } \times \left(\frac{p_{5} ^{+}}{p_{15}^{+}}\right)^{2} \frac{{\widetilde v}_{({15})5}}{{\widetilde v}_{15}{\widetilde v}_{5({15})}}    \\ 
+    \left(\frac{p_{1} ^{+}}{p_{2}^{+}}\right)^{2}
\frac{\widetilde{v}_{21}^{\star 4}}{\widetilde{v}_{15}^{\star}\widetilde{v}_{5\left(34\right)}^{\star}\widetilde{v}_{({34)}2}^{\star}  \widetilde{v}_{21}^{\star} } \times \left(\frac{p_{4} ^{+}}{p_{34}^{+}}\right)^{2} \frac{{\widetilde v}_{({34})3}}{{\widetilde v}_{43}{\widetilde v}_{3({34})}}  \Bigg]\, .
 \label{eq:5g_MHVbar}
\end{multline}
In the on-shell limit, we verified that the above expression reduces to the following compact form
\begin{equation}
    \mathcal{A}(1^-,2^-,3^-,4^+,5^+) =
 g^3 \left(\frac{p_{4} ^{+}}{p_{5}^{+}}\right)^{2}
\frac{\widetilde{v}_{54}^{4}}{\widetilde{v}_{15}\widetilde{v}_{54}\widetilde{v}_{43}  \widetilde{v}_{32}\widetilde{v}_{21} } \, ,
 \label{eq:5G_MHVbar_onshell}
\end{equation}
which is in agreement with Eq.~\eqref{eq:MHVbar_ampl}.

\subsection{6-point amplitudes}

\begin{figure}[h]
    \centering
 \includegraphics[width=16cm]{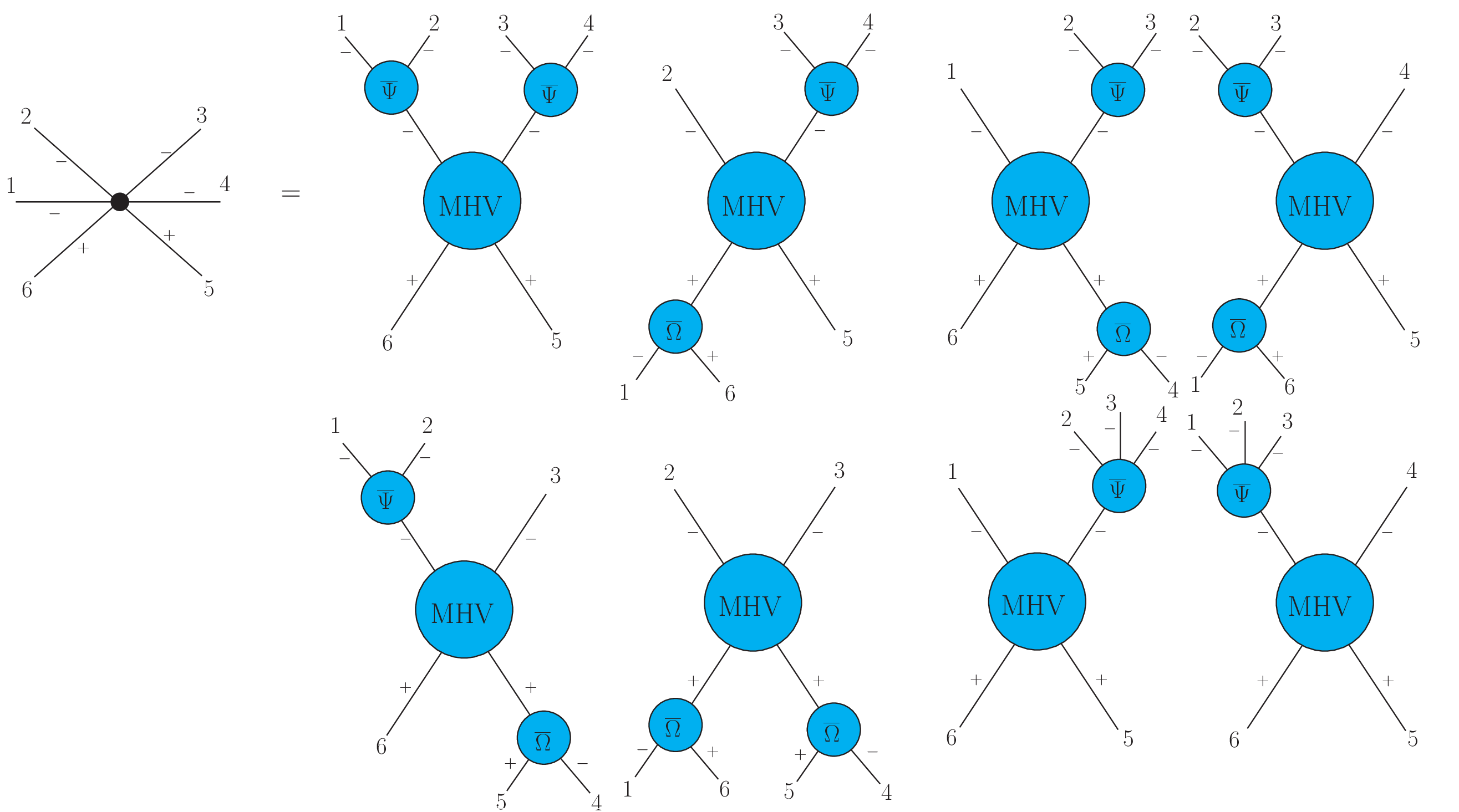}\\
 \includegraphics[width=6.8cm]{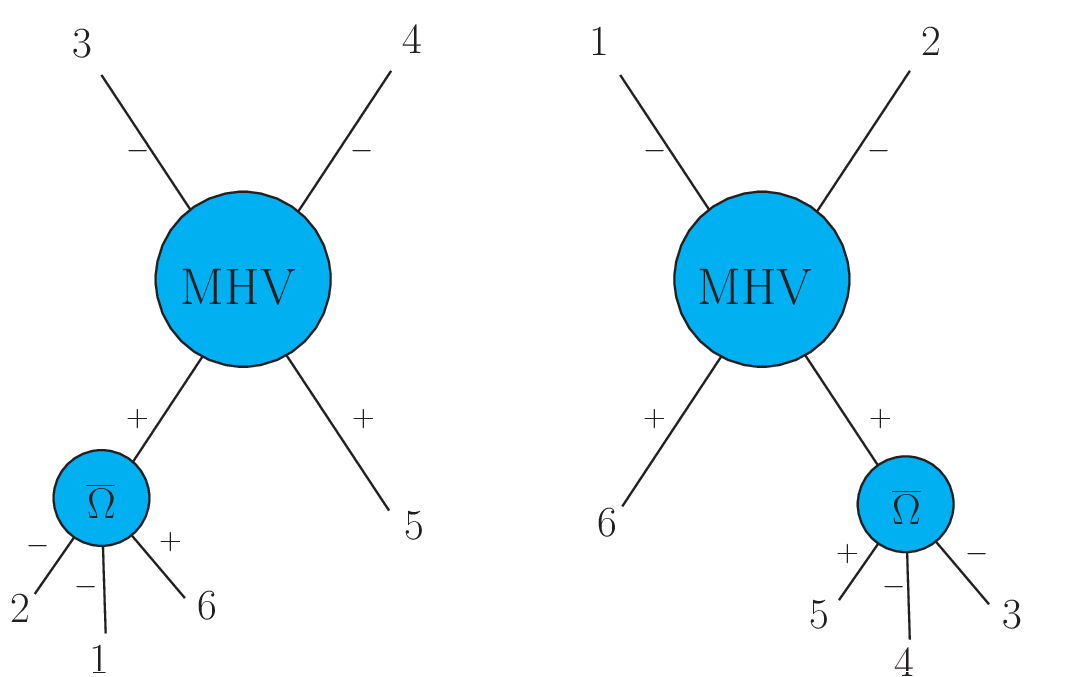}
    \caption{\small 
    LHS: The 6-point $\overline{\mathrm{MHV}}$ vertex $\mathcal{U}(1^-,2^-,3^-,4^-,5^+, 6^+)$ in our action. RHS: The contributions when deriving it via substituting $\widetilde{B}^{\star}_a[{Z}^{\star}](x^+;\mathbf{P})$ and $\widetilde{B}^{\bullet}_a[{Z}^{\star}, {Z}^{\bullet}](x^+;\mathbf{P})$ to the MHV action. This image was modified from our paper \cite{Kakkad:2021uhv}.}
    \label{fig:6gmhv_b}
\end{figure}

For 6-point, there are three configurations: MHV $(- - + + + +)$, $\overline{\mathrm{MHV}}$ $(+ + - - - -)$, and NMHV $(- - - + + +)$, for which the tree-level amplitude is non-vanishing. Since the MHV case is trivial, we do not repeat the discussion. The color-ordered $\overline{\mathrm{MHV}}$ amplitude $\mathcal{A}(1^-,2^-,3^-,4^-,5^+,6^+)$, can be obtained using the vertex $\mathcal{U}(1^-,2^-,3^-,4^-,5^+,6^+)$ Eq.~\eqref{eq:Z_gen_ker}. The contributions (also shown diagrammatically in Figure \ref{fig:6gmhv_b}) read 
\begin{multline}
 \mathcal{U}(1^-,2^-,3^-,4^-,5^+,6^+) = g^4
  \Bigg[  \Bigg( \left(\frac{p_{12} ^{+}}{p_{34}^{+}}\right)^{2}
\frac{\widetilde{v}_{(34)({12})}^{\star 4}}{\widetilde{v}_{({12}){6}}^{\star}\widetilde{v}_{{6}5}^{\star}\widetilde{v}_{{5}(34)}^{\star}  \widetilde{v}_{({34})({12})}^{\star} } \times  \frac{{\widetilde v}_{({12})1}}{{\widetilde v}_{21}{\widetilde v}_{1({12})}}\times  \frac{{\widetilde v}_{({34})3}}{{\widetilde v}_{43}{\widetilde v}_{3({34})}} \Bigg) \nonumber
\end{multline}
\begin{equation}
+ \Bigg(  \left(\frac{p_{2} ^{+}}{p_{34}^{+}}\right)^{2}
\frac{\widetilde{v}_{(34){2}}^{\star 4}}{\widetilde{v}_{2({16})}^{\star}\widetilde{v}_{{(16)5}}^{\star}\widetilde{v}_{{5}(34)}^{\star}  \widetilde{v}_{({34}){2}}^{\star} } \times \frac{{\widetilde v}_{({34})3}}{{\widetilde v}_{43}{\widetilde v}_{3({34})}} \times \left(\frac{p_{6} ^{+}}{p_{16}^{+}}\right)^{2} \frac{{\widetilde v}_{({16})6}}{{\widetilde v}_{16}{\widetilde v}_{6({16})}}   \Bigg)\nonumber
\end{equation}
\begin{equation}
+ \Bigg(  \left(\frac{p_{1} ^{+}}{p_{23}^{+}}\right)^{2}
\frac{\widetilde{v}_{(23){1}}^{\star 4}}{\widetilde{v}_{{16}}^{\star}\widetilde{v}_{{6}\left(45\right)}^{\star}\widetilde{v}_{({45)}(23)}^{\star}  \widetilde{v}_{({23}){1}}^{\star} } \times \frac{{\widetilde v}_{({23})2}}{{\widetilde v}_{32}{\widetilde v}_{2({23})}} \times \left(\frac{p_{5} ^{+}}{p_{45}^{+}}\right)^{2} \frac{{\widetilde v}_{({45})4}}{{\widetilde v}_{54}{\widetilde v}_{4({45})}}   \Bigg)\nonumber
\end{equation}
\begin{equation}
+ \Bigg( \left(\frac{p_{23} ^{+}}{p_{4}^{+}}\right)^{2}
\frac{\widetilde{v}_{4({23})}^{\star 4}}{\widetilde{v}_{({23})({16})}^{\star}\widetilde{v}_{({16})5}^{\star}\widetilde{v}_{{54}}^{\star}  \widetilde{v}_{{4}({23})}^{\star} } \times\frac{{\widetilde v}_{({23})2}}{{\widetilde v}_{32}{\widetilde v}_{2({23})}}  \times \left(\frac{p_{6} ^{+}}{p_{16}^{+}}\right)^{2} \frac{{\widetilde v}_{({16})6}}{{\widetilde v}_{16}{\widetilde v}_{6({16})}}   \Bigg) \nonumber
\end{equation}
\begin{equation}
+ \Bigg(  \left(\frac{p_{12} ^{+}}{p_{3}^{+}}\right)^{2}
\frac{\widetilde{v}_{3({12})}^{\star 4}}{\widetilde{v}_{({12}){6}}^{\star}\widetilde{v}_{{6}\left(45\right)}^{\star}\widetilde{v}_{({45)}3}^{\star}  \widetilde{v}_{{3}({12})}^{\star} } \times  \left(\frac{p_{5} ^{+}}{p_{45}^{+}}\right)^{2} \frac{{\widetilde v}_{({45})4}}{{\widetilde v}_{54}{\widetilde v}_{4({45})}} \times \frac{{\widetilde v}_{({12})1}}{{\widetilde v}_{21}{\widetilde v}_{1({12})}}  \Bigg) \nonumber
\end{equation}
\begin{equation}
+ \Bigg(  \left(\frac{p_{2} ^{+}}{p_{3}^{+}}\right)^{2}
\frac{\widetilde{v}_{32}^{\star 4}}{\widetilde{v}_{2({16})}^{\star}\widetilde{v}_{{(16)(45)}}^{\star}\widetilde{v}_{(45)3}^{\star}  \widetilde{v}_{{32}}^{\star} } \left(\frac{p_{5} ^{+}}{p_{45}^{+}}\right)^{2} \frac{{\widetilde v}_{({45})4}}{{\widetilde v}_{54}{\widetilde v}_{4({45})}} \left(\frac{p_{6} ^{+}}{p_{16}^{+}}\right)^{2} \frac{{\widetilde v}_{({16})6}}{{\widetilde v}_{16}{\widetilde v}_{6({16})}}   \Bigg)\nonumber
\end{equation}
\begin{equation}
- \Bigg(  \left(\frac{p_{1} ^{+}}{p_{234}^{+}}\right)^{2}
\frac{\widetilde{v}_{(234){1}}^{\star 4}}{\widetilde{v}_{{1}{6}}^{\star}\widetilde{v}_{{65}}^{\star}\widetilde{v}_{{5}(234)}^{\star}  \widetilde{v}_{({234}){1}}^{\star} } \times  \frac{{\widetilde v}_{({234})2}}{{\widetilde v}_{43}{\widetilde v}_{32}{\widetilde v}_{2({234})}}  \Bigg) \nonumber
\end{equation}
\begin{equation}
- \Bigg( \left(\frac{p_{123} ^{+}}{p_{4}^{+}}\right)^{2}
\frac{\widetilde{v}_{4({123})}^{\star 4}}{\widetilde{v}_{({123}){6}}^{\star}\widetilde{v}_{{65}}^{\star}\widetilde{v}_{{54}}^{\star}  \widetilde{v}_{{4}({123})}^{\star} } \times  \frac{{\widetilde v}_{({123})1}}{{\widetilde v}_{32}{\widetilde v}_{21}{\widetilde v}_{1({123})}} \Bigg) \nonumber
\end{equation}
\begin{equation}
- \Bigg(  \left(\frac{p_{3} ^{+}}{p_{4}^{+}}\right)^{2}
\frac{\widetilde{v}_{43}^{\star 4}}{\widetilde{v}_{3{(612)}}^{\star}\widetilde{v}_{{(612)5}}^{\star}\widetilde{v}_{54}^{\star}  \widetilde{v}_{43}^{\star} } \times \left(\frac{p_{6} ^{+}}{p_{612}^{+}}\right)^{2} \frac{{\widetilde v}_{({612})6}}{{\widetilde v}_{21}{\widetilde v}_{16}{\widetilde v}_{6({612})}}   \Bigg)\nonumber
\end{equation}
\begin{equation}
- \Bigg( \left(\frac{p_{1} ^{+}}{p_{2}^{+}}\right)^{2}
\frac{\widetilde{v}_{21}^{\star 4}}{\widetilde{v}_{{1}{6}}^{\star}\widetilde{v}_{{6}\left(345\right)}^{\star}\widetilde{v}_{({345)}2}^{\star}  \widetilde{v}_{21}^{\star} } \times \left(\frac{p_{5} ^{+}}{p_{345}^{+}}\right)^{2} \frac{{\widetilde v}_{({345})3}}{{\widetilde v}_{54}{\widetilde v}_{43}{\widetilde v}_{3({345})}}   \Bigg) \Bigg] \, .
 \label{eq:6g_MHVbar_Z}
\end{equation}

We verified numerically that in the on-shell limit the sum of all the terms reproduces
\begin{equation}
    \mathcal{A}(1^-,2^-,3^-,4^-,5^+,6^+) =
 g^4 \left(\frac{p_{5} ^{+}}{p_{6}^{+}}\right)^{2}
\frac{\widetilde{v}_{65}^{4}}{\widetilde{v}_{16}\widetilde{v}_{65}\widetilde{v}_{54}  \widetilde{v}_{43}\widetilde{v}_{32}\widetilde{v}_{21} } \, .
 \label{eq:6G_MHVbar_onshell}
\end{equation}

\begin{figure}[h]
    \centering
 \includegraphics[width=13cm]{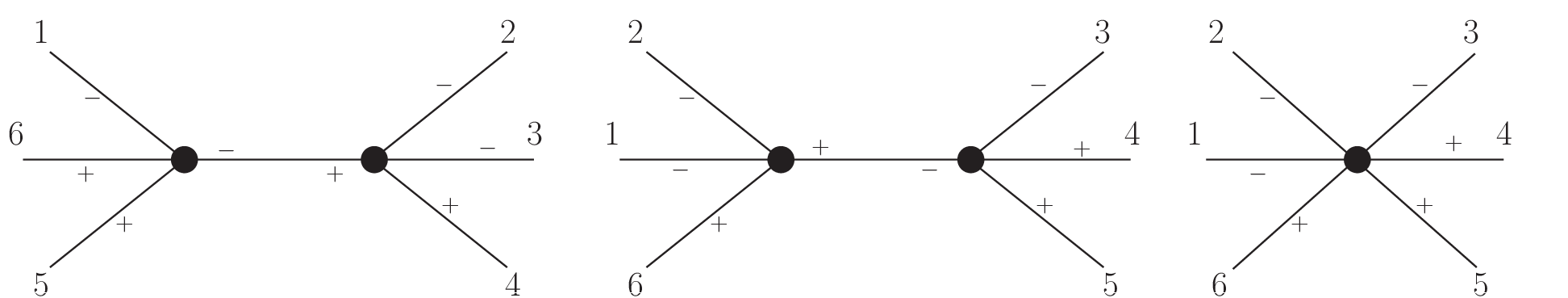}
    \caption{\small 
    Feynman diagrams contributing to the tree-level split-helicity 6-point NMHV $(- - - + + +)$ amplitude in our action $S\left[Z^{\bullet},Z^{\star}\right]$. This image was taken from our paper \cite{Kakkad:2021uhv}.}
    \label{fig:NMHV6}
\end{figure}

The 6-point NMHV $(- - - + + +)$ is the first case that receives contributions beyond a single vertex in our new action $S\left[Z^{\bullet},Z^{\star}\right]$ Eq.~\eqref{eq:Z_action1}. In total, there are three contributions to this amplitude shown in Figure \ref{fig:NMHV6}. Two of these contributions consist of two 4-point MHV vertices connected via a scalar propagator. The third is just the NMHV vertex $\mathcal{U}(1^-,2^-,3^-,4^+,5^+,6^+)$ from the action. We verified that the sum of all three contributions in the on-shell limit reproduces the known result \cite{Kosower1990}.

\subsection{7-point amplitudes}

\begin{figure}[h]
    \centering
 \includegraphics[width=13cm]{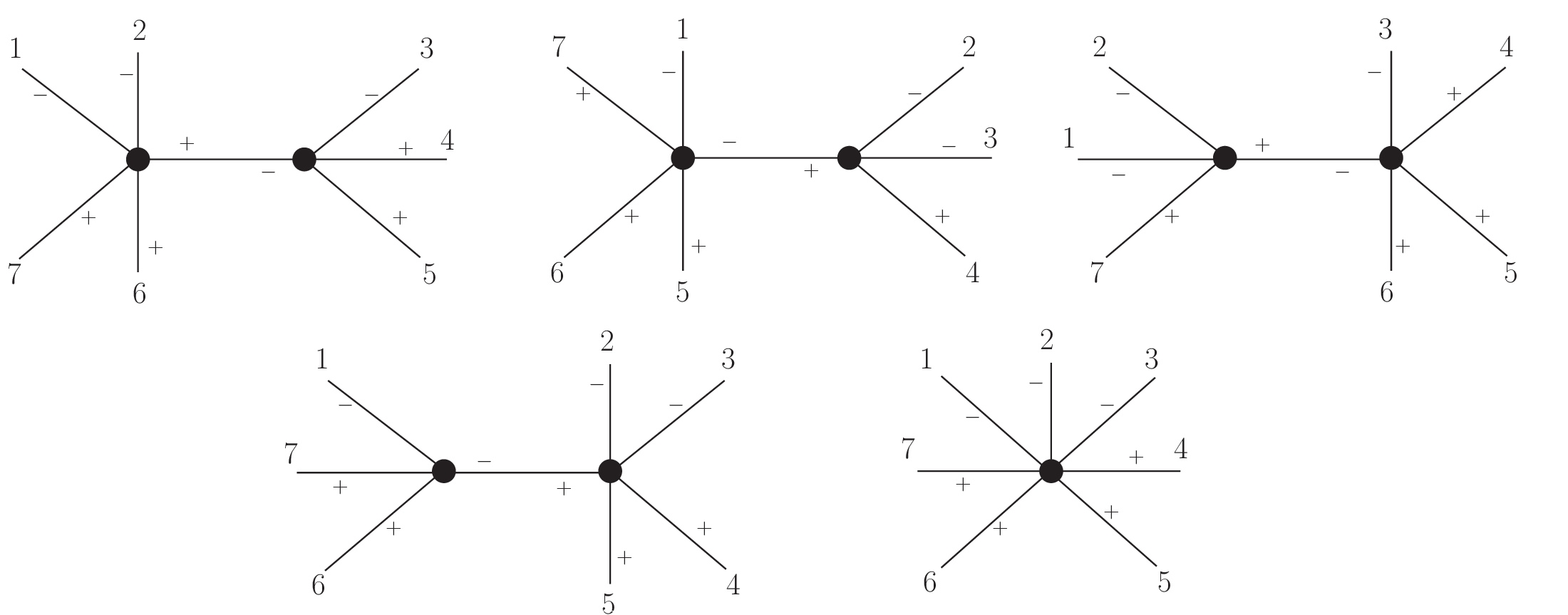}
    \caption{\small 
    Feynman diagrams contributing to the tree-level split-helicity 7-point NMHV $(- - - + + + +)$ amplitude in our action $S\left[Z^{\bullet},Z^{\star}\right]$. This image was taken from our paper \cite{Kakkad:2021uhv}.}
    \label{fig:NMHV7}
\end{figure}

For 7-point, there are four configurations: MHV $(- - + + + + +)$, $\overline{\mathrm{MHV}}$ $(+ + - - - - -)$, NMHV $(- - - + + + +)$, and NNMHV $(- - - - + + +)$ for which the tree-level amplitude is non-vanishing.  The MHV $(- - + + + + +)$ and $\overline{\mathrm{MHV}}$ $(+ + - - - - -)$ amplitudes can be computed from the corresponding vertices $\mathcal{L}_{- - + + + + +}$ and $\mathcal{L}_{+ + - - - - -}$ respectively in the on-shell limit.

For 7-point NMHV $(- - - + + + +)$, there are just five contributions in our action. These are shown in Figure \ref{fig:NMHV7}. Four of these contributions consist of two MHV vertices, one 4-point, and one 5-point, connected via the scalar propagator. The fifth is just the NMHV vertex $\mathcal{U}(1^-,2^-,3^-,4^+,5^+,6^+,7^+)$ from the action. Their explicit expressions read (in the order: from top left in Figure \ref{fig:NMHV7} to the right and then the second line) 
\begin{equation}
    D_1=i\, \mathcal{U} \left(1^-,2^-,[3,4,5]^+,6^+,7^+\right) \times \frac{i}{p^2_{6712}}\times i\,\mathcal{U} \left([6,7,1,2]^-,3^-,4^+,5^+\right) \, 
    \end{equation}
    \begin{equation}
    D_2= i\,\mathcal{U} \left(1^-,[2,3,4]^-,5^+,6^+,7^+\right) \times \frac{i}{p^2_{234}}\times i\,\mathcal{U} \left(2^-,3^-,4^+,[5,6,7,1]^+\right) \, 
    \end{equation}
    \begin{equation}
    D_3=i\, \mathcal{U} \left(1^-,2^-,[3,4,5,6]^+,7^+\right) \times \frac{i}{p^2_{3456}}\times i\,\mathcal{U} \left([7,1,2]^-,3^-,4^+,5^+,6^+\right) \, 
    \end{equation}
    \begin{equation}
    D_4= i\,\mathcal{U} \left(1^-,[2,3,4,5]^-,6^+,7^+\right) \times \frac{i}{p^2_{2345}}\times i\,\mathcal{U}\left(2^-,3^-,4^+,5^+,[6,7,1]^+\right) 
    \end{equation}
    \begin{equation}
    D_5= i\,\mathcal{U} \left(1^-,2^-,3^-,4^+,5^+,6^+,7^+\right)
    \,.
    \label{eq:7nmhv_z}
\end{equation}
where $i/p^2_{i_1\dots i_m}=i/(p_{i_1}+\dots+p_{i_m})^2$ is the scalar propagator in our new action $S\left[Z^{\bullet},Z^{\star}\right]$ Eq.~\eqref{eq:Z_action1}. The above terms combine as following
\begin{equation}
    D_1+D_2+D_3+D_4+\frac{1}{2}D_5 \,,
    \label{eq:sum_d5}
\end{equation}
where the factor of $1/2$ associated with $D_5$ is just the symmetry factor. It originates because of the fact that there are two color orderings contributing to each of the four terms $D_1$ to $D_4$. From this amplitude onwards \footnote{Recall, for 6-point $\mathrm{\overline{MHV}}$ $\mathcal{A}(1^-,2^-,3^-,4^-,5^+,6^+)$ we said we verified it numerically. There we first generated on-shell momenta in Mathematica. Then we gave those values as input both to $\mathcal{U}(1^-,2^-,3^-,4^-,5^+,6^+)$ Eq.~\eqref{eq:6g_MHVbar_Z} and the well-known form for $\mathcal{A}(1^-,2^-,3^-,4^-,5^+,6^+)$ $\mathrm{\overline{MHV}}$ amplitude  Eq.~\eqref{eq:6G_MHVbar_onshell} and found that the two results agreed.}, in order to validate our results, we used the following strategy. We computed the numerical values for the expression Eq.~\eqref{eq:sum_d5} using a set of on-shell momenta and compared it against the numerical results for the same amplitude obtained from the well-known \texttt{GGT} Mathematica package \cite{Dixon:2010ik} together with the \texttt{S@M} package \cite{Maitre2007}. The latter is used to generate numerical values for spinors corresponding to on-shell momenta. For the current amplitude, that is 7-point NMHV $(- - - + + + +)$, we found the expression Eq.~\eqref{eq:sum_d5} numerically agrees with the result from \texttt{GGT} Mathematica package up to an overall constant (integer multiple of $\sqrt{2}$) which originates from the differences in $\left\{\widetilde{v}_{ij}, \widetilde{v}^{\star}_{ij} \right\}$ and $\left\{\langle ij\rangle, [ij] \right\}$.

\begin{figure}[h]
    \centering
 \includegraphics[width=13cm]{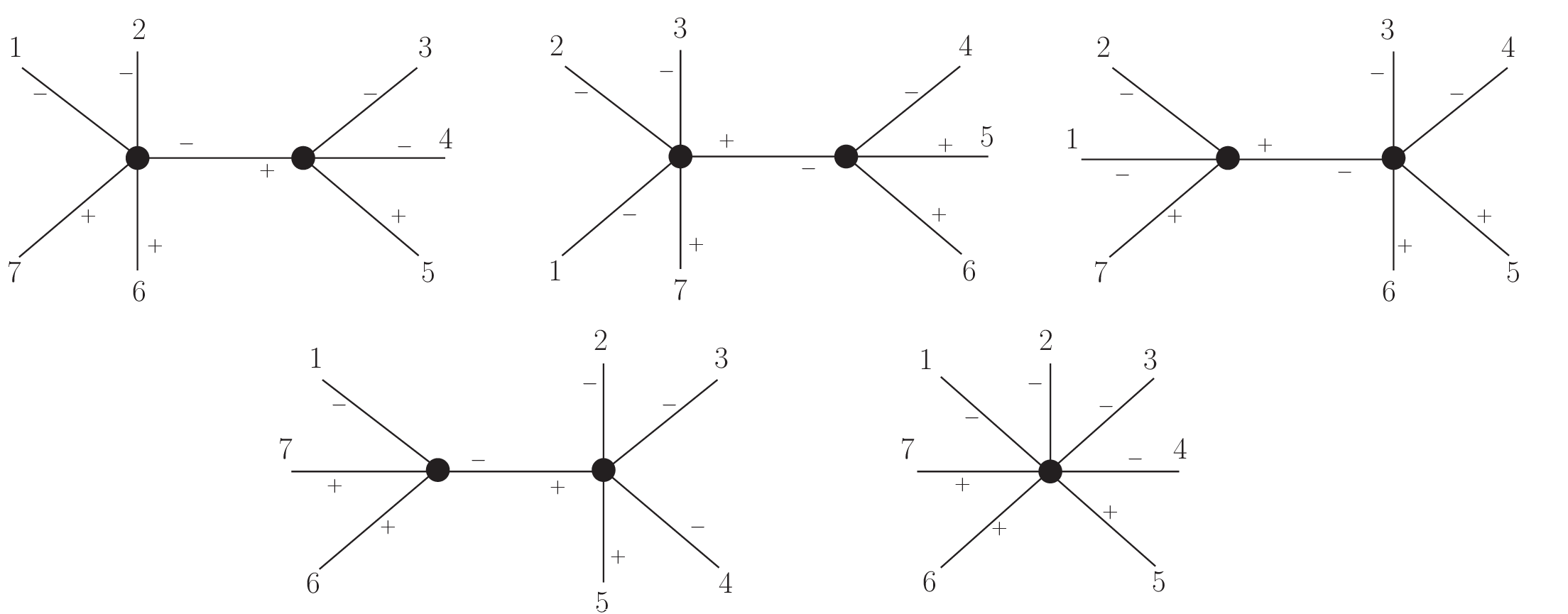}
    \caption{\small 
    Feynman diagrams contributing to the tree-level split-helicity 7-point NNMHV $(- - - - + + +)$ amplitude in our action $S\left[Z^{\bullet},Z^{\star}\right]$. This image was taken from our paper \cite{Kakkad:2021uhv}.}
    \label{fig:NNMHV7}
\end{figure}

We used the same strategy to compute 7-point NNMHV $(- - - - + + +)$ amplitude. In this case too we had exactly five contributions shown in Figure \ref{fig:NNMHV7}. Notice, this is the first case that contains contributions originating from the combination of the new vertices in our action (i.e. 5-point $\overline{\mathrm{MHV}}$ $(- - - + +)$ which are not present in the MHV action) with the MHV vertices. We verified that in the on-shell limit, the sum of these contributions agrees numerically with the results obtained from the \texttt{GGT} Mathematica package.

\subsection{8-point amplitudes}

In order to put our action through a rigorous test, we computed a couple of 8-point tree amplitudes as well. 

\begin{figure}[h]
    \centering
 \includegraphics[width=13cm]{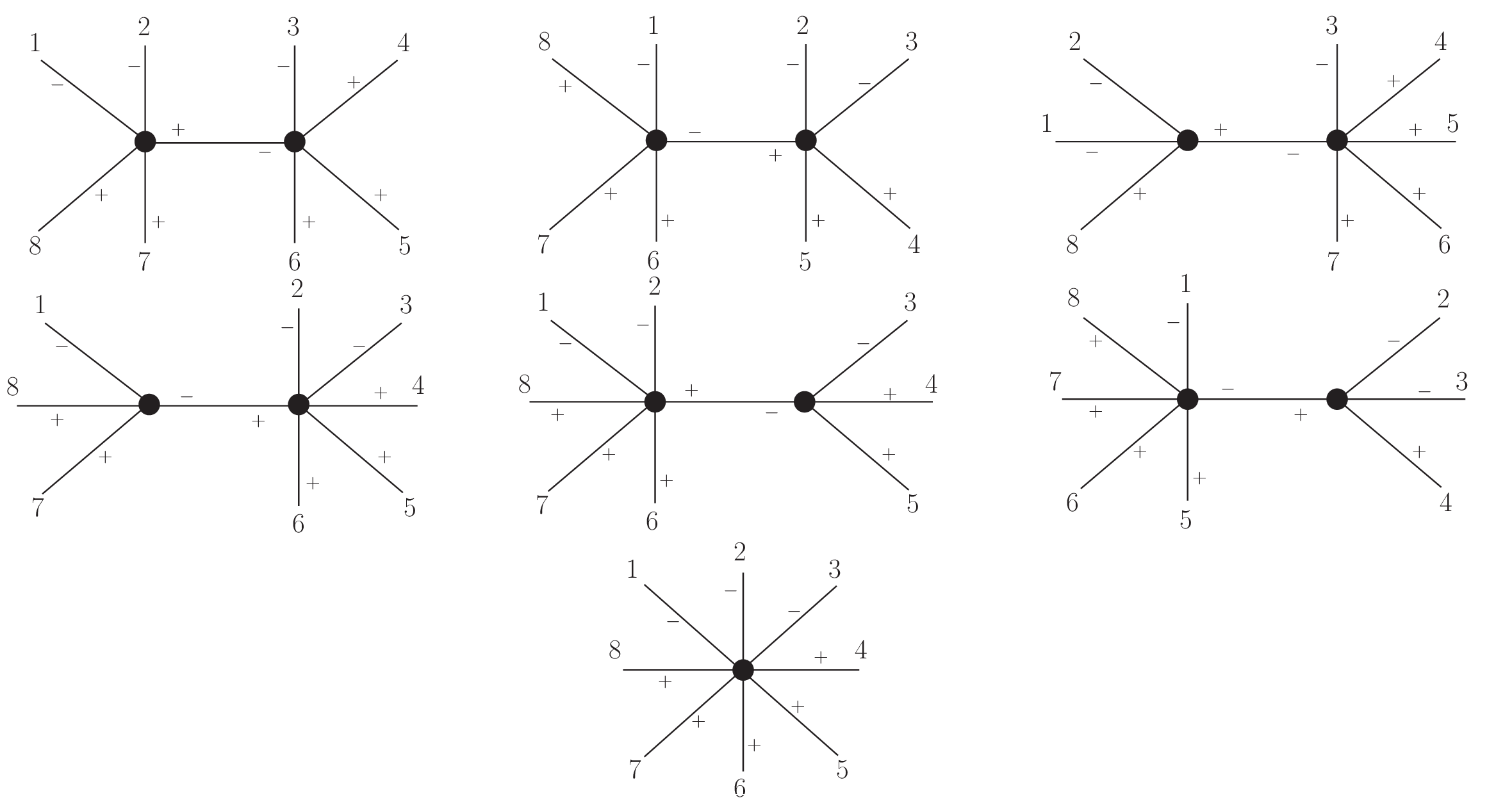}
    \caption{\small 
    Feynman diagrams contributing to the tree-level split-helicity 8-point NMHV $(- - - + + + + +)$ amplitude in our action $S\left[Z^{\bullet},Z^{\star}\right]$. This image was taken from our paper \cite{Kakkad:2021uhv}.}
    \label{fig:NMHV8}
\end{figure}

The first non-trivial case we considered was 8-point NMHV $(- - - + + + + +)$. For this, we had only seven contributions depicted in Figure \ref{fig:NMHV8}. Except for the NNMHV vertex, the remaining contributions consisted of a combination of a pair of MHV vertices. We checked that the sum of these, in the on-shell limit, agreed with the numerical results from the \texttt{GGT} Mathematica package.

\begin{figure}[h]
    \centering
 \includegraphics[width=13cm]{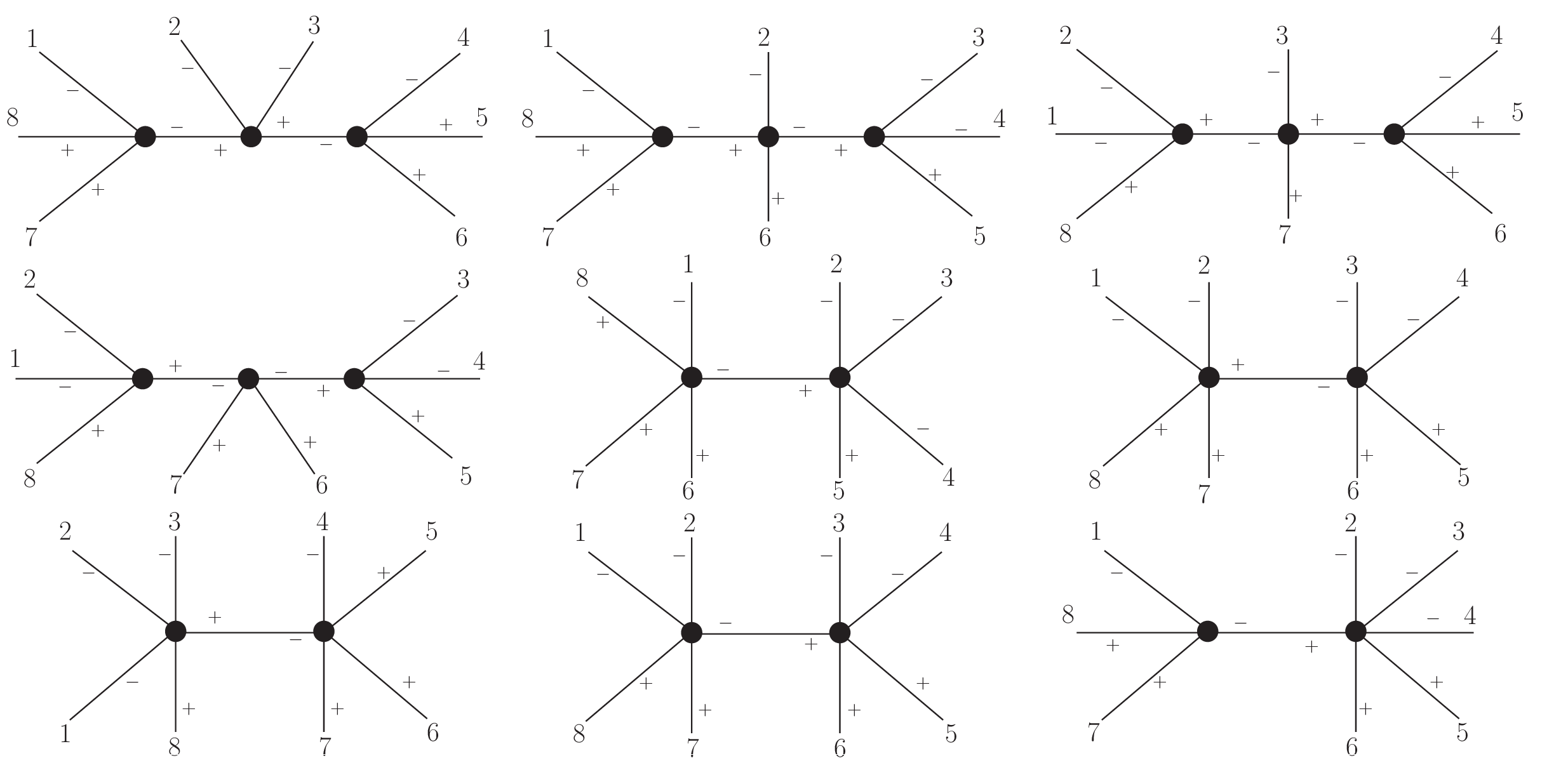}\\
 \includegraphics[width=13cm]{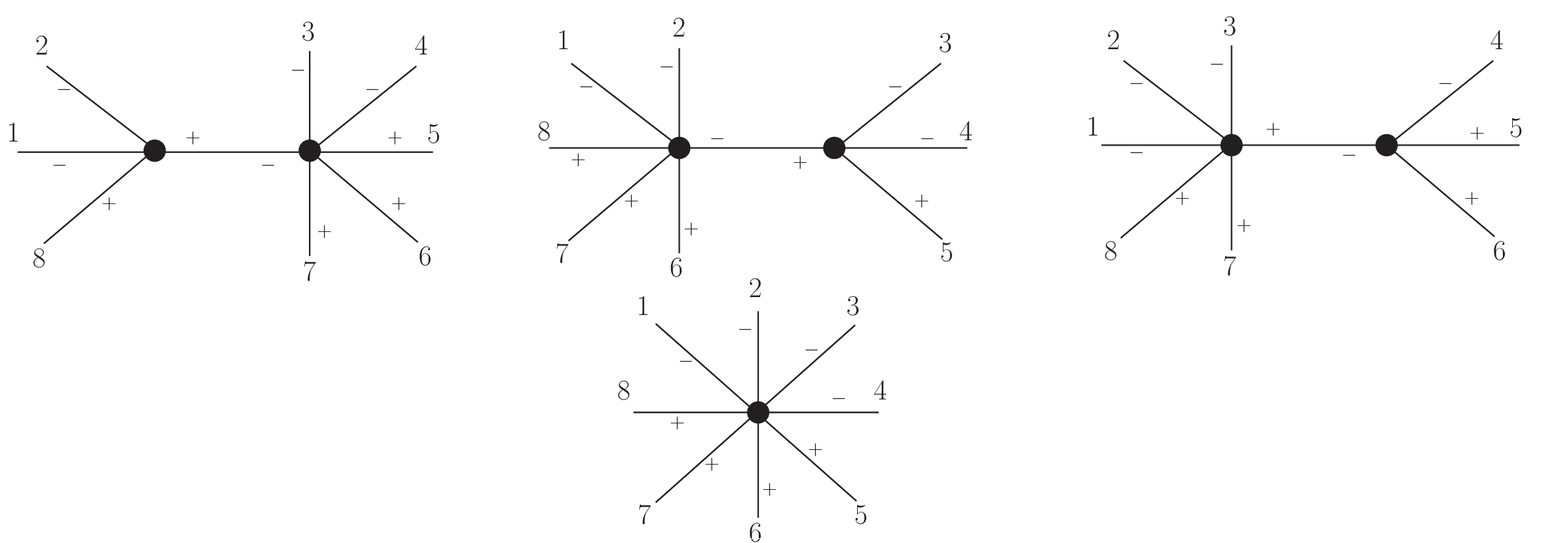}
    \caption{\small 
    Feynman diagrams contributing to the tree-level split-helicity 8-point NNMHV $(- - - - + + + +)$ amplitude in our action $S\left[Z^{\bullet},Z^{\star}\right]$. This image was taken from our paper \cite{Kakkad:2021uhv}.}
    \label{fig:NNMHV8}
\end{figure}

Finally, we computed the 8-point NNMHV $(- - - - + + + +)$ amplitude. There were 13 contributions to this amplitude. We represent these diagrammatically in Figure \ref{fig:NNMHV8}. This example turned out to be most crucial because the contributions consisted of non-trivial combinations of the interaction vertices from our action: three 4-point MHV vertices connected via two scalar propagators, higher point new interaction vertices in our action like 6-point NMHV connected with 4-point MHV vertex, and 5-point MHV connected with 5-point $\overline{\mathrm{MHV}}$. The explicit expressions for these contributions read (in the order: from top left in Figure \ref{fig:NNMHV8} to the right and then the next line and so on) 
\begin{multline}
    D_1 = i\,\mathcal{U} \left([7,8,1,2,3]^-, 4^-,5^+, 6^+\right) \times \frac{i}{p^2_{456}}\times i\,\mathcal{U} \left(1^-, [2,3,4,5,6]^-,7^+,8^+\right) \\ 
    \times \frac{i}{p^2_{781}} 
    \times i\,\mathcal{U} \left(2^-, 3^-, [4,5,6]^+,[7,8,1]^+\right) \, ,
\end{multline}
\vspace{-1cm}
\begin{multline}
    D_2 = i\,\mathcal{U} \left(3^-,4^-,5^+,[6,7,8,1,2]^+\right) \times \frac{i}{p^2_{67812}}\times i\,\mathcal{U} \left(1^-, [2,3,4,5,6]^-,7^+,8^+\right) \\
     \times \frac{i}{p^2_{781}}\times i\,\mathcal{U} \left(2^-, [3,4,5,]^-, 6^+,[7,8,1]^+\right) \, ,
\end{multline}   
\vspace{-1cm}
\begin{multline}
    D_3= i\,\mathcal{U} \left([7,8,1,2,3]^-,4^-,5^+, 6^+\right\} \times \frac{i}{p^2_{456}}\times i\,\mathcal{U} \left(1^-, 2^-, [3,4,5,6,7]^+,8^+\right)\\
     \times \frac{i}{p^2_{812}}\times i\,\mathcal{U} \left([8,1,2]^-,3^-,[4,5,6]^+, 7^+\right) \,, 
\end{multline}   
\vspace{-1cm}
\begin{multline}    
    D_4= i\,\mathcal{U} \left(3^-,4^-,5^+,[6,7,8,1,2]^+\right) \times \frac{i}{p^2_{67812}}\times i\,\mathcal{U} \left(1^-, 2^-, [3,4,5,6,7]^+,8^+\right)\\  \times \frac{i}{p^2_{812}}\times i\,\mathcal{U} \left([8,1,2]^-,[3,4,5]^-,6^+, 7^+\right) \,,
\end{multline} 
\begin{equation}    
    D_5= i\,\mathcal{U} \left(1^-,[2,3,4,5]^-,6^+,7^+, 8^+\right) \times \frac{i}{p^2_{2345}}\times i\,\mathcal{U} \left(2^-, 3^-,4^-,5^+,[6,7,8,1]\right)\,,
\end{equation}    
\begin{equation}    
    D_6= i\,\mathcal{U} \left(1^-,2^-,[3,4,5,6]^+,7^+, 8^+\right) \times \frac{i}{p^2_{3456}}\times i\,\mathcal{U} \left([7,8,1,2]^-,3^-,4^-,5^+,6^+\right)\,,
\end{equation}    
\begin{equation}    
    D_7= i\,\mathcal{U} \left(1^-,2^-,3^-,[4,5,6,7]^+,8^+\right) \times \frac{i}{p^2_{4567}}\times i\,\mathcal{U} \left([8,1,2,3]^-,4^-,5^+,6^+,7^+\right)\, ,
\end{equation}    
\begin{equation}    
    D_8= i\,\mathcal{U} \left(1^-,2^-,[3,4,5,6]^-,7^+,8^+\right) \times \frac{i}{P^2_{3456}}\times i\,\mathcal{U} \left(3^-, 4^-,5^+,6^+,[7,8,1,2]^+\right)\,,
\end{equation}    
\begin{equation}    
    D_9= i\,\mathcal{U} \left(1^-,[2,3,4,5,6]^-,7^+,8^+\right) \times \frac{i}{p^2_{23456}}\times i\,\mathcal{U} \left(2^-, 3^-,4^-,5^+,6^+,[7,8,1]^+\right)\,, 
\end{equation}   
\begin{equation}    
    D_{10}= i\,\mathcal{U} \left(1^-,2^-,[3,4,5,6,7]^+, 8^+\right\} \times \frac{i}{p^2_{34567}}\times i\,\mathcal{U} \left([8,1,2]^-,3^-,4^-,5^+,6^+, 7^+\right)\,, 
\end{equation}   
\begin{equation}    
    D_{11}= i\,\mathcal{U} \left(3^-,4^-,5^+,[6,7,8,1,2]^+\right) \times \frac{i}{p^2_{67812}}\times i\,\mathcal{U} \left(1^-, 2^-,[3,4,5]^-,6^+,7^+,8^+\right)\,,
\end{equation}    
\begin{equation}    
    D_{12}= i\,\mathcal{U} \left([7,8,1,2,3]^-,4^-,5^+, 6^+\right) \times \frac{i}{p^2_{456}}\times i\,\mathcal{U} \left(1^-,2^-,3^-,[4,5,6]^+,7^+, 8^+\right)\, , 
\end{equation}    
\begin{equation}    
    D_{13}= i\,\mathcal{U} \left(1^-,2^-,3^-,4^-,5^+,6^+,7^+,8^+\right) \, .
\end{equation}
These terms combine as follows
\begin{equation}
    D_1+D_2+D_3+D_4+\frac{1}{2}\left(D_5+D_6+D_7+D_8+D_9+D_{10}+D_{11}+D_{12}\right) + \frac{1}{4} D_{13} \,.
    \label{eq:8nnmhv_exp}
\end{equation}
Notice the factor of $1/2$ and $1/4$. These originate because there are four color orderings contributing to $D_1 - D_4$ and $4 \times 1/2 = 2$ color ordering contributing to $D_5 - D_{12}$. The expression Eq.~\eqref{eq:8nnmhv_exp} was evaluated numerically and we found that it agreed with the numerical results for the 8-point NNMHV amplitude from the \texttt{GGT} Mathematica package.

So far, we computed several split-helicity tree-level pure gluonic amplitudes using our new action $S\left[Z^{\bullet},Z^{\star}\right]$ Eq.~\eqref{eq:Z_action1} to demonstrate two important points. First, our action, as well as the master formula for the general vertex Eq.~\eqref{eq:Z_gen_ker}, is operational. Second, it allows for even more efficient computation of the amplitudes in terms of the number of diagrams when compared to the MHV rules (or equivalently the MHV action), needless to say, far more efficient when compared to the Yang-Mills action. Consider for instance the $n$-point split-helicity NMHV tree amplitudes. These require $2(n-3)$ diagrams in the MHV rules whereas $2(n-5)+1$ diagrams ($n\geq 5$) in our new action $S\left[Z^{\bullet},Z^{\star}\right]$ Eq.~\eqref{eq:Z_action1}. It is, however, important to note that the above comparison, in terms of the number of diagrams, should not be straightforwardly extrapolated to the number of contributing terms. This is because the vertex in our action consists of a sum of a few terms (i.e. we do not claim it to be a single term). However, what we would like to stress is that the master formula for the general vertex Eq.~\eqref{eq:Z_gen_ker}, treated as a single diagram, is perfectly operational and simple to execute. Finally, let us also point out that while computing amplitudes, we imposed the on-shell limit at the very end. One could, instead, also use the action for computing off-shell Green's function. 

\subsection{Delannoy numbers}
\label{subsec:del_num}

\begin{table}
\centering
\renewcommand{\arraystretch}{1.4}
\begin{tabular}{cccc}
\cline{1-3}
\multicolumn{1}{|c|}{\# legs}             & \multicolumn{1}{c|}{helicity} & \multicolumn{1}{c|}{\# diagrams} &  \\ 
\hhline{===}
\multicolumn{1}{|c|}{\multirow{2}{*}{4 point}} & \multicolumn{1}{c|}{$\mathrm{MHV}$}           & \multicolumn{1}{c|}{1}                                                                         &  \\ \cline{2-3}
\multicolumn{1}{|c|}{}                         & \multicolumn{1}{c|}{$\mathrm{\overline{MHV}}$}           & \multicolumn{1}{c|}{1}                                                                         &  \\ \cline{1-3}
\multicolumn{1}{|c|}{\multirow{2}{*}{5 point}} & \multicolumn{1}{c|}{$\mathrm{MHV}$}           & \multicolumn{1}{c|}{1}                                                                         &  \\ \cline{2-3}
\multicolumn{1}{|c|}{}                         & \multicolumn{1}{c|}{$\mathrm{\overline{MHV}}$}           & \multicolumn{1}{c|}{1}                                                                         &  \\ \cline{1-3}
\multicolumn{1}{|c|}{\multirow{3}{*}{6 point}} & \multicolumn{1}{c|}{$\mathrm{MHV}$}           & \multicolumn{1}{c|}{1}                                                                         &  \\ \cline{2-3}
\multicolumn{1}{|c|}{}                         & \multicolumn{1}{c|}{$\mathrm{NMHV}$}           & \multicolumn{1}{c|}{3}                                                                         &  \\ \cline{2-3}
\multicolumn{1}{|c|}{}                         & \multicolumn{1}{c|}{$\mathrm{\overline{MHV}}$}           & \multicolumn{1}{c|}{1}                                                                         &  \\ \cline{1-3}
\multicolumn{1}{|c|}{\multirow{4}{*}{7 point}} & \multicolumn{1}{c|}{$\mathrm{MHV}$}           & \multicolumn{1}{c|}{1}                                                                         &  \\ \cline{2-3}
\multicolumn{1}{|c|}{}                         & \multicolumn{1}{c|}{$\mathrm{NMHV}$}           & \multicolumn{1}{c|}{5}                                                                         &  \\ \cline{2-3}
\multicolumn{1}{|c|}{}                         & \multicolumn{1}{c|}{$\mathrm{NNMHV}$}           & \multicolumn{1}{c|}{5}                                                                         &  \\ \cline{2-3}
\multicolumn{1}{|c|}{}                         & \multicolumn{1}{c|}{$\mathrm{\overline{MHV}}$}           & \multicolumn{1}{c|}{1}                                                                         &  \\ \cline{1-3}
\multicolumn{1}{|c|}{\multirow{4}{*}{8 point}} & \multicolumn{1}{c|}{$\mathrm{MHV}$}           & \multicolumn{1}{c|}{1}                                                                         &  \\ \cline{2-3}
\multicolumn{1}{|c|}{}                         & \multicolumn{1}{c|}{$\mathrm{NMHV}$}           & \multicolumn{1}{c|}{7}                                                                         &  \\ \cline{2-3}
\multicolumn{1}{|c|}{}                         & \multicolumn{1}{c|}{$\mathrm{NNMHV}$}           & \multicolumn{1}{c|}{13}                                                                         &  \\ \cline{2-3}
\multicolumn{1}{|c|}{}                         & \multicolumn{1}{c|}{$\mathrm{NNNMHV}$}           & \multicolumn{1}{c|}{7}                                                                        &  \\ \cline{2-3}
\multicolumn{1}{|c|}{}                         & \multicolumn{1}{c|}{$\mathrm{\overline{MHV}}$}           & \multicolumn{1}{c|}{1}                                                                         &  \\ \cline{1-3}
                                               &                                 &                                                                                                & 
\end{tabular}
\caption{\small
The number of Feynman diagrams contributing to the different tree-level split-helicity amplitudes in our action $S\left[Z^{\bullet},Z^{\star}\right]$. This Table was taken from our paper \cite{Kakkad:2021uhv}.}
\label{table:Z_th_ampl}
\end{table}

\begin{figure}
    \centering
    \includegraphics[width=4.3cm]{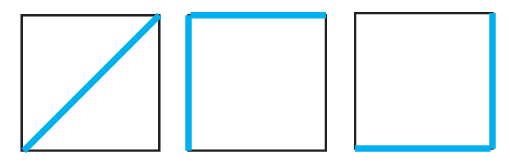}
    \caption{
    \small  The 3 paths to go from the origin $(0,0)$ of a 2D lattice to $(1,1)$ using just three moves: east $\rightarrow$, north $\uparrow$, and north-east $\nearrow \,$.}
    \label{fig:11DN}
\end{figure}

\begin{figure}
    \centering
    \includegraphics[width=9cm]{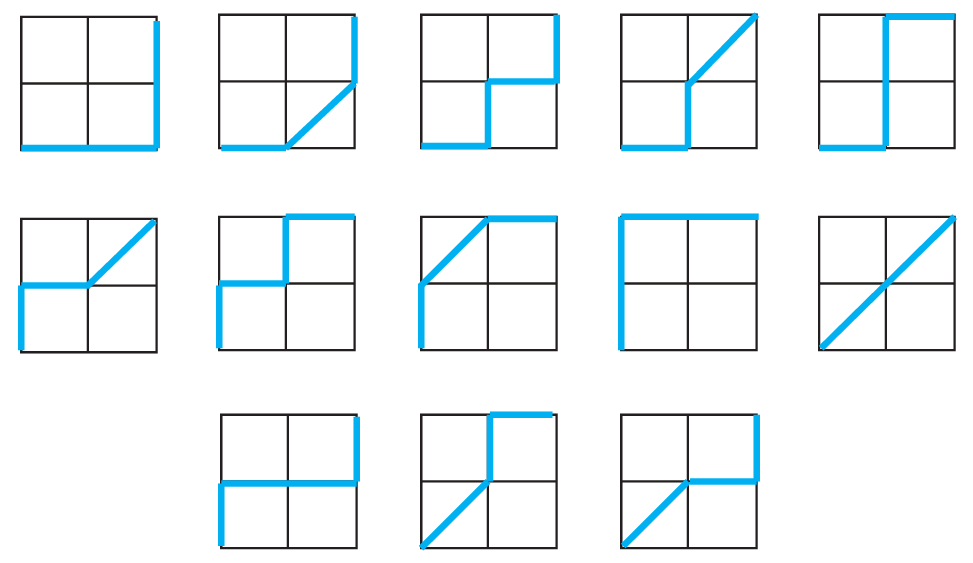}
    \caption{
    \small 
   The 13 paths to go from the origin $(0,0)$ of a 2D lattice to $(2,2)$ using just three moves: east $\rightarrow$, north $\uparrow$, and north-east $\nearrow \,$. }
    \label{fig:22DN}
\end{figure}

Recently, we discovered that the number of diagrams we encountered when computing the  different split-helicity tree-level amplitudes 
using our new action $S\left[Z^{\bullet},Z^{\star}\right]$ Eq.~\eqref{eq:Z_action1} (tabulated in Table \ref{table:Z_th_ampl}) follows the \textit{Delannoy numbers} 
$D(n,m)$. In order to understand these numbers consider a 2D lattice. Now, assume that one can use only the following three moves: east $\rightarrow$, north $\uparrow$, and north-east $\nearrow \,$; to go from a given lattice point to the other. These numbers $D(n,m)$ represent the total number of possible lattice paths one can use to go from the origin $(0,0)$ of this 2D lattice to a point $(n,m)$ such that each path is made up of a combination of these three moves only. For example, $D(1,1)$ and $D(2,2)$ represent the total possible paths, developed using the above-stated three moves, which one could use to go from the origin $(0,0)$ to $(1,1)$ and $(2,2)$ respectively. There are 3 and 13 paths for the former and latter respectively. We show them diagrammatically in Figures \ref{fig:11DN} and \ref{fig:22DN}. 

For a generic point $(n,m)$, one can compute $D(n,m)$ using the following identity
\begin{equation}
    D(n,m)= \sum_{i=0}^{n} \binom{m}{i} \binom{n+m-i}{m} = \sum_{i=0}^{n} 2^i \binom{m}{i} \binom{n}{i}\,.
    \label{eq:delannoy}
\end{equation}


We computed the numbers for the first few cases (see Table \ref{tab:del_num}). 

Let $n+2$ be the number of plus helicity legs and $m+2$ be the minus helicity legs in the split-helicity tree level pure gluonic amplitude $\mathcal{A}_{\underbrace{-\,\cdots\,-}_{m+2}\underbrace{+ \,\cdots\, +}_{n+2}}$. Notice, the number of diagrams in Table \ref{table:Z_th_ampl} for various values of $(n,m)$ (in $\mathcal{A}_{\underbrace{-\,\cdots\,-}_{m+2}\underbrace{+ \,\cdots\, +}_{n+2}}$ with $n,m \geq 0$) is exactly equal to the number of paths $D(n,m)$, for the same values of $(n,m)$, in Table \ref{tab:del_num}. Although it may seem like a coincidence, the appearance of Delannoy numbers when computing split-helicity tree amplitudes using our action $S\left[Z^{\bullet},Z^{\star}\right]$ Eq.~\eqref{eq:Z_action1} can be understood as follows. Let us first begin with realizing the 2D lattice. This is achieved using the vertices in our action $S\left[Z^{\bullet},Z^{\star}\right]$. Keeping the 4-point MHV at the origin $(0,0)$, one can increase the plus helictites by one along the horizontal axis and the minus helicity by one along the vertical axis. As a result, the MHV vertices sit along the horizontal axis whereas the $\mathrm{\overline{MHV}}$ vertices are along the vertical axis. The rest of the vertices are in the bulk between the two axes. This forms the 2D lattice. As for the three moves in Delannoy numbers, there are two ways of realizing this in amplitude computation using our action. The first way is what we call the \textit{bottom-up} approach. When computing the amplitudes, one contribution comes solely from the vertex with exactly the same helicity configuration as the amplitude. This corresponds to one move. The remaining contributions originate by connecting vertices via the scalar propagator. Given the structure of the propagator $(+ -)$ in our action, there are two ways of connecting any pair of vertices: the plus helicity of the first vertex connected to the minus helicity of the other and vice versa. These are the remaining two moves. This is in fact the approach we used for developing the contributions when computing amplitudes in the previous Subsections. The second way of realizing these three moves is the \textit{top-down} approach. Let us begin with a given tree amplitude. We know in our action one contribution is exactly a vertex. To develop remaining contributions, we can break the amplitude into a pair of sub-trees (tree-level sub-diagrams) connected in two ways via the scalar propagator. Now each of these sub-trees can be further decomposed into three ways. One vertex and a pair of sub-trees connected in two ways. Iterating this procedure we can reproduce the Delannoy number. Note the iteration ends when the sub-trees are exactly the vertices of our action that could not be further decomposed. This also implies that if one uses a different action, or a different set of building blocks, say the Yang-Mills action or the MHV action, the iteration would go on further, and then the number series would be completely different and not the Delannoy numbers. Finally, in the context of amplitudes, each Delannoy path must be thought of as a diagram contributing to the amplitude, and the diagram is made up of vertices, representing the lattice points, connected in a unique fashion. 

\begin{table}
\centering
\begin{tabular}{c|cccc}
(n,m) & 0 & 1 & 2  & 3  \\ \hline
0     & 1 & 1 & 1  & 1  \\
1     & 1 & 3 & 5  & 7  \\
1     & 1 & 5 & 13 & 25 \\
3     & 1 & 7 & 25 & 63
\end{tabular}
\caption{\small 
The Table represents the total number of paths $D(n,m)$ to go from the origin $(0,0)$ of a 2D lattice to a point $(n,m)$ using just three moves: east $\rightarrow$, north $\uparrow$, and north-east $\nearrow \,$.}
\label{tab:del_num}
\end{table}

Before we end this section, let us point out that the outcome of the appearance of Delannoy numbers in the context of computing split-helicity tree amplitudes using our action is that modifying Eq.~\eqref{eq:delannoy} one can simply predict the number of diagrams required to compute these amplitudes. By doing this, the number of contributions reads
\begin{align}
     \# \,\,\mathrm{contribution}\,\, \mathcal{A}_{\underbrace{-\,\cdots\,-}_{m+2}\underbrace{+ \,\cdots\, +}_{n+2}}^{\left(n+m+4 \, \mathrm{Tree}\right)} =& D(n+2,m+2) \nonumber\\
     =& \sum_{i=0}^{n} \binom{m}{i} \binom{n+m-i}{m} = \sum_{i=0}^{n} 2^i \binom{m}{i} \binom{n}{i}\,.
    \label{eq:delannoy_Z}
\end{align}
Note, the above discussion is restricted to the split-helicity tree amplitudes only. And therefore one cannot extrapolate these numbers for computing tree amplitudes, for the same number of plus and minus helicity legs, but with other helicity configurations.

\section{\texorpdfstring{$S[Z^{\bullet}, Z^{\star}]$}{ZAC} as a twistor space prescription}
\label{sec:Zac_TS}

Recall from Section \ref{sec:MHV_rules}, one of the main ideas of \cite{Witten2004} was that one can use a set of $(n_{-} - 1)$ degree one curves in twistor space connected via propagators to compute a tree-level pure gluonic amplitude with $n_{-}$
negative helicity legs. In terms of gauge theory amplitudes, these curves of degree one in twistor space were shown to be the MHV amplitudes. This in turn gave birth to the MHV rules \cite{Cachazo2004}. However, in the same work \cite{Witten2004}, it was also pointed out that it is not necessary to use only degree one curves. The same amplitude could also be computed using instead curves of a higher degree. In fact, in a series  of papers  \cite{Roiban1,Roiban2,Roiban3}, the authors demonstrated that the $\mathrm{\overline{MHV}}$ and the NMHV amplitudes could be computed using a single curve of degree $d = n_{-} - 1$. This implies there are two ways of computing the same amplitude: one is the completely "disconnected" approach which uses a set of disconnected $(n_{-} - 1)$ degree one curves (the MHV rules); and the "connected" approach which uses a single connected curve of degree $d = n_{-} - 1$ in twistor space.

\begin{figure}
    \centering
    \includegraphics[width=14cm]{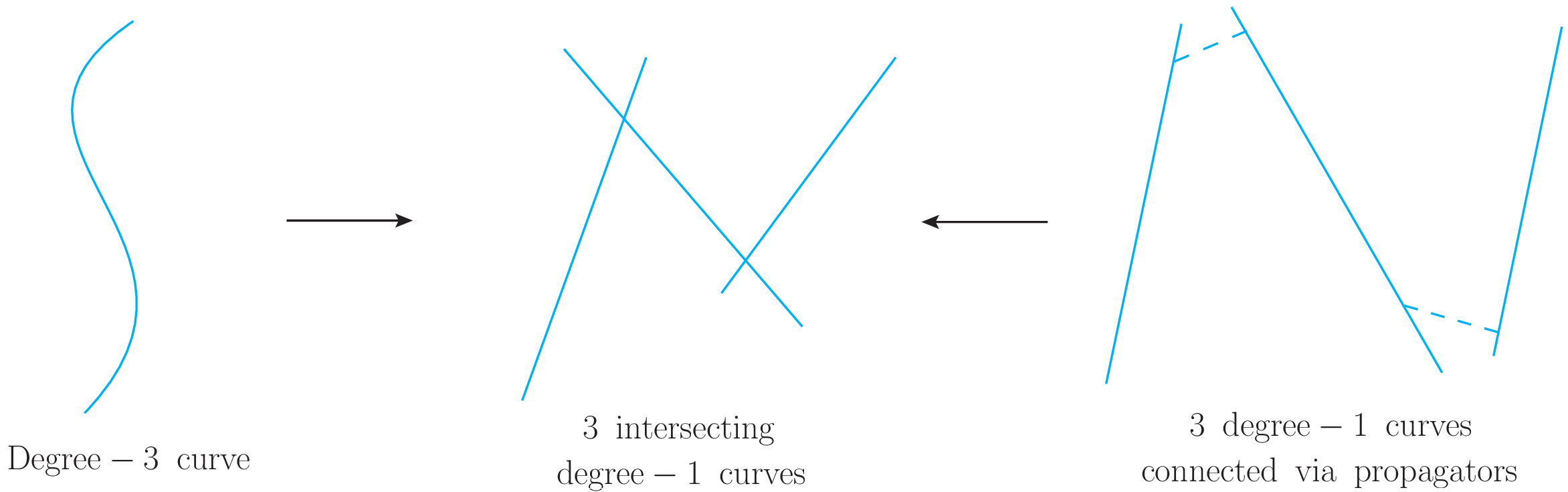}
    \caption{
    \small  
    The degree-3 curve (on the left) in twistor space can be shown to degenerate, over a locus, into 3 intersecting degree-one curves (in the middle), and on the same locus the propagators (represented via dashed line) connecting 3 degree-one curves (on the right) shrink to zero and therefore they too reduce to 3 intersecting degree-one curves.}
    \label{fig:TS_deg}
\end{figure}

In a later work \cite{gukov2004equivalence}, the authors demonstrated that the above-stated two approaches are indeed equivalent. They argued that there exists a locus over which the degree $d = n_{-} - 1$ curve in the connected approach can be shown to degenerate into intersecting $(n_{-} - 1)$ degree one curves and on the same locus the propagators connecting $(n_{-} - 1)$ degree one curves in the disconnected approach shrinks to zero thus they too reduce to intersecting $(n_{-} - 1)$ degree one curves (we represent this diagrammatically in Figure \ref{fig:TS_deg}). In the process, however, they conjectured that there exist intermediate prescriptions  where one could simply use a set of curves of different degrees $d_i$ such that their sum is $d = \sum d_i = n_{-} - 1$. The extreme situations consist of either the disconnected approach when $d_i = 1, \forall i$ (all curves are of degree one) or the connected approach if $i=1$ (single curve of degree $d$).

Motivated by the idea of intermediate twistor space prescriptions utilizing curves of different degrees, Bena et al. in \cite{Bena_2005} developed "NMHV vertices" which provided an explicit realization of this idea. What they call the "NMHV vertices", are essentially NMHV amplitudes computed using the MHV rules (the disconnected approach) and then continued off-shell using the CSW prescription Eq.~\eqref{eq:CSW_offshell}. Since imposing the on-shell limit to these vertices give the corresponding amplitudes, they should localize on $d = n_{-} - 1$ curve in twistor space. However, since they were computed using the disconnected approach they should also represent $(n_{-} - 1)$ degree one curves connected via propagators thereby unifying the two ideas. They further went on to develop a recursive approach using these NMHV vertices to compute gauge theory amplitudes thus providing an explicit example of an intermediate twistor space prescription (for further details see \cite{Bena_2005}).

The above set of ideas indicates that our action $S\left[Z^{\bullet},Z^{\star}\right]$ Eq.~\eqref{eq:Z_action1} is an explicit Minkowski-space realization of yet a different intermediate twistor space prescription for computing gauge theory amplitudes. Recall, our action has MHV vertices that correspond to the MHV amplitudes, just like the MHV action. These vertices, therefore, localize on degree one curves in twistor space. Then we have the $\mathrm{\overline{MHV}}$ vertices $(- \dots - + +)$. These too correspond to $\mathrm{\overline{MHV}}$ amplitudes in the on-shell limit and can thus be thought of as localizing either on connected curves of degree $d = n_{-} - 1$ or on a set of $(n_{-} - 1)$ degree one curves connected via propagators. To see this, consider, for the sake of simplicity, the five-point $\mathrm{\overline{MHV}}$ amplitude as computed using the MHV action
\begin{center}
\includegraphics[width=15.6cm]{Chapter_1/5gmhvbarcsw2.eps}
\end{center}
Above, the L.H.S. represents five-point $\mathrm{\overline{MHV}}$ amplitude which would localize on a degree 2 curve whereas the R.H.S. represents all possible degeneration of a  degree 2 curve in terms of a pair of degree one curves. In our action, we get similar terms where the 3-point MHV gets replaced by $\overline{ \Omega}_2 $ and $\overline{ \Psi}_2$ kernels (\emph{cf.} Figure \ref{fig:MHVbar5_vertex}). There are two things to understand. First, these kernels resum the tree-level connections one could develop using the $(--+)$ triple gluon vertex (or the 3-point MHV). Second, the degenerations on the R.H.S. above are all enclosed in the five-point $\mathrm{\overline{MHV}}$ vertex $\mathcal{L}_{---++}$ in our action. As a result in the on-shell limit, the vertex reproduces the amplitude. This suggests that the $\mathrm{\overline{MHV}}$ vertices in our action localize on degree $d = n_{-} - 1$ curves in twistor space. 

Apart from MHV and $\mathrm{\overline{MHV}}$ vertices, the story for the other vertices in our action is not as straightforward. Let us begin by pointing out that a generic vertex (with $d+1$ negative helicity legs) in our action does not correspond to a degree $d$ curve. This is because imposing the on-shell limit to the vertices doesn't give the corresponding amplitudes. However, as pointed out above, the kernels $\overline{ \Omega}_n $ and $\overline{ \Psi}_n$ resum the tree contributions of the 3-point MHV. These kernels, therefore represent a set of degree one curves connected via the propagator. Finally, recall, each vertex consists of a sum of terms where each term contains an MHV vertex substituted with these kernels (\emph{cf.} Figure \ref{fig:ZTH_vertex_gen}). Thus we can conclude that each vertex in our action consists of a sub-set (not the complete set) of possible degeneration of a degree $d$ curve in terms of degree one curves (MHV vertices). Consider again, for example, the 6-point NMHV $(- - - + + +)$ amplitude in our action. The contributions are shown below
\begin{center}
 \includegraphics[width=13cm]{Chapter_3/6nmhv.eps}
\end{center}
Since the sum of these in the on-shell limit give the amplitude, these terms represent all possible degeneration of a degree 2 curve. Now the first two terms represent two possible degenerations in terms of a pair of 4-point MHV vertices and the remaining are all encapsulated in the 6-point NMHV $(- - - + + +)$ vertex in our action. These account for the degeneration involving the 4-point MHV connected with the 3-point MHV.

All of the above suggests that our action consists of objects that correspond to either curve of different degrees in twistor space (MHV and $\mathrm{\overline{MHV}}$ vertices) or a collection of certain degenerations of a higher degree curve in terms of degree one curves (any other vertex). Thus, it provides the Minkowski-space action realization of an intermediate twistor space prescription for computing gauge theory amplitudes as pointed out in \cite{gukov2004equivalence}. Note, however, the proof of this correspondence is beyond the scope of this project and is left for the future.

\chapter{Quantum corrections to the MHV action}
\label{QMHV-chapter}

In this chapter, we summarize the work we did in \cite{Kakkad_2022} where we derived the one-loop effective MHV action. The MHV action, as shown in Chapter \ref{Chapt2}, allows for efficient computation of tree amplitudes when compared with the Yang-Mills action. But, when computing loop amplitudes, there are missing loop contributions originating either solely from the $(+ + -)$ triple gluon vertex or the mixing of this vertex with the MHV vertices (jointly, we call such contributions as the non-MHV contributions). Recall, this is exactly the vertex that is eliminated in Mansfield's transformation. Therefore, in order to systematically develop loop corrections to the MHV action, we used the one-loop effective action approach. We, however, argue that the one-loop effective MHV action should not be derived starting with the MHV action itself. One should rather start with the one-loop effective action for the Yang-Mills theory and then apply Mansfield's transformation. In this way, the non-MHV contributions are not missing. We verify this by computing the 4-point $(+ + + +)$ and $(+ ++ -)$ one-loop amplitudes and show that these match with the known results. This work lays the foundation for systematically developing the loop corrections to our new Wilson line-based action, discussed in Chapter \ref{WLAc-chapter}. We discuss this in the next chapter.

\section{The Problem}
\label{sec:prob}

With the success of our new action $S\left[Z^{\bullet},Z^{\star}\right]$ Eq.~\eqref{eq:Z_action1} (we shall call it the \emph{Z-field action} from now onwards) in efficiently computing tree-level pure gluonic amplitudes, the next natural step is to consider \emph{quantum corrections} i.e. the loop amplitudes. However, we have an immediate problem. Just as in the case of the MHV action, the Z-field action has missing loop contributions. To see this, let us first consider the MHV action. It consists of only the MHV vertices $(+ \dots + --)$ using which one cannot compute amplitudes where either all the gluons have plus helicities $(+ \dots + +)$ or if one has a minus helicity $(+ \dots + -)$ \cite{Brandhuber2007a,Fu_2009, Boels_2008, Ettle2007,Brandhuber2007,Elvang_2012}. As a result, they are naturally zero in the MHV action. At the tree level, for on-shell momenta, this is indeed the case (\emph{cf.} Eqs.~\eqref{eq:plus_tree_amp}) but the loop amplitudes of these types are in general non-zero. In fact, the all-plus $(+ \dots + +)$ and single-minus $(+ \dots + -)$ one-loop amplitudes are rational functions of spinor products. This implies that there are missing loop contributions when computing loop amplitudes using the MHV action Eq.~\eqref{eq:MHV_action}. Recall, the MHV action was derived from the Yang-Mills action Eq.~\eqref{eq:YM_LC_action} via Mansfield's transformation that maps the Self-Dual part of the Yang-Mills to a free theory, thereby eliminating the $(+ + -)$ triple gluon vertex. This vertex contributes to the all-plus $(+ \dots + +)$ and single-minus $(+ \dots + -)$ one-loop amplitudes. In fact, the former is the sole quantum correction to the Self-Dual theory. In the case of the Z-field action, this issue escalates because to derive it we eliminated both the triple gluon vertices from the Yang-Mills action. As a consequence, amplitudes of the type all-plus $(+ \dots + +)$, single-minus $(+ \dots + -)$, all-minus $(- \dots - -)$ and  single-plus $(- \dots - +)$ are all zero to all orders, which is not the case in Yang-Mills theory.

In the context of the MHV action, a few approaches have been developed to successfully recover the all-plus one-loop amplitude. For instance, in \cite{Brandhuber2007a}, these amplitudes were obtained from the Jacobian arising from a  holomorphic change of variables mapping the Self-Dual Yang-Mills (SDYM) theory to a free theory (we will discuss this in detail towards the end of the next Section). Then, in \cite{Ettle2007}, the violation of the S-matrix equivalence theorem was shown to bring back the contributions to the all-plus one-loop amplitude. In \cite{Brandhuber2007} yet another approach, based on the world-sheet regularization of \cite{CQT1,CQT2}, was used to introduce all-plus one-loop vertex to the MHV action. Finally, in \cite{Boels_2008}, the authors used the massive MHV rules (i.e. the MHV rules for Yang-Mills gauge theory coupled to a massive colored scalar) following which, $n$-point all-plus and single-minus one-loop amplitudes were developed in \cite{Elvang_2012}. In \cite{Kakkad_2022}, we employed a yet different approach, based on the One-Loop Effective action technique, that allowed us to systematically develop quantum corrections to the MHV action. This approach is, however, fairly generic and can be used for other actions (say the Z-field action), equivalent to the Yang-Mills action, derived via non-linear field transformations. In this chapter, we will summarize the work we did in \cite{Kakkad_2022} where we focused on using the MHV action as a toy model to test our approach. In the following chapter, we will discuss the extension of our approach to developing quantum corrections to the Z-field action.

\section{One-Loop Effective Action for SDYM Theory}
\label{sec:SDYM}

In this section, our focus is to review the one-loop effective action approach. To do this, we use the simple case of the Self-Dual Yang-Mills (SDYM) theory. Recall, from Eq.~\eqref{eq:SDactionLC}, that the Self-Dual action reads
\begin{equation}
S_{\mathrm{SD}}\left[A^{\bullet},A^{\star}\right]=\int dx^{+}\left(\mathcal{L}_{+-}+\mathcal{L}_{++-}\right)\,,
\end{equation}
\begin{equation}
    S_{\mathrm{SD}} [A^{\bullet}, A^{\star}] = \int\! dx^{+} d^{3}\mathbf{x}\, \left[
    -\mathrm{Tr}\,\hat{A}^{\bullet}\square\hat{A}^{\star} 
     -2ig\,\mathrm{Tr}\,\partial_{-}^{-1}\partial_{\bullet} \hat{A}^{\bullet}\left[\partial_{-}\hat{A}^{\star},\hat{A}^{\bullet}\right]
    \right]\, .
    \label{eq:actionSD}
\end{equation}
In order to derive the one-loop effective action, the starting point is the generating functional for the Green's function
\begin{equation}
    Z_{\mathrm{SD}}[J]=\int[dA]\, e^{i\left(S_{\mathrm{SD}}[A] + \int\!d^4x\, \Tr \hat{J}_j(x) \hat{A}^j(x)\right) } \,.
    \label{eq:ZG_SD}
\end{equation}
Above, $x \equiv (x^{+}, \mathbf{x})$; the index $j=\bullet,\star$ and $\hat{J}$ represents the external source (auxiliary) coupled to the Self-Dual action. Usually, there is an overall normalization $\mathcal{N}$ in the definition of the generating functional. We suppress it for the sake of simplicity. 

We now expand the terms in the exponent in Eq.~\eqref{eq:ZG_SD} around the classical solution ${\hat A}_{c}[J]=({\hat A}_{c}^{\bullet}[J],{\hat A}_{c}^{\star}[J])$. These are the solutions of the classical EOM
\begin{equation}
   \left. \frac{\delta S_{\mathrm{SD}}[A]}{\delta {\hat A}^j(x)}\right|_{{\hat A}={\hat A}_{c}}+{\hat J}_j(x)=0 \,.
    \label{eq:SD_EOM0}
\end{equation}
Recall, from Subsection \ref{subsec:SDYM_EOM}, for $j=\star$, we get the Self-Dual equation of motion. Substituting Eq.~\eqref{eq:actionSD} to Eq.~\eqref{eq:SD_EOM0} for $j=\star$, it reads 
\begin{equation}
    \Box {\hat{A}}^{\bullet} + 2ig{\partial}_{-} \left[ ({\partial}_{-}^{-1} {\partial}_{\bullet} {\hat{A}}^{\bullet}), {\hat{A}}^{\bullet} \right] + {\hat J}^{\bullet} = 0\, .
    \label{eq:SD_EOM1}
\end{equation}
In the same Subsection, we showed that the solution $ {\hat A}^{\bullet}_c[J^{\bullet}]$ of the equation above, when brought to the constant light-cone time $x^+$, was exactly equal to the solution ${\hat A}^{\bullet}[B^{\bullet}] $ of the Mansfield's transformation which, in turn, led to the interpretation of the solution as the inverse of the Wilson line $B^{\bullet}_a(x) \equiv \square^{-1}  {J}^{\bullet}(x) \equiv \mathcal{W}_{(+)}^a[A](x^+;\mathbf{x})$. 

Expanding the terms in the exponent around ${\hat A}_{c}[J]$ we get
\begin{multline}
    S_{\mathrm{SD}}[A] + \int\!d^4x\, \Tr \hat{J}_i(x) \hat{A}^i(x)  
    = S_{\mathrm{SD}}[A_c] + \int\!d^4x\, \Tr \hat{J}_i(x)\hat{A}_c^i(x) \\ + \int\!d^4x\,\Tr\left(\hat{A}^i(x)-\hat{A}_c^i(x)\right)
    \left(\frac{\delta S_{\mathrm{SD}}[A_c]}{\delta \hat{A}^i(x)}+{\hat J}_i(x)\right) \\
    +\frac{1}{2}\int\!d^4xd^4y\,\Tr\left(\hat{A}^i(x)-\hat{A}_c^i(x)\right)\frac{\delta^2 S_{\mathrm{SD}}[A_c]}{\delta\hat{A}^i(x)\delta\hat{A}^j(y)}\left(\hat{A}^j(y)-\hat{A}_c^j(y)\right) + \dots
    \label{eq:expansion_2nd}
\end{multline}
To derive the one-loop effective action, we need terms only up to the second order in the expansion above. Now, the first two terms on the R.H.S. above factor out of the path integral. The linear term vanishes owing to the classical EOMS Eq.~\eqref{eq:SD_EOM0}. As a result, we are left with the following Gaussian integral
\begin{equation}
    Z_{\mathrm{SD}}[J] \approx  e^{i\left(S_{\mathrm{SD}}[A_c] + \int\!d^4x\, \Tr \hat{J}_i(x)\hat{A}_c^i(x)\right)} \int[dA]\, e^{\frac{i}{2}\int\!d^4xd^4y\,\Tr\left(\hat{A}^i(x)-\hat{A}_c^i(x)\right)\left(\frac{\delta^2 S_{\mathrm{SD}}[A_c]}{\delta\hat{A}^i(x)\delta\hat{A}^j(y)}\right)\left(\hat{A}^j(y)-\hat{A}_c^j(y)\right) } \,,
    \label{eq:gen_YM1}
\end{equation}
where $\approx$ represents the truncation of the generating functional up to one loop. The standard technique for evaluating the above integral is through diagonalization of the matrix in the exponent. Since the Self-Dual action is linear in the $A^{\star}$ field, the matrix is simple (see Eq.~\eqref{eq:MAT_SD} below) and therefore we avoid the details of this procedure for now. Later when considering the Yang-Mills action in the following Section, we shall elaborate more on this. Upon integration we get
\begin{equation}
    Z_{\mathrm{SD}}[J]\approx \left[\det
    \left( 
    \frac{\delta^2 S_{\mathrm{SD}}[A_c]}
    {\delta \hat{A}^i(x)\delta \hat{A}^j(y)}
    \right)\right]^{-\frac{1}{2}}
    \exp\left\{i\left(S_{\mathrm{SD}}[A_c] 
    + \int\!d^4x\, \Tr \hat{J}_i(x) \hat{A}_c^i(x) \right) \right\} \,.
\end{equation}
The determinant is over both the color and the position degrees of freedom. The matrix in the determinant has the following explicit form
\begin{equation}
   \left(\begin{matrix}
     \frac{\delta^2 S_{\mathrm{SD}}[A_c]}
    {\delta\hat{A}^{\bullet}(x)\delta\hat{A}^{\star} (y)} 
     & \frac{\delta^2 S_{\mathrm{SD}}[A_c]}
    {\delta\hat{A}^{\bullet}(x)\delta\hat{A}^{\bullet} (y)} \\ \\
\frac{\delta^2 S_{\mathrm{SD}}[A_c]}
    {\delta\hat{A}^{\star}(x)\delta\hat{A}^{\star} (y)} & \frac{\delta^2 S_{\mathrm{SD}}[A_c]}
    {\delta\hat{A}^{\star}(x)\delta\hat{A}^{\bullet} (y)}    
\end{matrix}\right) = \left(\begin{matrix}
     \frac{\delta^2 S_{\mathrm{SD}}[A_c]}
    {\delta\hat{A}^{\bullet}(x)\delta\hat{A}^{\star} (y)} 
     & \frac{\delta^2 S_{\mathrm{SD}}[A_c]}
    {\delta\hat{A}^{\bullet}(x)\delta\hat{A}^{\bullet} (y)} \\ \\
\mathbb{0} & \frac{\delta^2 S_{\mathrm{SD}}[A_c]}
    {\delta\hat{A}^{\star}(x)\delta\hat{A}^{\bullet} (y)}    
\end{matrix}\right)\,.
\label{eq:MAT_SD}
\end{equation}
Owing to the simplicity of the above matrix, we can write
\begin{equation}
    \sqrt{\det
    \left( 
    \frac{\delta^2 S_{\mathrm{SD}}[A_c]}
    {\delta \hat{A}^i(x)\delta \hat{A}^j(y)}
    \right)}
    =  \det
    \left( 
    \frac{\delta^2 S_{\mathrm{SD}}[A_c]}
    {\delta \hat{A}^{\star}(x)\delta \hat{A}^{\bullet}(y)}
    \right)
     \,.
\end{equation}
The functional determinant on the R.H.S. of the equation above can be rewritten as the exponential of the trace of a logarithm. By doing this we get
\begin{equation}
    Z_{\mathrm{SD}}[J]\approx 
    \exp\left\{ iS_{\mathrm{SD}}[A_c] 
    + i\int\!d^4x\, \Tr \hat{J}_i(x) \hat{A}_c^i(x) 
    - \Tr\ln\left(\frac{\delta^2 S_{\mathrm{SD}}[A_c]}
    {\delta \hat{A}^{\star}(x)\delta \hat{A}^{\bullet}(y)} \right) 
    \right\}
    \label{eq:Partition1}
    \,.
\end{equation}
When computing amplitudes, we require only the connected contributions. Therefore it is rather useful to work with the generating functional for the connected Green's function, which is defined as
\begin{equation}
    W_{\mathrm{SD}}[J] = -i \ln Z_{\mathrm{SD}}[J]\,.
\end{equation}
Substituting Eq.~\eqref{eq:Partition1}, we get
\begin{equation}
   W_{\mathrm{SD}}[J] = S_{\mathrm{SD}}[A_c] 
    + \int\!d^4x\, \Tr \hat{J}_i(x) \hat{A}_c^i(x) 
    + i\, \Tr\ln\left(\frac{\delta^2 S_{\mathrm{SD}}[A_c]}
    {\delta \hat{A}^{\star}(x)\delta \hat{A}^{\bullet}(y)} \right) \,.
    \label{eq:SD_CONN_GF}
\end{equation}
In order to demonstrate that the above result is indeed one loop approximation of the generating functional for the connected Green's function, let us substitute the Self-dual action Eq.~\eqref{eq:actionSD} in the argument of log term. By doing this we get 
\begin{equation}
   \frac{\delta^2 S_{\mathrm{SD}}[A_c]}
    {\delta A^{\star I}\delta A^{\bullet J}} 
    = -\square_{IJ}-\left(V_{-++}\right)_{IJK}A_{c}^{\bullet K}
    \label{eq:d2SdAdA}
    \,.
\end{equation}
Above, we use the collective indices $I,J,K \dots$. These include the position, color, etc associated with the fields. Repeated indices are summed over. Since these indices account for both discrete and continuous variables, the summation implies a literal summation for the discrete variable and integration for the continuous. The triple gluon vertex in Eq.~\eqref{eq:d2SdAdA} reads 
\begin{equation}
    V_{-++}^{abc}(x,y,z)=2gf^{abc}\delta^4(x-y)
    \delta^4(x-z)\Big[\partial^{-1}_{-}(y)\partial_{\bullet}(y)-\partial^{-1}_{-}(z)\partial_{\bullet}(z)\Big]\partial_{-}(x) \, .
    \label{eq:v3_position}
\end{equation}
Notice, an additional factor of 2 in the definition of the vertex above as compared to the same vertex in the Yang-Mills action Eq.~\eqref{eq:YM_LC_action}. This originates from the second order functional derivative of this vertex with respect to the fields. Throughout this chapter, we will always associate such factors (originating from the functional derivative) with the corresponding vertices.

The inverse propagator $(\square_{IJ})$ in Eq.~\eqref{eq:d2SdAdA} can be factored out of the log term in the partition function Eq.~\eqref{eq:Partition1} as $\det(\square)$. This is a field-independent infinite constant and can therefore be absorbed in the overall normalization. This also "equips" the differentiated leg in the triple-gluon vertex with a propagator $(\square_{IJ}^{-1})$ as shown below
\begin{equation}
    W_{\mathrm{SD}}[J]=S_{\mathrm{SD}}[A_c]+J_I A_c^{\bullet I}
    +i\,\Tr \ln \Big[\mathbb{1} + \square^{-1}_{IJ}\left(V_{-++}\right)_{JKL}A_{c}^{\bullet L}\Big] \,.
    \label{eq:W_functional_0}
\end{equation}
Note, in the expression above, $\mathbb{1}$ appears due to the factoring of the inverse propagator and is essentially a Kronecker delta $\delta_{IJ}$. The log can be expanded into a series using
\begin{equation}
    \ln \left(\mathbb{1} +x\right)= \sum_{k=1}^{\infty} \frac{(-1)^{k+1}}{k} x^k\,.
    \label{eq:LOG_exp}
\end{equation}
Thus we can write
\begin{equation}
    W_{\mathrm{SD}}[J]=S_{\mathrm{SD}}[A_c]+J_I A_c^{\bullet I}
    +i\,\Tr \sum_{k=1}^{\infty} \frac{(-1)^{k+1}}{k} 
    \left[\square^{-1}_{IJ}\left(V_{-++}\right)_{JKL}A_{c}^{\bullet L}\right]^k \,.
    \label{eq:W_functional}
\end{equation}
Finally, recall the fields ${\hat A}_{c}[J]=({\hat A}_{c}^{\bullet}[J],{\hat A}_{c}^{\star}[J])$ are functionals of the source. Performing a Legendre transform, we can define an object that depends exclusively on the "fields" and not on the sources. This is done as follows
\begin{equation}
    \Gamma_{\mathrm{SD}}[A_c]=W_{\mathrm{SD}}[J]-J_IA_c^{\bullet I} \,.
\end{equation}
As a result we get this new functional $\Gamma_{\mathrm{SD}}[A_c]$ which depends only on the fields
\begin{align}
    \Gamma_{\mathrm{SD}}[A_c] =& S_{\mathrm{SD}}[A_c]
    +i \, \Tr \sum_{k=1}^{\infty} \frac{(-1)^{k+1}}{k} 
    \left[\square^{-1}_{IJ} \left(V_{-++}\right)_{JKL} A_c^{\bullet L}\right]^k \nonumber\\
    =&- A_c^{\star I}\square_{IJ}A_c^{\bullet J} 
    - \frac{1}{2}\left(V_{-++}\right)_{IJK}A_c^{\star I}A_c^{\bullet J}A_c^{\bullet K} \nonumber\\
    &+i\, \square^{-1}_{IJ}\left(V_{-++}\right)_{JIK} A_c^{\bullet K}
    -i\, \frac{1}{2} \square^{-1}_{I_1J_1}\left(V_{-++}\right)_{J_1I_2K_1}
    \square^{-1}_{I_2J_2}\left(V_{-++}\right)_{J_2I_1K_2} 
    A_c^{\bullet K_1} A_c^{\bullet K_2} + \dots \,.
    \label{eq:OLEA_SD}
\end{align}
\begin{figure}
    \centering
    \includegraphics[width=12cm]{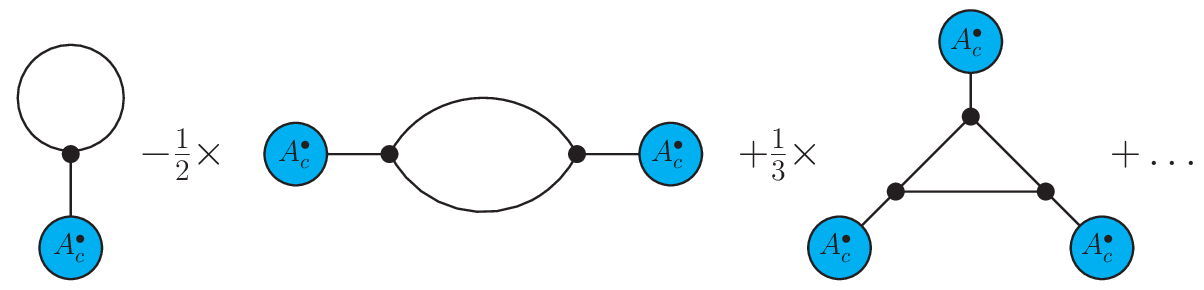}
    \caption{
    \small
   The one-loop corrections (up to 3-point) to the Self-Dual sector of the Yang-Mills theory originating from the log term in the one-loop effective Self-Dual action Eq.~\eqref{eq:OLEA_SD}. The first term is the $(+ + -)$ tadpole, the second is $(+ +)$ gluon self-energy and the third is the $\Delta^{+ + + }$ one-loop triangle contribution. This image was modified from our paper \cite{Kakkad_2022}.}
    \label{fig:SD_oneloopEffaction}
\end{figure}
The middle line above is just the Self-Dual action $S_{\mathrm{SD}}[A_c]$. The terms in the last line are obtained after taking the trace in the second term of the first line. Notice, the first term in the last line is a \textit{tadpole}. The second term is a \textit{bubble}. The dots represent the higher multiplicity one-loop terms like the \textit{triangle}, \textit{box}, and so on. We represent the first few of these in Figure~\ref{fig:SD_oneloopEffaction}. Thus we see that indeed the log term originating from the second-order expansion in Eq.~\eqref{eq:expansion_2nd} accounts for the one loop terms. The functional $\Gamma_{\mathrm{SD}}[A_c]$, therefore, consists of the classical action $S_{\mathrm{SD}}[A_c]$ and the one-loop corrections and is, thus, referred to as the one-loop effective action. Also, notice, all the one-loop corrections to SDYM shown in Figure \ref{fig:SD_oneloopEffaction} are all-plus $(+ + \dots +)$ one-loop terms. Finally, since one-loop are the only quantum correction to the Self-Dual sector (the higher loop corrections are zero), the action Eq.~\eqref{eq:OLEA_SD} should provide a full description of SDYM theory.

At this point, we would like to briefly highlight the motivation following which we decided to employ the one-loop effective action approach to develop quantum corrections to the action (MHV or Z-field actions) related to the Yang-Mills action via field transformation. Notice, the one-loop effective action separates the classical action from the loop contributions (the log term). This indicates that if one starts with the one-loop effective action and then applies the field transformations, their effects on the classical and quantum contributions can be considered separately. Indeed, as we see in the later sections that this segregation allows us to retain one-loop contributions. 

For now, as an immediate example, we will use the one-loop effective action for the SDYM theory to provide systematic proof for the technique employed in \cite{Brandhuber2007a} for computing the all-plus one-loop amplitudes. In this work, the authors performed a \textit{holomorphic} field redefinition that mapped the two fields $\left\{{\hat A}^{\bullet},  {\hat A}^{\star} \right\}$ in the Self-Dual sector of the Yang-Mills to a new pair $\left\{{\hat B}^{\bullet},  {\hat B}^{\star} \right\}$. The transformation reads
\begin{gather}
    {\hat A}^{\bullet} \longrightarrow {\hat B}^{\bullet}={\hat B}^{\bullet}[A^{\bullet}] \,, \nonumber  \\
    {\hat A}^{\star} \longrightarrow {\hat B}^{\star}={\hat A}^{\star} \,,
    \label{eq:BrandhubeTransf}
\end{gather}
where the mapping ${\hat A}^{\bullet} \longrightarrow {\hat B}^{\bullet}$ above is exactly the same as in Mansfield's transformation Eq.~\eqref{eq:AtoB_CT_def}. But, unlike in the case of Mansfield's transformation, they left the negative helicity field ${\hat A}^{\star}$ unaltered (hence the name \textit{holomorphic} field redefinition). The transformation maps the Self-Dual action to a free action
\begin{equation}
\mathcal{L}_{+-}\left[A^{\bullet},A^{\star}\right] +\mathcal{L}_{++-}\left[A^{\bullet},A^{\star}\right] \rightarrow \mathcal{L}_{+-}\left[B^{\bullet},B^{\star}\right] \,.
    \label{eq:Hol_Transf1}
\end{equation}
However, given that ${\hat B}^{\star}={\hat A}^{\star}$, these fields can be eliminated from the equation above to get
\begin{equation}
\Box {\hat{A}}^{\bullet} + 2ig{\partial}_{-} \left[ ({\partial}_{-}^{-1} {\partial}_{\bullet} {\hat{A}}^{\bullet}), {\hat{A}}^{\bullet} \right] = \Box\, \hat{B}^{\bullet} \, .
\label{eq:Hol_Transf2}
\end{equation}
Owing to the type of transformation, the Jacobian  is no longer field independent. It reads (in the collective index notation)
\begin{equation}
    \mathcal{J}  = \det \begin{vmatrix}
     \frac{\delta A^{\bullet I}}{\delta B^{\bullet J}} 
     &\mathbb{0} \\ 
\mathbb{0} 
     &\mathbb{1} 
\end{vmatrix} \,.
\label{eq:MT_jac_hol}
\end{equation}
Performing the above transformation to the generating functional Eq.\eqref{eq:ZG_SD}, we get
\begin{equation}
    Z_{\mathrm{SD}}[J] = \int[dB] \det\left(\frac{\delta A^{\bullet I}}{\delta B^{\bullet J}}\right) \exp\left\{ iS_{\mathrm{SD}}[B^{\bullet},B^{\star}] 
    + i\,A^{\bullet I}[B^{\bullet}]J_{\bullet I}
    + i\,B^{\star K}J_{\star K}\right\} \,,
    \label{eq:BrandhuberZ}
\end{equation}
where $S_{\mathrm{SD}}[B^{\bullet},B^{\star}]$ is just the free action and $A^{\bullet I}[B^{\bullet}]$ is exactly the same as in the case of Mansfield's transformation. Using Eq.~\eqref{eq:BrandhuberZ}, the authors in \cite{Brandhuber2007a} showed that the all-plus one-loop amplitude originates solely from the Jacobian. 

The result Eq.~\eqref{eq:BrandhuberZ}, can be proved using the one-loop effective action approach. To see this, let us differentiate the Self-Dual equation of motion Eq.~\eqref{eq:SD_EOM0}
\begin{equation}
    \frac{\delta^2 S[A_c]}{\delta {\hat A}^{\bullet}(y)\delta {\hat A}^{\star}(x)}=\frac{\delta {\hat J}^{\bullet}[A_c^{\bullet}](x)}{\delta {\hat A}^{\bullet}(y)} \,.
\end{equation}
Substituting this to Eq.~\eqref{eq:Partition1} we get
\begin{multline}
    Z_{\mathrm{SD}}[J]\approx \left[\det\left( \frac{\delta {\hat J}^{\bullet}[A_c^{\bullet}](x)}{\delta {\hat A}^{\bullet}(y)} \right)\right]^{-1} 
     \exp\left\{ iS_{\mathrm{SD}}[A_c] 
    + i\int\!d^4x\, \Tr \hat{J}_i(x) \hat{A}_c^i(x) \right\}  \\
    = \det\left( \frac{\delta {\hat A}_c^{\bullet}[J](y)}{\delta {\hat B}^{\bullet}(x)} \right)
     \exp\left\{ iS_{\mathrm{SD}}[A_c] 
    + i\int\!d^4x\, \Tr \hat{J}_i(x) \hat{A}_c^i(x) \right\} \,.
    \label{eq:SD_Brand}
\end{multline}
Above, in going from the first expression to the second we used $B^{\bullet}_a(x) \equiv \square^{-1}  {J}^{\bullet}(x)$. Then factored out the inverse propagator as $\det(\square)$ (and eventually discarded it). With this, we see that indeed the one-loop amplitudes obtained from the Eq.~\eqref{eq:SD_Brand} would match with those obtained from Eq.~\eqref{eq:BrandhuberZ}.

\section{One-loop Effective action for Yang-Mills theory}
\label{sec:Yang-Mills_OLEA}

In this section, we derive the one-loop effective action for the Yang-Mills action on the light cone. This, in turn, will lay the foundation for developing quantum corrections to the MHV action.

The starting point is again the generating functional for the full Green's function 
\begin{equation}
    Z[J]_{\mathrm{YM}}=\int[dA]\, e^{i\left(S_{\mathrm{YM}}[A] + \int\!d^4x\, \Tr \hat{J}_j(x) \hat{A}^j(x)\right) } \,,
    \label{eq:gen_YM0}
\end{equation}
where, the subscript $\mathrm{YM}$ stands for Yang-Mills. Above, $S_{\mathrm{YM}}[A]$ is the light-cone Yang-Mills action Eq.~\eqref{eq:YM_LC_action}
\begin{equation}
S_{\mathrm{YM}}\left[A^{\bullet},A^{\star}\right]=\int dx^{+}\left(\mathcal{L}_{+-}+\mathcal{L}_{++-}+\mathcal{L}_{+--}+\mathcal{L}_{++--}\right)\,,\label{eq:actionLC_YM1}
\end{equation}
and $\hat{J}_j$ represents the auxiliary source coupled to the action and the index $j=\bullet,\star$. 

As before, we expand the terms in the exponent around the classical solutions defined using the classical EOM
\begin{equation}
   \left. \frac{\delta S_{\mathrm{YM}}[A]}{\delta {\hat A}^j(x)}\right|_{{\hat A}={\hat A}_{c}}+{\hat J}_j(x)=0 \,,
    \label{eq:YMEOM0}
\end{equation}
where ${\hat A}_{c} = \left\{{\hat A}_{c}^{\bullet}(x), {\hat A}_{c}^{\star}(x) \right\}$.

Up to the second order in fields, the expansion reads
\begin{multline}
    S_{\mathrm{YM}}[A] + \int\!d^4x\, \Tr \hat{J}_i(x) \hat{A}^i(x)  
    = S_{\mathrm{YM}}[A_c] + \int\!d^4x\, \Tr \hat{J}_i(x)\hat{A}_c^i(x) \\ + \int\!d^4x\,\Tr\left(\hat{A}^i(x)-\hat{A}_c^i(x)\right)
    \left(\frac{\delta S_{\mathrm{YM}}[A_c]}{\delta \hat{A}^i(x)}+{\hat J}_i(x)\right) \\
    +\frac{1}{2}\int\!d^4xd^4y\,\Tr\left(\hat{A}^i(x)-\hat{A}_c^i(x)\right)\left(\frac{\delta^2 S_{\mathrm{YM}}[A_c]}{\delta\hat{A}^i(x)\delta\hat{A}^j(y)}\right)\left(\hat{A}^j(y)-\hat{A}_c^j(y)\right) + \dots
\end{multline}
The first two terms factor out of the path integral. The linear term vanishes owing to the classical EOM Eq.~\eqref{eq:YMEOM0}. This leaves us with the following integral
\begin{equation}
    Z[J]_{\mathrm{YM}} \approx  e^{i\left(S_{\mathrm{YM}}[A_c] + \int\!d^4x\, \Tr \hat{J}_i(x)\hat{A}_c^i(x)\right)} \int[dA]\, e^{\frac{i}{2}\int\!d^4xd^4y\,\Tr\left(\hat{A}^i(x)-\hat{A}_c^i(x)\right)\left(\frac{\delta^2 S_{\mathrm{YM}}[A_c]}{\delta\hat{A}^i(x)\delta\hat{A}^j(y)}\right)\left(\hat{A}^j(y)-\hat{A}_c^j(y)\right) } \,,
    \label{eq:gen_YMZJ}
\end{equation}
where the matrix in the exponent reads
\begin{equation}
   \mathrm{M}^{\mathrm{YM}}_{IJ}= \left(\begin{matrix}
     \frac{\delta^2 S_{\mathrm{YM}}[A_c]}
    {\delta A^{\bullet I}\delta A^{\star J}} 
     & \frac{\delta^2 S_{\mathrm{YM}}[A_c]}
    {\delta A^{\bullet I}\delta A^{\bullet J}} \\ \\
\frac{\delta^2 S_{\mathrm{YM}}[A_c]}
    {\delta A^{\star I}\delta A^{\star J}} & \frac{\delta^2 S_{\mathrm{YM}}[A_c]}
    {\delta A^{\star I}\delta A^{\bullet J}}    
\end{matrix}\right) = \left(\begin{matrix}
      S''_{\bullet\star}[A_c]
     & S''_{\bullet\bullet}[A_c] \\ \\
S''_{\star\star}[A_c] & S''_{\star\bullet}[A_c]    
\end{matrix}\right)\,.
\label{eq:MAT_YM1}
\end{equation}
We denote the matrix as $\mathrm{M}^{\mathrm{YM}}_{IJ}$. $I,J,\dots$ are the collective indices introduces earlier. On the R.H.S. we introduced a compact representation for the block matrices in $\mathrm{M}^{\mathrm{YM}}_{IJ}$. In order to perform the Gaussian integral Eq.~\eqref{eq:gen_YMZJ}, we must diagonalize the matrix $\mathrm{M}^{\mathrm{YM}}_{IJ}$. This can be done as shown below
\begin{multline}
    \left(\begin{matrix}
      S''_{\bullet\star}[A_c]
     & S''_{\bullet\bullet}[A_c] \\ \\
S''_{\star\star}[A_c] & S''_{\star\bullet}[A_c]    
\end{matrix}\right)= \left(\begin{matrix}
      \mathbb{1} 
     & \mathbb{0} \\ \\
S''_{\star\star}[A_c]\left(S''_{\bullet\star}[A_c]\right)^{-1} & \mathbb{1}     
\end{matrix}\right)\\
\times \left(\begin{matrix}
      S''_{\bullet\star}[A_c]
     & \mathbb{0} \\ \\
\mathbb{0} & S''_{\star\bullet}[A_c] - S''_{\star\star}[A_c]\left(S''_{\bullet\star}[A_c]\right)^{-1} S''_{\bullet\bullet}[A_c]
\end{matrix}\right) \\
\times \left(\begin{matrix}
      \mathbb{1}
     & \left(S''_{\bullet\star}[A_c]\right)^{-1}S''_{\bullet\bullet}[A_c] \\ \\
\mathbb{0} & \mathbb{1}    
\end{matrix}\right)\,.
\label{eq:MATdia}
\end{multline}
Notice, the determinant of the first and the third matrices above is one. Thus using it we can define a new set of auxiliary fields $\hat{C}^i(x)$ where $i = \{\bullet, \star \}$.
The Jacobian for the transformation $\hat{A}^i(x) \longrightarrow \hat{C}^i(x)$ is one. And, the Gaussian integral in Eq.~\eqref{eq:gen_YMZJ} reduces to
\begin{equation}
    \int[dC]\,\exp \left[ \frac{i}{2}\int\!d^4xd^4y\,\Tr\hat{C}^i(x)\left(\frac{\delta^2 S^{\mathrm{Diag}}_{\mathrm{YM}}[A_c]}{\delta\hat{A}^i(x)\delta\hat{A}^j(y)}\right)\hat{C}^j(y)\right]  \,,
\end{equation}
where the matrix in the exponent is the diagonalized matrix on the second line in Eq.~\eqref{eq:MATdia}. In collective index notation, it reads
\begin{equation}
    \mathrm{M}^{\mathrm{YM-Diag}}_{IJ} = \left(\begin{matrix}
      \frac{\delta^2 S_{\mathrm{YM}}[A_c]}
    {\delta A^{\bullet I}\delta A^{\star J}}
     & \mathbb{0} \\ \\
\mathbb{0} & \frac{\delta^2 S_{\mathrm{YM}}[A_c]}
    {\delta A^{\star I}\delta A^{\bullet J}} - \frac{\delta^2 S_{\mathrm{YM}}[A_c]}
    {\delta A^{\star I}\delta A^{\star K}}\left(\frac{\delta^2 S_{\mathrm{YM}}[A_c]}
    {\delta A^{\bullet K}\delta A^{\star L}}\right)^{-1} \frac{\delta^2 S_{\mathrm{YM}}[A_c]}
    {\delta A^{\bullet L}\delta A^{\bullet J}}
\end{matrix}\right)\,,
\end{equation}
where "Diag" stands for diagonalized. Thus, upon integration we get
\begin{equation}
    \left[ \det \left( \frac{\delta^2 S_{\mathrm{YM}}[A_c]}
    {\delta A^{\bullet I}\delta A^{\star J}}\right)\, \det \left(\frac{\delta^2 S_{\mathrm{YM}}[A_c]}
    {\delta A^{\star I}\delta A^{\bullet J}} - \frac{\delta^2 S_{\mathrm{YM}}[A_c]}
    {\delta A^{\star I}\delta A^{\star K}}\left(\frac{\delta^2 S_{\mathrm{YM}}[A_c]}
    {\delta A^{\bullet K}\delta A^{\star L}}\right)^{-1} \frac{\delta^2 S_{\mathrm{YM}}[A_c]}
    {\delta A^{\bullet L}\delta A^{\bullet J}}\right)\right]^{-\frac{1}{2}} \,.
    \label{eq:det_YM_full}
\end{equation}
Already at this level, one can cross-check the result for the Self-Dual Yang-Mills case. In that case, the above expression reduces to (given that $S''_{\star\star}[A_c] = 0$)
\begin{equation}
    \left[ \det \left( \frac{\delta^2 S_{\mathrm{SD}}[A_c]}
    {\delta A^{\bullet I}\delta A^{\star J}}\right)\, \det \left(\frac{\delta^2 S_{\mathrm{SD}}[A_c]}
    {\delta A^{\star I}\delta A^{\bullet J}} \right)\right]^{-\frac{1}{2}} \equiv \det
    \left( 
    \frac{\delta^2 S_{\mathrm{SD}}[A_c]}
    {\delta A^{\star I}\delta A^{\bullet J}}
    \right)
     \,.
\end{equation}
This is exactly what we had in Eq.~\eqref{eq:Partition1}.

\begin{figure}
    \centering
    \includegraphics[width=14.5cm]{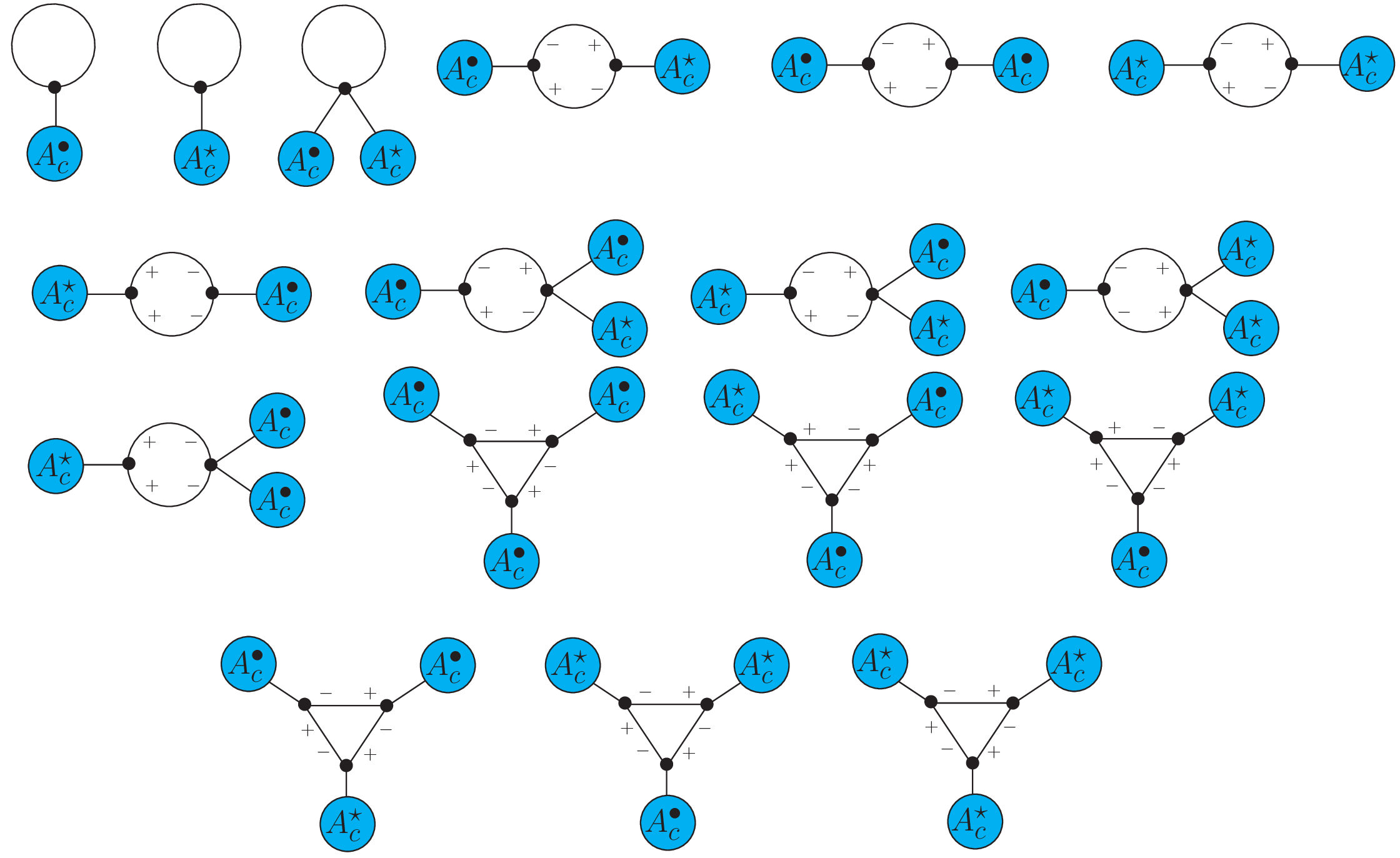}
    \caption{
    \small
    One-loop contributions with different helicity configurations up to 3-point originating from the log term in the Yang-Mills one-loop partition function Eq.~\eqref{eq:Partition_YM}. These consist of tadpoles: first three terms on the first line, self-energy corrections: last three terms on the first line and first term on the second line, swordfish: last three on the second line and first term on the third line, and triangles: the last three terms on the third line and the whole of the fourth line. This image was modified from our paper \cite{Kakkad_2022}.
    }
    \label{fig:YM_oneloopEffaction}
\end{figure}

Using Eq.~\eqref{eq:det_YM_full} for the generating fucntional Eq.~\eqref{eq:gen_YMZJ}, we get
\begin{multline}
   Z[J]_{\mathrm{YM}} \approx  \left[  \det \left\{ \frac{\delta^2 S_{\mathrm{YM}}[A_c]}
    {\delta A^{\star I}\delta A^{\bullet K}} \, \frac{\delta^2 S_{\mathrm{YM}}[A_c]}
    {\delta A^{\star K}\delta A^{\bullet J}} \right.\right.\\
   \left.\left. - \frac{\delta^2 S_{\mathrm{YM}}[A_c]}
    {\delta A^{\star I}\delta A^{\bullet K}} \, \frac{\delta^2 S_{\mathrm{YM}}[A_c]}
    {\delta A^{\star K}\delta A^{\star L}} \left( \frac{\delta^2 S_{\mathrm{YM}}[A_c]}
    {\delta A^{\bullet L}\delta A^{\star M}} \right)^{-1} \frac{\delta^2 S_{\mathrm{YM}}[A_c]}
    {\delta A^{\bullet M}\delta A^{\bullet J}}\right\}\right]^{-\frac{1}{2}} \\
   \exp\left\{i\left(S_{\mathrm{YM}}[A_c] 
    + \int\!d^4x\, \Tr \hat{J}_i(x) \hat{A}_c^i(x) \right) \right\} \,.
    \label{eq:det_ZYM}
\end{multline}
Rewriting the determinant as exponential of a trace of logarithm we get
\begin{multline}
    Z_{\mathrm{YM}}[J]\approx 
    \exp\left\{ iS_{\mathrm{YM}}[A_c] 
    + i\int\!d^4x\, \Tr \hat{J}_i(x) \hat{A}_c^i(x) \right.\\
   \left. - \frac{1}{2} \Tr\ln \left[ \frac{\delta^2 S_{\mathrm{YM}}[A_c]}
    {\delta A^{\star I}\delta A^{\bullet K}} \, \frac{\delta^2 S_{\mathrm{YM}}[A_c]}
    {\delta A^{\star K}\delta A^{\bullet J}} - \frac{\delta^2 S_{\mathrm{YM}}[A_c]}
    {\delta A^{\star I}\delta A^{\bullet K}} \, \frac{\delta^2 S_{\mathrm{YM}}[A_c]}
    {\delta A^{\star K}\delta A^{\star L}} \left( \frac{\delta^2 S_{\mathrm{YM}}[A_c]}
    {\delta A^{\bullet L}\delta A^{\star M}} \right)^{-1} \frac{\delta^2 S_{\mathrm{YM}}[A_c]}
    {\delta A^{\bullet M}\delta A^{\bullet J}}\right]
    \right\}
    \label{eq:Partition_YM}
    \,,
\end{multline}
where the repeated indices are summed over. Notice, further, that only the opposite helicity fields are contracted. The block matrices in the log term read
\begin{equation}
   \frac{\delta^2 S_{\mathrm{YM}}[A_c]}
    {\delta A^{\star I}\delta A^{\bullet J}} 
    = -\square_{IJ}-\left(V_{-++}\right)_{IJK}A_{c}^{\bullet K} -\left(V_{--+}\right)_{KIJ}A_{c}^{\star K} -\left(V_{--++}\right)_{LIJK}A_{c}^{\star L} A_{c}^{\bullet K}
    \label{eq:S+-}
    \,,
\end{equation}
\begin{equation}
   \frac{\delta^2 S_{\mathrm{YM}}[A_c]}
    {\delta A^{\bullet I}\delta A^{\bullet J}} 
    = -\left(V_{-++}\right)_{KIJ}A_{c}^{\star K} -\left(V_{--++}\right)_{KLIJ}A_{c}^{\star K} A_{c}^{\star L}
    \label{eq:S++}
    \,,
\end{equation}
\begin{equation}
   \frac{\delta^2 S_{\mathrm{YM}}[A_c]}
    {\delta A^{\star I}\delta A^{\star J}} 
    = -\left(V_{--+}\right)_{IJK}A_{c}^{\bullet K} -\left(V_{--++}\right)_{IJKL}A_{c}^{\bullet K} A_{c}^{\bullet L}
    \label{eq:S--}
    \,,
\end{equation}
The one-loop term in the Yang-Mills case Eq.~\eqref{eq:Partition_YM} is much more involved than that in the Self-Dual case. The treatment, however, proceeds in exactly the same fashion. We begin with factoring out the inverse propagator $(\square_{IJ})$ from the log term. At first sight, it may appear that one should get $(\square^2)$ as evident from the first term. This is indeed the case but the factor of $1/2$ outside the log compensates it. A similar factor of $(\square^2)$ originates also from the second term in the log. As before, we will discard this as $\det(\square)$ from the partition function. With this, the log term reduces to
\begin{multline}
     - \frac{1}{2} \Tr\ln \Bigg[  \Bigg\{ \Big( \mathbb{1} + \left(\square^{-1}V_{-++}\right)_{IKP}A_{c}^{\bullet P} +\left(\square^{-1}V_{--+}\right)_{PIK}A_{c}^{\star P} +\left(\square^{-1}V_{--++}\right)_{PIKQ}A_{c}^{\star P} A_{c}^{\bullet Q}\Big) \\
    \times \Big( \mathbb{1} + \left(\square^{-1}V_{-++}\right)_{KJR}A_{c}^{\bullet R} +\left(\square^{-1}V_{--+}\right)_{RKJ}A_{c}^{\star R} +\left(\square^{-1}V_{--++}\right)_{RKJS}A_{c}^{\star R} A_{c}^{\bullet S}\Big) \Bigg\}\\
     - \Bigg\{ \Big(\mathbb{1} + \left(\square^{-1}V_{-++}\right)_{IKP}A_{c}^{\bullet P} +\left(\square^{-1}V_{--+}\right)_{PIK}A_{c}^{\star P} +\left(\square^{-1}V_{--++}\right)_{PIKQ}A_{c}^{\star P} A_{c}^{\bullet Q}\Big) \\
     \times \Big(\left(\square^{-1}V_{--+}\right)_{KLR}A_{c}^{\bullet R} +\left(\square^{-1}V_{--++}\right)_{KLRS}A_{c}^{\bullet R} A_{c}^{\bullet S}\Big)\\
     \times \Big( \mathbb{1} + \left(\square^{-1}V_{-++}\right)_{MLT}A_{c}^{\bullet T} +\left(\square^{-1}V_{--+}\right)_{TML}A_{c}^{\star T} +\left(\square^{-1}V_{--++}\right)_{TMLU}A_{c}^{\star T} A_{c}^{\bullet U}\Big)^{-1}\\ 
     \times \Big(\left(\square^{-1}V_{-++}\right)_{VMJ}A_{c}^{\star V} +\left(\square^{-1}V_{--++}\right)_{WVMJ}A_{c}^{\star W} A_{c}^{\star V}\Big)\Bigg\}\Bigg]
    \label{eq:Partition_log}
    \,.
\end{multline}
Using the series expansion for the log term Eq.~\eqref{eq:LOG_exp}, 
up to the second order in fields, the expansion reads
\begin{align}
   -\frac{1}{2}\, \Bigg[2\, \left(\square^{-1}V_{-++}\right)_{IIP} A_c^{\bullet P} &+2\, \left(\square^{-1}V_{--+}\right)_{PII} A_c^{\star P} +2\, \left(\square^{-1}V_{--++}\right)_{PIIQ} A_c^{\star P}A_c^{\bullet Q} \nonumber\\
    &+\left(\square^{-1}V_{-++}\right)_{IKP}
    \left(\square^{-1}V_{-++}\right)_{KIR} 
    A_c^{\bullet P} A_c^{\bullet R}\nonumber\\
    &+\left(\square^{-1}V_{--+}\right)_{PIK}
    \left(\square^{-1}V_{--+}\right)_{RKI} 
    A_c^{\star P} A_c^{\star R}\nonumber\\
    &+\,\left(\square^{-1}V_{-++}\right)_{IKP}
    \left(\square^{-1}V_{--+}\right)_{RKI} 
    A_c^{\bullet P} A_c^{\star R}\nonumber\\
    &+\,\left(\square^{-1}V_{--+}\right)_{PIK}
    \left(\square^{-1}V_{-++}\right)_{KIR} 
    A_c^{\star P} A_c^{\bullet R}\nonumber\\
    &-\,\left(\square^{-1}V_{--+}\right)_{IMR}
    \left(\square^{-1}V_{-++}\right)_{VMI} 
    A_c^{\bullet R} A_c^{\star V}+ \dots \,\nonumber\\
    &-\,\frac{1}{2} \Big\{ 4\,\left(\square^{-1}V_{-++}\right)_{IKP}
    \left(\square^{-1}V_{-++}\right)_{KIR} 
    A_c^{\bullet P} A_c^{\bullet R}\nonumber\\
    &+ 4\,\left(\square^{-1}V_{--+}\right)_{PIK}
    \left(\square^{-1}V_{--+}\right)_{RKI} 
    A_c^{\star P} A_c^{\star R}\nonumber\\
    &+4\,\left(\square^{-1}V_{-++}\right)_{IKP}
    \left(\square^{-1}V_{--+}\right)_{RKI} 
    A_c^{\bullet P} A_c^{\star R}\nonumber\\
    &+4\,\left(\square^{-1}V_{--+}\right)_{PIK}
    \left(\square^{-1}V_{-++}\right)_{KIR} 
    A_c^{\star P} A_c^{\bullet R} + \dots \, \Big\}
    \Bigg]\,.
    \label{eq:2p_1loopexp}
\end{align}
Notice, the trace has already been taken in the terms above. The first line represents the different possible tadpoles using the vertices in the Yang-Mills action. All the remaining terms are bubbles with different helicity configurations. At this point, one may wonder if so many bubble configurations are even possible. As visible, the last four terms have exactly the same configuration as similar terms above those. These just have different overall numeric factors. The aim here was to keep the logarithmic expansion explicit due to which we did not combine them. The dots represent the higher point one-loop terms like triangles, boxes, etc. The one-loop contributions up to 3-point are shown in Figure \ref{fig:YM_oneloopEffaction}. 

Finally, using Eq.~\eqref{eq:Partition_YM}, one can obtain the generating functional for the connected Green's function via $\quad W_{\mathrm{YM}}[J] = -i \ln \left[ Z_{\mathrm{YM}}[J]\right]$. The one-loop effective action can then be derived as the Legendre transform as
\begin{equation}
   \Gamma_{\mathrm{YM}}[A_c] = W_{\mathrm{YM}}[J] - \int\!d^4x\, \Tr \hat{J}_i(x) \hat{A}_c^i(x) \,.
\end{equation}
By doing this, we get
\begin{multline}
   \Gamma_{\mathrm{YM}}[A_c] = S_{\mathrm{YM}}[A_c] \\
    + \frac{i}{2} \Tr\ln \left[ \frac{\delta^2 S_{\mathrm{YM}}[A_c]}
    {\delta A^{\star I}\delta A^{\bullet K}} \, \frac{\delta^2 S_{\mathrm{YM}}[A_c]}
    {\delta A^{\star K}\delta A^{\bullet J}} - \frac{\delta^2 S_{\mathrm{YM}}[A_c]}
    {\delta A^{\star I}\delta A^{\bullet K}} \, \frac{\delta^2 S_{\mathrm{YM}}[A_c]}
    {\delta A^{\star K}\delta A^{\star L}} \left( \frac{\delta^2 S_{\mathrm{YM}}[A_c]}
    {\delta A^{\bullet L}\delta A^{\star M}} \right)^{-1} \frac{\delta^2 S_{\mathrm{YM}}[A_c]}
    {\delta A^{\bullet M}\delta A^{\bullet J}}\right]\,.
    \label{eq:OLEA_YM}
\end{multline}
Above, it is understood that the inverse propagator has been factored out.
\section{One-loop Effective MHV action}
\label{sec:MHV_OLEA}

In this section, we summarize the main results of our paper \cite{Kakkad_2022}, where we used the one-loop effective action approach to systematically develop quantum corrections to the MHV action. The strategy we used was to start with the Yang-Mills one-loop partition function Eq.~\eqref{eq:Partition_YM}
\begin{equation}
    Z_{\mathrm{YM}}[J]\approx 
    \exp\left\{ iS_{\mathrm{YM}}[A_c] 
    + i\int\!d^4x\, \Tr \hat{J}_i(x) \hat{A}_c^i(x) - \frac{1}{2} \Tr\ln \left( 
    \frac{\delta^2 S_{\mathrm{YM}}[A_c]}
    {\delta \hat{A}^i(x)\delta \hat{A}^j(y)}
    \right)
    \right\}
    \label{eq:Partition_YMre}
    \,,
\end{equation}
then transform the classical fields $\left\{{\hat A}_{c}^{\bullet}(x), {\hat A}_{c}^{\star}(x) \right\} \longrightarrow \left\{{\hat B}_{c}^{\bullet}(x), {\hat B}_{c}^{\star}(x) \right\}$ via Mansfield's transformation
\begin{equation}
\mathrm{Tr}\,\hat{A}^{\bullet}_c\square\hat{A}^{\star}_c
+2ig\,\mathrm{Tr}\,\partial_{-}^{-1}\partial_{\bullet} \hat{A}_c^{\bullet}\left[\partial_{-}\hat{A}_c^{\star},\hat{A}_c^{\bullet}\right]
\,\, \longrightarrow \,\,
\mathrm{Tr}\,\hat{B}_c^{\bullet}\square\hat{B}_c^{\star}
\,.\label{eq:MansfieldTransf2}
\end{equation}
The log term in Eq.~\eqref{eq:Partition_YMre} is exactly the same as in Eq.~\eqref{eq:Partition_YM}. In the former, we represent it in a compact fashion for the sake of simplicity. Notice, we perform the transformation after the integration with respect to the fields in the partition function has been carried out.  

\begin{figure}
    \centering
    \includegraphics[width=8cm]{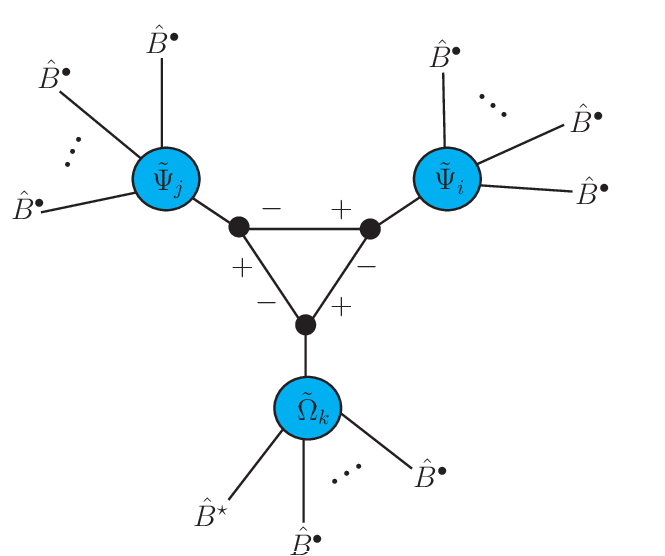}
    \caption{
    \small
    The $\hat{A}_c^i$ fields in the $\Delta^{+ + -}$ one-loop triangle contribution in Yang-Mills one-loop partition function Eq.~\eqref{eq:Partition_YM} transforms to $\hat{A}_c^i[B_c]$ following Mansfield's transformation. This substitution accounts for all the tree level connections using the $(+ + -)$ triple gluon vertex. This image was modified from our paper \cite{Kakkad_2022}.}
    \label{fig:loop_3V}
\end{figure}

Substituting the solutions $\hat{A}_c^i=\hat{A}_c^i[B_c]$ using Eq.~\eqref{eq:A_bull_solu}-\eqref{eq:A_star_solu} to Eq.~\eqref{eq:Partition_YMre} we see that the Yang-Mills action should transform to the MHV action
\begin{equation}
    S_{\mathrm{YM}}[A_c] \longrightarrow S_{\mathrm{MHV}}[B_c]\,,
\end{equation}
whereas for the remaining two terms the classical fields will simply get replaced by the functionals $\hat{A}_c^i=\hat{A}_c^i[B_c]$. As a result, the Yang-Mills partition function transforms to
\begin{equation}
    Z_{\mathrm{MHV}}[J]\approx 
    \exp\left\{ iS_{\mathrm{MHV}}[B_c] 
    + i\int\!d^4x\, \Tr \hat{J}_i(x) \hat{A}_c^i[B_c](x) - \frac{1}{2} \Tr\ln \left( 
    \frac{\delta^2 S_{\mathrm{YM}}[A_c [B_c]]}
    {\delta \hat{A}^i(x)\delta \hat{A}^j(y)}
    \right)
    \right\}
    \label{eq:PartitionMHV}
    \,.
\end{equation}
Let us elaborate on the log term in the above expression. Firstly, notice, it is the same as the log term in the Yang-Mills one-loop partition function Eq.~\eqref{eq:Partition_YM} but with the classical $\hat{A}_c^i$ fields replaced by the functionals $ \hat{A}_c^i[B_c]$. This has interesting consequences in the context of computing amplitudes. Consider, for example, the one-loop $(+ + -)$ triangle graph in the Yang-Mills one-loop partition function Eq.~\eqref{eq:Partition_YM}. There are two terms of this type with different routing of the helicities inside the loop. Therefore, consider specifically the first term in the final line of Figure \ref{fig:YM_oneloopEffaction}. Following Mansfield's transformation the $\hat{A}_c^i$ fields in this term  gets replaced by the functionals $\hat{A}_c^i[B_c]$ (see Figure \ref{fig:loop_3V}) which could, in turn, be expanded to any order. Recall, the latter accounts for all the tree level contributions originating from the $(+ + -)$ triple gluon vertex. As a result a term like this in the one-loop MHV partition function Eq.~\eqref{eq:PartitionMHV} accounts for several diagrams one would require to develop via tree level connection between the $(+ + -)$ triple gluon vertex in the classical Yang-Mills action $S_{\mathrm{YM}}[A_c]$ and the one-loop $(+ + -)$ triangle graph originating from the log term in the Yang-Mills one-loop partition function Eq.~\eqref{eq:Partition_YM} when computing amplitudes (following the rules summarized in Appendix \ref{sec:app_A6}) using Eq.~\eqref{eq:Partition_YM}. This implies, it is natural to expect that the number of contributions required to compute one-loop amplitudes using one-loop MHV partition function Eq.~\eqref{eq:PartitionMHV} is certainly lower as compared to those in the Yang-Mills one-loop partition function Eq.~\eqref{eq:Partition_YM}.

Another crucial aspect of the substitution $\hat{A}_c^i = \hat{A}_c^i[B_c]$  to the log term is that we get both the MHV as well as the non-MHV one-loop contributions. Let us try to first understand this via the nature of Mansfield's transformation itself. After that, we shall demonstrate this explicitly via an example. The way Mansfield's transformation Eq.~\eqref{eq:MansfieldTransf2} work upon substitution of $\hat{A}_c^i = \hat{A}_c^i[B_c]$ is as follows. The first order expansion of $\hat{A}_c^i[B_c]$ fields substituted to the kinetic terms on the L.H.S. of Eq.~\eqref{eq:MansfieldTransf2} gives the kinetic term on the R.H.S. Substitution of the higher order terms to the kinetic terms on the L.H.S. of Eq.~\eqref{eq:MansfieldTransf2} give rise to contributions which are explicitly canceled, term by term, against similar contributions originating from the substitution of $\hat{A}_c^i[B_c]$ to the $(+ + -)$ triple gluon vertex on the L.H.S. of Eq.~\eqref{eq:MansfieldTransf2}. As a result we are left with just the kinetic term on the R.H.S. Such cancellations will, however, not take place when $\hat{A}_c^i = \hat{A}_c^i[B_c]$ is substituted to the log term. This is simply because the block matrices in the log term read (in the collective index notation)
\begin{multline}
   \frac{\delta^2 S_{\mathrm{YM}}[A_c[B_c]]}
    {\delta A^{\star I}\delta A^{\bullet J}} 
    = -\square_{IJ}-\left(V_{-++}\right)_{IJK}A_{c}^{\bullet K}[B_c] -\left(V_{--+}\right)_{KIJ}A_{c}^{\star K}[B_c] \\
    -\left(V_{--++}\right)_{LIJK}A_{c}^{\star L}[B_c] A_{c}^{\bullet K}[B_c]
    \label{eq:S+-B}
    \,,
\end{multline}
\begin{equation}
   \frac{\delta^2 S_{\mathrm{YM}}[A_c[B_c]]}
    {\delta A^{\bullet I}\delta A^{\bullet J}} 
    = -\left(V_{-++}\right)_{KIJ}A_{c}^{\star K}[B_c] -\left(V_{--++}\right)_{KLIJ}A_{c}^{\star K}[B_c] A_{c}^{\star L}[B_c]
    \label{eq:S++B}
    \,,
\end{equation}
\begin{equation}
   \frac{\delta^2 S_{\mathrm{YM}}[A_c[B_c]]}
    {\delta A^{\star I}\delta A^{\star J}} 
    = -\left(V_{--+}\right)_{IJK}A_{c}^{\bullet K}[B_c] -\left(V_{--++}\right)_{IJKL}A_{c}^{\bullet K}[B_c] A_{c}^{\bullet L}[B_c]
    \label{eq:S--B}
    \,.
\end{equation}
Notice, only the first expression, above, has the (inverse) propagator term. However, since both the fields in the propagator term have been differentiated, upon substitution $\hat{A}_c^i = \hat{A}_c^i[B_c]$, it does not generate contributions that can cancel out similar contributions from the $(+ + -)$ triple gluon vertex. Due to this, the $(+ + -)$ Self-Dual vertex and, therefore, the one-loop contributions originating from the $(+ + -)$ Self-Dual vertex survives. Hence, the log term will generate both the MHV as well as the non-MHV one-loop contributions which would be missing if one simply used the MHV action instead to compute one-loop amplitudes.
It is also worth mentioning that these non-MHV contributions would still be missing if one derived the one-loop effective MHV action starting with the MHV action instead of starting with the Yang-Mills one-loop partition function followed by the Mansfield's transformation.

In order to demonstrate the above claim, concerning the non-MHV one-loop contributions not missing, let us compute the two point $(+ -)$ amputated connected Green's function using MHV one-loop effective action. This can be obtained from the one-loop MHV partition function Eq.~\eqref{eq:PartitionMHV} in two steps. First, we need the generating functional for the connected Green's function which reads
\begin{equation}
    W_{\mathrm{MHV}}[J]= S_{\mathrm{MHV}}[B_c[J]] 
    + \int\!d^4x\, \Tr \hat{J}_i(x) \hat{A}_c^i[B_c[J]](x) +i\, \frac{1}{2} \Tr\ln \left( 
    \frac{\delta^2 S_{\mathrm{YM}}[A_c [B_c[J]]]}
    {\delta \hat{A}^i(x)\delta \hat{A}^j(y)}
    \right)\,.
    \label{eq:WJ_MHV}
\end{equation}
Performing the Legendre transform we get
\begin{equation}
   \Gamma_{\mathrm{MHV}}[B_c]= S_{\mathrm{MHV}}[B_c]  +i\, \frac{1}{2} \Tr\ln \left( 
    \frac{\delta^2 S_{\mathrm{YM}}[A_c [B_c]]}
    {\delta \hat{A}^i(x)\delta \hat{A}^j(y)}
    \right)\,.
    \label{eq:G_MHV}
\end{equation}
For the example at hand, one can use either Eq.~\eqref{eq:WJ_MHV} or Eq.~\eqref{eq:G_MHV}. For the sake of completeness, we shall consider both. First, let us focus on the contributions using $\Gamma_{\mathrm{MHV}}[B_c]$. Essentially, there are two sources of contributions (below we shall also consider tadpole contributions but when computing amplitudes, later in Section \ref{sec:One-loop-OLEAMHV}, we will discard the tadpoles)
\begin{itemize}
    \item Following the rules for computing the amputated connected Green's function using the one-loop effective action discussed in Appendix \ref{sec:app_A6}, the first type of contribution originates from the functional derivative of $\Gamma_{\mathrm{MHV}}[B_c]$. For the two-point $(+ -)$, it reads
    \begin{equation}
     \frac{i}{2}\Bigg[\frac{\delta^2}
    {\delta B_c^{\bullet K_1} \delta B_c^{\star K_2}} \left\{ \Tr\ln \left( 
    \frac{\delta^2 S_{\mathrm{YM}}[A_c [B_c]]}
    {\delta \hat{A}^i(x)\delta \hat{A}^j(y)}
    \right)
    \right\}\Bigg]_{B_c= 0}\,.
   \end{equation}
   Notice, we did not include the classical action $S_{\mathrm{MHV}}[B_c]$ in the expression above because that would give  tree level contributions and not one-loop. The explicit one-loop contributions read
   \begin{multline}
   \frac{i}{2}\, \Bigg[  2\, \square^{-1}_{IJ}\left(V_{-+-}\right)_{JIK} \Omega_2^{KK_2K_1}  +2\, \square^{-1}_{IJ}\left(V_{-++-}\right)_{JIK_1K_2} \\
   +\,\square^{-1}_{I_1J_1}\left(V_{-++}\right)_{J_1I_2K_1}
    \square^{-1}_{I_2J_2}\left(V_{-+-}\right)_{J_2I_1K_2} \\
    +\,\square^{-1}_{I_1J_1}\left(V_{-+-}\right)_{J_1I_2K_2}
    \square^{-1}_{I_2J_2}\left(V_{-++}\right)_{J_2I_1K_1} \\
    -\,\square^{-1}_{I_1J_1}\left(V_{++-}\right)_{J_1I_2K_2}
    \square^{-1}_{I_2J_2}\left(V_{--+}\right)_{J_2I_1K_1} \\
    -\,\frac{1}{2} \Big\{ 4\,\,\square^{-1}_{I_1J_1}\left(V_{-++}\right)_{J_1I_2K_1}
    \square^{-1}_{I_2J_2}\left(V_{-+-}\right)_{J_2I_1K_2} \\
    +4\,\,\square^{-1}_{I_1J_1}\left(V_{-+-}\right)_{J_1I_2K_2}
    \square^{-1}_{I_2J_2}\left(V_{-++}\right)_{J_2I_1K_1}  \, \Big\}
    \Bigg]\,.
    \label{eq:2p_loopmhv}
\end{multline}
Diagrammatically, these contributions are
\begin{center}
\includegraphics[width=12cm]{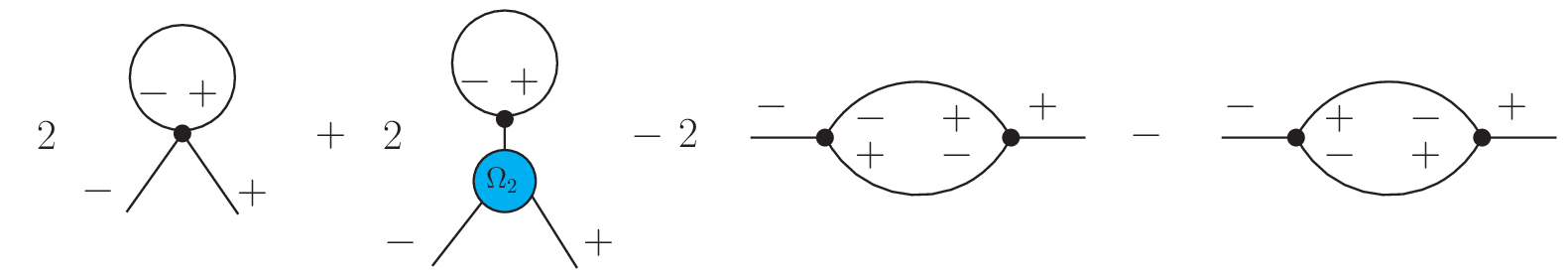}
\end{center}
 
\item  The second type of contribution arises via the tree level connection between the vertices in the classical action $S_{\mathrm{MHV}}[B_c]$ and the one-loop vertices in the log term in $\Gamma_{\mathrm{MHV}}[B_c]$ Eq.~\eqref{eq:G_MHV}. To the  two point $(+ -)$ one-loop case, there is just one such contribution originating from the tree level connection between the $(+ - -)$ triple gluon vertex in $S_{\mathrm{MHV}}[B_c]$ and the one-loop tadpole originating from the $(+ + -)$ triple gluon vertex in the log term. It reads
\begin{equation}
        -\frac{i}{2}\, \Bigg[2\, \square^{-1}_{IJ}\left(V_{-++}\right)_{JIL} \,\square^{-1}_{LK}\left(V^{\mathrm{MHV}}_{--+}\right)_{KK_1K_2}\,\Bigg]\,.
        \label{eq:mhv_tree_tad}
   \end{equation}
Diagrammatically, it is
   \begin{center}
\includegraphics[width=2.5cm]{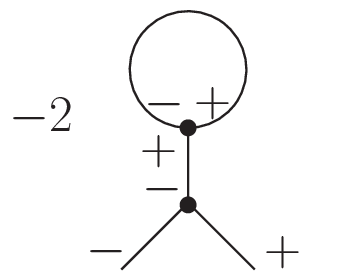}
\label{fig:Y}
\end{center}
\end{itemize}
Now, let us look at the above contributions using $W_{\mathrm{MHV}}[J]$ Eq.~\eqref{eq:WJ_MHV}. Any generic $n$-point connected Green's function can be obtained  from $W_{\mathrm{MHV}}[J]$ via functional derivatives with respect to the sources. In this case, the tree level connections to the external legs of the one-loop contributions are generated by the expansion of the classical fields in terms of the sources $\hat{B}_c^i[J]$. Therefore, for two point $(+ -)$ one-loop case we have just one source of contribution originating from the functional derivative of the log term with respect to the sources
\begin{equation}
       \Bigg[\frac{\delta^2 W_{\mathrm{MHV}}[J]}
    {\delta J^{\bullet K_1} \delta J^{\star K_2}}\Bigg]_{J= 0} = \frac{i}{2}\Bigg[\frac{\delta^2}
    {\delta J^{\bullet K_1} \delta J^{\star K_2}} \left\{ \Tr\ln \left( 
    \frac{\delta^2 S_{\mathrm{YM}}[A_c [B_c[J]]]}
    {\delta \hat{A}^i(x)\delta \hat{A}^j(y)}
    \right)
    \right\}\Bigg]_{J= 0}\,.
\end{equation}
The two sets of contributions we saw earlier when using $\Gamma_{\mathrm{MHV}}[B_c]$ Eq.~\eqref{eq:G_MHV} originate from the single source above. The first type of contributions shown in Eq.~\eqref{eq:2p_loopmhv} are the ones where the fields $\hat{B}_c^i[J]$ are expanded to first order in terms of the sources. The second type of contribution shown in Eq.~\eqref{eq:mhv_tree_tad} is the one where the field has been expanded to second order in source. Thus, in both cases (when using $\Gamma_{\mathrm{MHV}}[B_c]$ or $W_{\mathrm{MHV}}[J]$), the contributions to the two point $(+ -)$ one-loop amputated Green's function are exactly the same.
\begin{figure}
    \centering
    \includegraphics[width=15cm]{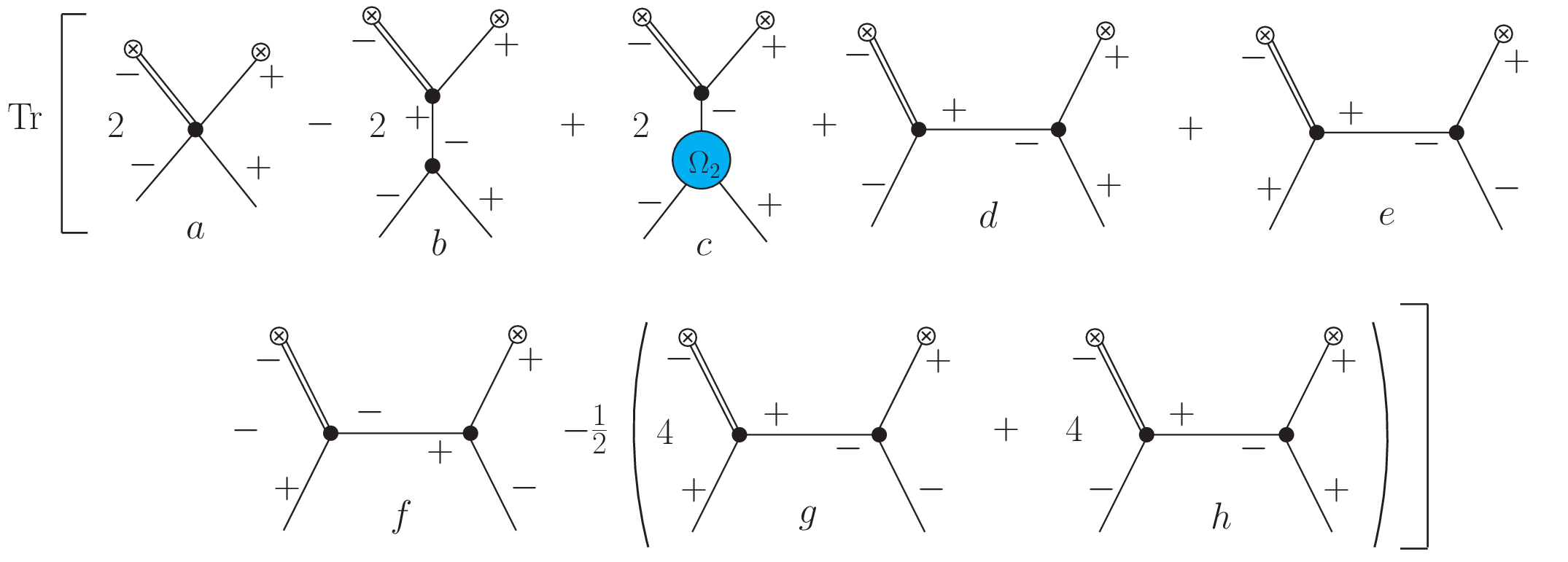}
    \caption{
    \small 
    The "un-traced" contributions to two point $(+ -)$ one-loop amputated connected Green's function originating from the one-loop effective MHV action Eq.~\eqref{eq:G_MHV}. Joining back the crossed circles we get the one-loop contributions. Notice, the double line on one of the legs in each term. This represents a propagator. This image was modified from our paper \cite{Kakkad_2022}.}
    \label{fig:2point}
\end{figure}

In order to demonstrate that these contributions Eq.~\eqref{eq:2p_loopmhv}-\eqref{eq:mhv_tree_tad} indeed consist of both MHV and non-MHV vertices in the loop, we un-trace the loops as shown in Figure \ref{fig:2point}. That is we cut open the loops and represent the cut legs via crossed circles in Figure \ref{fig:2point}. Also, notice that one of the cut legs is double lined. This represents a propagator. Combining the crossed circles we get back the one-loop contributions to the two point $(+ -)$ one-loop amputated connected Green's function. It turns out, the sum of the contributions a, b, c, and h in Figure~\ref{fig:2point} results in an un-traced 4-point MHV vertex. We show this diagrammatically in Figure~\ref{fig:4pointMHV}. The proof of this identity is collected in Appendix \ref{sec:app_A7}. Thus we have an MHV vertex in the loop. Note, further, this is the only contribution one would get when computing the two point $(+ -)$ one-loop amputated connected Green's function using just the MHV action. The remaining contributions i.e. d, e, f,and g would be missing. Furthermore, these contributions arise from the mixing of the $(+ + -)$ triple gluon vertex with the $(+ - -)$ triple-point MHV vertex. Thus, we see that the one-loop effective MHV action Eq.~\eqref{eq:G_MHV} does indeed consist of both the MHV and non-MHV contributions. 

\begin{figure}
    \centering
    \includegraphics[width=11cm]{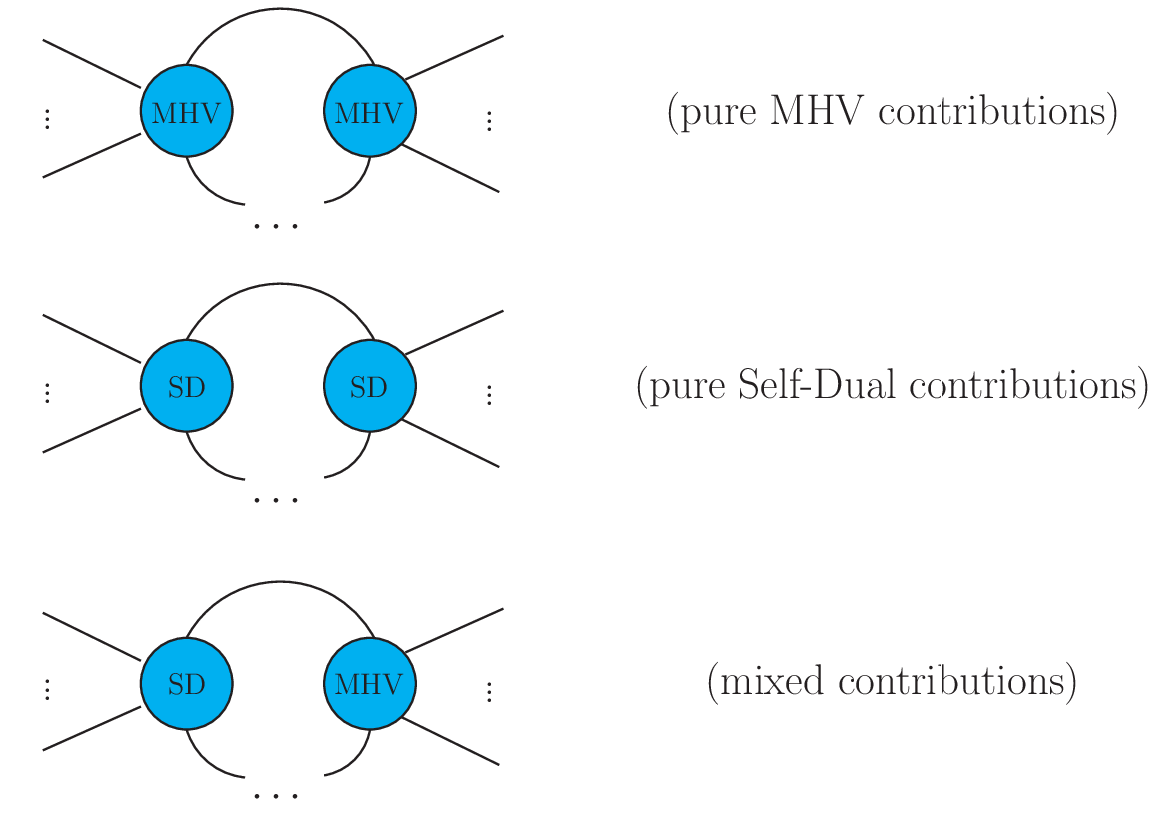}
    \caption{
    \small
     The one-loop contributions originating from the one-loop effective MHV action Eq.~\eqref{eq:G_MHV} can be categorized into three classes. The first class is the "pure MHV contributions". These consist of only the MHV vertices in the loop. The second class is the "pure Self-Dual contributions". These consist of only the $(+ + -)$ vertices in the loop. Finally, we have the "mixed contribution". These consist of a mixture of $(+ + -)$ vertices and MHV vertices $(+ \dots + + - -)$. The last two classes of contributions are what we collectively call the "non-MHV" contributions.   This image was modified from our paper \cite{Kakkad_2022}.}
    \label{fig:MHVloop_generic}
\end{figure}

In fact, this observation can be further generalized. In addition to purely MHV one-loop contributions, the one-loop effective MHV action Eq.~\eqref{eq:G_MHV} also consists of one-loop contributions where the vertices in the loop are either exclusively the Self-Dual vertices $(+ + -)$ or a mixture of the Self-Dual vertices $(+ + -)$ with the MHV vertices  $(+ + \dots + - -)$. The latter two would be missing if one computed one-loop amplitudes using just the MHV action. We show these three types of contributions in Figure \ref{fig:MHVloop_generic}.
\begin{figure}
    \centering
    \includegraphics[width=15cm]{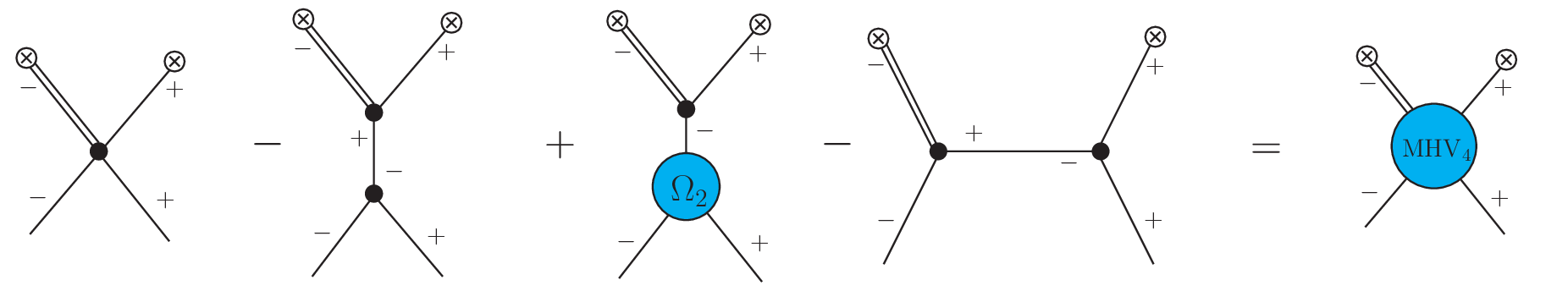}
    \caption{
    \small
    The sum of un-traced one-loop contributions a, b, c, and h in Figure \ref{fig:2point} is equal to the un-traced four point MHV vertex. The details of this identity are in Appendix \ref{sec:app_A7}. This image was modified from our paper \cite{Kakkad_2022}.} 
    \label{fig:4pointMHV}
\end{figure}

The three types of contributions shown in Figure \ref{fig:MHVloop_generic} indicate that there are more missing one-loop contributions when computing one-loop amplitudes using just the MHV action in four dimensions. Recall, any one-loop amplitude can be expressed in a basis consisting of a one-loop box, triangle, bubble, and rational terms. There are coefficients associated with the first three topologies which could be determined systematically in 4D using the unitarity cuts. These are therefore known as the cut constructible parts of the one-loop amplitudes. The rational terms can, however, not be determined using 4D unitarity cuts. For these one needs to perform D dimensional cuts \cite{Bern1994,Bern2007,Brandhuber2008,Perkins2009,Bern2011, DU1, DU2, DU3, DU4, DU5, DU6}. In 4D, MHV action is well-known to reproduce only the cut constructible parts of the one-loop amplitudes \cite{Bedford_cut}. These correspond to the contributions, termed "purely MHV" in Figure \ref{fig:MHVloop_generic}. This implies MHV action misses not only the all-plus $(+ \dots + +)$ and single-minus $(+ \dots + -)$ one-loop contributions, but it also misses the one-loop contributions necessary to compute the rational terms of the other one-loop amplitudes in 4D. These we believe must originate from the remaining two types of contributions in Figure~\ref{fig:MHVloop_generic}.  Consider first the "purely Self-Dual" contributions. These involve only the $(+ + -)$ triple gluon vertex in the loop and therefore give rise to all-plus $(+ \dots + +)$ one-loop terms. These are purely rational because the corresponding tree-level amplitudes vanish. They can contribute to the rational terms of the other one-loop amplitudes via tree-level connection with the MHV vertices in $S_{\mathrm{MHV}}[B_c]$ following the rules discussed in Appendix \ref{sec:app_A6}. The remaining contribution i.e. the "mixed contributions" in Figure \ref{fig:MHVloop_generic} can give rise to both rational and divergent terms. However, since the cut constructible pieces originate exclusively from the MHV contributions, it is expected that the divergent terms in the former should cancel out when all the "mixed contributions" are summed over thereby resulting in purely rational terms. In the following section, we use one-loop effective MHV action Eq.~\eqref{eq:G_MHV} to compute $(+ + + +)$ and $(+ + + -)$ one-loop amplitudes, and for the latter, we shall demonstrate the cancellations of the divergent terms originating from the mixed contributions resulting in a purely rational result.
\section{One-loop amplitudes}
\label{sec:One-loop-OLEAMHV}

So far, the discussion was focused on showing that the one-loop effective MHV action Eq.~\eqref{eq:G_MHV} does indeed consist of the non-MHV one-loop contributions that are missing in the MHV action Eq.~\eqref{eq:MHV_action}. However, this does not validate the one-loop effective MHV action Eq.~\eqref{eq:G_MHV} simply because the mere presence of non-MHV contributions does not prove that it is one-loop complete (no missing loop contributions). Therefore, in order to validate Eq.~\eqref{eq:G_MHV}, in this section, we will compute the all-plus $(+ \dots + +)$ and single-minus $(+ \dots + -)$ one-loop amplitudes using the one-loop effective MHV action Eq.~\eqref{eq:G_MHV}. As stated previously, these amplitudes are zero, to all orders, in the MHV action Eq.~\eqref{eq:MHV_action} \cite{Brandhuber2007a,Fu_2009, Boels_2008, Ettle2007,Brandhuber2007,Elvang_2012}. At the tree level, these are indeed zero in the on-shell limit (\emph{cf.} Eq.\eqref{eq:plus_tree_amp}) but are non-zero in general \cite{Bern_1994}. At one loop, both of these amplitudes are rational functions of spinor variables. Consider for instance the all-plus $(+ \dots + +)$ one-loop amplitude. It reads
\begin{equation}
       \mathcal{A}^{\mathrm{one-loop}}_n(+ + \dots +) = g^n\sum_{1 \leq i < j < k < l \leq n} \frac{\widetilde{v}_{ij}^{\star}\widetilde{v}_{jk}\widetilde{v}_{kl}^{\star}\widetilde{v}_{li}}{\widetilde{v}_{1n}^{\star}\widetilde{v}_{n\left(n-1\right)}^{\star}\widetilde{v}_{\left(n-1\right)\left(n-2\right)}^{\star}\dots\widetilde{v}_{21}^{\star}} \, ,
       \label{eq:all_plus_one_loop}
\end{equation}
where we suppressed the momentum conserving delta and the color factor. The above result is for the leading trace (the single-trace result) in color decomposition (\emph{cf.} Sub-section \ref{subsec:Col_dec}). In fact, throughout this section, when computing one-loop amplitudes, we will work in the 't~Hooft large $N_c$ limit (also known as the planar limit) \cite{tHooft:1973alw} for which only the leading trace partial amplitude in the color decomposition dominates. 

The computation of loop amplitudes is well-known to involve divergences that need to be regularized. Therefore, before we begin with the actual computation of amplitudes, let us briefly review the regularization scheme that we will employ.

\subsection{CQT regularization scheme}
\label{subsec:CQTreview}

The interaction vertices in the MHV action Eq.~\eqref{eq:MHV_action} have a compact single-term holomorphic expression only in 4D. In any other dimensions, this simplicity ceases to exist. In fact, in \cite{Ettle2007}, the authors explicitly demonstrated this when working with, more conventional, dimensional regularization scheme. This holds true also for the kernels in the solution ${\widetilde A}^{\bullet}[B^{\bullet}]$ and ${\widetilde A}^{\star}[B^{\bullet}, B^{\star}]$ Eq.~\eqref{eq:A_bull_solu}-\eqref{eq:A_star_solu}. That is, the simple holomorphic expressions for the kernels in these solutions are also a property of 4D. Therefore, to keep this simplicity intact we choose to work with the 4D world-sheet regularization scheme of Chakrabarti, Qiu, and Thorn (CQT)  \cite{CQT1,CQT2}. This scheme was originally introduced to explore the connections between quantum field theory and string theory. It, however, turned out to be useful, as we will see below, for the computation of loop amplitudes in light cone variables in 4D.  

\begin{figure}
    \centering
    \includegraphics[width=6cm]{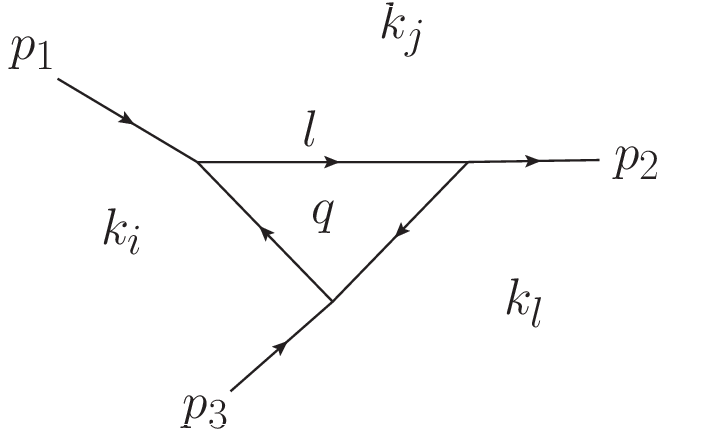}
    \caption{\small
     A sample planar one-loop diagram to demonstrate the assignment of region momenta in the CQT scheme. The bounded region inside the loop is assigned a region momentum $q$, where the un-bounded regions outside the loop are assigned region momenta $k_i$, $k_j$, and $k_l$. Using these, the line momenta can be expressed as $p_1= k_i-k_j$, $p_2= k_l-k_j$, $p_3= k_l-k_i$, and $l=q-k_j$. This image was taken from our paper \cite{Kakkad_2022}.}
    \label{fig:reg_mom}
\end{figure}

One of the key features of this scheme is that the loop integrand is expressed in terms of the region momenta as opposed to the more conventional line momenta associated with the internal and external lines in the Feynman diagram. This allows one to explore the world-sheet description of the scattering amplitudes in the planar limit. As a result, these region momenta are sometimes also dubbed as the "dual momenta" \cite{HOOFT1974461,Bardakci_2002}. To see how this works, consider a one-loop planar diagram shown in Figure \ref{fig:reg_mom}. The internal and external lines in the diagram divide the plane into bounded and unbounded regions. To the bounded region enclosed inside the loop, one associates a region momenta $q$ whereas to the unbounded regions, extending to infinity, outside the loops one assigns a region momenta $k_j$. Using these, one can express the line momenta as follows
\begin{equation}
    \mathrm{Line\,\, Momentum} =  \mathrm{Region\,\, Momentum\,\,on\,\, right} -  \mathrm{Region\,\, Momentum\,\,on\,\, left}\,,
    \label{eq:LM_con}
\end{equation}
Above, the notion of left and right is defined with respect to the orientation of the given line momentum. For instance, the line momenta in Figure \ref{fig:reg_mom} read
\begin{equation}
    p_1= k_i-k_j \,,\quad p_2= k_l-k_j\,, \quad p_3= k_l-k_i\,, \quad l=q-k_j \,.
    \label{eq:line_mom}
\end{equation}

With this, one can express the loop integrand entirely in terms of the region momenta. As a result, the integration over the loop momenta reduces to the integral over the loop region momenta $q$. Note, this assignment of region momentum works only for planar loop diagrams. As a result, the CQT scheme is suitable only for deriving the leading trace color ordered one-loop partial amplitudes. This limitation, however, does not concern us because, as mentioned above, we work in the 't~Hooft large $N_c$ limit where only the leading trace dominates.

In order to regularize the divergences originating from the different regions of integration over the loop region momenta $q$, the CQT scheme employs the following approach.
\begin{itemize}
    \item It uses an ultraviolet exponential cut-off to cut the integration over the loop region momenta $q$ along the transverse direction. Consider a one-loop Feynman diagram. To the loop integrand for such a Feynman diagram, in the CQT scheme, one introduces
    \begin{equation}
    \exp \Big( -\delta \mathbf{q}^2 \Big) \,, \quad \mathrm{where} \quad \mathbf{q}^2 = 2q^{\bullet}q^{\star}\,,
    \label{eq:cqt_exp}
\end{equation}
  where $\delta$ is the CQT regularization parameter. Throughout the computation, it is assumed to be positive ($\delta > 0$) and in the end, it is set to zero ($\delta \longrightarrow 0$). Note, since the ultraviolet cut-off Eq.~\eqref{eq:cqt_exp} uses only the transverse components of the loop region momenta ($\mathbf{q}^2 = 2q^{\bullet}q^{\star}$), it violates Lorentz invariance. Due to this, certain loop contributions that violate Lorentz invariance are non-zero in this scheme and must be canceled by introducing explicit counterterms. We will encounter some of these when computing one-loop amplitudes below.
    \item It uses discretization of the plus component of the loop momenta $q^{+}$ to regulate infrared divergences originating from the integration over certain intervals where the loop momentum is very small. The discretization converts the integral over the plus component of the loop momentum to a sum and the continuum limit is taken at the very end when computing physical quantities. There are however situations, as we will see when computing amplitudes below, where there are no such divergences. In that case, one can keep the plus component of the loop momentum continuous throughout the computation.
\end{itemize}

The counterterms mentioned above can be introduced systematically in the one-loop effective action approach. To do this, one starts by identifying all the necessary counterterms at the level of Yang-Mills. These are then introduced first in the Yang-Mills one-loop partition function as shown below
\begin{multline}
    Z_{\mathrm{YM}}[J]\approx 
    \exp\Bigg\{ iS_{\mathrm{YM}}[A_c] 
    + i\int\!d^4x\, \Tr \hat{J}_i(x) \hat{A}_c^i(x) - \frac{1}{2} \Tr\ln \left( 
    \frac{\delta^2 S_{\mathrm{YM}}[A_c]}
    {\delta \hat{A}^i(x)\delta \hat{A}^j(y)}
    \right) \\
    + i\Delta S_\mathrm{YM}^\mathrm{CQT}[A_c]
    \Bigg\}
    \label{eq:PartitionYM_CT}
    \,,
\end{multline}
where $\Delta S_\mathrm{YM}^\mathrm{CQT}[A_c]$ represents the set of all the counterterms required in the CQT scheme. We did not include a separate term representing the other counterterms, for instance, those required via renormalization. These could be thought of as included in the same set. We now perform  Mansfield's transformation Eq.~\eqref{eq:MansfieldTransf2} to Eq.~\eqref{eq:PartitionYM_CT} to get
\begin{multline}
    Z_{\mathrm{MHV}}[J]\approx 
    \exp\Bigg\{ iS_{\mathrm{MHV}}[B_c] 
    + i\int\!d^4x\, \Tr \hat{J}_i(x) \hat{A}_c^i[B_c](x) - \frac{1}{2} \Tr\ln \left( 
    \frac{\delta^2 S_{\mathrm{YM}}[A_c [B_c]]}
    {\delta \hat{A}^i(x)\delta \hat{A}^j(y)}
    \right)\\
    + i\Delta S_\mathrm{YM}^\mathrm{CQT}[A_c[B_c]]
    \Bigg\}
    \label{eq:PartitionMHV_CT}
    \,,
\end{multline}
where $\Delta S_\mathrm{YM}^\mathrm{CQT}[A_c[B_c]]$ represents the set of infinite counterterms originating from the substitution of the solutions of Mansfield's transformation ${\widetilde A}^{\bullet}[B^{\bullet}]$ and ${\widetilde A}^{\star}[B^{\bullet}, B^{\star}]$ Eq.~\eqref{eq:A_bull_solu}-\eqref{eq:A_star_solu} to $\Delta S_\mathrm{YM}^\mathrm{CQT}[A_c]$ in Eq.~\eqref{eq:PartitionYM_CT}. These are necessary to cancel out similar contributions originating from the log term in Eq.~\eqref{eq:PartitionMHV_CT}. Thus, in reality, we shall be using the CQT regularized one-loop effective MHV action obtained from Eq.~\eqref{eq:PartitionMHV_CT} to compute the one-loop amplitudes (and not simply Eq.~\eqref{eq:G_MHV}). 

It is worth pointing out that in \cite{Brandhuber2007}, the authors used canonically transformed (via Mansfield's transformation) counterterm in the CQT scheme to generate the missing contributions in the MHV action necessary to compute the all-plus $(+ + \dots +)$ one-loop amplitudes. Although we need these counterterms to compute the one-loop amplitudes correctly in this 4D scheme, these do not give rise to the missing non-MHV contribution in the MHV action Eq.~\eqref{eq:MHV_action}. In our approach, all the necessary one-loop contributions originate from the log term in Eq.~\eqref{eq:PartitionMHV_CT} and the counterterms merely cancel out the unwanted contributions that violate Lorentz invariance in the CQT scheme. We shall discuss the details of the work \cite{Brandhuber2007} below when computing the amplitudes.

Before we begin computing the one-loop amplitudes, let us point out the strategy that we shall follow
\begin{itemize}
    \item Using the CQT regularized one-loop effective MHV action, following the rules discussed in Appendix \ref{sec:app_A6}, we obtain all the Feynman diagrams contributing to the given amplitude with the necessary symmetry factors.
    \item The 1PI (one-particle irreducible) one-loop sub-diagrams in the above contributions (obtained via amputating the external tree level connections and/or the kernels from the solution ${\widetilde A}^{\bullet}[B^{\bullet}]$ and ${\widetilde A}^{\star}[B^{\bullet}, B^{\star}]$ Eq.~\eqref{eq:A_bull_solu}-\eqref{eq:A_star_solu}) will be exactly same as the one-loop 1PI contributions obtained from the log term in the Yang-Mills one-loop effective action Eq.~\eqref{eq:OLEA_YM}. The latter were explicitly computed in \cite{CQT1,CQT2}. Although we recompute some of these in Appendix \ref{sec:app_A8}, for the others we will highlight the important details and use the final results from their paper.
    \item To the above one-loop 1PI results, we substitute back the external tree-level connections and/or the kernels from the solution ${\widetilde A}^{\bullet}[B^{\bullet}]$ and ${\widetilde A}^{\star}[B^{\bullet}, B^{\star}]$ Eq.~\eqref{eq:A_bull_solu}-\eqref{eq:A_star_solu} with appropriate symmetry factors to get the final results for each diagram.
    \item The sum of all the diagrams must correspond to the final one-loop amplitude. Note, however, we shall focus only on the leading trace color-ordered amplitude.
\end{itemize}
\subsection{\texorpdfstring{$(+ + + +)$}{allplus} one-loop amplitude}
\label{sub:4plus}

In this Subsection, we will compute the leading trace color-ordered four-point $(+ + + +)$ one-loop amplitude using the CQT regularized one-loop effective MHV action. This amplitude is zero to all orders in the MHV action Eq.~\eqref{eq:MHV_action}. However, at one loop these are known to be a rational function of spinor products \cite{Bern:1991aq,Kunszt_1994}. Essentially, the all-plus one-loop amplitudes are the quantum corrections to the Self-Dual sector of the Yang-Mills theory. Since Mansfield's transformation maps this sector to a free theory when deriving the MHV action, the necessary contributions to compute this amplitude go missing. By computing this amplitude, we will demonstrate that these contributions are not missing in the one-loop effective MHV action.
\begin{figure}
    \centering
    \includegraphics[width=16cm]{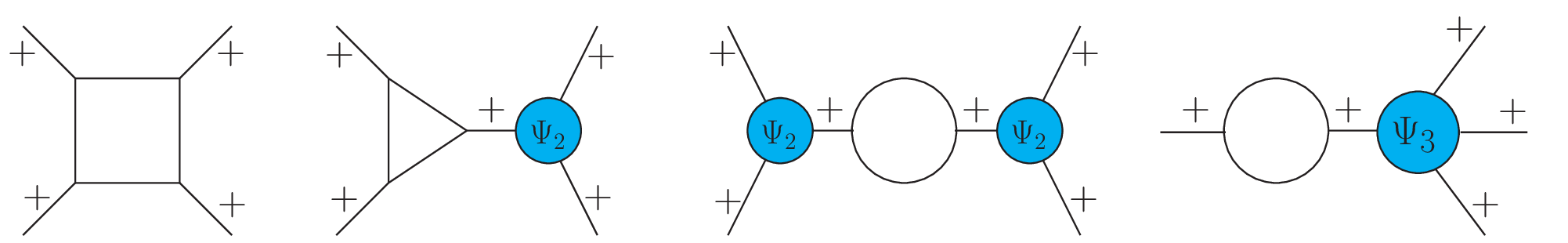}
    \caption{\small
The first three diagrams represent the one-loop contributions originating from the log term in the CQT regularized one-loop effective MHV action, following the rules in Appendix \ref{sec:app_A6}, to the $(+ + + +)$ one-loop amplitude. The last term is included to discuss an identity which states the sum of all four terms is zero. Note, there is no propagator connecting the kernels $\widetilde{\Psi}_n$ to the 1PI one-loop sub-diagrams. This image was modified from our paper \cite{Kakkad_2022}.}
    \label{fig:4plus_EA}
\end{figure}

Let us begin with the first step in the strategy mentioned in the previous Sub-section. Using the rules discussed in Appendix \ref{sec:app_A6}, the diagrams contributing to the $(+ + + +)$ one-loop amplitude are the first three shown in Figure \ref{fig:4plus_EA}. These contributions originate exclusively from the log term in Eq.~\eqref{eq:G_MHV}. In the Figure, we suppressed the symmetry factors for the sake of simplicity. The fourth diagram in Figure \ref{fig:4plus_EA} does not contribute because it consists of a self-energy term on the external leg. Upon amputation, this diagram will result in a tree-level contribution. We, however, still included it in Figure \ref{fig:4plus_EA} to discuss an interesting identity. This identity could be used to compute the amplitude itself. The identity states that the sum of all the four contributions in Figure \ref{fig:4plus_EA} with appropriate symmetry factors equal to zero \footnote{In \cite{CQT1} this observation was attributed to Zvi Bern.}. Notice, the third and the fourth contribution both involve the $(++)$ gluon self-energy which in the CQT scheme is non-zero. In Appendix \ref{sec:App_A81}, we re-derive this result. Explicitly, we show that this bubble with the following assignment of momenta
 \begin{center}
\includegraphics[width=3.6cm]{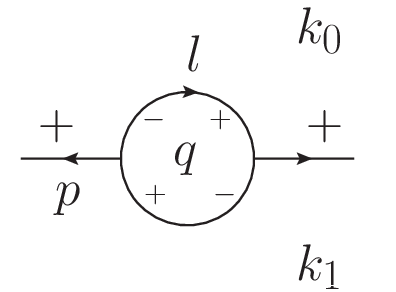}
\end{center}
reads
\begin{equation}
    \Pi^{+ +}= \frac{g^{2}}{12 \pi^{2}} \left[k_{0}^{\star 2}+k_{1}^{\star 2}+k_{0}^{\star} k_{1}^{\star}\right]\,.
    \label{eq:++GSEt}
\end{equation}
The non-zero result for the $(++)$ gluon self-energy, even on-shell, violates Lorentz invariance because it implies a gluon can flip its helicity. This must therefore be canceled by an explicit counterterm as shown below
\begin{center}
 \includegraphics[width=11cm]{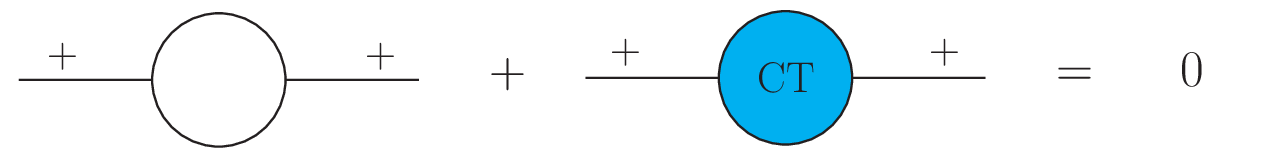}
\end{center}
Introducing this counterterm, we see that the last two contributions cancel out and the $(+ + + +)$ one-loop amplitude must then be equal to the sum of the first two contributions (this is what we will demonstrate below). But owing to the identity, the sum of the first two terms is equal to the negative of the last two terms. This implies one can compute the $(+ + + +)$ one-loop amplitude using just the counterterms. In fact, in \cite{Brandhuber2007}, the authors used exactly this approach. They first introduced the $(++)$ gluon self-energy counterterm Eq.~\eqref{eq:++GSEt} to the light-cone Yang-Mills action Eq.~\eqref{eq:YM_LC_action} and then performed Mansfield's transformation. This resulted in the MHV action plus an infinite series of all-plus $(+ + \dots +)$ one-loop vertices. For four point, these are exactly the last two terms in Figure \ref{fig:4plus_EA}. They showed that the sum of these two corresponds to the $(+ + + +)$ one-loop amplitude Eq.~\eqref{eq:all_plus_one_loop} in the on-shell limit.  

\begin{figure}
    \centering
    \includegraphics[width=16cm]{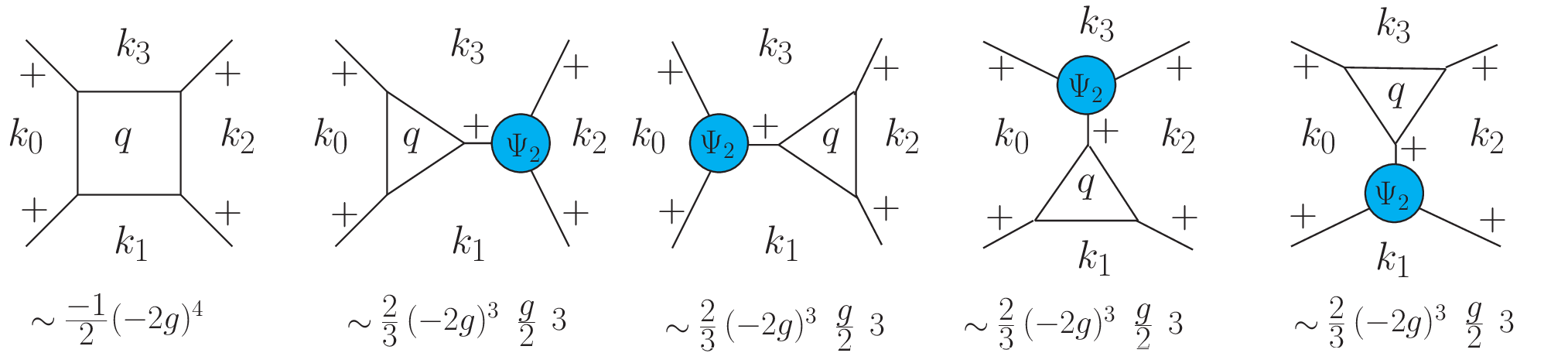}
    \caption{\small
    The explicit contributions to the $(+ + + +)$ one-loop amplitude with the symmetry factors as well as the assignment of region momenta. This image was modified from our paper \cite{Kakkad_2022}.}
    \label{fig:all_plus1}
\end{figure}

We will, however, follow the other way. After introducing the counter term we are left with just the first two contributions in Figure \ref{fig:4plus_EA}. The entire set of contributions originating from these two terms; the assignment of the line momenta and also the region momenta; and the appropriate symmetry factors are all shown explicitly in Figure \ref{fig:all_plus1}. 
Let us first understand the symmetry factors. Recall, the 1PI one-loop sub-diagrams in Figure \ref{fig:all_plus1} are all made up of $(+ + -)$ triple gluon vertex. This vertex has a factor of $(-2g)$ (see Eq.~\eqref{eq:v3_position}). This results in the factor of $(-2g)^4$ in the box diagram. The remaining factor of $(-1/2)$ arises from summing over all the box contributions originating from the log term in Eq.~\eqref{eq:G_MHV}. Similarly, for the triangular contributions (the remaining ones in Figure \ref{fig:all_plus1}), the $(-2g)^3$ originates from the $(+ + -)$ triple gluon vertices. The factor of $(2/3)$ arises from summing over all the 3-point 1PI one-loop triangle contributions $\Delta^{+ + + }$ originating from the log term in Eq.~\eqref{eq:G_MHV}. To each of the three legs in $\Delta^{+ + + }$, we need to substitute the solution ${\widetilde A}^{\bullet}[B^{\bullet}]$ Eq.~\eqref{eq:A_bull_solu}. Due to the symmetric structure of $\Delta^{+ + + }$, the substitution to all three legs is identical, and thus we get a factor of 3. The final factor of $(g/2)$ is due to the kernel $\widetilde{\Psi}_2$.

\begin{figure}
    \centering
    \includegraphics[width=13cm]{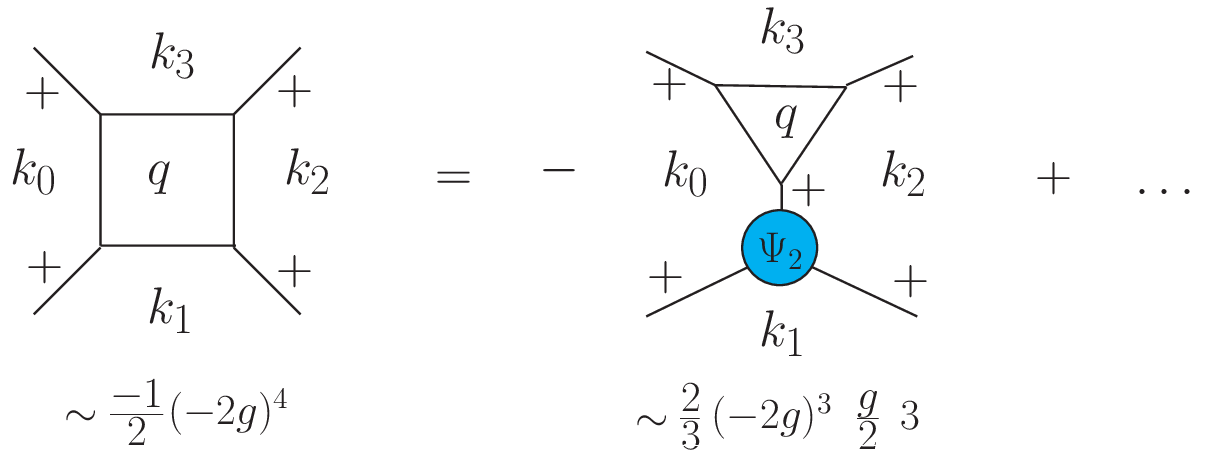}
    \caption{\small
    The box term $\square^{+ + + +}$ contributing to the $(+ + + +)$ one-loop amplitude can be reduced to negative times the triangular contribution $\Delta_{34}^{+ + + +}$ plus additional non-zero contribution.}
    \label{fig:box_tri}
\end{figure}

We denote the triangular contributions in Figure \ref{fig:all_plus1} as $\Delta_{ij}^{+ + + +}$ where the indices $i$ and $j$ denote the external on-shell legs attached to the triangle $\Delta^{+ + + }$. The triangular contributions in Figure~\ref{fig:all_plus1} are, therefore, $\Delta_{14}^{+ + + +}$, $\Delta_{23}^{+ + + +}$, $\Delta_{12}^{+ + + +}$ and $\Delta_{34}^{+ + + +}$ respectively. In Appendix~\ref{sec:App_A82}, we re-derived the expression for $\Delta_{12}^{+ + + +}$ in the CQT scheme starting first with the $\Delta^{+ + + }$. The expression reads
\begin{equation}
   \Delta_{12}^{+ + + +}= \frac{-g^{4}}{12 \pi^{2}} \frac{\left({\widetilde v}_{12}p_2^+\right)^{3}p_{3}^{+}}{p_{1}^{+} p_{2}^{+} p_{3}^{+}p_{4}^{+} p_{34}^{2}{\widetilde v}^{\star}_{34}}\,.
   \label{eq:tri_12}
\end{equation}
The remaining triangular contributions can be derived in a similar fashion. These read
\begin{equation}
    \Delta_{41}^{+ + + +}= \frac{-g^{4}}{12 \pi^{2}} \frac{\left({\widetilde v}_{41}p_1^+\right)^{3}p_{2}^{+}}{p_{1}^{+} p_{2}^{+} p_{3}^{+} p_{4}^{+} p_{14}^{2}{\widetilde v}^{\star}_{23}}\,,
    \label{eq:tri_41}
\end{equation}
\begin{equation}
   \Delta_{23}^{+ + + +}= \frac{-g^{4}}{12 \pi^{2}} \frac{ \left({\widetilde v}_{23}p_3^+\right)^{3}p_{4}^{+}}{p_{1}^{+} p_{2}^{+} p_{3}^{+} p_{4}^{+} p_{14}^{2}{\widetilde v}^{\star}_{41}}\,.
   \label{eq:tri_23}
\end{equation} 
 Above, we did not present the expression for $\Delta_{34}^{+ + + +}$. This is because it can be canceled against the box term \cite{CQT1}. In general, the box term $\square^{+ + + +}$ is more complicated to evaluate than the triangular contributions $\Delta_{ij}^{+ + + +}$. Thus, if possible, the strategy is to reduce the box term to the other, much simpler topology. For the box term $\square^{+ + + +}$ in Figure \ref{fig:all_plus1}, it turns out that these can be reduced to negative of the triangular contribution $\Delta_{34}^{+ + + +}$ plus some additional non-zero contribution. We represent this reduction diagrammatically in Figure~\ref{fig:box_tri} (the "$\dots$" in the Figure represents the non-vanishing additional contributions mentioned above). Thus, adding these together the triangular contribution $\Delta_{34}^{+ + + +}$ cancels out and we get
\begin{multline}
  \square^{+ + + +} \,+\, \Delta_{34}^{+ + + +} = \frac{-g^{4}}{12 \pi^{2}} \frac{p_{1}^{+}}{p_{1}^{+} p_{2}^{+} p_{3}^{+} p_{4}^{+} {\widetilde v}^{\star}_{12} p_{14}^{2}}\\
  \left[{\widetilde v}_{41}p_{1}^{+} {\widetilde v}_{23}p_{3}^{+}\left({\widetilde v}_{41}p_{1}^{+}+{\widetilde v}_{23}p_{3}^{+}\right)+{\widetilde v}_{34}p_{4}^{+}\left({\widetilde v}_{41}^{2}p_{1}^{+ 2}+{\widetilde v}_{23}^{2}p_{3}^{+ 2}\right)\right]\,.
    \label{eq:box+tri}
\end{multline}
The $(+ + + +)$ one-loop amplitude is the sum of all these contributions
\begin{equation}
    \mathcal{A}_{\mathrm{one-loop}}^{+ + + +} = \square^{+ + + +} \,+\, \Delta_{34}^{+ + + +} \,+\, \Delta_{12}^{+ + + +}\,+\, \Delta_{41}^{+ + + +}\,+\, \Delta_{23}^{+ + + +}\,.
\end{equation}
Substituting the expressions Eq.~\eqref{eq:tri_12}-\eqref{eq:box+tri}, after a bit of tedious algebra, in the on-shell limit we get
\begin{equation}
      \mathcal{A}_{\mathrm{one-loop}}^{+ + + +} = \frac{g^{4}}{24 \pi^{2}} \frac{{\widetilde v}_{21} {\widetilde v}_{43}}{{\widetilde v}^{\star}_{21} {\widetilde v}^{\star}_{43}} \, .
       \label{eq:4_plus_one_loop}
\end{equation}
The above result for leading trace  color ordered $(+ + + +)$ one-loop amplitude is in agreement with the known results \cite{Bern:1991aq,Kunszt_1994}, Eq.~\eqref{eq:all_plus_one_loop} (modulo an overall momentum conserving delta and a color factor).
\subsection{\texorpdfstring{$(+ + + -)$}{allplusm} one-loop amplitude}
\label{sub:3plus-minus}

\begin{figure}
    \centering
 \includegraphics[width=15.6cm]{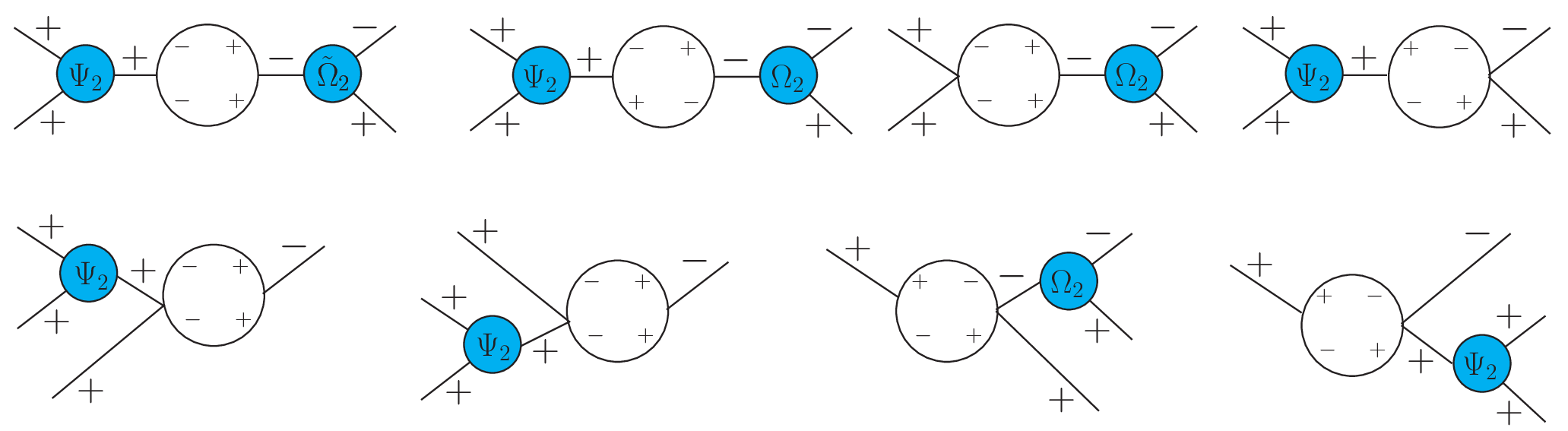}\\
 \includegraphics[width=15cm]{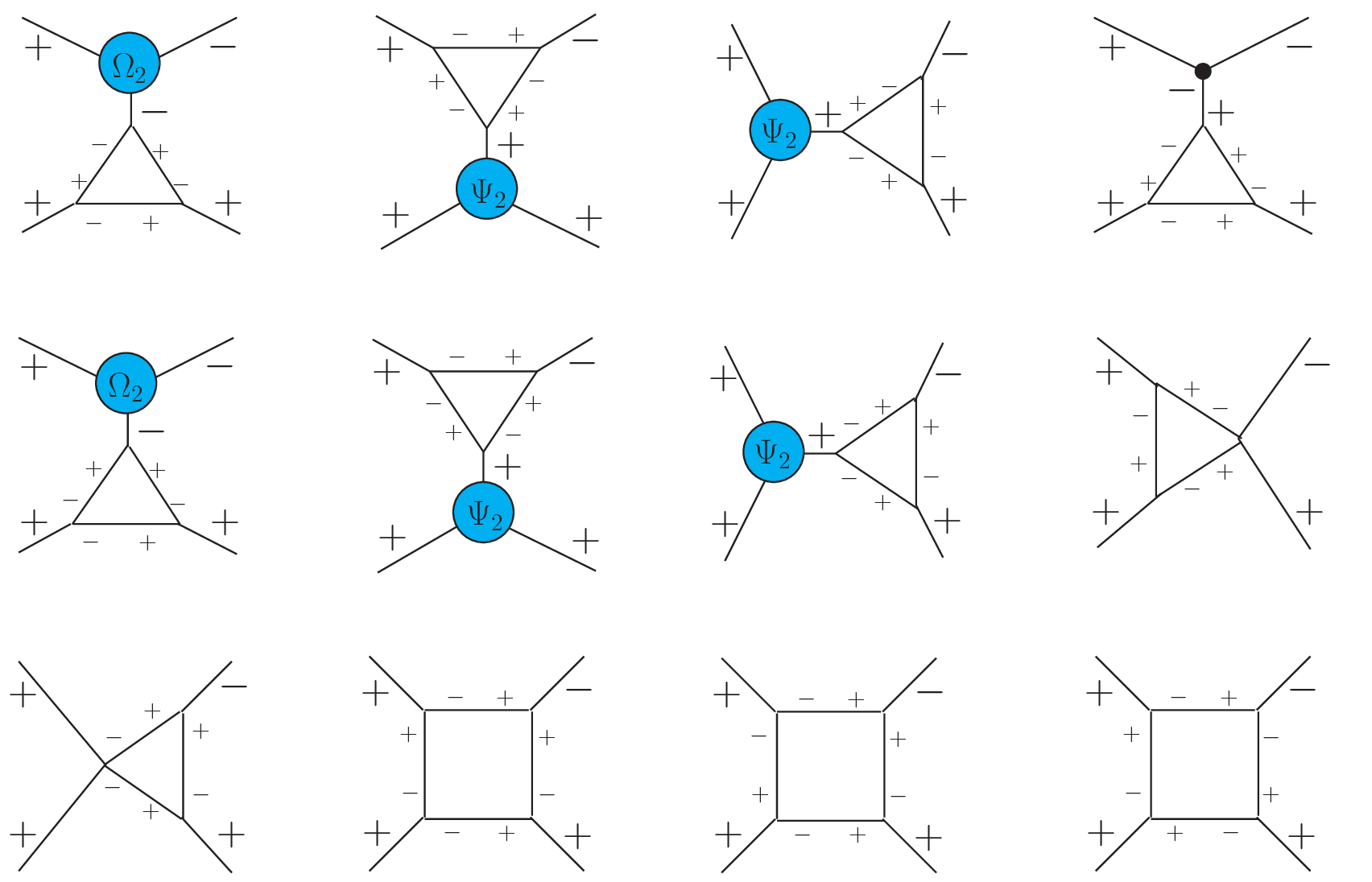} 
    \caption{\small 
    The one-loop contributions originating from the CQT regularized one-loop effective MHV action, following the rules in Appendix \ref{sec:app_A6}, to the $(+ + + -)$ one-loop amplitude. The last term on the third line represents the contribution involving tree level connection between the $(+ - -)$ MHV vertex in classical MHV action $S_{\mathrm{MHV}}[B_c]$ and the $\Delta^{+ + + }$ one-loop sub-diagram from the log term in Eq.~\eqref{eq:G_MHV}. We suppressed the symmetry factors and region momenta for the sake of simplicity. This image was modified from our paper \cite{Kakkad_2022}.}
    \label{fig:3plusminus}
\end{figure}

In this Subsection, we will compute the leading trace color-ordered four-point $(+ + + -)$ one-loop amplitude using the CQT regularized one-loop effective MHV action. Unlike in the previous case, this amplitude gets contribution both from the log term alone as well as the tree level connection between the MHV vertices in classical MHV action $S_{\mathrm{MHV}}[B_c]$ and the log term in Eq.~\eqref{eq:G_MHV}. The contributions are shown in Figure \ref{fig:3plusminus}. We did not include diagrams involving the $(+ +)$ gluon self-energy. This is because we need to introduce a $(+ +)$ counterterm in the CQT scheme and as a result, contributions involving the $(+ +)$ gluon self-energy get explicitly canceled by a similar contribution origination from $\Delta S_\mathrm{YM}^\mathrm{CQT}[A_c[B_c]]$. As before, we suppress the symmetry factors in the Figure for the sake of simplicity. We name the legs $(+ + + -)$ as 1, 2, 3, and 4 in the anticlockwise fashion and associate four momenta $p_i$ to each leg respectively. 

Let us begin with the first two terms in Figure \ref{fig:3plusminus}. Both of these diagrams involve the substitution of the second order expansion of the  solutions ${\widetilde A}^{\bullet}[B^{\bullet}]$ and ${\widetilde A}^{\star}[B^{\bullet}, B^{\star}]$ Eq.~\eqref{eq:A_bull_solu}-\eqref{eq:A_star_solu} to the two legs of $(+ -)$ gluon self-energy respectively. The latter can be derived in the CQT scheme in exactly the same way as $(+ +)$ gluon self-energy. Following the re-derivation of $(+ +)$ discussed in Appendix \ref{sec:App_A81}, the $(+ -)$ bubble reads
\begin{multline}
\Pi^{+ -} =\frac{g^{2}}{4 \pi^{2}} p^{2}\Bigg(\sum_{q^{+}}\left[\frac{1}{q^{+}}+\frac{1}{p^{+}-q^{+}}\right] \ln \left\{\frac{q^{+}\left(p^{+}-q^{+}\right)}{p^{+2}} p^{2} \delta e^{\gamma}\right\} \\
-\frac{11}{6} \ln \left(p^{2} \delta e^{\gamma}\right)+\frac{67}{18}\Bigg) + \frac{g^{2}}{4 \pi^{2}}\frac{1}{\delta}\sum_{q^{+}}\frac{1}{p^{+}}\Bigg(1 +\frac{p^{+ 2}}{q^{+ 2}} +\frac{p^{+ 2}}{(p^{+}-q^{+})^2} \Bigg)\,,
\label{eq:pi+-}
\end{multline}
where $p$ is the line momentum associated with the external legs of the bubble. $q$ is the region momentum associated with the loop. Notice, if we take the continuum limit, the sum over $q^+$ reduces to an integral that is divergent. As stated above, such divergences are regularized by the discretization of the plus component of the momentum. The discretization proceeds as follows. $q^+ = l P^+$, where $l=1,2,\dots,N$, and $p^+ = N P^+$. With this, $N$ represents the total number of steps in which the plus component of the external line momentum is discretized and $P^+$ represents the step length (in the world-sheet description, it, however, gets the interpretation of the "mass scale"). As a result, the summation can be written as $\sum_{q^{+}}= P^+\sum_{l=1}^{N-1}$. These divergences are, however, an outcome of the light-cone gauge choice (and are therefore spurious artifacts) and will get canceled by similar contributions originating from the other one-loop diagrams when computing the amplitudes. Finally, $\delta$ is the CQT transverse space regularization parameter and $\gamma$ is the Euler's constant.  

In expression Eq.~\eqref{eq:pi+-}, there are essentially two types of terms. Those in the first bracket have logarithmic divergence and the ones in the second bracket have $1/ \delta$. The latter were termed as "quadratic divergent" \footnote{Given the form of the exponential cut-off used in the CQT scheme Eq~\eqref{eq:cqt_exp}, these terms originate from contributions that are quadratic divergent.} in \cite{CQT1}
\begin{equation}
   \Pi^{+ -}_{\mathrm{QD}} = \frac{g^{2}}{4 \pi^{2}}\frac{1}{\delta}\sum_{q^{+}}\frac{1}{p^{+}}\Bigg(1 +\frac{p^{+ 2}}{q^{+ 2}} +\frac{p^{+ 2}}{(p^{+}-q^{+})^2} \Bigg)\,.
   \label{eq:PI_QD}
\end{equation}
Above "QD" stands for quadratic divergent. These were shown in \cite{CQT1} to be the artifacts of the regularization scheme and, in total, two counterterms were needed to get rid of $\Pi^{+ -}_{\mathrm{QD}}$. 
With this, the $(+ -)$ gluon self-energy reduces to
\begin{multline}
\Pi^{+ -}_{\mathrm{sub}} =\frac{g^{2}}{4 \pi^{2}} p^{2}\Bigg(\sum_{q^{+}}\left[\frac{1}{q^{+}}+\frac{1}{p^{+}-q^{+}}\right] \ln \left\{\frac{q^{+}\left(p^{+}-q^{+}\right)}{p^{+2}} p^{2} \delta e^{\gamma}\right\} \\
-\frac{11}{6} \ln \left(p^{2} \delta e^{\gamma}\right)+\frac{67}{18}\Bigg)\,,
\label{eq:pi+-_ct}
\end{multline}
where "sub" stands for subtracted. Notice, the above expression is still divergent in the continuum limit. As stated above, these divergences are the outcome of the light-cone gauge choice and do not require any new set of counterterms. As we will see below, they cancel against similar contributions originating from the other diagrams. Substituting ${\widetilde A}^{\bullet}[B^{\bullet}]$ and ${\widetilde A}^{\star}[B^{\bullet}, B^{\star}]$ Eq.~\eqref{eq:A_bull_solu}-\eqref{eq:A_star_solu}, to the above expression we get the first two diagrams in Figure~\ref{fig:3plusminus}. By doing this, we get
\begin{multline}
\mathcal{A}_{\mathrm{SE}}^{+ + + -} =\frac{ -g^{4}p_{4}^{+}}{2 \pi^{2} p_{1}^{+} p_{2}^{+} p_{3}^{+}}\Bigg[\frac{{\widetilde v}_{34}p_{4}^{+}p_{2}^{+}}{{\widetilde v}_{21}^{\star}}\\
\Bigg\{\sum_{q^{+}}\left[\frac{1}{q^{+}}+\frac{1}{p_{12}^{+}-q^{+}}\right] \ln \Bigg(\frac{q^{+}\left(p_{12}^{+}-q^{+}\right)}{p_{12}^{+ 2}} p_{12}^{2} e^{\gamma} \delta\Bigg)
-\frac{11}{6} \ln \left(p_{12}^{2} e^{\gamma} \delta\right)+\frac{67}{18}\Bigg\}\\
+\frac{{\widetilde v}_{41}p_{3}^{+}p_{1}^{+}}{{\widetilde v}_{32}^{\star}}\Bigg\{\sum_{q^{+}+p_4^+}\left[\frac{1}{q^{+}+p_4^+}+\frac{1}{p_1^{+}-q^{+}}\right] \ln \Bigg(\frac{(q^{+}+p_4^+)\left(p_1^{+}-q^{+}\right)}{p_{14}^{+2}} p_{14}^{2} e^{\gamma} \delta\Bigg)\\
-\frac{11}{6} \ln \left(p_{14}^{2} e^{\gamma} \delta\right)+\frac{67}{18}\Bigg\}\Bigg] \,,
\label{eq:self_contri_OL}
\end{multline}
where the subscript "SE" stands for self-energy.

Let us now consider the set of diagrams in Figure \ref{fig:3plusminus} which involve 3-point one-loop 1PI sub-diagrams. There are two topologies contributing to this type
\begin{center}
    \includegraphics[width=8cm]{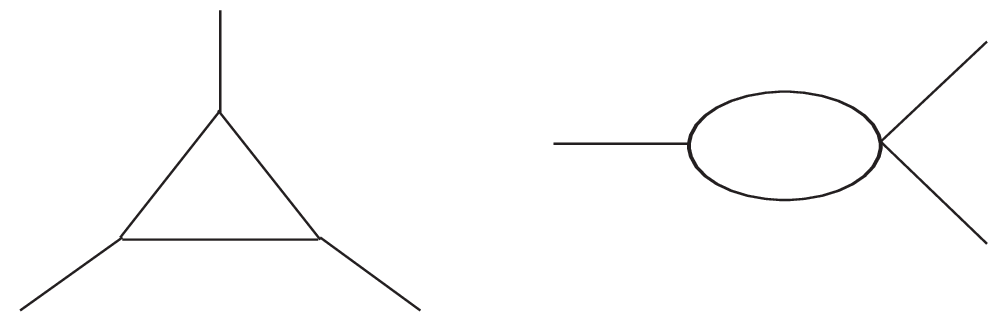}
\end{center}
The left one is what we call a triangle (as before) and the right one is what "is called" a "swordfish". In Figure \ref{fig:3plusminus}, the last two diagrams in the first line and the whole of the second line consists of the $(+ + -)$ swordfish sub-diagrams. On the other hand, the whole of the third line and the first three diagrams in the fourth line consists of the triangle sub-diagram. Notice, however, there are two types of triangle sub-diagrams: $\Delta^{+ + +}$ and $\Delta^{+ + -}$. 

The last diagram on the third line in Figure \ref{fig:3plusminus} is made up of tree level connection of the $(+ + -)$ MHV vertex in the classical MHV action $S_{\mathrm{MHV}}[B_c]$ with the triangle $\Delta^{+ + +}$ originating from the log term in Eq.~\eqref{eq:G_MHV}. The remaining contributions include the 3-point one-loop $(+ + -)$  sub-diagrams (triangle + swordfish). In the CQT scheme, these 3-point one-loop $(+ + -)$  sub-diagrams result in contributions that violate Lorentz invariance and therefore require an explicit counterterm. This counterterm reads \cite{CQT1}
\begin{center}
    \includegraphics[width=9cm]{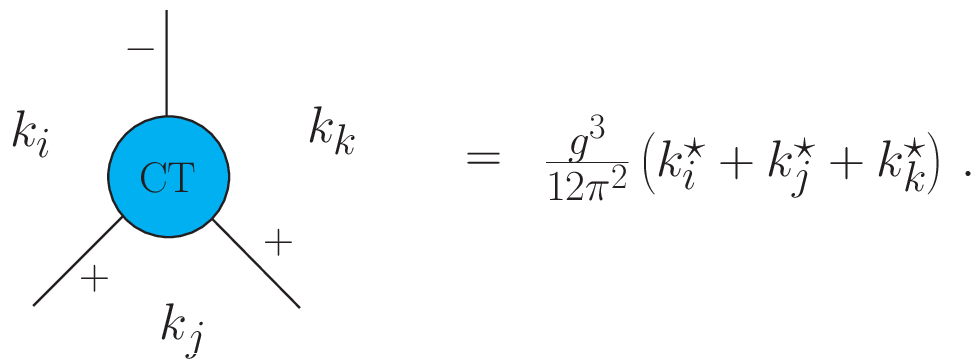}
\end{center}
Substituting ${\widetilde A}^{\bullet}[B^{\bullet}]$ and ${\widetilde A}^{\star}[B^{\bullet}, B^{\star}]$ Eq.~\eqref{eq:A_bull_solu}-\eqref{eq:A_star_solu} to the 3-point one-loop $(+ + -)$ and combining it with $\Delta^{+ + +}$ term following the tree connection, we get
\begin{multline} 
\mathcal{A}_{\mathrm{TS}}^{+ + + -}=\frac{-g^{4}p_{4}^{+}}{4 \pi^{2} p_{1}^{+} p_{2}^{+} p_{3}^{+}}\Bigg[\frac{{\widetilde v}_{34}p_{4}^{+}p_{2}^{+}}{{\widetilde v}_{21}^{\star}}\Bigg\{\frac{22}{3} \ln \left(p_{12}^{2} e^{\gamma} \delta\right)-\frac{140}{9}-S_{3}^{q^{+}}\left(p_{1}, p_{2}\right)-S_{3}^{q^{+}}\left(-p_{4},-p_{3}\right)\\
+\frac{p_{1}^{+} p_{2}^{+}}{3 p_{12}^{+2}}\Bigg\}
+\frac{{\widetilde v}_{41}p_{3}^{+}p_{1}^{+}}{{\widetilde v}_{32}^{\star}}\Bigg\{\frac{22}{3} \ln \left(p_{14}^{2} e^{\gamma} \delta\right)-\frac{140}{9}-S_{2}^{q^{+}}\left(-p_{4},-p_{23}\right)-S_{1}^{q^{+}+p_{4}^{+}}\left(p_{14}, p_{2}\right)+\frac{p_{2}^{+} p_{3}^{+}}{3 p_{14}^{+2}}\Bigg\}\Bigg] \\
+\frac{g^{4}p_{3}^{+}p_{2}^{+2}}{3 \pi^{2} p_{1}^{+}  p_{12}^{+2}} \frac{{\widetilde v}_{12}^3 {\widetilde v}_{34}^{\star}}{p_{12}^{4}}+\frac{g^{4}p_{1}^{+ 2}p_{3}^{+ 2}}{3 \pi^{2} p_{2}^{+} p_{4}^{+} p_{14}^{+2}} \frac{{\widetilde v}_{23}^3 {\widetilde v}_{41}^{\star}}{p_{14}^{4}}\,.
\label{eq:TS_contri_OL}
\end{multline}
Above, the subscript "TS" stands for triangle + swordfish. And, $S_{l}^{q^{+}}\!\left(p_{i}, p_{j}\right)$ is an infrared sensitive term. The explicit expression for this quantity can be found in Appendix \ref{sec:App_A83}.

Finally, we are left with the 4-point one-loop 1PI diagrams in Figure \ref{fig:3plusminus}. These consist of two different topologies shown below
\begin{center}
    \includegraphics[width=7cm]{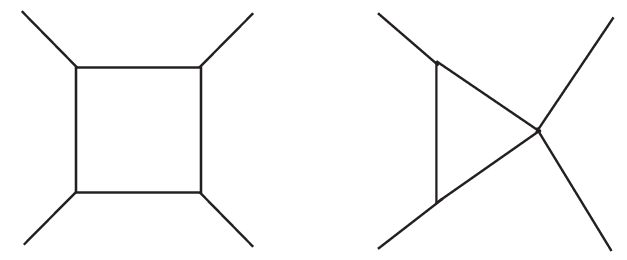}
\end{center}
On the left, we have the box diagram and on the right, we have the so-called "quartic" contribution (made by combining the triple gluon vertices with the quartic interaction vertex in the Yang-Mills action Eq.~\eqref{eq:YM_LC_action}). In Figure \ref{fig:3plusminus}, the last term on the fourth line and the first term on the last line are both quartic contributions. The remaining ones i.e. the last three terms on the final line are all box contributions. Just as in the case of $\square^{+ + + +}$, the box contributions here too can be reduced to triangular contributions. Since the 4-point one-loop 1PI diagrams (both box and the quartic) do not involve any tree level connections or the substitution of ${\widetilde A}^{\bullet}[B^{\bullet}]$ and ${\widetilde A}^{\star}[B^{\bullet}, B^{\star}]$ Eq.~\eqref{eq:A_bull_solu}-\eqref{eq:A_star_solu} to the external legs, we can simply use the result for these terms from \cite{CQT1}. We cross checked the symmetry factors to make sure they are correct. Their sum reads
\begin{multline}
    \mathcal{A}_{\mathrm{BQ}}^{ + + + -}=\frac{ g^{4} }{2 \pi^{2}}\Bigg[ {-\left\{\frac{p_{1}^{+}p^+_{2\overline{3}}-3 p_{3}^{+}p^+_{23}}{6 p_{12}^{+} p_{23}^{+} p_{4}^{+} {\widetilde v}_{14}{\widetilde v}_{41}^{\star}}+\frac{p_{3}^{+}(-3 p_{1}^{+} p_{12}^{+}+p_{3}^{+}p^+_{\overline{1}2})}{6 p_{1}^{+} p_{12}^{+2} p_{4}^{+} {\widetilde v}_{12}{\widetilde v}_{21}^{\star}}\right\} {\widetilde v}_{12}^2p_{2}^{+} } \\
+\left\{\frac{p_{1}^{+}p^+_{\overline{2}3}+p_{3}^{+} p_{23}^{+}}{6 p_{1}^{+} p_{3}^{+} p_{23}^{+2} {\widetilde v}_{12}{\widetilde v}_{21}^{\star}}-\frac{p_{1}^{+}p^+_{\overline{2}3}+3 p_{3}^{+} p_{23}^{+}}{6 p_{12}^{+} p_{3}^{+} p_{23}^{+} p_{4}^{+} {\widetilde v}_{14}{\widetilde v}_{41}^{\star}}\right\} {\widetilde v}_{12}{\widetilde v}_{34}p_2^+p_4^+\\
+\frac{p_{4}^{+}}{2p_{1}^{+} p_{2}^{+} p_{3}^{+}} \frac{{\widetilde v}_{34}p_{4}^{+}p_{2}^{+}}{{\widetilde v}_{21}^{\star}}\Bigg\{\frac{11}{3} \ln \left(\delta e^{\gamma} p_{12}^{2}\right)-\frac{11}{3} \ln \left(\delta e^{\gamma} p_{14}^{2}\right)\\
     -S_{3}^{q^{+}}\left(p_{1}, p_{2}\right)-S_{3}^{q^{+}}\left(-p_{4},-p_{3}\right)+S_{2}^{q^{+}}\left(-p_{4},-p_{23}\right) +S_{1}^{q^{+}+p_{4}^{+}}\left(p_{14}, p_{2}\right)\\
-2 \sum_{q^{+}}\left[\frac{1}{q^{+}}+\frac{1}{p_{12}^{+}-q^{+}}\right] \ln \Bigg(\frac{q^{+}\left(p_{12}^{+}-q^{+}\right)}{p_{12}^{+ 2}} p_{12}^{2} e^{\gamma} \delta\Bigg)\\
+2 \sum_{q^{+}+p_4^+}\left[\frac{1}{q^{+}+p_4^+}+\frac{1}{p_1^{+}-q^{+}}\right] \ln \Bigg(\frac{(q^{+}+p_4^+)\left(p_1^{+}-q^{+}\right)}{p_{14}^{+2}} p_{14}^{2} e^{\gamma} \delta\Bigg)\Bigg\}\Bigg]\,,
\label{eq:box_quar}
\end{multline}
where $p^+_{ij}=p_{i}^{+}+p_{j}^{+}$ and $p^+_{i\overline{j}}=p_{i}^{+}-p_{j}^{+}$. Above, the subscript "BQ" stands for box + quartic. 

The  $(+ + + -)$ one-loop amplitude is the sum of all the above contributions
\begin{equation}
    \mathcal{A}_{\mathrm{one-loop}}^{+ + + -} = \mathcal{A}_{\mathrm{SE}}^{+ + + -} + \mathcal{A}_{\mathrm{TS}}^{+ + + -} + \mathcal{A}_{\mathrm{BQ}}^{+ + + -}\,.
\end{equation}
Substituting Eqs.~\eqref{eq:self_contri_OL}-\eqref{eq:box_quar}, after a bit of tedious algebra in the on-shell limit we get
\begin{equation}
    \mathcal{A}_{\mathrm{one-loop}}^{+ + + -} = \frac{-g^{4}}{24 \pi^{2}} \frac{ p_3^{+}{\widetilde v}_{13}^2}{p_1^+{\widetilde v}_{14} {\widetilde v}_{43}{\widetilde v}^{\star}_{21} {\widetilde v}^{\star}_{32}} (p_{12}^2 + p_{14}^2 )\,.
    \label{eq:3plus-oneloop}
\end{equation}
Notice, the divergent pieces indeed cancel out, to give a rational result, as we claimed for the mixed contributions in Figure \ref{fig:MHVloop_generic}. The above results for the leading trace color ordered 4-point one-loop amplitude is in agreement with the results obtained in \cite{Bern:1991aq,Kunszt_1994} (modulo an overall momentum conserving delta and a color factor).

Thus, we see that the  one-loop effective MHV action Eq.~\eqref{eq:G_MHV} does contain the contributions necessary to compute the all-plus $(+ \dots + +)$ and single-minus $(+ \dots + -)$ one-loop amplitudes.
\chapter{Quantum corrections to the new action}
\label{QZth-chapter}

In this chapter, our focus is on two major goals. The first is to extend the one-loop effective action approach from the previous chapter to systematically develop quantum corrections to the new Wilson line-based action we developed in Chapter \ref{WLAc-chapter}. The one-loop action obtained this way will be one-loop complete with no missing loop contributions and will also allow for efficient computation of one-loop amplitudes as compared to both the one-loop effective Yang-Mills and MHV action. There is however a drawback in the one-loop action derived in the previous chapter where we start with the one-loop effective Yang-Mills action and then perform Mansfield's transformation. Recall, the main motive to develop the  MHV as well as the new Wilson line-based action was to switch over to "bigger" interaction vertices that would allow for efficient computation of pure gluonic amplitudes. On the contrary, the loop contributions originating from the log term in the one-loop effective MHV action use only the Yang-Mills vertices, instead of the MHV vertices, to form the loops. Therefore, the second goal is to derive one-loop effective action, both for the MHV as well as our new action, such that the new interaction vertices are explicit in the log term. We do this first for the MHV action where we start with the Yang-Mills partition function; perform Mansfield's transformation to the Yang-Mills action as well as the source terms and then derive the one-loop action by integrating the field fluctuations. Keeping the source term unaltered is crucial to this approach. The one-loop effective MHV action derived this way has MHV vertices explicit in the log term. We then demonstrate that the one-loop action is both one-loop complete as well as equal (modulo a field independent volume divergent factor irrelevant for amplitude computation) to the one-loop effective MHV action derived in the previous chapter. This approach is then extended to derive the one-loop action for our new action. Finally using it we compute several one-loop amplitudes. All the results discussed in this chapter are new and have not been published yet. The manuscript for this work is under preparation.

\section{One-loop Effective Z-field action}
\label{sec:Zth_OLEA}

In Chapter \ref{WLAc-chapter}, we derived a new Wilson line-based action Eq.~\eqref{eq:Z_action1} (we shall call it the "Z-field action" throughout this chapter) by exchanging the gluon fields in the light-cone Yang-Mills action Eq.~\eqref{eq:YM_LC_action} with Wilson line degrees of freedom via a canonical transformation. This indeed turned out to be beneficial in the context of computing pure gluonic tree amplitudes because the number of contributing diagrams reduced drastically (\emph{cf.} Section \ref{sec:Zac_TR_AMP}). The major reason for this multi-fold simplicity was that the above stated exchange of gluon fields with Wilson line degrees of freedom eliminated both the triple-point interaction vertices from the Yang-Mills action and resulted in an infinite number of interaction vertices with different helicity configurations. Although beneficial at the tree level, this elimination makes the Z-field action "quantum incomplete". Recall, in order to derive the MHV action Eq.~\eqref{eq:MHV_action} from the light-cone Yang-Mills action Eq.~\eqref{eq:YM_LC_action}, one eliminates the $(+ + -)$ triple gluon vertex via Mansfield's transformation Eq.~\eqref{eq:Man_Transf1}. Due to this, all the loop contributions originating either solely from this vertex or the mixing of this vertex with the MHV vertices are missing. This issue escalates in the Z-field action because all the loop contributions originating 
\begin{itemize}
    \item solely from the $(+ + -)$ triple point interaction vertex,
    \item solely from the $(+ - -)$ triple point interaction vertex,
    \item from the mixing of $(+ + -)$, $(+ - -)$ and the Z-field interaction vertices,
\end{itemize}
are all missing.  As a result, the all-plus $(+ \dots + +)$, single-minus $(+ \dots + -)$, all-minus $(- \dots - -)$ and  single-plus $(- \dots - +)$ helicity loop amplitudes cannot be computed using the Z-field action. Furthermore, restricting the discussion to 4D, the rational contributions to the other one-loop amplitudes will also be missing from the Z-field action. This is because, recall from Figure \ref{fig:ZTH_vertex_gen}, any generic vertex in the Z-field action can be derived from the substitution of $\widetilde{B}^{\star}_a[{Z}^{\star}](x^+;\mathbf{P})$ and $\widetilde{B}^{\bullet}_a[{Z}^{\star}, {Z}^{\bullet}](x^+;\mathbf{P})$ Eq.~\eqref{eq:BstarZ_exp}-\eqref{eq:BbulletZ_exp} to the vertices in the MHV action Eq.~\eqref{eq:MHV_action}. The kernels in $\widetilde{B}^{\star}_a[{Z}^{\star}](x^+;\mathbf{P})$ and $\widetilde{B}^{\bullet}_a[{Z}^{\star}, {Z}^{\bullet}](x^+;\mathbf{P})$ account for the tree level contributions originating from $(+ - -)$ triple point MHV interaction vertex. This implies any generic vertex in the Z-field action is essentially made up of only the MHV vertices. The latter is known to give only the cut-constructible parts of the one-loop amplitudes in 4D. Thus computing one-loop amplitudes using the Z-field action will also give only the cut-constructible result and not the full amplitude.

In the previous chapter, we used the one-loop effective action approach to overcome similar issues in the MHV action. We demonstrated that starting with the one-loop effective Yang-Mills action Eq.~\eqref{eq:OLEA_YM}, which is one-loop complete, one can derive the one-loop effective MHV action Eq.~\eqref{eq:G_MHV} via Mansfield's transformation Eq.~\eqref{eq:Man_Transf1}. This way, there are no missing one-loop contributions. In this section, we will extend this approach to systematically develop one-loop corrections to the Z-field action. 

In order to do this, we use the same strategy as in the case of deriving the one-loop effective MHV action Eq.~\eqref{eq:G_MHV}. That is, we start with the one-loop effective Yang-Mills action Eq.~\eqref{eq:OLEA_YM} and then perform the transformation that derives the Z-field action. Recall, however, from Figure \ref{fig:CT_paths} that there are two ways of deriving the Z-field action. One is the direct approach using the generating functional 
\begin{equation}
   \mathcal{G}(q,Q) \equiv \mathcal{G}[A^\bullet,Z^\star](x^+) =
    -\int\! d^3\mathbf{x}\,\,\,\Tr\,
     \hat{\mathcal{W}}^{\,-1}_{(-)}[Z](x)\,\,
     \partial_- \hat{\mathcal{W}}_{(+)}[A](x) \,.
    \label{eq:generatingfuncOL}
\end{equation}
The second is via a set of two consecutive canonical transformations. First maps the Self-Dual part of the Yang-Mills action to the kinetic term in the MHV action Eq.~\eqref{eq:SD_MT} and the second maps the Anti-Self-Dual part of the MHV action to the kinetic term in the Z-field action Eq.~\eqref{eq:BtoZtransform}.

Below, we argue that either of the two ways can be used to derive the same one-loop effective Z-field action. For the sake of simplicity, we start with the latter approach involving two consecutive canonical transformations. Note, however, since the derivation involves field transformations, it is preferable to first derive the one-loop effective Z-field partition function and then derive the effective action via the Legendre transform of the generating functional for the connected Green's function. Symbolically, the derivation of one-loop effective Z-field action in this approach involves
\begin{equation}
   Z_{\mathrm{YM}}[A_c[J]] \xrightarrow[]{\hat{A}^{\bullet}[{B}^{\bullet}]\,,\,\hat{A}^{\star}[{B}^{\bullet},{B}^\star]} Z_{\mathrm{MHV}}[B_c[J]] \xrightarrow[]{\hat{B}^{\bullet}[{Z}^{\bullet},{Z}^{\star}]\,,\,\hat{B}^{\star}[{Z}^\star]} Z[Z_c[J]]\longrightarrow W[Z_c[J]] \xrightarrow[]{\mathrm{LT}}\Gamma[Z_c]\,.
   \label{eq:OLEA_ABZ}
\end{equation}
Above, LT stands for the Legendre transform. The first step consists of deriving the one-loop effective MHV partition function from the one-loop effective Yang-Mills partition function Eq.~\eqref{eq:Partition_YM}. This was already achieved in the previous chapter. Thus the starting point for us is the one-loop effective MHV partition function Eq.~\eqref{eq:PartitionMHV}
\begin{equation}
    Z_{\mathrm{MHV}}[J]\approx 
    \exp\left\{ iS_{\mathrm{MHV}}[B_c] 
    + i\int\!d^4x\, \Tr \hat{J}_i(x) \hat{A}_c^i[B_c](x) - \frac{1}{2} \Tr\ln \left( 
    \frac{\delta^2 S_{\mathrm{YM}}[A_c [B_c]]}
    {\delta \hat{A}^i(x)\delta \hat{A}^j(y)}
    \right)
    \right\}
    \label{eq:PartitionMHVOL}
    \,.
\end{equation}
Now, we perform the second transformation shown below
\begin{equation}
\mathcal{L}_{-+}[B^{\bullet},B^{\star}]+\mathcal{L}_{--+}[B^{\bullet},B^{\star}]
\,\, \longrightarrow \,\,
\mathcal{L}_{-+}[Z^{\bullet},Z^{\star}]
\,.
\label{eq:BtoZtransformOL}
\end{equation}
In order to execute the above transformation, we need to substitute the expression for the classical fields $\left\{{\hat B}_{c}^{\bullet},{\hat B}_{c}^{\star}\right\} \longrightarrow \left\{{\hat B}_{c}^{\bullet}[{Z}_{c}^{\bullet}, {Z}_{c}^{\star}],{\hat B}_{c}^{\star}[{Z}_{c}^{\star}]\right\}$ using the position space version of the solutions $\widetilde{B}^{\star}_a[{Z}^{\star}](x^+;\mathbf{P})$ and $\widetilde{B}^{\bullet}_a[{Z}^{\star}, {Z}^{\bullet}](x^+;\mathbf{P})$ Eq.~\eqref{eq:BstarZ_exp}-\eqref{eq:BbulletZ_exp}. Making this substitution to the classical MHV action $S_{\mathrm{MHV}}[B_c]$, we get
\begin{equation}
    S_{\mathrm{MHV}}[B_c] \longrightarrow S[Z_c]\,.
    \label{eq:Cmhv2Cz}
\end{equation}
This was already discussed in detail in Subsection \ref{subsec:Zac_der}.
Finally, we substitute the solution $\hat{B}^{\star}_c[{Z}_c^{\star}]$ and $\hat{B}^{\bullet}_c[{Z}_c^{\star}, {Z}_c^{\bullet}]$ to the source term and the log term in Eq.~\eqref{eq:PartitionMHVOL}
\begin{multline}
     Z[J]\approx 
    \exp\Bigg\{ i\,S[Z_c] 
    + i\int\!d^4x\, \Tr \hat{J}_i(x) \hat{A}_c^i[B_c[Z_c]](x) 
    - \frac{1}{2} \Tr\ln \left( 
    \frac{\delta^2 S_{\mathrm{YM}}[A_c [B_c[Z_c]]]}
    {\delta \hat{A}^i(x)\delta \hat{A}^j(y)}
    \right)
    \Bigg\}
    \label{eq:PartitionZOL}
    \,.
\end{multline}
Notice the replacement $\hat{B}_c^i = \hat{B}_c^i[Z_c]$ (where $i= \bullet, \star$). The matrices in the log term read
\begin{multline}
   \frac{\delta^2 S_{\mathrm{YM}}[A_c[B_c[Z_c]]]}
    {\delta A^{\star I}\delta A^{\bullet J}} 
    = -\square_{IJ}-\left(V_{-++}\right)_{IJK}A_{c}^{\bullet K}[B_c[Z_c]] -\left(V_{--+}\right)_{KIJ}A_{c}^{\star K}[B_c[Z_c]] \\
    -\left(V_{--++}\right)_{LIJK}A_{c}^{\star L}[B_c[Z_c]] A_{c}^{\bullet K}[B_c[Z_c]]
    \label{eq:S+-Z}
    \,,
\end{multline}
\begin{equation}
   \frac{\delta^2 S_{\mathrm{YM}}[A_c[B_c[Z_c]]]}
    {\delta A^{\bullet I}\delta A^{\bullet J}} 
    = -\left(V_{-++}\right)_{KIJ}A_{c}^{\star K}[B_c[Z_c]] -\left(V_{--++}\right)_{KLIJ}A_{c}^{\star K}[B_c[Z_c]] A_{c}^{\star L}[B_c[Z_c]]
    \label{eq:S++Z}
    \,,
\end{equation}
\begin{equation}
   \frac{\delta^2 S_{\mathrm{YM}}[A_c[B_c[Z_c]]]}
    {\delta A^{\star I}\delta A^{\star J}} 
    = -\left(V_{--+}\right)_{IJK}A_{c}^{\bullet K}[B_c[Z_c]] -\left(V_{--++}\right)_{IJKL}A_{c}^{\bullet K}[B_c[Z_c]] A_{c}^{\bullet L}[B_c[Z_c]]
    \label{eq:S--Z}
    \,.
\end{equation}
Just as in the case of one-loop effective MHV partition function Eq.~\eqref{eq:PartitionMHVOL}, the inverse propagator term in Eq.~\eqref{eq:S+-Z} is field independent. As a result, it does not generate contributions that could cancel out similar contributions originating from the triple point interaction vertices $\left(V_{-++}\right)_{IJK}A_{c}^{\bullet K}[B_c[Z_c]] $ and $\left(V_{--+}\right)_{KIJ}A_{c}^{\star K}[B_c[Z_c]]$ in the log term. Therefore, the loop contributions involving the vertices $(+ + -)$ and $(- - +)$ will no longer be missing. The above result leads to the one-loop effective Z-field partition function. 

Before we proceed to derive the one-loop effective Z-field action, let us, briefly, consider using the 
second approach involving the generating functional Eq.~\eqref{eq:generatingfuncOL}. Symbolically, the derivation looks like the following
\begin{equation}
   Z_{\mathrm{YM}}[A_c[J]] \xrightarrow[]{\hat{A}^{\bullet}[{Z}^{\bullet},{Z}^\star]\,,\,\hat{A}^{\star}[{Z}^{\bullet},{Z}^\star]} Z[Z_c[J]]\longrightarrow W[Z_c[J]] \xrightarrow[]{\mathrm{LT}}\Gamma[Z_c]\,.
   \label{eq:OLEA_AZ}
\end{equation}
In this case, the one-loop effective Z-field partition function is first obtained directly from the one-loop effective Yang-Mills partition function 
Eq.~\eqref{eq:Partition_YM}. 
To achieve this, we substitute the classical fields $\left\{{\hat A}_{c}^{\bullet},{\hat A}_{c}^{\star}\right\} $ $ \longrightarrow $ $ \left\{{\hat A}_{c}^{\bullet}[{Z}_{c}^{\bullet}, {Z}_{c}^{\star}],{\hat A}_{c}^{\star}[{Z}_{c}^{\bullet},{Z}_{c}^{\star}]\right\}$ using  Eq.~\eqref{eq:Abullet_to_Z}-\eqref{eq:Astar_to_Z}. By doing this to the classical action $S_{\mathrm{YM}}[A_c]$, following the discussion in Subsection \ref{subsec:Zac_der} we get $S_{\mathrm{YM}}[A_c] \longrightarrow S[Z_c]$. In the same Subsection, we also demonstrated that the solution $\left\{{\hat A}_{c}^{\bullet}[{Z}_{c}^{\bullet}, {Z}_{c}^{\star}],{\hat A}_{c}^{\star}[{Z}_{c}^{\bullet},{Z}_{c}^{\star}]\right\}$ can be obtained via a two step procedure where we substitute the expressions for $\hat{B}^{\star}_c[{Z}_c^{\star}]$ and $\hat{B}^{\bullet}_c[{Z}_c^{\star}, {Z}_c^{\bullet}]$  into the expressions for $\hat{A}^{\bullet}_c[{B}_c^{\bullet}]$ and $\hat{A}^{\star}_c[{B}_c^{\bullet}, {B}_c^{\star}]$. Thus, substituting the latter to the source term and the log term in the one-loop effective Yang-Mills partition function, we get exactly the same one-loop effective Z-field partition function Eq.~\eqref{eq:PartitionZOL}.
With this, we see that the  one-loop effective Z-field partition function Eq.~\eqref{eq:PartitionZOL} can  equivalently be derived following either of the two ways. 

\begin{figure}
    \centering
    \includegraphics[width=10cm]{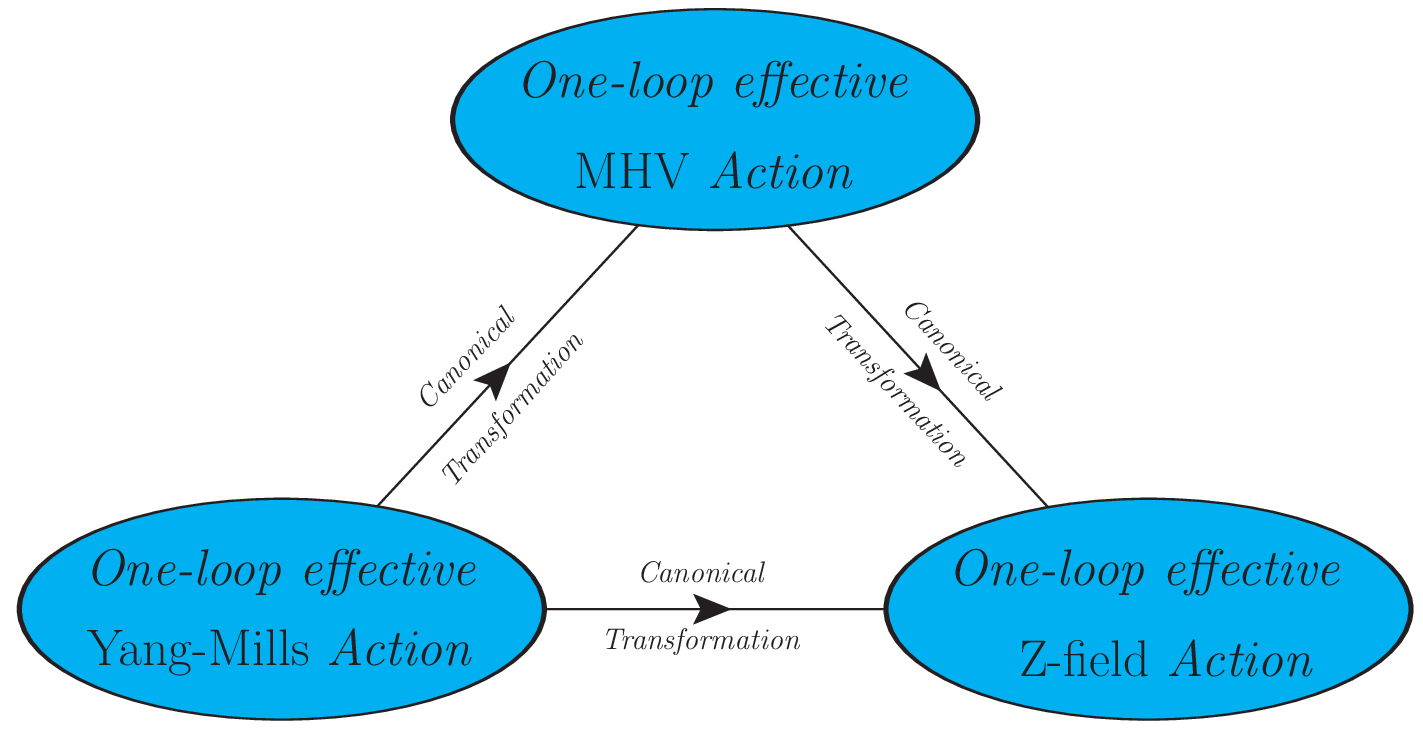}
    \caption{\small
    Two ways of deriving the one-loop effective Z-field action $\Gamma[Z_c]$. First is the direct approach $\Gamma_{\mathrm{YM}}[A_c] \longrightarrow \Gamma[Z_c]$ using the generating function Eq.~\eqref{eq:generatingfuncOL}. The second is via two consecutive transformations $\Gamma_{\mathrm{YM}}[A_c] \longrightarrow \Gamma_{\mathrm{MHV}}[B_c] \longrightarrow \Gamma[Z_c]$.}
    \label{fig:OLEApaths}
\end{figure}

Using Eq.~\eqref{eq:PartitionZOL}, the generating functional for the connected Green's function has the schematic form
\begin{equation}
    W[J]= S[Z_c] 
    + \int\!d^4x\, \Tr \hat{J}_i(x) \hat{A}_c^i[Z_c[J]](x) +i\, \frac{1}{2} \Tr\ln \left( 
    \frac{\delta^2 S_{\mathrm{YM}}[A_c [Z_c[J]]]}
    {\delta \hat{A}^i(x)\delta \hat{A}^j(y)}
    \right)\,.
    \label{eq:WJ_Zac}
\end{equation}
Performing the Legendre transform we get
\begin{equation}
   \Gamma[Z_c]= S[Z_c]  +i\, \frac{1}{2} \Tr\ln \left( 
    \frac{\delta^2 S_{\mathrm{YM}}[A_c [Z_c]]}
    {\delta \hat{A}^i(x)\delta \hat{A}^j(y)}
    \right)\,.
    \label{eq:G_Zac}
\end{equation}
Note, since the one-loop effective Z-field partition function obtained via the two approaches, is exactly the same; the one-loop effective Z-field action obtained using either of the two Eq.~\eqref{eq:OLEA_ABZ} or \eqref{eq:OLEA_AZ} must also be the same. Thus we can conclude that just like the Z-field action, the one-loop effective Z-field action can also be derived from the one-loop effective Yang-Mills action in two ways.  First is the direct approach using the generating function Eq.~\eqref{eq:generatingfuncOL}
\begin{equation}
   \Gamma_{\mathrm{YM}}[A_c] \xrightarrow[]{\hat{A}^{\bullet}[{Z}^{\bullet},{Z}^{\star}]\,,\hat{A}^{\star}[{Z}^{\bullet},{Z}^\star]} \Gamma[Z_c]\,.
   \label{eq:OLEA_dir}
\end{equation}
The second is via a set of two consecutive transformations
\begin{equation}
   \Gamma_{\mathrm{YM}}[A_c] \xrightarrow[]{\hat{A}^{\bullet}[{B}^{\bullet}]\,,\hat{A}^{\star}[{B}^{\bullet},{B}^\star]} \Gamma_{\mathrm{MHV}}[B_c] \xrightarrow[]{\hat{B}^{\bullet}[{Z}^{\bullet},{Z}^{\star}]\,,\hat{B}^{\star}[{Z}^\star]} \Gamma[Z_c]\,.
   \label{eq:OLEA_ind}
\end{equation}
We represent these two ways diagrammatically in Figure \ref{fig:OLEApaths}.

One may wonder, what is the actual gain from deriving the one-loop effective Z-field action Eq.~\eqref{eq:G_Zac}. There are essentially two important gains. First is that the one-loop effective Z-field action Eq.~\eqref{eq:G_Zac} is one-loop complete i.e. there are no missing one-loop contributions. Second, the number of diagrams required to compute higher multiplicity one-loop amplitudes using the one-loop effective Z-field action Eq.~\eqref{eq:G_Zac} will be considerably less when compared to the one-loop effective MHV action Eq.~\eqref{eq:G_MHV} and way less when compared with the one-loop effective Yang-Mills action Eq.~\eqref{eq:OLEA_YM}. Since these aspects are crucial, let us elaborate. 

As previously stated, the major advantage of the one-loop effective action approach itself is that it provides a systematic way for deriving quantum corrections to actions (like the MHV action and the Z-field action) that are related to the Yang-Mills action via field transformation. Now, recall from the derivation of the one-loop effective MHV action Eq.~\eqref{eq:G_MHV} that we started with the one-loop effective Yang-Mills action Eq.~\eqref{eq:OLEA_YM}. The latter is, by construction, one-loop complete and all we achieve by performing Mansfield's transformation is a remarkable reduction in the number of diagrams for computing pure gluonic amplitudes up to one-loop. There are two major sources for this reduction. Following the rules for computing amplitudes using one-loop effective action discussed in Appendix \ref{sec:app_A6}, we see that for any given one-loop amplitude, either the log term can contribute independently or a one-loop term originating from the log term can undergo tree level connection with interaction vertices in the classical action to give a contribution (let us call it the tree-loop contribution). In going from one-loop effective Yang-Mills action Eq.~\eqref{eq:OLEA_YM} to one-loop effective MHV action Eq.~\eqref{eq:G_MHV}, we achieve a reduction in the number of diagrams in both of these sectors. The substitution of the solution of Mansfield's transformation to the log term accounts for all the tree level connections originating from the $(+ + -)$ triple-gluon vertex that one needs to do in the case of  one-loop effective Yang-Mills action. Secondly, the interaction vertices in the classical action get "bigger" (from three and four-point interaction vertices to the MHV vertices). Therefore the number of diagrams originating from the tree-loop contributions also reduces. These advantages grow further when going from one-loop effective MHV action Eq.~\eqref{eq:G_MHV} to one-loop effective Z-field action Eq.~\eqref{eq:G_Zac}. Lets us start first with the quantum completeness of the action. Just like in the case of one-loop effective MHV action Eq.~\eqref{eq:G_MHV}, since we derive the one-loop effective Z-field action Eq.~\eqref{eq:G_Zac} starting with an action that is already one-loop complete and has the classical and loop contribution separated, the resulting action following the transformation is guaranteed to be one-loop complete as well (however for the sake of completeness, we will demonstrate this explicitly by computing one-loop amplitudes in Section \ref{sec:loopamp_ZAC}). Regarding the efficiency in computing one-loop amplitudes using the one-loop effective Z-field action Eq.~\eqref{eq:G_Zac}, the classical action here has even bigger building blocks as compared to the MHV action. Furthermore, in the log term the substitution of $ \left\{{\hat A}_{c}^{\bullet}[{Z}_{c}^{\bullet}, {Z}_{c}^{\star}],{\hat A}_{c}^{\star}[{Z}_{c}^{\bullet},{Z}_{c}^{\star}]\right\}$, accounts for the tree level connection involving both the triple point interaction vertices $(+ + -)$ and $(+ - -)$. Both these factors further increase the efficiency when compared with the one-loop effective MHV action. Note, however, this efficiency is visible only when computing higher multiplicity one-loop amplitudes. For amplitudes say the 4-point one-loop there is practically not much difference between the one-loop effective MHV action Eq.~\eqref{eq:G_MHV} and the one-loop effective Z-field action Eq.~\eqref{eq:G_Zac}.

Although both the one-loop effective MHV action Eq.~\eqref{eq:G_MHV} and the one-loop effective Z-field action Eq.~\eqref{eq:G_Zac} provide a huge simplicity for computing pure gluonic amplitudes up to one-loop when compared with the one-loop effective Yang-Mills action Eq.~\eqref{eq:OLEA_YM}, they both have a major issue that restricts a further increase in the efficiency. In the following section, we discuss this at length.

\section{Issue with MHV/Z-field one-loop effective actions}
\label{sec:Issue_MHV_Zac}

\begin{figure}
    \centering
 \hspace*{1.5cm}\includegraphics[width=11.2cm]{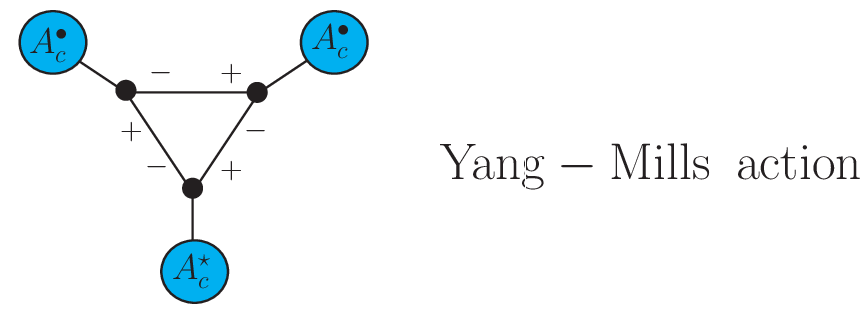}\\
 \includegraphics[width=13cm]{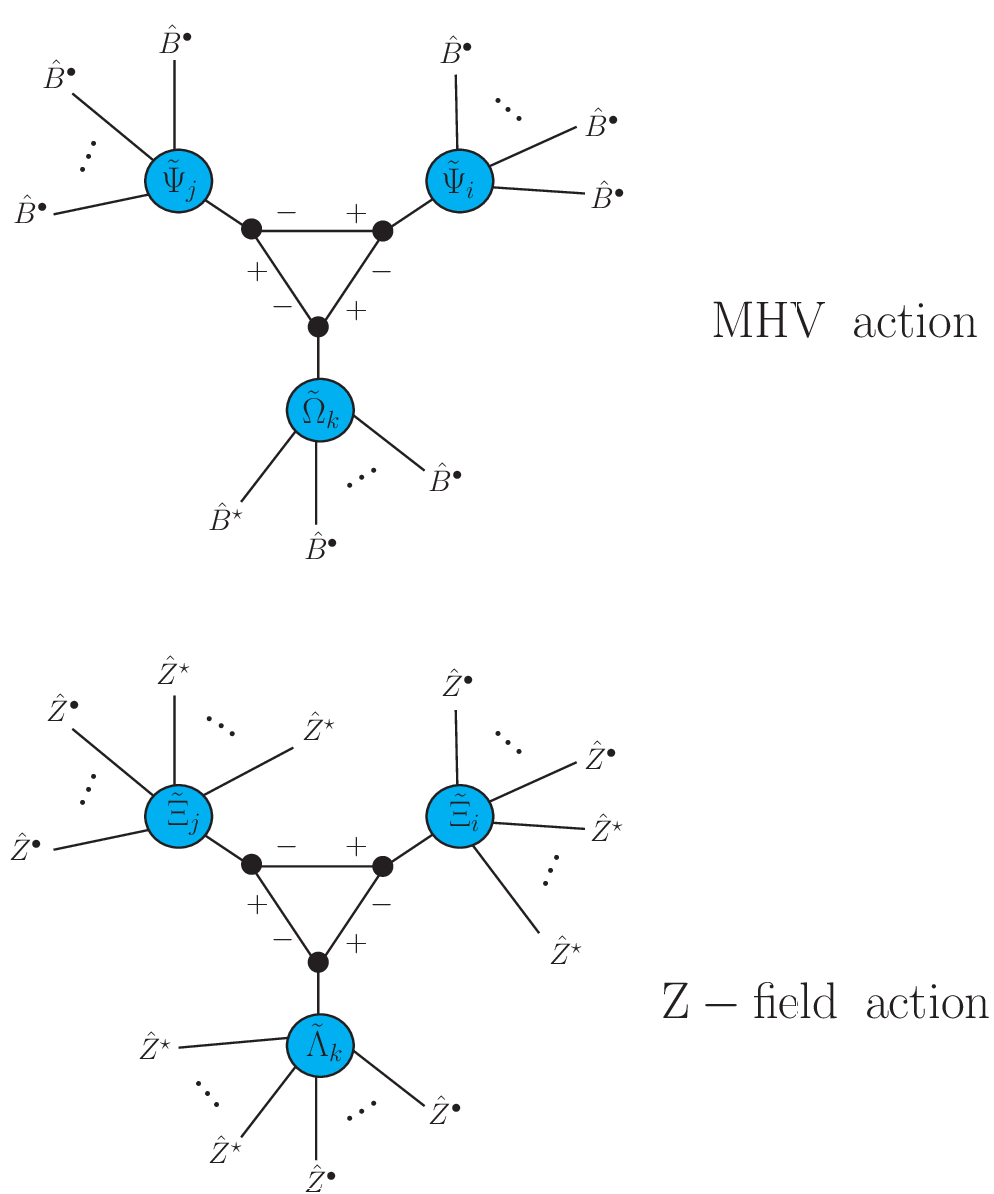} 
    \caption{\small 
    In going from top (a sample loop term in the one-loop effective Yang-Mills action Eq.~\eqref{eq:G_YM1}) to the bottom (the same loop term in the one-loop effective Z-field action Eq.~\eqref{eq:G_Zac1}), only the classical field outside the loop undergoes a field substitution $\hat{A}_c^i \longrightarrow \hat{A}_c^i[B_c] \longrightarrow \hat{A}_c^i[Z_c]$. The first substitution  accounts for all the tree-level connections one needs to make between the external legs of the loop term and the $(+ + -)$ triple gluon vertex whereas the second accounts for all such tree connections using both the triple gluon vertices $(+ + -)$ and $(+ - -)$. However, in all three cases, the interaction vertices participating in the loop formation are still the Yang-Mills vertices.}
    \label{fig:loopVer}
\end{figure}

The main idea behind deriving the MHV action Eq.~\eqref{eq:MHV_action} and the Z-field action Eq.~\eqref{eq:Z_action1} was to replace the triple and four-point interaction vertices of the Yang-Mills action Eq.~\eqref{eq:YM_LC_action} with bigger interaction vertices so as to enable efficient computation of the pure gluonic amplitudes. However, if we look at the general formulas for one-loop effective action for the three theories
\begin{equation}
   \Gamma_{\mathrm{YM}}[A_c]= S_{\mathrm{YM}}[A_c]  +i\, \frac{1}{2} \Tr\ln \left( 
    \frac{\delta^2 S_{\mathrm{YM}}[A_c ]}
    {\delta \hat{A}^i(x)\delta \hat{A}^j(y)}
    \right)\,,
    \label{eq:G_YM1}
\end{equation}
\begin{equation}
   \Gamma_{\mathrm{MHV}}[B_c]= S_{\mathrm{MHV}}[B_c]  +i\, \frac{1}{2} \Tr\ln \left( 
    \frac{\delta^2 S_{\mathrm{YM}}[A_c [B_c]]}
    {\delta \hat{A}^i(x)\delta \hat{A}^j(y)}
    \right)\,,
    \label{eq:G_MHV1}
\end{equation}
\begin{equation}
   \Gamma[Z_c]= S[Z_c]  +i\, \frac{1}{2} \Tr\ln \left( 
    \frac{\delta^2 S_{\mathrm{YM}}[A_c [Z_c]]}
    {\delta \hat{A}^i(x)\delta \hat{A}^j(y)}
    \right)\,,
    \label{eq:G_Zac1}
\end{equation}
in going from $S_{\mathrm{YM}}[A_c] \longrightarrow S_{\mathrm{MHV}}[B_c] \longrightarrow S[Z_c]$, at each step the interaction vertices are getting bigger and bigger. But this does not happen for the loop terms. In going from Eq.~\eqref{eq:G_YM1} to Eq.~\eqref{eq:G_Zac1}, only the classical field outside the loop in the log term undergoes a field substitution $\hat{A}_c^i \longrightarrow \hat{A}_c^i[B_c] \longrightarrow \hat{A}_c^i[Z_c]$. We represent this diagrammatically for a sample one-loop term in Figure \ref{fig:loopVer}. The first substitution  accounts for all the tree level connections one needs to make between the external legs of the loop term and the $(+ + -)$ triple gluon vertex whereas the second accounts for all such tree connections using both the triple gluon vertices $(+ + -)$ and $(+ - -)$. Notice, however, in all three cases the interaction vertices participating in the loop formation are still the Yang-Mills triple and four-point interaction vertices. Thus, if we could make the MHV vertices explicit in the log term of Eq.~\eqref{eq:G_MHV1} and the Z-field vertices in the log term of Eq.~\eqref{eq:G_Zac1}, given that these are much bigger building blocks as compared to the Yang-Mills vertices, the number of contributing diagrams to any given one-loop amplitudes could be further decreased. We demonstrate this fact below.

In the previous chapter, we showed that when all the one-loop contributions originating from the one-loop effective MHV action Eq.~\eqref{eq:G_MHV1} to the $(+ -)$ self-energy are  un-traced (cut-open), we get (diagrammatically)
\begin{center}
    \includegraphics[width=15cm]{Chapter_4/2P_EAT.eps}
\end{center}
Furthermore, the terms a, b, c, and h add up to give an un-traced 4-point MHV vertex shown below
\begin{center}
    \includegraphics[width=15cm]{appendix/mhv_4prt.eps}
\end{center}
A similar exercise can be done using the one-loop effective Z-field action Eq.~\eqref{eq:G_Zac1} instead. The only difference would be that the MHV vertex connected via a propagator to the un-traced $(+ + -)$ tadpole in the term b above will be replaced by $\overline{\Psi}_2$ (originating from second order expansion of ${\widetilde B}_{c}^{\bullet}[{Z}_{c}^{\bullet}, {Z}_{c}^{\star}]$ Eq.~\eqref{eq:B_star_Z}) connected to the un-traced $(+ + -)$ tadpole. The remaining terms would be exactly the same and the calculation would proceed in exactly the same fashion as shown in Appendix \ref{sec:app_A7}. This example demonstrates that, indeed, when all the contributions to a given one-loop case are put together, some of these can be combined to make the bigger building blocks (both in the MHV and Z-field theory) explicit in the loop. By doing this, we see that the number of diagrams indeed reduces. 

The above procedure of combining a sub-set of one-loop contributions to get the MHV vertices or the Z-field vertices explicit in the loop is simple only in the case of a lower point one-loop amplitude. In the following section, we show how to make MHV and Z-field vertices explicit in the loop. This will, in turn, allow for even more efficient computation of one-loop amplitudes.

\section{Revisiting the one-loop effective action}
\label{sec:OLEA_rev}

In this section, our target is to make the new interaction vertices, MHV vertices in the case of one-loop effective MHV action and Z-field vertices in the case of one-loop effective Z-field action, explicit in the log term of their corresponding one-loop effective action. To achieve this, we will re-derive the actions via a different approach. Recall, previously we started with the Yang-Mills generating functional Eq.~\eqref{eq:gen_YM0} and derived the one-loop effective Yang-Mills generating functional Eq.~\eqref{eq:Partition_YM}. Then we performed the transformations deriving the MHV or the Z-field one-loop effective partition functions Eqs.~\eqref{eq:PartitionMHV} and \eqref{eq:PartitionZOL} respectively. Finally, using the latter we obtained the corresponding one-loop effective actions. Now, we will change the order of operations. We start with the Yang-Mills generating functional Eq.~\eqref{eq:gen_YM0}. Then we perform the canonical transformations, crucially applying them to the source term as well. The Yang-Mills action transforms into the MHV or the Z-field action and the source term become non-linear in new fields (B-fields or the Z-fields). Using this newly derived generating functional (MHV or Z-field), we derive the MHV or the Z-field one-loop effective partition functions respectively, and then proceed to obtain the corresponding one-loop effective actions. In this approach, the source terms play a crucial role because owing to the non-linearity (in new fields) of the source term, it contributes to the log term. As we demonstrate below, the above outlined procedure does indeed make the new interaction vertices explicit in the log term. We start first with the one-loop effective MHV action and then, following the same steps, derive  the one-loop effective Z-field action.

\subsection{One-loop effective MHV action}
\label{subsec:OLEA_MHVrev}

Following the procedure stated above, the starting point is the Yang-Mills generating functional
\begin{equation}
    Z[J]=\int[dA]\, e^{i\left(S_{\mathrm{YM}}[A] + \int\!d^4x\, \Tr \hat{J}_j(x) \hat{A}^j(x)\right) } \,.
    \label{eq:gen_YM4olea}
\end{equation}
Next, we perform Mansfield's transformation Eq.~\eqref{eq:Man_Transf1}. By doing this, we get
\begin{equation}
    Z_{\mathrm{MHV}}[J]=\int[dB]\, e^{i\left(S_{\mathrm{MHV}}[B] + \int\!d^4x\, \Tr \hat{J}_j(x) \hat{A}^j[B](x)\right) } \,.
    \label{eq:MHV_genr}
\end{equation}
Notice, the current term was also transformed. Above, we suppressed the overall normalizations, and since the Jacobian for Mansfield's transformation Eq.~\eqref{eq:MT_jac} is field independent, we absorbed it in the normalization. At this point, one may wonder if what we are doing is contrary to the claims made in the previous chapter (\emph{cf.} the discussion in the second last paragraph in Section \ref{sec:MHV_OLEA}) where we argued that if one derives the one-loop effective MHV action starting with the MHV action itself, the loop contributions originating either solely from the Self-Dual vertex $(+ + -)$ or the mixing of this vertex with the MHV vertices will be missing in the log term. What we meant was, if we follow the standard procedure for defining a generating functional using the MHV action coupled with a "new" auxiliary source (say $J'$), as shown below
\begin{equation}
    Z_{\mathrm{MHV}}[J']=\int[dB]\, e^{i\left(S_{\mathrm{MHV}}[B] + \int\!d^4x\, \Tr \hat{J'}_j(x) \hat{B}^j(x)\right) } \,,
    \label{eq:MHV_genrrad}
\end{equation}
and then derive the one-loop effective MHV action, the log term will simply have the second order functional derivative of the MHV vertices (the source term being linear in B-fields would not contribute). The loop contributions originating from these will consist of only the MHV vertices and will, therefore,  only give the cut-constructible parts of any one-loop amplitude in 4D (and not the full one-loop amplitude). Notice, however, in going from Eq.~\eqref{eq:gen_YM4olea} to Eq.~\eqref{eq:MHV_genr} the source is unaltered. The field redefinition replaces $\hat{A}^i \longrightarrow \hat{A}^i[B]$ (where $i= \bullet, \star$) also in the source term 
\begin{equation}
    \int\!d^4x\, \Tr \hat{J}_j(x) \hat{A}^j(x) \longrightarrow \int\!d^4x\, \Tr \hat{J}_j(x) \hat{A}^j[B](x)\,,
\end{equation}
due to which the source term becomes non-linear in the fields. Now, when we derive the one-loop effective action, the source term will also have a non-vanishing contribution to the log term. Already at the level of the generating functional, a straightforward way of seeing this is to consider the evaluation of the $n$-point correlator using Eq.~\eqref{eq:MHV_genr} and Eq.~\eqref{eq:MHV_genrrad}. For the former we get
\begin{equation}
   \langle0|\mathcal{T} \{{A}^{j_1\, a_1}[B](x_1) {A}^{j_2\, a_2}[B](x_2) \dots {A}^{j_n\, a_n}[B](x_n) e^{i\,S_{\mathrm{MHV}}[B]}\} |0\rangle\,,
   \label{eq:nCORR_R}
\end{equation}
whereas for the latter we get
\begin{equation}
   \langle0|\mathcal{T} \{{B}^{j_1\, a_1}(x_1) {B}^{j_2\, a_2}(x_2) \dots {B}^{j_n\, a_n}(x_n) e^{i\,S_{\mathrm{MHV}}[B]}\} |0\rangle\,,
   \label{eq:nCORR_wr}
\end{equation}
where $a_i$ is the color index and $j_i= \bullet, \star$. In the case of Eq.~\eqref{eq:nCORR_wr}, the only interaction vertices are the MHV vertices (in the action in the exponent). Therefore, at the loop level, upon contracting these vertices with the external fields ${B}^{j_i\, a_i}(x_i)$ (via Wick contraction) we get loop diagrams where each interaction vertex is just the MHV vertex. Thus when computing amplitudes via amputating the connected contributions, we will get only the cut-constructible contributions in 4D. Whereas in the case of Eq.~\eqref{eq:nCORR_R}, the ${A}^{j_i\, a_i}[B](x_i)$ is first expanded in terms of ${B}^{j_i\, a_i}(x_i)$ fields which are then Wick contracted with the MHV vertices originating from the action in the exponent to  get loop diagrams. These diagrams therefore consist of the MHV vertices as well as the kernels ${\Psi}_{n}$ or ${\Omega}_{n}$, (\emph{cf.} Eqs.~\eqref{eq:psi_kernel}, \eqref{eq:omega_kernel}). These kernels account for the all tree level connections one can make using the $(+ + -)$ Self-Dual vertex. Thus, the loop diagrams involve both the $(+ + -)$ Self-Dual vertex as well as the MHV vertices. Therefore, we see that when computing the loop amplitudes via amputating the connected Green's functions evaluated using Eq.~\eqref{eq:nCORR_R}, one should get both the MHV as well as the non-MHV loop contributions.

In order to see this explicitly for the one-loop effective action, let us expand the terms in the exponent of Eq.~\eqref{eq:MHV_genr} around the classical $\hat{B}_c^i$ fields
\begin{multline}
    S_{\mathrm{MHV}}[B] + \int\!d^4x\, \Tr \hat{J}_i(x) \hat{A}^i[B](x)  
    = S_{\mathrm{MHV}}[B_c] + \int\!d^4x\, \Tr \hat{J}_i(x)\hat{A}^i[B_c](x) \\ + \int\!d^4x\,\Tr\left(\hat{B}^i(x)-\hat{B}_c^i(x)\right)
    \left(\frac{\delta S_{\mathrm{MHV}}[B_c]}{\delta \hat{B}^i(x)}+\int\!d^4y\,{\hat J}_k(y)\frac{\delta \hat{A}^k[B_c](y)}{\delta \hat{B}^i(x)}\right) \\
    +\frac{1}{2}\int\!d^4xd^4y\,\Tr\left(\hat{B}^i(x)-\hat{B}_c^i(x)\right)\left(\frac{\delta^2 S_{\mathrm{MHV}}[B_c]}{\delta\hat{B}^i(x)\delta\hat{B}^j(y)} \right. \\
    \left. +\int\!d^4z\,{\hat J}_k(z)\frac{\delta^2 \hat{A}^k[B_c](z)}{\delta \hat{B}^i(x)\delta\hat{B}^j(y)}\right)\left(\hat{B}^j(y)-\hat{B}_c^j(y)\right) \,.
    \label{eq:Der_olea_mhv}
\end{multline}
As before, we truncate the expansion to the second order. The classical $\hat{B}_c^i$ fields are defined as the solutions of the following classical EOM 
\begin{equation}
    \left.\frac{\delta S_{\mathrm{MHV}}[B^{\bullet}, B^{\star}]}{\delta \hat{B}^{\bullet}(x)}+\int\!d^4y\,\left({\hat J}_{\bullet}(y)\frac{\delta \hat{A}^{\bullet}[B^{\bullet}](y)}{\delta \hat{B}^{\bullet}(x)}+{\hat J}_{\star}(y)\frac{\delta \hat{A}^{\star}[B^{\bullet}, B^{\star}](y)}{\delta \hat{B}^{\bullet}(x)}\right)\right|_{{\hat B}={\hat B}_{c}} =0\,.
    \label{eq:B_bul_EOM}
\end{equation}
\begin{equation}
    \left. \frac{\delta S_{\mathrm{MHV}}[B^{\bullet}, B^{\star}]}{\delta \hat{B}^{\star}(x)}+\int\!d^4y\,{\hat J}_{\star}(y)\frac{\delta \hat{A}^{\star}[B^{\bullet}, B^{\star}](y)}{\delta \hat{B}^{\star}(x)}\right|_{{\hat B}={\hat B}_{c}} =0\,.
    \label{eq:B_star_EOM}
\end{equation}
We shall dub the above as the "MHV classical EOM". The use of $\hat{B}_c^i$ for the solution of the MHV classical EOM may appear a misuse of notation. This is because we used exactly the same notation in the previous approach (used in the previous chapter). Recall, what we termed $\hat{B}_c^i$ in the previous approach were the fields obtained via Mansfield's  transformation of the classical $\hat{A}_c^i$ fields. The latter were the solution of the Yang-Mills classical EOM Eq.~\eqref{eq:YMEOM0}. Here, what we call the $\hat{B}_c^i$ fields are solutions of the MHV classical EOM. These could, in principle, be different objects. However, it turns out they are not. In Appendix \ref{sec:App91} we show that these are exactly the same fields. In fact, we also demonstrate that Mansfield's transformation applied to the Yang-Mills classical EOM Eq.~\eqref{eq:YMEOM0} gives the MHV classical EOM Eq.~\eqref{eq:B_bul_EOM}-\eqref{eq:B_star_EOM}. This also provides us a consistency check for defining the classical $\hat{B}_c^i$ fields via the MHV classical EOM Eq.~\eqref{eq:B_bul_EOM}-\eqref{eq:B_star_EOM}.

Owing to Eq.~\eqref{eq:B_bul_EOM}-\eqref{eq:B_star_EOM}, the linear terms in Eq.\eqref{eq:Der_olea_mhv} vanishes. The first two terms factor out of the path integral. This leaves us with the Gaussian integral which can be performed following the procedure in Section  \ref{sec:Yang-Mills_OLEA}. By doing this, we get
\begin{multline}
    Z_{\mathrm{MHV}}[J] \approx  \left[ \det \left(\frac{\delta^2 S_{\mathrm{MHV}}[B_c]}{\delta\hat{B}^i(x)\delta\hat{B}^k(y)} +\int\!d^4z\,{\hat J}_l(z)\frac{\delta^2 \hat{A}^l[B_c](z)}{\delta \hat{B}^i(x)\delta\hat{B}^k(y)}\right)\right]^{-\frac{1}{2}} \\
   \exp\left\{i\left(S_{\mathrm{MHV}}[B_c] 
    + \int\!d^4x\, \Tr \hat{J}_l(x)\hat{A}^l[B_c](x) \right) \right\} \,.
    \label{eq:det_MHV}
\end{multline}
The matrix in the determinant above reads (using the collective indices $I=\left\{x, a\right\}$, $K=\left\{y, b\right\}$ where $x$, $y$ are position and $a$, $b$ are the color degrees of freedom) 
\begin{equation}
 \mathrm{M}^{\mathrm{MHV}}_{IK}  =  \left(\begin{matrix}
     \frac{\delta^2 S_{\mathrm{MHV}}[B_c]}{\delta B^{\bullet I}\delta B^{\star K}}+ J_{\star L}\frac{\delta^2 A^{\star L}[B_c]}{\delta B^{\bullet I}\delta B^{\star K}} 
     \quad &\frac{\delta^2 S_{\mathrm{MHV}}[B_c]}{\delta B^{\bullet I}\delta B^{\bullet K}}+ J_{\star L}\frac{\delta^2 A^{\star L}[B_c]}{\delta B^{\bullet I}\delta B^{\bullet K}} +  J_{\bullet L}\frac{\delta^2 A^{\bullet L}[B_c]}{\delta B^{\bullet I}\delta B^{\bullet K}}\\ \\
\frac{\delta^2 S_{\mathrm{MHV}}[B_c]}{\delta B^{\star I}\delta B^{\star K}}
     \quad &\frac{\delta^2 S_{\mathrm{MHV}}[B_c]}{\delta B^{\star I}\delta B^{\bullet K}}+ J_{\star L}\frac{\delta^2 A^{\star L}[B_c]}{\delta B^{\star I}\delta B^{\bullet K}} 
    \label{eq:M_MHV_J}
\end{matrix}\right) \,.
\end{equation}
Notice, the matrix consists of non-commuting blocks. Thus the usual formulae for the $2\times2$ determinant do not apply. For the sake of simplicity let us rewrite the matrix in a very general form consisting of four blocks as shown below
\begin{equation}
 \mathrm{M}^{\mathrm{MHV}}_{IK}  = \left(\begin{matrix}     \left(\mathrm{M}^{\mathrm{MHV}}_{IK}\right)_{11} &\left(\mathrm{M}^{\mathrm{MHV}}_{IK}\right)_{12}\\ \\
\left(\mathrm{M}^{\mathrm{MHV}}_{IK}\right)_{21} &\left(\mathrm{M}^{\mathrm{MHV}}_{IK}\right)_{22} \end{matrix}\right)\,.
\end{equation}
Following the derivation in Section  \ref{sec:Yang-Mills_OLEA}, the determinant reads (\emph{cf.} Eq.~\eqref{eq:det_YM_full})
\begin{equation}
    \det \left[\mathrm{M}^{\mathrm{MHV}}_{IK} \right]  = \det \left[ \left(\mathrm{M}^{\mathrm{MHV}}_{IK}\right)_{11}\left(\mathrm{M}^{\mathrm{MHV}}_{IK}\right)_{22} - \left(\mathrm{M}^{\mathrm{MHV}}_{IK}\right)_{11}\left(\mathrm{M}^{\mathrm{MHV}}_{IK}\right)_{21}\left(\mathrm{M}^{\mathrm{MHV}}_{IK}\right)_{11}^{-1}\left(\mathrm{M}^{\mathrm{MHV}}_{IK}\right)_{12}\right]\,.
    \label{eq:OLgnMHV}
\end{equation}
Using the above, the MHV one-loop partition function Eq.~\eqref{eq:det_MHV} reads
\begin{equation}
    Z_{\mathrm{MHV}}[J] \approx  \left[ \det \left(\mathrm{M}^{\mathrm{MHV}}_{IK} \right)\right]^{-\frac{1}{2}}
   \exp\left\{i\left(S_{\mathrm{MHV}}[B_c] 
    + \int\!d^4x\, \Tr \hat{J}_l(x)\hat{A}^l[B_c](x) \right) \right\} \,.
    \label{eq:det_MHVag}
\end{equation}
From this point, deriving the one-loop effective MHV action may seem straightforward but there is a subtle point. Instead of the usual linear Legendre transform, we have to apply
 \begin{equation}
   \Gamma_{\mathrm{MHV}}[B_c] = W_{\mathrm{MHV}}[J] - \int\!d^4x\, \Tr \hat{J}_i(x) \hat{A}^i[B_c](x) \,, 
   \label{eq:LT_mhv1}
\end{equation}
where $W_{\mathrm{MHV}}[J] = -i \ln \left[ Z_{\mathrm{MHV}}[J]\right]$, we get
\begin{equation}
    \Gamma_{\mathrm{MHV}}[B_c] = S_{\mathrm{MHV}}[B_c] 
      +i\, \frac{1}{2} \Tr\ln \left( 
     \mathrm{M}^{\mathrm{MHV}}_{IK}
     \right)\,.
     \label{eq:G_MHVnew}
\end{equation}

Notice, however, that the matrix $\mathrm{M}^{\mathrm{MHV}}_{IK}$ in the expression above has source dependent terms as evident from Eq.~\eqref{eq:M_MHV_J}. For the one-loop generating functional Eq.~\eqref{eq:det_MHVag}, which is a functional of the source $J$, the form of the matrix $\mathrm{M}^{\mathrm{MHV}}_{IK}$ in Eq.~\eqref{eq:M_MHV_J}  requires no changes. But the one-loop effective action Eq.~\eqref{eq:G_MHVnew} is a functional exclusively of the classical fields. Thus we need to exchange the source in terms of fields. This can be done using the following identities derived in Appendix \ref{sec:App91}
\begin{equation}
  \frac{\delta S_{\mathrm{YM}}[A[B_c]]}{\delta A^{\star L}}=-J_{\star L}\,, \quad\quad
    \frac{\delta S_{\mathrm{YM}}[A[B_c]]}{\delta A^{\bullet L}}=- J_{\bullet L} \,.
    \label{eq:J_curr}
\end{equation}
Substituting the above in Eq.~\eqref{eq:M_MHV_J} we get
\begin{multline}
\mathrm{M}^{\mathrm{MHV}}_{IK}  =  \left(\begin{matrix}
     \frac{\delta^2 S_{\mathrm{MHV}}[B_c]}{\delta B^{\bullet I}\delta B^{\star K}} &\frac{\delta^2 S_{\mathrm{MHV}}[B_c]}{\delta B^{\bullet I}\delta B^{\bullet K}}\\ \\
\frac{\delta^2 S_{\mathrm{MHV}}[B_c]}{\delta B^{\star I}\delta B^{\star K}}
      &\frac{\delta^2 S_{\mathrm{MHV}}[B_c]}{\delta B^{\star I}\delta B^{\bullet K}} 
\end{matrix}\right) \\
+ \left(\begin{matrix}
    - \frac{\delta S_{\mathrm{YM}}[A[B_c]]}{\delta A^{\star L}}\frac{\delta^2 A^{\star L}[B_c]}{\delta B^{\bullet I}\delta B^{\star K}} 
     &- \frac{\delta S_{\mathrm{YM}}[A[B_c]]}{\delta A^{\star L}}\frac{\delta^2 A^{\star L}[B_c]}{\delta B^{\bullet I}\delta B^{\bullet K}} -  \frac{\delta S_{\mathrm{YM}}[A[B_c]]}{\delta A^{\bullet L}}\frac{\delta^2 A^{\bullet L}[B_c]}{\delta B^{\bullet I}\delta B^{\bullet K}}\\ \\
\mathbb{0}
      &- \frac{\delta S_{\mathrm{YM}}[A[B_c]]}{\delta A^{\star L}}\frac{\delta^2 A^{\star L}[B_c]}{\delta B^{\star I}\delta B^{\bullet K}} 
    \label{eq:M_MHV}
\end{matrix}\right)\,.
\end{multline}
The above result has quite a few interesting aspects. The first matrix on the R.H.S. above consists of exclusively the MHV vertices. In fact, this is the only matrix one will get (the second matrix consisting of the source terms vanishes) if one derives the one-loop effective MHV action starting with the naive generating functional in Eq.~\eqref{eq:MHV_genrrad}. Therefore we can conclude that the one-loop contributions originating either solely from the determinant of the first matrix in Eq.~\eqref{eq:M_MHV} or via tree connections between the one-loop terms originating from the determinant of this matrix with the interaction vertices in the classical MHV action $S_{\mathrm{MHV}}[B_c]$ will consist of MHV vertices only and will thus give only the cut-constructible parts of the one-loop amplitudes in 4D. This implies the remaining one-loop contribution to the one-loop amplitudes should originate from the second matrix in Eq.~\eqref{eq:M_MHV}. To see this lets us consider just the second matrix
\begin{equation}
\left(\begin{matrix}
    - \frac{\delta S_{\mathrm{YM}}[A[B_c]]}{\delta A^{\star L}}\frac{\delta^2 A^{\star L}[B_c]}{\delta B^{\bullet I}\delta B^{\star K}} 
     &- \frac{\delta S_{\mathrm{YM}}[A[B_c]]}{\delta A^{\star L}}\frac{\delta^2 A^{\star L}[B_c]}{\delta B^{\bullet I}\delta B^{\bullet K}} -  \frac{\delta S_{\mathrm{YM}}[A[B_c]]}{\delta A^{\bullet L}}\frac{\delta^2 A^{\bullet L}[B_c]}{\delta B^{\bullet I}\delta B^{\bullet K}}\\ \\
\mathbb{0}
      &- \frac{\delta S_{\mathrm{YM}}[A[B_c]]}{\delta A^{\star L}}\frac{\delta^2 A^{\star L}[B_c]}{\delta B^{\star I}\delta B^{\bullet K}} 
    \label{eq:M_MHV2nd}
\end{matrix}\right)\,,
\end{equation}
where
\begin{multline}
   \frac{\delta S_{\mathrm{YM}}[A[B_c]]}
    {\delta A^{\star L}} 
    = -\square_{LJ}A^{\bullet J}[B_c]-\left(V_{-++}\right)_{LJK}A^{\bullet J}[B_c] A^{\bullet K}[B_c] -\left(V_{--+}\right)_{KLJ}A^{\star K}[B_c] A^{\bullet J}[B_c] \\
    -\left(V_{--++}\right)_{ILJK}A^{\star I}[B_c] A^{\bullet J}[B_c] A^{\bullet K}[B_c]
    \label{eq:S+-mhv}
    \,,
\end{multline}
\begin{multline}
   \frac{\delta S_{\mathrm{YM}}[A[B_c]]}
    {\delta A^{\bullet L}} 
    = -\square_{LJ}A^{\star J}[B_c]-\left(V_{-++}\right)_{JLK}A^{\star J}[B_c] A^{\bullet K}[B_c] -\left(V_{--+}\right)_{KJL}A^{\star K}[B_c] A^{\star J}[B_c] \\
    -\left(V_{--++}\right)_{IJLK}A^{\star I}[B_c] A^{\star J}[B_c] A^{\bullet K}[B_c]
    \label{eq:S+-mhv2}
    \,,
\end{multline}
\begin{equation}
\frac{\delta^2 A^{\bullet L}[B_c]}{\delta B^{\bullet I}\delta B^{\bullet K}} = 
\sum_{n=2}^{\infty} 
     \, \widetilde{\Psi}_n^{L\{I K J_1\dots J_n\}} {B}_c^{\bullet J_1} \dots {B}_c^{\bullet J_n}\,,
    \label{eq:A_bull_soluCL}
\end{equation}
\begin{equation}
\frac{\delta^2 A^{\star L}[B_c]}{\delta B^{\star I}\delta B^{\bullet K}} =  \sum_{n=2}^{\infty} 
    \, {\widetilde \Omega}_{n}^{L I \left \{ K J_1 \dots J_n \right \} } {B}_c^{\bullet J_1}\dots {B}_c^{\bullet J_n}\, ,
    \label{eq:A_star_soluCL}
\end{equation}
\begin{equation}
\frac{\delta^2 A^{\star L}[B_c]}{\delta B^{\bullet I}\delta B^{\bullet K}} =  \sum_{n=2}^{\infty} 
    \, {\widetilde \Omega}_{n}^{L J_1 \left \{I K J_3 \dots J_n \right \} } {B}_c^{\star J_1}{B}_c^{\bullet J_3}\dots {B}_c^{\bullet J_n}\, .
    \label{eq:A_star_soluCL2}
\end{equation}
Above, the last three relations were obtained via the functional derivative of the solution of Mansfield's transformation Eq.~\eqref{eq:A_bull_solu}-\eqref{eq:A_star_solu}. A word of caution. Recall, Mansfield's transformation and its solution were all defined over the constant light-cone time $x^+$. The above functional derivatives are over full 4D.  However, since the kernels of the solution of Mansfield's transformation are independent of the light-cone time. In going from former to latter (3D to 4D) we only get  an additional delta for the light-cone time. 

For the sake of simplicity, let us first consider the substitution of only the lowest power of B-fields in the above expressions Eq.~\eqref{eq:S+-mhv}-\eqref{eq:A_star_soluCL2} to the matrix Eq.~\eqref{eq:M_MHV2nd} (we discuss the consequences of the substitution of higher power terms below). By doing that we get 
\begin{equation}
\left(\begin{matrix}
    \square_{LJ}{\widetilde \Omega}_{2}^{L K \left \{ I \right \}} B_c^{\bullet J} \quad
     &\square_{LJ} {\widetilde \Omega}_{2}^{L J_1 \left \{I K \right \} }B_c^{\bullet J}{B}_c^{\star J_1} +  \square_{LJ}\widetilde{\Psi}_2^{L\{I K \}}B_c^{\star J}\\ \\
\mathbb{0} \quad
      &\square_{LJ}{\widetilde \Omega}_{2}^{L I \left \{ K \right \} }B_c^{\bullet J}
    \label{eq:M_MHV3rd}
\end{matrix}\right)\,,
\end{equation}
where the inverse propagator $\square_{LJ}$ originates from considering the lowest power term in Eq.~\eqref{eq:S+-mhv}-\eqref{eq:S+-mhv2}. Let us  focus only on the determinant of the matrix above. In that case, the off-diagonal term can be put to zero. This will further simplify our discussion. With this we have
\begin{equation}
 \det \begin{vmatrix}
    \square_{LJ}{\widetilde \Omega}_{2}^{L K \left \{ I \right \}} B_c^{\bullet J} \quad
     &\mathbb{0}\\ \\
\mathbb{0} \quad
      &\square_{LJ}{\widetilde \Omega}_{2}^{L I \left \{ K \right \} }B_c^{\bullet J}
    \label{eq:M_MHV4th}
\end{vmatrix}\,.
\end{equation}
In momentum space, substituting the expression for ${\widetilde \Omega}_{2}$ using Eq.~\eqref{eq:omega_kernel}, with a bit of algebra one can show (\emph{cf.} Eq.~\eqref{eq:2gam})
\begin{equation}
    P_L^2 {\widetilde \Omega}_{2}^{L K \left \{ I \right \}} = \left[\left(\frac{p_K^+}{p_L^+}\right)^2 + \left(\frac{p_I^+}{p_L^+}\right)^2\right]\left(V_{-++}\right)_{KIL}\,,
\end{equation}
where $p_L^+$ is the plus component of the momentum associated with the collective index $L$. Substituting the above in Eq.~\eqref{eq:M_MHV4th}, we get
\begin{equation}
 \det \begin{vmatrix}
    \left[\left(\frac{p_K^+}{p_L^+}\right)^2 + \left(\frac{p_I^+}{p_L^+}\right)^2\right]\left(V_{-++}\right)_{KIL} B_c^{\bullet L} \quad
     &\mathbb{0}\\ \\
\mathbb{0} \quad
      &\left[\left(\frac{p_K^+}{p_L^+}\right)^2 + \left(\frac{p_I^+}{p_L^+}\right)^2\right]\left(V_{-++}\right)_{IKL}B_c^{\bullet L}
    \label{eq:M_MHV5th}
\end{vmatrix}\,.
\end{equation}
Notice, the vertices in the matrix above are exactly the Self-Dual vertex $(+ + -)$ and the determinant of the above matrix will give the one-loop terms consisting solely of the Self-Dual vertex. Let us consider now the terms with a higher power in fields step by step. Instead of ${\widetilde \Omega}_{2}$, if we had ${\widetilde \Omega}_{n}$ the contributions would still involve  purely the Self-Dual vertex because these kernels account for only chains  (tree level connections) of $(+ + -)$. Thus, we see that the one-loop terms involving only the Self-Dual vertex are present in the second matrix in Eq.~\eqref{eq:M_MHV}. Now, let us consider the higher power  terms (in fields) in Eq.~\eqref{eq:S+-mhv}-\eqref{eq:S+-mhv2}. These include both the  triple gluon vertices and the four gluon vertex. Combining these with the ${\widetilde \Omega}_{n}$ kernels we get some of the mixed contributions of the Self-dual vertex $(+ + -)$ with the 3-point MHV $(+ - -)$. In order to exhaust the set of all mixed contributions we need to consider both the matrices in Eq.~\eqref{eq:M_MHV}. Note, as long as we consider the determinant of just the second matrix in Eq.~\eqref{eq:M_MHV}, the off-diagonal term is useless. It contributes only when the first matrix (consisting of purely the MHV vertices) is also considered. The main point to conclude from the above discussion is that the determinant of the matrix Eq.~\eqref{eq:M_MHV} does account for all the three types of contributions shown below
\begin{center}
    \includegraphics[width=10cm]{Chapter_4/OnleLoopEffAction_MHV_generic_1.eps}
\end{center}

The above discussion only ascertains the fact that there are both the MHV as well non-MHV one-loop contributions in the one-loop effective MHV action Eq.~\eqref{eq:G_MHVnew} derived in the new approach. 

This, however, does not guarantee the quantum completeness (up to one loop) of the action. In the previous chapter, we explicitly computed one-loop amplitudes to demonstrate quantum completeness of the effective action obtained there. Below, we show that the above effective action Eq.~\eqref{eq:G_MHVnew} is equal, up to a field independent factor, to the former.

The one-loop effective action derived in the previous chapter reads
\begin{equation}
   \Gamma_{\mathrm{MHV}}[B_c] = S_{\mathrm{MHV}}[B_c] 
     +i\, \frac{1}{2} \Tr\ln \left( 
    \mathrm{N}^{\mathrm{MHV}}_{IK}
    \right)\,,
    \label{eq:G_MHVold}
\end{equation}
where the matrix $\mathrm{N}^{\mathrm{MHV}}_{IK}$ had the following form
\begin{equation}
  \mathrm{N}^{\mathrm{MHV}}_{IK} =   \left(\begin{matrix}
     \frac{\delta^2 S_{\mathrm{YM}}[A[B_c]]}{\delta A^{\bullet I}\delta A^{\star K}} &\frac{\delta^2 S_{\mathrm{YM}}[A[B_c]]}{\delta A^{\bullet I}\delta A^{\bullet K}}\\ \\
\frac{\delta^2 S_{\mathrm{YM}}[A[B_c]]}{\delta A^{\star I}\delta A^{\star K}}
      &\frac{\delta^2 S_{\mathrm{YM}}[A[B_c]]}{\delta A^{\star I}\delta A^{\bullet K}} 
\end{matrix}\right)\,.
\label{eq:detMHVold}
\end{equation}
By computing one-loop amplitudes, we showed that the above action is indeed one-loop complete. The above action differs from the one we derived via the new approach Eq.~\eqref{eq:G_MHVnew} only in the log term. In Appendix \ref{sec:A101}, we demonstrate that the two log terms or equivalently the determinant of the matrices in the two actions are proportional to each other up to a field-independent volume divergent factor which does not contribute to one-loop amplitudes and can therefore be discarded. Precisely, we prove that
\begin{equation}
\det\mathrm{M}^{\mathrm{MHV}}_{IK} = \mathcal{N} \Big[ \mathcal{J}_{\mathrm{MT}}\Big]^2 \times \, \det\begin{vmatrix}
     \frac{\delta^2 S_{\mathrm{YM}}[A[B_c]]}{\delta A^{\bullet Q}\delta A^{\star P}} &\frac{\delta^2 S_{\mathrm{YM}}[A[B_c]]}{\delta A^{\bullet Q}\delta A^{\bullet P}}\\ \\
\frac{\delta^2 S_{\mathrm{YM}}[A[B_c]]}{\delta A^{\star Q}\delta A^{\star P}}
      &\frac{\delta^2 S_{\mathrm{YM}}[A[B_c]]}{\delta A^{\star Q}\delta A^{\bullet P}} 
\end{vmatrix}\,,
\label{eq:M_mhv12}
\end{equation}
where $\mathcal{N} = \Big[\det \delta(Q^+-I^+)\det \delta(P^+-K^+)\Big] $ and $I^+$ represents the plus component of the position 4-vector $y^+$ associated with the collective index $I=\left\{\left(y^{+},y^{-},y^{\bullet},y^{\star}\right); a\right\}$. This factor is field-independent as well as volume divergent. $\mathcal{J}_{\mathrm{MT}}$ is the Jacobian of Mansfield's transformation Eq.~\eqref{eq:MT_jac} which is also field-independent. On the L.H.S. of Eq.~\eqref{eq:M_mhv12}, we have the determinant of the matrix that appears in the one-loop effective MHV action derived via the new approach Eq.~\eqref{eq:G_MHVnew}, whereas on the R.H.S. we have the determinant of the matrix Eq.~\eqref{eq:detMHVold} that appears in the one-loop effective MHV action derived in the previous chapter Eq.~\eqref{eq:G_MHVold}. These are related via the product of $\mathcal{N}$  and the Jacobian $\mathcal{J}_{\mathrm{MT}}$ both of which do not contribute to amplitudes and therefore can be factored out of the partition function and absorbed in the overall normalization. Thus, we can conclude that the one-loop amplitudes computed using both of these actions must be identical. Furthermore, since the former was shown to be one-loop complete, the latter must be one-loop complete as well.
\subsection{One-loop effective Z-field action}
\label{subsec:OLEA_ZACrev}

The one-loop effective MHV action Eq.~\eqref{eq:G_MHVnew} derived in the previous Subsection does fulfill the goals we had in mind. Firstly, it is one-loop complete and secondly, it has the bigger interaction vertices -- the MHV vertices -- explicit in the log term. It will, therefore, allow for even more efficient computation of higher multiplicity one-loop amplitudes when compared with the one-loop effective MHV action derived in the previous chapter. In this Subsection, we extend our new approach to derive the one-loop effective Z-field action so that the Z-field interaction vertices are explicit in the loop.

As of now, we have repeated the procedure for deriving a one-loop effective action quite a few times. Therefore, in this Subsection we will be brief and highlight only the important steps in the derivation. In the new approach we must start with the Yang-Mills generating functional and then perform the canonical transformation  Eq.~\eqref{eq:AtoZ_ct1}-\eqref{eq:AtoZ_ct2}. By doing this we get
\begin{equation}
    Z[J]=\int[dA]\, e^{i\left(S_{\mathrm{YM}}[A] + \int\!d^4x\, \Tr \hat{J}_j(x) \hat{A}^j(x)\right) } \longrightarrow \int[dZ]\, e^{i\left(S[Z] + \int\!d^4x\, \Tr \hat{J}_j(x) \hat{A}^j[Z](x)\right) } \,.
   \label{eq:SZ_genr}
\end{equation}
Above, $S[Z]$ is the Z-field action Eq.~\eqref{eq:Z_action1} and $\hat{A}^j[Z](x)$ are the solutions Eq.~\eqref{eq:Abullet_to_Z}-\eqref{eq:Astar_to_Z} of the canonical transformation  Eq.~\eqref{eq:AtoZ_ct1}-\eqref{eq:AtoZ_ct2}. Notice, the source term is also transformed.

Expanding the terms in the exponent around the classical Z-fields $\hat{Z}_c^i$ fields up to second order, we get
\begin{multline}
    S[Z] + \int\!d^4x\, \Tr \hat{J}_i(x) \hat{A}^i[Z](x)  
    = S[Z_c] + \int\!d^4x\, \Tr \hat{J}_i(x)\hat{A}^i[Z_c](x) \\ + \int\!d^4x\,\Tr\left(\hat{Z}^i(x)-\hat{Z}_c^i(x)\right)
    \left(\frac{\delta S[Z_c]}{\delta \hat{Z}^i(x)}+\int\!d^4y\,{\hat J}_k(y)\frac{\delta \hat{A}^k[Z_c](y)}{\delta \hat{Z}^i(x)}\right) \\
    +\frac{1}{2}\int\!d^4xd^4y\,\Tr\left(\hat{Z}^i(x)-\hat{Z}_c^i(x)\right)\left(\frac{\delta^2 S[Z_c]}{\delta\hat{Z}^i(x)\delta\hat{Z}^j(y)}\right. \\
    \left.+\int\!d^4z\,{\hat J}_k(z)\frac{\delta^2 \hat{A}^k[Z_c](z)}{\delta \hat{Z}^i(x)\delta\hat{Z}^j(y)}\right)\left(\hat{Z}^j(y)-\hat{Z}_c^j(y)\right) \,.
    \label{eq:Der_olea_sz}
\end{multline}
We define the classical Z-fields $\hat{Z}_c^i$ as the solution of the following equations
\begin{equation}
    \left.\frac{\delta S[Z_c^{\bullet}, Z_c^{\star}]}{\delta \hat{Z}^{\bullet}(x)}+\int\!d^4y\,\left[{\hat J}_{\bullet}(y)\frac{\delta \hat{A}^{\bullet}[Z_c^{\bullet}, Z_c^{\star}](y)}{\delta \hat{Z}^{\bullet}(x)}+{\hat J}_{\star}(y)\frac{\delta \hat{A}^{\star}[Z_c^{\bullet}, Z_c^{\star}](y)}{\delta \hat{Z}^{\bullet}(x)}\right]\right|_{{\hat Z}={\hat Z}_{c}} =0\,,
    \label{eq:Z_bul_EOM}
\end{equation}
\begin{equation}
   \left.\frac{\delta S[Z_c^{\bullet}, Z_c^{\star}]}{\delta \hat{Z}^{\star}(x)}+\int\!d^4y\,\left[{\hat J}_{\bullet}(y)\frac{\delta \hat{A}^{\bullet}[Z_c^{\bullet}, Z_c^{\star}](y)}{\delta \hat{Z}^{\star}(x)}+{\hat J}_{\star}(y)\frac{\delta \hat{A}^{\star}[Z_c^{\bullet}, Z_c^{\star}](y)}{\delta \hat{Z}^{\star}(x)}\right]\right|_{{\hat Z}={\hat Z}_{c}} =0\,.
    \label{eq:Z_star_EOM}
\end{equation}
We dub these as the "Z-field classical EOM". In Appendix \ref{sec:App92} we check the consistency of these equations and demonstrate that these can be derived from the Yang-Mills classical EOM Eq.~\eqref{eq:YMEOM0} via the transformation  Eq.~\eqref{eq:AtoZ_ct1}-\eqref{eq:AtoZ_ct2}.

Owing to Eq.~\eqref{eq:Z_bul_EOM}-\eqref{eq:Z_star_EOM}, the linear terms in Eq.~\eqref{eq:Der_olea_sz} vanish, the first two terms factor out of path integral. Finally, by performing the Gaussian integral we get
\begin{equation}
   Z[J] \approx  \Big[ \det\,\mathrm{M}^{\mathrm{Z}}_{IK}\Big]^{-\frac{1}{2}}
   \exp\left\{i\left(S[Z_c] 
    + \int\!d^4x\, \Tr \hat{J}_i(x)\hat{A}^i[Z_c](x) \right) \right\} \,.
    \label{eq:det_SZ}
\end{equation}
Where the matrix $\mathrm{M}^{\mathrm{Z}}_{IK}$ reads
 \begin{multline}
\mathrm{M}^{\mathrm{Z}}_{IK}  =  \left(\begin{matrix}
     \frac{\delta^2 S[Z_c]}{\delta Z^{\bullet I}\delta Z^{\star K}} &\frac{\delta^2 S[Z_c]}{\delta Z^{\bullet I}\delta Z^{\bullet K}}\\ \\
\frac{\delta^2 S[Z_c]}{\delta Z^{\star I}\delta Z^{\star K}}
      &\frac{\delta^2 S[Z_c]}{\delta Z^{\star I}\delta Z^{\bullet K}} 
\end{matrix}\right) 
+ \left(\begin{matrix}
    J_{\star L}\frac{\delta^2 A^{\star L}[Z_c]}{\delta Z^{\bullet I}\delta Z^{\star K}} + J_{\bullet L}\frac{\delta^2 A^{\bullet L}[Z_c]}{\delta Z^{\bullet I}\delta Z^{\star K}} 
     & J_{\star L}\frac{\delta^2 A^{\star L}[Z_c]}{\delta Z^{\bullet I}\delta Z^{\bullet K}} +  J_{\bullet L}\frac{\delta^2 A^{\bullet L}[Z_c]}{\delta Z^{\bullet I}\delta Z^{\bullet K}}\\ \\
 J_{\star L}\frac{\delta^2 A^{\star L}[B_c]}{\delta Z^{\star I}\delta Z^{\star K}} + J_{\bullet L}\frac{\delta^2 A^{\bullet L}[B_c]}{\delta Z^{\star I}\delta Z^{\star K}}
      &J_{\star L}\frac{\delta^2 A^{\star L}[B_c]}{\delta Z^{\star I}\delta Z^{\bullet K}} + J_{\bullet L}\frac{\delta^2 A^{\bullet L}[B_c]}{\delta Z^{\star I}\delta Z^{\bullet K}}
\end{matrix}\right)
\label{eq:MZ_Col}
\end{multline}
The superscript $Z$ on the matrix $\mathrm{M}^{\mathrm{Z}}_{IK}$ indicates the Z-field. The matrix in this case is more involved as compared to the one-loop effective MHV action. The first matrix above consists of purely the Z-field vertices. The one-loop contributions originating either solely from this matrix or via tree level connections with the classical Z-field action $S[Z_c]$  should account for only the cut-constructible parts of one-loop amplitudes in 4D. This is because, as stated previously, the vertices in the Z-field action are ultimately made up of MHV vertices only. This implies, just as in the case of one-loop effective MHV action Eq.~\eqref{eq:G_MHVnew}, the purely Self-Dual loop contributions, as well as some of the mixed (Self-Dual with MHV vertices), should originate from the second matrix consisting of the source terms. The sources in the second matrix above can be exchanged with fields using the following identities derived in Appendix~\ref{sec:App92}
\begin{equation}
  \frac{\delta S_{\mathrm{YM}}[A[Z_c]]}{\delta A^{\star L}}=-J_{\star L}\,, \quad\quad
    \frac{\delta S_{\mathrm{YM}}[A[Z_c]]}{\delta A^{\bullet L}}=- J_{\bullet L} \,.
    \label{eq:J_currZ}
\end{equation}
Substituting these to Eq.~\eqref{eq:MZ_Col} followed by some simple manipulations using the basic functional calculus we prove in Appendix \ref{sec:A102} that 
\begin{equation}
\det\mathrm{M}^{\mathrm{Z}}_{IK}  =\mathcal{N} \Big[ \mathcal{J}\Big]^2   \times\, \det\left.\Big(\mathrm{M}^{\mathrm{MHV}}_{SR}\right|_{\hat{B}_c^i = \hat{B}_c^i[Z_c]} \Big)\,.
\label{eq:MZ_MHV}
\end{equation}
Above, $\mathcal{N} = \Big[\det \delta(S^+-I^+)\det \delta(R^+-K^+)\Big]$; $\mathcal{J}$ is the Jacobian Eq.~\eqref{eq:BZ_jac} for the canonical transformation mapping the Anti-Self-Dual sector of the MHV action to the kinetic term in the Z-field action. Both these factors are field-independent. The determinant $\det\mathrm{M}^{\mathrm{Z}}_{IK}$ on the L.H.S. is exactly the one that appears in the one-loop Z-field partition function Eq.~\eqref{eq:det_SZ} whereas the determinant that appears on the R.H.S. is a bit more tricky. It is the determinant of the matrix obtained via substituting the solutions $\hat{B}_c^i = \hat{B}_c^i[Z_c]$ to the matrix that appears in the one-loop effective MHV action Eq.~\eqref{eq:G_MHVnew}. That is
\begin{multline}
\det\left.\Big(\mathrm{M}^{\mathrm{MHV}}_{SR}\right|_{\hat{B}_c^i = \hat{B}_c^i[Z_c]} \Big)  =  \det \left|\left(\begin{matrix}
     \frac{\delta^2 S_{\mathrm{MHV}}[B_c]}{\delta B^{\bullet I}\delta B^{\star K}} &\frac{\delta^2 S_{\mathrm{MHV}}[B_c]}{\delta B^{\bullet I}\delta B^{\bullet K}}\\ \\
\frac{\delta^2 S_{\mathrm{MHV}}[B_c]}{\delta B^{\star I}\delta B^{\star K}}
      &\frac{\delta^2 S_{\mathrm{MHV}}[B_c]}{\delta B^{\star I}\delta B^{\bullet K}} 
\end{matrix}\right)_{\hat{B}_c^i = \hat{B}_c^i[Z_c]} \right.\\
\left. + \left( \begin{matrix}
    - \frac{\delta S_{\mathrm{YM}}[A[B_c]]}{\delta A^{\star L}}\frac{\delta^2 A^{\star L}[B_c]}{\delta B^{\bullet I}\delta B^{\star K}} 
     &- \frac{\delta S_{\mathrm{YM}}[A[B_c]]}{\delta A^{\star L}}\frac{\delta^2 A^{\star L}[B_c]}{\delta B^{\bullet I}\delta B^{\bullet K}} -  \frac{\delta S_{\mathrm{YM}}[A[B_c]]}{\delta A^{\bullet L}}\frac{\delta^2 A^{\bullet L}[B_c]}{\delta B^{\bullet I}\delta B^{\bullet K}}\\ \\
\mathbb{0}
      &- \frac{\delta S_{\mathrm{YM}}[A[B_c]]}{\delta A^{\star L}}\frac{\delta^2 A^{\star L}[B_c]}{\delta B^{\star I}\delta B^{\bullet K}} 
\end{matrix}\right)_{\hat{B}_c^i = \hat{B}_c^i[Z_c]} \right|\,.
\end{multline}
Using Eq.~\eqref{eq:M_mhv12}, we can rewrite it as
\begin{equation}
\det\left.\Big(\mathrm{M}^{\mathrm{MHV}}_{SR}\right|_{\hat{B}_c^i = \hat{B}_c^i[Z_c]} \Big) = \mathcal{N} \Big[ \mathcal{J}_{\mathrm{MT}}\Big]^2 \times \, \det \begin{vmatrix}
     \frac{\delta^2 S_{\mathrm{YM}}[A[B_c]]}{\delta A^{\bullet Q}\delta A^{\star P}} &\frac{\delta^2 S_{\mathrm{YM}}[A[B_c]]}{\delta A^{\bullet Q}\delta A^{\bullet P}}\\ \\
\frac{\delta^2 S_{\mathrm{YM}}[A[B_c]]}{\delta A^{\star Q}\delta A^{\star P}}
      &\frac{\delta^2 S_{\mathrm{YM}}[A[B_c]]}{\delta A^{\star Q}\delta A^{\bullet P}} 
\end{vmatrix}_{\hat{B}_c^i = \hat{B}_c^i[Z_c]}\,,
\label{eq:M_mhv123}
\end{equation}
or
\begin{equation}
\det\left.\Big(\mathrm{M}^{\mathrm{MHV}}_{SR}\right|_{\hat{B}_c^i = \hat{B}_c^i[Z_c]} \Big) = \mathcal{N} \Big[ \mathcal{J}_{\mathrm{MT}}\Big]^2 \times \, \det \begin{vmatrix}
     \frac{\delta^2 S_{\mathrm{YM}}[A[B[Z_c]]]}{\delta A^{\bullet Q}\delta A^{\star P}} &\frac{\delta^2 S_{\mathrm{YM}}[A[B[Z_c]]]}{\delta A^{\bullet Q}\delta A^{\bullet P}}\\ \\
\frac{\delta^2 S_{\mathrm{YM}}[A[B[Z_c]]]}{\delta A^{\star Q}\delta A^{\star P}}
      &\frac{\delta^2 S_{\mathrm{YM}}[A[B[Z_c]]]}{\delta A^{\star Q}\delta A^{\bullet P}} 
\end{vmatrix}\,.
\label{eq:M_mhv1234}
\end{equation}
Substituting  the above in Eq.~\eqref{eq:MZ_MHV} and discarding all the field-independent as well as the volume divergent factors, we get
\begin{equation}
\det\mathrm{M}^{\mathrm{Z}}_{IK}  =\det \begin{vmatrix}
     \frac{\delta^2 S_{\mathrm{YM}}[A[B[Z_c]]]}{\delta A^{\bullet Q}\delta A^{\star P}} &\frac{\delta^2 S_{\mathrm{YM}}[A[B[Z_c]]]}{\delta A^{\bullet Q}\delta A^{\bullet P}}\\ \\
\frac{\delta^2 S_{\mathrm{YM}}[A[B[Z_c]]]}{\delta A^{\star Q}\delta A^{\star P}}
      &\frac{\delta^2 S_{\mathrm{YM}}[A[B[Z_c]]]}{\delta A^{\star Q}\delta A^{\bullet P}} 
\end{vmatrix}\,.
\label{eq:MZ_MHVre}
\end{equation}
The R.H.S. is exactly the determinant we had in the one-loop effective Z-field action Eq.~\eqref{eq:G_Zac} we derived in the previous approach. 

\begin{figure}
    \centering
    \includegraphics[width=10cm]{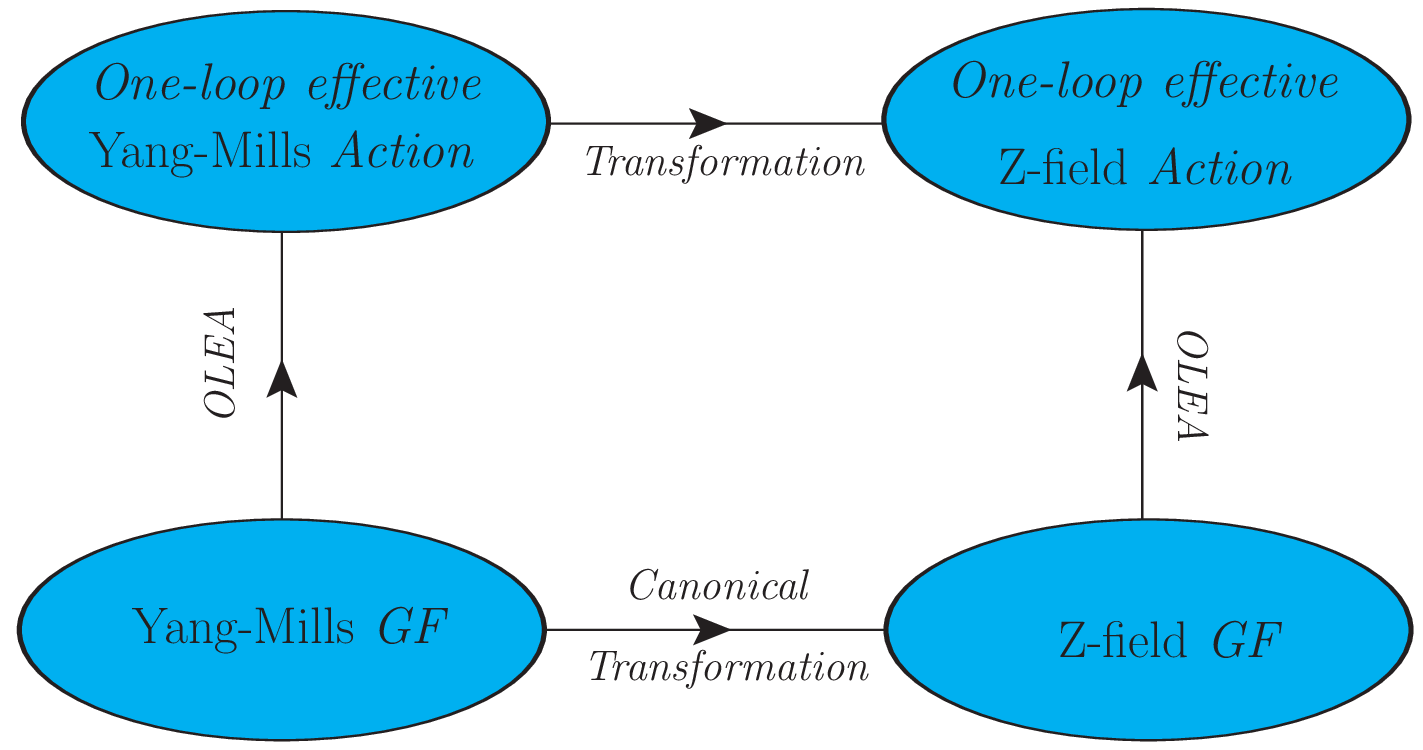}
    \caption{\small
    The one-loop effective Z-field action can be derived in two ways. First involves performing the canonical transformation Eq.~\eqref{eq:AtoZ_ct1}-\eqref{eq:AtoZ_ct2} to the Yang-Mills generating functional (keeping the source terms unaltered) and then deriving the one-loop effective action (dubbed as OLEA) via integration of the field fluctuations. The second involves deriving the one-loop effective Yang-Mills action starting with the Yang-Mills generating functional and then performing the transformations Eq.~\eqref{eq:AtoZ_ct1}-\eqref{eq:AtoZ_ct2}. In the figure, \textit{GF} stands for generating functional.}
    \label{fig:OLEAZacpaths}
\end{figure}

The one-loop effective action in the new approach obtained from Eq.~\eqref{eq:det_SZ} reads
\begin{equation}
   \Gamma[Z_c] = S[Z_c] 
     +i\, \frac{1}{2} \Tr\ln \left( 
    \mathrm{M}^{\mathrm{Z}}_{IK}
    \right)\,.
    \label{eq:G_Zacnew}
\end{equation}
Owing to Eq.~\eqref{eq:MZ_MHVre}, we see it is equal (in the context of computing one-loop amplitudes) to the one-loop effective Z-field action Eq.~\eqref{eq:G_Zac} derived previously in Section \ref{sec:Zth_OLEA}. We represent this equivalence diagrammatically in Figure \ref{fig:OLEAZacpaths}. Note, however, this equivalence stems from the fact that the source term introduced in the Yang-Mills partition function (coupled with the A-fields) remains unaltered throughout the process.

In summary, the one-loop effective actions (MHV as well as the Z-field) in the new approach are exactly the same as the ones we derived in the previous approach and are one-loop complete. However, in the new approach, the bigger interaction vertices are explicit in the log term and is therefore more suitable for computing higher multiplicity one-loop amplitudes.


\section{Amplitudes}
\label{sec:loopamp_ZAC}

In this section, we use the one-loop effective Z-field action to compute $(+ + + +)$, $(+ + + -)$, $(- - - -)$, $(- - - +)$ and $(- - + +)$ one-loop amplitudes. As before, we will focus on the leading trace color-ordered amplitude. Computing the first four types will ascertain that these are no longer missing as in the case of Z-field action. Recall, however, that the Z-field action also misses rational contributions to any one-loop amplitude in 4D. To make sure this is no longer the case, we compute the 4-point MHV $(- - + +)$ one-loop amplitude in 4D using the one-loop effective Z-field action. It is thus preferable to continue using the CQT scheme introduced in Section \ref{sec:One-loop-OLEAMHV}. However, the CQT scheme, in general, requires introducing additional counterterms which considerably complicates its practical application in Z-theory. We elaborate on this below.

Let us assume that we use the one-loop effective Z-field action Eq.~\eqref{eq:G_Zacnew} derived in Section \ref{sec:OLEA_rev}. Consider computing 4-point MHV $(- - + +)$ one-loop amplitudes using this action. It has Z-field vertices explicit in the log term. Thus a contribution to this amplitude will originate form joining the opposite helicity legs of the 6-point NMHV $(+ + + - - -)$ vertex in the Z-field action. We show this diagrammatically below 
\begin{center}
    \includegraphics[width=4cm]{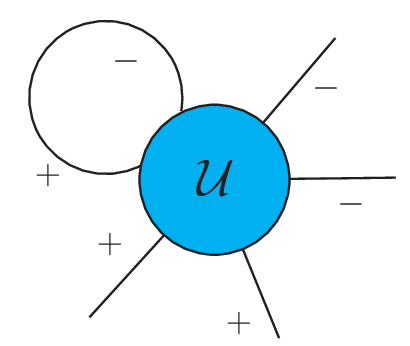}
\end{center}

Following the computation of one-loop amplitudes in the previous chapter, we know that quite a few one-loop terms involve contributions that violate Lorentz invariance in the CQT scheme and therefore require an explicit counterterm. In the previous chapter, we already encountered counterterms for the $(+ +)$, $(+ -)$, $(- -)$ self-energy terms; $(+ + -)$ one-loop term including both the triangle and swordfish topologies. It is natural to expect that some of these will be present in the above contribution. This is because the 6-point NMHV $(+ + + - - -)$ vertex in the Z-field action is made up of a 4-point MHV combined with a 3-point MHV. The former can be further decomposed into a tree connection of Yang-Mills vertices. When joining the opposite helicity legs of the 6-point NMHV $(+ + + - - -)$ vertex, we get a result consisting of contributions originating from tadpoles and also those that violate Lorentz invariance. The problem is all such contributions are intertwined and apriori looking at the result we get for one-loop $(- - + +)$ coming from  6-point NMHV $(+ + + - - -)$ vertex, it is hard to determine what the unwanted pieces are. These could be a combination of the above mentioned one-loop contributions for which the counterterms are known in the CQT. Furthermore, not all unwanted pieces require counterterms. Some of them are artifacts of the gauge choice and must cancel out in the amplitude computations. All these require an explicit study of contributions originating from a higher multiplicity vertex to a given one-loop amplitude in the CQT and this is beyond the scope of the present work.

For now, we avoid those issues using the one-loop effective Z-field action Eq.~\eqref{eq:G_Zac} derived in the previous approach discussed in Section \ref{sec:Zth_OLEA}. This is because the vertices in the log term in that approach are Yang-Mills vertices. Thus we can control the contributions that require counterterms explicitly. All the counterterms required in the Yang-Mills action were already uncovered in \cite{CQT1,CQT2}. Finally, recall, when computing lower multiplicity amplitudes like the 4-point ones considered in this section, both the action are equally favorable. The action derived via the new approach will provide, upon determining the CQT counterterms, a more efficient computation of higher multiplicity amplitudes. Thus, keeping the above in mind we use one-loop effective Z-field action Eq.~\eqref{eq:G_Zac}.
\subsection{\texorpdfstring{$(+ + + +)$}{allplus} and \texorpdfstring{$(- - - -)$}{allminus} one-loop amplitudes}
\label{sub:4plusminus}

In this Subsection, we will compute the leading trace color-ordered four-point $(+ + + +)$ and $(- - - -)$ one-loop amplitudes using the one-loop effective Z-field action Eq.~\eqref{eq:G_Zac} in the CQT scheme. Both these amplitudes are zero in the naive Z-field action Eq.~\eqref{eq:Z_action1} because they receive contributions solely from the Self-Dual $(+ + -)$ and the Anti-Self-Dual $(+ - -)$ vertices, respectively, both of which were eliminated via the canonical transformation Eq.~\eqref{eq:AtoZ_ct1}-\eqref{eq:AtoZ_ct2}. However, as we saw in the previous chapter, the $(+ + + +)$  one-loop amplitude is non-zero and is a rational function of the spinor products. The same applies to its conjugate $(+ \leftrightarrow -)$, i.e. the $(- - - -)$ one-loop amplitude, as we will see below.

Let us begin with the $(+ + + +)$ one-loop amplitude. The computation of this amplitude using the one-loop effective Z-field action Eq.~\eqref{eq:G_Zac} proceeds in the same way as what we saw in the previous chapter where we used the one-loop effective MHV action Eq.~\eqref{eq:G_MHV} instead. Thus, we do not repeat the entire calculation here. We simply highlight the main points necessary to assemble the final result. Following the rules discussed in Appendix \ref{sec:app_A6}, the contributions to $(+ + + +)$ one-loop amplitude originate only from the log term in Eq.~\eqref{eq:G_Zac}. These turn out to be identical to the ones in Figure \ref{fig:4plus_EA}. The contributions involving the $(+ +)$ one-loop bubble get canceled by similar contributions originating for the $(+ +)$ counterterm. With this, we are left with the following  
\begin{center}
    \includegraphics[width=16cm]{Chapter_4/all_plus1T.eps}
\end{center}
The triangular contribution $\Delta_{12}^{+ + + +}$ was computed in Appendix \ref{sec:App_A82}. Following the same procedure we can compute the other triangular contributions as well. 
 The box contribution $\square^{+ + + +}$, on the other hand, is computed by reducing it to the triangular contribution $\Delta_{34}^{+ + + +}$ plus additional terms. The explicit expressions for these can be found in Eqs.~\eqref{eq:tri_12}-\eqref{eq:box+tri}. Summing these over, in the on-shell limit we reproduce the known result
 \begin{equation}
      \mathcal{A}_{\mathrm{one-loop}}^{+ + + +} = \frac{g^{4}}{24 \pi^{2}} \frac{{\widetilde v}_{21} {\widetilde v}_{43}}{{\widetilde v}^{\star}_{21} {\widetilde v}^{\star}_{43}} \, .
       \label{eq:4_plus_one_loopre}
\end{equation}

The $(- - - -)$ one-loop amplitude is conjugate to $(+ + + +)$ one-loop amplitude in the sense that all the $+$ helicity gluons are exchanged with the $-$ helicity gluons. This conjugation replaces the Self-Dual vertices $(+ + -)$ in the diagrams contributing to the latter with the Anti-Self-Dual vertices $(+ - -)$ (as we will see below). And, in terms of momentum, the conjugation exchanges the transverse components $\bullet \leftrightarrow \star$. Due to this, the computation of this amplitude proceeds in a fashion similar to the $(+ + + +)$ one-loop amplitude but with the momentum components $\bullet \leftrightarrow \star$ exchanged.

\begin{figure}
    \centering
    \includegraphics[width=16cm]{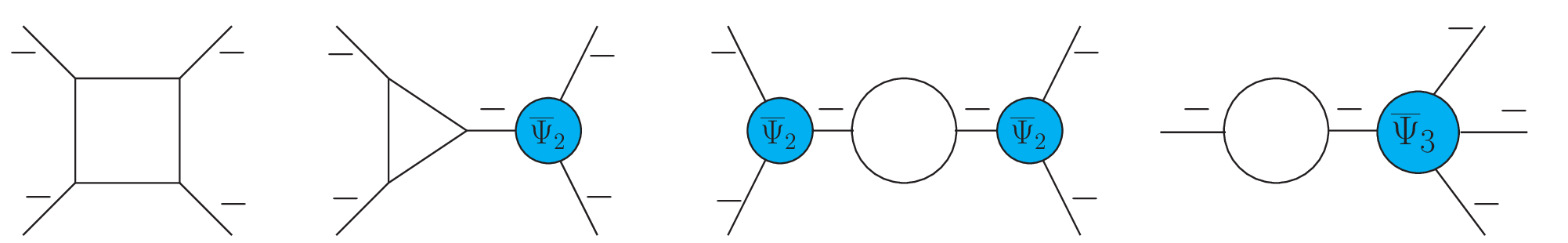}
    \caption{\small
    The first three diagrams represent the one-loop contributions originating from the log term in the one-loop effective Z-field action, following the rules in Appendix \ref{sec:app_A6}, to the $(- - - -)$ one-loop amplitude. The last term is included to discuss an identity which states the sum of all four terms is zero. Note, there is no propagator connecting the kernels $\overline{{\Psi}}_n$ to the 1PI one-loop sub-diagrams.}
    \label{fig:4Mct}
\end{figure}

\begin{figure}
    \centering
    \includegraphics[width=16cm]{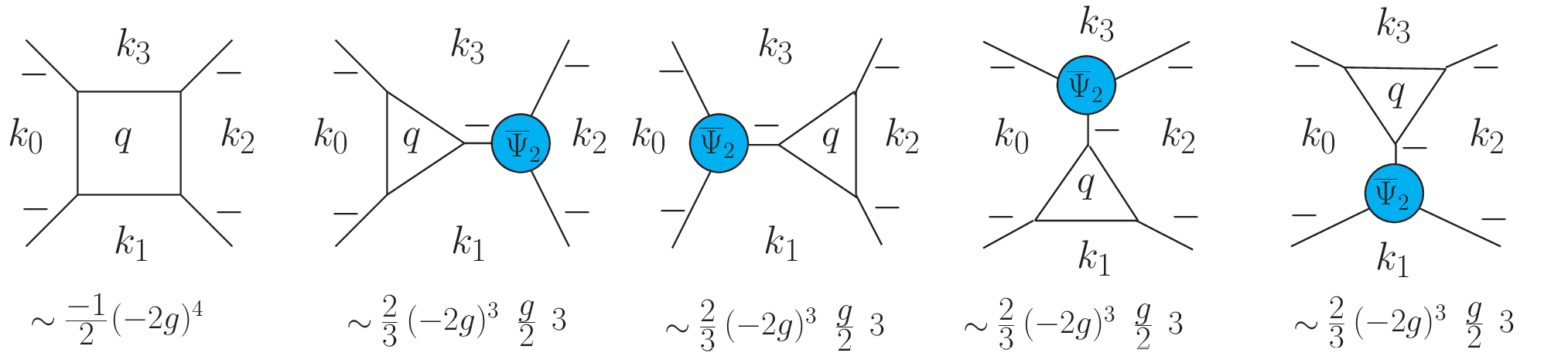}
    \caption{\small
    The explicit contributions to the $(- - - -)$ one-loop amplitude with the symmetry factors as well as the assignment of region momenta.}
    \label{fig:4Mall}
\end{figure}

Following the rules discussed in Appendix \ref{sec:app_A6}, the contributions to $(- - - -)$ one-loop amplitude also originate solely from the log term in Eq.~\eqref{eq:G_Zac}. These are the first three diagrams shown in Figure \ref{fig:4Mct}. Notice, all the diagrams in this figure can be obtained from those in Figure \ref{fig:4plus_EA} (contributing to $(+ + + +)$ one-loop amplitude) via the conjugation $+ \leftrightarrow -$. This replaces the kernels ${\Psi }_n$ in the latter with $\overline{{\Psi}}_n$. The last two diagrams in Figure \ref{fig:4Mct} involve the $(- -)$ gluon self-energy. Just like the $(+ +)$ gluon self-energy, this too is none-zero in the CQT scheme and thus requires an explicit counterterm. For the line/region momenta assignment shown below
\begin{center}
    \includegraphics[width=4cm]{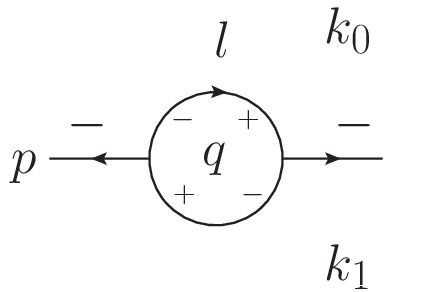}
\end{center}

the $(- -)$ gluon self-energy reads (\emph{cf.} Eq.~\eqref{eq:++GSE_ap} from Appendix \ref{sec:App_A81})
\begin{equation}
    \Pi^{--}= \frac{g^{2}}{12 \pi^{2}} \left[k_{0}^{\bullet 2}+k_{1}^{\bullet 2}+k_{0}^{\bullet} k_{1}^{\bullet}\right]\,.
    \label{eq:--GSE_bub}
\end{equation}
Introducing the above counterterm, the last two diagrams in Figure \ref{fig:4Mct} cancel out against similar contributions from the counterterm. Note, however, as stated above, the rules discussed in Appendix \ref{sec:app_A6} allow for only the first three diagrams shown in Figure \ref{fig:4Mct}. The last one was included to highlight an identity (similar to the one we encountered for the $(+ + + +)$ one-loop case in Subsection \ref{sub:4plus}). According to this identity, the sum of all four contributions in Figure \ref{fig:4Mct} with appropriate symmetry factors is zero. Therefore one can simply compute the $(- - - -)$ one-loop amplitude using just the $(- -)$ counterterm.

After introducing the $(- -)$ counterterm, we are left with the contributions shown in Figure \ref{fig:4Mall}. The computation of these are step by step identical (modulo the exchange of the transverse components $\bullet \leftrightarrow \star$ of the line/region momenta) to the calculation we did for the $(+ + + +)$ one-loop case in Sub-section \ref{sub:4plus}. Therefore, we do not repeat the details of the calculation and simply quote the final expression we obtained. These read
\begin{equation}
   \Delta_{12}^{- - - -}= \frac{-g^{4}}{12 \pi^{2}} \frac{\left({\widetilde v}^{\star}_{12}p_2^+\right)^{3}p_{3}^{+}}{p_{1}^{+} p_{2}^{+} p_{3}^{+}p_{4}^{+} p_{34}^{2}{\widetilde v}_{34}}\,,
\end{equation}
\begin{equation}
    \Delta_{41}^{- - - }= \frac{-g^{4}}{12 \pi^{2}} \frac{\left({\widetilde v}^{\star}_{41}p_1^+\right)^{3}p_{2}^{+}}{p_{1}^{+} p_{2}^{+} p_{3}^{+} p_{4}^{+} p_{14}^{2}{\widetilde v}_{23}}\,,
\end{equation}
\begin{equation}
   \Delta_{23}^{- - - -}= \frac{-g^{4}}{12 \pi^{2}} \frac{ \left({\widetilde v}^{\star}_{23}p_3^+\right)^{3}p_{4}^{+}}{p_{1}^{+} p_{2}^{+} p_{3}^{+} p_{4}^{+} p_{14}^{2}{\widetilde v}_{41}}\,,
\end{equation} 
\begin{multline}
  \square^{- - - -} \,+\, \Delta_{34}^{- - - -} = \frac{-g^{4}}{12 \pi^{2}} \frac{p_{1}^{+}}{p_{1}^{+} p_{2}^{+} p_{3}^{+} p_{4}^{+} {\widetilde v}_{12} p_{14}^{2}}\\
  \left[{\widetilde v}^{\star}_{41}p_{1}^{+} {\widetilde v}^{\star}_{23}p_{3}^{+}\left({\widetilde v}^{\star}_{41}p_{1}^{+}+{\widetilde v}^{\star}_{23}p_{3}^{+}\right)+{\widetilde v}^{\star}_{34}p_{4}^{+}\left({\widetilde v}^{\star 2}_{41}p_{1}^{+ 2}+{\widetilde v}^{\star 2}_{23}p_{3}^{+ 2}\right)\right]\,.
    \label{eq:box+tri4minus}
\end{multline}
Notice, the above results are exactly conjugate ($\widetilde{v}_{ij} \leftrightarrow \widetilde{v}^{\star}_{ij}$) to the expression Eqs.~\eqref{eq:tri_12}-\eqref{eq:box+tri}. This is simply because we assigned the line/region momenta to the terms in Figure~\ref{fig:4Mall} in exactly the same fashion as the terms contributing to $(+ + + +)$ one-loop case shown in Figure~\ref{fig:all_plus1}. 

The sum of all the above contributions, after a bit of tedious algebra, give $(- - - -)$ one-loop amplitude 
\begin{equation}
      \mathcal{A}_{\mathrm{one-loop}}^{- - - -} = \frac{g^{4}}{24 \pi^{2}} \frac{{\widetilde v}^{\star}_{21} {\widetilde v}^{\star}_{43}}{{\widetilde v}_{21} {\widetilde v}_{43}}     \, .
       \label{eq:4_minus_one_loop}
\end{equation}
Indeed, the above result is conjugate to $\mathcal{A}_{\mathrm{one-loop}}^{+ + + +}$ Eq.~\eqref{eq:4_plus_one_loopre}.

\subsection{\texorpdfstring{$(+ + + -)$}{allplus} and \texorpdfstring{$(- - - +)$}{allminus} one-loop amplitudes}
\label{sub:3plus+minus}

In this Subsection, we will compute the leading trace color-ordered four-point $(+ + + -)$ and $(- - - +)$ one-loop amplitudes using the one-loop effective Z-field action Eq.~\eqref{eq:G_Zac} in the CQT scheme. These too cannot be computed in the naive classical Z-field action. 

Let us start with the four-point $(+ + + -)$ one-loop amplitude. The diagrams contributing to this amplitude, using the rules from Appendix \ref{sec:app_A6}, are almost (except for one difference that we discuss below) all identical to the ones we had in Figure \ref{fig:3plusminus} where we used the one-loop effective MHV action Eq.~\eqref{eq:G_MHV}. The only difference is that when using the one-loop effective Z-field action Eq.~\eqref{eq:G_Zac} all the contributions originate solely from the log term. Whereas in one-loop effective MHV action Eq.~\eqref{eq:G_MHV}, there was a contribution involving tree level connection between the $(+ - -)$ MHV vertex in classical MHV action $S_{\mathrm{MHV}}[B_c]$ and the $\Delta^{+ + + }$ one-loop sub-diagram from the log term in Eq.~\eqref{eq:G_MHV}. This was the last term on the third line of Figure \ref{fig:3plusminus}. When deriving the one-loop effective Z-field action Eq.~\eqref{eq:G_Zac}, the $(+ - -)$ MHV vertex gets eliminated. As a result, in the case of one-loop effective Z-field action Eq.~\eqref{eq:G_Zac}, this diagram in Figure \ref{fig:3plusminus} gets replaced by a similar term where the tree level connection with the $(+ - -)$ MHV vertex in the former is replaced by the second order expansion of the $\widetilde{B}^{\bullet}_a[{Z}^{\star}, {Z}^{\bullet}](x^+;\mathbf{P})$ field Eq.~\eqref{eq:BbulletZ_exp} which results in the kernel $\overline{\widetilde \Omega}\,^{a b_1 \left \{b_2 \right \}}_{2}(\mathbf{P}; \mathbf{p}_{1}, \left \{ \mathbf{p}_{2} \right \} )$ shown below
\begin{center}
    \includegraphics[width=9cm]{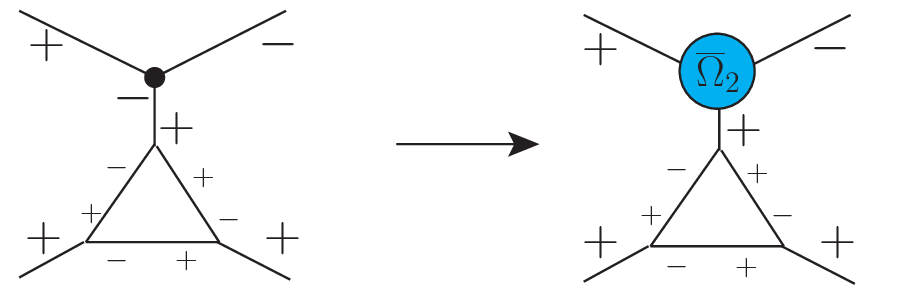}
\end{center}

Except for this difference, the entire calculation proceeds in exactly the same fashion as shown in Subsection \ref{sub:3plus-minus}. We, therefore, do not repeat it here. We checked that assigning the line/region momenta in exactly the same fashion as we did in Subsection \ref{sub:3plus-minus}, we could reproduce the expressions for the Self-Energy contributions $\mathcal{A}_{\mathrm{SE}}^{+ + + -}$ Eq.~\eqref{eq:self_contri_OL}, the Triangle-Swordfish contributions $\mathcal{A}_{\mathrm{TS}}^{+ + + -}$ Eq.~\eqref{eq:TS_contri_OL} and the Box-Quartic contributions $\mathcal{A}_{\mathrm{BQ}}^{+ + + -}$ Eq.~\eqref{eq:box_quar}. Summing which, in the on-shell limit, we reproduced the result Eq.~\eqref{eq:3plus-oneloop}
\begin{equation}
    \mathcal{A}_{\mathrm{one-loop}}^{+ + + -} = \frac{-g^{4}}{24 \pi^{2}} \frac{ p_3^{+}{\widetilde v}_{13}^2}{p_1^+{\widetilde v}_{14} {\widetilde v}_{43}{\widetilde v}^{\star}_{21} {\widetilde v}^{\star}_{32}} (p_{12}^2 + p_{14}^2 )\,.
    \label{eq:3plus-oneloopre}
\end{equation}

\begin{figure}
    \centering
 \includegraphics[width=15.8cm]{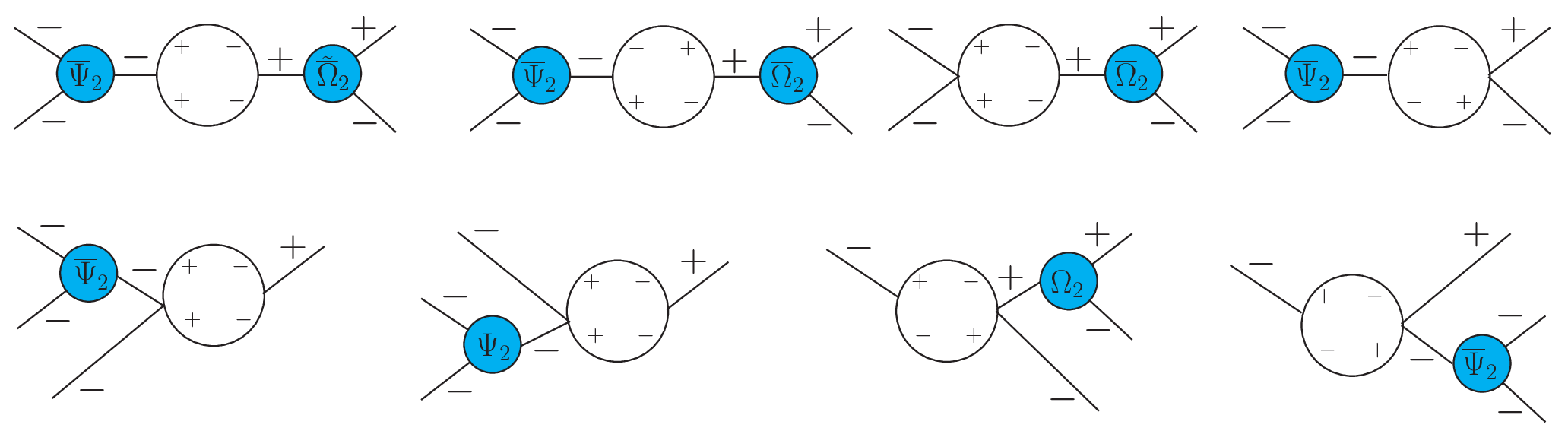}\\
 \includegraphics[width=16cm]{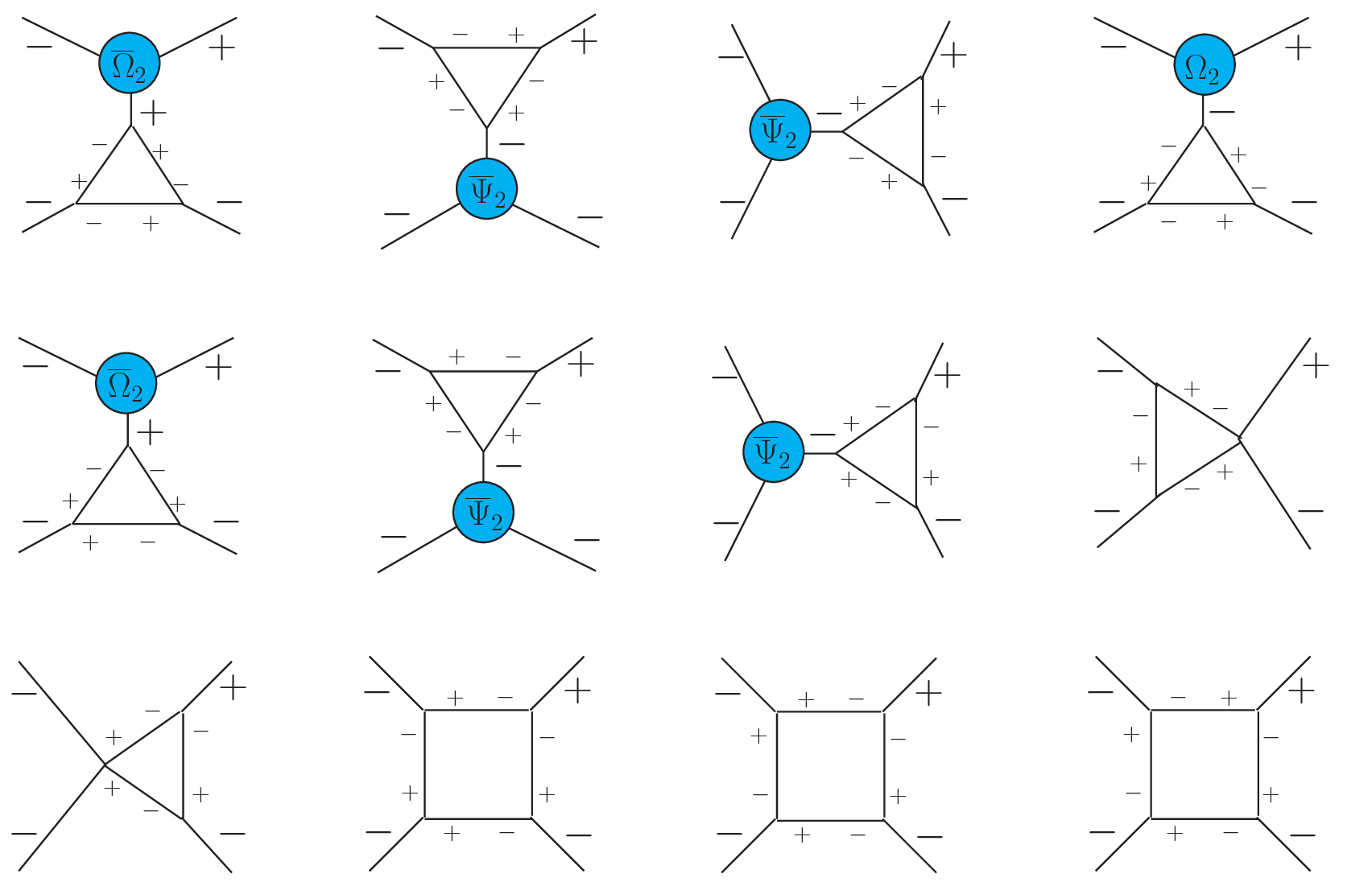} 
    \caption{\small 
    The one-loop contributions originating from the one-loop effective Z-field action, following the rules in Appendix \ref{sec:app_A6}, to the $(- - - +)$ one-loop amplitude. We suppressed the symmetry factors and region momenta for the sake of simplicity.}
    \label{fig:3minusplus}
\end{figure}

Now, let us consider the $(- - - +)$ one-loop amplitude. In this case too, the contributions originate solely from the log term in the one-loop effective Z-field action Eq.~\eqref{eq:G_Zac}. These are shown in Figure \ref{fig:3minusplus}. Notice, the diagrams in Figure \ref{fig:3minusplus} can be directly obtained, instead, from those in Figure \ref{fig:3plusminus} by simply flipping the helicities $+ \leftrightarrow -$ everywhere. In terms of vertices, this implies replacing the Self-Dual vertex $(+ + -)$ with the Anti-Self-Dual $(- - +)$, and for the kernels, we have $\Big\{ {\widetilde \Psi}_{2}^{a \left \{b_i b_j \right \} }(\mathbf{p}; \left \{\mathbf{p}_{i}, \mathbf{p}_{j} \right \}), {\widetilde \Omega}_{2}^{c \left \{b_k b_l \right \} }(\mathbf{q}; \left \{\mathbf{p}_{k}, \mathbf{p}_{l} \right \})\Big\} \rightarrow \Big\{ \overline{\widetilde \Psi}\,^{ a \left \{b_i b_j \right \} }_{2}(\mathbf{p}; \left \{\mathbf{p}_{i}, \mathbf{p}_{j} \right \}), \overline{\widetilde \Omega}\,^{c \left \{b_k b_l \right \} }_{2}(\mathbf{q}; \left \{\mathbf{p}_{k}, \mathbf{p}_{l} \right \})\Big\}$ respectively. This interchanges the spinor products $\widetilde{v}_{ij} \leftrightarrow \widetilde{v}^{\star}_{ij}$. Furthermore, we checked that the symmetry factors in both sets of diagrams, related via $+ \leftrightarrow -$, are identical. As a result, already at the level of diagrams, we have a cross-check that the contributions necessary to compute $(- - - +)$ one-loop amplitude are not missing in Eq.~\eqref{eq:G_Zac}. 

For the computation of the analytic expression, we assign the line/region momenta to each of these diagrams in exactly the same fashion as for the contribution in Figure \ref{fig:3plusminus} for the $(+ + + -)$ one-loop amplitude. That is we assign $(p_1, p_2, p_3, p_4)$ to the external legs $(- - - +)$ in an anti-clockwise fashion. The main reason for doing this is that it provides an intermediate cross-check. Precisely, if we assign the momenta in exactly the same fashion then we know that the intermediate results we get for the contributions, in the current computation, should be the conjugate of those we had in $(+ + + -)$ one-loop case in Subsection \ref{sub:3plus-minus}.

Let us proceed with the first two terms involving the $(+ -)$ gluon self-energy Eq.~\eqref{eq:pi+-_ct}. Substituting the second order expansion of $\widetilde{B}^{\star}_a[{Z}^{\star}](x^+;\mathbf{P})$ and $\widetilde{B}^{\bullet}_a[{Z}^{\star}, {Z}^{\bullet}](x^+;\mathbf{P})$ Eq.~\eqref{eq:BstarZ_exp}-\eqref{eq:BbulletZ_exp}, we get
\begin{multline}
\mathcal{A}_{\mathrm{SE}}^{- - - +} =\frac{ -g^{4}p_{4}^{+}}{2 \pi^{2} p_{1}^{+} p_{2}^{+} p_{3}^{+}}\Bigg[\frac{{\widetilde v}^{\star}_{34}p_{4}^{+}p_{2}^{+}}{{\widetilde v}_{21}}\\
\Bigg\{\sum_{q^{+}}\left[\frac{1}{q^{+}}+\frac{1}{p_{12}^{+}-q^{+}}\right] \ln \Bigg(\frac{q^{+}\left(p_{12}^{+}-q^{+}\right)}{p_{12}^{+ 2}} p_{12}^{2} e^{\gamma} \delta\Bigg)
-\frac{11}{6} \ln \left(p_{12}^{2} e^{\gamma} \delta\right)+\frac{67}{18}\Bigg\}\\
+\frac{{\widetilde v}^{\star}_{41}p_{3}^{+}p_{1}^{+}}{{\widetilde v}_{32}}\Bigg\{\sum_{q^{+}+p_4^+}\left[\frac{1}{q^{+}+p_4^+}+\frac{1}{p_1^{+}-q^{+}}\right] \ln \Bigg(\frac{(q^{+}+p_4^+)\left(p_1^{+}-q^{+}\right)}{p_{14}^{+2}} p_{14}^{2} e^{\gamma} \delta\Bigg)\\
-\frac{11}{6} \ln \left(p_{14}^{2} e^{\gamma} \delta\right)+\frac{67}{18}\Bigg\}\Bigg] \,,
\label{eq:self_contri_OL3m}
\end{multline}
where "SE" stands for self-energy. Notice, the above expression is indeed conjugate to Eq.~\eqref{eq:self_contri_OL}.

Next, we consider all the contributions that consist of the one-loop triangle and swordfish sub-diagrams. Notice, in Figure \ref{fig:3minusplus}, the last term on the third line consists of $\Delta^{- - -}$ one-loop sub-diagram connected with  ${\widetilde \Omega}_{2}$. The $\Delta^{- - -}$ can be computed exactly in the same fashion as $\Delta^{+ + +}$. The latter was re-derived in Appendix \ref{sec:App_A82}. Except for this term, all the others consist of $(- - +)$ one-loop sub-diagram (triangle + swordfish topology). This requires an explicit counterterm in the CQT scheme shown below \cite{CQT1,CQT2}
\begin{center}
    \includegraphics[width=8cm]{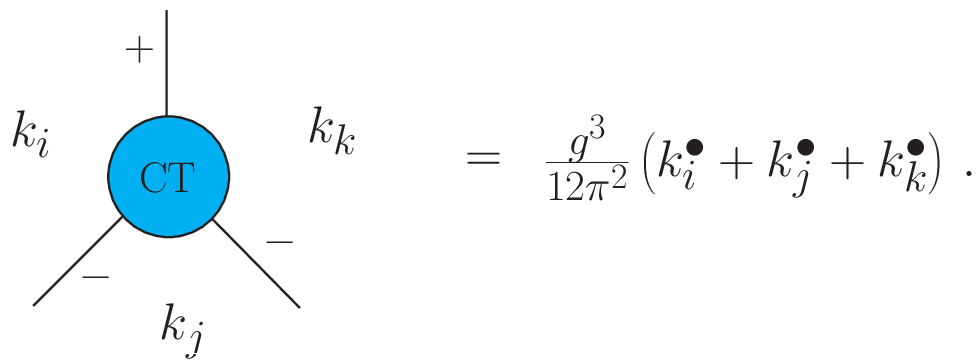}
\end{center}

Substituting the second order expansion of $\widetilde{B}^{\star}_a[{Z}^{\star}](x^+;\mathbf{P})$ and $\widetilde{B}^{\bullet}_a[{Z}^{\star}, {Z}^{\bullet}](x^+;\mathbf{P})$ Eq.~\eqref{eq:BstarZ_exp}-\eqref{eq:BbulletZ_exp} to the $(- - +)$ one-loop sub-diagram  and combining it with the $\Delta^{- - -}$- ${\widetilde \Omega}_{2}$ term, we get
\begin{multline}
\mathcal{A}_{\mathrm{TS}}^{- - - +}=\frac{-g^{4}p_{4}^{+}}{4 \pi^{2} p_{1}^{+} p_{2}^{+} p_{3}^{+}}\Bigg[\frac{{\widetilde v}^{\star}_{34}p_{4}^{+}p_{2}^{+}}{{\widetilde v}_{21}}\Bigg\{\frac{22}{3} \ln \left(p_{12}^{2} e^{\gamma} \delta\right)-\frac{140}{9}-S_{3}^{q^{+}}\left(p_{1}, p_{2}\right)-S_{3}^{q^{+}}\left(-p_{4},-p_{3}\right)\\
+\frac{p_{1}^{+} p_{2}^{+}}{3 p_{12}^{+2}}\Bigg\}
+\frac{{\widetilde v}^{\star}_{41}p_{3}^{+}p_{1}^{+}}{{\widetilde v}_{32}}\Bigg\{\frac{22}{3} \ln \left(p_{14}^{2} e^{\gamma} \delta\right)-\frac{140}{9}-S_{2}^{q^{+}}\left(-p_{4},-p_{23}\right)-S_{1}^{q^{+}+p_{4}^{+}}\left(p_{14}, p_{2}\right)+\frac{p_{2}^{+} p_{3}^{+}}{3 p_{14}^{+2}}\Bigg\}\Bigg] \\
+\frac{g^{4}p_{3}^{+}p_{2}^{+2}}{3 \pi^{2} p_{1}^{+}  p_{12}^{+2}} \frac{{\widetilde v}_{12}^{\star 3} {\widetilde v}_{34}}{p_{12}^{4}}+\frac{g^{4}p_{1}^{+ 2}p_{3}^{+ 2}}{3 \pi^{2} p_{2}^{+} p_{4}^{+} p_{14}^{+2}} \frac{{\widetilde v}_{23}^{\star 3} {\widetilde v}_{41}}{p_{14}^{4}}\,,
\label{eq:TS_contri_OL3m}
\end{multline}
where "TS" represents triangle + swordfish and the expression for $S_{l}^{q^{+}}\!\left(p_{i}, p_{j}\right)$ can be found in Appendix \ref{sec:App_A83}.

Finally, we have the 4-point one-loop terms. These consist of the box and the quartic topology. These do not require any substitution of $\widetilde{B}^{\star}_a[{Z}^{\star}](x^+;\mathbf{P})$ and $\widetilde{B}^{\bullet}_a[{Z}^{\star}, {Z}^{\bullet}](x^+;\mathbf{P})$ Eq.~\eqref{eq:BstarZ_exp}-\eqref{eq:BbulletZ_exp}. Their sum reads
\begin{multline}
    \mathcal{A}_{\mathrm{BQ}}^{- - - +}=\frac{ g^{4} }{2 \pi^{2}}\Bigg[ {-\left\{\frac{p_{1}^{+}p^+_{2\overline{3}}-3 p_{3}^{+}p^+_{23}}{6 p_{12}^{+} p_{23}^{+} p_{4}^{+} {\widetilde v}^{\star}_{14}{\widetilde v}_{41}}+\frac{p_{3}^{+}(-3 p_{1}^{+} p_{12}^{+}+p_{3}^{+}p^+_{\overline{1}2})}{6 p_{1}^{+} p_{12}^{+2} p_{4}^{+} {\widetilde v}^{\star}_{12}{\widetilde v}_{21}}\right\} {\widetilde v}_{12}^{\star 2}p_{2}^{+} } \\
+\left\{\frac{p_{1}^{+}p^+_{\overline{2}3}+p_{3}^{+} p_{23}^{+}}{6 p_{1}^{+} p_{3}^{+} p_{23}^{+2} {\widetilde v}^{\star}_{12}{\widetilde v}_{21}}-\frac{p_{1}^{+}p^+_{\overline{2}3}+3 p_{3}^{+} p_{23}^{+}}{6 p_{12}^{+} p_{3}^{+} p_{23}^{+} p_{4}^{+} {\widetilde v}^{\star}_{14}{\widetilde v}_{41}}\right\} {\widetilde v}^{\star}_{12}{\widetilde v}^{\star}_{34}p_2^+p_4^+\\
+\frac{p_{4}^{+}}{2p_{1}^{+} p_{2}^{+} p_{3}^{+}} \frac{{\widetilde v}^{\star}_{34}p_{4}^{+}p_{2}^{+}}{{\widetilde v}_{21}}\Bigg\{\frac{11}{3} \ln \left(\delta e^{\gamma} p_{12}^{2}\right)-\frac{11}{3} \ln \left(\delta e^{\gamma} p_{14}^{2}\right)\\
     -S_{3}^{q^{+}}\left(p_{1}, p_{2}\right)-S_{3}^{q^{+}}\left(-p_{4},-p_{3}\right)+S_{2}^{q^{+}}\left(-p_{4},-p_{23}\right) +S_{1}^{q^{+}+p_{4}^{+}}\left(p_{14}, p_{2}\right)\\
-2 \sum_{q^{+}}\left[\frac{1}{q^{+}}+\frac{1}{p_{12}^{+}-q^{+}}\right] \ln \Bigg(\frac{q^{+}\left(p_{12}^{+}-q^{+}\right)}{p_{12}^{+ 2}} p_{12}^{2} e^{\gamma} \delta\Bigg)\\
+2 \sum_{q^{+}+p_4^+}\left[\frac{1}{q^{+}+p_4^+}+\frac{1}{p_1^{+}-q^{+}}\right] \ln \Bigg(\frac{(q^{+}+p_4^+)\left(p_1^{+}-q^{+}\right)}{p_{14}^{+2}} p_{14}^{2} e^{\gamma} \delta\Bigg)\Bigg\}\Bigg]\,,
\label{eq:BQ---+}
\end{multline}
where "BQ" represents box + quartic. Notice, all the three expressions above Eqs.~\eqref{eq:self_contri_OL3m}-\eqref{eq:BQ---+} are conjugate of $\mathcal{A}_{\mathrm{SE}}^{+ + + -}$ Eq.~\eqref{eq:self_contri_OL}, $\mathcal{A}_{\mathrm{TS}}^{+ + + -}$ Eq.~\eqref{eq:TS_contri_OL} and  $\mathcal{A}_{\mathrm{BQ}}^{+ + + -}$ Eq.~\eqref{eq:box_quar}. Thus already at this level, we can ascertain that all the one-loop contributions necessary to compute $(- - - +)$ one-loop amplitude are present in Eq.~\eqref{eq:G_Zac}. Summing these, after a bit of tedious algebra, in the on-shell limit we get
\begin{equation}
    \mathcal{A}_{\mathrm{one-loop}}^{- - - +} = \frac{-g^{4}}{24 \pi^{2}} \frac{ p_3^{+}{\widetilde v}^{\star 2}_{13}}{p_1^+{\widetilde v}^{\star}_{14} {\widetilde v}^{\star}_{43}{\widetilde v}_{21} {\widetilde v}_{32}} (p_{12}^2 + p_{14}^2 )\,.
    \label{eq:3plus-oneloop--}
\end{equation}
The above result is conjugate to Eq.~\eqref{eq:3plus-oneloopre}.

\subsection{\texorpdfstring{$(+ + - -)$}{mhv} one-loop amplitude}
\label{sub:oneloopMHV}

\begin{figure}
    \centering
 \includegraphics[width=14.3cm]{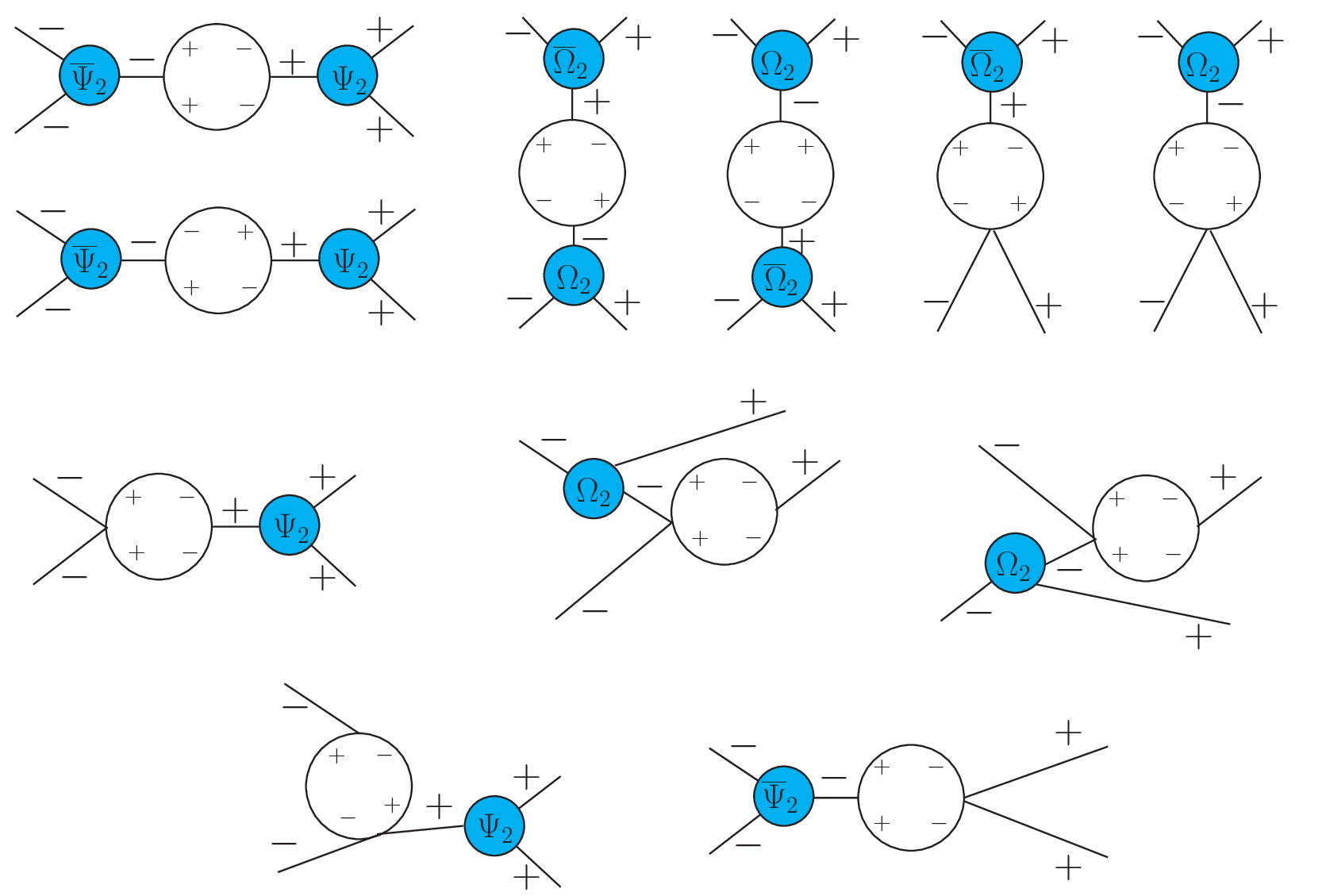}\\
 \includegraphics[width=13.4cm]{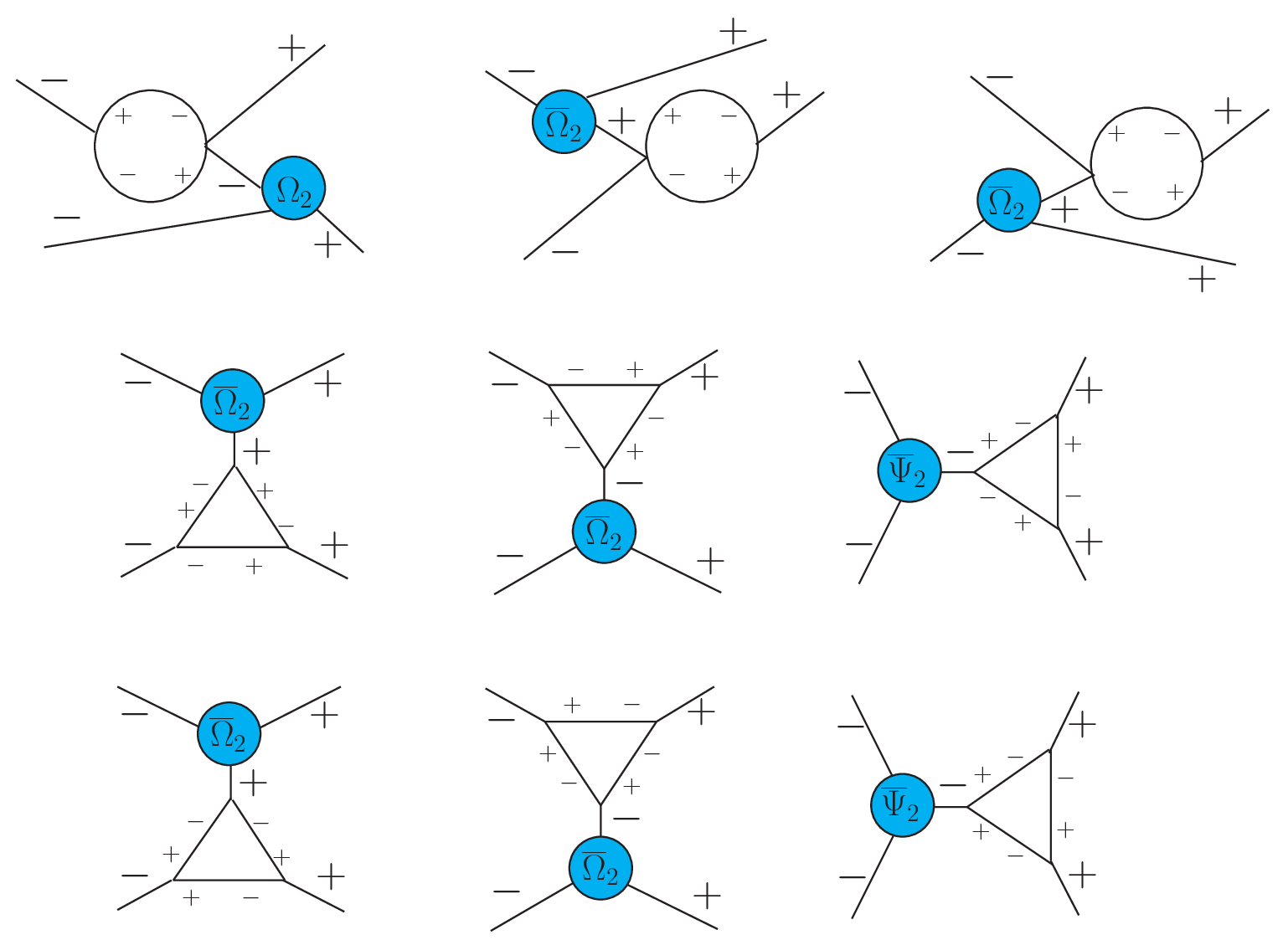} \\
 \includegraphics[width=8cm]{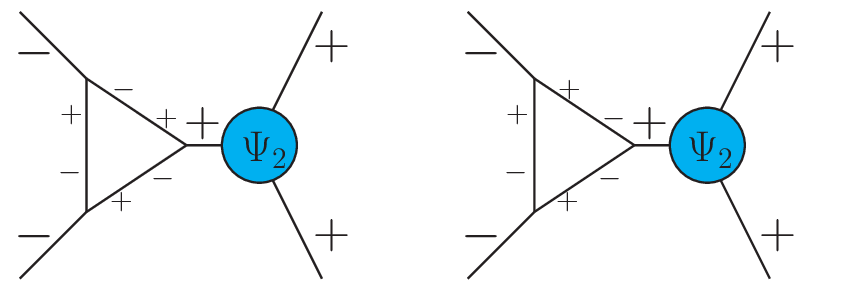} 
    \caption{\small 
    The one-loop contributions, containing 2-point and the 3-point 1PI one-loop sub-diagrams, originating from the one-loop effective Z-field action, following the rules in Appendix \ref{sec:app_A6}, to the $(- - + +)$ one-loop amplitude. These include
bubbles (first four), triangles (last eight), and swordfish (remaining) topologies. We suppressed the symmetry factors and region momenta for brevity.}
    \label{fig:4MHVOL1}
\end{figure}

\begin{figure}
    \centering
 \includegraphics[width=14cm]{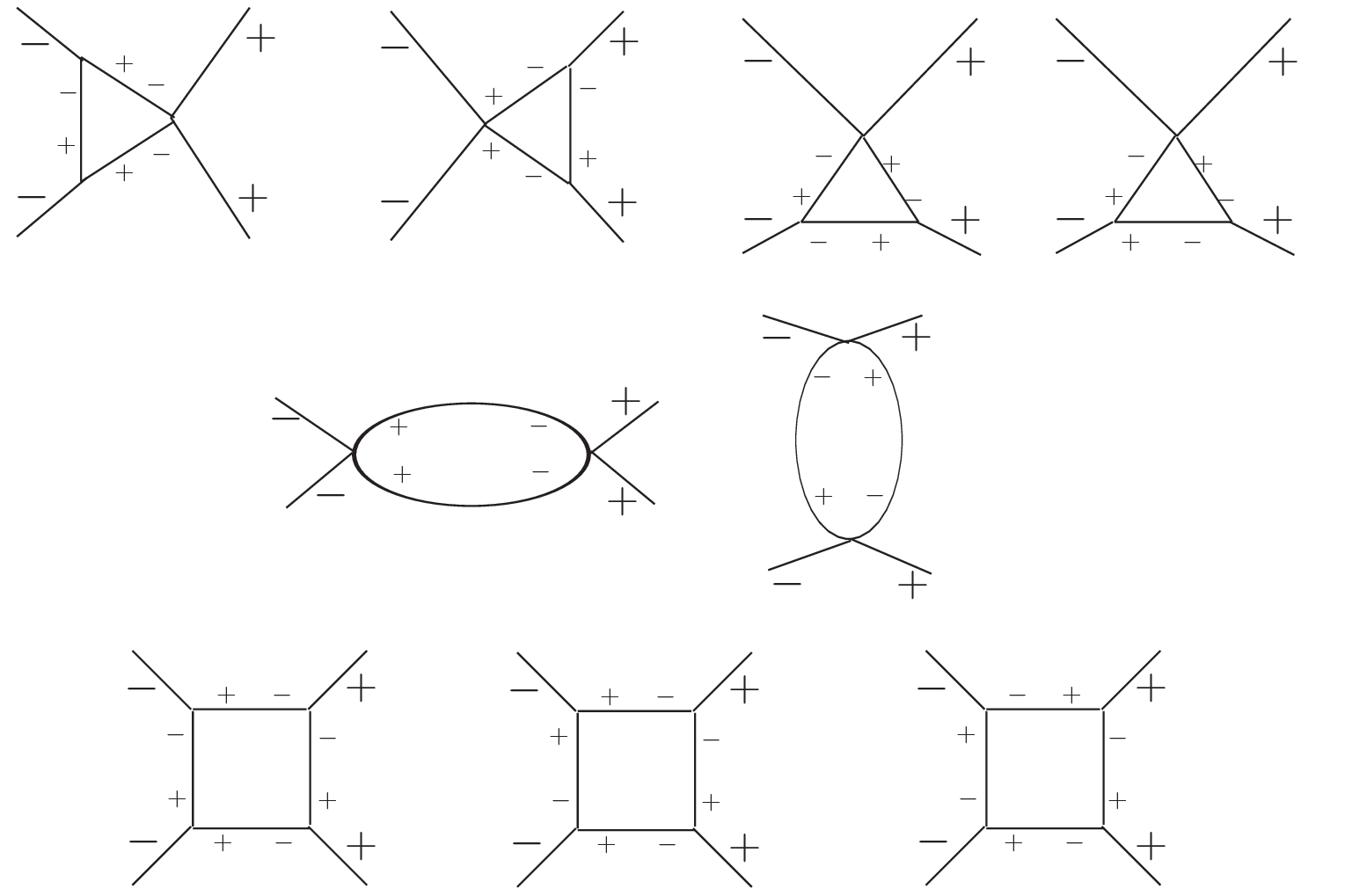}
    \caption{\small 
    The one-loop contributions, containing 4-point 1PI one-loop sub-diagrams, originating from the one-loop effective Z-field action, following the rules in Appendix \ref{sec:app_A6}, to the $(- - + +)$ one-loop amplitude. These include
quartic (first four), box (last three), and double-quartic (remaining two in the middle) topologies. We suppressed the symmetry factors and region momenta for brevity.}
    \label{fig:4MHVOL2}
\end{figure}

In this Subsection, we will compute the leading trace color-ordered four-point $(+ + - -)$ one-loop amplitude using the one-loop effective Z-field action Eq.~\eqref{eq:G_Zac} in the CQT scheme. All the contributions to this amplitude originate solely from the log term in Eq.~\eqref{eq:G_Zac}. Unlike the previous cases, we do not show the contributions altogether in a single figure. We rather show them separately in two Figures \ref{fig:4MHVOL1}-\ref{fig:4MHVOL2}. The former represents all the contributions involving the 2-point and the 3-point 1PI one-loop sub-diagrams. These include bubbles, triangles, and swordfish topologies. The latter represents all the 4-point 1PI one-loop contributions to $(+ + - -)$ one-loop amplitude. These include quartic, double-quartic (this naming will be explained below), and box topologies. For this computation, we name the external legs $(+ + - -)$ as $(p_1, p_2, p_3, p_4)$ in an anti-clockwise fashion. As before, we focus on the leading trace and suppress the color for the sake of simplicity.

Let us begin with the contributions in Figure \ref{fig:4MHVOL1}. These include bubbles, triangles, and swordfish one-loop sub-diagrams. Notice, we avoided the contributions involving the $(+ +)$ and $(- -)$ gluon self-energy because these would get canceled by similar contributions originating from the $(+ +)$ and $(- -)$ counterterms. We already encountered each of the 2-point and the 3-point one-loop sub-diagrams in Figure \ref{fig:4MHVOL1} separately in the previous computations. The only thing that changes here is the substitution of fields on the external legs. Therefore, we do not reiterate the results and the associated counterterms for these here. Substituting the second order expansion of $\widetilde{A}^{\bullet}_a[{B}^{\bullet}](x^+;\mathbf{P})$ and $\widetilde{A}^{\star}_a[{B}^{\bullet}, {B}^{\star}](x^+;\mathbf{P})$ Eq.~\eqref{eq:A_bull_solu}-\eqref{eq:A_star_solu} as well as $\widetilde{B}^{\star}_a[{Z}^{\star}](x^+;\mathbf{P})$ and $\widetilde{B}^{\bullet}_a[{Z}^{\star}, {Z}^{\bullet}](x^+;\mathbf{P})$ Eq.~\eqref{eq:BstarZ_exp}-\eqref{eq:BbulletZ_exp}, we get
\begin{multline}
    \mathcal{A}_{\mathrm{SE+TS}}^{+ + - -}=  -\frac{1}{32 \pi^2}\frac{ {\widetilde v}_{12}{\widetilde v}^{\star}_{34} p_{12}^{+2} }{p_1^{+}  p_3^{+}  p_{12}^2}\Bigg[\frac{11}{3} \ln \left(p_{12}^2 e^\gamma \delta\right)-\frac{73}{9}+\frac{p_1^{+} p_2^{+}+p_3^{+} p_4^{+}}{3 p_{12}^{+2}} -S_3\left(p_1, p_2\right) \\  -S_3\left(-p_4,-p_3\right)+  
    \sum_{q^{+}}\left[\frac{1}{q^{+}}+\frac{1}{p_{12}^{+}-q^{+}}\right] \ln \left\{\frac{q^{+}\left(p_{12}^{+}-q^{+}\right)}{p_{12}^{+2}} p_{12}^{2} \delta e^{\gamma}\right\}\Bigg]\\
    -\frac{1}{32 \pi^2}\left(\frac{p_1^{+ 2} p_3^{+ 2}}{p_2^{+} p_4^{+}} \frac{{\widetilde v}_{23} {\widetilde v}^{\star}_{41}}{p_{14}^{+2} p_{14}^2}+ \frac{p_2^{+} p_4^{+}{\widetilde v}^{\star}_{23}{\widetilde v}_{41}}{p_{14}^{+2} p_{14}^2}\right)\Bigg[\frac{11}{3} \ln \left(p_{14}^2 e^\gamma \delta\right)-\frac{73}{9} -S_2\left(-p_4,-p_{23}\right) \\  -S_1\left(p_{14}, p_2\right)+
    \sum_{q^{+}}\left[\frac{1}{q^{+}}+\frac{1}{p_{14}^{+}-q^{+}}\right] \ln \left\{\frac{q^{+}\left(p_{14}^{+}-q^{+}\right)}{p_{14}^{+2}} p_{14}^{2} \delta e^{\gamma}\right\}\Bigg]\,,
    \label{eq:SETS}
\end{multline}
where "SE+TS" stands for self-energy + triangle-swordfish.

Finally, we are left with the 4-point one-loop 1PI contributions. These are shown in Figure~\ref{fig:4MHVOL2}. The contributions consist of three types: box, quartic, and double-quartic. The first two types have already been introduced. The double-quartic are the ones that are made up of two $4\mbox{-}$point interaction vertex (quartic vertex) in the Yang-Mills action as shown below
\begin{center}
    \includegraphics[width=5cm]{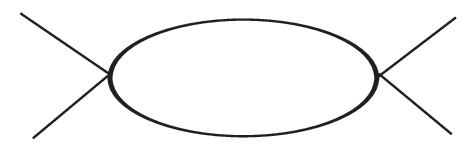}
\end{center}

As before, since the 4-point one-loop 1PI contributions do not involve any field substitution on the external legs, we can use the results for these from \cite{CQT2}. Below we just highlight the necessary details associated with computing these. 

Out of the three types of contribution in Figure~\ref{fig:4MHVOL2}, the most difficult ones to compute are the box diagrams. The quartic and the double-quartic are relatively simple. Notice, the former consists of the $(+ - -)$ and $(+ + -)$ triple gluon vertices connected to the quartic vertex $(+ + - -)$ via three propagator. Thus the numerator of the former consists of the product ${\widetilde v}^{\star}_{ij}{\widetilde v}_{kl}$ originating from the  triple gluon vertices times a rational function of the plus component of momenta. The latter on the other hand consists of two quartic vertices connected via a pair of propagators and thus the numerator is simply a rational function of the plus component of momenta. Both these sets of terms in the CQT scheme give rise to infrared divergent terms. These, however, cancel out against similar contributions originating from the other terms. The $(+ + - -)$ one-loop box terms are much more complicated.  Below, we represent the generic form of the integrand originating from the box diagrams in terms of the region momenta
\begin{equation}
   \mathcal{I}_{\mathrm{box}}^{+ + - -} = \frac{1}{(2 \pi)^4} \frac{\mathrm{Rational} \times \mathrm{N}}{p_1^{+} p_2^{+} p_3^{+} p_4^{+}\left(q-k_0\right)^2\left(q-k_1\right)^2\left(q-k_2\right)^2\left(q-k_3\right)^2}\,.
\end{equation}
where "Rational" represents a rational function of only the plus component of the momenta. Whereas the $\mathrm{N}$ (numerator) is a function of the spinor products. Given the box contributions in Figure \ref{fig:4MHVOL2}, $\mathrm{N}$ takes one of the following two forms
\begin{equation}
    \mathrm{N}_1 = {\widetilde v}^{\star}_{ij}\,{\widetilde v}^{\star}_{kl}\,{\widetilde v}_{mn}\,{\widetilde v}_{pq}\,, \quad\quad \mathrm{N}_2 = {\widetilde v}^{\star}_{ij}\,{\widetilde v}_{kl}\,{\widetilde v}^{\star}_{mn}\,{\widetilde v}_{pq}\,.
\end{equation}
The box integrands where the numerator is of the type $\mathrm{N}_1$ can be completely reduced into triangular contributions plus additional non-vanishing terms just as in the case of $\square^{+ + + +}$. These do not give rise to any additional collinear divergences. However, these do contain poles in $q^+$ which give rise to complicated infrared divergent terms. As stated previously in Subsection \ref{subsec:CQTreview}, such divergences are dealt via discretization which reduces the integral over $q^+$ to a sum. The remaining box integrands where the numerator is of the type $\mathrm{N}_2$ are much more complicated. These cannot be completely reduced in terms of triangular contribution as the former (with numerator $\mathrm{N}_1$) and each of these box diagrams further introduce additional collinear divergent terms. A careful treatment of the reduction procedure revealed \cite{CQT2} that these collinear divergences cancel against similar terms originating from the  reduction of other box diagrams with the similar type of numerator $\mathrm{N}_2$ (there are more than one box diagrams with a numerator of the type $\mathrm{N}_2$). However, the final results does include complicated infrared divergent pieces. Due to this, we refrain from quoting the results for these 4-point one-loop 1PI contributions here. Interestingly, when these are all summed together, most, but not all, of these divergences cancel out. And the result, including Eq.~\eqref{eq:SETS} reads \cite{CQT2}
\begin{multline}
    \mathcal{A}_{\mathrm{one-loop}}^{+ + - -}= \frac{g^2}{8 \pi^2}\left[\log ^2 \frac{p_{12}^2}{p_{14}^2}+\pi^2+\frac{11}{3} \log \left[p_{14}^2 \delta e^\gamma\right]-\frac{73}{9}\right] \mathcal{V}_{\mathrm{tree}}\left(1^+,2^+,3^-,4^-\right) \\ +\frac{g^4}{12 \pi^2} \left[1 -\left(\frac{p_{1}^{+}p_{3}^{+}+p_{2}^{+}p_{4}^{+}}{\left(p_{1}^{+}+p_{4}^{+}\right)^{2}}\right)\right]+\mathcal{A}_{\mathrm{IR}}^{+ + - -}\,,
    \label{eq:4++--amp}
\end{multline} 
where $\mathcal{V}_{\mathrm{tree}}\left(1^+,2^+,3^-,4^-\right)$ is the color-ordered tree level MHV amplitude Eq.~\eqref{eq:MHV_vertex}  and $\mathcal{A}_{\mathrm{IR}}^{+ + - -}$ is the infrared divergent factor. The explicit expression for the latter can be found in Appendix \ref{sec:App_A84}. Notice, the terms in the square bracket on the second line above. These do not have a factor of color-ordered tree level MHV amplitude multiplying to it. Furthermore, the term in the bracket is proportional (modulo $g^2/24\pi^2$) to the 4-point interaction vertex in the Yang-Mills action Eq.~\eqref{eq:YM_LC_action}. These require explicit counterterm in the CQT scheme \cite{CQT2}. Including these, we get
\begin{multline}
    \mathcal{A}_{\mathrm{one-loop}}^{+ + - -}= \frac{g^2}{8 \pi^2}\left[\log ^2 \frac{p_{12}^2}{p_{14}^2}+\pi^2+\frac{11}{3} \log \left[p_{14}^2 \delta e^\gamma\right]-\frac{67}{9}\right] \mathcal{V}_{\mathrm{tree}}\left(1^+,2^+,3^-,4^-\right) \\ +\mathcal{A}_{\mathrm{IR}}^{+ + - -}\,,
    \label{eq:4++--ampct}
\end{multline} 
which is consistent with the known result \cite{Kunszt_1994} (modulo an overall momentum conserving delta and a color factor). Notice, that the consistency check for singular amplitudes is non-trivial as the exact form of divergent one-loop results depends on regularization. However, amplitudes should possess a proper collinear and soft structure. In \cite{CQT2}, the authors computed cross sections using the above results where all the infrared divergent terms cancel out and showed that the results they obtain are in exact agreement with the standard results. 

With this, we conclude that the one-loop effective Z-field action Eq.~\eqref{eq:G_Zac} is indeed one-loop complete with no missing one-loop contributions. 


\chapter{Conclusion}
\label{chap:concl}

In this work, we derived a new classical action for \textit{gluodynmaics} by performing two consecutive canonical transformations to the Yang-Mills action on the light cone. The first transformation maps the Yang-Mills action to the so-called MHV action and in the process eliminates the $(+ + -)$ triple point interaction vertex. The second transformation maps the MHV action to the new action and it eliminates the other triple point interaction vertex i.e. $(+ - -)$. We, however, showed that the new action could also be derived from the Yang-Mills action via a single canonical transformation that eliminates both the triple point interaction vertices at once. Due to this, the lowest multiplicity interaction vertex in the new action is the 4-point MHV $(+ + - -)$. Higher multiplicity interaction vertices include $\mathrm{N}^k\mathrm{MHV}$ where $1\leq k \leq n-4$ with $n$ being the multiplicity. All the vertices are easy to compute because they have a closed analytic form. Using this we computed several split-helicity tree level pure gluonic amplitudes up to 8-point. The highest number of planar diagrams we had were 13. Due to the lack of triple point interaction vertices, the new action allowed for even more efficient computation of pure gluonic tree amplitudes when compared with the MHV action.  Furthermore, when tabulated, we discovered that the number of diagrams we got for the split-helicity tree amplitudes using our action followed the \textit{Delannoy} number series. This allows us both to predict the number of diagrams required to compute these amplitudes using our action and to quantitatively compare against the number of diagrams required in other formalisms, say the MHV action.

An interesting aspect of the new action is that the fields in the action turn out to be Wilson line degrees of freedom. Precisely, the fields in the new action are given by straight infinite Wilson lines of the fields in the MHV action. These lines lie on a plane spanned by $\varepsilon_\perp^-$ and $\eta$, where $\eta={1}/{\sqrt{2}}\left(1,0,0,-1\right)$ and $\varepsilon_{\perp}^{\pm}={1}/{\sqrt{2}}\left(0,1,\pm i,0\right)$. In fact, the exploration of these Wilson lines was triggered by the existence of similar structures for the fields in the MHV action itself. It turns out that the latter is given by straight infinite Wilson lines of the gauge fields in the Yang-Mills action. These, however, live on the perpendicular plane spanned by $\varepsilon_\perp^+$ and $\eta$. Thus, in terms of gauge fields, the field in the new action is "Wilson lines of Wilson lines".  Both the triple point interaction vertices that got eliminated, when deriving the new action, are resummed inside these geometric objects. These, therefore, encode the Self-Dual and the Anti-Self-Dual sector of Yang-Mills. Finally, in the context of scattering amplitudes, going from the Yang-Mills action to the new action the fields get exchanged by the Wilson lines which in turn leads to a new set of bigger building blocks that provide remarkable efficiency in computing tree level amplitudes. This indicates that the Wilson line degrees of freedom are probably better as compared to the fundamental gauge fields in the context of scattering amplitudes.

At the loop level, however, our action is incomplete. There are missing loop contributions which originate either solely from $(+ + -)$ triple point vertex or solely from $(+ - -)$ triple point vertex or the intermixing of these vertices with the other interaction vertices of our action. This is simply reflected by the fact that pure gluonic amplitudes where either all the external gluons have the same helicity or one of them has a different helicity as compared to the rest are all zero in our action to all orders. These amplitudes are indeed zero (in the on-shell limit) at the tree level but are non-zero in general. Similar issues exist for the MHV action as well. Therefore to systematically develop quantum corrections, we use the one-loop effective action approach where we start with the one-loop effective Yang-Mills action and then we perform the field transformations. The one-loop effective Yang-Mills action is by construction one-loop complete. Furthermore, substituting the solution of the transformation takes care of tree level connection involving the triple gluon vertex eliminated via the transformation ($(+ + -)$ in case of MHV action, and both $(+ + -)$ and $(+ - -)$ in the case of our new action).  As a result, the number of diagrams contributing to a given one-loop amplitude is the least in the case of our new action. We used this approach first to develop quantum corrections to the MHV action and demonstrated that there are indeed no missing loop contributions. In fact, we validated the one-loop effective MHV action by explicitly computing the $(+ + + +)$ and $(+ + + -)$ one-loop amplitudes which could not be computed in the MHV theory. After that, we extended it to develop the loop corrections to our new action. To do this, we started with the one-loop effective Yang-Mills action and then performed the field transformations to obtain our new classical action plus one-loop corrections. We validated this action by computing $(+ + + +)$, $(+ + + -)$, $(- - - -)$, $(- - - +)$, and $(+ + - -)$ one-loop amplitudes.

The major issue with the above approach is that the vertices participating in the loop formation are the Yang-Mills vertices. That is the MHV vertices or the interaction vertices of our new action are not manifest in the loop. We, however, demonstrated that when all the one-loop contributions are summed over, these vertices could indeed be realized in the loop. Therefore to make the new interaction vertices manifest in the loop we re-derived the one-loop effective action via a different approach where we first perform the canonical transformation to the Yang-Mills partition function transforming also the current dependent term, and then derive the corresponding one-loop effective actions. We showed that this way the new vertices are manifest and the action is one-loop complete with no missing contributions. We did this both for the MHV action as well as our new action.  The explicit presence of the bigger interaction vertices in the loop makes the computation of higher multiplicity one-loop amplitudes even more efficient when compared with the one-loop effective action derived in the previous where these vertices were not manifest in the loop. Finally, given that the interaction vertices in our action are bigger than those in the MHV action, our one-loop corrected new action should enable much more efficient computation of pure gluonic amplitudes up to one-loop.

\chapter{Outlook}
\label{chap:outlook}

In this section, we highlight some possible long term future research directions emanating from the work done in this thesis
\begin{itemize}
   \item \textit{Supersymmetric extension of our action}: Given that there are no triple point vertices in our action, it allowed for even more efficient computation of higher multiplicity tree amplitudes when compared to the MHV action. However, just like the MHV action, our action had missing loop contributions. $( + + \dots +)$, $(- + \dots +)$, $( - - \dots -)$ and $(- - \dots - +)$ helicity amplitudes are all zero in our action but are not zero in QCD. To overcome this, we used the one-loop effective action approach. However, note that the above stated missing loop amplitudes are zero both at the tree as well as the loop level in $\mathcal{N} = 4$ supersymmetric Yang-Mills (SYM) theory. This indicates that our action is a natural candidate for supersymmetric gauge theory in which case it will be quantum complete with no missing contributions. Exploring different formulations of scattering amplitudes in $\mathcal{N} = 4$ SYM, in the past, has unraveled hidden structures and symmetries underlying particle interactions. In fact, the analytic expressions for tree amplitudes, used for numerical predictions of cross sections in QCD, were first obtained in SYM using the Yangian symmetry. At the loop level, although the amplitudes in SYM and QCD are not equal, the former provides a simple arena to test new ideas and techniques which could later be used to perform similar calculations in QCD. 
    \item \textit{Geometric formulation of scattering amplitudes using the Wilson lines}: One of the key features of our action is that the fields in the action turn out to be Wilson lines extending over the Self-Dual and the Anti-Self-Dual planes. These Wilson lines encode the eliminated Self-Dual and the Anti-Self-Dual sectors. It, therefore, creates a possibility of describing the amplitudes geometrically in terms of these Wilson lines. 
    \item \textit{Our action as a twistor space prescription}: In section \ref{sec:Zac_TS} we presented an intuitive discussion that some of the vertices in our action should localize on curves of different degrees ($d \geq 1$) in the twistor space whereas the remaining ones consist of a set of degeneration of a higher degree curve in terms of lines in the twistor space. This indicates our action goes beyond the MHV action (consisting of degree one curves in twistor space) and provides a new prescription involving higher degree curves. However, the entire discussion was intuitive. Establishing the exact correspondence requires using the whole twistor space machinery, but seems feasible.
     \item \textit{Higher loop corrections to our action}: In this work, we developed the one-loop corrections to our new action using the one-loop effective action approach. When deriving the latter we started with the partition function and then expand the terms in the exponent (action plus the source term) around the classical configuration up to the second order. Considering higher order terms, we can systematically introduce higher loop correction to our action. However, there are challenges already at the one-loop level which must be addressed first. First, the log term in the one-loop effective action is a bit complicated. Due to this, we have not been able to develop a closed analytic form (a master equation) for a generic one-loop contribution in the one-loop action. It is natural to expect that this complication will further increase for higher loop corrections. Second, if we intend to use the CQT scheme, we must first determine the necessary counterterms both for one-loop contributions originating from higher multiplicity vertices and the higher-loop contributions.  
\end{itemize}


\bibliographystyle{JHEP}
\justifying
\bibliography{bibs/sample}

\appendix
\renewcommand{\thechapter}{A\arabic{chapter}}
\chapter{Light-cone Yang-Mills action}
\label{sec:app_A1}

In this appendix, we re-derive the expression for the Yang-Mills action using the "double-null" coordinates in the light-cone gauge \cite{Kotko2017}. We start with the fully covariant form of the $SU(3)$ Yang-Mills action on the Minkowski space Eq.~\eqref{eq:YM_cov}
\begin{equation}
S_{\mathrm{YM}}=\int d^{4}x\,\mathrm{Tr}\left\{ -\frac{1}{4}\hat{F}_{\mu\nu}\hat{F}^{\mu\nu}\right\} \,.
\label{eq:YM_MC}
\end{equation}
Substituting $\hat{F}^{\mu\nu}=\partial^{\mu}\hat{A}^{\nu}-\partial^{\nu}\hat{A}^{\mu}-ig\left[\hat{A}^{\mu},\hat{A}^{\nu}\right]$, we get Eq.~\eqref{eq:YM_action_cov}
\begin{multline}
S_{\mathrm{YM}}=\int d^{4}x\,\mathrm{Tr}\Bigg\{-\frac{1}{2}\left(\partial^{\mu}\hat{A}_{\nu}\right)^{2}+\frac{1}{2}\partial^{\mu}\hat{A}^{\nu}\partial_{\nu}\hat{A}_{\mu}+ig\partial^{\mu}\hat{A}^{\nu}\left[\hat{A}_{\mu},\hat{A}_{\nu}\right]\\
+\frac{1}{4}g^{2}\left[\hat{A}^{\mu},\hat{A}^{\nu}\right]\left[\hat{A}_{\mu},\hat{A}_{\nu}\right]\Bigg\}\,.\label{eq:YM_action_explicit}
\end{multline}
We switch to the double-null coordinates, in which the field components read
\begin{gather} 
 {\hat A}^{+}=\frac{1}{\sqrt{2}} \left({\hat A}^0+{\hat A}^3\right) \,, \, {\hat A}^{-}=\frac{1}{\sqrt{2}} \left({\hat A}^0-{\hat A}^3\right) \,, \nonumber\\
 {\hat A}^{\bullet}=-\frac{1}{\sqrt{2}} \left({\hat A}^1+i{\hat A}^2\right) \,,\,{\hat A}^{\star}=-\frac{1}{\sqrt{2}} \left({\hat A}^1-i{\hat A}^2\right) \,,
\end{gather}
Furthermore, we impose the light-cone gauge ${\hat A}\cdot\eta={\hat A}^{+}=0$. The action reads
\begin{multline}
S_{\mathrm{YM}} = \int\! dx^{+} d^{3}\mathbf{x}\,\,\mathrm{Tr}\Bigg\{-\hat{A}^{\bullet}\square\hat{A}^{\star} - \frac{1}{2}\hat{A}^{-}\partial_{-}^{2}\hat{A}^{-} -\hat{A}^{-}\partial_{-}\partial_{\bullet}\hat{A}^{\bullet}-\hat{A}^{-}\partial_{-}\partial_{\star}\hat{A}^{\star}\\ - \frac{1}{2}\hat{A}^{\bullet}\partial_{\bullet}^{2}\hat{A}^{\bullet}- \frac{1}{2}\hat{A}^{\star}\partial_{\star}^{2}\hat{A}^{\star}
-\hat{A}^{\star}\partial_{\bullet}\partial_{\star}\hat{A}^{\bullet} -ig\,\partial_{-}\hat{A}^{\bullet}\left[\hat{A}^{-},\hat{A}^{\star}\right]-ig\,\partial_{-}\hat{A}^{\star}\left[\hat{A}^{-},\hat{A}^{\bullet}\right]\\ - ig\,\partial_{\bullet}\hat{A}^{\bullet}\left[\hat{A}^{\bullet},\hat{A}^{\star}\right]-ig\,\partial_{\star}\hat{A}^{\star}\left[\hat{A}^{\star},\hat{A}^{\bullet}\right]
+\frac{1}{2}g^{2}\left[\hat{A}^{\bullet},\hat{A}^{\star}\right]\left[\hat{A}^{\star},\hat{A}^{\bullet}\right]\Bigg\}\,,
\end{multline}
where $\square=2\left(\partial_{+}\partial_{-}-\partial_{\bullet}\partial_{\star}\right)$. The action is quadratic in $\hat{A}^-$ field and therefore it can be integrated out of the partition function \cite{Scherk1975}. To see this explicitly, let us collect the terms involving $\hat{A}^-$
\begin{multline}
-\frac{1}{2}\hat{A}^{-}\partial_{-}^{2}\hat{A}^{-}-\hat{A}^{-}\partial_{-}\partial_{\bullet}\hat{A}^{\bullet}-\hat{A}^{-}\partial_{-}\partial_{\star}\hat{A}^{\star}
-ig\left(\partial_{-}A^{\bullet}\left[\hat{A}^{-},\hat{A}^{\star}\right]+\partial_{-}A^{\star}\left[\hat{A}^{-},\hat{A}^{\bullet}\right]\right)\,.
\end{multline}
Notice, except for the first term, all the remaining terms are linear in $\hat{A}^-$. Thus, we can rewrite the above as
\begin{equation}
-\frac{1}{2}\hat{A}^{-}\partial_{-}^{2}\hat{A}^{-}-\hat{G}[A^{\bullet}, A^{\star}]\,\hat{A}^{-}\,,
\label{eq:A_minus}
\end{equation}
where the functional $\hat{G}[A^{\bullet}, A^{\star}]$ is defined as
\begin{equation}
G^{a}[A^{\bullet}, A^{\star}]=\partial_{-}\partial_{\bullet}A^{\bullet a}+\partial_{-}\partial_{\star}A^{\star a}-gf^{abc}\left(\partial_{-}A_{c}^{\bullet}A_{b}^{\star}+\partial_{-}A_{b}^{\bullet}A_{c}^{\star}\right)\,.
\label{eq:GAbulAstar}
\end{equation}
Using $\hat{G}[A^{\bullet}, A^{\star}]$, we perform a field redefinition $\hat{A}^{-}=\hat{K}-\partial_{-}^{-2}\hat{G}[A^{\bullet}, A^{\star}]\,$ that reduces expression \eqref{eq:A_minus} to
\begin{equation}
-\frac{1}{2}\hat{K}\partial_{-}^{2}\hat{K}+\frac{1}{2}\hat{G}[A^{\bullet}, A^{\star}]\,\partial_{-}^{-2}\,\hat{G}[A^{\bullet}, A^{\star}]\,.
\end{equation}
The Jacobian for this redefinition is field-independent and can therefore be absorbed in the normalization of the path integral. The first term above results in a Gaussian integral 
\begin{equation}
\int\left[dK\right]\,\exp\left[\int\! dx^{+} d^{3}\mathbf{x}\,\,\,\mathrm{Tr}\Bigg\{ i\frac{1}{2}\hat{K}\partial_{-}^{2}\hat{K}\Bigg\}\right]=\left(\det\partial_{-}^{2}\right)^{-1/2}\,,
\end{equation}
which too gives rise to a field-independent determinant and can therefore be ignored. The second term survives and the action reduces to the following form
\begin{multline}
S_{\mathrm{YM}}=\int\! dx^{+} d^{3}\mathbf{x}\,\mathrm{Tr}\Bigg\{-\hat{A}^{\bullet}\square\hat{A}^{\star}-\frac{1}{2}\hat{A}^{\bullet}\partial_{\bullet}^{2}\hat{A}^{\bullet}-\frac{1}{2}\hat{A}^{\star}\partial_{\star}^{2}\hat{A}^{\star}-\hat{A}^{\star}\partial_{\bullet}\partial_{\star}\hat{A}^{\bullet}
-ig\partial_{\bullet}\hat{A}^{\bullet}\left[\hat{A}^{\bullet},\hat{A}^{\star}\right]\\
+ig\partial_{\star}\hat{A}^{\star}\left[\hat{A}^{\star},\hat{A}^{\bullet}\right]
+\frac{1}{2}g^{2}\left[\hat{A}^{\bullet},\hat{A}^{\star}\right]\left[\hat{A}^{\star},\hat{A}^{\bullet}\right]+\frac{1}{2}\hat{G}[A^{\bullet}, A^{\star}]\,\partial_{-}^{-2}\,\hat{G}[A^{\bullet}, A^{\star}]\Bigg\}\,.\label{eq:Action_LC}
\end{multline}
Notice, the action depends only on the two transverse field components $\left(\hat{A}^{\bullet},\hat{A}^{\star}\right)$. The above expression, after substituting Eq.~\eqref{eq:GAbulAstar}, can be further simplified via integration by parts with the assumption that the field components vanish at infinity. By doing this, after a bit of massaging, we can compactly re-express the action as 
\begin{equation}
    S_{\mathrm{YM}}=\int\! dx^{+} \left( \mathcal{L}_{+ -} + \mathcal{L}_{+ + -} + \mathcal{L}_{- - +} + \mathcal{L}_{+ + - -}\right) \, ,
    \label{eq:SYM_com}
\end{equation}
where the kinetic term reads
\begin{equation}
\mathcal{L}_{+ -}=\int d^{3}\mathbf{x}\,\mathrm{Tr}\Big\{-\hat{A}^{\bullet}\square\hat{A}^{\star}\Big\}\,.
\label{eq:YM_kin}
\end{equation}
The triple gluon vertices read
\begin{equation}
\mathcal{L}_{+ + -}=-2ig\,\int d^{3}\mathbf{x}\,\mathrm{Tr}\Big\{\partial_{-}^{-1}\partial_{\bullet}\hat{A}^{\bullet}\left[\partial_{-}\hat{A}^{\star},\hat{A}^{\bullet}\right]\Big\}\,,
\label{eq:YM_sdv}
\end{equation}
\begin{equation}
\mathcal{L}_{- - +}=-2ig\,\int d^{3}\mathbf{x}\,\mathrm{Tr}\Big\{\partial_{-}^{-1}\partial_{\star}\hat{A}^{\star}\left[\partial_{-}\hat{A}^{\bullet},\hat{A}^{\star}\right]\Big\}\,.
\end{equation}
Note, in our convention, $\bullet$ represents a plus helicity and $\star$ represents a minus helicity. Finally, the four gluon vertex reads
\begin{equation}
\mathcal{L}_{++--}=-g^{2}\int d^{3}\mathbf{x}\,\mathrm{Tr}\left\{ \left[\partial_{-}\hat{A}^{\bullet},\hat{A}^{\star}\right]\partial_{-}^{-2}\left[\partial_{-}\hat{A}^{\star},\hat{A}^{\bullet}\right]\right\} \,.
\end{equation}

\chapter{Re-deriving the MHV action}
\label{sec:app_A2}

The MHV action can be derived starting with the light-cone Yang-Mills action \eqref{eq:SYM_com} via a canonical field transformation that maps the Yang-Mills fields $\left( {\hat A}^{\bullet}, {\hat A}^{\star} \right)$ to a new pair of fields $\left( {\hat B}^{\bullet}, {\hat B}^{\star} \right)$ \cite{Mansfield2006}. In this appendix, we re-derive the MHV action step by step. In Section \ref{sec:CFR} we discuss the field transformation proposed in \cite{Mansfield2006}. Thereafter in Section \ref{sec:SoFT}, we solve these transformations to obtain solutions which are then used in Section \ref{sec:mhv_action} to explicitly compute the MHV vertices.

\section{The canonical field redefinition}
\label{sec:CFR}
The transformation $\left( {\hat A}^{\bullet}, {\hat A}^{\star} \right) \longrightarrow \left( {\hat B}^{\bullet}, {\hat B}^{\star} \right)$ is defined such that it maps the kinetic term $(+ -)$ and the $(+ + -)$ triple-gluon vertex in the Yang-Mills action \eqref{eq:SYM_com} to the kinetic term in the new action
\begin{equation}
\mathcal{L}_{+ -}\left[A^{\bullet},A^{\star}\right]+\mathcal{L}_{+ + -}\left[A^{\bullet},A^{\star}\right]=\mathcal{L}_{+ -}\left[B^{\bullet},B^{\star}\right]\,.
\label{eq:Mans_tra}
\end{equation}
Note, the field redefinition is on the surface of constant light-cone time $x^+$. It is further assumed to be canonical satisfying the following 
\begin{equation}
B^{\bullet}_a(x^+;\mathbf{x})=B^{\bullet}_a\left[A^{\bullet}\right](x^+;\mathbf{x})\,, \quad \quad
\partial_{-}A_{a}^{\star}(x^+;\mathbf{x})=\int d^{3}\mathbf{y}\,\frac{\delta B_{c}^{\bullet}(x^+;\mathbf{y})}{\delta A_{a}^{\bullet}(x^+;\mathbf{x})}\partial_{-}B_{c}^{\star}(x^+;\mathbf{y})\,,
\label{eq:B_CT_def}
\end{equation}
where the first relation above implies that the ${\hat B}^{\bullet}$ field is a functional of just ${\hat A}^{\bullet}$ field. This in turn renders ${\hat A}^{\star}$ field linear in ${\hat B}^{\star}$ and vice versa. Substituting \eqref{eq:YM_kin} and \eqref{eq:YM_sdv} on the left side of \eqref{eq:Mans_tra} we get
\begin{equation}
\int d^{3}\mathbf{x}\,\mathrm{Tr}\Big\{-\hat{A}^{\bullet}\square\hat{A}^{\star}-2ig\partial_{-}^{-1}\partial_{\bullet}\hat{A}^{\bullet}\left[\partial_{-}\hat{A}^{\star},\hat{A}^{\bullet}\right]\Big\}=\int d^{3}\mathbf{x}\,\mathrm{Tr}\Big\{-\hat{B}^{\bullet}\square\hat{B}^{\star}\Big\}\,.
\label{eq:Transf1}
\end{equation}
Using $\square=2\left(\partial_{+}\partial_{-}-\partial_{\bullet}\partial_{\star}\right)$ and integration by parts, again with the assumption that the field components vanish at infinity, the above expression can be rewritten as 
\begin{multline}
\int d^{3}\mathbf{x}\,\mathrm{Tr}\Big\{2\left(\partial_{+}-\partial_{\bullet}\partial_{\star}\partial_{-}^{-1}\,\right)\hat{A}^{\bullet}(x^+;\mathbf{x})\,\partial_{-}\hat{A}^{\star}(x^+;\mathbf{x})\,\\
-2ig\left[\hat{A}^{\bullet}(x^+;\mathbf{x}),\partial_{-}^{-1}\partial_{\bullet}\hat{A}^{\bullet}(x^+;\mathbf{x})\right]\partial_{-}\hat{A}^{\star}(x^+;\mathbf{x})\Big\} \\
=\int d^{3}\mathbf{x}\,\mathrm{Tr}\Big\{2\left(\partial_{+}-\partial_{\bullet}\partial_{\star}\partial_{-}^{-1}\,\right)\hat{B}^{\bullet}(x^+;\mathbf{x})\,\partial_{-}\hat{B}^{\star}(x^+;\mathbf{x})\,\Big\}\,.
\end{multline}
Substituting for the $\partial_{-}\hat{A}^{\star}(x^+;\mathbf{x})$ using \eqref{eq:B_CT_def} we get
\begin{multline}
\int d^{3}\mathbf{x}\, d^{3}\mathbf{y}\,\mathrm{Tr}\Big\{\left(2\left(\partial_{+}-\partial_{\bullet}\partial_{\star}\partial_{-}^{-1}\right)\hat{A}^{\bullet}(x^+;\mathbf{x})-2ig\left[\hat{A}^{\bullet}(x^+;\mathbf{x}),\partial_{-}^{-1}\partial_{\bullet}\hat{A}^{\bullet}(x^+;\mathbf{x})\right]\right)t^{c}\Big\}\\
\times\frac{\delta B_{a}^{\bullet}(x^+;\mathbf{y})}{\delta A_{c}^{\bullet}(x^+;\mathbf{x})}\partial_{-}B_{a}^{\star}(x^+;\mathbf{y})
=\int d^{3}\mathbf{x}\,\mathrm{Tr}\Big\{2\left(\partial_{+}-\partial_{\bullet}\partial_{\star}\partial_{-}^{-1}\right)\hat{B}^{\bullet}(x^+;\mathbf{x})\partial_{-}\hat{B}^{\star}(x^+;\mathbf{x})\Big\}\,.
\end{multline}
$\partial_{-}{B}_a^{\star}(x^+;\mathbf{x})$ can be factored out of the above expression via renaming the dummy variables $\mathbf{x} \leftrightarrow \mathbf{y}$ on the L.H.S. Furthermore, since ${B}_a^{\star}(x^+;\mathbf{x})$ is arbitrary we get
\begin{multline}
\int d^{3}\mathbf{y}\,\mathrm{Tr}\Big\{\left(2\left(\partial_{+}-\partial_{\bullet}\partial_{\star}\partial_{-}^{-1}\right)\hat{A}^{\bullet}(x^+;\mathbf{y})-2ig\left[\hat{A}^{\bullet}(x^+;\mathbf{y}),\partial_{-}^{-1}\partial_{\bullet}\hat{A}^{\bullet}(x^+;\mathbf{y})\right]\right)t^{c}\Big\}\\
\times\frac{\delta B_{a}^{\bullet}(x^+;\mathbf{x})}{\delta A_{c}^{\bullet}(x^+;\mathbf{y})}
=2\left(\partial_{+}-\partial_{\bullet}\partial_{\star}\partial_{-}^{-1}\right)B_a^{\bullet}(x^+;\mathbf{x})\,.
\label{eq:exp_in}
\end{multline}
As stated previously, expression \eqref{eq:B_CT_def} implies $\hat{B}^{\bullet}$ is a functional of just $\hat{A}^{\bullet}$. Due to this any dependence of $\hat{B}^{\bullet}$ on $x^+$ is implicit through $\hat{A}^{\bullet}$. Thus we have
\begin{equation}
 \partial_{+}B_{a}^{\bullet}(x^+;\mathbf{x}) = \int d^{3}\mathbf{y}\,\frac{\delta B_{a}^{\bullet}(x^+;\mathbf{x})}{\delta A_{c}^{\bullet}(x^+;\mathbf{y})}\partial_{+}A_{c}^{\bullet}(x^+;\mathbf{y})\,.
\end{equation}
These terms cancel out in expression \eqref{eq:exp_in} and we get the final expression
\begin{multline}   \partial_{\bullet}\partial_{\star}\partial_{-}^{-1}B_a^{\bullet}(x^+;\mathbf{x}) = \int d^{3}\mathbf{y}\,\mathrm{Tr}\Big\{\left(\partial_{\bullet}\partial_{\star}\partial_{-}^{-1}\hat{A}^{\bullet}(x^+;\mathbf{y})\right.\\
  \left. +ig\left[\hat{A}^{\bullet}(x^+;\mathbf{y}),\partial_{-}^{-1}\partial_{\bullet}\hat{A}^{\bullet}(x^+;\mathbf{y})\right]\right)t^{c}\Big\}\frac{\delta B_{a}^{\bullet}(x^+;\mathbf{x})}{\delta A_{c}^{\bullet}(x^+;\mathbf{y})}
\,. \label{eq:MT_abulb}
\end{multline}
Equations \eqref{eq:B_CT_def} and \eqref{eq:MT_abulb} govern the field transformation $\left( {\hat A}^{\bullet}, {\hat A}^{\star} \right) \longrightarrow \left( {\hat B}^{\bullet}, {\hat B}^{\star} \right)$.

\section{Solution of the field transformation}
\label{sec:SoFT}

In this appendix, we re-derive the solution for the field transformations \eqref{eq:B_CT_def} and \eqref{eq:MT_abulb} in the momentum space following closely \cite{Ettle2006b}.
\subsection{Solution for \texorpdfstring{${\hat A}^{\bullet}[B^{\bullet}]$}{Abullet}}
\label{subsec:A_bul_deri}

In order to obtain the solution for ${\hat A}^{\bullet}[B^{\bullet}]$, we need to solve 
\begin{multline}   \partial_{\bullet}\partial_{\star}\partial_{-}^{-1}B_a^{\bullet}(x^+;\mathbf{x}) = \int d^{3}\mathbf{y}\,\mathrm{Tr}\Big\{\left(\partial_{\bullet}\partial_{\star}\partial_{-}^{-1}\hat{A}^{\bullet}(x^+;\mathbf{y})\right.\\
  \left. +ig\left[\hat{A}^{\bullet}(x^+;\mathbf{y}),\partial_{-}^{-1}\partial_{\bullet}\hat{A}^{\bullet}(x^+;\mathbf{y})\right]\right)t^{c}\Big\}\frac{\delta B_{a}^{\bullet}(x^+;\mathbf{x})}{\delta A_{c}^{\bullet}(x^+;\mathbf{y})}
\,, \label{eq:MT_abulb1}
\end{multline}
keeping in mind that $\hat{B}^{\bullet}$ is a functional of just $\hat{A}^{\bullet}$ and vice versa. The above expression can be rewritten as
\begin{multline}   \int d^{3}\mathbf{x}\,\partial_{\bullet}\partial_{\star}\partial_{-}^{-1}B_a^{\bullet}(x^+;\mathbf{x})\frac{\delta A_{c}^{\bullet}(x^+;\mathbf{y})}{\delta B_{a}^{\bullet}(x^+;\mathbf{x})} = \partial_{\bullet}\partial_{\star}\partial_{-}^{-1}A^{\bullet}_c(x^+;\mathbf{y})\\
   -gf^{abc}A^{\bullet}_a(x^+;\mathbf{y})\,\partial_{-}^{-1}\partial_{\bullet}A^{\bullet}_b(x^+;\mathbf{y})
\,,
\end{multline}
which in momentum space reads
\begin{multline}   i\int d^{3}\mathbf{q}\,\,\hat{q}\,\frac{\delta {\widetilde A}_{c}^{\bullet}(x^+;\mathbf{P})}{\delta {\widetilde B}_{a}^{\bullet}(x^+;\mathbf{q})} {\widetilde B}_a^{\bullet}(x^+;\mathbf{q}) = i\,\hat{P}\,{\widetilde A}^{\bullet}_c(x^+;\mathbf{P})\\
   +\frac{g}{2}f^{abc}\int d^{3}\mathbf{q}_1\,d^{3}\mathbf{q}_2\, \delta^3(\mathbf{q}_1 + \mathbf{q}_2 - \mathbf{P})\Big(\frac{q_1^{\star}}{q_1^+} -\frac{q_2^{\star}}{q_2^+}\Big){\widetilde A}^{\bullet}_a(x^+;\mathbf{q}_1)\,{\widetilde A}^{\bullet}_b(x^+;\mathbf{q}_2)
\,, \label{eq:MT_abulb2}
\end{multline}
where
\begin{equation}
    \hat{q} = q^{\bullet} q^{\star}/ q^{+}\,.
\end{equation}
It is straightforward to see that at first order, assuming $\hat{P} \neq 0$, we have  ${\widetilde A}^{\bullet}_c(x^+;\mathbf{P}) = {\widetilde B}^{\bullet}_c(x^+;\mathbf{P})$. However, more general solutions where the first-order term is independent of ${\widetilde B}^{\bullet}_c(x^+;\mathbf{P})$  are possible. A detailed discussion of this can be found in \cite{Ettle2006b}. For current purposes, such terms are not necessary and we can postulate the following solution
\begin{multline}
\widetilde{A}^{\bullet}_a(x^+;\mathbf{P}) = \widetilde{B}^{\bullet}_a(x^+;\mathbf{P})\\
+\sum_{n=2}^{\infty} 
    \int d^3\mathbf{p}_1\dots d^3\mathbf{p}_n \, \widetilde{\Psi}_n^{a\{b_1\dots b_n\}}(\mathbf{P};\{\mathbf{p}_1,\dots ,\mathbf{p}_n\}) \prod_{i=1}^n\widetilde{B}^{\bullet}_{b_i}(x^+;\mathbf{p}_i)\,.
    \label{eq:A_bull_exp}
\end{multline}
At this point, the curly braces for momentum and color in the kernel 
$\widetilde{\Psi}_n^{a\{b_1\dots b_n\}}(\mathbf{P};\{\mathbf{p}_1,\dots ,\mathbf{p}_n\})$ may appear ambiguous. This will become clear later. 

We need to solve for the kernels $\widetilde{\Psi}_n^{a\{b_1\dots b_n\}}(\mathbf{P};\{\mathbf{p}_1,\dots ,\mathbf{p}_n\})$ in order to determine the solution Eq.~\eqref{eq:A_bull_exp}. This is achieved by substituting Eq.~\eqref{eq:A_bull_exp} to Eq.~\eqref{eq:MT_abulb2} and then equating the same order (in fields) terms on the L.H.S and R.H.S.  The first order term is trivial, as stated above. For the second order, we have
\begin{multline}   i\,\int d^3\mathbf{p}_1 d^3\mathbf{p}_2 \,(\hat{p}_1 + \hat{p}_2)\widetilde{\Psi}_2^{a\{b_1b_2\}}(\mathbf{P};\{\mathbf{p}_1,\mathbf{p}_2\}) \widetilde{B}^{\bullet}_{b_1}(x^+;\mathbf{p}_1) {\widetilde B}_{b_2}^{\bullet}(x^+;\mathbf{p}_2) \\
= i\int d^3\mathbf{p}_1 d^3\mathbf{p}_2 \,\,\hat{P}\,\widetilde{\Psi}_2^{a\{b_1b_2\}}(\mathbf{P};\{\mathbf{p}_1,\mathbf{p}_2\}) \widetilde{B}^{\bullet}_{b_1}(x^+;\mathbf{p}_1) {\widetilde B}_{b_2}^{\bullet}(x^+;\mathbf{p}_2)\\
   +\frac{g}{2}f^{ab_1 b_2}\int d^{3}\mathbf{p}_1\,d^{3}\mathbf{p}_2\, \delta^3(\mathbf{p}_1 + \mathbf{p}_2 - \mathbf{P})\Big(\frac{p_1^{\star}}{p_1^+} -\frac{p_2^{\star}}{p_2^+}\Big){\widetilde B}^{\bullet}_{b_1}(x^+;\mathbf{p}_1)\,{\widetilde B}^{\bullet}_{b_2}(x^+;\mathbf{p}_2)
\,. \label{eq:abulb2}
\end{multline}
Using the identity $if^{ab_1 b_2}=\mathrm{Tr}(t^a t^{b_1} t^{b_2})-\mathrm{Tr}(t^a t^{b_2} t^{b_1})$ we get
\begin{equation}
    \widetilde{\Psi}_2^{a\{b_1b_2\}}(\mathbf{P};\{\mathbf{p}_1,\mathbf{p}_2\})= -\frac{g}{2}\,\big[\mathrm{Tr}(t^a t^{b_1} t^{b_2})-\mathrm{Tr}(t^a t^{b_2} t^{b_1})\big]\frac{\delta^3(\mathbf{p}_1 + \mathbf{p}_2 - \mathbf{P})}{\hat{p}_1 + \hat{p}_2 -\hat{P}}\frac{\widetilde{v}_{21}}{p_2^+}\,.
\end{equation}
Rewriting the denominator $p_{12}^+(\hat{p}_1 + \hat{p}_2 -\hat{P}) = - \widetilde{v}_{21}\widetilde{v}^{\star}_{12} $ we get
\begin{multline}
    \widetilde{\Psi}_2^{a\{b_1b_2\}}(\mathbf{P};\{\mathbf{p}_1,\mathbf{p}_2\})= -\frac{g}{2}\,\Bigg( \frac{\delta^3(\mathbf{p}_1 + \mathbf{p}_2 - \mathbf{P})}{\widetilde{v}^{\star}_{21}}\frac{p_{12}^+}{p_1^+}\mathrm{Tr}(t^a t^{b_1} t^{b_2}) \\
    + \frac{\delta^3(\mathbf{p}_1 + \mathbf{p}_2 - \mathbf{P})}{\widetilde{v}^{\star}_{12}}\frac{p_{12}^+}{p_2^+}\mathrm{Tr}(t^a t^{b_2} t^{b_1})\Bigg)\,.
    \label{eq:sym_psi2}
\end{multline}
Note, since the kernel above is integrated with respect to the 3-momenta of the associated fields in the solution Eq.~\eqref{eq:A_bull_exp}, it is possible to rename the 3-momenta and the color of fields in such a way that the two terms above collapse into one. For instance, if we rename $2\leftrightarrow 1$ in the second term above we get
\begin{equation}
    \widetilde{\Psi}_2^{a\{b_1b_2\}}(\mathbf{P};\{\mathbf{p}_1,\mathbf{p}_2\})= -g\, \frac{\delta^3(\mathbf{p}_1 + \mathbf{p}_2 - \mathbf{P})}{\widetilde{v}^{\star}_{21}}\frac{p_{12}^+}{p_1^+}\mathrm{Tr}(t^a t^{b_1} t^{b_2})\,.
    \label{eq:psi_2_od}
\end{equation}
On the contrary, the single term above can be symmetrized with respect to the momentum and color to obtain Eq.~\eqref{eq:sym_psi2}. We use curly braces in the definition of the kernels to indicate this symmetry with respect to the interchange of momentum and color of the associated fields.

For the third order in fields, substituting Eq.~\eqref{eq:A_bull_exp} to Eq.~\eqref{eq:MT_abulb2} we get
\begin{multline}   
i\int d^3\mathbf{p}_1 d^3\mathbf{p}_2 d^3\mathbf{p}_3\,(\hat{p}_1 + \hat{p}_2+ \hat{p}_3 -\hat{P})\widetilde{\Psi}_3^{a\{b_1b_2b_3\}}(\mathbf{P};\{\mathbf{p}_1,\mathbf{p}_2,\mathbf{p}_3\}) \widetilde{B}^{\bullet}_{b_1}(x^+;\mathbf{p}_1) \dots {\widetilde B}_{b_3}^{\bullet}(x^+;\mathbf{p}_3) \\
= \frac{g}{2}f^{ac_1 c_2}\int d^{3}\mathbf{q}_1\,d^{3}\mathbf{q}_2\, \delta^3(\mathbf{q}_1 + \mathbf{q}_2 - \mathbf{P})\Big(\frac{q_1^{\star}}{q_1^+} -\frac{q_2^{\star}}{q_2^+}\Big)\Big[\int d^3\mathbf{p}_1 d^3\mathbf{p}_2 \widetilde{\Psi}_2^{c_1\{b_1b_2\}}(\mathbf{q}_1;\{\mathbf{p}_1,\mathbf{p}_2\}) \\
\widetilde{B}^{\bullet}_{b_1}(x^+;\mathbf{p}_1) {\widetilde B}_{b_2}^{\bullet}(x^+;\mathbf{p}_2) {\widetilde B}^{\bullet}_{c_2}(x^+;\mathbf{q}_2)+{\widetilde B}^{\bullet}_{c_1}(x^+;\mathbf{q}_1)\int d^3\mathbf{p}_2 d^3\mathbf{p}_3 \,\widetilde{\Psi}_2^{c_2\{b_2b_3\}}(\mathbf{q}_2;\{\mathbf{p}_2,\mathbf{p}_3\}) \\
\widetilde{B}^{\bullet}_{b_2}(x^+;\mathbf{p}_2) {\widetilde B}_{b_3}^{\bullet}(x^+;\mathbf{p}_3)\Big]\,. 
\label{eq:abulb3}
\end{multline}
At this point, we can either derive the fully symmetrized kernel or just the ordered kernel (with a particular ordering of color and momenta). For former, one needs to fully symmetrize the R.H.S expression above. To achieve this, we need to rename the color and momenta in two more ways by introducing an overall factor of $1/3$. For instance, the kernel $\widetilde{\Psi}_2^{c_1\{b_1b_2\}}(\mathbf{q}_1;\{\mathbf{p}_1,\mathbf{p}_2\})$ in the first term on R.H.S can be renamed as $\widetilde{\Psi}_2^{c_1\{b_2b_3\}}(\mathbf{q}_1;\{\mathbf{p}_2,\mathbf{p}_3\})$ or $\widetilde{\Psi}_2^{c_1\{b_3b_1\}}(\mathbf{q}_1;\{\mathbf{p}_3,\mathbf{p}_1\})$ and similarly for the second term. Just for the sake of simplicity, below we derive the expression for the ordered kernel. Renaming $\left\{\mathbf{q}_2, c_2\right\}\longrightarrow \left\{\mathbf{p}_3, b_3\right\}$ and $\left\{\mathbf{q}_1, c_1\right\}\longrightarrow \left\{\mathbf{p}_1, b_1\right\}$ respectively to the first and the second term on the R.H.S of the expression above we get
\begin{multline}
    \widetilde{\Psi}_3^{a\{b_1b_2b_3\}}(\mathbf{P};\{\mathbf{p}_1,\mathbf{p}_2,\mathbf{p}_3\}) = g^2\frac{\delta^3(\mathbf{p}_{123} - \mathbf{P})}{(\hat{p}_1 + \hat{p}_2+ \hat{p}_3 -\hat{P})}\Bigg[
    \frac{\widetilde{v}_{3(12)} p_{12}^+}{p_3^+ \widetilde{v}^{\star}_{21} p_1^+}+\frac{\widetilde{v}_{(23)1}}{ \widetilde{v}^{\star}_{32} p_2^+} \Bigg]\mathrm{Tr}(t^a t^{b_1} t^{b_2} t^{b_3})\,,
\end{multline}
where $\mathbf{p}_{123} = \mathbf{p}_1 + \mathbf{p}_2 + \mathbf{p}_3$. With a bit of algebra, the above expression can be rewritten as
\begin{multline}
    \widetilde{\Psi}_3^{a\{b_1b_2b_3\}}(\mathbf{P};\{\mathbf{p}_1,\mathbf{p}_2,\mathbf{p}_3\}) = g^2\frac{\delta^3(\mathbf{p}_{123} - \mathbf{P})}{(\hat{p}_1 + \hat{p}_2+ \hat{p}_3 -\hat{P})}\\
    \frac{p_{123}^+}{p_1^+}\frac{1}{\widetilde{v}^{\star}_{32}\widetilde{v}^{\star}_{21}}\Bigg[
    \frac{\widetilde{v}_{12(123)} \widetilde{v}^{\star}_{23}}{p_2^+ }+\frac{\widetilde{v}_{1(123)}\widetilde{v}^{\star}_{12}}{p_1^+} \Bigg]\mathrm{Tr}(t^a t^{b_1} t^{b_2} t^{b_3})\,.
    \label{eq:psi_3_in}
\end{multline}
Using the following identity
\cite{Motyka2009}:
\begin{equation}
(\hat{p}_1 + \hat{p}_2+ \dots \hat{p}_n -\hat{p}_{1\dots n})=\sum_{i=1}^{n-1}\frac{\widetilde{v}^{\star}_{i\left(i+1\right)}}{p_{i}^{+}}\,\widetilde{v}_{\left(1\dots i\right)\left(1\dots n\right)}\,,\label{eq:MS_ID}
\end{equation}
expression Eq.~\eqref{eq:psi_3_in} simplifies to
\begin{equation}
    \widetilde{\Psi}_3^{a\{b_1b_2b_3\}}(\mathbf{P};\{\mathbf{p}_1,\mathbf{p}_2,\mathbf{p}_3\}) = g^2\delta^3(\mathbf{p}_{123} - \mathbf{P})
    \frac{p_{123}^+}{p_1^+}\frac{1}{\widetilde{v}^{\star}_{32}\widetilde{v}^{\star}_{21}}\mathrm{Tr}(t^a t^{b_1} t^{b_2} t^{b_3})\,.
    \label{eq:psi_3}
\end{equation}

The above result for ordered (in color and momentum) kernels can be generalized as 
\begin{multline}
    \widetilde{\Psi}_n^{a\{b_1b_2 \dots b_n\}}(\mathbf{P};\{\mathbf{p}_1,\mathbf{p}_2, \dots \mathbf{p}_n\}) = (-g)^{n-1}\delta^3(\mathbf{p}_{12 \dots n} - \mathbf{P})\\
    \frac{p_{12 \dots n}^+}{p_1^+}\frac{1}{\widetilde{v}^{\star}_{n\, n-1} \dots \widetilde{v}^{\star}_{32}\widetilde{v}^{\star}_{21}}\mathrm{Tr}(t^a t^{b_1} t^{b_2} t^{b_3})\,.
    \label{eq:psi_n}
\end{multline}
Owing to the integration over the momentum of associated fields in the solution Eq.~\eqref{eq:A_bull_exp}, the kernels can be rewritten in fully symmetrized form as
\begin{multline}
    \widetilde{\Psi}_n^{a\{b_1b_2 \dots b_n\}}(\mathbf{P};\{\mathbf{p}_1,\mathbf{p}_2, \dots \mathbf{p}_n\}) = \frac{(-g)^{n-1}}{n!}\delta^3(\mathbf{p}_{12 \dots n} - \mathbf{P})\\
    \!\!\sum_{\text{\scriptsize permutations}}
    \frac{p_{12 \dots n}^+}{p_1^+}\frac{1}{\widetilde{v}^{\star}_{n\, n-1} \dots \widetilde{v}^{\star}_{32}\widetilde{v}^{\star}_{21}}\mathrm{Tr}(t^a t^{b_1} t^{b_2} t^{b_3})\,.
    \label{eq:psi_kernel_der}
\end{multline}
where the sum is over all the permutations of $(1,2 \dots, n)$.
\subsection{Solution for \texorpdfstring{${\hat A}^{\star}[B^{\bullet}, B^{\star}]$}{Astar}}
\label{subsec:A_star_deri}

In order to obtain a solution for ${\hat A}^{\star}[B^{\bullet}, B^{\star}]$, we need to solve the following 
\begin{equation}
\partial_{-}A_{a}^{\star}(x^+;\mathbf{x})=\int d^{3}\mathbf{y}\,\frac{\delta B_{c}^{\bullet}(x^+;\mathbf{y})}{\delta A_{a}^{\bullet}(x^+;\mathbf{x})}\partial_{-}B_{c}^{\star}(x^+;\mathbf{y})\,.
\end{equation}
The above expression can be rearranged such that we have a functional derivative of the ${\widetilde A}^{\bullet}[B^{\bullet}]$ field. Doing this, in momentum space we get 
\begin{equation}
P^+{\widetilde B}_{a}^{\star}(x^+;\mathbf{P})=\int d^{3}\mathbf{q}\,\frac{\delta {\widetilde A}_{c}^{\bullet}(x^+;\mathbf{q})}{\delta {\widetilde B}_{a}^{\bullet}(x^+;\mathbf{P})}q^+{\widetilde A}_{c}^{\star}(x^+;\mathbf{q})\,.
\label{eq:B_star_def}
\end{equation}
In the previous subsection, we derived
the solution for ${\widetilde A}^{\bullet}[B^{\bullet}]$ field Eq.~\eqref{eq:A_bull_exp}. At first order in fields we had ${\widetilde A}^{\bullet}_c(x^+;\mathbf{q}) = {\widetilde B}^{\bullet}_c(x^+;\mathbf{q})$. Substituting this in  Eq.~\eqref{eq:B_star_def} we obtain
\begin{equation}
P^+{\widetilde B}_{a}^{\star}(x^+;\mathbf{P})=\int d^{3}\mathbf{q}\,\delta^a_c \,\delta^3(\mathbf{q} - \mathbf{P})q^+{\widetilde A}_{c}^{\star}(x^+;\mathbf{q})\,.
\end{equation}
Therefore, we see that even for ${\widetilde A}^{\star}[B^{\bullet}, B^{\star}]$ at first order we have ${\widetilde A}^{\star}_a(x^+;\mathbf{P}) = {\widetilde B}^{\star}_a(x^+;\mathbf{P})$. Since ${\widetilde A}^{\bullet}[B^{\bullet}]$ is a series only in ${\widetilde B}^{\bullet}$ fields, Eq.~\eqref{eq:B_star_def} implies ${\widetilde A}^{\star}[B^{\bullet}, B^{\star}]$ must be linear in ${\widetilde B}^{\star}$ fields. As a result, we can postulate the following series solution in momentum space
\begin{multline}
\widetilde{A}^{\star}_a(x^+;\mathbf{P}) = {\widetilde B}^{\star}_a(x^+;\mathbf{P})\\
+ \sum_{n=2}^{\infty} 
    \int d^3\mathbf{p}_1\dots d^3\mathbf{p}_n \, {\widetilde \Omega}_{n}^{a b_1 \left \{b_2 \cdots b_n \right \} }(\mathbf{P}; \mathbf{p}_1 ,\left \{ \mathbf{p}_2 , \dots ,\mathbf{p}_n \right \}) \widetilde{B}^{\star}_{b_1}(x^+;\mathbf{p}_1)\prod_{i=2}^n\widetilde{B}^{\bullet}_{b_i}(x^+;\mathbf{p}_i)\, .
    \label{eq:A_star_exp1}
\end{multline}
Notice the use of curly braces in ${\widetilde \Omega}_{n}^{a b_1 \left \{b_2 \cdots b_n \right \} }(\mathbf{P}; \mathbf{p}_1 ,\left \{ \mathbf{p}_2 , \dots ,\mathbf{p}_n \right \})$ only for certain momentum and color indices. As stated in the previous subsection, these braces represent symmetrization with respect to the interchange of the momentum and color indices. However, as we shall see, unlike the $\widetilde{\Psi}_n^{a\{b_1b_2 \dots b_n\}}(\mathbf{P};\{\mathbf{p}_1,\mathbf{p}_2, \dots \mathbf{p}_n\})$ kernels, now this symmetry is no longer applicable for all the indices.  Furthermore, we fix the $\widetilde{B}^{\star}$ field to $\widetilde{B}^{\star}_{b_1}(x^+;\mathbf{p}_1)$ for all orders in the series. This is not a necessity. We do this just for the sake of simplicity. 

In order to determine the kernels ${\widetilde \Omega}_{n}^{a b_1 \left \{b_2 \cdots b_n \right \} }(\mathbf{P}; \mathbf{p}_1 ,\left \{ \mathbf{p}_2 , \dots ,\mathbf{p}_n \right \})$, we substitute Eq.~\eqref{eq:A_bull_exp} and Eq.~\eqref{eq:A_star_exp1} to Eq.~\eqref{eq:B_star_def} and equate terms with same number of fields. For the second order, we have
\begin{multline}
\int d^3\mathbf{P} d^3\mathbf{p}_1 d^3\mathbf{p}_2 \,\widetilde{\Psi}_2^{a\{b_1b_2\}}(\mathbf{P};\{\mathbf{p}_1,\mathbf{p}_2\}) \Big[\delta^c_{b_1} \,\delta^3(\mathbf{q} - \mathbf{p}_1) {\widetilde B}_{b_2}^{\bullet}(x^+;\mathbf{p}_2)\\
+\widetilde{B}^{\bullet}_{b_1}(x^+;\mathbf{p}_1)\delta^c_{b_2} \,\delta^3(\mathbf{q} - \mathbf{p}_2)\Big]P^+{\widetilde B}_{a}^{\star}(x^+;\mathbf{P})\\
+\int d^3\mathbf{p}_1 d^3\mathbf{p}_2 \, q^+{\widetilde \Omega}_{2}^{c b_1 \left\{b_2\right\}}(\mathbf{q}; \mathbf{p}_1 ,\left\{\mathbf{p}_2\right\}) \widetilde{B}^{\star}_{b_1}(x^+;\mathbf{p}_1)\widetilde{B}^{\bullet}_{b_2}(x^+;\mathbf{p}_2)=0\,.
\end{multline}
The curly brace over just one index is redundant and can therefore be dropped. Simplifying the above expression, we get
\begin{multline}
\int d^3\mathbf{P} d^3\mathbf{p}_2 \,\widetilde{\Psi}_2^{a\{c b_2\}}(\mathbf{P};\{\mathbf{q},\mathbf{p}_2\})  {\widetilde B}_{b_2}^{\bullet}(x^+;\mathbf{p}_2)P^+{\widetilde B}_{a}^{\star}(x^+;\mathbf{P})\\
+\int d^3\mathbf{P} d^3\mathbf{p}_1  \,\widetilde{\Psi}_2^{a\{b_1c\}}(\mathbf{P};\{\mathbf{p}_1,\mathbf{q}\})\widetilde{B}^{\bullet}_{b_1}(x^+;\mathbf{p}_1)P^+{\widetilde B}_{a}^{\star}(x^+;\mathbf{P})\\
+\int d^3\mathbf{p}_1 d^3\mathbf{p}_2 \, q^+{\widetilde \Omega}_{2}^{c b_1 b_2}(\mathbf{q}; \mathbf{p}_1 ,\mathbf{p}_2) \widetilde{B}^{\star}_{b_1}(x^+;\mathbf{p}_1)\widetilde{B}^{\bullet}_{b_2}(x^+;\mathbf{p}_2)=0\,.
\end{multline}
Substituting for $\widetilde{\Psi}_2$ using Eq.~\eqref{eq:psi_2_od} we get
\begin{multline}
\int d^3\mathbf{P} d^3\mathbf{p}_2 \,\left( -g\, \frac{\delta^3(\mathbf{q} + \mathbf{p}_2 - \mathbf{P})}{\widetilde{v}^{\star}_{2q}}\frac{p_{q2}^+}{q^+}\mathrm{Tr}(t^a t^{c} t^{b_2})\right) {\widetilde B}_{b_2}^{\bullet}(x^+;\mathbf{p}_2)P^+{\widetilde B}_{a}^{\star}(x^+;\mathbf{P})\\
+\int d^3\mathbf{P} d^3\mathbf{p}_1  \,\left(-g\, \frac{\delta^3(\mathbf{p}_1 + \mathbf{q} - \mathbf{P})}{\widetilde{v}^{\star}_{q1}}\frac{p_{1q}^+}{p_1^+}\mathrm{Tr}(t^a t^{b_1} t^{c})\right)\widetilde{B}^{\bullet}_{b_1}(x^+;\mathbf{p}_1)P^+{\widetilde B}_{a}^{\star}(x^+;\mathbf{P})\\
+\int d^3\mathbf{p}_1 d^3\mathbf{p}_2 \, q^+{\widetilde \Omega}_{2}^{c b_1 b_2}(\mathbf{q}; \mathbf{p}_1 ,\mathbf{p}_2) \widetilde{B}^{\star}_{b_1}(x^+;\mathbf{p}_1)\widetilde{B}^{\bullet}_{b_2}(x^+;\mathbf{p}_2)=0\,.
\end{multline}
Renaming the terms such that we have $\widetilde{B}^{\star}_{b_1}(x^+;\mathbf{p}_1)\widetilde{B}^{\bullet}_{b_2}(x^+;\mathbf{p}_2)$ in all of them, the above expression can be rewritten as
\begin{multline}
-\int d^3\mathbf{p}_1 d^3\mathbf{p}_2 \,\left( -g\, \frac{\delta^3(\mathbf{q} - \mathbf{p}_2 - \mathbf{p}_1)}{p_2^+\widetilde{v}^{\star}_{12}}\frac{p_1^+ p_{12}^+}{p_{12}^+}\mathrm{Tr}(t^{c} t^{b_2} t^{b_1})\right) {\widetilde B}_{b_2}^{\bullet}(x^+;\mathbf{p}_2)p_1^+{\widetilde B}_{b_1}^{\star}(x^+;\mathbf{p}_1)\\
-\int d^3\mathbf{p}_1 d^3\mathbf{p}_2  \,\left(-g\, \frac{\delta^3( \mathbf{q} -\mathbf{p}_2 - \mathbf{p}_1)}{\widetilde{v}^{\star}_{21}}\frac{p_1^+ p_{12}^+}{p_1^+ p_{12}^+}\mathrm{Tr}(t^{c}t^{b_1} t^{b_2} )\right)\widetilde{B}^{\bullet}_{b_2}(x^+;\mathbf{p}_2)p_1^+{\widetilde B}_{b_1}^{\star}(x^+;\mathbf{p}_1)\\
+\int d^3\mathbf{p}_1 d^3\mathbf{p}_2 \, p_{12}^+{\widetilde \Omega}_{2}^{c b_1 b_2}(\mathbf{q}; \mathbf{p}_1 ,\mathbf{p}_2) \widetilde{B}^{\star}_{b_1}(x^+;\mathbf{p}_1)\widetilde{B}^{\bullet}_{b_2}(x^+;\mathbf{p}_2)=0\,.
\end{multline}
Thus, we get the following relation
\begin{multline}
    {\widetilde \Omega}_{2}^{c b_1 b_2}(\mathbf{q}; \mathbf{p}_1 ,\mathbf{p}_2)= -g\,\left(\frac{p_1^+}{p_{12}^+}\right)^2 \Bigg[  \frac{\delta^3(\mathbf{q} - \mathbf{p}_2 - \mathbf{p}_1)}{\widetilde{v}^{\star}_{12}}\frac{ p_{12}^+}{p_{2}^+}\mathrm{Tr}(t^{c} t^{b_2} t^{b_1})\\
    +  \frac{\delta^3( \mathbf{q} -\mathbf{p}_2 - \mathbf{p}_1)}{\widetilde{v}^{\star}_{21}}\frac{ p_{12}^+}{p_1^+}\mathrm{Tr}(t^{c}t^{b_1} t^{b_2} )\Bigg]\,.
    \label{eq:Omega_2_der}
\end{multline}
Comparing the terms in the square brackets above with those in Eq.~\eqref{eq:sym_psi2}, it is tempting to rewrite them as $ \widetilde{\Psi}_2^{a\{b_1b_2\}}(\mathbf{P};\{\mathbf{p}_1,\mathbf{p}_2\})$. This will, however, be a misuse of notation because the kernel ${\widetilde \Omega}_{2}^{c b_1 b_2}(\mathbf{q}; \mathbf{p}_1,\mathbf{p}_2)$ is not symmetric under the renaming $2\leftrightarrow 1$ as evident from Eq.~\eqref{eq:Omega_2_der}. Therefore, we introduce the following notation
\begin{multline}
    \widetilde{\Psi}_n^{a b_1b_2 \dots b_n}(\mathbf{P};\mathbf{p}_1,\mathbf{p}_2, \dots \mathbf{p}_n) = \frac{(-g)^{n-1}}{n!}\delta^3(\mathbf{p}_{12 \dots n} - \mathbf{P})\\
    \!\!\sum_{\text{\scriptsize permutations}}
    \frac{p_{12 \dots n}^+}{p_1^+}\frac{1}{\widetilde{v}^{\star}_{n\, n-1} \dots \widetilde{v}^{\star}_{32}\widetilde{v}^{\star}_{21}}\mathrm{Tr}(t^a t^{b_1} t^{b_2}\dots t^{b_n})\,.
    \label{eq:psi_kernel_not}
\end{multline}
Comparing the above with Eq.~\eqref{eq:psi_kernel_der} may generate confusion. Let us clarify. The kernel  $\widetilde{\Psi}_n^{a\{b_1b_2 \dots b_n\}}(\mathbf{P};\{\mathbf{p}_1,\mathbf{p}_2, \dots \mathbf{p}_n\})$ makes explicit the fact that there is a symmetry with respect to the interchange of the indices enclosed in the braces. As a result, they can be expressed either in a fully symmetric form as shown in Eq.~\eqref{eq:psi_kernel_der} or, equivalently, ordered as shown in Eq.~\eqref{eq:psi_n}. This equivalence exists due to the integral over the momentum of the associated fields.  The notation Eq.~\eqref{eq:psi_kernel_not}, on the other hand, explicitly represents a fully symmetric kernel. Thus, using the above notation in Eq.~\eqref{eq:Omega_2_der} we get
\begin{equation}
    {\widetilde \Omega}_{2}^{c b_1 b_2}(\mathbf{q}; \mathbf{p}_1 ,\mathbf{p}_2)= 2\,\left(\frac{p_1^+}{p_{12}^+}\right)^2 \widetilde{\Psi}_2^{c b_1b_2}(\mathbf{q};\mathbf{p}_1,\mathbf{p}_2)\,.
    \label{eq:Omega_2_psi}
\end{equation}

Substituting Eq.~\eqref{eq:A_bull_exp} and Eq.~\eqref{eq:A_star_exp1} to Eq.~\eqref{eq:B_star_def}, for the third order, we have
\begin{multline}
    \int d^3\mathbf{p}_1\dots d^3\mathbf{p}_3 \,q^+\, {\widetilde \Omega}_{3}^{c b_1 \left \{b_2 b_3 \right \} }(\mathbf{q}; \mathbf{p}_1 ,\left \{ \mathbf{p}_2 , \mathbf{p}_3 \right \}) \widetilde{B}^{\star}_{b_1}(x^+;\mathbf{p}_1)\widetilde{B}^{\bullet}_{b_2}(x^+;\mathbf{p}_2)\widetilde{B}^{\bullet}_{b_3}(x^+;\mathbf{p}_3)\\
    +\int d^3\mathbf{P} d^3\mathbf{p}_1 d^3\mathbf{p}_2 \,\widetilde{\Psi}_2^{a\{b_1b_2\}}(\mathbf{P};\{\mathbf{p}_1,\mathbf{p}_2\}) \Big[\delta^c_{b_1} \,\delta^3(\mathbf{q} - \mathbf{p}_1) {\widetilde B}_{b_2}^{\bullet}(x^+;\mathbf{p}_2)\\
+\widetilde{B}^{\bullet}_{b_1}(x^+;\mathbf{p}_1)\delta^c_{b_2} \,\delta^3(\mathbf{q} - \mathbf{p}_2)\Big]P^+\int d^3\mathbf{t}_1 d^3\mathbf{t}_2 \,{\widetilde \Omega}_{2}^{a e_1 e_2}(\mathbf{P}; \mathbf{t}_1 ,\mathbf{t}_2) \widetilde{B}^{\star}_{e_1}(x^+;\mathbf{t}_1)\widetilde{B}^{\bullet}_{e_2}(x^+;\mathbf{t}_2)\\
+ \int d^3\mathbf{P} d^3\mathbf{p}_1 \dots d^3\mathbf{p}_3 \,\widetilde{\Psi}_3^{a\{b_1b_2 b_3\}}(\mathbf{P};\{\mathbf{p}_1,\mathbf{p}_2,\mathbf{p}_3\}) \Big[\delta^c_{b_1} \,\delta^3(\mathbf{q} - \mathbf{p}_1) {\widetilde B}_{b_2}^{\bullet}(x^+;\mathbf{p}_2){\widetilde B}_{b_3}^{\bullet}(x^+;\mathbf{p}_3)\\
+\widetilde{B}^{\bullet}_{b_1}(x^+;\mathbf{p}_1)\delta^c_{b_2} \,\delta^3(\mathbf{q} - \mathbf{p}_2){\widetilde B}_{b_3}^{\bullet}(x^+;\mathbf{p}_3)+ \widetilde{B}^{\bullet}_{b_1}(x^+;\mathbf{p}_1){\widetilde B}_{b_2}^{\bullet}(x^+;\mathbf{p}_2)\delta^c_{b_3} \,\delta^3(\mathbf{q} - \mathbf{p}_3)\Big]\\
P^+{\widetilde B}_{a}^{\star}(x^+;\mathbf{P})=0\,.
\end{multline}
Integrating out the deltas, we get
\begin{multline}
    \int d^3\mathbf{p}_1\dots d^3\mathbf{p}_3 \,q^+\, {\widetilde \Omega}_{3}^{c b_1 \left \{b_2 b_3 \right \} }(\mathbf{q}; \mathbf{p}_1 ,\left \{ \mathbf{p}_2 , \mathbf{p}_3 \right \}) \widetilde{B}^{\star}_{b_1}(x^+;\mathbf{p}_1)\widetilde{B}^{\bullet}_{b_2}(x^+;\mathbf{p}_2)\widetilde{B}^{\bullet}_{b_3}(x^+;\mathbf{p}_3)\\
    +\int  d^3\mathbf{p}_2  d^3\mathbf{t}_1 d^3\mathbf{t}_2\,\widetilde{\Psi}_2^{a\{c b_2\}}(\mathbf{t}_{12};\{\mathbf{q},\mathbf{p}_2\}) 
t_{12}^+ \,{\widetilde \Omega}_{2}^{a e_1 e_2}(\mathbf{t}_{12}; \mathbf{t}_1 ,\mathbf{t}_2) {\widetilde B}_{b_2}^{\bullet}(x^+;\mathbf{p}_2)\widetilde{B}^{\star}_{e_1}(x^+;\mathbf{t}_1)\widetilde{B}^{\bullet}_{e_2}(x^+;\mathbf{t}_2)\\
    +\int d^3\mathbf{p}_1 d^3\mathbf{t}_1 d^3\mathbf{t}_2 \, \,\widetilde{\Psi}_2^{a\{b_1c\}}(\mathbf{t}_{12};\{\mathbf{p}_1,\mathbf{q}\})
t_{12}^+{\widetilde \Omega}_{2}^{a e_1 e_2}(\mathbf{t}_{12}; \mathbf{t}_1 ,\mathbf{t}_2) \widetilde{B}^{\bullet}_{b_1}(x^+;\mathbf{p}_1)\widetilde{B}^{\star}_{e_1}(x^+;\mathbf{t}_1)\widetilde{B}^{\bullet}_{e_2}(x^+;\mathbf{t}_2)\\
+ \int d^3\mathbf{P} d^3\mathbf{p}_2 d^3\mathbf{p}_3 \,\widetilde{\Psi}_3^{a\{cb_2 b_3\}}(\mathbf{P};\{\mathbf{q},\mathbf{p}_2,\mathbf{p}_3\}) {\widetilde B}_{b_2}^{\bullet}(x^+;\mathbf{p}_2){\widetilde B}_{b_3}^{\bullet}(x^+;\mathbf{p}_3)
P^+{\widetilde B}_{a}^{\star}(x^+;\mathbf{P})\\
+ \int d^3\mathbf{P} d^3\mathbf{p}_1 d^3\mathbf{p}_3 \,\widetilde{\Psi}_3^{a\{b_1 c b_3\}}(\mathbf{P};\{\mathbf{p}_1,\mathbf{q},\mathbf{p}_3\}) 
\widetilde{B}^{\bullet}_{b_1}(x^+;\mathbf{p}_1)\delta^c_{b_2} {\widetilde B}_{b_3}^{\bullet}(x^+;\mathbf{p}_3)
P^+{\widetilde B}_{a}^{\star}(x^+;\mathbf{P})\\
+ \int d^3\mathbf{P} d^3\mathbf{p}_1 d^3\mathbf{p}_2 \,\widetilde{\Psi}_3^{a\{b_1b_2 c\}}(\mathbf{P};\{\mathbf{p}_1,\mathbf{p}_2,\mathbf{q}\})  \widetilde{B}^{\bullet}_{b_1}(x^+;\mathbf{p}_1){\widetilde B}_{b_2}^{\bullet}(x^+;\mathbf{p}_2)
P^+{\widetilde B}_{a}^{\star}(x^+;\mathbf{P})=0\,.
\end{multline}
The expression above can be rewritten, by renaming the terms such that each of them has $\widetilde{B}^{\star}_{b_1}(x^+;\mathbf{p}_1)\widetilde{B}^{\bullet}_{b_2}(x^+;\mathbf{p}_2)\widetilde{B}^{\bullet}_{b_3}(x^+;\mathbf{p}_3)$, as follows
\begin{multline}
    \int d^3\mathbf{p}_1\dots d^3\mathbf{p}_3 \,q^+\, {\widetilde \Omega}_{3}^{c b_1 \left \{b_2 b_3 \right \} }(\mathbf{q}; \mathbf{p}_1 ,\left \{ \mathbf{p}_2 , \mathbf{p}_3 \right \}) \widetilde{B}^{\star}_{b_1}(x^+;\mathbf{p}_1)\widetilde{B}^{\bullet}_{b_2}(x^+;\mathbf{p}_2)\widetilde{B}^{\bullet}_{b_3}(x^+;\mathbf{p}_3)\\
    =\int  d^3\mathbf{p}_1 d^3\mathbf{p}_2d^3\mathbf{p}_3\,\Big[\widetilde{\Psi}_2^{a\{c b_3\}}(\mathbf{p}_{12};\{\mathbf{q},-\mathbf{p}_3\}) 
p_{12}^+ \,{\widetilde \Omega}_{2}^{a b_1 b_2}(\mathbf{p}_{12}; \mathbf{p}_1 ,\mathbf{p}_2)\\ 
    +\,\widetilde{\Psi}_2^{a\{b_2c\}}(\mathbf{p}_{13};\{-\mathbf{p}_2,\mathbf{q}\})
p_{13}^+{\widetilde \Omega}_{2}^{a b_1 b_3}(\mathbf{p}_{13}; \mathbf{p}_1 ,\mathbf{p}_3)
- \,\widetilde{\Psi}_3^{b_1\{cb_2 b_3\}}(\mathbf{p}_1;\{\mathbf{q},-\mathbf{p}_2,-\mathbf{p}_3\}) p_1^+\\
- \,\widetilde{\Psi}_3^{b_1\{b_2 c b_3\}}(\mathbf{p}_1;\{-\mathbf{p}_2,\mathbf{q},-\mathbf{p}_3\}) p_1^+ - \,\widetilde{\Psi}_3^{b_1\{b_2b_3 c\}}(\mathbf{p}_1;\{-\mathbf{p}_2,-\mathbf{p}_3,\mathbf{q}\})p_1^+\Big]\\
{\widetilde B}_{b_1}^{\star}(x^+;\mathbf{p}_1){\widetilde B}_{b_2}^{\bullet}(x^+;\mathbf{p}_2){\widetilde B}_{b_3}^{\bullet}(x^+;\mathbf{p}_3)\,.
\end{multline}
Thus, we obtain the following relation for the kernel
\begin{multline}
    p_{123}^+\, {\widetilde \Omega}_{3}^{c b_1 \left \{b_2 b_3 \right \} }(\mathbf{q}; \mathbf{p}_1 ,\left \{ \mathbf{p}_2 , \mathbf{p}_3 \right \}) 
    =\Big[\widetilde{\Psi}_2^{a\{c b_3\}}(\mathbf{p}_{12};\{\mathbf{q},-\mathbf{p}_3\}) 
p_{12}^+ \,{\widetilde \Omega}_{2}^{a b_1 b_2}(\mathbf{p}_{12}; \mathbf{p}_1 ,\mathbf{p}_2)\\ 
    +\,\widetilde{\Psi}_2^{a\{b_2c\}}(\mathbf{p}_{13};\{-\mathbf{p}_2,\mathbf{q}\})
p_{13}^+{\widetilde \Omega}_{2}^{a b_1 b_3}(\mathbf{p}_{13}; \mathbf{p}_1 ,\mathbf{p}_3)
- \,\widetilde{\Psi}_3^{b_1\{cb_2 b_3\}}(\mathbf{p}_1;\{\mathbf{q},-\mathbf{p}_2,-\mathbf{p}_3\}) p_1^+\\
- \,\widetilde{\Psi}_3^{b_1\{b_2 c b_3\}}(\mathbf{p}_1;\{-\mathbf{p}_2,\mathbf{q},-\mathbf{p}_3\}) p_1^+ - \,\widetilde{\Psi}_3^{b_1\{b_2b_3 c\}}(\mathbf{p}_1;\{-\mathbf{p}_2,-\mathbf{p}_3,\mathbf{q}\})p_1^+\Big]\,.
\label{eq:omega_3_int}
\end{multline}
After substituting the expression for $\widetilde{\Psi}_2$, $\widetilde{\Omega}_2$ and $\widetilde{\Psi}_3$, the R.H.S can be decomposed into six terms each of which has a certain ordering of color and momentum $\{(\mathbf{p}_1, b_1),(\mathbf{p}_2, b_2),(\mathbf{p}_3, b_3)\}$. Below we explicitly compute the final expression for one of the six orderings. The remaining can be computed in a similar fashion. Let us consider the ordering $\mathrm{Tr}\left(t^{a}t^{b_{1}}t^{b_{2}}t^{b_{3}}\right)$. The first and the fifth term on the R.H.S of Eq.~\eqref{eq:omega_3_int} contribute to this. Substituting for the kernels we get
\begin{equation}
    \frac{g^2}{2}p_1^{+ 2}\Big[ -\frac{p_{12}^+}{p_3^+ p_1^+ \widetilde{v}^{\star}_{(123)3}\widetilde{v}^{\star}_{21}}-\frac{1}{p_2^+  \widetilde{v}^{\star}_{(123)3}\widetilde{v}^{\star}_{32}}\Big]\delta^3(\mathbf{p}_{123} - \mathbf{q})
    \mathrm{Tr}(t^c t^{b_1} t^{b_2} t^{b_3})\,.
    \label{eq:exp_r}
\end{equation}
With a bit of algebra, we obtain the following identity
\begin{equation}
    -\frac{p_{12}^+}{p_1^+ \widetilde{v}^{\star}_{(123)3}\widetilde{v}^{\star}_{21}}+\frac{1}{ \widetilde{v}^{\star}_{(123)3}\widetilde{v}^{\star}_{23}}= \frac{1}{\widetilde{v}^{\star}_{12}\widetilde{v}^{\star}_{23}}\,.
\end{equation}
Substituting this in Eq.~\eqref{eq:exp_r} and massaging the terms a bit we get
\begin{equation}
    \frac{g^2}{2}\frac{p_1^{+ 2}}{p_{123}^+}\Big[ \frac{1}{\widetilde{v}^{\star}_{21}\widetilde{v}^{\star}_{32}}\frac{p_{123}^+}{p_1^+}\Big]\delta^3(\mathbf{p}_{123} - \mathbf{q})
    \mathrm{Tr}(t^c t^{b_1} t^{b_2} t^{b_3})\,.
    \label{eq:exp_ri}
\end{equation}
Following the same procedure for the other orderings, we get
\begin{multline}
    {\widetilde \Omega}_{3}^{c b_1 \left \{b_2 b_3 \right \} }(\mathbf{q}; \mathbf{p}_1 ,\left \{ \mathbf{p}_2 , \mathbf{p}_3 \right \}) 
    = \frac{g^2}{2}\left(\frac{p_1^{+}}{p_{123}^+}\right)^2\delta^3(\mathbf{p}_{123} - \mathbf{q})
    \\
    \Big[ \frac{1}{\widetilde{v}^{\star}_{21}\widetilde{v}^{\star}_{32}}\frac{p_{123}^+}{p_1^+} \mathrm{Tr}(t^c t^{b_1} t^{b_2} t^{b_3}) +\frac{1}{\widetilde{v}^{\star}_{23}\widetilde{v}^{\star}_{31}}\frac{p_{123}^+}{p_1^+} \mathrm{Tr}(t^c t^{b_1} t^{b_3} t^{b_2})\\ +\frac{1}{\widetilde{v}^{\star}_{12}\widetilde{v}^{\star}_{31}}\frac{p_{123}^+}{p_2^+} \mathrm{Tr}(t^c t^{b_2} t^{b_1} t^{b_3})+ \frac{1}{\widetilde{v}^{\star}_{13}\widetilde{v}^{\star}_{32}}\frac{p_{123}^+}{p_2^+} \mathrm{Tr}(t^c t^{b_2} t^{b_3} t^{b_1})\\
    + \frac{1}{\widetilde{v}^{\star}_{21}\widetilde{v}^{\star}_{13}}\frac{p_{123}^+}{p_3^+} \mathrm{Tr}(t^c t^{b_3} t^{b_1} t^{b_2}) +\frac{1}{\widetilde{v}^{\star}_{12}\widetilde{v}^{\star}_{23}}\frac{p_{123}^+}{p_3^+} \mathrm{Tr}(t^c t^{b_3} t^{b_2} t^{b_1})\Big]\,.
    \label{eq:omega_3_full}
\end{multline}
Comparing the above with Eq.~\eqref{eq:psi_kernel_not}, we get
\begin{equation}
\widetilde{\Omega}_{3}^{cb_{1}\left\{ b_{2}b_{3}\right\} }\left(\mathbf{q};\mathbf{p}_{1},\left\{ \mathbf{p}_{2},\mathbf{p}_{3}\right\} \right)=3\left(\frac{p_{1}^{+}}{p_{123}^{+}}\right)^{2}\widetilde{\Psi}_{3}^{cb_{1}b_{2}b_{3}}\left(\mathbf{q};\mathbf{p}_{1},\mathbf{p}_{2},\mathbf{p}_{3}\right)\,.
\label{eq:omega_3_psi}
\end{equation}
Notice, when put together with fields
\begin{equation}
     \int d^3\mathbf{p}_1\dots d^3\mathbf{p}_3 \,q^+\, {\widetilde \Omega}_{3}^{c b_1 \left \{b_2 b_3 \right \} }(\mathbf{q}; \mathbf{p}_1 ,\left \{ \mathbf{p}_2 , \mathbf{p}_3 \right \}) \widetilde{B}^{\star}_{b_1}(x^+;\mathbf{p}_1)\widetilde{B}^{\bullet}_{b_2}(x^+;\mathbf{p}_2)\widetilde{B}^{\bullet}_{b_3}(x^+;\mathbf{p}_3)\,.
\end{equation}
the expression can be seen to remain unaltered under the renaming $2\leftrightarrow3$ of the $\widetilde{B}^{\bullet}$ fields because under this renaming half of the terms in the kernel Eq.~\eqref{eq:omega_3_full} change into the other half and vice versa thereby keeping the whole expression for the kernel unaltered. Hence the braces enclose only the momentum and color associated with bullet fields in ${\widetilde \Omega}_{3}^{c b_1 \left \{b_2 b_3 \right \} }(\mathbf{q}; \mathbf{p}_1 ,\left \{ \mathbf{p}_2, \mathbf{p}_3 \right \})$ and not all.

The results Eq.~\eqref{eq:omega_3_psi} can be generalized as follows
\begin{equation}
\widetilde{\Omega}_{n}^{cb_{1}\left\{ b_{2}\dots b_{n}\right\} }\left(\mathbf{q};\mathbf{p}_{1},\left\{ \mathbf{p}_{2},\dots,\mathbf{p}_{n}\right\} \right)=n\left(\frac{p_{1}^{+}}{p_{1\dots n}^{+}}\right)^{2}\widetilde{\Psi}_{n}^{cb_{1}\dots b_{n}}\left(\mathbf{P};\mathbf{p}_{1},\dots,\mathbf{p}_{n}\right)\,.
\label{eq:omega_n_psi}
\end{equation}

\section{The MHV action}
\label{sec:mhv_action}

To derive the MHV action, we need to substitute the solution for ${\hat A}^{\bullet}[B^{\bullet}]$ and ${\hat A}^{\star}[B^{\bullet}, B^{\star}]$ to the Yang-Mills action 
\begin{multline}
S_{\mathrm{YM}}\left[A^{\bullet},A^{\star}\right]=\int dx^{+}\int d^{3}\mathbf{x}\,\,\Bigg\{ 
-\mathrm{Tr}\,\hat{A}^{\bullet}\square\hat{A}^{\star}
-2ig\,\mathrm{Tr}\,\partial_{-}^{-1}\partial_{\bullet} \hat{A}^{\bullet}\left[\partial_{-}\hat{A}^{\star},\hat{A}^{\bullet}\right] \\
-2ig\,\mathrm{Tr}\,\partial_{-}^{-1}\partial_{\star}\hat{A}^{\star}\left[\partial_{-}\hat{A}^{\bullet},\hat{A}^{\star}\right]
-2g^{2}\,\mathrm{Tr}\,\left[\partial_{-}\hat{A}^{\bullet},\hat{A}^{\star}\right]\partial_{-}^{-2}\left[\partial_{-}\hat{A}^{\star},\hat{A}^{\bullet}\right]
\Bigg\}
\,.\label{eq:YM_LC_action1}
\end{multline}
From Mansfield's transformation, we know that \begin{equation}
\int d^{3}\mathbf{x}\,\mathrm{Tr}\Big\{-\hat{A}^{\bullet}\square\hat{A}^{\star}-2ig\partial_{-}^{-1}\partial_{\bullet}\hat{A}^{\bullet}\left[\partial_{-}\hat{A}^{\star},\hat{A}^{\bullet}\right]\Big\}=\int d^{3}\mathbf{x}\,\mathrm{Tr}\Big\{-\hat{B}^{\bullet}\square\hat{B}^{\star}\Big\}\,.
\label{eq:MTransf1}
\end{equation}
Therefore, we are left with just the last two terms i.e., $\mathcal{L}_{+--}\left[A^{\bullet},A^{\star}\right]$ and $\mathcal{L}_{++--}\left[A^{\bullet},A^{\star}\right]$ in Eq.~\eqref{eq:YM_LC_action1} to which we need to substitute the solutions ${\hat A}^{\bullet}[B^{\bullet}]$ and ${\hat A}^{\star}[B^{\bullet}, B^{\star}]$ to obtain the new action. Since the latter was derived in momentum space in the previous section, in order to perform the substitution, it is necessary that we rewrite $\mathcal{L}_{+--}\left[A^{\bullet},A^{\star}\right]$ and $\mathcal{L}_{++--}\left[A^{\bullet},A^{\star}\right]$ in momentum space as well. They read
\begin{multline}
\mathcal{L}_{--+}\left[A^{\bullet},A^{\star}\right]=\int d^{3}\mathbf{p}_{1}d^{3}\mathbf{p}_{2}d^{3}\mathbf{p}_{3}\,\delta^{3}\left(\mathbf{p}_{1}+\mathbf{p}_{2}+\mathbf{p}_{3}\right)\widetilde{V}_{--+}^{b_1b_2b_3}\left(\mathbf{p}_{1},\mathbf{p}_{2},\mathbf{p}_{3}\right)\,\\
\widetilde{A}^{\star}_{b_1}(x^+;\mathbf{p}_1)\widetilde{A}^{\star}_{b_2}(x^+;\mathbf{p}_2)\widetilde{A}^{\bullet}_{b_3}(x^+;\mathbf{p}_3)\,,
\label{eq:YM3p}
\end{multline}
\begin{multline}
\mathcal{L}_{--++}\left[A^{\bullet},A^{\star}\right]=\int d^{3}\mathbf{p}_{1}d^{3}\mathbf{p}_{2}d^{3}\mathbf{p}_{3}d^{3}\mathbf{p}_{4}\delta^{3}\left(\mathbf{p}_{1}+\mathbf{p}_{2}+\mathbf{p}_{3}+\mathbf{p}_{4}\right)\,\widetilde{V}_{--++}^{b_{1}b_{2}b_{3}b_{4}}\left(\mathbf{p}_{1},\mathbf{p}_{2},\mathbf{p}_{3},\mathbf{p}_{4}\right)\\
\widetilde{A}^{\star}_{b_1}(x^+;\mathbf{p}_1)\widetilde{A}^{\star}_{b_2}(x^+;\mathbf{p}_2)\widetilde{A}^{\bullet}_{b_3}(x^+;\mathbf{p}_3)\widetilde{A}^{\bullet}_{b_4}(x^+;\mathbf{p}_4)\,,
\label{eq:YM4p}
\end{multline}
where the vertices have the following form
\begin{equation}
\widetilde{V}_{--+}^{b_1b_2b_3}\left(\mathbf{p}_{1},\mathbf{p}_{2},\mathbf{p}_{3}\right)=-igf^{b_1b_2b_3}\left(\frac{p_{1}^{\bullet}}{p_{1}^{+}}-\frac{p_{2}^{\bullet}}{p_{2}^{+}}\right)p_{3}^{+}\,,
\label{eq:vertex3g}
\end{equation}
and 
\begin{multline}
    \widetilde{V}_{--++}^{b_{1}b_{2}b_{3}b_{4}}\left(\mathbf{p}_{1},\mathbf{p}_{2},\mathbf{p}_{3},\mathbf{p}_{4}\right)=\frac{g^{2}}{2}\Bigg[f^{ab_{4}b_{1}}f^{ab_{2}b_{3}}\frac{p_{1}^{+}p_{3}^{+}+p_{2}^{+}p_{4}^{+}}{\left(p_{1}^{+}+p_{4}^{+}\right)^{2}}+f^{ab_{4}b_{2}}f^{ab_{1}b_{3}}\frac{p_{1}^{+}p_{4}^{+}+p_{2}^{+}p_{3}^{+}}{\left(p_{1}^{+}+p_{3}^{+}\right)^{2}}\Bigg]\,.\label{eq:vertex4g}
\end{multline}

Substituting solutions for ${\hat A}^{\bullet}[B^{\bullet}]$ and ${\hat A}^{\star}[B^{\bullet}, B^{\star}]$ using Eq.~\eqref{eq:A_bull_exp}-\eqref{eq:A_star_exp1} to Eq.~\eqref{eq:YM3p}, we see that the lowest-order vertex in the new action is a 3-point vertex obtained from the first order expansion of the fields as shown below
\begin{multline}
\mathcal{L}_{--+}\left[B^{\bullet},B^{\star}\right]
=\int d^{3}\mathbf{p}_{1}d^{3}\mathbf{p}_{2}d^{3}\mathbf{p}_{3}\,\delta^{3}\left(\mathbf{p}_{1}+\mathbf{p}_{2}+\mathbf{p}_{3}\right)\widetilde{V}_{--+}^{b_{1}b_{2}b_{3}}\left(\mathbf{p}_{1},\mathbf{p}_{2},\mathbf{p}_{3}\right)\,\\
\widetilde{B}^{\star}_{b_1}(x^+;\mathbf{p}_1)\widetilde{B}^{\star}_{b_2}(x^+;\mathbf{p}_2)\widetilde{B}^{\bullet}_{b_3}(x^+;\mathbf{p}_3)\,,
\end{multline}
where the vertex $\widetilde{V}_{--+}^{b_{1}b_{2}b_{3}}\left(\mathbf{p}_{1},\mathbf{p}_{2},\mathbf{p}_{3}\right)$ is exactly the same as in Eq.~\eqref{eq:vertex3g}. Using the identity $if^{b_1 b_2b_3}=\mathrm{Tr}( t^{b_1} t^{b_2} t^{b_3})-\mathrm{Tr}(t^{b_1} t^{b_3} t^{b_2})$, we can rewrite the vertex as follows
\begin{equation}
\widetilde{V}_{--+}^{b_{1}b_{2}b_{3}}\left(\mathbf{p}_{1},\mathbf{p}_{2},\mathbf{p}_{3}\right)=\mathrm{Tr}\left(t^{b_1}t^{b_2} t^{b_3}\right)
 \mathcal{V}\left(1^-,2^-,3^+\right)+\mathrm{Tr}\left(t^{b_1}t^{b_3} t^{b_2}\right)
 \mathcal{V}\left(1^-,3^+,2^-\right)\,.
\end{equation}
where 
\begin{equation}
\mathcal{V}\left(1^-,2^-,3^+\right)=-g\,v_{12}^{\star}p_{3}^{+}=-g\,\left(\frac{p_{1}^{+}}{p_{2}^{+}}\right)^{2}\frac{\tilde{v}_{21}^{\star 4}}{\tilde{v}_{13}^{\star}\tilde{v}_{32}^{\star}\tilde{v}_{21}^{\star}}\,.
\end{equation}
\begin{equation}
\mathcal{V}\left(1^-,3^+,2^-\right)=-g\,\left(\frac{p_{2}^{+}}{p_{1}^{+}}\right)^{2}\frac{\tilde{v}_{12}^{\star 4}}{\tilde{v}_{31}^{\star}\tilde{v}_{23}^{\star}\tilde{v}_{12}^{\star}}\,.
\end{equation}
In the color ordered vertex above, we use numbers to represent the momentum associated with a given leg. Furthermore, we put the helicity of a given leg explicitly on the number. 

Next, starting with the 4-point vertex $(- - + +)$, all the higher point vertices will receive contribution both from the triple and four-gluon vertex in the Yang-Mills action via the substitution of ${\hat A}^{\bullet}[B^{\bullet}]$ and ${\hat A}^{\star}[B^{\bullet}, B^{\star}]$. For the 4-point vertex $(- - + +)$ we have

\begin{multline}
\int d^{3}\mathbf{p}_{1}d^{3}\mathbf{p}_{2}d^{3}\mathbf{p}_{3}d^{3}\mathbf{q}_{1}d^{3}\mathbf{q}_{2}\,\delta^{3}\left(\mathbf{p}_{123}\right)\widetilde{V}_{--+}^{b_{1}b_{2}b_{3}}\left(\mathbf{p}_{1},\mathbf{p}_{2},\mathbf{p}_{3}\right)\,\\ \Big[\widetilde{\Omega}_{2}^{b_{1}e_{1}e_{2}}\left(\mathbf{p}_{1};\mathbf{q}_{1},\mathbf{q}_{2}\right)\widetilde{B}_{e_{1}}^{\star}(x^+;\mathbf{q}_1)\widetilde{B}_{e_{2}}^{\bullet}(x^+;\mathbf{q}_2)\widetilde{B}_{b_{2}}^{\star}(x^+;\mathbf{p}_2)\widetilde{B}_{b_{3}}^{\bullet}(x^+;\mathbf{p}_3)\\
+\widetilde{B}_{b_{1}}^{\star}(x^+;\mathbf{p}_1)\,\widetilde{\Omega}_{2}^{b_{2}e_{1}e_{2}}\left(\mathbf{p}_{2};\mathbf{q}_{1},\mathbf{q}_{2}\right)\widetilde{B}_{e_{1}}^{\star}(x^+;\mathbf{q}_1)\widetilde{B}_{e_{2}}^{\bullet}(x^+;\mathbf{q}_2)\widetilde{B}_{b_{3}}^{\bullet}(x^+;\mathbf{p}_3)\\
+\widetilde{B}_{b_{1}}^{\star}(x^+;\mathbf{p}_1)\widetilde{B}_{b_{2}}^{\star}(x^+;\mathbf{p}_2)\widetilde{\Psi}_{2}^{b_{3}\left\{ e_{1}e_{2}\right\} }\left(\mathbf{p}_{3};\left\{ \mathbf{q}_{1},\mathbf{q}_{2}\right\} \right)\widetilde{B}_{e_{1}}^{\bullet}(x^+;\mathbf{q}_1)\widetilde{B}_{e_{2}}^{\bullet}(x^+;\mathbf{q}_2)\Big]\\
+ \int d^{3}\mathbf{p}_{1}d^{3}\mathbf{p}_{2}d^{3}\mathbf{p}_{3}d^{3}\mathbf{p}_{4}\delta^{3}\left(\mathbf{p}_{1234}\right)\,\widetilde{V}_{--++}^{b_{1}b_{2}b_{3}b_{4}}\left(\mathbf{p}_{1},\mathbf{p}_{2},\mathbf{p}_{3},\mathbf{p}_{4}\right)\\
\widetilde{B}^{\star}_{b_1}(x^+;\mathbf{p}_1)\widetilde{B}^{\star}_{b_2}(x^+;\mathbf{p}_2)\widetilde{B}^{\bullet}_{b_3}(x^+;\mathbf{p}_3)\widetilde{B}^{\bullet}_{b_4}(x^+;\mathbf{p}_4)\,.
\end{multline}
where the first three terms originate from the $(--+)$ triple gluon vertex and the last term is from the four gluon vertex $(--++)$ in the Yang-Mills action. Renaming each of the terms such that we have $\widetilde{B}^{\star}_{b_1}(x^+;\mathbf{p}_1)\widetilde{B}^{\star}_{b_2}(x^+;\mathbf{p}_2)\widetilde{B}^{\bullet}_{b_3}(x^+;\mathbf{p}_3)\widetilde{B}^{\bullet}_{b_4}(x^+;\mathbf{p}_4)$ in all, we get
\begin{multline}
\int d^{3}\mathbf{p}_{1}d^{3}\mathbf{p}_{2}d^{3}\mathbf{p}_{3}d^{3}\mathbf{p}_{4}\,\,\delta^{3}\left(\mathbf{p}_{1234}\right)\,\widetilde{B}^{\star}_{b_1}(x^+;\mathbf{p}_1)\widetilde{B}^{\star}_{b_2}(x^+;\mathbf{p}_2)\widetilde{B}^{\bullet}_{b_3}(x^+;\mathbf{p}_3)\widetilde{B}^{\bullet}_{b_4}(x^+;\mathbf{p}_4)\,\\
\Bigg[\frac{1}{2}\widetilde{V}_{--+}^{ab_{2}b_{3}}\left(\mathbf{p}_{14},\mathbf{p}_{2},\mathbf{p}_{3}\right)\widetilde{\Omega}_{2}^{ab_{1}b_{4}}\left(\mathbf{p}_{14};\mathbf{p}_{1},\mathbf{p}_{4}\right)
+\frac{1}{2}\widetilde{V}_{--+}^{ab_{2}b_{4}}\left(\mathbf{p}_{13},\mathbf{p}_{2},\mathbf{p}_{4}\right)\widetilde{\Omega}_{2}^{ab_{1}b_{3}}\left(\mathbf{p}_{13};\mathbf{p}_{1},\mathbf{p}_{3}\right)\\
+\frac{1}{2}\widetilde{V}_{--+}^{b_{1}ab_{3}}\left(\mathbf{p}_{1},\mathbf{p}_{24},\mathbf{p}_{3}\right)\widetilde{\Omega}_{2}^{ab_{2}b_{4}}\left(\mathbf{p}_{24};\mathbf{p}_{2},\mathbf{p}_{4}\right)
+\frac{1}{2}\widetilde{V}_{--+}^{b_{1}ab_{4}}\left(\mathbf{p}_{1},\mathbf{p}_{23},\mathbf{p}_{3}\right)\widetilde{\Omega}_{2}^{ab_{2}b_{3}}\left(\mathbf{p}_{23};\mathbf{p}_{2},\mathbf{p}_{3}\right)\\
+\widetilde{V}_{--+}^{b_{1}b_{2}a}\left(\mathbf{p}_{1},\mathbf{p}_{2},\mathbf{p}_{34}\right)\widetilde{\Psi}_{2}^{a\{b_{3}b_{4}\}}\left(\mathbf{p}_{34};\{\mathbf{p}_{3},\mathbf{p}_{4}\}\right)+\widetilde{V}_{--++}^{b_{1}b_{2}b_{3}b_{4}}\left(\mathbf{p}_{1},\mathbf{p}_{2},\mathbf{p}_{3},\mathbf{p}_{4}\right)\Bigg]\,.
\label{eq:MHV_4_ym}
\end{multline}
Notice, we symmetrized the terms involving $\widetilde{\Omega}_{2}$ kernels in the first and the second line above. Substituting the vertices $\widetilde{V}_{--+}$ and $\widetilde{V}_{--++}$ as wells as the $\widetilde{\Omega}_{2}$ kernels, the expression above can be rewritten as follows
\begin{multline}
    \int d^{3}\mathbf{p}_{1}d^{3}\mathbf{p}_{2}d^{3}\mathbf{p}_{3}d^{3}\mathbf{p}_{4}\,\,\delta^{3}\left(\mathbf{p}_{1}+\mathbf{p}_{2}+\mathbf{p}_{3}+\mathbf{p}_{4}\right)\, \widetilde{\mathcal{V}}_{--++}^{b_{1} b_{2}b_{3}b_{4}}\left(\mathbf{p}_{1},\mathbf{p}_{2},\mathbf{p}_{3},\mathbf{p}_{4}\right)\\
\,\widetilde{B}^{\star}_{b_1}(x^+;\mathbf{p}_1)\widetilde{B}^{\star}_{b_2}(x^+;\mathbf{p}_2)\widetilde{B}^{\bullet}_{b_3}(x^+;\mathbf{p}_3)\widetilde{B}^{\bullet}_{b_4}(x^+;\mathbf{p}_4)\,,    
\end{multline}
where
\begin{equation}
\widetilde{\mathcal{V}}_{--++}^{b_{1} b_{2}b_{3}b_{4}}\left(\mathbf{p}_{1},\mathbf{p}_{2},\mathbf{p}_{3},\mathbf{p}_{4}\right)= \!\!\sum_{\underset{\text{\scriptsize permutations}}{\text{noncyclic}}}
 \mathrm{Tr}\left(t^{b_1}t^{b_2} t^{b_3}t^{b_4}\right)
 \mathcal{V}\left(1^-,2^-,3^+,4^+\right)
\,.
\label{eq:MHV_vertex1}
\end{equation}
Below, we explicitly compute the vertex with $\mathrm{Tr}\left(t^{b_1}t^{b_2} t^{b_3}t^{b_4}\right)$ color ordering. The remaining color ordered vertices can be computed in a similar fashion. From the expression Eq.~\eqref{eq:MHV_4_ym}, it is straightforward to see that the first, fourth, fifth and sixth terms contribute to this ordering. Collecting them together we get
\begin{multline}
   \mathcal{V}\left(1^-,2^-,3^+,4^+\right)= \frac{1}{2}g^{2}\,\Bigg[\frac{p_{1}^{+ 2}p_{3}^{+}}{p_{2}^{+}p_{4}^{+}p_{14}^{+}}\,\frac{\widetilde{v}_{2\left(23\right)}^{\star}}{\widetilde{v}_{14}^{\star}}
    + \frac{p_{2}^{+}p_{4}^{+}}{p_{14}^{+}p_{23}^{+}}\,\frac{\widetilde{v}_{\left(14\right)1}^{\star}}{\widetilde{v}_{32}^{\star}}\\
-\frac{p_{12}^{+}p_{34}^{+}}{p_{2}^{+}p_{3}^{+}}\,\frac{\widetilde{v}_{21}^{\star}}{\widetilde{v}_{43}^{\star}}
-\frac{p_{1}^{+}p_{3}^{+}+p_{2}^{+}p_{4}^{+}}{\left(p_{1}^{+}+p_{4}^{+}\right)^{2}}\Bigg]\,.
\label{eq:Mhv_4_4}
\end{multline}
With a bit of tedious algebra, the R.H.S of the expression above can be reduced to the following
\begin{equation}
\mathcal{V}\left(1^-,2^-,3^+,4^+\right)= \frac{g^2}{2}\left(\frac{p_{1}^{+}}{p_{2}^{+}}\right)^{2}\frac{\widetilde{v}_{21}^{\star 4}}{\widetilde{v}_{14}^{\star}\widetilde{v}_{43}^{\star}\widetilde{v}_{32}^{\star}\widetilde{v}_{21}^{\star}}\,.
\end{equation}
The above expression was also cross-checked against Eq.~\eqref{eq:Mhv_4_4} for different sets of off-shell momenta and we found agreement between the two. Similarly, the higher point vertices can be derived and shown to satisfy the following
\begin{equation}
\mathcal{V}\left(1^-,2^-,3^+,\dots,n^+\right)= 
\frac{(-g)^{n-2}}{(n-2)!}  \left(\frac{p_{1}^{+}}{p_{2}^{+}}\right)^{2}
\frac{\widetilde{v}_{21}^{\star 4}}{\widetilde{v}_{1n}^{\star}\widetilde{v}_{n\left(n-1\right)}^{\star}\widetilde{v}_{\left(n-1\right)\left(n-2\right)}^{\star}\dots\widetilde{v}_{21}^{\star}}
\,.
\label{eq:MHV_colororder}
\end{equation}
In fact, the five-point computation was done explicitly in \cite{Ettle2006b}.

Thus we conclude that the solution to Mansfield's transformation when substituted to the triple gluon vertex $(--+)$ and the four gluon vertex $(--++)$ in the Yang-Mills action results in a series of MHV vertices
\begin{multline}
\mathcal{L}_{+--}\left[A^{\bullet}[B^{\bullet}],A^{\star}[B^{\bullet}, B^{\star}]\right] +\mathcal{L}_{++--}\left[A^{\bullet}[B^{\bullet}],A^{\star}[B^{\bullet}, B^{\star}]\right]\\
=\mathcal{L}_{--+}\left[B^{\bullet},B^{\star}\right]+\mathcal{L}_{--++}\left[B^{\bullet},B^{\star}\right]+\mathcal{L}_{--+++}\left[B^{\bullet},B^{\star}\right]+\dots\,.
\end{multline}

\begin{figure}
    \centering
    \includegraphics[width=16.2cm]{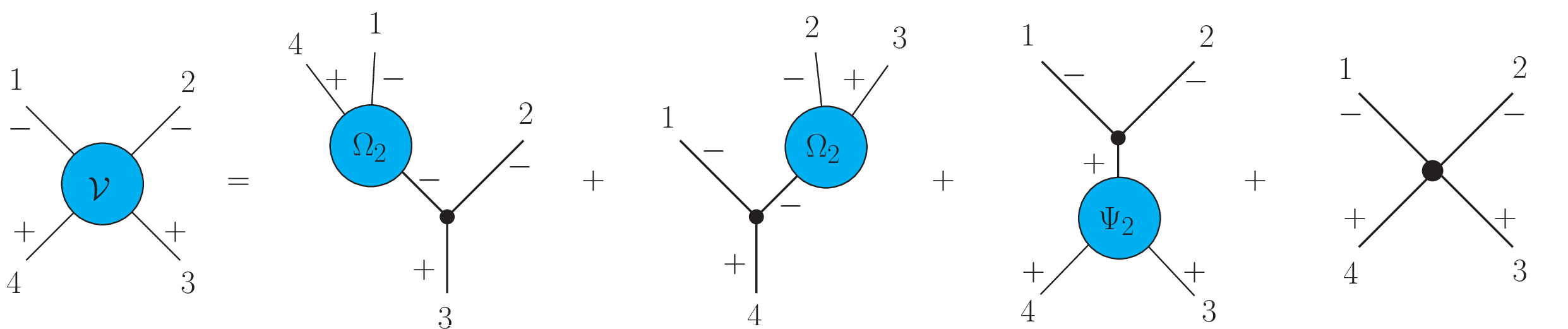}
    \caption{\small 
    LHS: 4-point color-ordered split-helicity vertex $\mathcal{V}\left(1^-,2^-,3^+,4^+\right)$ in the MHV action. RHS: The contributions originating from the substitution of the solutions for ${\hat A}^{\bullet}[B^{\bullet}]$ and ${\hat A}^{\star}[B^{\bullet}, B^{\star}]$ Eq.~\eqref{eq:A_bull_exp}-\eqref{eq:A_star_exp1} to the triple gluon $(- - +)$ and the four gluon $(- - + +)$ vertices in the Yang-Mills action.
}
    \label{fig:mhv_ver_4p}
\end{figure}
\chapter{\texorpdfstring{${\hat B}^{\bullet}[A^{\bullet}]:$}{Bbullet} Solution of SD EOM}
\label{sec:app_A3}

The Self-Dual equation reads Eq.~\eqref{eq:SD_EOM_j}
\begin{equation}
    \Box {\hat{A}}^{\bullet} + 2ig{\partial}_{-} \left[ ({\partial}_{-}^{-1} {\partial}_{\bullet} {\hat{A}}^{\bullet}), {\hat{A}}^{\bullet} \right] - \Box\, \hat{B}^{\bullet} = 0\, .
    \label{eq:SD_EOM_j1}
\end{equation}
It is preferable to re-write the above equation in momentum space since we know the kernels of ${\widetilde B}^{\bullet}[A^{\bullet}]$ in it (\emph{cf.} Eq.\eqref{eq:Bbullet_sol1}). Doing so we obtain
\begin{multline}
    -P^{2} \widetilde{A}^{\bullet}_a (P) + ig f^{a b c} \int\!d^{4}p_{1}d^{4}p_{2} \,{\delta}^{4}(p_{1}+p_{2}-P) \left \{ \frac{p_{12}^{+}}{p_{1}^{+}}  {\widetilde v}_{12} \right \}  \widetilde{A}^{\bullet }_b (p_{1})\widetilde{A}^{\bullet }_c (p_{2})\\ + P^{2} \widetilde{B}^{\bullet}_a (P) = 0\, .
    \label{eq:eomft}
\end{multline}
The aim is to show that the series ${\widetilde B}^{\bullet}[A^{\bullet}]$ Eq.~\eqref{eq:Bbullet_sol1}, with $\widetilde{\Gamma}_n^{a\{b_1\dots b_n\}}(\mathbf{P};\{\mathbf{p}_1,\dots ,\mathbf{p}_n\}) \neq 0$ for $n\geq 3$,  obtained by inverting ${\widetilde A}^{\bullet}[B^{\bullet}]$ satisfies the Self-Dual equation Eq.~\eqref{eq:eomft}.
Let us begin by writing ${\widetilde B}^{\bullet}[A^{\bullet}]$ in 4D so that we can substitute it in Eq.~\eqref{eq:eomft}
\begin{equation}
    \widetilde{B}^{\bullet}_a(P) = \sum_{n=1}^{\infty} 
    \int d^4p_1\dots d^4p_n \, \widetilde{\Gamma}_n^{a\{b_1\dots b_n\}}(P;\{p_1,\dots ,p_n\}) \widetilde{A^{\bullet}}_{b_1}(p_1)\dots \widetilde{A^{\bullet}}_{b_n}(p_n)\,,
    \label{eq:j[A]1}
\end{equation}
 where
\begin{equation}
    \widetilde{\Gamma}_n^{a\{b_1\dots b_n\}} (P;\{p_1,\dots,p_n\}) = (-g)^{n-1} \frac{\delta^4\left(p_1+\dots+p_n-P\right)\, \mathrm{Tr} \left(t^{a} t^{b_{1}} \cdots t^{b_{n}}\right)}
    {\widetilde{v}^{\star}_{1(1\cdots n)} \widetilde{v}^{\star}_{(12)(1\cdots n)} \cdots \widetilde{v}^{\star}_{(1 \cdots n-1)(1\cdots n)}} \, .
    \label{eq:Gamma_n1}
\end{equation}
Essentially, going from $\widetilde{B}_a^{\bullet}\left[A^{\bullet}\right](x^+;\mathbf{P})$ to $\widetilde{B}^{\bullet}_a\left[A^{\bullet}\right](P)$ we get an extra delta for the minus component of the momenta because the kernels $\widetilde{\Gamma}_n^{a\{b_1\dots b_n\}}(\mathbf{P};\{\mathbf{p}_1,\dots ,\mathbf{p}_n\}) $ depend only on the 3-momenta. Now, we substitute $\widetilde{B}^{\bullet}_a\left[A^{\bullet}\right](P)$ to Eq.\eqref{eq:eomft} under the constraint $P^2\longrightarrow 0$. This is because the solution ${\widetilde A}^{\bullet}[B^{\bullet}]$ was obtained with the current (which is $B^{\bullet}$) having support on the light cone. 

For the first order, we get $P^{2} \widetilde{A}^{\bullet}_a (P)$ which essentially cancels out the first term in Eq.\eqref{eq:eomft}. For the second order, we have
\begin{align}
    \widetilde{\Gamma}_2^{a\{b_1 b_2\}}(P;\{p_1 , p_2\}) =&\, -g \frac{{\delta}^{4}(p_{1}+p_{2}-P)}{{\widetilde v}^{\star}_{1(12)}} \mathrm{Tr}(t^{a} t^{b_{1}} t^{b_{2}})\,, \nonumber\\
    =&\,\frac{ig}{2}\frac{{\delta}^{4}(p_{1}+p_{2}-P)}{{\widetilde v}_{21}{\widetilde v}^{\star}_{12}} f^{a b_{1}b_{2}}\left \{ \frac{p_{12}^{+}}{p_{1}^{+}} {\widetilde v}_{12} \right \} \, ,
\end{align}
where the second expression is obtained by symmetrizing the first. Recall, $\widetilde{\Gamma}_2 = -\widetilde{\Psi}_2$ (\emph{cf.} Eq.\eqref{eq:psi2_gamma2}). From the latter, we know that the two final state gluons are on-shell therefore, given that $P=p_1+p_2$, we can write $P^2 = -2{\widetilde v}_{21}{\widetilde v}^{\star}_{12}$. Substituting this to the expression above we get
\begin{equation}
    P^2 \widetilde{\Gamma}_2^{a\{b_1 b_2\}}(P;\{p_1 , p_2\}) = -igf^{a b_{1}b_{2}}\left \{ \frac{p_{12}^{+}}{p_{1}^{+}} \, {\widetilde v}_{12} \right \}{\delta}^{4}(p_{1}+p_{2}-P) \, .
     \label{eq:2gam}
\end{equation}
This exactly cancels out the triple gluon vertex term in Eq.\eqref{eq:eomft}. Note, until this point, we didn't have to impose $P^2\longrightarrow 0$. However, imposing this on $P^2 \widetilde{\Gamma}_2^{a\{b_1 b_2\}}(P;\{p_1 , p_2\})$ would give a triple gluon vertex with all the three gluons on-shell which must be zero (for real momenta). This would happen only via ${\widetilde v}_{12} \rightarrow 0$, as evident from Eq.\eqref{eq:2gam}, implying 
\begin{equation}
  P^2 = p_{12}^2\rightarrow 0 \hspace{0.5cm}\Leftrightarrow \hspace{0.5cm} \widetilde{v}_{12}\longrightarrow 0  \, .
  \label{eq:v12}
\end{equation}
This can be further generalized. Consider $P=p_1+\dots +p_n$, where all the $n$ gluons are on-shell. Thus we can write
\begin{eqnarray}
    P^2 = -\sum_{i,j=1}^{n} \widetilde{v}_{ij}\widetilde{v}^{\star}_{ji} = \sum_{i,j=1}^{n} \frac{p_j^+}{p_i^+}\widetilde{v}_{ij}\widetilde{v}^{\star}_{ij} \, .
    \label{eq:pode}
\end{eqnarray}
From the rightmost expression above we see that each term in the sum is positive definite therefore
\begin{equation}
  P^2 \rightarrow 0 \hspace{0.5cm}\Rightarrow \hspace{0.5cm} \widetilde{v}_{ij}\widetilde{v}^{\star}_{ji} \longrightarrow 0 \hspace{0.5cm} \forall i,j \, .
  \label{eq:vvstarz}
\end{equation}
Following Eq.\eqref{eq:v12}, we can write
\begin{equation}
    P^2 \rightarrow 0 \hspace{0.5cm}\Leftrightarrow \widetilde{v}_{ij}\rightarrow 0\, \hspace{0.5cm} \forall i,j.
    \label{eq:vtildeto0}
\end{equation}
This generalization will be useful below. 

After the cancellation of both the terms in the self-dual equation Eq.~\eqref{eq:eomft}, we are left to show 
\begin{equation}
     \sum_{n=3}^{\infty}\left [\int\!d^{4}p_{1}\cdots d^{4}p_{n}\, P^2 \widetilde{\Gamma}_n^{a\{b_1\dots b_n\}}(P;\{p_1,\dots ,p_n\}) \widetilde{A^{\bullet}}_{b_1}(p_1)\dots \widetilde{A^{\bullet}}_{b_n}(p_n) \right ]_{P^2 \rightarrow 0} = 0 \, .
     \label{eq:hotz}
\end{equation}
Since the final state gluons are on-shell we can write
\begin{equation}
    P^2  \widetilde{\Gamma}_n^{a\{b_1\dots b_n\}}(P;\{p_1,\dots ,p_n\})    = 
     -P^+ D_{1 \cdots n} \widetilde{\Gamma}_n^{a\{b_1\dots b_n\}}(P;\{p_1,\dots ,p_n\})\, .
    \label{eq:pd}
\end{equation}
From \cite{Kotko2017}, we can re-write the kernels as
\begin{multline}
     \widetilde{\Gamma}_n^{a\{b_1\dots b_n\}}(P;\{p_1,\dots ,p_n\})  = (-g)^{n-1} \delta^{4} (p_{1} + \cdots +p_{n} - P) \frac{-2}{D_{1 \cdots n}} \\
    \times\!\frac{v_{n(n-1)}\widetilde{v}^{\star}_{(1 \cdots n-1)(1\cdots n)} + v_{(n-1)(n-2)} \widetilde{v}^{\star}_{(1 \cdots n-2)(1\cdots n)} + \cdots v_{21}\widetilde{v}^{\star}_{1(1\cdots n)} }{\widetilde{v}^{\star}_{1(1\cdots n)} \widetilde{v}^{\star}_{(12)(1\cdots n)} \cdots \widetilde{v}^{\star}_{(1 \cdots n-1)(1\cdots n)}} \, .
    \label{eq:gfac}
\end{multline}
Substituting this in Eq.~\eqref{eq:pd}, we get
\begin{multline}
     P^2 \widetilde{\Gamma}_n^{a\{b_1\dots b_n\}}(P;\{p_1,\dots ,p_n\}) = (-g)^{n-1}\delta^{4} (p_{1} + \cdots +p_{n} - P) 2P^+ \\
    \times\!\frac{v_{n(n-1)}\widetilde{v}^{\star}_{(1 \cdots n-1)(1\cdots n)} + v_{(n-1)(n-2)} \widetilde{v}^{\star}_{(1 \cdots n-2)(1\cdots n)} + \cdots v_{21}\widetilde{v}^{\star}_{1(1\cdots n)} }{\widetilde{v}^{\star}_{1(1\cdots n)} \widetilde{v}^{\star}_{(12)(1\cdots n)} \cdots \widetilde{v}^{\star}_{(1 \cdots n-1)(1\cdots n)}} \, ,
\end{multline}
which simplifies to
\begin{multline}
     P^2 \widetilde{\Gamma}_n^{a\{b_1\dots b_n\}}(P;\{p_1,\dots ,p_n\}) = (-g)^{n-1}\delta^{4} (p_{1} + \cdots +p_{n} - P) 2P^+  \\
    \times \left [ \frac{v_{n(n-1)}}{\widetilde{v}^{\star}_{1(1\cdots n)} \widetilde{v}^{\star}_{(12)(1\cdots n)} \cdots \widetilde{v}^{\star}_{(1 \cdots n-2)(1\cdots n)}} \right. \\
    \left. + \frac{v_{(n-1)(n-2)}}{\widetilde{v}^{\star}_{1(1\cdots n)} \widetilde{v}^{\star}_{(12)(1\cdots n)} \cdots \widetilde{v}^{\star}_{(1 \cdots n-3)(1\cdots n)}\widetilde{v}^{\star}_{(1 \cdots n-1)(1\cdots n)}} + \cdots \right] \, .
\end{multline}
Using $\widetilde{v}_{\left(q\right)\left(p\right)}=q^{+}v_{\left(p\right)\left(q\right)}$ in the numerator of each term, we see that owing to Eq.\eqref{eq:vtildeto0} each term on the R.H.S above vanishes in the limit $P^2\longrightarrow 0$. Therefore Eq.~\eqref{eq:hotz} is indeed true.
\chapter{\texorpdfstring{${\widetilde B}^{\star}[A^{\bullet},A^{\star}]$}{Bstar} in momentum space}
\label{sec:app_A4}

In this appendix, we re-derive the momentum space expression for ${\widetilde B}^{\star}[A^{\bullet},A^{\star}]$ field first by "inverting" the ${\widetilde A}^{\star}[B^{\bullet},B^{\star}]$ and then, in the next section, we show that the kernels match with those obtained by Fourier transforming the straight infinite Wilson line based expression for ${\hat B}^{\star}[A^{\bullet},A^{\star}]$. Both of these calculations were done in \cite{Kakkad2020}.

\section{\texorpdfstring{${\widetilde B}^{\star}[A^{\bullet},A^{\star}]$}{Bstar} from \texorpdfstring{${\widetilde A}^{\star}[B^{\bullet},B^{\star}]$}{Bstar} and \texorpdfstring{${\widetilde A}^{\bullet}[B^{\bullet}]$}{Bstar}}
\label{subsec:app_A41}

Let us begin by proposing the following expansion for the ${\widetilde B}^{\star}[A^{\bullet},A^{\star}]$ in momentum space
\begin{multline}
    {\widetilde{B}}_a^\star (x^+;\mathbf{P}) = {\widetilde A}^\star_{a} (x^+;\mathbf{P}) + \sum_{n=2}^{\infty} \int\!d^{3}\mathbf{p}_{1}\cdots d^{3}\mathbf{p}_{n} {\widetilde \Upsilon}_{n}^{a b_1 \left \{b_2 \cdots b_n \right \} }(\mathbf{P}; \mathbf{p}_{1} ,\left \{ \mathbf{p}_{2} , \cdots \mathbf{p}_{n} \right \})  \\
     {\widetilde A}^\star_{b_1} (x^+;\mathbf{p}_{1}){\widetilde A}^\bullet_{b_2} (x^+;\mathbf{p}_{2}) \cdots {\widetilde A}^\bullet_{b_n} (x^+;\mathbf{p}_{n}) \, .
   \label{eq:Bstar1}
\end{multline}
In order to obtain the kernels ${\widetilde \Upsilon}_{n}^{a b_1 \left \{b_2 \cdots b_n \right \} }(\mathbf{P}; \mathbf{p}_{1} ,\left \{ \mathbf{p}_{2} , \cdots \mathbf{p}_{n} \right \})$ we need to substitute the expression for ${\widetilde A}^{\bullet}[B^{\bullet}]$ and ${\widetilde A}^{\star}[B^{\bullet},B^{\star}]$ using Eq.~\eqref{eq:A_bull_solu}-\eqref{eq:A_star_solu} on the R.H.S of the expression above and equate terms with equal order in fields. 

The expression for ${\widetilde A}^{\bullet}[B^{\bullet}]$ and ${\widetilde A}^{\star}[B^{\bullet},B^{\star}]$ reads
\begin{multline}
\widetilde{A}^{\bullet}_a(x^+;\mathbf{P}) = \widetilde{B}^{\bullet}_a(x^+;\mathbf{P})\\
+\sum_{n=2}^{\infty} 
    \int d^3\mathbf{p}_1\dots d^3\mathbf{p}_n \, \widetilde{\Psi}_n^{a\{b_1\dots b_n\}}(\mathbf{P};\{\mathbf{p}_1,\dots ,\mathbf{p}_n\}) \prod_{i=1}^n\widetilde{B}^{\bullet}_{b_i}(x^+;\mathbf{p}_i)\,,
    \label{eq:A_bull_solu2}
\end{multline}
\begin{multline}
\widetilde{A}^{\star}_a(x^+;\mathbf{P}) = {\widetilde B}^{\star}_a(x^+;\mathbf{P})\\
+ \sum_{n=2}^{\infty} 
    \int d^3\mathbf{p}_1\dots d^3\mathbf{p}_n \, {\widetilde \Omega}_{n}^{a b_1 \left \{b_2 \cdots b_n \right \} }(\mathbf{P}; \mathbf{p}_1 ,\left \{ \mathbf{p}_2 , \dots ,\mathbf{p}_n \right \}) \widetilde{B}^{\star}_{b_1}(x^+;\mathbf{p}_1)\prod_{i=2}^n\widetilde{B}^{\bullet}_{b_i}(x^+;\mathbf{p}_i)\, ,
    \label{eq:A_star_solu2}
\end{multline}
where
\begin{equation}
    {\widetilde \Psi}_{n}^{a \left \{b_1 \cdots b_n \right \} }(\mathbf{P}; \left \{\mathbf{p}_{1},  \dots ,\mathbf{p}_{n} \right \}) =- (-g)^{n-1} \,\,  
    \frac{{\widetilde v}^{\star}_{(1 \cdots n)1}}{{\widetilde v}^{\star}_{1(1 \cdots n)}} \, 
    \frac{\delta^{3} (\mathbf{p}_{1} + \cdots +\mathbf{p}_{n} - \mathbf{P})\,\,  \mathrm{Tr} (t^{a} t^{b_{1}} \cdots t^{b_{n}})}{{\widetilde v}^{\star}_{21}{\widetilde v}^{\star}_{32} \cdots {\widetilde v}^{\star}_{n(n-1)}}  
      \, ,
    \label{eq:psi_kernel2}
\end{equation}
\begin{equation}
    {\widetilde \Omega}_{n}^{a b_1 \left \{b_2 \cdots b_n \right \} }(\mathbf{P}; \mathbf{p}_{1} , \left \{ \mathbf{p}_{2} , \dots ,\mathbf{p}_{n} \right \} ) = n \left(\frac{p_1^+}{p_{1\cdots n}^+}\right)^2 {\widetilde \Psi}_{n}^{a b_1 \cdots b_n }(\mathbf{P};  \mathbf{p}_{1},  \dots , \mathbf{p}_{n}) \, .
    \label{eq:omega_kernel2}
\end{equation}
When substituting Eq.~\eqref{eq:Abullet_sol2}-\eqref{eq:A_star_solu2} to Eq.~\eqref{eq:Bstar1}, we will be using different color and momentum indices for each of these series to avoid confusions. The first order term is trivial. For the second order in fields, we get 
\begin{multline}
    0 = \int\!d^{3}\mathbf{q}_{1}d^{3}\mathbf{q}_{2} {\widetilde \Upsilon}_{2}^{a c_1 \left \{c_2 \right \} }(\mathbf{P}; \mathbf{q}_{1} ,\left \{ \mathbf{q}_{2} \right \}) {\widetilde B}^\star_{c_1} (x^+;\mathbf{q}_{1}){\widetilde B}^\bullet_{c_2} (x^+;\mathbf{q}_{2})  \\
 +  \int\!d^{3}\mathbf{p}_{1}d^{3}\mathbf{p}_{2} {\widetilde \Omega}_{2}^{a b_1 \left \{b_2 \right \} }(\mathbf{P}; \mathbf{p}_{1} ,\left \{ \mathbf{p}_{2} \right \}) {\widetilde B}^{\star}_{b_1} (x^+;\mathbf{p}_{1}) {\widetilde B}^{\bullet}_{b_2} (x^+;\mathbf{p}_{2})\, .
  \label{eq:2ndterm}
\end{multline}
Although the curly braces (on just a single index) in the expression above are redundant, we keep them for consistency in the symbolic representation of the kernel. Substituting for ${\widetilde \Omega}_{2}^{a b_1 \left \{b_2 \right \} }(\mathbf{P}; \mathbf{p}_{1} ,\left \{ \mathbf{p}_{2} \right \})$ using Eq.~\eqref{eq:omega_kernel2} we get
\begin{multline}
    {\widetilde \Upsilon}_{2}^{a b_1 \left \{b_2 \right \} }(\mathbf{P}; \mathbf{p}_{1} ,\left \{ \mathbf{p}_{2} \right \})  = - 2 \left(\frac{p_1^+}{p_{12}^+}\right )^2 \left[\frac{1}{2}\widetilde{\Psi}_2^{a\{b_1 b_2\}}(\mathbf{P};\{\mathbf{p}_1,\mathbf{p}_2\}) \right. \\
    \left. + \frac{1}{2}\widetilde{\Psi}_2^{a\{b_2 b_1\}}(\mathbf{P};\{\mathbf{p}_2,\mathbf{p}_1\})   \right] \, .
    \label{eq:uppsi}
\end{multline}
Recalling \cite{Kotko2017}
\begin{center}
\includegraphics[width=5.5cm]{Chapter_2/psi_2.eps}
\end{center}
\begin{multline}
    {\widetilde \Upsilon}_{2}^{a b_1 \left \{b_2 \right \} }(\mathbf{P}; \mathbf{p}_{1} ,\left \{ \mathbf{p}_{2} \right \})  =  2 \left(\frac{p_1^+}{p_{12}^+}\right )^2 \left[\frac{1}{2}\widetilde{\Gamma}_2^{a\{b_1 b_2\}}(\mathbf{P};\{\mathbf{p}_1,\mathbf{p}_2\}) \right. \\
    \left. + \frac{1}{2}\widetilde{\Gamma}_2^{a\{b_2 b_1\}}(\mathbf{P};\{\mathbf{p}_2,\mathbf{p}_1\})   \right] \, .
    \label{eq:uppsi2}
\end{multline}
Just like for the $\widetilde\Psi$ kernels in Eq.~\eqref{eq:psi_kernel_not}, we introduce similar notation for the $\widetilde\Gamma$ kernels as shown below
\begin{multline}
    \widetilde{\Gamma}_n^{a b_1b_2 \dots b_n}(\mathbf{P};\mathbf{p}_1,\mathbf{p}_2, \dots \mathbf{p}_n) = \frac{(-g)^{n-1}}{n!}\delta^3(\mathbf{p}_{12 \dots n} - \mathbf{P})\\
    \!\!\sum_{\text{\scriptsize permutations}}
    \frac{1}{\widetilde{v}_{1\left(1\dots n\right)}^{\star}\widetilde{v}_{\left(12\right)\left(1\dots n\right)}^{\star}\dots\widetilde{v}_{\left(1\dots n-1\right)\left(1\dots n\right)}^{\star}}\mathrm{Tr}(t^a t^{b_1} t^{b_2}\dots  t^{b_n})\,.
    \label{eq:gamma_kernel_not}
\end{multline}
Using this we get
\begin{equation}
    {\widetilde \Upsilon}_{2}^{a b_1 \left \{b_2 \right \} }(\mathbf{P}; \mathbf{p}_{1} ,\left \{ \mathbf{p}_{2} \right \})  =  2 \left(\frac{p_1^+}{p_{12}^+}\right )^2  {\widetilde \Gamma}_{2}^{a b_1 b_2 }(\mathbf{P}; \mathbf{p}_{1} , \mathbf{p}_{2} ) \, .
    \label{eq:r2ob}
\end{equation}

For the third order in fields, we have
\begin{multline}
     0 = \int\!d^{3}\mathbf{q}_{1}d^{3}\mathbf{q}_{2} d^{3}\mathbf{q}_{3} {\widetilde \Upsilon}_{3}^{a c_1 \left \{c_2 c_3 \right \} }(\mathbf{P}; \mathbf{q}_{1} ,\left \{ \mathbf{q}_{2} , \mathbf{q}_{3} \right \}) {\widetilde B}^\star_{c_1} (x^+;\mathbf{q}_{1}){\widetilde B}^\bullet_{c_2} (x^+;\mathbf{q}_{2})
{\widetilde B}^\bullet_{c_3} (x^+;\mathbf{q}_{3}) \\
+ \int\!d^{3}\mathbf{p}_{1}d^{3}\mathbf{p}_{2} d^{3}\mathbf{p}_{3} {\widetilde \Omega}_{3}^{a b_1 \left \{b_2 b_3 \right \} }(\mathbf{P}; \mathbf{p}_{1} ,\left \{ \mathbf{p}_{2} , \mathbf{p}_{3} \right \})  {\widetilde B}^{\star}_{b_1} (x^+;\mathbf{p}_{1})  
 {\widetilde B}^{\bullet}_{b_2} (x^+;\mathbf{p}_{2}) {\widetilde B}^{\bullet}_{b_3}(x^+;\mathbf{p}_{3}) \\
 + \int\!d^{3}\mathbf{q}_{1}d^{3}\mathbf{q}_{2} {\widetilde \Upsilon}_{2}^{a c_1 \left \{c_2 \right \} }(\mathbf{P} ; \mathbf{q}_{1} ,\left \{ \mathbf{q}_{2} \right \})  
 \Bigg[\int\!d^{3}\mathbf{p}_{1}d^{3}\mathbf{p}_{2} {\widetilde \Omega}_{2}^{c_1 b_1 \left \{b_2 \right \} }(\mathbf{q}_{1}; \mathbf{p}_{1},\left \{ \mathbf{p}_{2}  \right \}) {\widetilde B}^\star_{b_1} (x^+;\mathbf{p}_{1})  \\
  {\widetilde B}^\bullet_{b_2} (x^+;\mathbf{p}_{2}){\widetilde B}^{\bullet}_{c_2} (x^+;\mathbf{q}_{2})  + {\widetilde B}^{\star}_{c_1} (x^+;\mathbf{q}_{1})\int\!d^{3}\mathbf{s}_{1} d^{3}\mathbf{s}_{2} {\widetilde \Psi}_{2}^{c_2  \left \{d_1 d_2 \right \} }(\mathbf{q}_{2}; \left \{ \mathbf{s}_{1} , \mathbf{s}_{2} \right \}) \\
  {\widetilde B}^\bullet_{d_1} (x^+;\mathbf{s}_{1}){\widetilde B}^\bullet_{d_2} (x^+;\mathbf{s}_{2})    \Bigg]\,.
\end{multline}
Renaming the terms such we have ${\widetilde B}^\star_{e_1} (x^+;\mathbf{t}_{1}){\widetilde B}^\bullet_{e_2} (x^+;\mathbf{t}_{2}){\widetilde B}^\bullet_{e_3} (x^+;\mathbf{t}_{3})$ in all of the terms above we get
\begin{multline}
    0 = \int\!d^{3}\mathbf{t}_{1}d^{3}\mathbf{t}_{2} d^{3}\mathbf{t}_{3}  {\widetilde B}^\star_{e_1} (x^+;\mathbf{t}_{1}){\widetilde B}^\bullet_{e_2} (x^+;\mathbf{t}_{2}){\widetilde B}^\bullet_{e_3} (x^+;\mathbf{t}_{3})  
 \left[ {\widetilde \Upsilon}_{3}^{a e_1 \left \{e_2 e_3 \right \} }(\mathbf{P}; \mathbf{t}_{1} ,\left \{ \mathbf{t}_{2} , \mathbf{t}_{3} \right \}) \right. \\
 \left. + {\widetilde \Omega}_{3}^{a e_1 \left \{e_2 e_3 \right \} }(\mathbf{P}; \mathbf{t}_{1} ,\left \{ \mathbf{t}_{2} , \mathbf{t}_{3} \right \})  + {\widetilde \Upsilon}_{2}^{a c_1 \left \{e_3 \right \} }(\mathbf{P} ; \mathbf{t}_{1}+\left\{ \mathbf{t}_{2}\right\} ,\left \{ \mathbf{t}_{3} \right \}){\widetilde \Omega}_{2}^{c_1 e_1 \left \{e_2 \right \} }(\mathbf{q}_{1}; \mathbf{t}_{1},\left \{ \mathbf{t}_{2}  \right \})\right. \\
 \left.  + {\widetilde \Upsilon}_{2}^{a e_1 \left \{c_2 \right \} }(\mathbf{P} ; \mathbf{t}_{1} ,\left \{ \mathbf{t}_{2}+\mathbf{t}_{3} \right \}) {\widetilde \Psi}_{2}^{c_2  \left \{e_2 e_3 \right \} }(\mathbf{q}_{2}; \left \{ \mathbf{t}_{2} , \mathbf{t}_{3} \right \})
 \right]\,.
\end{multline}
Form which we get the following relation among the kernels
\begin{multline}
    {\widetilde \Upsilon}_{3}^{a e_1 \left \{e_2 e_3 \right \} }(\mathbf{P}; \mathbf{t}_{1} ,\left \{ \mathbf{t}_{2} , \mathbf{t}_{3} \right \}) = - \left[{\widetilde \Omega}_{3}^{a e_1 \left \{e_2 e_3 \right \} }(\mathbf{P}; \mathbf{t}_{1} ,\left \{ \mathbf{t}_{2} , \mathbf{t}_{3} \right \})  \right.  \\
 \left.  + {\widetilde \Upsilon}_{2}^{a c_1 \left \{e_3 \right \} }(\mathbf{P} ; \mathbf{t}_{1}+\left\{ \mathbf{t}_{2}\right\} ,\left \{ \mathbf{t}_{3} \right \}) {\widetilde \Omega}_{2}^{c_1 e_1 \left \{e_2 \right \} }(\mathbf{q}_{1}; \mathbf{t}_{1},\left \{ \mathbf{t}_{2}  \right \})  \right.  \\
 \left. + {\widetilde \Upsilon}_{2}^{a e_1 \left \{c_2 \right \} }(\mathbf{P} ; \mathbf{t}_{1} ,\left \{ \mathbf{t}_{2}+\mathbf{t}_{3} \right \}){\widetilde \Psi}_{2}^{c_2  \left \{e_2 e_3 \right \} }(\mathbf{q}_{2}; \left \{ \mathbf{t}_{2} , \mathbf{t}_{3} \right \})
 \right] \,.
 \label{eq:up3lon}
\end{multline}
The first term on the R.H.S above is fully symmetric, however, the remaining two terms can be symmetrized with respect to the ${\widetilde B}^{\bullet}$ fields. By doing that we get,
\begin{multline}
    {\widetilde \Upsilon}_{3}^{a e_1 \left \{e_2 e_3 \right \} }(\mathbf{P}; \mathbf{t}_{1} ,\left \{ \mathbf{t}_{2} , \mathbf{t}_{3} \right \}) = - 3 \left(\frac{t_1^+}{t^+_{123}}\right)^2\frac{1}{3!}\\
    \Bigg[ \widetilde{\Psi}_3^{a\{e_1 e_2 e_3\}}(\mathbf{P};\{\mathbf{t}_1,\mathbf{t}_2,\mathbf{t}_3\}) + \widetilde{\Psi}_2^{a\{b e_3\}}(\mathbf{P};\{\mathbf{t}_{12},\mathbf{t}_3\})\widetilde{\Gamma}_2^{b\{e_1 e_2\}}(\mathbf{t}_{12};\{\mathbf{t}_1,\mathbf{t}_2\})\\
    + \widetilde{\Psi}_2^{a\{e_1 b\}}(\mathbf{P};\{\mathbf{t}_1,\mathbf{t}_{23}\})\widetilde{\Gamma}_2^{b\{e_2 e_3\}}(\mathbf{t}_{23};\{\mathbf{t}_2,\mathbf{t}_3\})\\
   + \widetilde{\Psi}_3^{a\{e_1 e_3 e_2\}}(\mathbf{P};\{\mathbf{t}_1,\mathbf{t}_3,\mathbf{t}_2\}) + \widetilde{\Psi}_2^{a\{b e_2\}}(\mathbf{P};\{\mathbf{t}_{13},\mathbf{t}_2\})\widetilde{\Gamma}_2^{b\{e_1 e_3\}}(\mathbf{t}_{13};\{\mathbf{t}_1,\mathbf{t}_3\})\\
    + \widetilde{\Psi}_2^{a\{e_1 b\}}(\mathbf{P};\{\mathbf{t}_1,\mathbf{t}_{23}\})\widetilde{\Gamma}_2^{b\{e_3 e_2\}}(\mathbf{t}_{23};\{\mathbf{t}_3,\mathbf{t}_2\})\\
    + \widetilde{\Psi}_3^{a\{e_2 e_1 e_3\}}(\mathbf{P};\{\mathbf{t}_2,\mathbf{t}_1,\mathbf{t}_3\}) + \widetilde{\Psi}_2^{a\{b e_3\}}(\mathbf{P};\{\mathbf{t}_{12},\mathbf{t}_3\})\widetilde{\Gamma}_2^{b\{e_2 e_1\}}(\mathbf{t}_{12};\{\mathbf{t}_2,\mathbf{t}_1\})\\
    + \widetilde{\Psi}_2^{a\{e_2 b\}}(\mathbf{P};\{\mathbf{t}_2,\mathbf{t}_{13}\})\widetilde{\Gamma}_2^{b\{e_1 e_3\}}(\mathbf{t}_{13};\{\mathbf{t}_1,\mathbf{t}_3\})\\
    + \widetilde{\Psi}_3^{a\{e_2 e_3 e_1\}}(\mathbf{P};\{\mathbf{t}_2,\mathbf{t}_3,\mathbf{t}_1\}) + \widetilde{\Psi}_2^{a\{b e_1\}}(\mathbf{P};\{\mathbf{t}_{23},\mathbf{t}_1\})\widetilde{\Gamma}_2^{b\{e_2 e_3\}}(\mathbf{t}_{23};\{\mathbf{t}_2,\mathbf{t}_3\})\\
    + \widetilde{\Psi}_2^{a\{e_2 b\}}(\mathbf{P};\{\mathbf{t}_2,\mathbf{t}_{13}\})\widetilde{\Gamma}_2^{b\{e_3 e_1\}}(\mathbf{t}_{13};\{\mathbf{t}_3,\mathbf{t}_1\})\\
    + \widetilde{\Psi}_3^{a\{e_3 e_1 e_2\}}(\mathbf{P};\{\mathbf{t}_3,\mathbf{t}_1,\mathbf{t}_2\}) + \widetilde{\Psi}_2^{a\{b e_2\}}(\mathbf{P};\{\mathbf{t}_{13},\mathbf{t}_2\})\widetilde{\Gamma}_2^{b\{e_3 e_1\}}(\mathbf{t}_{13};\{\mathbf{t}_3,\mathbf{t}_1\})\\
    + \widetilde{\Psi}_2^{a\{e_3 b\}}(\mathbf{P};\{\mathbf{t}_3,\mathbf{t}_{12}\})\widetilde{\Gamma}_2^{b\{e_1 e_2\}}(\mathbf{t}_{12};\{\mathbf{t}_1,\mathbf{t}_2\})\\
    +\widetilde{\Psi}_3^{a\{e_3 e_2 e_1\}}(\mathbf{P};\{\mathbf{t}_3,\mathbf{t}_2,\mathbf{t}_1\}) + \widetilde{\Psi}_2^{a\{b e_1\}}(\mathbf{P};\{\mathbf{t}_{23},\mathbf{t}_1\})\widetilde{\Gamma}_2^{b\{e_3 e_2\}}(\mathbf{t}_{23};\{\mathbf{t}_3,\mathbf{t}_2\})\\
    + \widetilde{\Psi}_2^{a\{e_3 b\}}(\mathbf{P};\{\mathbf{t}_3,\mathbf{t}_{12}\})\widetilde{\Gamma}_2^{b\{e_2 e_1\}}(\mathbf{t}_{12};\{\mathbf{t}_2,\mathbf{t}_1\})\Bigg]\,.
\end{multline}
However, using the following identity \cite{Kotko2017}
\begin{center}
\includegraphics[width=12.3cm]{Chapter_2/psi_3.eps}
\end{center}
we get
\begin{equation}
    {\widetilde \Upsilon}_{3}^{a e_1 \left \{e_2 e_3 \right \} }(\mathbf{P}; \mathbf{t}_{1} ,\left \{ \mathbf{t}_{2} , \mathbf{t}_{3} \right \})
 =  3\left(\frac{t_1^+}{t^+_{123}}\right)^2  {\widetilde \Gamma}_3^{a e_1 e_2 e_3}\left(\mathbf{P}; \mathbf{t}_{1} , \mathbf{t}_{2} , \mathbf{t}_{3} \right ) \, .
\end{equation}
The above results can be generalized for any $n$ as follows
\begin{equation}
    {\widetilde \Upsilon}_{n}^{a b_1 \left \{b_2 \cdots b_n \right \} }(\mathbf{P}; \mathbf{p}_1 ,\left \{ \mathbf{p}_2 , \cdots \mathbf{p}_n \right \}) = n\left(\frac{p_1^+}{p_{1\cdots n}^+}\right )^2 {\widetilde \Gamma}_{n}^{a b_1 \cdots b_n }\left(\mathbf{P}; \mathbf{p}_{1}  \cdots \mathbf{p}_{n} \right) \,.
    \label{eq:Upsilon_n}
\end{equation}

\section{\texorpdfstring{${\widetilde B}^{\star}[A^{\bullet},A^{\star}]$}{Bstar} from Wilson line \texorpdfstring{${\hat B}^{\star}[A^{\bullet},A^{\star}]$}{Astar}}
\label{subsec:app_A42}

In this section we Fourier transform the Wilson line ${\hat B}^{\star}[A^{\bullet},A^{\star}](x^+;\mathbf{x})$ to obtain the momentum space expression  ${\widetilde B}^{\star}[A^{\bullet},A^{\star}](x^+;\mathbf{P})$. To do this we start with the following definition 
\begin{multline}
    B_a^{\star}(x) = 
    \int_{-\infty}^{+\infty}\! d\alpha\,\, 
    \mathrm{Tr} \Big\{
    \frac{1}{2\pi g} t^a \partial_-^{-1} 
    \int\! d^4y \,
     \left[\partial_-^2 {A}^{\star}_c (y)\right]  \\
     \frac{\delta}{\delta {A}^{\bullet}_c (y)} \,
    \mathbb{P} \exp {\left[ig \int_{-\infty}^{+ \infty}\! ds \,
    \hat{A}^{\bullet}(x+s\varepsilon_{\alpha}^+)\right]  } 
    \Big\} \,.
    \label{eq:Bstar_WL1}
\end{multline}
In order to conveniently manage the computation, we separate the straight infinite Wilson line from the $\alpha$ integral as follows
\begin{equation}
    B_a^{\star}(x) = 
    \int_{-\infty}^{+\infty}\! d\alpha\,\, 
    \mathrm{Tr} \Big\{
    \frac{1}{2\pi g} t^a \partial_-^{-1} G(x)\Big\} \,,
    \label{eq:bstar2}
\end{equation}
where we encode the derivative of the Wilson line in $G(x)$ as shown below
\begin{equation}
    G(x) = \int\! d^4y \,
     \left[\partial_-^2 {A}^{\star}_c (y)\right]  
     \frac{\delta}{\delta {A}^{\bullet}_c (y)} \,
    \mathbb{P} \exp {\left[ig \int_{-\infty}^{+ \infty}\! ds \,
    \hat{A}^{\bullet}(x+s\varepsilon_{\alpha}^+)\right]  } \, .
    \label{eq:gdef}
\end{equation}
The path-ordered exponential above can be expanded into a series. The functional derivative of this series with respect to ${A}^{\bullet}_c (y)$  will introduce a Dirac delta $\delta^4 (x+s_i \varepsilon_{\alpha}^+- y)$ relating the position co-ordinate of the former with that of the differentiated ${A}^{\bullet}_{b_i} (x+s_i \varepsilon_{\alpha})$ field and a Kronecker delta $\delta^c_{b_i}$ relating their color. After solving for these deltas we get
\begin{multline}
    G(x)= ig\int_{-\infty}^{+ \infty}\! ds_1\,  \partial_-^2 {\hat{A}}^\star (x+s_1 \varepsilon_{\alpha}^+)\\
    +  (ig)^2\int_{-\infty}^{+ \infty} \! ds_1\,\int_{-\infty}^{s_1}\!  ds_2\, \left \{{\hat{A}}^\bullet (x+s_1 \varepsilon_{\alpha}^+) \partial_-^2 {\hat{A}}^\star(x+s_2 \varepsilon_{\alpha}^+) + \partial_-^2 {\hat{A}}^\star(x+s_1 \varepsilon_{\alpha}^+){\hat{A}}^\bullet (x+s_2 \varepsilon_{\alpha}^+) \right \}\\ + (ig)^3\int_{-\infty}^{+ \infty} \! ds_1\,
    \int_{-\infty}^{s_1} \! ds_2\, \int_{-\infty}^{s_2} \! ds_3\,  \left \{ {\hat{A}}^\bullet (x+s_1 \varepsilon_{\alpha}^+){\hat{A}}^\bullet (x+s_2 \varepsilon_{\alpha}^+) \partial_-^2 {\hat{A}}^\star(x+s_3 \varepsilon_{\alpha}^+) \right.   \\
    \left. + {\hat{A}}^\bullet (x+s_1 \varepsilon_{\alpha}^+)\partial_-^2 {\hat{A}}^\star(x+s_2 \varepsilon_{\alpha}^+){\hat{A}}^\bullet (x+s_3 \varepsilon_{\alpha}^+)  \right.   \\
    \left. +\partial_-^2 {\hat{A}}^\star(x+s_1 \varepsilon_{\alpha}^+) {\hat{A}}^\bullet (x+s_2 \varepsilon_{\alpha}^+){\hat{A}}^\bullet (x+s_3 \varepsilon_{\alpha}^+)\right\} + \cdots 
\end{multline}
The R.H.S. of the expression above can be re-written as
\begin{multline}
    G(x)= ig\int_{-\infty}^{+\infty}\! ds_1\,\int_{-\infty}^{+\infty}\!d^3\, \mathbf{p}_1(-i p_1^+)^2 e^{-i s_1 \mathbf{e}_{\alpha}\cdot \mathbf{p}_1}  e^{-i\mathbf{x}\cdot\mathbf{p}_1}{\widetilde A}_{b_1}^\star (x^+;\mathbf{p}_1) t^{b_1} \\
+(ig)^2 \int_{-\infty}^{+\infty}\! ds_1\, \int_{-\infty}^{s_1} \! ds_2\, \int_{-\infty}^{+\infty}\! d^3\,\mathbf{p}_1\, d^3 \mathbf{p}_2\,e^{-i s_1 \mathbf{e}_{\alpha}\cdot \mathbf{p}_1}e^{-i s_2 \mathbf{e}_{\alpha}\cdot \mathbf{p}_2} e^{-i\mathbf{x}\cdot(\mathbf{p}_1+\mathbf{p}_2)}  \\
 \left\{ {\widetilde A}_{b_1}^\bullet (x^+;\mathbf{p_1})t^{b_1}(-i p_2^+)^2{\widetilde A}_{b_2}^\star (x^+;\mathbf{p_2}) t^{b_2} +    (-i p_1^+)^2{\widetilde A}_{b_1}^\star (x^+;\mathbf{p_1})t^{b_1} {\widetilde A}_{b_2}^\bullet (x^+;\mathbf{p_2})t^{b_2}\right\} \\
 + (ig)^3\int_{-\infty}^{+ \infty}\! ds_1\,\int_{-\infty}^{s_1} \! ds_2\,\int_{-\infty}^{s_2} \! ds_3\,\int_{-\infty}^{+\infty}\!d^3\,\mathbf{p}_1 d^3 \mathbf{p}_2\, d^3 \mathbf{p}_3\, e^{-i s_1 \mathbf{e}_{\alpha}\cdot \mathbf{p}_1 -i s_2 \mathbf{e}_{\alpha}\cdot \mathbf{p}_2 -i s_3 \mathbf{e}_{\alpha}\cdot \mathbf{p}_3} \\
 e^{-i\mathbf{x}\cdot(\mathbf{p}_1+\mathbf{p}_2+\mathbf{p}_3)}
\left\{ {\widetilde A}_{b_1}^\bullet (x^+;\mathbf{p}_1)t^{b_1} {\widetilde A}_{b_2}^\bullet (x^+;\mathbf{p}_2)t^{b_2}(-i p_3^+)^2{\widetilde A}_{b_3}^\star (x^+;\mathbf{p}_3)t^{b_3}\right. \\
\left.  +  {\widetilde A}_{b_1}^\bullet (x^+;\mathbf{p}_1)t^{b_1}  (-i p_2^+)^2{\widetilde A}_{b_2}^\star (x^+;\mathbf{p}_2)t^{b_2}{\widetilde A}_{b_3}^\bullet (x^+;\mathbf{p}_3)t^{b_3}\right.  \\
\left. + (-i p_1^+)^2{\widetilde A}_{b_1}^\star (x^+;\mathbf{p}_1)t^{b_1} {\widetilde A}_{b_2}^\bullet (x^+;\mathbf{p}_2)t^{b_2}{\widetilde A}_{b_3}^\bullet (x^+;\mathbf{p}_3)t^{b_3}\right\} + \cdots
\label{eq:gdcop}
\end{multline}
Above we used $\mathbf{e}_{\alpha} \equiv (-\alpha, 0, -1)$ to represent  $(x^-, x^\bullet, x^\star)$ coordinates of  $\varepsilon_{\alpha}^+$. For the ordered integrals over the $s_i$ ($i = 1, \dots, n$), we have the following relation
\begin{multline}
     \int_{-\infty}^{+\infty}\! ds_1\, \cdots \int_{-\infty}^{s_{n-1}} \!ds_n \, e^{-i s_1 \mathbf{e}_{\alpha}\cdot \mathbf{p}_1}\cdots e^{-i s_n \mathbf{e}_{\alpha}\cdot \mathbf{p}_n} = 2\pi\delta(\mathbf{e}_{\alpha}\cdot \mathbf{p}_{1\cdots n}) \\
    \times \frac{i^{n-1}}{(\mathbf{e}_{\alpha}\cdot \mathbf{p}_{2\cdots n} + i\epsilon)(\mathbf{e}_{\alpha}\cdot \mathbf{p}_{3\cdots n} + i\epsilon) \cdots (\mathbf{e}_{\alpha}\cdot \mathbf{p}_{n} + i\epsilon)} \, .
    \label{eq:odintre}
\end{multline}
The delta $\delta(\mathbf{e}_{\alpha}\cdot \mathbf{p}_{1\cdots n})$ on the R.H.S above allows to easily perform the $\alpha$ integration in Eq.\eqref{eq:bstar2}. Fourier transforming the expression Eq.\eqref{eq:bstar2} to 3D momentum space (keeping $x^+$ fixed) and integrating out $\alpha$ we get (below we suppress the $i\epsilon$)
\begin{multline}
    {\widetilde{B}}_a^\star (x^+;\mathbf{P})
= {\widetilde A}^\star_{a} (x^+;\mathbf{P}) + (-g)\int_{-\infty}^{+\infty}\! d^3\,\mathbf{p}_1 d^3\, \mathbf{p}_{2}\frac{\delta^3 (\mathbf{P} - \mathbf{p}_{{12}})}{{\widetilde v}^{\star}_{1(12)}} \\ \left\{ {\widetilde A}^\bullet_{b_1} (x^+;\mathbf{p}_{1}) \left(\frac{p_2^+}{p_{12}^+}\right )^2{\widetilde A}^\star_{b_2} (x^+;\mathbf{p}_{2}) +  \left(\frac{p_1^+}{p_{12}^+}\right )^2  {\widetilde A}^\star_{b_1} (x^+;\mathbf{p}_{1})  {\widetilde A}^\bullet_{b_2} (x^+;\mathbf{p}_{2}) \right\} \mathrm{Tr}(t^a t^{b_1} t^{b_2}) \\
+ g^2 \int_{-\infty}^{+\infty}\! d^3\,\mathbf{p}_{1} d^3\, \mathbf{p}_{2} d^3\, \mathbf{p}_{3} \frac{\delta^3 (\mathbf{P} - \mathbf{p}_{{123}})}{\widetilde{v}^{\star}_{1(123)}\widetilde{v}^{\star}_{(12)(123)}}   \left\{{\widetilde A}^\bullet_{b_1} (x^+;\mathbf{p}_{1}){\widetilde A}^\bullet_{b_2} (x^+;\mathbf{p}_{2}) \left(\frac{p_3^+}{p_{123}^+}\right )^2{\widetilde A}^\star_{b_3} (x^+;\mathbf{p}_{3}) \right.  \\
\left. +  {\widetilde A}^\bullet_{b_1} (x^+;\mathbf{p}_{1})   \left(\frac{p_2^+}{p_{123}^+}\right )^2{\widetilde A}^\star_{b_2} (x^+;\mathbf{p}_{2}){\widetilde A}^\bullet_{b_3} (x^+;\mathbf{p}_{3}) \right. \\
 \left.  + \left(\frac{p_1^+}{p_{123}^+}\right )^2{\widetilde A}^\star_{b_1} (x^+;\mathbf{p}_{1}){\widetilde A}^\bullet_{b_2} (x^+;\mathbf{p}_{2}){\widetilde A}^\bullet_{b_3} (x^+;\mathbf{p}_{3}) \right\}\mathrm{Tr}(t^a t^{b_1} t^{b_2} t^{b_3})+ \cdots 
\label{eq:bstarmom}
\end{multline}
Let us deal with the above expression order by order. The first order is trivial. Identifying that the kernel in the second order term is $\widetilde{\Gamma}_2^{a\{b_1 b_2\}}(\mathbf{P};\{\mathbf{p}_1,\mathbf{p}_2\})$, we can re-write it as
\begin{multline}
    \left[{\widetilde{B}}_a^\star (x^+;\mathbf{P})\right]_{2nd} = 2 \int_{-\infty}^{+\infty}\! d^3\,\mathbf{p}_{1}\, d^3\, \mathbf{p}_{2}\, \frac{1}{2}\widetilde{\Gamma}_2^{a\{b_1 b_2\}}(\mathbf{P};\{\mathbf{p}_1,\mathbf{p}_2\})\\
    \left\{ {\widetilde A}^\bullet_{b_1} (x^+;\mathbf{p}_{1})\left(\frac{p_2^+}{p_{12}^+}\right )^2{\widetilde A}^\star_{b_2} (x^+;\mathbf{p}_{2}) +     {\widetilde A}^\star_{b_1} (x^+;\mathbf{p}_{1}) \left(\frac{p_1^+}{p_{12}^+}\right )^2 {\widetilde A}^\bullet_{b_2} (x^+;\mathbf{p}_{2}) \right\} \, .
\end{multline}
Renaming $2 \leftrightarrow 1$ in the first term above we get
\begin{multline}
    \left[{\widetilde{B}}_a^\star (x^+;\mathbf{P})\right]_{2nd} = 2 \int_{-\infty}^{+\infty}\! d^3\,\mathbf{p}_{1} d^3\,  \mathbf{p}_{2}\,{\widetilde A}^\star_{b_1} (x^+;\mathbf{p}_{1}){\widetilde A}^\bullet_{b_2} (x^+;\mathbf{p}_{2}) \\
\times \left(\frac{p_1^+}{p_{12}^+}\right )^2  \left[\frac{1}{2}\widetilde{\Gamma}_2^{a\{b_1 b_2\}}(\mathbf{P};\{\mathbf{p}_1,\mathbf{p}_2\}) + \frac{1}{2}\widetilde{\Gamma}_2^{a\{b_2 b_1\}}(\mathbf{P};\{\mathbf{p}_2,\mathbf{p}_1\})   \right] \, .
\label{eq:gam_2sta}
\end{multline}
Using the notation defined in Eq.~\eqref{eq:gamma_kernel_not}, we can re-write Eq.~\eqref{eq:gam_2sta} as
\begin{multline}
     \left[{\widetilde{B}}_a^\star (x^+;\mathbf{P})\right]_{2nd} = \int_{-\infty}^{+\infty}\! d^3\,\mathbf{p}_{1}\, d^3\, \mathbf{p}_{2}\,{\widetilde A}^\star_{b_1} (x^+;\mathbf{p}_{1}){\widetilde A}^\bullet_{b_2} (x^+;\mathbf{p}_{2}) \\
    \left[ 2  \left(\frac{p_1^+}{p_{12}^+}\right )^2 {\widetilde \Gamma}_2^{a b_1 b_2}(\mathbf{P}; \mathbf{p}_{1}, \mathbf{p}_{2} )     \right] \, .
    \label{eq:bsttwo}
\end{multline}
For the third order, we have
\begin{multline}
    \left[{\widetilde{B}}_a^\star (x^+;\mathbf{P})\right]_{3rd} = (3!) \int_{-\infty}^{+\infty}\! d^3\,\mathbf{p}_{1} \,d^3\, \mathbf{p}_{2} \,d^3\, \mathbf{p}_{3} \,\frac{1}{3!}\widetilde{\Gamma}_3^{a\{b_1 b_2 b_3\}}(\mathbf{P};\{\mathbf{p}_1,\mathbf{p}_2,\mathbf{p}_3\})\\
    \left\{{\widetilde A}^\bullet_{b_1} (x^+;\mathbf{p}_{1}){\widetilde A}^\bullet_{b_2} (x^+;\mathbf{p}_{2}) \left(\frac{p_3^+}{p_{123}^+}\right )^2{\widetilde A}^\star_{b_3} (x^+;\mathbf{p}_{3})+  {\widetilde A}^\bullet_{b_1} (x^+;\mathbf{p}_{1})   \left(\frac{p_2^+}{p_{123}^+}\right )^2{\widetilde A}^\star_{b_2} (x^+;\mathbf{p}_{2})\right.  \\
 \left. {\widetilde A}^\bullet_{b_3} (x^+;\mathbf{p}_{3})  + \left(\frac{p_1^+}{p_{123}^+}\right )^2{\widetilde A}^\star_{b_1} (x^+;\mathbf{p}_{1}){\widetilde A}^\bullet_{b_2} (x^+;\mathbf{p}_{2}){\widetilde A}^\bullet_{b_3} (x^+;\mathbf{p}_{3}) \right\} \, .
\end{multline}
Renaming the terms such that we have ${\widetilde A}^\star_{b_1} (x^+;\mathbf{p}_{1}){\widetilde A}^\bullet_{b_2} (x^+;\mathbf{p}_{2}){\widetilde A}^\bullet_{b_3} (x^+;\mathbf{p}_{3})$ in each of the terms above, we get
\begin{multline}
    \left[{\widetilde{B}}_a^\star (x^+;\mathbf{P})\right]_{3rd} = (3!) \int_{-\infty}^{+\infty}\!d^3\, \mathbf{p}_{1} d^3\, \mathbf{p}_{2} d^3\, \mathbf{p}_{3}\,{\widetilde A}^\star_{b_1} (x^+;\mathbf{p}_{1}){\widetilde A}^\bullet_{b_2} (x^+;\mathbf{p}_{2}) 
 {\widetilde A}^\bullet_{b_3} (x^+;\mathbf{p}_{3}) 
 \left(\frac{p_1^+}{p_{123}^+}\right )^2 \\
\frac{1}{3!} \left[ \widetilde{\Gamma}_3^{a\{b_1 b_2 b_3\}}(\mathbf{P};\{\mathbf{p}_1,\mathbf{p}_2,\mathbf{p}_3\}) + \widetilde{\Gamma}_3^{a\{b_2 b_1 b_3\}}(\mathbf{P};\{\mathbf{p}_2,\mathbf{p}_1,\mathbf{p}_3\}) +\widetilde{\Gamma}_3^{a\{b_3 b_2 b_1\}}(\mathbf{P};\{\mathbf{p}_3,\mathbf{p}_2,\mathbf{p}_1\})    \right] \, .
\end{multline}
The three terms above can be symmetrized with respect to the ${\widetilde A}^\bullet$ fields. This can be done as shown below
\begin{multline}
    \left[{\widetilde{B}}_a^\star (x^+;\mathbf{P})\right]_{3rd} = (3!) \int_{-\infty}^{+\infty}\! d^3\,\mathbf{p}_{1} d^3\, \mathbf{p}_{2} d^3\, \mathbf{p}_{3}\,{\widetilde A}^\star_{b_1} (x^+;\mathbf{p}_{1}){\widetilde A}^\bullet_{b_2} (x^+;\mathbf{p}_{2})
{\widetilde A}^\bullet_{b_3} (x^+;\mathbf{p}_{3}) \\
\left(\frac{p_1^+}{p_{123}^+}\right )^2
 \frac{1}{3!} \left[ \frac{1}{2}\widetilde{\Gamma}_3^{a\{b_1 b_2 b_3\}}(\mathbf{P};\{\mathbf{p}_1,\mathbf{p}_2,\mathbf{p}_3\}) +\frac{1}{2}\widetilde{\Gamma}_3^{a\{b_1 b_3 b_2\}}(\mathbf{P};\{\mathbf{p}_1,\mathbf{p}_3,\mathbf{p}_2\}) \right.  \\
+ \left. \frac{1}{2}\widetilde{\Gamma}_3^{a\{b_2 b_1 b_3\}}(\mathbf{P};\{\mathbf{p}_2,\mathbf{p}_1,\mathbf{p}_3\})+ \frac{1}{2}\widetilde{\Gamma}_3^{a\{b_3 b_1 b_2\}}(\mathbf{P};\{\mathbf{p}_3,\mathbf{p}_1,\mathbf{p}_2\})\right.  \\
+\left. \frac{1}{2}\widetilde{\Gamma}_3^{a\{b_3 b_2 b_1\}}(\mathbf{P};\{\mathbf{p}_3,\mathbf{p}_2,\mathbf{p}_1\}) +\frac{1}{2}\widetilde{\Gamma}_3^{a\{b_2 b_3 b_1\}}(\mathbf{P};\{\mathbf{p}_2,\mathbf{p}_3,\mathbf{p}_1\})    \right] \, .
\end{multline}
Using the notation Eq.~\eqref{eq:gamma_kernel_not}, the above expression can be re-written as
\begin{multline}
     \left[{\widetilde{B}}_a^\star (x^+;\mathbf{P})\right]_{3rd} = \int_{-\infty}^{+\infty}\! d^3\,\mathbf{p}_{1} d^3\, \mathbf{p}_{2} d^3\, \mathbf{p}_{3}\,{\widetilde A}^\star_{b_1} (x^+;\mathbf{p}_{1}){\widetilde A}^\bullet_{b_2} (x^+;\mathbf{p}_{2})
    {\widetilde A}^\bullet_{b_3} (x^+;\mathbf{p}_{3})\\
    \left[ 3  \left(\frac{p_1^+}{p_{123}^+}\right )^2 {\widetilde \Gamma}_3^{a b_1 b_2 b_3}(\mathbf{P}; \mathbf{p}_{1}, \mathbf{p}_{2}, \mathbf{p}_{3} )     \right]\,.
    \label{eq:bstth}
\end{multline}
This procedure can be generalized for any $n$ as follows. For order $n$ in fields we start with the expression below
\begin{multline}
    \left[{\widetilde{B}}_a^\star (x^+;\mathbf{P})\right]_{n^{th}} = n! \int_{-\infty}^{+\infty}\! d^3\, \mathbf{p}_{1} \cdots d^3 \mathbf{p}_{n}\,  \frac{1}{n!}\widetilde{\Gamma}_{n}^{a\{b_{1}\dots b_{n}\}}\left(\mathbf{P};\{\mathbf{p}_{1},\dots,\mathbf{p}_{n}\}\right)\\
\left\{{\widetilde A}^\bullet_{b_1} (x^+;\mathbf{p}_{1})\cdots{\widetilde A}^\bullet_{b_{n-1}} (x^+;\mathbf{p}_{n-1}) \left(\frac{p_n^+}{p_{1\cdots n}^+}\right )^2{\widetilde A}^\star_{b_n} (x^+;\mathbf{p}_{n})  \right.  \\
 \left. +  {\widetilde A}^\bullet_{b_1} (x^+;\mathbf{p}_{1}) \cdots  \left(\frac{p_{n-1}^+}{p_{1\cdots n}^+}\right )^2{\widetilde A}^\star_{b_{n-1}} (x^+;\mathbf{p}_{n-1}){\widetilde A}^\bullet_{b_n} (x^+;\mathbf{p}_{n}) \right.  \\
 \left.+ \cdots + \left(\frac{p_1^+}{p_{1\cdots n}^+}\right )^2{\widetilde A}^\star_{1} (x^+;\mathbf{p}_{1}){\widetilde A}^\bullet_{b_2} (x^+;\mathbf{p}_{2})\cdots{\widetilde A}^\bullet_{b_n} (x^+;\mathbf{p}_{n}) \right\} \, .
 \label{eq:bnexp}
\end{multline}
Renaming the terms such that we have ${\widetilde A}^\star_{1} (x^+;\mathbf{p}_{1}){\widetilde A}^\bullet_{b_2} (x^+;\mathbf{p}_{2})\cdots{\widetilde A}^\bullet_{b_n} (x^+;\mathbf{p}_{n})$ in all the terms we get
\begin{multline}
    \left[{\widetilde{B}}_a^\star (x^+;\mathbf{P})\right]_{n^{th}} = n! \int_{-\infty}^{+\infty}\! d^3\,\mathbf{p}_{1} \cdots d^3\, \mathbf{p}_{n}\, {\widetilde A}^\star_{1} (x^+;\mathbf{p}_{1}){\widetilde A}^\bullet_{b_2} (x^+;\mathbf{p}_{2})\cdots{\widetilde A}^\bullet_{b_n} (x^+;\mathbf{p}_{n})\,  \\
\left(\frac{p_1^+}{p_{1\cdots n}^+}\right )^2  \frac{1}{n!} \bigg[ \widetilde{\Gamma}_{n}^{a\{b_{1}\dots b_{n}\}}\left(\mathbf{P};\{\mathbf{p}_{1},\dots,\mathbf{p}_{n}\}\right)\, + \,\widetilde{\Gamma}_{n}^{a\{b_{2}b_{1}b_{3}\dots b_{n}\}}\left(\mathbf{P};\{\mathbf{p}_{2},\mathbf{p}_{1},\mathbf{p}_{3},\dots,\mathbf{p}_{n}\}\right) +\cdots  \\
   + \widetilde{\Gamma}_{n}^{a\{b_{n-1}b_{2}\dots b_{n-2} b_1 b_{n}\}}\left(\mathbf{P};\{\mathbf{p}_{n-1},\mathbf{p}_{2},\dots ,\mathbf{p}_{n-2},\mathbf{p}_{1},\mathbf{p}_{n}\}\right)  \\
  + \widetilde{\Gamma}_{n}^{a\{b_{n}b_{2}\dots b_{n-1} b_1 \}}\left(\mathbf{P};\{\mathbf{p}_{n},\mathbf{p}_{2},\dots ,\mathbf{p}_{n-1},\mathbf{p}_{1}\}\right)     \bigg]\, ,
\end{multline}
where each term can be symmetrized with respect to the ${\widetilde A}^\bullet$ fields. This is done by renaming each of the $n$ terms in $(n-1)!$ ways. This introduces a factor of $1/(n-1)!$ which cancels against the $n!$ leaving an overall factor of $n$. By doing this we obtain
\begin{multline}
     \left[{\widetilde{B}}_a^\star (x^+;\mathbf{P})\right]_{n^{th}} = \int_{-\infty}^{+\infty}\! d^3\,\mathbf{p}_{1} \cdots d^3 \mathbf{p}_{n} \,{\widetilde A}^\star_{b_1} (x^+;\mathbf{p}_{1}) {\widetilde A}^\bullet_{b_2} (x^+;\mathbf{p}_{2})
    \cdots
    {\widetilde A}^\bullet_{b_n} (x^+;\mathbf{p}_{n})\\
    \left[ n\left(\frac{p_1^+}{p_{1\cdots n}^+}\right )^2 {\widetilde \Gamma}_{n}^{a b_1 \cdots b_n }(\mathbf{P}; \mathbf{p}_{1}  \cdots \mathbf{p}_{n} )    \right] \, .
    \label{eq:bstn}
\end{multline}

The above results match with the one we derived in the previous section Eq.~\eqref{eq:Upsilon_n}.
\chapter{Elimination of triple-gluon vertices from Yang-Mills action}
\label{sec:app_A5}

In this appendix, we demonstrate that the transformation
\begin{equation}
    \left\{\hat{A}^{\bullet},\hat{A}^{\star}\right\} \rightarrow \Big\{\hat{Z}^{\bullet}\big[{A}^{\bullet},{A}^{\star}\big],\hat{Z}^{\star}\big[{A}^{\bullet},{A}^{\star}\big]\Big\} \, ,
    \label{eq:general_transf_app}
\end{equation}
indeed eliminates both the triple-gluon vertices from the Yang-Mills action Eq.~\eqref{eq:actionLC_YM}. Recall, however, that this transformation can be executed in two ways (\emph{cf.} Figure \ref{fig:CT_paths}). For the current purpose, we will use the amalgamation of both approaches. Firstly, we use the second approach which involves two consecutive canonical transformations: Yang-Mills $\longrightarrow$  MHV action $\longrightarrow$ new action $S[{Z}^{\bullet}, {Z}^{\star}]$, to obtain the solutions $\widetilde{A}^{\bullet}_a[{Z}^{\bullet}, {Z}^{\star}](x^+;\mathbf{P})$ and $\widetilde{A}^{\star}_a[{Z}^{\bullet}, {Z}^{\star}](x^+;\mathbf{P})$. To achieve this, we begin by substituting the expressions for $\widetilde{B}^{\star}_a[{Z}^{\star}](x^+;\mathbf{P})$ and $\widetilde{B}^{\bullet}_a[{Z}^{\star}, {Z}^{\bullet}](x^+;\mathbf{P})$ Eq.~\eqref{eq:BstarZ_exp}-\eqref{eq:BbulletZ_exp} into the expressions for $\widetilde{A}^{\bullet}_a[{B}^{\bullet}](x^+;\mathbf{P})$ and $\widetilde{A}^{\star}_a[{B}^{\bullet}, {B}^{\star}](x^+;\mathbf{P})$ Eq.~\eqref{eq:A_bull_solu}-\eqref{eq:A_star_solu} to obtain $\widetilde{A}^{\bullet}_a[{Z}^{\bullet}, {Z}^{\star}](x^+;\mathbf{P})$ and $\widetilde{A}^{\star}_a[{Z}^{\bullet}, {Z}^{\star}](x^+;\mathbf{P})$. After that, we substitute these directly to the Yang-Mills action Eq.~\eqref{eq:actionLC_YM} and show that both the triple-gluon vertices cancel out.

In order to demonstrate the cancellation of the triple-gluon vertices, we need $\widetilde{A}^{\bullet}_a[{Z}^{\bullet}, {Z}^{\star}](x^+;\mathbf{P})$ and $\widetilde{A}^{\star}_a[{Z}^{\bullet}, {Z}^{\star}](x^+;\mathbf{P})$ only up to second order in $Z$ fields. Making the above stated substitutions, we get
\begin{multline}
    \Big[{\widetilde {A}}^{\bullet}_{a} (x^+;\mathbf{P})\Big]_{2nd} =\int\!d^{3}\mathbf{p}_{1} d^{3}\mathbf{p}_{2}\, \overline{\widetilde \Omega}\,^{a b_1 \left \{b_2 \right \}}_{2}(\mathbf{P}; \mathbf{p_1} ,\left \{ \mathbf{p_2} \right \}) {\widetilde Z}^{\bullet}_{b_1} (x^+;\mathbf{p}_{1}) {\widetilde Z}^{\star}_{b_2} (x^+;\mathbf{p}_{2}) \\
    + \int\!d^{3}\mathbf{p}_{1}\, d^{3}\mathbf{p}_{2} {\widetilde \Psi}_{2}^{a \left \{b_1 b_2 \right \} }(\mathbf{P}; \left \{\mathbf{p}_{1}, \mathbf{p}_{2} \right \}) {\widetilde Z}^{\bullet}_{b_1} (x^+;\mathbf{p}_{1}){\widetilde Z}^{\bullet}_{b_2} (x^+;\mathbf{p}_{2}) \, ,
    \label{eq:A_bull_2}
\end{multline}
and
\begin{multline}
    \Big[{\widetilde A}^{\star}_{a} (x^+;\mathbf{P})\Big]_{2nd} = \int\!d^{3}\mathbf{p}_{1}\, d^{3}\mathbf{p}_{2}\, \overline{\Psi}\,^{a\{b_1 b_2\}}_2(\mathbf{P};\{\mathbf{p}_1,\mathbf{p}_2\})
     {\widetilde Z}^{\star}_{b_1} (x^+;\mathbf{p}_{1}){\widetilde Z}^{\star}_{b_2} (x^+;\mathbf{p}_{2}) \\ + \int\!d^{3}\mathbf{p}_{1}\, d^{3}\mathbf{p}_{2}\, {\widetilde \Omega}_{2}^{a b_1 \left \{b_2  \right \} }(\mathbf{P}; \mathbf{p}_{1} ,\left \{ \mathbf{p}_{2}  \right \}) {\widetilde Z}^{\star}_{b_1} (x^+;\mathbf{p}_{1}) {\widetilde Z}^{\bullet}_{b_2} (x^+;\mathbf{p}_{2}) \, .
    \label{eq:astar_2}
\end{multline}
The kernels in the above expression correspond to the momentum space versions of the kernels $\Xi_{1,1}^{ab_1 b_{2}}(\mathbf{x};\mathbf{y}_1, \mathbf{y}_{2})$,  $\Xi_{2,0}^{ab_1 b_{2}}(\mathbf{x};\mathbf{y}_1, \mathbf{y}_{2})$, $\Lambda_{2,0}^{ab_1 b_{2}}(\mathbf{x};\mathbf{y}_1, \mathbf{y}_{2})$ and $\Lambda_{1,1}^{ab_1 b_{2}}(\mathbf{x};\mathbf{y}_1, \mathbf{y}_{2})$, respectively, considered in Eq.~\eqref{eq:Abullet_to_Z}-\eqref{eq:Astar_to_Z}.

In momentum space, the kinetic term and both the triple-gluon vertices in the Yang-Mills action read
\begin{equation}
    \mathcal{L}_{+-}=\int d^{4}{p}_{1}d^{4}{p}_{2}\,\delta^{4}\left({p}_{1}+{p}_{2}\right)\, {p}_{1}^{2}\,\,
\widetilde{A}_{a}^{\bullet}\left({p}_{1}\right)\widetilde{A}_{a}^{\star}\left({p}_{2}\right)\,,
\label{eq:kin_sym}
\end{equation}
\begin{multline}
\mathcal{L}_{++-}=\int d^{3}\mathbf{p}_{1}d^{3}\mathbf{p}_{2}d^{3}\mathbf{p}_{3}\,\delta^{3}\left(\mathbf{p}_{1}+\mathbf{p}_{2}+\mathbf{p}_{3}\right)\widetilde{V}_{++-}^{abc}\left(\mathbf{p}_{1},\mathbf{p}_{2},\mathbf{p}_{3}\right)\,\\
\widetilde{A}_{a}^{\bullet}\left(x^+;\mathbf{p}_{1}\right)\widetilde{A}_{b}^{\bullet}\left(x^+;\mathbf{p}_{2}\right)\widetilde{A}_{c}^{\star}\left(x^+;\mathbf{p}_{3}\right)\,,
\label{eq:3g++-}
\end{multline}
where
\begin{equation}
\widetilde{V}_{++-}^{abc}\left(\mathbf{p}_{1},\mathbf{p}_{2},\mathbf{p}_{3}\right)=-igf^{abc}\left(\frac{p_{1}^{\star}}{p_{1}^{+}}-\frac{p_{2}^{\star}}{p_{2}^{+}}\right)p_{3}^{+}\,,
\label{eq:3g++-V}
\end{equation}
and
\begin{multline}
    \mathcal{L}_{--+}=\int d^{3}\mathbf{p}_{1}d^{3}\mathbf{p}_{2}d^{3}\mathbf{p}_{3}\,\delta^{3}\left(\mathbf{p}_{1}+\mathbf{p}_{2}+\mathbf{p}_{3}\right)\widetilde{V}_{--+}^{abc}\left(\mathbf{p}_{1},\mathbf{p}_{2},\mathbf{p}_{3}\right)\, \\
\widetilde{A}_{a}^{\star}\left(x^+;\mathbf{p}_{1}\right)\widetilde{A}_{b}^{\star}\left(x^+;\mathbf{p}_{2}\right)\widetilde{A}_{c}^{\bullet}\left(x^+;\mathbf{p}_{3}\right)\,,
\label{eq:3g+--}
\end{multline}
with
\begin{equation}
\widetilde{V}_{--+}^{abc}\left(\mathbf{p}_{1},\mathbf{p}_{2},\mathbf{p}_{3}\right)=-igf^{abc}\left(\frac{p_{1}^{\bullet}}{p_{1}^{+}}-\frac{p_{2}^{\bullet}}{p_{2}^{+}}\right)p_{3}^{+}\,.\label{eq:3g+--V}
\end{equation}
An important point to notice is that the kernels in Eqs.~\eqref{eq:A_bull_2}-\eqref{eq:astar_2} as well as the triple-gluon vertices in Eqs.~\eqref{eq:3g++-V}-\eqref{eq:3g+--V} are obtained via the 3D Fourier transform of the corresponding expression in $\mathbf{x}\equiv\left(x^{-},x^{\bullet},x^{\star}\right)$ (3D position space) keeping the light-cone time $x^+$ fixed. Since the kernels as well as the vertices do not depend on $x^+$, a 4D Fourier transform will only introduce an extra delta 
$\delta\left(p_1^- + \dots + p_n^-\right)$ conserving the minus component of the momenta. The expression for the kinetic term Eq.~\eqref{eq:kin_sym}, on the other hand, is in 4D momentum space because the inverse propagator ($\square=2(\partial_+\partial_- - \partial_{\bullet}\partial_{\star})$) contains the $\partial_+$ (derivative with respect to light-cone time $x^+$) operator. And for that matter, using $\mathcal{L}_{+-}$ is in some sense misuse of notation because the Lagrangian is an integral over $\mathbf{x}\equiv\left(x^{-},x^{\bullet},x^{\star}\right)$, action $S$, on the other hand, is an integral over 4D position $x=(x^+,\mathbf{x})$. But for now, we will overlook this detail and keep using $\mathcal{L}_{+-}$.

Substituting the 4D versions of Eqs.~\eqref{eq:A_bull_2}-\eqref{eq:astar_2} to the kinetic term Eq.~\eqref{eq:kin_sym}, we get
\begin{multline}
    \mathcal{L}_{+-}=\int d^{4}{p}_{1}d^{4}{p}_{2}\,\delta^{4}\left({p}_{1}+{p}_{2}\right)\, {p}_{1}^{2}\,\,
\Bigg\{\Bigg[\int\!d^{4}{q}_{1} d^{4}{q}_{2}\,\, \overline{\widetilde \Omega}\,^{a c_1 \left \{c_2 \right \} }_2({p}_{1}; {q}_{1} ,\left \{ {q}_{2} \right \}) \\
\times {\widetilde Z}^{\bullet}_{c_1} ({q}_{1}) {\widetilde Z}^{\star}_{c_2} ({q}_{2})     +  {\widetilde \Psi}_{2}^{a \left \{c_1 c_2 \right \} }({p}_{1}; \left \{{q}_{1}, {q}_{2} \right \}) {\widetilde Z}^{\bullet}_{c_1} ({q}_{1}){\widetilde Z}^{\bullet}_{c_2} ({q}_{2}) \,\Bigg]\widetilde{Z}_{a}^{\star}\left({p}_{2}\right)\, \\ +\widetilde{Z}_{a}^{\bullet}\left({p}_{1}\right) 
\times \Bigg[ \int\!d^{4}{q}_{1}\, d^{4}{q}_{2}\,\, \overline{\Psi}\,^{a \left \{c_1 c_2 \right \} }_2({p}_{2}; \left \{{q}_{1},   {q}_{2} \right \})      {\widetilde Z}^{\star}_{c_1} ({q}_{1}){\widetilde Z}^{\star}_{c_2} ({q}_{2})\\
     +  {\widetilde \Omega}_{2}^{a c_1 \left \{c_2  \right \} }({p}_{2}; {q}_{1} ,\left \{ {q}_{2}  \right \}) {\widetilde Z}^{\star}_{c_1} ({q}_{1}) {\widetilde Z}^{\bullet}_{c_2} ({q}_{2})\Bigg]\Bigg\} \,.
     \label{eq:kin_subs}
\end{multline}
The above expression has two types of field configurations: ${\widetilde{Z}^\star}{\widetilde{Z}^\bullet}{\widetilde{Z}^\bullet}$ and ${\widetilde{Z}^\star}{\widetilde{Z}^\star}{\widetilde{Z}^\bullet}$. For the moment let us focus on the latter (which we denote as $\mathcal{T}_{--+}$)
\begin{multline}
    \mathcal{T}_{--+}=\int d^{4}{p}_{1}d^{4}{p}_{2}\,\delta^{4}\left({p}_{1}+{p}_{2}\right)\, {p}_{1}^{2}\,\, \Bigg[\int\!d^{4}{q}_{1} d^{4}{q}_{2} \,\, \overline{\widetilde \Omega}\,^{a c_1 \left \{c_2 \right \} }_2({p}_{1}; {q}_{1} ,\left \{ {q}_{2} \right \})\\
\times {\widetilde Z}^{\bullet}_{c_1} ({q}_{1}) {\widetilde Z}^{\star}_{c_2} ({q}_{2}) \widetilde{Z}_{a}^{\star}\left({p}_{2}\right)\,  +\widetilde{Z}_{a}^{\bullet}\left({p}_{1}\right)
    \int\!d^{4}{q}_{1}\, d^{4}{q}_{2}\,\, \overline{\widetilde{\Psi}}\,^{a \left \{c_1 c_2 \right \} }_2({p}_{2}; \left \{{q}_{1},   {q}_{2} \right \}) {\widetilde Z}^{\star}_{c_1} ({q}_{1}){\widetilde Z}^{\star}_{c_2} ({q}_{2}) \Bigg] \, .
     \label{eq:v--+_tr}
\end{multline}
Integrating out $p_1$, $p_2$
in the first, second term respectively and then renaming the variables such that we have ${\widetilde Z}^{\star}_{b_1} ({p}_{1}){\widetilde Z}^{\star}_{b_2} ({p}_{2})\widetilde{Z}_{b_3}^{\bullet}\left({p}_{3}\right)$ in both the terms, we get
\begin{multline}
   \mathcal{T}_{--+}=\int d^{4}{p}_{1}d^{4}{p}_{2}d^{4}{p}_{3}\, \Bigg[ {p}_{2}^{2}\,\,\overline{\widetilde \Omega}\,^{b_2 b_3 \left \{b_1 \right \} }_2(-{p}_{2}; {p}_{3} ,\left \{ {p}_{1} \right \}) \\ +  {p}_{3}^{2}\overline{\widetilde{\Psi}}\,^{b_3 \left \{b_1 b_2 \right \} }_2(-{p}_{3}; \left \{{p}_{1},   {p}_{2} \right \}) \Bigg]{\widetilde Z}^{\star}_{b_1} ({p}_{1}){\widetilde Z}^{\star}_{b_2} ({p}_{2})\widetilde{Z}_{b_3}^{\bullet}\left({p}_{3}\right) \, .
   \label{eq:v--t_com}
\end{multline}
Now, the aim is to show that the above term would cancel out the $\widetilde{V}_{--+}$ triple-gluon vertex. But for that, we need to bring the above expression on the fixed light-cone time $x^+$ hypersurface. In order to do that we introduce a set of new auxiliary fields
\begin{equation}
    {\widetilde Z}_{b_i}(p_i) = \frac{{\widetilde \kappa}_{b_i}(p_i) }{p_i^2} \, ,
    \label{eq:kappa}
\end{equation}
both for $\widetilde{Z}^\bullet$ and $\widetilde{Z}^\star$ fields. Substituting Eq.~\eqref{eq:kappa} to Eq.~\eqref{eq:v--t_com} we get
\begin{multline}
   \mathcal{T}_{--+}=\int d^{4}{p}_{1}d^{4}{p}_{2}d^{4}{p}_{3}\, \Bigg[ {p}_{2}^{2}\,\,\overline{\widetilde \Omega}\,^{b_2 b_3 \left \{b_1 \right \} }_2(-{p}_{2}; {p}_{3} ,\left \{ {p}_{1} \right \}) \\ +  {p}_{3}^{2}\overline{\widetilde{\Psi}}\,^{b_3 \left \{b_1 b_2 \right \} }_2(-{p}_{3}; \left \{{p}_{1},   {p}_{2} \right \}) \Bigg]\frac{{\widetilde \kappa}^{\star}_{b_1}(p_1) }{p_1^2}\frac{{\widetilde \kappa}^{\star}_{b_2}(p_2) }{p_2^2}\frac{{\widetilde \kappa}^{\bullet}_{b_3}(p_3) }{p_3^2} \, .
   \label{eq:v--t_kap}
\end{multline}
For the sake of simplicity, let us focus on just the first term above. It can be re-written as
\begin{multline}
    \int d^{3}\mathbf{p}_{1}d^{3}\mathbf{p}_{2}d^{3}\mathbf{p}_{3}\,\,\int dp_1^-\, dp_2^-\ dp_3^- dP^-\, \int dz^+ \int dy^+
    e^{iz^+(P^- - p_2^-)} \,\,    e^{iy^+(P^- +p_1^- + p_3^-)}  \\  (-p_{13}^+)(P^- + \hat{p}_{13})
    \overline{\widetilde \Omega}\,^{b_2 b_3 \left \{b_1 \right \} }_2(-\mathbf{p}_2; \mathbf{p}_{3} ,\left \{ \mathbf{p}_{1} \right \}) \frac{1}{2p_1^+ [p_1^- - \hat{p}_{1} + i\epsilon]}\frac{1}{2p_2^+ [p_2^- - \hat{p}_{2} + i\epsilon]}\\
    \frac{1}{2p_3^+ [p_3^- - \hat{p}_{3} + i\epsilon]}{\widetilde \kappa}^{\star}_{b_1}(p_1){\widetilde \kappa}^{\star}_{b_2}(p_2){\widetilde \kappa}^{\star}_{b_3}(p_3) \, .
    \label{eq:v--t1}
\end{multline}
where 
\begin{equation}
    \hat{p}_i = \frac{p_i^\bullet p_i^\star}{p_i^+} \, .
    \label{eq:hat_def}
\end{equation}
Notice the change from $\overline{\widetilde \Omega}\,^{b_2 b_3 \left \{b_1 \right \} }_2(-{p}_{2}; {p}_{3} ,\left \{ {p}_{1} \right \}) $ $ \longrightarrow $ $ \overline{\widetilde \Omega}\,^{b_2 b_3 \left \{b_1 \right \} }_2(-\mathbf{p}_2; \mathbf{p}_{3} ,\left \{ \mathbf{p}_{1} \right \})$ in going from Eq.~\eqref{eq:v--t_kap} to Eq.~\eqref{eq:v--t1}. Essentially, we used the $\delta\left(p_1^- + p_2^- + p_3^-\right)$ in the former and re-expressed the ${p}_{2}^{2} = {p}_{13}^{2}$ using a dummy variable $P$ as
\begin{multline}
 \int dp_1^-\, dp_2^-\ dp_3^-  {p}_{13}^{2} \delta\left(p_1^- + p_2^- + p_3^-\right) =  \int dp_1^-\, dp_2^-\ dp_3^- dP^-\, \int dz^+ \int dy^+
    e^{iz^+(P^- - p_2^-)} \,\, \\   e^{iy^+(P^- +p_1^- + p_3^-)}    (-p_{13}^+)(P^- + \hat{p}_{13})\,.
\end{multline}
Also, we make the $i\epsilon$ prescription explicit in Eq.~\eqref{eq:v--t1}. Integrating out the minus component of the momenta $p_1$, $p_2$, and $p_3$ we get
\begin{multline}
    \int d^{3}\mathbf{p}_{1}d^{3}\mathbf{p}_{2}d^{3}\mathbf{p}_{3}\,\,\,p_{13}^+(\hat{p}_{1} + \hat{p}_{3} -\hat{p}_{13})\,\,
    \overline{\widetilde \Omega}\,^{b_2 b_3 \left \{b_1 \right \} }_2(-\mathbf{p}_2; \mathbf{p}_{3} ,\left \{ \mathbf{p}_{1} \right \}) \frac{{(i\pi)}^3}{p_1^+ p_2^+ p_3^+} \\ \times \Bigg[ \prod_{i=1}^{3}\Theta (-p_i^+) \frac{i}{\hat{p}_{1}+\hat{p}_{2} + \hat{p}_{3} + i\epsilon}
    +\prod_{i=1}^{3}\Theta (p_i^+) {(-1)}^3 \frac{-i}{\hat{p}_{1}+\hat{p}_{2} + \hat{p}_{3} + i\epsilon} \Bigg]\\
    \times {\widetilde \kappa}^{\star}_{b_1}\left(\hat{p}_{1};\mathbf{p}_{1}\right){\widetilde \kappa}^{\star}_{b_2}\left(\hat{p}_{2};\mathbf{p}_{2}\right){\widetilde \kappa}^{\star}_{b_3}\left(\hat{p}_{3};\mathbf{p}_{3}\right) \, ,
    \label{eq:v--+1_cl}
    \end{multline}
where $\Theta (p_i^+)$ is Heaviside step function. Since $\left\{p_{13}^+(\hat{p}_{1} + \hat{p}_{3} -\hat{p}_{13})\,\,
    \overline{\widetilde \Omega}\,^{b_2 b_3 \left \{b_1 \right \} }_2(-\mathbf{p}_2; \mathbf{p}_{3} ,\left \{ \mathbf{p}_{1} \right \})\right\}$ is independent of the minus component of the momentum, the above expression can be re-written as
\begin{multline}
  \int dx^+\, \int d^{3}\mathbf{p}_{1}d^{3}\mathbf{p}_{2}d^{3}\mathbf{p}_{3}\,\,\,p_{13}^+(\hat{p}_{1} + \hat{p}_{3} -\hat{p}_{13})\,\,
  \overline{\widetilde \Omega}\,^{b_2 b_3 \left \{b_1 \right \} }_2(-\mathbf{p}_2; \mathbf{p}_{3} ,\left \{ \mathbf{p}_{1} \right \}) \\ 
  \times {\widetilde Z}^{\star}_{b_1} (x^+;\mathbf{p}_{1}){\widetilde Z}^{\star}_{b_2} (x^+;\mathbf{p}_{2})\widetilde{Z}_{b_3}^{\bullet}\left(x^+;\mathbf{p}_{3}\right) \, ,
  \label{eq:v--+vcl}
\end{multline}
where, in going from Eq.~\eqref{eq:v--+1_cl}
 to Eq.~\eqref{eq:v--+vcl}, we used the following 
\begin{multline}
     \int dx^+\, \int d^{3}\mathbf{p}_{1}\cdots d^{3}\mathbf{p}_{n}\,\,\,
  {\widetilde f}(\mathbf{p}_1 \cdots \mathbf{p}_{n})
   {\widetilde Z}^{\star}_{b_1} (x^+;\mathbf{p}_{1}) \cdots \widetilde{Z}_{b_n}^{\bullet}\left(x^+;\mathbf{p}_{n}\right)\\ = 
  \int d^{3}\mathbf{p}_{1}\cdots d^{3}\mathbf{p}_{n}\,\,\, {\widetilde f}(\mathbf{p}_1 \cdots \mathbf{p}_{n})\frac{{(i\pi)}^n}{p_1^+ \cdots p_n^+} \Bigg[ \prod_{i=1}^{n}\Theta (-p_i^+) \frac{i}{\hat{p}_{1}+\cdots + \hat{p}_{n} + i\epsilon} \\
    +\prod_{i=1}^{n}\Theta (p_i^+) {(-1)}^n \frac{-i}{\hat{p_{1}}+\cdots + \hat{p_{n}} + i\epsilon} \Bigg]{\widetilde \kappa}^{\star}_{b_1}\left(\hat{p}_{1};\mathbf{p}_{1}\right) \cdots {\widetilde \kappa}^{\star}_{b_n}\left(\hat{p}_{n};\mathbf{p}_{n}\right) \, .
\end{multline}
Above, ${\widetilde f}(\mathbf{p}_1 \cdots \mathbf{p}_{n})$ represents a generic function of only the three momenta $\mathbf{p}_i$. With this, we have the expression Eq.~\eqref{eq:v--+vcl} on the constant light-cone time $x^+$.

Repeating the above steps for the second term in Eq.~\eqref{eq:v--t_kap}, we get
\begin{multline}
  \int dx^+\, \int d^{3}\mathbf{p}_{1}d^{3}\mathbf{p}_{2}d^{3}\mathbf{p}_{3}\,\,\,p_{12}^+(\hat{p}_{1} + \hat{p}_{2} -\hat{p}_{12})
  \overline{\widetilde{\Psi}}\,^{b_3 \left \{b_1 b_2 \right \} }_2(-\mathbf{p}_3; \left \{\mathbf{p}_{1},   \mathbf{p}_{2} \right \}) \\
  \times {\widetilde Z}^{\star}_{b_1} (x^+;\mathbf{p}_{1}) {\widetilde Z}^{\star}_{b_2} (x^+;\mathbf{p}_{2})\widetilde{Z}_{b_3}^{\bullet}\left(x^+;\mathbf{p}_{3}\right) \, .
  \label{eq:v--+vcl2}
\end{multline}
Adding the two expressions, Eq.~\eqref{eq:v--+vcl}-\eqref{eq:v--+vcl2}, we get
\begin{multline}
    \mathcal{T}_{--+}=\int dx^+\, \int d^{3}\mathbf{p}_{1}d^{3}\mathbf{p}_{2}d^{3}\mathbf{p}_{3}\,
    \Bigg[p_{13}^+(\hat{p}_{1} + \hat{p}_{3} -\hat{p}_{13}) 
  \,\, \overline{\widetilde \Omega}\,^{b_2 b_3 \left \{b_1 \right \} }_2(-\mathbf{p}_2; \mathbf{p}_{3} ,\left \{ \mathbf{p}_{1} \right \}) \\ + p_{12}^+(\hat{p}_{1} + \hat{p}_{2} -\hat{p}_{12})
  \,\, \overline{\widetilde{\Psi}}\,^{b_3 \left \{b_1 b_2 \right \} }_2(-\mathbf{p}_3; \left \{\mathbf{p}_{1},   \mathbf{p}_{2} \right \})\Bigg] {\widetilde Z}^{\star}_{b_1} (x^+;\mathbf{p}_{1}){\widetilde Z}^{\star}_{b_2} (x^+;\mathbf{p}_{2})\widetilde{Z}_{b_3}^{\bullet}\left(x^+;\mathbf{p}_{3}\right) \, .
  \label{eq:v--t_can}
\end{multline}
Substituting $\overline{\widetilde \Omega}\,_2$ and $\overline{\widetilde{\Psi}}\,_2$ using Eq.~\eqref{eq:omegaBar_kernel}-\eqref{eq:psiBar_kernel} and using the identity $p_{ij}^+(\hat{p_{i}} + \hat{p_{j}} -\hat{p_{ij}}) =-{\widetilde v}_{(i)(j)}{\widetilde v}^{\ast}_{(j)(i)}$, the above expression can be simplified to
\begin{multline}
    \mathcal{T}_{--+}=\int dx^+\, \int d^{3}\mathbf{p}_{1}d^{3}\mathbf{p}_{2}d^{3}\mathbf{p}_{3}\,\delta^{3}\left(\mathbf{p}_{1}+\mathbf{p}_{2}+\mathbf{p}_{3}\right) \\
    \times \Big( ig f^{b_1 b_2 b_3} p_3^+ {v}^{\ast}_{12}  \Big) 
  \times {\widetilde Z}^{\star}_{b_1} (x^+;\mathbf{p}_{1}){\widetilde Z}^{\star}_{b_2} (x^+;\mathbf{p}_{2})\widetilde{Z}_{b_3}^{\bullet}\left(x^+;\mathbf{p}_{3}\right) \, .
  \label{eq:v--t_can1}
\end{multline}
Comparing the above with the expression for $\widetilde{V}_{--+}^{abc}\left(\mathbf{p}_{1},\mathbf{p}_{2},\mathbf{p}_{3}\right)$ in Eq.~\eqref{eq:3g+--V}, we can write
\begin{multline}
    \mathcal{T}_{--+}=\int dx^+\, \int d^{3}\mathbf{p}_{1}d^{3}\mathbf{p}_{2}d^{3}\mathbf{p}_{3}\,\delta^{3}\left(\mathbf{p}_{1}+\mathbf{p}_{2}+\mathbf{p}_{3}\right) \\
    \times \Big( - \widetilde{V}_{--+}^{b_1 b_2 b_3}\left(\mathbf{p}_{1},\mathbf{p}_{2},\mathbf{p}_{3}\right) \Big) 
  \times {\widetilde Z}^{\star}_{b_1} (x^+;\mathbf{p}_{1}){\widetilde Z}^{\star}_{b_2} (x^+;\mathbf{p}_{2})\widetilde{Z}_{b_3}^{\bullet}\left(x^+;\mathbf{p}_{3}\right) \, .
  \label{eq:v--t_comp}
\end{multline}
This will exactly cancel out the $\widetilde{V}_{--+}^{abc}\left(\mathbf{p}_{1},\mathbf{p}_{2},\mathbf{p}_{3}\right)$ term originating from the substitution of the first order expansion of $A$ fields
\begin{equation}
    A_a^{\bullet}(x^+;\mathbf{P})= Z_{a}^{\bullet}(x^+;\mathbf{P})+\dots \,\,, \qquad
    A_a^{\star}(x^+;\mathbf{P})=
    Z_{a}^{\star}(x^+;\mathbf{P})+\dots \,\,.
    \label{eq:A_to_Z_zeroth_mom}
\end{equation}
to the $(+ - -)$ triple gluon vertex in the Yang-Mills action and will thus eliminate it.

The cancellation of the $(+ + -)$ triple gluon vertex can be demonstrated in exactly the same way. We shall not repeat it here. With this, we conclude that the transformation Eq.~\eqref{eq:general_transf_app} indeed eliminates both the triple-gluon vertices in the Yang-Mills action.


\chapter{Computing amplitudes from One-Loop Effective Action}
\label{sec:app_A6}

In this appendix, we briefly recall the standard textbook discussion (that can be found in say \cite{Peskin:1995ev}) for computing scattering amplitudes using the one-loop effective action.

Let us start with the definition of a generic one-loop effective action $\Gamma[A_c]$ as a Legendre transform of the generating functional for the connected Green's function $W[J]$ shown below
\begin{equation}
    \Gamma[A_c]=W[J]-J_IA_c^{I} \,.
    \label{eq:gam_gen}
\end{equation}
Using the above relation, we can write
\begin{equation}
     \frac{\delta \Gamma[A_c]}
    {\delta A_c^{I}} = -J_I \, , \quad \frac{\delta W[J]}
    {\delta J_I} = A_c^{I} \, .
    \label{eq:EOM_EA}
\end{equation}
Thus, we see that the source has an interpretation of the functional derivative of the one-loop effective action with respect to the classical field whereas the classical field is the functional derivative of the generating functional for the connected Green's function with respect to the source. Above (as well as throughout this appendix) we use the collective indices.

Differentiating the second relation in Eq.~\eqref{eq:EOM_EA} with respect to the classical field we get
\begin{equation}
    \frac{\delta}
    {\delta A_c^{J}}\left(\frac{\delta W[J]}
    {\delta J_I} \right)=\frac{\delta J_K}
    {\delta A_c^{J}}\left(\frac{\delta^2 W[J]}
    {\delta J_K \delta J_I} \right) = - \frac{\delta^2 \Gamma[A_c]}
    {\delta A_c^{J} \delta A_c^{K}}\left(\frac{\delta^2 W[J]}
    {\delta J_K \delta J_I} \right) = \delta^{I}_J \, .
    \label{eq:EOM_2}
\end{equation}
The second-order functional derivative of $W[J]$ with respect to the sources represents the 2-point connected Green's function. Up to one loop, it consists of the propagator and the self-energy (or the bubble) terms. From the above expression, we see that it is inverse to the second-order functional derivative of $\Gamma[A_c]$ with respect to the fields. In fact, we can rearrange the above expression as
\begin{figure}
    \centering
    \includegraphics[width=7cm]{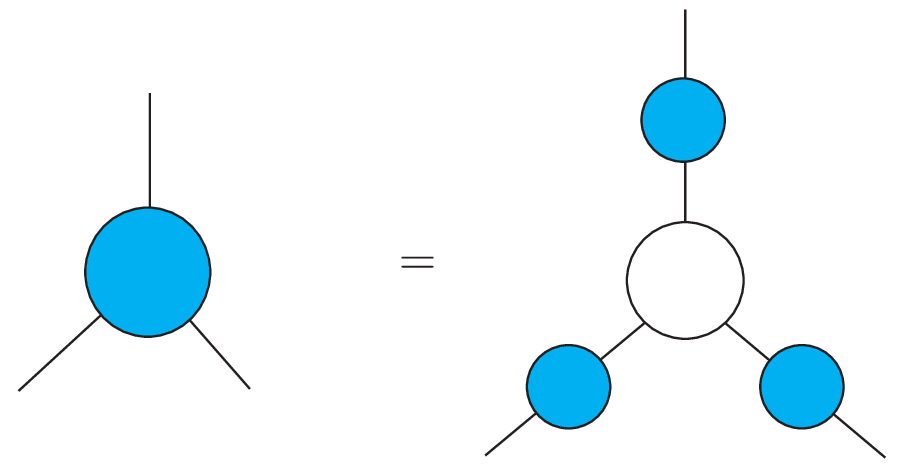}
    \caption{\small
   Each of the blue blobs in the figure represents the connected Green's function. The white blob, on the other hand, is the third order functional derivative of $\Gamma[A_c]$ with respect to the classical fields. It can be obtained by amputating the two-point connected Green's function. Due to this, it gets the interpretation of the 3-point amputated connected Green's function. This image was modified from our paper \cite{Kakkad_2022}.}
    \label{fig:3_point_FR}
\end{figure}
\begin{equation}
    \frac{\delta^2 W[J]}
    {\delta J_K \delta J_I} = - \left(\frac{\delta^2 \Gamma[A_c]}
    {\delta A_c^{I} \delta A_c^{K}} \right)^{-1} \, .
    \label{eq:w2_g2}
\end{equation}
Owing to this, the second-order functional derivative of $\Gamma[A_c]$ is quite commonly referred to as the \textit{inverse propagator}.

Differentiating Eq.~\eqref{eq:w2_g2}, one more time we get
\begin{multline}
    \frac{\delta^3 W[J]}
    {\delta J_L \delta J_K \delta J_I} = - \frac{\delta A_c^{M}}
    {\delta J_L} \frac{\delta}
    {\delta A_c^{M}} \left[ \left(\frac{\delta^2 \Gamma[A_c]}
    {\delta A_c^{I} \delta A_c^{K}} \right)^{-1} \right] \\
    =- \left(\frac{\delta^2 \Gamma[A_c]}
    {\delta A_c^{M} \delta A_c^{L}} \right)^{-1}  \left(\frac{\delta^2 \Gamma[A_c]}
    {\delta A_c^{N} \delta A_c^{K}} \right)^{-1} 
    \left(\frac{\delta^2 \Gamma[A_c]} 
    {\delta A_c^{P} \delta A_c^{I}} \right)^{-1} 
    \frac{\delta^3 \Gamma[A_c]}
    {\delta A_c^{M} \delta A_c^{N} \delta A_c^{P}} \, .
    \label{eq:w3_g3}
\end{multline}
The quantity on L.H.S. is the 3-point connected Green's function. It can be amputated using the three 2-point connected Green's function on the R.H.S. to obtain the third-order derivative of $\Gamma[A_c]$. Due to this, the latter can be interpreted as the amputated 3-point connected Green's function where the amputation accounts for both the propagators and bubbles on the external legs. We represent this diagrammatically in Figure \ref{fig:3_point_FR}.

\begin{figure}
    \centering
    \includegraphics[width=13.5cm]{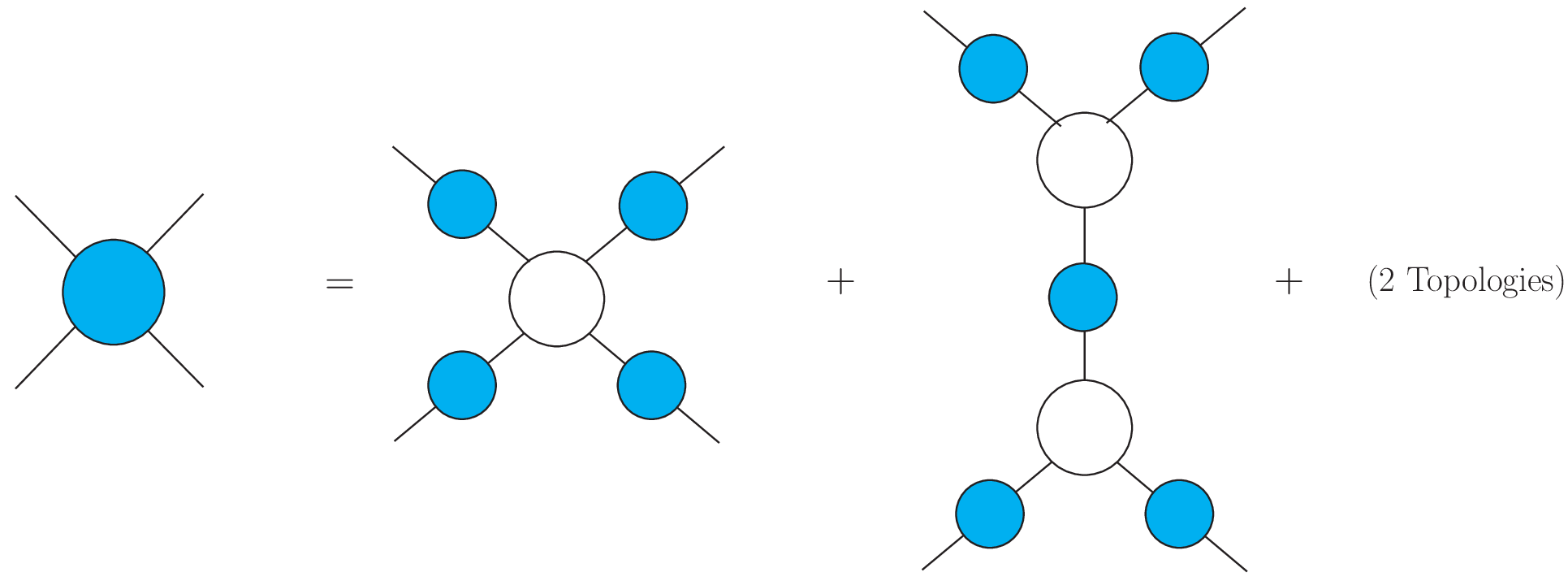}
    \caption{\small
    Each of the blue blobs in the figure represents the connected Green's function. The white blobs are the functional derivative of $\Gamma[A_c]$ with respect to the classical fields. This image was modified from our paper \cite{Kakkad_2022}.}
    \label{fig:4_point_FR}
\end{figure}

In a similar way, differentiating one more time we get
\begin{multline}
    \frac{\delta^4 W[J]}
    {\delta J_Q \delta J_L \delta J_K \delta J_I} = \frac{\delta A_c^{R}}
    {\delta J_Q} \frac{\delta}
    {\delta A_c^{N}} \left( \frac{\delta^3 W[J]}
    {\delta J_L \delta J_K \delta J_I} \right) \\
    =-\left(\frac{\delta^2 \Gamma[A_c]}
    {\delta A_c^{R} \delta A_c^{Q}} \right)^{-1} \left(\frac{\delta^2 \Gamma[A_c]}
    {\delta A_c^{M} \delta A_c^{L}} \right)^{-1} \frac{\delta^3 \Gamma[A_c]}
    {\delta A_c^{R} \delta A_c^{M} \delta A_c^{T}} \left(\frac{\delta^2 \Gamma[A_c]}     {\delta A_c^{S} \delta A_c^{T}} \right)^{-1}  \left(\frac{\delta^2 \Gamma[A_c]}
    {\delta A_c^{N} \delta A_c^{K}} \right)^{-1} \\ \times \frac{\delta^3 \Gamma[A_c]}
    {\delta A_c^{S} \delta A_c^{N} \delta A_c^{P}} \left(\frac{\delta^2 \Gamma[A_c]}     {\delta A_c^{P} \delta A_c^{I}} \right)^{-1}
   + \mathrm{2 \,\, Topologies\,\,}\\ - \left(\frac{\delta^2 \Gamma[A_c]}
    {\delta A_c^{R} \delta A_c^{Q}} \right)^{-1} \left(\frac{\delta^2 \Gamma[A_c]}
    {\delta A_c^{M} \delta A_c^{L}} \right)^{-1} \frac{\delta^4 \Gamma[A_c]}
    {\delta A_c^{R} \delta A_c^{M} \delta A_c^{N}\delta A_c^{P}} \left(\frac{\delta^2 \Gamma[A_c]}
    {\delta A_c^{N} \delta A_c^{K}} \right)^{-1}  \left(\frac{\delta^2 \Gamma[A_c]} 
    {\delta A_c^{P} \delta A_c^{I}} \right)^{-1}\,.
    \label{eq:w4_g4}
\end{multline}
This time, the amputated 4-point connected Green's function consists of two terms. First is the fourth-order derivative of $\Gamma[A_c]$. Second, two third-order derivatives of $\Gamma[A_c]$ connected via a 2-point connected Green's function (see Figure \ref{fig:4_point_FR}). Since the amputated connected Green's function corresponds to the amplitude (with the same number of external legs), these are the only contributions to a 4-point amplitude.

\begin{figure}
    \centering
    \includegraphics[width=14cm]{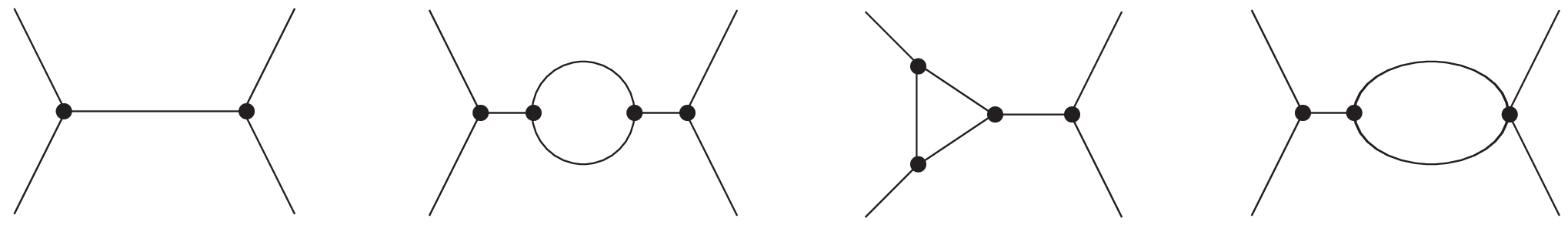}
    \caption{\small
    The contributions to a four point amplitude, up to one loop, originating from joining two third-order derivatives of $\Gamma[A_c]$ with respect to the classical fields via a two-point connected Green's function which up to one loop includes a propagator and a bubble. Each 
$\bullet$ represents an interaction vertex in the action. This image was taken from our paper \cite{Kakkad_2022}.}
    \label{fig:gamma_3VC}
\end{figure}

\begin{figure}
    \centering
    \includegraphics[width=14cm]{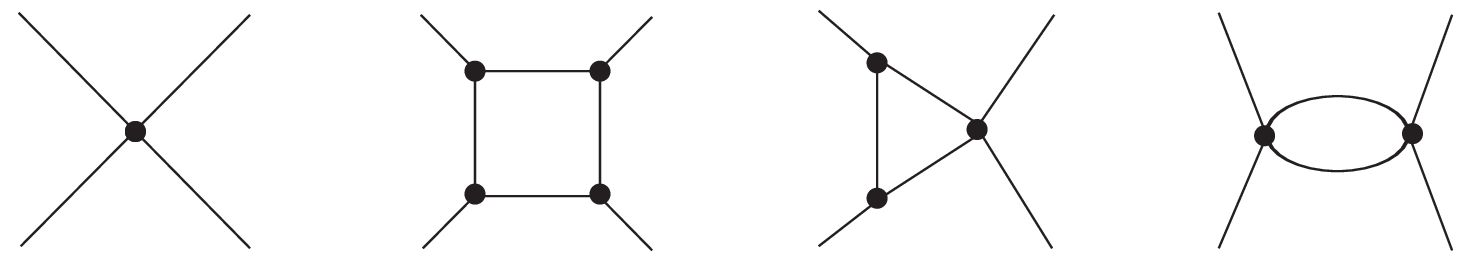}
    \caption{\small
    The contributions to a four point amplitude, up to one loop, originating from the fourth-order derivative of $\Gamma[A_c]$ with respect to the classical fields. Each 
$\bullet$ represents an interaction vertex in the action. This image was taken from our paper \cite{Kakkad_2022}.}
    \label{fig:gamma_4V}
\end{figure}

To demonstrate that this indeed accounts for all the contributions necessary to compute an amplitude, consider, for the sake of simplicity, a four-point amplitude (up to one-loop) in a theory consisting of triple and four-point interaction vertices. The first type of contributions originating from the fourth-order derivative of $\Gamma[A_c]$ are shown in Figure \ref{fig:gamma_4V}.
The second type originating from two third-order derivatives of $\Gamma[A_c]$ connected via a 2-point connected Green's function are shown in Figure \ref{fig:gamma_3VC}. Notice, these two sets, indeed, exhaust all the possible geometries that can contribute to a four-point amplitude up to one loop.

The higher-order derivatives can be computed in exactly the same way. By doing that one can see that the above results generalize as follows
\begin{itemize}
   \item The $n$-th order functional derivative of $\Gamma[A_c]$ gives the vertices.
   \begin{equation}
       \Bigg[\frac{\delta^n \Gamma[A_c]}
    {\delta A_c^{I} \delta A_c^{J} \dots \delta A_c^{K}}\Bigg]_{A_c= 0} \,.
   \end{equation}
   These, in fact, correspond to the $n$-point \textit{one-particle irreducible } (1PI) contributions to an $n$-point amplitude.
   \item These vertices could be joined together using 2-point connected Green's functions (propagator and bubble up to one loop) to obtain a higher multiplicity contribution. 
\end{itemize}

\chapter{Four point MHV}
\label{sec:app_A7}

In this appendix, we prove the identity shown diagrammatically below
\begin{center}
    \includegraphics[width=13cm]{appendix/mhv_4prt.eps}
\end{center}
The above contributions are obtained by un-tracing (cutting open) all one-loop diagrams. That is, by joining (or tracing over) the two crossed circles in each term we obtain the one-loop contribution. The double lines represent a propagator associated with the given leg.

The legs $(- - + +)$ are named 1, 2, 3, and 4 anticlockwise respectively. The momentum and the color index associated with each leg is, therefore, $\left\{b_i, {p}_{i}\right\}$. For the sake of simplicity, we will focus on the color-ordered case with $(t^{b_1}t^{b_2}t^{b_3}t^{b_4})$ order. The first contribution on the L.H.S. is the four-point interaction vertex in the Yang-Mills action. In momentum space, it reads Eq.~\eqref{eq:vertex4g}
\begin{equation}
    D_1 = -\frac{1}{2}\frac{4}{p_1^2} g^2 \frac{p_1^+ p_3^+ + p_2^+ p_4^+}{(p_1^+ + p_4^+)^2} \,\, \mathrm{Tr}(t^{b_1}t^{b_2}t^{b_3}t^{b_4}) \,.
\end{equation}
Notice, the propagator $1/p_1^2$ on the external leg named 1 and an additional factor of 4 above. It originates due to the second order derivative of the interaction vertex with respect to the plus and the minus helicity fields in the log term Eq.~\eqref{eq:PartitionMHV}.

The second term on the L.H.S. is obtained from the tree level connection between the un-traced $(- + +)$ tadpole from the log term and the $(- - +)$ triple gluon vertex from $S_{MHV}[B]$ in Eq.~\eqref{eq:PartitionMHV} following the rules discussed in Appendix \ref{sec:app_A6}. Substituting the expressions for both the types of triple gluon vertices, involved in this contribution, we get
\begin{equation}
    D_2 = -\frac{1}{p_1^2}g^2 2\Bigg(\frac{p^{\star}}{p^+} - \frac{p_4^{\star}}{p_4^+}\Bigg)p_1^+ \frac{1}{p^2} \,2\, \Bigg(\frac{p^{\bullet}}{p^+} - \frac{p_2^{\bullet}}{p_2^+}\Bigg)p_3^+\,\, \mathrm{Tr}(t^{b_1}t^{b_2}t^{b_3}t^{b_4}) \,.
\end{equation}
The above expression can be further simplified using the identities $p=p_1+p_4=-(p_2+p_3)$, and $p^2=2(p^+p^- - p^{\bullet}p^{\star})$. Substituting these above we get
\begin{equation}
    D_2 = \frac{4}{p_1^2}g^2 \frac{1}{2} \frac{(p_1^+)^2 p_3^+ {\tilde v}^{\star}_{2(23)}}{p_{14}^+ p_4^+ p_2^+ {\tilde v}^{\star}_{14}}\,\, \mathrm{Tr}(t^{b_1}t^{b_2}t^{b_3}t^{b_4}) \,.
\end{equation}

The third term originates exclusively from the log term in Eq.~\eqref{eq:PartitionMHV}. It involves the un-traced $(- - +)$ tadpole where the minus helicity field $\hat{A}_c^{\star}[B^{\bullet}, B^{\star}]$ is expanded to second order in the $\hat{B}_c^{i}$ fields. Substituting the expression for ${\widetilde \Omega}_{2}^{a b_2 \left \{b_3 \right \} }(\mathbf{p}; \mathbf{p}_2 ,\left \{ \mathbf{p}_3 \right \})$ using Eq.~\eqref{eq:omega_kernel} we get
\begin{equation}
    D_3 = \frac{4}{p_1^2}g^2 \frac{1}{2} \frac{p_2^+ p_4^+ {\tilde v}^{\star}_{(14)1}}{p_{14}^+ p_{23}^+ {\tilde v}^{\star}_{32}}\,\, \mathrm{Tr}(t^{b_1}t^{b_2}t^{b_3}t^{b_4}) \,.
\end{equation}

The final contribution is the un-traced $(- +)$ bubble from the log term in Eq.~\eqref{eq:PartitionMHV}. It is made up of the two triple gluon vertices. Substituting the expressions for the triple gluon vertices we get
\begin{equation}
    D_4 = -\frac{1}{p_1^2}g^2 2\Bigg(\frac{p_4^{\star}}{p_4^+} - \frac{p_3^{\star}}{p_3^+}\Bigg)(-p^+) \frac{1}{p^2} \,2\, \Bigg(\frac{p_2^{\bullet}}{p_2^+} - \frac{p_1^{\bullet}}{p_1^+}\Bigg)p^+\,\, \mathrm{Tr}(t^{b_1}t^{b_2}t^{b_3}t^{b_4}) \,.
\end{equation}
As before, using the identities $p=p_1+p_2=-(p_3+p_4)$, and $p^2=2(p^+p^- - p^{\bullet}p^{\star})$, we can simplify the above expression to
\begin{equation}
    D_4 = -\frac{4}{p_1^2}g^2 \frac{1}{2} \frac{p_{12}^+ p_{34}^+ {\tilde v}^{\star}_{21}}{p_2^+ p_3^+ {\tilde v}^{\star}_{43}}\,\, \mathrm{Tr}(t^{b_1}t^{b_2}t^{b_3}t^{b_4}) \,.
\end{equation}
Note, the four terms above are exactly the same (modulo the numeric factor of 4 the propagator $1/p_1^2$ ) as the ones we had earlier in Eq.~\eqref{eq:Mhv_4_4} where we derived the 4-point MHV vertex in the MHV action. As a result we can write
\begin{equation}
    D_1 +  D_2 +  D_3 + D_4 = \frac{4}{p_1^2}\frac{g^2}{2}  \left(\frac{p_{1}^{+}}{p_{2}^{+}}\right)^{2}
\frac{\widetilde{v}_{21}^{*4}}{\widetilde{v}_{14}^{*}\widetilde{v}_{43}^{*}\widetilde{v}_{32}^{*}\widetilde{v}_{21}^{*}}\,\, \mathrm{Tr}(t^{b_1}t^{b_2}t^{b_3}t^{b_4}) \,.
\end{equation}
The R.H.S. of the above expression represents the un-traced 4-point color ordered MHV vertex with a factor of 4 and a propagator on the leg named 1. The factor of 4 would originate from the second order derivative of the 4-point MHV vertex in the MHV action with respect to the plus and the minus helicity fields. This completes the proof.

\chapter{One-loop calculations in CQT regularisation scheme}
\label{sec:app_A8}

In this appendix, we re-derive some of the one-loop results in the CQT regularisation scheme. 

\section{Gluon Self-Energy \texorpdfstring{$\Pi^{+ +}$}{GSE}}
\label{sec:App_A81}

\begin{figure}
    \centering
 \includegraphics[width=3.5cm]{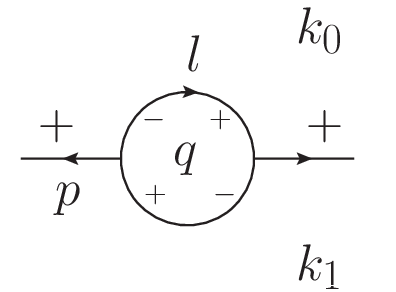}
    \caption{\small 
    The $(+ +)$ gluon self-energy contribution. $l$ is the loop line momentum and $p$ is the line momentum of the external leg. $q$ is the region momentum of the bounded region inside the loop and $k_0$, and $k_1$ are the region momenta of the exterior regions outside the loop. This image was taken from our paper \cite{Kakkad_2022}.}
    \label{fig:2_ct}
\end{figure}

In this section, we re-derive the expression for the $(+ +)$ gluon self-energy, or equivalently the bubble, graph $\Pi^{++}$ shown in Figure \ref{fig:2_ct}. Both the vertices in the $\Pi^{+ +}$ self-energy diagram are $(- + +)$ triple gluon vertex. In momentum space, the vertex reads

\begin{equation}
\widetilde{V}_{++-}^{b_1b_2b_3}\left({p}_{1},{p}_{2},{p}_{3}\right)=-igf^{b_1b_2b_3}\left(\frac{p_{1}^{\star}}{p_{1}^{+}}-\frac{p_{2}^{\star}}{p_{2}^{+}}\right)p_{3}^{+}\, \delta^{4}\left({p}_{1}+{p}_{2}+{p}_{3}\right).
\label{eq:vertex3g_aga}
\end{equation}
Note, however, the above expression is the momentum space expression for the $(- + +)$ triple gluon vertex in the Yang-Mills action. It does not account for an additional factor of 2 that will originate from the second order derivative with respect to the fields in the log term in Eq.~\eqref{eq:PartitionMHV}. So, precisely, what we need is the Fourier transform of the position space vertex in Eq.~\eqref{eq:v3_position}. It reads
\begin{equation}
    V_{-++}^{abc}(x,y,z)=2gf^{abc}\delta^4(x-y)
    \delta^4(x-z)\Big[\partial^{-1}_{-}(y)\partial_{\bullet}(y)-\partial^{-1}_{-}(z)\partial_{\bullet}(z)\Big]\partial_{-}(x) \, .
    \label{eq:v3_positionag}
\end{equation}
In momentum space we get
\begin{equation}
\widetilde{V}_{++-}^{b_1b_2b_3}\left({p}_{1},{p}_{2},{p}_{3}\right)=-2igf^{b_1b_2b_3}\left(\frac{p_{1}^{\star}}{p_{1}^{+}}-\frac{p_{2}^{\star}}{p_{2}^{+}}\right)p_{3}^{+}\, \delta^{4}\left({p}_{1}+{p}_{2}+{p}_{3}\right).
\label{eq:vertex3g_ag}
\end{equation}
Using the above expression, the diagram in Figure \ref{fig:2_ct} can be expressed as
\begin{align}
\Pi^{++} &=8 g^{2} \int \frac{\mathrm{d}^{4} l}{(2 \pi)^{4}}\Bigg(\frac{p^{\star}}{p^+} - \frac{l^{\star}}{l^+}\Bigg)(p+l)^+ \frac{1}{l^2} \,\frac{1}{{(p+l)}^2}\, \Bigg(\frac{p^{\star}}{p^+} - \frac{{(p+l)}^{\star}}{{(p+l)}^+}\Bigg)l^+ \,, \\
&=\frac{g^{2}}{2 \pi^{4}} \int \mathrm{d}^{4} l \frac{1}{\left(p^{+}\right)^{2}}\left(p^{+} l^{\star}-l^{+} p^{\star}\right)\left(p^{+}\left(p^{\star}+l^{\star}\right)-\left(p^{+}+l^{+}\right) p^{\star}\right) \frac{1}{l^{2}(p+l)^{2}} .
\label{eq:pi_--1}
\end{align}
We suppressed the color factors for brevity. Recall, in CQT scheme, the $1/l^+$ singularities are regulated via discretization of the plus component of the momenta. Notice, however, in going from the first step to the second step above, the $1/l^+$ singularities cancel out as a result we can keep $l^+$ continuous throughout the calculation. One may worry about the $l^+$ in the propagator, but as we show below in Eq.~\eqref{eq:sch_prop}, the propagators are exponentiated in the CQT scheme.
Next, we need to replace the line momenta in terms of the region momenta and introduce the exponential cutoff Eq.~\eqref{eq:cqt_exp}. Using Eq.~\eqref{eq:LM_con} for the diagram in Figure \ref{fig:2_ct}, we get
\begin{equation}
    p= k_0-k_1 \,,\quad \quad l=q-k_0 \,.
    \label{eq:line_def2p}
\end{equation}
For the propagators, we use the Schwinger representation as shown below
\begin{equation}
    \frac{e^{-\delta \boldsymbol{q}^{2}}}{\left(q-k_{0}\right)^{2}\left(q-k_{1}\right)^{2}}=\int_0^{\infty} d t_{1} d t_{2} e^{-t_{1}\left(q-k_{0}\right)^{2}-t_{2}\left(q-k_{1}\right)^{2}-\delta \mathbf{q}^{2}} \,.
    \label{eq:sch_prop}
\end{equation}
Note, in CQT scheme we work with the region momenta. These are not Wick rotated and therefore in some sense such exponentiation is a prescription. Substituting Eq.~\eqref{eq:line_def2p}-\eqref{eq:sch_prop} in Eq.~\eqref{eq:pi_--1}, we get
\begin{align}
\Pi^{++}=& \frac{g^{2}}{2 \pi^{4}} \int_{0}^{\infty} \mathrm{d} t_{1} \mathrm{~d} t_{2} \int \mathrm{d}^{4} q \frac{1}{\left(k_{0}^{+}\right)^{2}} e^{-t_{1}(q-k_0)^{2}-t_{2}\left(q-k_1\right)^{2}-\delta \mathbf{q}^{2}}  \\
& \times\left[k_{0}^{+}\left(q^{\star}-k_0^{\star}\right)-\left(q^{+}-k_{0}^{+}\right)\left(k_{0}^{\star}-k_{1}^{\star}\right)\right]\left[k_{0}^{+}\left(q^{\star}-k_1^{\star}\right)-q^{+}\left(k_{0}^{\star}-k_1^{\star}\right)\right] \,,
\end{align}
where using the translational invariance along the plus component of the region momenta we put $k_{1}^{+}=0$. Notice, the minus component $q^-$ is only in the exponent. Therefore, it can be integrated out to obtain a delta: $\pi \delta\left(\left(t_{1}+t_{2}\right) q^{+}-t_{2} p^{+}\right)$ which could then be used to integrate out the plus component of the loop region momenta $q^+$. This leaves us with the transverse components which can be integrated out by completing the square in the exponent. Performing all the integrals and changing the variables as $T=t_{1}+t_{2}, \alpha=t_{1} /\left(t_{1}+t_{2}\right)$, we get
\begin{equation}
   \Pi^{++}=\frac{g^{2}}{2 \pi^{2}} \int_{0}^{1} \mathrm{~d} \alpha \int_{0}^{\infty} \mathrm{d} T \frac{ \delta^{2}\,\left[\alpha k_1^{\star}+(1-\alpha) k_0^{\star}\right]^{2}}{(T+\delta)^{3}} e^{-T \alpha(1-\alpha) p^{2}-\frac{\delta T}{T+\delta}\left(\alpha \mathbf{k}_0+(1-\alpha) \mathbf{k}_1\right)^{2}}  \,.
\end{equation}
The $\delta$ in the expression above represents the parameter used in the cutoff Eq.~\eqref{eq:sch_prop}. Performing the integral over $T$, in the limit of the regularization parameter $\delta \longrightarrow 0$ we get:
\begin{equation}
    \Pi^{++}= \frac{g^{2}}{4 \pi^{2}} \int_{0}^{1} \mathrm{~d} \alpha \left[\alpha k_1^{\star}+(1-\alpha) k_0^{\star}\right]^{2}\,,
\end{equation}
which, after integrating out the $\alpha$ gives
\begin{equation}
    \Pi^{++}= \frac{g^{2}}{12 \pi^{2}} \left[k_{0}^{\star 2}+k_{1}^{\star 2}+k_{0}^{\star} k_{1}^{\star}\right]\,.
    \label{eq:--GSE_ap}
\end{equation}
A non-zero value for $\Pi^{++}$ implies that a gluon can flip its helicity and therefore violates the Loretz invariance. Such a contribution must be canceled via a counterterm. The other two types of gluon self-energies $\Pi^{--}$ and $\Pi^{-+}$ can be calculated in exactly the same fashion. For the former, the calculation is exactly the same, step by step, as for $\Pi^{++}$ with the interchange of $\star \leftrightarrow \bullet$. As a result, we get
\begin{equation}
    \Pi^{--}= \frac{g^{2}}{12 \pi^{2}} \left[k_{0}^{\bullet 2}+k_{1}^{\bullet 2}+k_{0}^{\bullet} k_{1}^{\bullet}\right]\,.
    \label{eq:++GSE_ap}
\end{equation}
The result, for $\Pi^{-+}$, however, involves two additional terms which require the inclusion of new counterterms. We discuss this in detail in the main text.

\section{Triangle contribution \texorpdfstring{$\Delta^{+ + + +}_{ij}$}{Tri}}
\label{sec:App_A82}

\begin{figure}
    \centering
 \includegraphics[width=3.8cm]{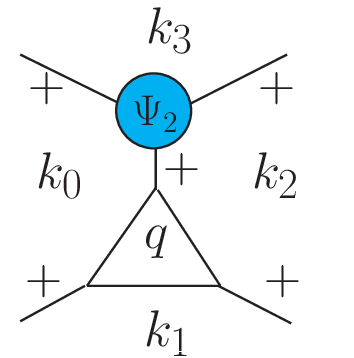}
    \caption{\small 
The $\Delta^{+ + + +}_{12}$ triangular contribution. $q$ is the region momentum of the bounded region inside the loop. $k_0$, $k_1$, $k_2$, and $k_3$ are the region momenta of the exterior regions outside the loop. There is no propagator connecting the kernel $\widetilde{\Psi}_2$ with the $\Delta^{+ + +}$ sub-diagram. This image was modified from our paper \cite{Kakkad_2022}.}
    \label{fig:4_tri}
\end{figure}

\begin{figure}
    \centering
 \includegraphics[width=4cm]{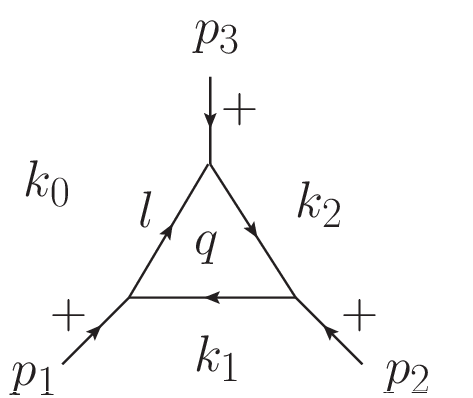}
    \caption{\small 
    The $\Delta^{+ + +}$ triangle contribution. $p_1$, $p_2$, and $p_3$ are the line momenta associated with the external legs. $q$ is the region momentum of the bounded region inside the loop. $k_0$, $k_1$, and $k_2$ are the region momenta of the exterior regions outside the loop. This image was taken from our paper \cite{Kakkad_2022}.}
    \label{fig:3_tri}
\end{figure}

 In this section, we re-derive the expression for the $\Delta^{+ + + +}_{ij}$ which essentially consists of a one-loop triangle $\Delta^{+ + +}$ convoluted with second order expansion of $\widetilde{A}^{\bullet}[B^{\bullet}]$. Specifically, we will obtain the expression for $\Delta^{+ + + +}_{12}$ shown in Figure \ref{fig:4_tri}.  In order to do it step by step, let us first derive the expression for the  one-loop triangle $\Delta^{+ + +}$ shown in Figure \ref{fig:3_tri}. All three vertices in the triangle are $(- + +)$ triple gluon vertex. Using Eq.~\eqref{eq:vertex3g_ag}, we can write the following explicit expression for the diagram shown in Figure \ref{fig:3_tri}
 
\begin{multline}
\Delta^{+ + +} = -g^{3}16 \int \frac{\mathrm{d}^{4} l}{(2 \pi)^{4}}\Bigg(\frac{p_1^{\star}}{p_1^+} - \frac{l^{\star}}{l^+}\Bigg)(-p_1+l)^+ \frac{1}{l^2}\,\frac{1}{{(-p_1+l)}^2} \,\frac{1}{{(p_3+l)}^2}\,\\ \Bigg(\frac{p_3^{\star}}{p_3^+} - \frac{{(p_3+l)}^{\star}}{{(p_3+l)}^+}\Bigg)l^+ \Bigg(\frac{p_2^{\star}}{p_2^+} - \frac{{(-p_1+l)}^{\star}}{{(-p_1+l)}^+}\Bigg)(l+p_3)^+ \,.
\end{multline}
Again, we suppressed the color factors for simplicity. The factors depending on $l^+$ in the denominator can be canceled to obtain
\begin{multline}
  \Delta^{+ + +}  =-\frac{g^{3}}{\pi^{4}} \int \mathrm{d}^{4} l \frac{1}{p_1^{+}p_2^{+}p_3^{+}}\left(p_1^{\bullet} l^{+}-l^{\star} p_1^{+}\right)\left(p_2^{\star}\left(-p_1^{+}+l^{+}\right)-\left(-p_1^{\star}+l^{\star}\right) p_2^{+}\right) \\
 \left(p_3^{\star}\left(p_3^{+}+l^{+}\right)-\left(p_3^{\star}+l^{\star}\right) p_3^{+}\right) \frac{1}{l^{2}{(-p_1+l)}^2(p_3+l)^{2}} .
 \label{eq:d---}
\end{multline}
As a result, we can keep $l^+$ continuous throughout the calculation. Again, we do not worry about the $l^+$ in the propagators because the propagators will get exponentiated following Eq.~\eqref{eq:sch_prop}. Using Eq.~\eqref{eq:LM_con}, the line momenta in Figure \ref{fig:3_tri} can be expressed in terms of the region momenta as follows
\begin{equation}
    p_1= k_1-k_0 \,,\quad p_2= k_2-k_1\,, \quad p_3= k_0-k_2\,, \quad l=q-k_0 \,.
    \label{eq:line_def}
\end{equation}
Substituting Eq.~\eqref{eq:line_def} in Eq.~\eqref{eq:d---} followed by inserting the CQT cutoff, we get
\begin{multline}
  \Delta^{+ + +}  =-\frac{g^{3}}{\pi^{4}} \int \mathrm{d}^{4} q \frac{1}{p_1^{+}p_2^{+}p_3^{+}}\left((k_1-k_0)^{\star} \left(q-k_0\right)^{+}-(q-k_0)^{\star} (k_1-k_0)^{+}\right)\\
  \left((k_2-k_1)^{\star}\left(q-k_1\right)^+-\left(q-k_1\right)^{\star} (k_2-k_1)^{+}\right) \\
 \left((k_0-k_2)^{\star}\left(q-k_2\right)^+-\left(q-k_2\right)^{\star} (k_0-k_2)^{+}\right)\\ \int_0^{\infty} d t_{1} d t_{2} d t_{3} e^{-t_{1}\left(q-k_{0}\right)^{2}-t_{2}\left(q-k_{1}\right)^{2}-t_{3}\left(q-k_{2}\right)^{2}-\delta \mathbf{q}^{2}}\, .
\end{multline}
In order to integrate out the loop region momenta $q$, we follow the same steps as in the case of gluon self-energy: first, integrate the $q^-$ appearing only in the exponent to get a delta in $q^+$, then integrate out $q^+$ using the delta. Finally, integrate the transverse components by completing the square in the exponent. By doing this, we get
\begin{align}
\Delta^{+++}=&-\frac{g^{3}}{\pi^{2}} \int_{0}^{\infty} \frac{d t_{1} d t_{2} d t_{3}}{p_1^{+}p_2^{+}p_3^{+}t_{123}\left(t_{123}+\delta\right)} \exp \left(-\frac{\delta\left(\boldsymbol{k}_{0} t_{3}+\boldsymbol{k}_{1} t_{1}+\boldsymbol{k}_{2} t_{2}\right)^{2}}{t_{123}\left(t_{123}+\delta\right)}\right) \nonumber\\
& \exp \left(-\frac{t_{1} t_{3}\left(k_{1}-k_{0}\right)^{2}+t_{1} t_{2}\left(k_{1}-k_{2}\right)^{2}+t_{2} t_{3}\left(k_{2}-k_{0}\right)^{2}}{t_{123}}\right) \nonumber\\
& \left(\frac{t_{1} {\widetilde v}_{12}p_2^+}{ t_{123}}-\frac{\delta\left(t_{1} k_{1}^{\star}+t_{2} k_{2}^{\star}+t_{3} k_{0}^{\star}\right)}{t_{123}\left(\delta+t_{123}\right)}\right) \left(\frac{t_{2} {\widetilde v}_{12}p_2^+}{ t_{123}}-\frac{\delta\left(t_{1} k_{1}^{\star}+t_{2} k_{2}^{\star}+t_{3} k_{0}^{\star}\right)}{t_{123}\left(\delta+t_{123}\right)}\right)\nonumber\\
&\left(\frac{t_{3} {\widetilde v}_{12}p_2^+}{ t_{123}}-\frac{\delta\left(t_{1} k_{1}^{\star}+t_{2} k_{2}^{\star}+t_{3} k_{0}^{\star}\right)}{t_{123}\left(\delta+t_{123}\right)}\right)\,.
\end{align}
Above, $t_{123}=t_{1}+t_{2}+t_{3}$ and $\delta$ represents the parameter in the CQT scheme. Setting $\delta \rightarrow 0$, the expression reduces to
\begin{multline}
\Delta^{+++}  =-\frac{g^{3}}{\pi^{2}} \int_{0}^{\infty} \frac{d t_{1} d t_{2} d t_{3}}{p_{1}^{+} p_{2}^{+} p_{3}^{+} t_{123}^{5}} t_{1} t_{2} t_{3}\left({\widetilde v}_{12}p_2^+\right)^{3} \\
\exp \left(-\frac{t_{1} t_{3}\left(k_{1}-k_{0}\right)^{2}+t_{1} t_{2}\left(k_{1}-k_{2}\right)^{2}+t_{2} t_{3}\left(k_{2}-k_{0}\right)^{2}}{t_{123}}\right) \,.
\end{multline}
Renaming the variables as $T=t_{1}+t_{2}+t_{3}$,  $\alpha=t_{1} /\left(t_{1}+t_{2}+t_{3}\right)$, $\beta=t_{2} /\left(t_{1}+t_{2}+t_{3}\right)$, and integrating with respect to $T$, we get:
\begin{multline}
    \Delta^{+++}  =-\frac{g^{3}}{\pi^{2}} \frac{\left({\widetilde v}_{12}p_2^+\right)^{3}}{p_{1}^{+} p_{2}^{+} p_{3}^{+}}\\ \int_{\alpha+\beta<1} d \alpha \,\,d \beta \frac{\alpha \beta(1-\alpha-\beta)}{\alpha(1-\alpha-\beta)\left(k_{1}-k_{0}\right)^{2}+\alpha \beta\left(k_{1}-k_{2}\right)^{2}+\beta(1-\alpha-\beta)\left(k_{2}-k_{0}\right)^{2}} \,.
\end{multline}
The denominator of the expression above consists of the square of the momenta of the external legs of the triangle. In its current form, the above results hold for all three legs of the triangle off-shell. It can, however, be simplified if we put some of the legs on-shell. Recall, we want to compute $\Delta^{+++}$ to finally obtain $\Delta^{++++}_{ij}$ where the legs $p_{i}^{2}=0 $, $p_{j}^{2}=0$ are both on-shell. Specifically, for $\Delta^{++++}_{12 }$ we can put $p_{1}^{2}=0 $, $p_{2}^{2}=0$ on-shell. By doing this, the above expression simplifies and we obtain 
\begin{equation}
    \left.\Delta^{+++}\right|_{p_{1}^{2}=0, p_{2}^{2}=0}  =-\frac{g^{3}}{6 \pi^{2}} \frac{\left({\widetilde v}_{12}p_2^+\right)^{3}}{p_{1}^{+} p_{2}^{+} p_{3}^{+} p_3^{2}}\,.
\end{equation}
Above, $p_{3}$ is the off-shell leg. To avoid confusion in what follows, let us rename it to $P$.
\begin{equation}
    \left.\Delta^{+++}\right|_{p_{1}^{2}=0, p_{2}^{2}=0}  =-\frac{g^{3}}{6 \pi^{2}} \frac{\left({\widetilde v}_{12}p_2^+\right)^{3}}{p_{1}^{+} p_{2}^{+} P^{+} P^{2}}\,.
    \label{eq:tri_+++P}
\end{equation}
In order to obtain $\Delta^{++++}_{12 }$ (see Figure \ref{fig:4_tri}), we need to substitute 
${\widetilde \Psi}_{2}(\mathbf{P}; \left \{\mathbf{p}_{3},  \mathbf{p}_{4} \right \})$ (we suppressed the color dependence of the kernel)  using Eq.~\eqref{eq:psi_kernel}  to the off-shell leg $P$ in Eq.\eqref{eq:tri_+++P}. By doing this we get
\begin{equation}
    \Delta^{+ + + +}_{12}  =-\frac{g^{4}}{12 \pi^{2}} \frac{\left({\widetilde v}_{12}p_2^+\right)^{3}p_{3}^{+}}{p_{1}^{+} p_{2}^{+} p_{3}^{+}p_{4}^{+} p_{34}^{2}{\widetilde v}^{\star}_{34}}\,.
\end{equation}

\section{Expression for \texorpdfstring{$S_{l}^{q^{+}}\left(p_{i}, p_{j}\right)$}{Sre}}
\label{sec:App_A83}

In this section we list the explicit expression for $S_{l}^{q^{+}}\left(p_{i}, p_{j}\right)$ with $i=1$ and $j=2$, derived in \cite{CQT1,CQTnotes}. It reads
\begin{multline}
S_{1}^{q^{+}}\left(p_{1}, p_{2}\right)=\sum_{q^{+}<p_{1}^{+}}\left\{\left[\frac{2}{q^{+}}+\frac{1}{p_{12}^{+}-q^{+}}+\frac{1}{p_{1}^{+}-q^{+}}\right]\left(\ln \left(\delta p_{1}^{2} e^{\gamma}\right)+\ln \frac{q^{+}}{p_{1}^{+}}\right)\right.\\
\left.+\left[\frac{2}{p_{1}^{+}-q^{+}}-\frac{1}{p_{12}^{+}-q^{+}}+\frac{1}{q^{+}}\right] \ln \frac{p_{1}^{+}-q^{+}}{p_{1}^{+}}\right\} \\
+\sum_{q^{+}>p_{1}^{+}}\left\{\left[\frac{1}{q^{+}}+\frac{2}{p_{12}^{+}-q^{+}}+\frac{1}{q^{+}-p_{1}^{+}}\right]\left(\ln \left(\delta p_{1}^{2} e^{\gamma}\right)+\ln \frac{p_{12}^{+}-q^{+}}{p_{2}^{+}}\right)\right. \\
+\sum_{q^{+} \neq p_{1}^{+}}\left[\frac{1}{q^{+}}+\frac{2}{p_{12}^{+}-q^{+}}+\frac{1}{q^{+}-p_{1}^{+}}\right] \ln \frac{p_{12}^{+}-q^{+}}{p_{12}^{+}}\,,
\end{multline}

\begin{multline}
    S_{2}^{q^{+}}\left(p_{1}, p_{2}\right)=\sum_{q^{+} \neq p_{1}^{+}}\left[\frac{2}{q^{+}}+\frac{1}{p_{12}^{+}-q^{+}}+\frac{1}{p_{1}^{+}-q^{+}}\right] \ln \frac{q^{+}}{p_{12}^{+}}\\
+\sum_{q^{+}<p_{1}^{+}}\left\{\left[\frac{2}{q^{+}}+\frac{1}{p_{12}^{+}-q^{+}}+\frac{1}{p_{1}^{+}-q^{+}}\right]\left(\ln \left(\delta p_{2}^{2} e^{\gamma}\right)+\ln \frac{q^{+}}{p_{1}^{+}}\right)\right\}\\
+\sum_{q^{+}>p_{1}^{+}}\left\{\left[\frac{1}{q^{+}}+\frac{2}{p_{12}^{+}-q^{+}}+\frac{1}{q^{+}-p_{1}^{+}}\right]\left(\ln \left(\delta p_{2}^{2} e^{\gamma}\right)+\ln \frac{p_{12}^{+}-q^{+}}{p_{2}^{+}}\right)\right.
\\
\left. +\left[\frac{2}{q^{+}-p_{1}^{+}}+\frac{1}{p_{12}^{+}-q^{+}}-\frac{1}{q^{+}}\right] \ln \frac{q^{+}-p_{1}^{+}}{p_{2}^{+}}\right\}\,,
\end{multline}

\begin{multline}
    S_{3}^{q^{+}}\left(p_{1}, p_{2}\right)=\sum_{q^{+}<p_{1}^{+}}\left\{\left[\frac{2}{q^{+}}+\frac{1}{p_{12}^{+}-q^{+}}+\frac{1}{p_{1}^{+}-q^{+}}\right]\left(\ln \left(\delta p_{12}^{2} e^{\gamma}\right)+\ln \frac{q^{+}}{p_{12}^{+}}\right)\right.\\
+\left[\frac{1}{q^{+}}+\frac{2}{p_{12}^{+}-q^{+}}+\frac{1}{q^{+}-p_{1}^{+}}\right] \ln \frac{p_{12}^{+}-q^{+}}{p_{12}^{+}}\\
\left.+\left[\frac{2}{p_{1}^{+}-q^{+}}-\frac{1}{p_{12}^{+}-q^{+}}+\frac{1}{q^{+}}\right] \ln \frac{p_{1}^{+}-q^{+}}{p_{1}^{+}}\right\}\\
+\sum_{q^{+}>p_{1}^{+}}\left\{\left[\frac{1}{q^{+}}+\frac{2}{p_{12}^{+}-q^{+}}+\frac{1}{q^{+}-p_{1}^{+}}\right]\left(\ln \left(\delta p_{12}^{2} e^{\gamma}\right)+\ln \frac{p_{12}^{+}-q^{+}}{p_{12}^{+}}\right)\right.\\
+\left[\frac{2}{q^{+}}+\frac{1}{p_{12}^{+}-q^{+}}+\frac{1}{p_{1}^{+}-q^{+}}\right] \ln \frac{q^{+}}{p_{12}^{+}}\\
\left.+\left[\frac{2}{q^{+}-p_{1}^{+}}+\frac{1}{p_{12}^{+}-q^{+}}-\frac{1}{q^{+}}\right] \ln \frac{q^{+}-p_{1}^{+}}{p_{2}^{+}}\right\}\,.
\end{multline}
\section{Expression for \texorpdfstring{$\mathcal{A}_{\mathrm{IR}}^{+ + - -}$}{AIR}}
\label{sec:App_A84}

In this section we represent the explicit expression for $\mathcal{A}_{\mathrm{IR}}^{+ + - -}$ appearing in the result of the 4-point $(+ + - -)$ one-loop amplitude 
$\mathcal{A}_{\mathrm{one-loop}}^{+ + - -}$ Eq.~\eqref{eq:4++--amp}. It reads \cite{CQT2}

\begin{equation}
     \mathcal{A}_{\mathrm{IR}}^{+ + - -}= -\frac{g^2}{8 \pi^2}  \mathcal{V}_{\mathrm{tree}}\left(1^+,2^+,3^-,4^-\right)\times \mathcal{F}_{\mathrm{IR}}^{+ + - -}[q^+]\,,
\end{equation}

where $\mathcal{V}_{\mathrm{tree}}\left(1^+,2^+,3^-,4^-\right)$ is the tree level MHV amplitude Eq.~\eqref{eq:MHV_vertex}  and $\mathcal{F}_{\mathrm{IR}}^{+ + - -}[q^+]$ is the divergent factor depending on the range of $q^+$. For $k_0^{+}<q^{+}<k_3^{+}$, 

\begin{align} 
\mathcal{F}_{\mathrm{IR}}^{+ + - -}[q^+] = & {\left[\frac{1}{q^{+}-k_3^{+}}+\frac{1}{q^{+}-k_1^{+}}\right] \log \frac{\left(k_2^{+}-q^{+}\right)\left(-k_0^{+}+q^{+}\right) p_{12}^2 \delta e^\gamma}{\left(k_0^{+}-k_2^{+}\right)^2}} \nonumber\\ 
& -\left[\frac{2}{q^{+}-k_1^{+}}\right] \log \frac{\left(k_1^{+}-q^{+}\right)\left(q^{+}-k_0^{+}\right) p_{12}^2 \delta e^\gamma}{\left(k_1^{+}-k_0^{+}\right)\left(k_2^{+}-k_0^{+}\right)} \nonumber\\ 
& -\left[\frac{2}{q^{+}-k_3^{+}}\right] \log \frac{\left(k_3^{+}-q^{+}\right)\left(-k_0^{+}+q^{+}\right) p_{12}^2 \delta e^\gamma}{\left(k_3^{+}-k_0^{+}\right)\left(k_2^{+}-k_0^{+}\right)} \nonumber \\ 
& +\left[\frac{2}{q^{+}-k_0^{+}}\right] \log \frac{\left(-k_0^{+}+q^{+}\right)^2 p_{14}^2 \delta e^\gamma}{\left(k_1^{+}-k_0^{+}\right)\left(k_3^{+}-k_0^{+}\right)} \,.
\end{align}
For $k_3^{+}<q^{+}<k_1^{+}$,
\begin{align}
\mathcal{F}_{\mathrm{IR}}^{+ + - -}[q^+]= & -\left[\frac{1}{q^{+}-k_3^{+}}-\frac{1}{q^{+}-k_1^{+}}\right] \log \frac{\left(k_2^{+}-q^{+}\right)\left(-k_0^{+}+q^{+}\right) p_{12}^2 \delta e^\gamma}{\left(k_0^{+}-k_2^{+}\right)^2} \nonumber \\ 
& -\left[\frac{1}{q^{+}-k_0^{+}}-\frac{1}{q^{+}-k_2^{+}}\right] \log \frac{\left(k_3^{+}-q^{+}\right)\left(q^{+}-k_1^{+}\right) p_{14}^2 \delta e^\gamma}{\left(-k_1^{+}+k_3^{+}\right)^2} \nonumber \\ 
& -\left[\frac{2}{q^{+}-k_1^{+}}\right] \log \frac{\left(k_0^{+}-q^{+}\right)\left(q^{+}-k_1^{+}\right) p_{12}^2 \delta e^\gamma}{\left(-k_1^{+}+k_0^{+}\right)\left(k_0^{+}-k_2^{+}\right)} \nonumber \\ 
& +\left[\frac{2}{q^{+}-k_3^{+}}\right] \log \frac{\left(q^{+}-k_3^{+}\right)\left(k_2^{+}-q^{+}\right) p_{12}^2 \delta e^\gamma}{\left(k_2^{+}-k_0^{+}\right)\left(k_2^{+}-k_3^{+}\right)} \nonumber \\ 
& -\left[\frac{2}{q^{+}-k_2^{+}}\right] \log \frac{\left(-k_2^{+}+q^{+}\right)\left(k_3^{+}-q^{+}\right) p_{14}^2 \delta e^\gamma}{\left(k_3^{+}-k_2^{+}\right)\left(-k_1^{+}+k_3^{+}\right)} \nonumber \\ 
& +\left[\frac{2}{q^{+}-k_0^{+}}\right] \log \frac{\left(k_0^{+}-q^{+}\right)\left(q^{+}-k_1^{+}\right) p_{14}^2 \delta e^\gamma}{\left(-k_1^{+}+k_0^{+}\right)\left(-k_1^{+}+k_3^{+}\right)} \,.
\end{align}
And, for $k_1^{+}<q^{+}<k_2^{+}$,
\begin{align}
\mathcal{F}_{\mathrm{IR}}^{+ + - -}[q^+] =
& -\left[\frac{1}{q^{+}-k_3^{+}}+\frac{1}{q^{+}-k_1^{+}}\right] \log \frac{\left(k_2^{+}-q^{+}\right)\left(-k_0^{+}+q^{+}\right) p_{12}^2 \delta e^\gamma}{\left(k_0^{+}-k_2^{+}\right)^2} \nonumber \\
& +\left[\frac{2}{q^{+}-k_1^{+}}\right] \log \frac{\left(q^{+}-k_1^{+}\right)\left(-k_2^{+}+q^{+}\right) p_{12}^2 \delta e^\gamma}{\left(k_2^{+}-k_1^{+}\right)\left(k_0^{+}-k_2^{+}\right)} \nonumber \\
& +\left[\frac{2}{q^{+}-k_3^{+}}\right] \log \frac{\left(q^{+}-k_3^{+}\right)\left(k_2^{+}-q^{+}\right) p_{12}^2 \delta e^\gamma}{\left(k_2^{+}-k_0^{+}\right)\left(k_2^{+}-k_3^{+}\right)} \nonumber \\
& -\left[\frac{2}{q^{+}-k_2^{+}}\right] \log \frac{\left(-k_2^{+}+q^{+}\right)^2 p_{14}^2 \delta e^\gamma}{\left(k_2^{+}-k_1^{+}\right)\left(k_2^{+}-k_3^{+}\right)}\,.
\end{align}

Notice, the infrared divergent terms contain the CQT ultraviolet regulator $\delta$. This intermixes the infrared divergences with the ultraviolet ones. These get canceled against similar terms originating from the self-energy corrections to the external legs in the one-loop amplitude.
\chapter{Consistency check for the classical EOM}
\label{sec:app_A9}

In this appendix, we demonstrate that
\begin{itemize}
    \item The MHV classical EOM Eq.~\eqref{eq:B_bul_EOM}-\eqref{eq:B_star_EOM} and the Z-field classical EOM Eq.~\eqref{eq:Z_bul_EOM}-\eqref{eq:Z_star_EOM} can be obtained from the  Yang-Mills classical EOM  Eq.~\eqref{eq:YMEOM0} via the canonical transformations that derive the corresponding action i.e. Eq.~\eqref{eq:Man_Transf1} (Mansfield's transformation) and Eq.~\eqref{eq:AtoZ_ct1}-\eqref{eq:AtoZ_ct2} respectively. This provides a consistency check between the classical EOMs.
    \item The solutions  ${\hat B}^i_{c}[J]$ of the MHV classical EOM Eq.~\eqref{eq:B_bul_EOM}-\eqref{eq:B_star_EOM} and ${\hat Z}^i_{c}[J]$ of the Z-field classical EOM Eq.~\eqref{eq:Z_bul_EOM}-\eqref{eq:Z_star_EOM} can be obtained from the solution ${\hat A}^i_{c}[J]$ of the Yang-Mills classical EOM Eq.~\eqref{eq:YMEOM0} via the canonical transformations that derives the corresponding actions.
\end{itemize}

\section{MHV classical EOM}
\label{sec:App91}

In this section, we demonstrate the above for the MHV classical EOM Eq.~\eqref{eq:B_bul_EOM}-\eqref{eq:B_star_EOM} and its solution ${\hat B}^i_{c}[J]$. Precisely, we start with the MHV classical EOM, and then through simple manipulation using functional calculus we relate them to the Yang-Mills classical EOM. The MHV classical EOM read
\begin{equation}
    \frac{\delta S_{\mathrm{MHV}}[B_c^{\bullet}, B_c^{\star}]}{\delta \hat{B}^{\bullet}(x)}+\int\!d^4y\,\left({\hat J}_{\bullet}(y)\frac{\delta \hat{A}^{\bullet}[B_c^{\bullet}](y)}{\delta \hat{B}^{\bullet}(x)}+{\hat J}_{\star}(y)\frac{\delta \hat{A}^{\star}[B_c^{\bullet}, B_c^{\star}](y)}{\delta \hat{B}^{\bullet}(x)}\right) =0\,.
    \label{eq:B_bul_EOM1}
\end{equation}
\begin{equation}
    \frac{\delta S_{\mathrm{MHV}}[B_c^{\bullet}, B_c^{\star}]}{\delta \hat{B}^{\star}(x)}+\int\!d^4y\,{\hat J}_{\star}(y)\frac{\delta \hat{A}^{\star}[B_c^{\bullet}, B_c^{\star}](y)}{\delta \hat{B}^{\star}(x)} =0\,.
    \label{eq:B_star_EOM1}
\end{equation}
Before proceeding further, let us establish the relation for the inverse of the functional derivatives of the A-fields with respect to B-fields as shown below
\begin{equation}
    \int\!d^4x\,\frac{\delta \hat{A}^{\bullet}[B_c^{\bullet}](y)}{\delta \hat{B}^{\bullet}(x)}\left(\frac{\delta \hat{A}^{\bullet}[B_c^{\bullet}](z)}{\delta \hat{B}^{\bullet}(x)}\right)^{-1} = \,\delta^4(y-z)\,.
    \label{eq:abul_inv}
\end{equation}
\begin{equation}
    \int\!d^4x\,\frac{\delta \hat{A}^{\star}[B_c^{\bullet}, B_c^{\star}](y)}{\delta \hat{B}^{\star}(x)}\left(\frac{\delta \hat{A}^{\star}[B_c^{\bullet}, B_c^{\star}](z)}{\delta \hat{B}^{\star}(x)}\right)^{-1} = \,\delta^4(y-z)\,.
    \label{eq:astar_inv}
\end{equation}
These two are the only relations we need for current purposes. 

Using Eq.~\eqref{eq:astar_inv} in  Eq.~\eqref{eq:B_star_EOM1} we get
\begin{equation}
   \int\!d^4x\, \frac{\delta S_{\mathrm{MHV}}[B_c^{\bullet}, B_c^{\star}]}{\delta \hat{B}^{\star}(x)}\left(\frac{\delta \hat{A}^{\star}[B_c^{\bullet}, B_c^{\star}](z)}{\delta \hat{B}^{\star}(x)}\right)^{-1}=-{\hat J}_{\star}(z)\,.
    \label{eq:B_star_EOM2}
\end{equation}
Recall, the MHV action is obtained by substituting the solution of Mansfield's transformation $A^{\bullet}[B_c^{\bullet}], A^{\star}[B_c^{\bullet}, B_c^{\star}]$ Eq.~\eqref{eq:A_bull_solu}-\eqref{eq:A_star_solu} to the Yang-Mills action. Thus, we can write
\begin{equation}
    S_{\mathrm{MHV}}[B_c^{\bullet}, B_c^{\star}] = S_{\mathrm{YM}}[A^{\bullet}[B_c^{\bullet}], A^{\star}[B_c^{\bullet}, B_c^{\star}]]\,.
    \label{eq:MHV_YMac}
\end{equation}
Differentiating both sides with respect to $\hat{B}^{\star}(x)$ field and using the chain rule for functional derivatives in the R.H.S of the above expression, we get
\begin{equation}
    \frac{\delta S_{\mathrm{MHV}}[B_c^{\bullet}, B_c^{\star}]}{\delta \hat{B}^{\star}(x)} =\int\!d^4y\, \frac{\delta S_{\mathrm{YM}}[A^{\bullet}[B_c^{\bullet}], A^{\star}[B_c^{\bullet}, B_c^{\star}]]}{\delta \hat{A}^{\star}(y)}\,\frac{\delta \hat{A}^{\star}[B_c^{\bullet}, B_c^{\star}](y)}{\delta \hat{B}^{\star}(x)}\,.
\end{equation}
Substituting this to Eq.~\eqref{eq:B_star_EOM2}, we obtain
\begin{equation}
   \int\!d^4x\, \!d^4y\, \frac{\delta S_{\mathrm{YM}}[A^{\bullet}[B_c^{\bullet}], A^{\star}[B_c^{\bullet}, B_c^{\star}]]}{\delta \hat{A}^{\star}(y)}\,\frac{\delta \hat{A}^{\star}[B_c^{\bullet}, B_c^{\star}](y)}{\delta \hat{B}^{\star}(x)}\left(\frac{\delta \hat{A}^{\star}[B_c^{\bullet}, B_c^{\star}](z)}{\delta \hat{B}^{\star}(x)}\right)^{-1}=-{\hat J}_{\star}(z)\,.
\end{equation}
Owing to Eq.~\eqref{eq:astar_inv}, the above expression reduces to
\begin{equation}
   \frac{\delta S_{\mathrm{YM}}[A^{\bullet}[B_c^{\bullet}], A^{\star}[B_c^{\bullet}, B_c^{\star}]]}{\delta \hat{A}^{\star}(z)}=-{\hat J}_{\star}(z)\,.
   \label{eq:Jstar_eom}
\end{equation}
The above expression is essentially
\begin{equation}
   \left.\frac{\delta S_{\mathrm{YM}}[A^{\bullet}, A^{\star}]}{\delta \hat{A}^{\star}(z)}\right|_{{\hat A}={\hat A}[B_{c}]}=-{\hat J}_{\star}(z)\,.
   \label{eq:YM_dumeom}
\end{equation}
Comparing it with the Yang-Mills classical EOM
\begin{equation}
   \left.\frac{\delta S_{\mathrm{YM}}[A^{\bullet}, A^{\star}]}{\delta \hat{A}^{\star}(z)}\right|_{{\hat A}={\hat A}_c}=-{\hat J}_{\star}(z)\,,
   \label{eq:YM_eom1}
\end{equation}
we conclude that the MHV classical EOM Eq.~\eqref{eq:B_star_EOM1} can be obtained from the Yang-Mills classical EOM Eq.~\eqref{eq:YM_eom1} via Mansfield's transformation. Furthermore, the solution ${\hat B}^i_{c}[J]$ obtained from the MHV classical EOM Eq.~\eqref{eq:Jstar_eom} should be identical to the solution obtained from inverting ${\hat A}^i[B_{c}[J]]$ obtained from Eq.~\eqref{eq:YM_dumeom}. The latter is basically the solution ${\hat A}^i_{c}[J]$ for the Yang-Mills classical EOM followed by Mansfield's transformation. Thus we can also conclude that the solution ${\hat B}^i_{c}[J]$ of the MHV classical EOM is related to the solution ${\hat A}^i_{c}[J]$ of the Yang-Mills classical EOM via Mansfield's transformation.

Let us verify the above claims for the second EOM Eq.~\eqref{eq:B_bul_EOM1}. Substituting Eq.~\eqref{eq:abul_inv} to Eq.~\eqref{eq:B_bul_EOM1}, we get
\begin{multline}
    \int\!d^4x\,\frac{\delta S_{\mathrm{MHV}}[B_c^{\bullet}, B_c^{\star}]}{\delta \hat{B}^{\bullet}(x)} \left(\frac{\delta \hat{A}^{\bullet}[B_c^{\bullet}](z)}{\delta \hat{B}^{\bullet}(x)}\right)^{-1}\\
   +\int\!d^4y\,d^4x\left({\hat J}_{\star}(y)\frac{\delta \hat{A}^{\star}[B_c^{\bullet}, B_c^{\star}](y)}{\delta \hat{B}^{\bullet}(x)}\right)\left(\frac{\delta \hat{A}^{\bullet}[B_c^{\bullet}](z)}{\delta \hat{B}^{\bullet}(x)}\right)^{-1} =- {\hat J}_{\bullet}(z)\,.
    \label{eq:B_bul_EOM2}
\end{multline}
Differentiating both sides of Eq.~\eqref{eq:MHV_YMac} with respect to $\hat{B}^{\bullet}(x)$ field and using the chain rule for functional derivatives in the R.H.S, we get
\begin{multline}
    \frac{\delta S_{\mathrm{MHV}}[B_c^{\bullet}, B_c^{\star}]}{\delta \hat{B}^{\bullet}(x)} =\int\!d^4y\, \frac{\delta S_{\mathrm{YM}}[A^{\bullet}[B_c^{\bullet}], A^{\star}[B_c^{\bullet}, B_c^{\star}]]}{\delta \hat{A}^{\star}(y)}\,\frac{\delta \hat{A}^{\star}[B_c^{\bullet}, B_c^{\star}](y)}{\delta \hat{B}^{\bullet}(x)}\\
    + \int\!d^4y\, \frac{\delta S_{\mathrm{YM}}[A^{\bullet}[B_c^{\bullet}], A^{\star}[B_c^{\bullet}, B_c^{\star}]]}{\delta \hat{A}^{\bullet}(y)}\,\frac{\delta \hat{A}^{\bullet}[B_c^{\bullet}](y)}{\delta \hat{B}^{\bullet}(x)}\,.
    \label{eq:SMHV_bul}
\end{multline}
Substituting Eq.~\eqref{eq:Jstar_eom} and \eqref{eq:SMHV_bul} to Eq.~\eqref{eq:B_bul_EOM2}, we get
\begin{multline}
    \int\!d^4x\,\left\{ \int\!d^4y\, \frac{\delta S_{\mathrm{YM}}[A^{\bullet}[B_c^{\bullet}], A^{\star}[B_c^{\bullet}, B_c^{\star}]]}{\delta \hat{A}^{\star}(y)}\,\frac{\delta \hat{A}^{\star}[B_c^{\bullet}, B_c^{\star}](y)}{\delta \hat{B}^{\bullet}(x)}\right.\\
    \left.+ \int\!d^4y\, \frac{\delta S_{\mathrm{YM}}[A^{\bullet}[B_c^{\bullet}], A^{\star}[B_c^{\bullet}, B_c^{\star}]]}{\delta \hat{A}^{\bullet}(y)}\,\frac{\delta \hat{A}^{\bullet}[B_c^{\bullet}](y)}{\delta \hat{B}^{\bullet}(x)}\right\}\left(\frac{\delta \hat{A}^{\bullet}[B_c^{\bullet}](z)}{\delta \hat{B}^{\bullet}(x)}\right)^{-1}\\
   +\int\!d^4y\,d^4x\left(-\frac{\delta S_{\mathrm{YM}}[A^{\bullet}[B_c^{\bullet}], A^{\star}[B_c^{\bullet}, B_c^{\star}]]}{\delta \hat{A}^{\star}(y)}\frac{\delta \hat{A}^{\star}[B_c^{\bullet}, B_c^{\star}](y)}{\delta \hat{B}^{\bullet}(x)}\right)\left(\frac{\delta \hat{A}^{\bullet}[B_c^{\bullet}](z)}{\delta \hat{B}^{\bullet}(x)}\right)^{-1} =- {\hat J}_{\bullet}(z)\,.
    \label{eq:B_bul_EOM3}
\end{multline}
The first and the third term on the L.H.S. of the expression above cancel out. Using Eq.~\eqref{eq:abul_inv} for the middle term we get
\begin{equation}
    \frac{\delta S_{\mathrm{YM}}[A^{\bullet}[B_c^{\bullet}], A^{\star}[B_c^{\bullet}, B_c^{\star}]]}{\delta \hat{A}^{\bullet}(z)}=- {\hat J}_{\bullet}(z) = \left.\frac{\delta S_{\mathrm{YM}}[A^{\bullet}, A^{\star}]}{\delta \hat{A}^{\bullet}(z)}\right|_{{\hat A}={\hat A}[B_{c}]}
\end{equation}
Comparing the above with the second Yang-Mills classical EOM
\begin{equation}
   \left.\frac{\delta S_{\mathrm{YM}}[A^{\bullet}, A^{\star}]}{\delta \hat{A}^{\bullet}(z)}\right|_{{\hat A}={\hat A}_c}=-{\hat J}_{\bullet}(z)\,.
   \label{eq:YM_eom2}
\end{equation}
we can make exactly the same conclusions made previously. This verifies the claims.
\section{Z-field classical EOM}
\label{sec:App92}

In this section, we perform the consistency check between the Z-field classical EOM Eq.~\eqref{eq:Z_bul_EOM}-\eqref{eq:Z_star_EOM} and the  Yang-Mills classical EOM  Eq.~\eqref{eq:YMEOM0} and demonstrate that the two sets of EOM as well as their solutions are related via the canonical transformations  Eq.~\eqref{eq:AtoZ_ct1}-\eqref{eq:AtoZ_ct2} that derives the Z-field action.

The Z-field classical EOM Eq.~\eqref{eq:Z_bul_EOM}-\eqref{eq:Z_star_EOM} read
\begin{multline}
    \frac{\delta S[Z_c^{\bullet}, Z_c^{\star}]}{\delta \hat{Z}^{\bullet}(x)}+\int\!d^4y_1d^4y_2\,\left[{\hat J}_{\bullet}(y_1)\frac{\delta \hat{A}^{\bullet}[B^{\bullet}](y_1)}{\delta \hat{B}^{\bullet}(y_2)}\frac{\delta \hat{B}^{\bullet}[Z_c^{\bullet}, Z_c^{\star}](y_2)}{\delta \hat{Z}^{\bullet}(x)}\right.\\
    \left.+{\hat J}_{\star}(y_1)\frac{\delta \hat{A}^{\star}[B^{\bullet}, B^{\star}](y_1)}{\delta \hat{B}^{\bullet}(y_2)}\frac{\delta \hat{B}^{\bullet}[Z_c^{\bullet}, Z_c^{\star}](y_2)}{\delta \hat{Z}^{\bullet}(x)}\right] =0\,.
    \label{eq:Z_bul_EOM1}
\end{multline}
\begin{multline}
    \frac{\delta S[Z_c^{\bullet}, Z_c^{\star}]}{\delta \hat{Z}^{\star}(x)}+\int\!d^4y_1d^4y_2\,\left[{\hat J}_{\star}(y_1)\left(\frac{\delta \hat{A}^{\star}[B^{\bullet}, B^{\star}](y_1)}{\delta \hat{B}^{\star}(y_2)}\frac{\delta \hat{B}^{\star}[Z_c^{\star}](y_2)}{\delta \hat{Z}^{\star}(x)}\right. \right.\\
    \left.\left.+\frac{\delta \hat{A}^{\star}[B^{\bullet}, B^{\star}](y_1)}{\delta \hat{B}^{\bullet}(y_2)}\frac{\delta \hat{B}^{\bullet}[Z_c^{\bullet}, Z_c^{\star}](y_2)}{\delta \hat{Z}^{\star}(x)}\right)+{\hat J}_{\bullet}(y_1)\frac{\delta \hat{A}^{\bullet}[B^{\bullet}](y_1)}{\delta \hat{B}^{\bullet}(y_2)}\frac{\delta \hat{B}^{\bullet}[Z_c^{\bullet}, Z_c^{\star}](y_2)}{\delta \hat{Z}^{\star}(x)}\right] =0\,.
    \label{eq:Z_star_EOM1}
\end{multline}
Throughout this section, we shall use the following relation for the inverse of the functional derivatives of the B-fields with respect to Z-fields
\begin{equation}
    \int\!d^4x\,\frac{\delta \hat{B}^{\star}[Z_c^{\star}](y)}{\delta \hat{Z}^{\star}(x)}\left(\frac{\delta \hat{B}^{\star}[Z_c^{\star}](z)}{\delta \hat{Z}^{\star}(x)}\right)^{-1} = \,\delta^4(y-z)\,.
    \label{eq:bstar_inv}
\end{equation}
\begin{equation}
    \int\!d^4x\,\frac{\delta \hat{B}^{\bullet}[Z_c^{\bullet}, Z_c^{\star}](y)}{\delta \hat{Z}^{\bullet}(x)}\left(\frac{\delta \hat{B}^{\bullet}[Z_c^{\bullet}, Z_c^{\star}](z)}{\delta \hat{Z}^{\bullet}(x)}\right)^{-1} = \,\delta^4(y-z)\,.
    \label{eq:bbul_inv}
\end{equation}

Substituting Eq.~\eqref{eq:bbul_inv} to  Eq.~\eqref{eq:Z_bul_EOM1} we get
\begin{multline}
    \int\!d^4x\frac{\delta S[Z_c^{\bullet}, Z_c^{\star}]}{\delta \hat{Z}^{\bullet}(x)}\left(\frac{\delta \hat{B}^{\bullet}[Z_c^{\bullet}, Z_c^{\star}](z)}{\delta \hat{Z}^{\bullet}(x)}\right)^{-1}\\
    +\int\!d^4y_1\,\left[{\hat J}_{\bullet}(y_1)\frac{\delta \hat{A}^{\bullet}[B^{\bullet}](y_1)}{\delta \hat{B}^{\bullet}(z)}+{\hat J}_{\star}(y_1)\frac{\delta \hat{A}^{\star}[B^{\bullet}, B^{\star}](y_1)}{\delta \hat{B}^{\bullet}(z)}\right] =0\,.
    \label{eq:Z_bul_EOM2}
\end{multline}
Recall, the Z-field action Eq.~\eqref{eq:Z_action1} can be derived from the light-cone Yang-Mills action Eq.~\eqref{eq:YM_LC_action} via two consecutive canonical transformations  $S_{\mathrm{YM}}\left[A^{\bullet},A^{\star}\right] \longrightarrow S_{\mathrm{MHV}}\left[B^{\bullet},B^{\star}\right] \longrightarrow S\left[Z^{\bullet},Z^{\star}\right]$. Thus we can write
\begin{equation}
    S[Z_c^{\bullet}, Z_c^{\star}] = S_{\mathrm{YM}}[A^{\bullet}[B^{\bullet}[Z_c^{\bullet}, Z_c^{\star}]], A^{\star}[B^{\bullet}[Z_c^{\bullet}, Z_c^{\star}], B^{\star}[Z_c^{\star}]]]\,.
    \label{eq:ZacYM}
\end{equation}
Differentiating both sides of the expression above with respect to $\hat{Z}^{\bullet}(x)$ field and using the chain rule for functional derivatives in the R.H.S, we get 
\begin{multline}
\frac{\delta S[Z_c^{\bullet}, Z_c^{\star}]}{\delta \hat{Z}^{\bullet}(x)} =\int\!d^4y_1d^4y_2\,  \frac{\delta S_{\mathrm{YM}}[A^{\bullet}[B^{\bullet}], A^{\star}[B^{\bullet}, B^{\star}]]}{\delta \hat{A}^{\star}(y_1)}\,\frac{\delta \hat{A}^{\star}[B^{\bullet}, B^{\star}](y_1)}{\delta \hat{B}^{\bullet}(y_2)}\frac{\delta \hat{B}^{\bullet}[Z_c^{\bullet}, Z_c^{\star}](y_2)}{\delta \hat{Z}^{\bullet}(x)}\\
    + \int\!d^4y_1d^4y_2\, \frac{\delta S_{\mathrm{YM}}[A^{\bullet}[B^{\bullet}], A^{\star}[B^{\bullet}, B^{\star}]]}{\delta \hat{A}^{\bullet}(y_1)}\,\frac{\delta \hat{A}^{\bullet}[B^{\bullet}](y_1)}{\delta \hat{B}^{\bullet}(y_2)}\frac{\delta \hat{B}^{\bullet}[Z_c^{\bullet}, Z_c^{\star}](y_2)}{\delta \hat{Z}^{\bullet}(x)}\,.
    \label{eq:Sz_SYM}
   \end{multline}
Substituting the above to Eq.~\eqref{eq:Z_bul_EOM2} and using Eq.~\eqref{eq:bbul_inv} we get
\begin{multline}
   \int\!d^4y_1\, \left[\frac{\delta S_{\mathrm{YM}}[A^{\bullet}[B^{\bullet}], A^{\star}[B^{\bullet}, B^{\star}]]}{\delta \hat{A}^{\star}(y_1)}\,\frac{\delta \hat{A}^{\star}[B^{\bullet}, B^{\star}](y_1)}{\delta \hat{B}^{\bullet}(z)}\right.\\
    \left.
    + \frac{\delta S_{\mathrm{YM}}[A^{\bullet}[B^{\bullet}], A^{\star}[B^{\bullet}, B^{\star}]]}{\delta \hat{A}^{\bullet}(y_1)}\,\frac{\delta \hat{A}^{\bullet}[B^{\bullet}](y_1)}{\delta \hat{B}^{\bullet}(z)} \right.\\
    \left. +{\hat J}_{\bullet}(y_1)\frac{\delta \hat{A}^{\bullet}[B^{\bullet}](y_1)}{\delta \hat{B}^{\bullet}(z)}+{\hat J}_{\star}(y_1)\frac{\delta \hat{A}^{\star}[B^{\bullet}, B^{\star}](y_1)}{\delta \hat{B}^{\bullet}(z)}\right] =0\,.
    \label{eq:J_b_s}
\end{multline}
Using the above equation, the second classical EOM Eq.~\eqref{eq:Z_star_EOM1} can be reexpressed as
\begin{multline}
    \frac{\delta S[Z_c^{\bullet}, Z_c^{\star}]}{\delta \hat{Z}^{\star}(x)}+\int\!d^4y_1d^4y_2\,\left[{\hat J}_{\star}(y_1)\frac{\delta \hat{A}^{\star}[B^{\bullet}, B^{\star}](y_1)}{\delta \hat{B}^{\star}(y_2)}\frac{\delta \hat{B}^{\star}[Z_c^{\star}](y_2)}{\delta \hat{Z}^{\star}(x)}\right. \\
    \left.- \left(\frac{\delta S_{\mathrm{YM}}[A^{\bullet}[B^{\bullet}], A^{\star}[B^{\bullet}, B^{\star}]]}{\delta \hat{A}^{\star}(y_1)}\,\frac{\delta \hat{A}^{\star}[B^{\bullet}, B^{\star}](y_1)}{\delta \hat{B}^{\bullet}(y_2)}
    + \frac{\delta S_{\mathrm{YM}}[A^{\bullet}[B^{\bullet}], A^{\star}[B^{\bullet}, B^{\star}]]}{\delta \hat{A}^{\bullet}(y_1)}\,\frac{\delta \hat{A}^{\bullet}[B^{\bullet}](y_1)}{\delta \hat{B}^{\bullet}(y_2)} \right)\right.\\
   \left.\frac{\delta \hat{B}^{\bullet}[Z_c^{\bullet}, Z_c^{\star}](y_2)}{\delta \hat{Z}^{\star}(x)}\right] =0\,.
   \label{eq:J_star_z}
\end{multline}
Again, differentiating both sides of the Eq.~\eqref{eq:ZacYM} with respect to $\hat{Z}^{\star}(x)$ field and using the chain rule for functional derivatives in the R.H.S, we get
\begin{multline}
\frac{\delta S[Z_c^{\bullet}, Z_c^{\star}]}{\delta \hat{Z}^{\star}(x)} =\int\!d^4y_1d^4y_2\,  \frac{\delta S_{\mathrm{YM}}[A^{\bullet}[B^{\bullet}], A^{\star}[B^{\bullet}, B^{\star}]]}{\delta \hat{A}^{\star}(y_1)}\,\frac{\delta \hat{A}^{\star}[B^{\bullet}, B^{\star}](y_1)}{\delta \hat{B}^{\star}(y_2)}\frac{\delta \hat{B}^{\star}[Z_c^{\bullet}, Z_c^{\star}](y_2)}{\delta \hat{Z}^{\star}(x)}\\
+\int\!d^4y_1d^4y_2\,  \frac{\delta S_{\mathrm{YM}}[A^{\bullet}[B^{\bullet}], A^{\star}[B^{\bullet}, B^{\star}]]}{\delta \hat{A}^{\star}(y_1)}\,\frac{\delta \hat{A}^{\star}[B^{\bullet}, B^{\star}](y_1)}{\delta \hat{B}^{\bullet}(y_2)}\frac{\delta \hat{B}^{\bullet}[Z_c^{\bullet}, Z_c^{\star}](y_2)}{\delta \hat{Z}^{\star}(x)}\\
    + \int\!d^4y_1d^4y_2\, \frac{\delta S_{\mathrm{YM}}[A^{\bullet}[B^{\bullet}], A^{\star}[B^{\bullet}, B^{\star}]]}{\delta \hat{A}^{\bullet}(y_1)}\,\frac{\delta \hat{A}^{\bullet}[B^{\bullet}](y_1)}{\delta \hat{B}^{\bullet}(y_2)}\frac{\delta \hat{B}^{\bullet}[Z_c^{\bullet}, Z_c^{\star}](y_2)}{\delta \hat{Z}^{\star}(x)}\,.
    \label{eq:Sz_SYM1}
\end{multline}
Substituting Eq.~\eqref{eq:Sz_SYM1} in Eq.~\eqref{eq:J_star_z}, we see that the last two terms in the latter cancel out and we are left with
\begin{multline}
    \int\!d^4y_1d^4y_2\,\left[\frac{\delta S_{\mathrm{YM}}[A^{\bullet}[B^{\bullet}], A^{\star}[B^{\bullet}, B^{\star}]]}{\delta \hat{A}^{\star}(y_1)}\,\frac{\delta \hat{A}^{\star}[B^{\bullet}, B^{\star}](y_1)}{\delta \hat{B}^{\star}(y_2)}\frac{\delta \hat{B}^{\star}[Z_c^{\bullet}, Z_c^{\star}](y_2)}{\delta \hat{Z}^{\star}(x)} \right.\\
    \left. +{\hat J}_{\star}(y_1)\frac{\delta \hat{A}^{\star}[B^{\bullet}, B^{\star}](y_1)}{\delta \hat{B}^{\star}(y_2)}\frac{\delta \hat{B}^{\star}[Z_c^{\star}](y_2)}{\delta \hat{Z}^{\star}(x)} \right] =0\,.
\end{multline}
The above expression implies
\begin{equation}
   \frac{\delta S_{\mathrm{YM}}[A^{\bullet}[Z^{\bullet}, Z^{\star}], A^{\star}[Z^{\bullet}, Z^{\star}]]}{\delta \hat{A}^{\star}(y_1)}=-{\hat J}_{\star}(y_1)\,.
   \label{eq:Jstar_eom1}
\end{equation}
Substituting the above in Eq.~\eqref{eq:J_b_s} we get
\begin{equation}
   \int\!d^4y_1\, \left[\frac{\delta S_{\mathrm{YM}}[A^{\bullet}[B^{\bullet}], A^{\star}[B^{\bullet}, B^{\star}]]}{\delta \hat{A}^{\bullet}(y_1)}\,\frac{\delta \hat{A}^{\bullet}[B^{\bullet}](y_1)}{\delta \hat{B}^{\bullet}(z)}  +{\hat J}_{\bullet}(y_1)\frac{\delta \hat{A}^{\bullet}[B^{\bullet}](y_1)}{\delta \hat{B}^{\bullet}(z)}\right] =0\,.
\end{equation}
which simplifies to
\begin{equation}
    \frac{\delta S_{\mathrm{YM}}[A^{\bullet}[Z^{\bullet}, Z^{\star}], A^{\star}[Z^{\bullet}, Z^{\star}]]}{\delta \hat{A}^{\bullet}(y_1)}=- {\hat J}_{\bullet}(y_1)\,.
    \label{eq:Zac_eom2}
\end{equation}

Eqs.~\eqref{eq:Jstar_eom1} and \eqref{eq:Zac_eom2} can be rewritten as
\begin{equation}
   \left.\frac{\delta S_{\mathrm{YM}}[A^{\bullet}, A^{\star}]}{\delta \hat{A}^{\star}(y_1)}\right|_{{\hat A}={\hat A}[Z_c]}=-{\hat J}_{\star}(y_1)\,,\quad \quad \left.\frac{\delta S_{\mathrm{YM}}[A^{\bullet}, A^{\star}]}{\delta \hat{A}^{\bullet}(y_1)}\right|_{{\hat A}={\hat A}[Z_c]}=-{\hat J}_{\bullet}(y_1)\,.
\end{equation}
Thus, we see that indeed the Z-field classical EOM Eq.~\eqref{eq:Z_bul_EOM}-\eqref{eq:Z_star_EOM} and the  Yang-Mills classical EOM  Eq.~\eqref{eq:YMEOM0} are related via the canonical transformations  Eq.~\eqref{eq:AtoZ_ct1}-\eqref{eq:AtoZ_ct2} that derives the Z-field action. Furthermore, the solutions ${\hat Z}^i_{c}[J]$ and ${\hat A}^i_{c}[J]$ are also related via the same transformation. 

\chapter{\texorpdfstring{$\,\,\,\,$}{sp}Determinant of matrix \texorpdfstring{$\mathrm{M}_{IK}$}{M}}
\label{sec:app_A10}

We derived the one-loop effective MHV  action in two ways in Chapter \ref{QMHV-chapter} and Chapter \ref{QZth-chapter} respectively. Similarly, the one-loop Z-field action was also derived using these two ways. The two one-loop effective actions, both for the MHV theory as well as the Z-field theory, differ only in the log term. In this appendix, we demonstrate that the log terms and therefore the two actions derived using two different ways are identical up to a redundant factor (for amplitude computation). Precisely, we show that the determinant of the two matrices $\mathrm{M}^{\mathrm{MHV}}_{IK}$ appearing in the two one-loop effective MHV action derived in Chapter \ref{QMHV-chapter} and Chapter \ref{QZth-chapter} respectively are identical up to a redundant factor. We demonstrate the same for the determinant of the matrices $\mathrm{M}^{\mathrm{Z}}_{IK}$ in the one-loop effective Z-field action. Throughout this appendix, we use the collective indices.
\section{\texorpdfstring{$\,\,\,\,$}{sp}One-loop effective MHV action}
\label{sec:A101}

In this section, we focus on the determinant of the matrix $\mathrm{M}^{\mathrm{MHV}}_{IK}$ in the one-loop effective MHV action derived via the new approach in Chapter \ref{QZth-chapter}. It reads Eq.~\eqref{eq:M_MHV}
\begin{multline}
\mathrm{M}^{\mathrm{MHV}}_{IK}  =  \left(\begin{matrix}
     \frac{\delta^2 S_{\mathrm{MHV}}[B_c]}{\delta B^{\bullet I}\delta B^{\star K}} &\frac{\delta^2 S_{\mathrm{MHV}}[B_c]}{\delta B^{\bullet I}\delta B^{\bullet K}}\\ \\
\frac{\delta^2 S_{\mathrm{MHV}}[B_c]}{\delta B^{\star I}\delta B^{\star K}}
      &\frac{\delta^2 S_{\mathrm{MHV}}[B_c]}{\delta B^{\star I}\delta B^{\bullet K}} 
\end{matrix}\right) \\
+ \left(\begin{matrix}
    - \frac{\delta S_{\mathrm{YM}}[A[B_c]]}{\delta A^{\star L}}\frac{\delta^2 A^{\star L}[B_c]}{\delta B^{\bullet I}\delta B^{\star K}} 
     &- \frac{\delta S_{\mathrm{YM}}[A[B_c]]}{\delta A^{\star L}}\frac{\delta^2 A^{\star L}[B_c]}{\delta B^{\bullet I}\delta B^{\bullet K}} -  \frac{\delta S_{\mathrm{YM}}[A[B_c]]}{\delta A^{\bullet L}}\frac{\delta^2 A^{\bullet L}[B_c]}{\delta B^{\bullet I}\delta B^{\bullet K}}\\ \\
\mathbb{0}
      &- \frac{\delta S_{\mathrm{YM}}[A[B_c]]}{\delta A^{\star L}}\frac{\delta^2 A^{\star L}[B_c]}{\delta B^{\star I}\delta B^{\bullet K}} 
    \label{eq:M_1}
\end{matrix}\right)\,.
\end{multline}
Recall from Eq.~\eqref{eq:MHV_YMac} that
\begin{equation}
    S_{\mathrm{MHV}}[B_c^{\bullet}, B_c^{\star}] = S_{\mathrm{YM}}[A^{\bullet}[B_c^{\bullet}], A^{\star}[B_c^{\bullet}, B_c^{\star}]]\,.
    \label{eq:MHV_YMac1}
\end{equation}
Differentiating both sides with respect to B-fields while using the chain rule on the R.H.S., we can write the following relations
\begin{multline}
    \frac{\delta^2 S_{\mathrm{MHV}}[B_c]}{\delta B^{\bullet I}\delta B^{\star K}} = \frac{\delta S_{\mathrm{YM}}[A[B_c]]}{\delta A^{\star P}}\frac{\delta^2 A^{\star P}[B_c]}{\delta B^{\bullet I}\delta B^{\star K}}\\
    +\frac{\delta A^{\star Q}[B_c]}{\delta B^{\bullet I}}\frac{\delta^2 S_{\mathrm{YM}}[A[B_c]]}{\delta A^{\star Q}\delta A^{\star P}}\frac{\delta A^{\star P}[B_c]}{\delta B^{\star K}}+\frac{\delta A^{\bullet Q}[B_c]}{\delta B^{\bullet I}}\frac{\delta^2 S_{\mathrm{YM}}[A[B_c]]}{\delta A^{\bullet Q}\delta A^{\star P}}\frac{\delta A^{\star P}[B_c]}{\delta B^{\star K}}\,.
\end{multline}
\begin{multline}
    \frac{\delta^2 S_{\mathrm{MHV}}[B_c]}{\delta B^{\bullet I}\delta B^{\bullet K}} = \frac{\delta S_{\mathrm{YM}}[A[B_c]]}{\delta A^{\bullet P}}\frac{\delta^2 A^{\bullet P}[B_c]}{\delta B^{\bullet I}\delta B^{\bullet K}}+\frac{\delta S_{\mathrm{YM}}[A[B_c]]}{\delta A^{\star P}}\frac{\delta^2 A^{\star P}[B_c]}{\delta B^{\bullet I}\delta B^{\bullet K}}\\
    +\frac{\delta A^{\star Q}[B_c]}{\delta B^{\bullet I}}\frac{\delta^2 S_{\mathrm{YM}}[A[B_c]]}{\delta A^{\star Q}\delta A^{\bullet P}}\frac{\delta A^{\bullet P}[B_c]}{\delta B^{\bullet K}}+\frac{\delta A^{\bullet Q}[B_c]}{\delta B^{\bullet I}}\frac{\delta^2 S_{\mathrm{YM}}[A[B_c]]}{\delta A^{\bullet Q}\delta A^{\bullet P}}\frac{\delta A^{\bullet P}[B_c]}{\delta B^{\bullet K}}\\
    +\frac{\delta A^{\star Q}[B_c]}{\delta B^{\bullet I}}\frac{\delta^2 S_{\mathrm{YM}}[A[B_c]]}{\delta A^{\star Q}\delta A^{\star P}}\frac{\delta A^{\star P}[B_c]}{\delta B^{\bullet K}}+\frac{\delta A^{\bullet Q}[B_c]}{\delta B^{\bullet I}}\frac{\delta^2 S_{\mathrm{YM}}[A[B_c]]}{\delta A^{\bullet Q}\delta A^{\star P}}\frac{\delta A^{\star P}[B_c]}{\delta B^{\bullet K}}\,.
\end{multline}
\begin{multline}
     \frac{\delta^2 S_{\mathrm{MHV}}[B_c]}{\delta B^{\star I}\delta B^{\star K}} = 
    \frac{\delta A^{\star Q}[B_c]}{\delta B^{\star I}}\frac{\delta^2 S_{\mathrm{YM}}[A[B_c]]}{\delta A^{\star Q}\delta A^{\star P}}\frac{\delta A^{\star P}[B_c]}{\delta B^{\star K}}\,. \hspace{6.5cm}
\end{multline}
\begin{multline}
    \frac{\delta^2 S_{\mathrm{MHV}}[B_c]}{\delta B^{\star I}\delta B^{\bullet K}} = \frac{\delta S_{\mathrm{YM}}[A[B_c]]}{\delta A^{\star P}}\frac{\delta^2 A^{\star P}[B_c]}{\delta B^{\star I}\delta B^{\bullet K}}\\
    +\frac{\delta A^{\star Q}[B_c]}{\delta B^{\star I}}\frac{\delta^2 S_{\mathrm{YM}}[A[B_c]]}{\delta A^{\star Q}\delta A^{\star P}}\frac{\delta A^{\star P}[B_c]}{\delta B^{\bullet K}}+\frac{\delta A^{\star Q}[B_c]}{\delta B^{\star I}}\frac{\delta^2 S_{\mathrm{YM}}[A[B_c]]}{\delta A^{\star Q}\delta A^{\bullet P}}\frac{\delta A^{\bullet P}[B_c]}{\delta B^{\bullet K}}\,.
\end{multline}    
Substituting these to Eq.~\eqref{eq:M_1} we get
\begin{multline}
\mathrm{M}^{\mathrm{MHV}}_{IK}  =   \left(\begin{matrix}
     \substack{\frac{\delta A^{\star Q}[B_c]}{\delta B^{\bullet I}}\frac{\delta^2 S_{\mathrm{YM}}[A[B_c]]}{\delta A^{\star Q}\delta A^{\star P}}\frac{\delta A^{\star P}[B_c]}{\delta B^{\star K}}\\+\frac{\delta A^{\bullet Q}[B_c]}{\delta B^{\bullet I}}\frac{\delta^2 S_{\mathrm{YM}}[A[B_c]]}{\delta A^{\bullet Q}\delta A^{\star P}}\frac{\delta A^{\star P}[B_c]}{\delta B^{\star K}}} & \substack{\frac{\delta A^{\star Q}[B_c]}{\delta B^{\bullet I}}\frac{\delta^2 S_{\mathrm{YM}}[A[B_c]]}{\delta A^{\star Q}\delta A^{\bullet P}}\frac{\delta A^{\bullet P}[B_c]}{\delta B^{\bullet K}}+\frac{\delta A^{\bullet Q}[B_c]}{\delta B^{\bullet I}}\frac{\delta^2 S_{\mathrm{YM}}[A[B_c]]}{\delta A^{\bullet Q}\delta A^{\bullet P}}\frac{\delta A^{\bullet P}[B_c]}{\delta B^{\bullet K}}\\+\frac{\delta A^{\star Q}[B_c]}{\delta B^{\bullet I}}\frac{\delta^2 S_{\mathrm{YM}}[A[B_c]]}{\delta A^{\star Q}\delta A^{\star P}}\frac{\delta A^{\star P}[B_c]}{\delta B^{\bullet K}}+\frac{\delta A^{\bullet Q}[B_c]}{\delta B^{\bullet I}}\frac{\delta^2 S_{\mathrm{YM}}[A[B_c]]}{\delta A^{\bullet Q}\delta A^{\star P}}\frac{\delta A^{\star P}[B_c]}{\delta B^{\bullet K}}}\\ \\
\frac{\delta A^{\star Q}[B_c]}{\delta B^{\star I}}\frac{\delta^2 S_{\mathrm{YM}}[A[B_c]]}{\delta A^{\star Q}\delta A^{\star P}}\frac{\delta A^{\star P}[B_c]}{\delta B^{\star K}}
      & \substack{\frac{\delta A^{\star Q}[B_c]}{\delta B^{\star I}}\frac{\delta^2 S_{\mathrm{YM}}[A[B_c]]}{\delta A^{\star Q}\delta A^{\star P}}\frac{\delta A^{\star P}[B_c]}{\delta B^{\bullet K}}+\frac{\delta A^{\star Q}[B_c]}{\delta B^{\star I}}\frac{\delta^2 S_{\mathrm{YM}}[A[B_c]]}{\delta A^{\star Q}\delta A^{\bullet P}}\frac{\delta A^{\bullet P}[B_c]}{\delta B^{\bullet K}}}
\end{matrix}\right) \\
+ \left(\begin{matrix}
     \frac{\delta S_{\mathrm{YM}}[A[B_c]]}{\delta A^{\star P}}\frac{\delta^2 A^{\star P}[B_c]}{\delta B^{\bullet I}\delta B^{\star K}} 
     & \frac{\delta S_{\mathrm{YM}}[A[B_c]]}{\delta A^{\star P}}\frac{\delta^2 A^{\star P}[B_c]}{\delta B^{\bullet I}\delta B^{\bullet K}} +  \frac{\delta S_{\mathrm{YM}}[A[B_c]]}{\delta A^{\bullet P}}\frac{\delta^2 A^{\bullet P}[B_c]}{\delta B^{\bullet I}\delta B^{\bullet K}}\\ \\
\mathbb{0}
      & \frac{\delta S_{\mathrm{YM}}[A[B_c]]}{\delta A^{\star P}}\frac{\delta^2 A^{\star P}[B_c]}{\delta B^{\star I}\delta B^{\bullet K}} 
\end{matrix}\right) \\
+ \left(\begin{matrix}
    - \frac{\delta S_{\mathrm{YM}}[A[B_c]]}{\delta A^{\star L}}\frac{\delta^2 A^{\star L}[B_c]}{\delta B^{\bullet I}\delta B^{\star K}} 
     &- \frac{\delta S_{\mathrm{YM}}[A[B_c]]}{\delta A^{\star L}}\frac{\delta^2 A^{\star L}[B_c]}{\delta B^{\bullet I}\delta B^{\bullet K}} -  \frac{\delta S_{\mathrm{YM}}[A[B_c]]}{\delta A^{\bullet L}}\frac{\delta^2 A^{\bullet L}[B_c]}{\delta B^{\bullet I}\delta B^{\bullet K}}\\ \\
\mathbb{0}
      &- \frac{\delta S_{\mathrm{YM}}[A[B_c]]}{\delta A^{\star L}}\frac{\delta^2 A^{\star L}[B_c]}{\delta B^{\star I}\delta B^{\bullet K}} 
    \label{eq:M_2}
\end{matrix}\right)\,.
\end{multline}
Notice, the last two matrices exactly cancel out. With this, we are left with the following
\begin{equation}
\mathrm{M}^{\mathrm{MHV}}_{IK}  =   \left(\begin{matrix}
     \substack{\frac{\delta A^{\star Q}[B_c]}{\delta B^{\bullet I}}\frac{\delta^2 S_{\mathrm{YM}}[A[B_c]]}{\delta A^{\star Q}\delta A^{\star P}}\frac{\delta A^{\star P}[B_c]}{\delta B^{\star K}}\\+\frac{\delta A^{\bullet Q}[B_c]}{\delta B^{\bullet I}}\frac{\delta^2 S_{\mathrm{YM}}[A[B_c]]}{\delta A^{\bullet Q}\delta A^{\star P}}\frac{\delta A^{\star P}[B_c]}{\delta B^{\star K}}} & \substack{\frac{\delta A^{\star Q}[B_c]}{\delta B^{\bullet I}}\frac{\delta^2 S_{\mathrm{YM}}[A[B_c]]}{\delta A^{\star Q}\delta A^{\bullet P}}\frac{\delta A^{\bullet P}[B_c]}{\delta B^{\bullet K}}+\frac{\delta A^{\bullet Q}[B_c]}{\delta B^{\bullet I}}\frac{\delta^2 S_{\mathrm{YM}}[A[B_c]]}{\delta A^{\bullet Q}\delta A^{\bullet P}}\frac{\delta A^{\bullet P}[B_c]}{\delta B^{\bullet K}}\\+\frac{\delta A^{\star Q}[B_c]}{\delta B^{\bullet I}}\frac{\delta^2 S_{\mathrm{YM}}[A[B_c]]}{\delta A^{\star Q}\delta A^{\star P}}\frac{\delta A^{\star P}[B_c]}{\delta B^{\bullet K}}+\frac{\delta A^{\bullet Q}[B_c]}{\delta B^{\bullet I}}\frac{\delta^2 S_{\mathrm{YM}}[A[B_c]]}{\delta A^{\bullet Q}\delta A^{\star P}}\frac{\delta A^{\star P}[B_c]}{\delta B^{\bullet K}}}\\ \\
\frac{\delta A^{\star Q}[B_c]}{\delta B^{\star I}}\frac{\delta^2 S_{\mathrm{YM}}[A[B_c]]}{\delta A^{\star Q}\delta A^{\star P}}\frac{\delta A^{\star P}[B_c]}{\delta B^{\star K}}
      & \substack{\frac{\delta A^{\star Q}[B_c]}{\delta B^{\star I}}\frac{\delta^2 S_{\mathrm{YM}}[A[B_c]]}{\delta A^{\star Q}\delta A^{\star P}}\frac{\delta A^{\star P}[B_c]}{\delta B^{\bullet K}}+\frac{\delta A^{\star Q}[B_c]}{\delta B^{\star I}}\frac{\delta^2 S_{\mathrm{YM}}[A[B_c]]}{\delta A^{\star Q}\delta A^{\bullet P}}\frac{\delta A^{\bullet P}[B_c]}{\delta B^{\bullet K}}}
\end{matrix}\right)\,. 
\end{equation}
The complicated looking matrix on the R.H.S. above can be decomposed into the product of three matrices as shown below
\begin{equation}
\mathrm{M}^{\mathrm{MHV}}_{IK}  = \left(\begin{matrix}
     \frac{\delta A^{\bullet Q}[B_c]}
    {\delta B^{\bullet I}} 
     & \frac{\delta A^{\star Q}[B_c]}
    {\delta B^{\bullet I}}\\ \\
 \mathbb{0}
     &\frac{\delta A^{\star Q}[B_c]}
    {\delta B^{\star I}} 
\end{matrix}\right) \left(\begin{matrix}
     \frac{\delta^2 S_{\mathrm{YM}}[A[B_c]]}{\delta A^{\bullet Q}\delta A^{\star P}} &\frac{\delta^2 S_{\mathrm{YM}}[A[B_c]]}{\delta A^{\bullet Q}\delta A^{\bullet P}}\\ \\
\frac{\delta^2 S_{\mathrm{YM}}[A[B_c]]}{\delta A^{\star Q}\delta A^{\star P}}
      &\frac{\delta^2 S_{\mathrm{YM}}[A[B_c]]}{\delta A^{\star Q}\delta A^{\bullet P}} 
\end{matrix}\right) \left(\begin{matrix}
     \frac{\delta A^{\star P}[B_c]}
    {\delta B^{\star K}}  
     & \frac{\delta A^{\star P}[B_c]}
    {\delta B^{\bullet K}}\\ \\
 \mathbb{0}
     &\frac{\delta A^{\bullet P}[B_c]}
    {\delta B^{\bullet K}}
\end{matrix}\right)\,.
\end{equation}
Thus the determinant reads
\begin{multline}
\det\mathrm{M}^{\mathrm{MHV}}_{IK}  = \det\begin{vmatrix}
     \frac{\delta A^{\bullet Q}[B_c]}
    {\delta B^{\bullet I}} 
     & \frac{\delta A^{\star Q}[B_c]}
    {\delta B^{\bullet I}}\\ \\
 \mathbb{0}
     &\frac{\delta A^{\star Q}[B_c]}
    {\delta B^{\star I}} 
\end{vmatrix} \times \, \det\begin{vmatrix}
     \frac{\delta^2 S_{\mathrm{YM}}[A[B_c]]}{\delta A^{\bullet Q}\delta A^{\star P}} &\frac{\delta^2 S_{\mathrm{YM}}[A[B_c]]}{\delta A^{\bullet Q}\delta A^{\bullet P}}\\ \\
\frac{\delta^2 S_{\mathrm{YM}}[A[B_c]]}{\delta A^{\star Q}\delta A^{\star P}}
      &\frac{\delta^2 S_{\mathrm{YM}}[A[B_c]]}{\delta A^{\star Q}\delta A^{\bullet P}} 
\end{vmatrix} \\
\times \, \det\begin{vmatrix}
     \frac{\delta A^{\star P}[B_c]}
    {\delta B^{\star K}}  
     & \frac{\delta A^{\star P}[B_c]}
    {\delta B^{\bullet K}}\\ \\
 \mathbb{0}
     &\frac{\delta A^{\bullet P}[B_c]}
    {\delta B^{\bullet K}}
\end{vmatrix}
\label{eq:M_4}
\end{multline}
The above result is interesting. Notice, the determinant we have on L.H.S. is of the matrix in the one-loop effective MHV action derived via the new approach in Chapter \ref{QZth-chapter}, whereas the determinant of the matrix in the middle of the R.H.S. above is exactly the same as the determinant we have in the one-loop effective MHV action Eq.~\eqref{eq:G_MHV} derived in Chapter \ref{QMHV-chapter}. And, the first and the third determinant resemble the Jacobian of Mansfield's transformation shown below (we make the position explicit just for the sake of simplicity)
\begin{equation}
    \mathcal{J}_{\mathrm{MT}}  = \det \begin{vmatrix}
     \frac{\delta {\hat A}^{\bullet} (x^+;\mathbf{x})}
    {\delta {\hat B}^{\bullet} (x^+;\mathbf{y})} 
     &\mathbb{0} \\ \\
\frac{\delta {\hat A}^{\star}(x^+;\mathbf{x})}
    {\delta {\hat B}^{\bullet}(x^+;\mathbf{y})} 
     &\frac{\delta {\hat A}^{\star}(x^+;\mathbf{x})}
    {\delta {\hat B}^{\star}(x^+;\mathbf{y})} 
\end{vmatrix} \,.
\label{eq:MT_jacag}
\end{equation}
Recall, however, that Mansfield's transformation as well as its solutions are both defined on the constant light-cone time $x^+$. Whereas, the functional derivatives in Eq.~\eqref{eq:M_4}  are over the full 4D space.  But these are related to the functional derivatives of the solution of Mansfield's transformation via an extra delta corresponding to the light-cone time because the kernels in the latter are independent of the light-cone time. Explicitly, we can write
\begin{equation}
  \int\!dx^+\, \delta(x^+-y^+)  \frac{\delta \hat{A}^{\star}[B_c^{\bullet}, B_c^{\star}](y)}{\delta \hat{B}^{\star}(x)} \equiv \frac{\delta \hat{A}^{\star}[B_c^{\bullet}, B_c^{\star}](y^+;\mathbf{y})}{\delta \hat{B}^{\star}(y^+;\mathbf{x})}
\end{equation}
and similarly for the other functional derivative terms in the determinant. As a result, the two determinants are related as follows
\begin{equation}
    \det \begin{vmatrix}
     \frac{\delta {\hat A}^{\bullet}[B_c](x)}
    {\delta {\hat B}^{\bullet}(y)} 
     & \frac{\delta {\hat A}^{\star}[B_c](x)}
    {\delta {\hat B}^{\bullet}(y)}\\ \\
 \mathbb{0}
     &\frac{\delta {\hat A}^{\star}[B_c](x)}
    {\delta {\hat B}^{\star}(y)} 
\end{vmatrix} \approx \mathcal{N} \times \det \begin{vmatrix}
     \frac{\delta {\hat A}^{\bullet}[B_c] (x^+;\mathbf{x})}
    {\delta {\hat B}^{\bullet} (x^+;\mathbf{y})} 
     & \frac{\delta {\hat A}^{\star}[B_c](x^+;\mathbf{x})}
    {\delta {\hat B}^{\bullet}(x^+;\mathbf{y})}\\ \\
 \mathbb{0}
     &\frac{\delta {\hat A}^{\star}[B_c](x^+;\mathbf{x})}
    {\delta {\hat B}^{\star}(x^+;\mathbf{y})} 
\end{vmatrix} \,.
\end{equation}
where $\approx$ implies the two determinants are equal up to a field independent volume divergent factor 
$\mathcal{N}$ originating from the determinant of a delta $\mathcal{N} \approx \det[\delta(y^+-x^+)]$.

Substituting the above in Eq.~\eqref{eq:M_4}, we get
\begin{equation}
\det\mathrm{M}^{\mathrm{MHV}}_{IK} = \mathcal{N} \Big[ \mathcal{J}_{\mathrm{MT}}\Big]^2 \times \, \det\begin{vmatrix}
     \frac{\delta^2 S_{\mathrm{YM}}[A[B_c]]}{\delta A^{\bullet Q}\delta A^{\star P}} &\frac{\delta^2 S_{\mathrm{YM}}[A[B_c]]}{\delta A^{\bullet Q}\delta A^{\bullet P}}\\ \\
\frac{\delta^2 S_{\mathrm{YM}}[A[B_c]]}{\delta A^{\star Q}\delta A^{\star P}}
      &\frac{\delta^2 S_{\mathrm{YM}}[A[B_c]]}{\delta A^{\star Q}\delta A^{\bullet P}} 
\end{vmatrix}\,.
\label{eq:M_5}
\end{equation}
where $\mathcal{N} = \Big[\det \delta(Q^+-I^+)\det \delta(P^+-K^+)\Big] $ and $I^+$ represents the plus component of the position 4-vector $y^+$ associated with the collective index $I=\left\{\left(y^{+},y^{-},y^{\bullet},y^{\star}\right); a\right\}$. Notice both the quantities $\mathcal{N}$ as well as the Jacobian $\mathcal{J}_{\mathrm{MT}}$ are field independent and therefore do not contribute to the computation of one-loop amplitudes. Furthermore, the former is a divergent factor that could be absorbed in the normalization of the partition function. As a result, we conclude that the determinant of the matrices appearing in the one-loop effective MHV action derived via two approaches in Chapter \ref{QZth-chapter} and Chapter \ref{QMHV-chapter} respectively are equal up to a redundant factor. Therefore, the one-loop effective MHV action derived in these two ways give exactly the same one-loop amplitudes.
\section{\texorpdfstring{$\,\,\,\,$}{sp}One-loop effective Z-field action}
\label{sec:A102}

In this section, we focus on the determinant of the matrix $\mathrm{M}^{\mathrm{Z}}_{IK}$ appearing in the one-loop effective Z-field action derived via the new approach in Chapter \ref{QZth-chapter}. The matrix reads
\begin{multline}
\mathrm{M}^{\mathrm{Z}}_{IK}  =  \left(\begin{matrix}
     \frac{\delta^2 S[Z_c]}{\delta Z^{\bullet I}\delta Z^{\star K}} &\frac{\delta^2 S[Z_c]}{\delta Z^{\bullet I}\delta Z^{\bullet K}}\\ \\
\frac{\delta^2 S[Z_c]}{\delta Z^{\star I}\delta Z^{\star K}}
      &\frac{\delta^2 S[Z_c]}{\delta Z^{\star I}\delta Z^{\bullet K}} 
\end{matrix}\right) \\
+ \left(\begin{matrix}
    J_{\star L}\frac{\delta^2 A^{\star L}[Z_c]}{\delta Z^{\bullet I}\delta Z^{\star K}} + J_{\bullet L}\frac{\delta^2 A^{\bullet L}[Z_c]}{\delta Z^{\bullet I}\delta Z^{\star K}} 
     & J_{\star L}\frac{\delta^2 A^{\star L}[Z_c]}{\delta Z^{\bullet I}\delta Z^{\bullet K}} +  J_{\bullet L}\frac{\delta^2 A^{\bullet L}[Z_c]}{\delta Z^{\bullet I}\delta Z^{\bullet K}}\\ \\
 J_{\star L}\frac{\delta^2 A^{\star L}[B_c]}{\delta Z^{\star I}\delta Z^{\star K}} + J_{\bullet L}\frac{\delta^2 A^{\bullet L}[B_c]}{\delta Z^{\star I}\delta Z^{\star K}}
      &J_{\star L}\frac{\delta^2 A^{\star L}[B_c]}{\delta Z^{\star I}\delta Z^{\bullet K}} + J_{\bullet L}\frac{\delta^2 A^{\bullet L}[B_c]}{\delta Z^{\star I}\delta Z^{\bullet K}}
    \label{eq:MZ_1}
\end{matrix}\right) \,.
\end{multline}
Recall the Z-field action can be obtained by substituting the solution $B^{\star}[Z_c^{\star}], B^{\bullet}[Z_c^{\bullet}, Z_c^{\star}]$ Eq.~\eqref{eq:BstarZ_exp}-\eqref{eq:BbulletZ_exp} to the MHV action. Thus, we can write
\begin{equation}
    S[Z_c^{\bullet}, Z_c^{\star}] = S_{\mathrm{MHV}}[ B^{\bullet}[Z_c^{\bullet}, Z_c^{\star}], B^{\star}[Z_c^{\star}]]\,.
    \label{eq:ZacYMag}
\end{equation}
Differentiating both sides with respect to Z-fields and using the chain rule on the R.H.S. we can develop the following relations
\begin{multline}
    \frac{\delta^2 S[Z_c]}{\delta Z^{\bullet I}\delta Z^{\star K}} = \frac{\delta S_{\mathrm{MHV}}[B[Z_c]]}{\delta B^{\bullet R}}\frac{\delta^2 B^{\bullet R}[Z_c]}{\delta Z^{\bullet I}\delta Z^{\star K}}\\
    +\frac{\delta B^{\bullet S}[Z_c]}{\delta Z^{\bullet I}}\frac{\delta^2 S_{\mathrm{MHV}}[B[Z_c]]}{\delta B^{\bullet S}\delta B^{\bullet R}}\frac{\delta B^{\bullet R}[Z_c]}{\delta Z^{\star K}}+\frac{\delta B^{\bullet S}[Z_c]}{\delta Z^{\bullet I}}\frac{\delta^2 S_{\mathrm{MHV}}[B[Z_c]]}{\delta B^{\bullet S}\delta B^{\star R}}\frac{\delta B^{\star R}[Z_c]}{\delta Z^{\star K}}\,.
    \label{eq:SZ_BS}
\end{multline}
\begin{multline}
     \frac{\delta^2 S[Z_c]}{\delta Z^{\bullet I}\delta Z^{\bullet K}} = 
    \frac{\delta B^{\bullet S}[Z_c]}{\delta Z^{\bullet I}}\frac{\delta^2 S_{\mathrm{MHV}}[B[Z_c]]}{\delta B^{\bullet S}\delta B^{\bullet R}}\frac{\delta B^{\bullet R}[Z_c]}{\delta Z^{\bullet K}}\,. \hspace{6.5cm}
    \label{eq:SZ_BB}
\end{multline}
\begin{multline}
    \frac{\delta^2 S[Z_c]}{\delta Z^{\star I}\delta Z^{\star K}} = \frac{\delta S_{\mathrm{MHV}}[B[Z_c]]}{\delta B^{\bullet R}}\frac{\delta^2 B^{\bullet R}[Z_c]}{\delta Z^{\star I}\delta Z^{\star K}}+\frac{\delta S_{\mathrm{MHV}}[B[Z_c]]}{\delta B^{\star R}}\frac{\delta^2 B^{\star R}[Z_c]}{\delta Z^{\star I}\delta Z^{\star K}}\\
    +\frac{\delta B^{\star S}[Z_c]}{\delta Z^{\star I}}\frac{\delta^2 S_{\mathrm{MHV}}[B[Z_c]]}{\delta B^{\star S}\delta B^{\bullet R}}\frac{\delta B^{\bullet R}[Z_c]}{\delta Z^{\star K}}  
    +\frac{\delta B^{\bullet S}[Z_c]}{\delta Z^{\star I}}\frac{\delta^2 S_{\mathrm{MHV}}[B[Z_c]]}{\delta B^{\bullet S}\delta B^{\bullet R}}\frac{\delta B^{\bullet R}[Z_c]}{\delta Z^{\star K}}\\    
    +\frac{\delta B^{\star S}[Z_c]}{\delta Z^{\star I}}\frac{\delta^2 S_{\mathrm{MHV}}[B[Z_c]]}{\delta B^{\star S}\delta B^{\star R}}\frac{\delta B^{\star R}[Z_c]}{\delta Z^{\star K}}    
    +\frac{\delta B^{\bullet S}[Z_c]}{\delta Z^{\star I}}\frac{\delta^2 S_{\mathrm{MHV}}[B[Z_c]]}{\delta B^{\bullet S}\delta B^{\star R}}\frac{\delta B^{\star R}[Z_c]}{\delta Z^{\star K}}\,.
    \label{eq:SZ_SS}
\end{multline}
\begin{multline}
    \frac{\delta^2 S[Z_c]}{\delta Z^{\star I}\delta Z^{\bullet K}} = \frac{\delta S_{\mathrm{MHV}}[B[Z_c]]}{\delta B^{\bullet R}}\frac{\delta^2 B^{\bullet R}[Z_c]}{\delta Z^{\star I}\delta Z^{\bullet K}}\\
    +\frac{\delta B^{\bullet S}[Z_c]}{\delta Z^{\star I}}\frac{\delta^2 S_{\mathrm{MHV}}[B[Z_c]]}{\delta B^{\bullet S}\delta B^{\bullet R}}\frac{\delta B^{\bullet R}[Z_c]}{\delta Z^{\bullet K}}+\frac{\delta B^{\star S}[Z_c]}{\delta Z^{\star I}}\frac{\delta^2 S_{\mathrm{MHV}}[B[Z_c]]}{\delta B^{\star S}\delta B^{\bullet R}}\frac{\delta B^{\bullet R}[Z_c]}{\delta Z^{\bullet K}}\,.
    \label{eq:SZ_SB}
\end{multline}    
Since the above expressions are a bit more complicated, we develop the following notation for the matrix $\mathrm{M}^{\mathrm{Z}}_{IK}$ 
\begin{equation}
    \mathrm{M}^{\mathrm{Z}}_{IK}  = \left(\begin{matrix}     \left(\mathrm{M}^{\mathrm{Z}}_{IK}\right)_{11} &\left(\mathrm{M}^{\mathrm{Z}}_{IK}\right)_{12}\\ \\
\left(\mathrm{M}^{\mathrm{Z}}_{IK}\right)_{21} &\left(\mathrm{M}^{\mathrm{Z}}_{IK}\right)_{22} 
\end{matrix}\right)\,,
\end{equation}
so that we can deal with the blocks separately. 

 In Eq.~\eqref{eq:MZ_1}, the block $\left(\mathrm{M}^{\mathrm{Z}}_{IK}\right)_{11} $ reads
\begin{align}
\left(\mathrm{M}^{\mathrm{Z}}_{IK}\right)_{11}  =&\, \frac{\delta^2 S[Z_c]}{\delta Z^{\bullet I}\delta Z^{\star K}} + J_{\star L}\frac{\delta^2 A^{\star L}[Z_c]}{\delta Z^{\bullet I}\delta Z^{\star K}} + J_{\bullet L}\frac{\delta^2 A^{\bullet L}[Z_c]}{\delta Z^{\bullet I}\delta Z^{\star K}}\,, \nonumber\\
    =&\,\frac{\delta^2 S[Z_c]}{\delta Z^{\bullet I}\delta Z^{\star K}} + J_{\star L}\frac{\delta A^{\star L}[B_c]}{\delta B^{\bullet R}}\frac{\delta^2 B^{\bullet R}[Z_c]}{\delta Z^{\bullet I}\delta Z^{\star K}}+ J_{\bullet L}\frac{\delta A^{\bullet L}[B_c]}{\delta B^{\bullet R}}\frac{\delta^2 B^{\bullet R}[Z_c]}{\delta Z^{\bullet I}\delta Z^{\star K}}\nonumber\\
    &\,+\frac{\delta B^{\bullet S}[Z_c]}{\delta Z^{\bullet I}}J_{\star L}\frac{\delta^2 A^{\star L}[B_c]}{\delta B^{\bullet S}\delta B^{\bullet R}}\frac{\delta B^{\bullet R}[Z_c]}{\delta Z^{\star K}}+\frac{\delta B^{\bullet S}[Z_c]}{\delta Z^{\bullet I}}J_{\star L}\frac{\delta^2 A^{\star L}[B_c]}{\delta B^{\bullet S}\delta B^{\star R}}\frac{\delta B^{\star R}[Z_c]}{\delta Z^{\star K}}\nonumber\\
    &\,+\frac{\delta B^{\bullet S}[Z_c]}{\delta Z^{\bullet I}}J_{\bullet L}\frac{\delta^2 A^{\bullet L}[B_c]}{\delta B^{\bullet S}\delta B^{\bullet R}}\frac{\delta B^{\bullet R}[Z_c]}{\delta Z^{\star K}}\,,
\end{align}
where in going from the first expression to the second we used $A^{i}[Z_c] = A^{i}[B_c [Z_c]]$ and then applied the chain rule to the functional derivatives. Finally, substituting Eq.~\eqref{eq:SZ_BS} we get
\begin{align}
\left(\mathrm{M}^{\mathrm{Z}}_{IK}\right)_{11}  
    =&\,\Bigg[\frac{\delta S_{\mathrm{MHV}}[B[Z_c]]}{\delta B^{\bullet R}} + J_{\star L}\frac{\delta A^{\star L}[B_c]}{\delta B^{\bullet R}}+ J_{\bullet L}\frac{\delta A^{\bullet L}[B_c]}{\delta B^{\bullet R}}\Bigg]\frac{\delta^2 B^{\bullet R}[Z_c]}{\delta Z^{\bullet I}\delta Z^{\star K}}\nonumber\\
    &\,+\frac{\delta B^{\bullet S}[Z_c]}{\delta Z^{\bullet I}}\Bigg[\frac{\delta^2 S_{\mathrm{MHV}}[B[Z_c]]}{\delta B^{\bullet S}\delta B^{\bullet R}}+ J_{\star L}\frac{\delta^2 A^{\star L}[B_c]}{\delta B^{\bullet S}\delta B^{\bullet R}}+J_{\bullet L}\frac{\delta^2 A^{\bullet L}[B_c]}{\delta B^{\bullet S}\delta B^{\bullet R}}\Bigg]\frac{\delta B^{\bullet R}[Z_c]}{\delta Z^{\star K}}\nonumber\\
    &\,+\frac{\delta B^{\bullet S}[Z_c]}{\delta Z^{\bullet I}}\Bigg[\frac{\delta^2 S_{\mathrm{MHV}}[B[Z_c]]}{\delta B^{\bullet S}\delta B^{\star R}}+ J_{\star L}\frac{\delta^2 A^{\star L}[B_c]}{\delta B^{\bullet S}\delta B^{\star R}}\Bigg]\frac{\delta B^{\star R}[Z_c]}{\delta Z^{\star K}}\,.
    \label{eq:m11}
\end{align}
The first term above vanishes owing to the MHV classical EOM Eq.~\eqref{eq:B_bul_EOM1}. In order to compactly represent the second and third terms, let us introduce the following notation for the matrix $\mathrm{M}^{\mathrm{MHV}}_{IK}$ Eq.~\eqref{eq:M_1}
\begin{equation}
    \mathrm{M}^{\mathrm{MHV}}_{IK}  = \left(\begin{matrix}     \left(\mathrm{M}^{\mathrm{MHV}}_{IK}\right)_{11} &\left(\mathrm{M}^{\mathrm{MHV}}_{IK}\right)_{12}\\ \\
\left(\mathrm{M}^{\mathrm{MHV}}_{IK}\right)_{21} &\left(\mathrm{M}^{\mathrm{MHV}}_{IK}\right)_{22} \end{matrix}\right)\,.
\label{eq:mat_bl}
\end{equation}
Comparing the second and the third term in Eq.~\eqref{eq:m11} with the blocks in the matrix $\mathrm{M}^{\mathrm{MHV}}_{IK}$ Eq.~\eqref{eq:M_1}, we get
\begin{multline}
 \left(\mathrm{M}^{\mathrm{Z}}_{IK}\right)_{11}  
    =\frac{\delta B^{\bullet S}[Z_c]}{\delta Z^{\bullet I}}\Big[ \left.\left(\mathrm{M}^{\mathrm{MHV}}_{SR}\right)_{12}\right|_{\hat{B}_c^i = \hat{B}_c^i[Z_c]}\Big]\frac{\delta B^{\bullet R}[Z_c]}{\delta Z^{\star K}} \\
    + \frac{\delta B^{\bullet S}[Z_c]}{\delta Z^{\bullet I}}\Big[\left.\left(\mathrm{M}^{\mathrm{MHV}}_{SR}\right)_{11}\right|_{\hat{B}_c^i = \hat{B}_c^i[Z_c]}\Big]\frac{\delta B^{\star R}[Z_c]}{\delta Z^{\star K}}\,,
\end{multline}
where $\left.\left(\mathrm{M}^{\mathrm{MHV}}_{SR}\right)_{jk}\right|_{\hat{B}_c^i = \hat{B}_c^i[Z_c]}$ represents the block $\left(\mathrm{M}^{\mathrm{MHV}}_{IK}\right)_{jk}$ in the matrix $\mathrm{M}^{\mathrm{MHV}}_{IK}$ Eq.~\eqref{eq:mat_bl} or equivalently \eqref{eq:M_1} appearing in the one-loop effective MHV action such that the ${B}_c^i $ fields in the matrix entries have been replaced with the solution ${B}_c^i[Z_c] = B^{\star}[Z_c^{\star}], B^{\bullet}[Z_c^{\bullet}, Z_c^{\star}]$ Eq.~\eqref{eq:BstarZ_exp}-\eqref{eq:BbulletZ_exp}. 

Similarly, for the remaining blocks, we have
\begin{multline}
 \left(\mathrm{M}^{\mathrm{Z}}_{IK}\right)_{22}  
    =\frac{\delta B^{\bullet S}[Z_c]}{\delta Z^{\star I}}\Big[ \left.\left(\mathrm{M}^{\mathrm{MHV}}_{SR}\right)_{12}\right|_{\hat{B}_c^i = \hat{B}_c^i[Z_c]}\Big]\frac{\delta B^{\bullet R}[Z_c]}{\delta Z^{\bullet K}}\\
    + \frac{\delta B^{\star S}[Z_c]}{\delta Z^{\star I}}\Big[\left.\left(\mathrm{M}^{\mathrm{MHV}}_{SR}\right)_{22}\right|_{\hat{B}_c^i = \hat{B}_c^i[Z_c]}\Big]\frac{\delta B^{\bullet R}[Z_c]}{\delta Z^{\bullet K}}\,.
\end{multline}
\begin{align}
  \left(\mathrm{M}^{\mathrm{Z}}_{IK}\right)_{12}  
    =&\,\frac{\delta^2 S[Z_c]}{\delta Z^{\bullet I}\delta Z^{\bullet K}}+J_{\star L}\frac{\delta^2 A^{\star L}[Z_c]}{\delta Z^{\bullet I}\delta Z^{\bullet K}} +  J_{\bullet L}\frac{\delta^2 A^{\bullet L}[Z_c]}{\delta Z^{\bullet I}\delta Z^{\bullet K}}\,, \nonumber\\
    =&\,\frac{\delta B^{\bullet S}[Z_c]}{\delta Z^{\bullet I}}\frac{\delta^2 S_{\mathrm{MHV}}[B[Z_c]]}{\delta B^{\bullet S}\delta B^{\bullet R}}\frac{\delta B^{\bullet R}[Z_c]}{\delta Z^{\bullet K}}+ \frac{\delta B^{\bullet S}[Z_c]}{\delta Z^{\bullet I}}J_{\star L}\frac{\delta^2 A^{\star L}[B_c]}{\delta B^{\bullet S}\delta B^{\bullet R}}\frac{\delta B^{\bullet R}[Z_c]}{\delta Z^{\bullet K}}\nonumber\\
    &\,\,\,+\frac{\delta B^{\bullet S}[Z_c]}{\delta Z^{\bullet I}}J_{\bullet L}\frac{\delta^2 A^{\bullet L}[B_c]}{\delta B^{\bullet S}\delta B^{\bullet R}}\frac{\delta B^{\bullet R}[Z_c]}{\delta Z^{\bullet K}}\,,\nonumber\\
    =&\,\frac{\delta B^{\bullet S}[Z_c]}{\delta Z^{\bullet I}}\Bigg[\frac{\delta^2 S_{\mathrm{MHV}}[B[Z_c]]}{\delta B^{\bullet S}\delta B^{\bullet R}}+J_{\star L}\frac{\delta^2 A^{\star L}[B_c]}{\delta B^{\bullet S}\delta B^{\bullet R}}+J_{\bullet L}\frac{\delta^2 A^{\bullet L}[B_c]}{\delta B^{\bullet S}\delta B^{\bullet R}}\Bigg]\frac{\delta B^{\bullet R}[Z_c]}{\delta Z^{\bullet K}}\,,\nonumber\\
    =&\,\frac{\delta B^{\bullet S}[Z_c]}{\delta Z^{\bullet I}}\Big[ \left.\left(\mathrm{M}^{\mathrm{MHV}}_{SR}\right)_{12}\right|_{\hat{B}_c^i = \hat{B}_c^i[Z_c]}\Big]\frac{\delta B^{\bullet R}[Z_c]}{\delta Z^{\bullet K}}\,.
\end{align}
\begin{align}
  \left(\mathrm{M}^{\mathrm{Z}}_{IK}\right)_{21}  
    =&\,\frac{\delta^2 S[Z_c]}{\delta Z^{\star I}\delta Z^{\star K}}+ J_{\star L}\frac{\delta^2 A^{\star L}[B_c]}{\delta Z^{\star I}\delta Z^{\star K}} + J_{\bullet L}\frac{\delta^2 A^{\bullet L}[B_c]}{\delta Z^{\star I}\delta Z^{\star K}}\,,\nonumber\\
    =&\,\frac{\delta^2 S[Z_c]}{\delta Z^{\star I}\delta Z^{\star K}} +J_{\star L}\frac{\delta A^{\star L}[B_c]}{\delta B^{\star R}}\frac{\delta^2 B^{\star R}[Z_c]}{\delta Z^{\star I}\delta Z^{\star K}}+\left(J_{\star L}\frac{\delta A^{\star L}[B_c]}{\delta B^{\bullet R}}+ J_{\bullet L}\frac{\delta A^{\bullet L}[B_c]}{\delta B^{\bullet R}}\right)\frac{\delta^2 B^{\bullet R}[Z_c]}{\delta Z^{\star I}\delta Z^{\star K}}\nonumber\\
    &\,+\frac{\delta B^{\bullet S}[Z_c]}{\delta Z^{\star I}}J_{\star L}\frac{\delta^2 A^{\star L}[B_c]}{\delta B^{\bullet S}\delta B^{\star R}}\frac{\delta B^{\star R}[Z_c]}{\delta Z^{\star K}}+\frac{\delta B^{\star S}[Z_c]}{\delta Z^{\star I}}J_{\star L}\frac{\delta^2 A^{\star L}[B_c]}{\delta B^{\star S}\delta B^{\bullet R}}\frac{\delta B^{\bullet R}[Z_c]}{\delta Z^{\star K}}\nonumber\\
    &\,+\frac{\delta B^{\bullet S}[Z_c]}{\delta Z^{\star I}}J_{\star L}\frac{\delta^2 A^{\star L}[B_c]}{\delta B^{\bullet S}\delta B^{\bullet R}}\frac{\delta B^{\bullet R}[Z_c]}{\delta Z^{\star K}}+\frac{\delta B^{\bullet S}[Z_c]}{\delta Z^{\star I}}J_{\bullet L}\frac{\delta^2 A^{\bullet L}[B_c]}{\delta B^{\bullet S}\delta B^{\bullet R}}\frac{\delta B^{\bullet R}[Z_c]}{\delta Z^{\star K}}\,.
\end{align}
Substituting Eq.~\eqref{eq:SZ_SS} above we get
\begin{align}
  \left(\mathrm{M}^{\mathrm{Z}}_{IK}\right)_{21}  
    =&\,\Bigg[\frac{\delta S_{\mathrm{MHV}}[B[Z_c]]}{\delta B^{\star R}} + J_{\star L}\frac{\delta A^{\star L}[B_c]}{\delta B^{\star R}}\Bigg]\frac{\delta^2 B^{\star R}[Z_c]}{\delta Z^{\star I}\delta Z^{\star K}}\nonumber\\
     &\,+\Bigg[\frac{\delta S_{\mathrm{MHV}}[B[Z_c]]}{\delta B^{\bullet R}} + J_{\star L}\frac{\delta A^{\star L}[B_c]}{\delta B^{\bullet R}}+ J_{\bullet L}\frac{\delta A^{\bullet L}[B_c]}{\delta B^{\bullet R}}\Bigg]\frac{\delta^2 B^{\bullet R}[Z_c]}{\delta Z^{\star I}\delta Z^{\star K}}\nonumber\\
    &\,+\frac{\delta B^{\bullet S}[Z_c]}{\delta Z^{\star I}}\Bigg[\frac{\delta^2 S_{\mathrm{MHV}}[B[Z_c]]}{\delta B^{\bullet S}\delta B^{\bullet R}}+ J_{\star L}\frac{\delta^2 A^{\star L}[B_c]}{\delta B^{\bullet S}\delta B^{\bullet R}}+J_{\bullet L}\frac{\delta^2 A^{\bullet L}[B_c]}{\delta B^{\bullet S}\delta B^{\bullet R}}\Bigg]\frac{\delta B^{\bullet R}[Z_c]}{\delta Z^{\star K}}\nonumber\\
    &\,+\frac{\delta B^{\bullet S}[Z_c]}{\delta Z^{\star I}}\Bigg[\frac{\delta^2 S_{\mathrm{MHV}}[B[Z_c]]}{\delta B^{\bullet S}\delta B^{\star R}}+ J_{\star L}\frac{\delta^2 A^{\star L}[B_c]}{\delta B^{\bullet S}\delta B^{\star R}}\Bigg]\frac{\delta B^{\star R}[Z_c]}{\delta Z^{\star K}}\nonumber\\
    &\,+\frac{\delta B^{\star S}[Z_c]}{\delta Z^{\star I}}\Bigg[\frac{\delta^2 S_{\mathrm{MHV}}[B[Z_c]]}{\delta B^{\star S}\delta B^{\bullet R}}+ J_{\star L}\frac{\delta^2 A^{\star L}[B_c]}{\delta B^{\star S}\delta B^{\bullet R}}\Bigg]\frac{\delta B^{\bullet R}[Z_c]}{\delta Z^{\star K}}\nonumber\\
    &\,+\frac{\delta B^{\star S}[Z_c]}{\delta Z^{\star I}}\Bigg[\frac{\delta^2 S_{\mathrm{MHV}}[B[Z_c]]}{\delta B^{\star S}\delta B^{\star R}}\Bigg]\frac{\delta B^{\star R}[Z_c]}{\delta Z^{\star K}}\,.
\end{align}
Owing to the MHV classical EOM Eq.~\eqref{eq:B_bul_EOM1}-\eqref{eq:B_star_EOM1}, the first two terms in the expression above vanish. The remaining terms can be rewritten as
\begin{multline}
  \left(\mathrm{M}^{\mathrm{Z}}_{IK}\right)_{21}  
    =\frac{\delta B^{\bullet S}[Z_c]}{\delta Z^{\star I}}\Big[\left. \left(\mathrm{M}^{\mathrm{MHV}}_{SR}\right)_{12}\right|_{\hat{B}_c^i = \hat{B}_c^i[Z_c]}\Big]\frac{\delta B^{\bullet R}[Z_c]}{\delta Z^{\star K}}\\
    + \frac{\delta B^{\bullet S}[Z_c]}{\delta Z^{\star I}}\Big[\left.\left(\mathrm{M}^{\mathrm{MHV}}_{SR}\right)_{11}\right|_{\hat{B}_c^i = \hat{B}_c^i[Z_c]}\Big]\frac{\delta B^{\star R}[Z_c]}{\delta Z^{\star K}}\\
    +  \frac{\delta B^{\star S}[Z_c]}{\delta Z^{\star I}}\Big[\left.\left(\mathrm{M}^{\mathrm{MHV}}_{SR}\right)_{22}\right|_{\hat{B}_c^i = \hat{B}_c^i[Z_c]}\Big]\frac{\delta B^{\bullet R}[Z_c]}{\delta Z^{\star K}}\\
    +\frac{\delta B^{\star S}[Z_c]}{\delta Z^{\star I}}\Big[ \left.\left(\mathrm{M}^{\mathrm{MHV}}_{SR}\right)_{21}\right|_{\hat{B}_c^i = \hat{B}_c^i[Z_c]}\Big]\frac{\delta B^{\star R}[Z_c]}{\delta Z^{\star K}}\,.
\end{multline}
Putting back all the blocks together, we get
\begin{equation}
\mathrm{M}^{\mathrm{Z}}_{IK}  =   \left(\begin{matrix}
     \substack{\frac{\delta B^{\bullet S}[Z_c]}{\delta Z^{\bullet I}}\Big[ \left(\mathrm{M}^{\mathrm{MHV}}_{SR}\right)_{12}\Big]\frac{\delta B^{\bullet R}[Z_c]}{\delta Z^{\star K}}\\
     + \frac{\delta B^{\bullet S}[Z_c]}{\delta Z^{\bullet I}}\Big[\left(\mathrm{M}^{\mathrm{MHV}}_{SR}\right)_{11}\Big]\frac{\delta B^{\star R}[Z_c]}{\delta Z^{\star K}}} &\quad 
     \substack{\frac{\delta B^{\bullet S}[Z_c]}{\delta Z^{\bullet I}}\Big[ \left(\mathrm{M}^{\mathrm{MHV}}_{SR}\right)_{12}\Big]\frac{\delta B^{\bullet R}[Z_c]}{\delta Z^{\bullet K}}}\\ \\
\substack{\frac{\delta B^{\bullet S}[Z_c]}{\delta Z^{\star I}}\Big[ \left(\mathrm{M}^{\mathrm{MHV}}_{SR}\right)_{12}\Big]\frac{\delta B^{\bullet R}[Z_c]}{\delta Z^{\star K}}\\
+ \frac{\delta B^{\bullet S}[Z_c]}{\delta Z^{\star I}}\Big[\left(\mathrm{M}^{\mathrm{MHV}}_{SR}\right)_{11}\Big]\frac{\delta B^{\star R}[Z_c]}{\delta Z^{\star K}}\\
    +  \frac{\delta B^{\star S}[Z_c]}{\delta Z^{\star I}}\Big[\left(\mathrm{M}^{\mathrm{MHV}}_{SR}\right)_{22}\Big]\frac{\delta B^{\bullet R}[Z_c]}{\delta Z^{\star K}}\\
    +\frac{\delta B^{\star S}[Z_c]}{\delta Z^{\star I}}\Big[ \left(\mathrm{M}^{\mathrm{MHV}}_{SR}\right)_{21}\Big]\frac{\delta B^{\star R}[Z_c]}{\delta Z^{\star K}}}
      & \quad \substack{\frac{\delta B^{\bullet S}[Z_c]}{\delta Z^{\star I}}\Big[ \left(\mathrm{M}^{\mathrm{MHV}}_{SR}\right)_{12}\Big]\frac{\delta B^{\bullet R}[Z_c]}{\delta Z^{\bullet K}}\\
      + \frac{\delta B^{\star S}[Z_c]}{\delta Z^{\star I}}\Big[\left(\mathrm{M}^{\mathrm{MHV}}_{SR}\right)_{22}\Big]\frac{\delta B^{\bullet R}[Z_c]}{\delta Z^{\bullet K}}}
\end{matrix} \right)_{\hat{B}_c^i = \hat{B}_c^i[Z_c]}
\end{equation}
The matrix on R.H.S. above can be decomposed into the product of three matrices as shown below
\begin{equation}
\mathrm{M}^{\mathrm{Z}}_{IK}  = \left(\begin{matrix}
     \frac{\delta B^{\bullet S}[Z_c]}
    {\delta Z^{\bullet I}} 
     & \mathbb{0} \\ \\
 \frac{\delta B^{\bullet S}[Z_c]}
    {\delta Z^{\star I}}
     &\frac{\delta B^{\star S}[Z_c]}
    {\delta Z^{\star I}} 
\end{matrix}\right) \left(\begin{matrix}
     \left(\mathrm{M}^{\mathrm{MHV}}_{SR}\right)_{11} &\left(\mathrm{M}^{\mathrm{MHV}}_{SR}\right)_{12}\\ \\
\left(\mathrm{M}^{\mathrm{MHV}}_{SR}\right)_{21}
      &\left(\mathrm{M}^{\mathrm{MHV}}_{SR}\right)_{22} 
\end{matrix}\right)_{\hat{B}_c^i = \hat{B}_c^i[Z_c]} \left(\begin{matrix}
     \frac{\delta B^{\star R}[Z_c]}
    {\delta Z^{\star K}} 
     & \mathbb{0} \\ \\
 \frac{\delta B^{\bullet R}[Z_c]}
    {\delta Z^{\star K}}
     &\frac{\delta B^{\bullet R}[Z_c]}
    {\delta Z^{\bullet K}} 
\end{matrix}\right)\,.
\end{equation}
Thus the determinant of the matrix reads
\begin{multline}
\det\mathrm{M}^{\mathrm{Z}}_{IK}  = \det\begin{vmatrix}
     \frac{\delta B^{\bullet S}[Z_c]}
    {\delta Z^{\bullet I}} 
     & \mathbb{0} \\ \\
 \frac{\delta B^{\bullet S}[Z_c]}
    {\delta Z^{\star I}}
     &\frac{\delta B^{\star S}[Z_c]}
    {\delta Z^{\star I}} 
\end{vmatrix} \times \, \det \begin{vmatrix}
     \left(\mathrm{M}^{\mathrm{MHV}}_{SR}\right)_{11} &\left(\mathrm{M}^{\mathrm{MHV}}_{SR}\right)_{12}\\ \\
\left(\mathrm{M}^{\mathrm{MHV}}_{SR}\right)_{21}
      &\left(\mathrm{M}^{\mathrm{MHV}}_{SR}\right)_{22} 
\end{vmatrix}_{\hat{B}_c^i = \hat{B}_c^i[Z_c]} \\ \times \, \det\begin{vmatrix}
     \frac{\delta B^{\star R}[Z_c]}
    {\delta Z^{\star K}} 
     & \mathbb{0} \\ \\
 \frac{\delta B^{\bullet R}[Z_c]}
    {\delta Z^{\star K}}
     &\frac{\delta B^{\bullet R}[Z_c]}
    {\delta Z^{\bullet K}} 
\end{vmatrix}
\end{multline}
Notice, the matrix in the middle of R.H.S. above is the matrix $\mathrm{M}^{\mathrm{MHV}}_{IK}$ Eq.~\eqref{eq:M_1} appearing in the one-loop effective MHV action (derived in Chapter \ref{QZth-chapter}, using the new approach where we start with the Yang-Mills partition function, perform Mansfield's transformation and then derive the one-loop effective MHV action) where the ${B}_c^i$ fields in the matrix entries have been replaced with the solution ${B}_c^i[Z_c] = B^{\star}[Z_c^{\star}], B^{\bullet}[Z_c^{\bullet}, Z_c^{\star}]$ Eq.~\eqref{eq:BstarZ_exp}-\eqref{eq:BbulletZ_exp}. 
And, the determinant of the first and the third matrices are related to the Jacobian Eq.~\eqref{eq:BZ_jac} for the canonical transformation mapping the Anti-Self-Dual sector of the MHV action to the kinetic term in the Z-field action via determinant of delta for the light-cone time. Explicitly, 
\begin{equation}
\det\mathrm{M}^{\mathrm{Z}}_{IK}  = \Big[\det \delta(S^+-I^+)\det \delta(R^+-K^+)\Big] \times\, \det\left.\Big(\mathrm{M}^{\mathrm{MHV}}_{SR}\right|_{\hat{B}_c^i = \hat{B}_c^i[Z_c]} \Big)\,.
\label{eq:M_z}
\end{equation}
where  $I^+$ represents the plus component of the position 4-vector $y^+$ associated with the collective index $I=\left\{\left(y^{+},y^{-},y^{\bullet},y^{\star}\right); a\right\}$. Substituting Eq.~\eqref{eq:M_5}, we get (avoiding the redundant factors)
\begin{equation}
\det\mathrm{M}^{\mathrm{Z}}_{IK}  \approx \det\begin{vmatrix}
     \frac{\delta^2 S_{\mathrm{YM}}[A[B[Z_c]]]}{\delta A^{\bullet Q}\delta A^{\star P}} &\frac{\delta^2 S_{\mathrm{YM}}[A[B[Z_c]]]}{\delta A^{\bullet Q}\delta A^{\bullet P}}\\ \\
\frac{\delta^2 S_{\mathrm{YM}}[A[B[Z_c]]]}{\delta A^{\star Q}\delta A^{\star P}}
      &\frac{\delta^2 S_{\mathrm{YM}}[A[B[Z_c]]]}{\delta A^{\star Q}\delta A^{\bullet P}} 
\end{vmatrix} \,,
\label{eq:M_z22}
\end{equation}
where $\approx$ implies up to a field-independent volume divergent factor. Above, on L.H.S. we have the determinant appearing in the one-loop effective Z-field action derived in Chapter \ref{QZth-chapter} via the new approach whereas on R.H.S. we have the determinant appearing in the one-loop effective Z-field action derived in the old approach (developed in Chapter \ref{QMHV-chapter}). Their equality implies that the two ways of deriving one-loop effective Z-field action give the same action. 

\end{document}